\documentclass[phd,12pt]{psuthesis}
\usepackage[T1]{fontenc}
\usepackage{lmodern}
\usepackage{textcomp}
\usepackage{microtype}

\usepackage{amsmath}
\usepackage{amssymb}
\usepackage{subfigure}
\usepackage{eqlist} % Makes for a nice list of symbols.
\usepackage[nosepfour,warning,np,debug,autolanguage]{numprint}
\usepackage{acro}
\usepackage{booktabs}
\usepackage[final]{graphicx}
\usepackage[dvipsnames]{color}
\DeclareGraphicsExtensions{.pdf, .jpg}

\usepackage{url}
\usepackage{cite}

\usepackage{titlesec}

\usepackage{hyperref}
\usepackage{enumitem}
\usepackage{epigraph}
\usepackage{fixmath}
\usepackage{booktabs}
\usepackage{amsmath}
\usepackage[mathscr]{euscript}
\usepackage{pst-solides3d}
\usepackage{siunitx}
\protected\def\numpi{\text{\ensuremath{\pi}}}
\usepackage{physics}
\usepackage{mathtools}
\usepackage{chemfig,chemmacros}
\usepackage{amsbsy}
\usepackage{enumitem}
\usepackage{mhchem}
\chemsetup{
  modules = {polymers} ,
  polymers/delimiters = ()
}
\usepackage{caption}
\usepackage[utf8]{inputenc}
\usepackage{cleveref}
\crefname{section}{§}{§§}
\Crefname{section}{§}{§§}
\usepackage[OT2,T1]{fontenc}
\DeclareSymbolFont{cyrletters}{OT2}{wncyr}{m}{n}
\DeclareMathSymbol{\Sha}{\mathalpha}{cyrletters}{"58}
\DeclareMathOperator{\sinc}{sinc}
\usepackage[normalem]{ulem}
\useunder{\uline}{\ul}{}
\DeclareMathAlphabet\mathbfcal{OMS}{cmsy}{b}{n}
\ProvideTextCommand{\DJ}{OT1}{\raisebox{0.25ex}{-}\kern-0.4em D}

\input{SupplementaryMaterial/UserDefinedCommands}

\title{Applied Nanofabrication for X-ray Grating Spectroscopy}

\author{Jake A.\ McCoy}
\dept{Astronomy \& Astrophysics}
\degreedate{May 2021}
\copyrightyear{2021}

\documenttype{Dissertation}
\submittedto{The Graduate School}

\collegesubmittedto{}

\numberofreaders{4}

\advisor[] %Associate Head for the Graduate Program
		{Randall McEntaffer \\ Dissertation Advisor, Chair of Committee, Head of the Graduate Program}
		{Professor of Astronomy \& Astrophysics, Physics, Materials Science \& Engineering}

\readerone[]
			{Abraham Falcone}
			{Research Professor of Astronomy \& Astrophysics} %Research Professor

\readertwo[]
			{Fabien Gris\'{e}}
			{Assistant Research Professor of Astronomy \& Astrophysics} %Assistant Research Professor

\readerthree[]
			{Suvrath Mahadevan}
			{Professor of Astronomy \& Astrophysics}

\readerfour[]
			{Susan Trolier-McKinstry}
			{Evan Pugh University Professor and Steward S. Flaschen Professor of Ceramic Science \& Engineering, Electrical Engineering} 

\definecolor{gray75}{gray}{0.75}
\newcommand{\hsp}{\hspace{15pt}}
\titleformat{\chapter}[display]{\fontsize{30}{30}\selectfont\bfseries\sffamily}{Chapter \thechapter\hsp\textcolor{gray75}{\raisebox{3pt}{|}}}{0pt}{}{}

\titleformat{\section}[block]{\Large\bfseries\sffamily}{\thesection}{12pt}{}{}
\titleformat{\subsection}[block]{\large\bfseries\sffamily}{\thesubsection}{12pt}{}{}

\usepackage{listings}
\begin{document}
\pagestyle{fancy}
\fancyhead[L,C,R]{}
\fancyfoot[L,R]{}
\fancyfoot[C]{\thepage}
\renewcommand{\headrulewidth}{0pt}
\renewcommand{\footrulewidth}{0pt}
%%%%%%%%%%%%%%%%%%%%%%%%
% Preliminary Material %
%%%%%%%%%%%%%%%%%%%%%%%%
% This command is needed to properly set up the frontmatter.
\frontmatter

%%%%%%%%%%%%%%%%%%%%%%%%%%%%%%%%%%%%%%%%%%%%%%%%%%%%%%%%%%%%%%
% IMPORTANT
%
% The following commands allow you to include all the
% frontmatter in your thesis. If you don't need one or more of
% these items, you can comment it out. Most of these items are
% actually required by the Grad School -- see the Thesis Guide
% for details regarding what is and what is not required for
% your particular degree.
%%%%%%%%%%%%%%%%%%%%%%%%%%%%%%%%%%%%%%%%%%%%%%%%%%%%%%%%%%%%%%
% !!! DO NOT CHANGE THE SEQUENCE OF THESE ITEMS !!!
%%%%%%%%%%%%%%%%%%%%%%%%%%%%%%%%%%%%%%%%%%%%%%%%%%%%%%%%%%%%%%

% Generates the title page based on info you have provided
% above.
\psutitlepage

% Generates the committee page -- this is bound with your
% thesis. If this is an baccalaureate honors thesis, then
% comment out this line.
\psucommitteepage

% Generates the abstract. The argument should point to the
% file containing your abstract. 
%
%\pagestyle{plain}
\chapter*{Abstract}
    \begin{singlespace}
        % Place abstract below.

\vspace{-0.3in}

Measuring the diffuse, highly-ionized baryonic content in galactic halos and the intergalactic medium through soft x-ray absorption spectroscopy of active galactic nuclei is a main scientific objective of the Lynx X-ray Observatory mission concept that can only be accomplished with a next-generation grating spectrometer. 
Realizing such an instrument using reflection grating technology requires thousands of custom blazed gratings that each perform with high diffraction efficiency to be manufactured and aligned to intercept radiation coming to a focus in a Wolter-I telescope. 
With Lynx performance requirements demanding significant improvements over Chandra and XMM-Newton in terms of both spectral sensitivity and spectral resolving power, previous approaches to blazed grating fabrication face limitations in their ability to meet these goals simultaneously. 

The aim of this thesis is to implement two recently-developed techniques in nanofabrication for this task, with an emphasis on beamline diffraction-efficiency testing for characterizing spectral sensitivity. 
In particular, thermally-activated selective topography equilibration (TASTE) is pursued as a means for fabricating a master grating with the key advantage that it enables blazed groove facets to be patterned in polymeric electron-beam resist over a non-parallel groove layout not limited by substrate crystal structure. 
Additionally, substrate-conformal imprint lithography (SCIL) is studied as a method for mass manufacturing high-fidelity grating replicas in a silica sol-gel resist while avoiding many of the detriments associated with large-area patterning in other nanoimprint techniques. 

Diffraction-efficiency testing of sub-micron grating prototypes coated with gold shows that TASTE is capable of meeting Lynx requirements for spectral sensitivity, with room for improvement at small groove periods, and that while SCIL offers a promising avenue for Lynx grating production, imprints suffer a small blaze-angle reduction due to resist shrinkage. 
With process development for TASTE established and equipment for SCIL recently installed at the Penn State Nanofabrication Laboratory, these findings will contribute to upcoming sounding-rocket experiments for x-ray spectroscopy. 
Accompanying this dissertation are appendices that outline physics fundamentals for x-ray spectral lines, x-ray optics, and diffraction gratings. 

    \end{singlespace}
\newpage

% Generates the Table of Contents
\thesistableofcontents

% Generates the List of Figures
\begin{singlespace}
\renewcommand{\listfigurename}{\sffamily\Huge List of Figures}
\setlength{\cftparskip}{\baselineskip}
\addcontentsline{toc}{chapter}{List of Figures}
%\fancypagestyle{plain}{%
%\fancyhf{} % clear all header and footer fields
%\fancyfoot[C]{\thepage}} % except the center
\listoffigures
\end{singlespace}
\clearpage

% Generates the List of Tables
\begin{singlespace}
\renewcommand{\listtablename}{\sffamily\Huge List of Tables}
\setlength{\cftparskip}{\baselineskip}
\addcontentsline{toc}{chapter}{List of Tables}
\listoftables
\end{singlespace}
\clearpage

% Generates the List of Symbols. The argument should point to
% the file containing your List of Symbols. 
%\thesislistofsymbols{SupplementaryMaterial/ListOfSymbols}

% Generates the Acknowledgments. The argument should point to
% the file containing your Acknowledgments. 
%
\chapter{Acknowledgments}
%\addcontentsline{toc}{chapter}{Acknowledgments}
\begin{singlespace}
    % Place acknowledgments below.

This research was supported by the National Aeronautics and Space Administration (NASA) Space Technology Research Fellowship (NSTRF) under grant No.\ NNX16AP92H and used resources of the Advanced Light Source (ALS), which is a U.S. Department of Energy (DOE) Office of Science User Facility under contract No.\ DE-AC02-05CH11231, the Nanofabrication Laboratory and the Materials Characterization Laboratory at the Penn State Materials Research Institute, the Quattrone Nanofabrication Facility at the Singh Center for Nanotechnology, University of Pennsylvania, and Philips SCIL Nanoimprint Solutions in Eindhoven, Netherlands. 
Any opinions, findings, and conclusions or recommendations expressed in this document do not necessarily reflect the view of NASA, the U.S. DOE, or any other agency. 

Firstly, I would like to express my gratitude to my advisor, Prof.\ Randy McEntaffer, whom I met during my time as a graduate student at the University of Iowa before my transfer to Penn State. 
Thank you, Randy, for always fostering a positive work environment, for giving me a chance as a researcher in various areas and for supporting me patiently through my time-consuming thesis-writing efforts. 
My sincere thanks also goes to the Iowa Department of Physics \& Astronomy for my education in physics and initial support as a graduate student; in particular, I would like to thank Hannah Marlowe Ayta\c{c}, Ryan Allured and Prof.\ Phil Kaaret for piquing my curiosity for x-ray instrumentation and hardware-related research. 
Additionally, I thank my fellow labmates in the McEntaffer Group for all their help and comradery throughout the years and for the fun times outside of work. 

Special thanks are owed to Chad Eichfeld, Michael Labella, Kathy Gehoski, Bangzhi Liu, Andy Fitzgerald, Tim Tighe and other staff at the Penn State Materials Research Institute for their help with electron-beam lithography, microscopy, wet chemistry and more. 
Moreover, I wish to extend my thanks to Marc Verschuuren at Philips SCIL Nanoimprint Solutions and Gerald Lopez at the Singh Center for Nanotechnology for their inspiration and help with my studies on nanoimprint lithography. 
I would also like to acknowledge the support of Eric Gullikson at the ALS as well as my NSTRF advisors: the late Misha Gubarev and Jessica Gaskin at NASA's Marshall Space Flight Center.  

Saving the most important for last, I graciously thank my parents, Jeffrey and Tina McCoy, for their love and support as well as their instillment for the importance of education and self-investment. I am also grateful for the rest of my extended family and my friends for their encouragement throughout my life.

\end{singlespace}

% Generates the Epigraph/Dedication. The first argument should
% point to the file containing your Epigraph/Dedication and
% the second argument should be the title of this page. 
\thesisdedication{SupplementaryMaterial/Dedication}{Dedication}

%%%%%%%%%%%%%%%%%%%%%%%%%%%%%%%%%%%%%%%%%%%%%%%%%%%%%%
% This command is needed to get the main part of the %
% document going.                                    %
%%%%%%%%%%%%%%%%%%%%%%%%%%%%%%%%%%%%%%%%%%%%%%%%%%%%%%
\thesismainmatter

%%%%%%%%%%%%%%%%%%%%%%%%%%%%%%%%%%%%%%%%%%%%%%%%%%
% This is an AMS-LaTeX command to allow breaking %
% of displayed equations across pages. Note the  %
% closing the "}" just before the bibliography.  %
%%%%%%%%%%%%%%%%%%%%%%%%%%%%%%%%%%%%%%%%%%%%%%%%%%
\allowdisplaybreaks{
%\pagestyle{fancy}
%\fancyhead{}
%
%%%%%%%%%%%%%%%%%%%%%%
% THE ACTUAL CONTENT %
%%%%%%%%%%%%%%%%%%%%%%
% Chapters
% !TEX root = ../McCoy-Dissertation.tex
\chapter[Astrophysical Motivation for Custom Blazed Gratings]{Astrophysical Motivation for \\Custom Blazed Gratings}\label{ch:introduction}
%%%%%%%%%%%%%%%%%%%%%%%%%%%%%%%%%%%%%%%%%--------------------------------------------------
The physical basis for the science of spectroscopy is that atomic electrons absorb and emit electromagnetic radiation at a set of discrete frequencies unique to the specific ion, atom or molecule to which they are bound. 
Measuring these chemical signatures across the electromagnetic spectrum in astronomy is crucial not only for identifying elemental compositions in Solar System objects, planetary atmospheres, stars, galaxies and more, but also for probing physical conditions in various cosmic plasmas through application of atomic physics. 
Home to a large number of spectral lines associated with inner-shell electrons in atoms with atomic number $\mathcal{Z} \geq 6$, the soft x-ray bandpass of the electromagnetic spectrum\footnote{As described in \cref{ap:x-ray_intro}, this spectral band is defined by wavelength on the order of a nanometer (\si{\nm}) with photon energy ranging from a few hundred of electronvolts (\si{\electronvolt}) to a few kiloelectronvolts (\si{\kilo\electronvolt}). For reference, photons of visible light are characteristic of $\sim$~\SI{2}{\electronvolt} while hard x-rays commonly used for radiography have photon energy on the order of tens of \si{\kilo\electronvolt}.} plays a special role in the study of how \emph{baryonic matter}\footnote{This refers to ordinary matter composed of protons and neutrons, which are in turn made up of \emph{quarks}, as opposed to \emph{dark matter} or \emph{dark energy} \cite{Liddle03}.} cycles through galaxies and how these processes play into galaxy evolution. %(with electrons and neutrinos being \emph{leptons})
While hydrogen and helium are virtually the only elements whose nuclei were produced in the early Universe during \emph{primordial nucleosynthesis} and are by far the most abundant elements in the cosmos, making up $\sim \SI{74}{\percent}$ and $\sim \SI{24}{\percent}$ of all baryonic matter, respectively \cite{Mo10,Liddle03,Schneider06}, the remaining $\sim \SI{2}{\percent}$ is dominated by atomic nuclei of higher $\mathcal{Z}$ (known as \emph{metals} in astronomy) that are generated during the life cycles of stars with more than about eight times the mass of the Sun with $M_{\odot} \approx \SI{2e30}{\kilogram}$ \cite[\emph{cf.\@} \cref{ap:quantum_spectral}]{Hoyle46,Rolfs88,Carroll07,Ryden10}. 
These elements of low-to-mid $\mathcal{Z}$, such as oxygen, carbon and neon, produce soft x-ray spectral lines in highly-ionized states characteristic of hot, diffuse plasmas, where hydrogen and helium may be completely stripped of their electrons, rendering them invisible from the standpoint of spectroscopy at longer wavelengths \cite{Kahn02}. 

The regions in between star systems in a typical star-forming galaxy (the \emph{interstellar medium}) are enriched with $\mathcal{Z} \geq 6$ atomic nuclei predominately by supernovae that result from the deaths of $\gtrapprox 8 M_{\odot}$ stars \cite{Schneider06,Carroll07};  % and stellar winds. 
such stellar feedback also disperses metals into the gravitationally-bound gas that surrounds a galaxy (the \emph{circumgalactic medium}) or pervades galaxy groups and clusters (\emph{intragroup} and \emph{intracluster media}), and additionally, regions in between isolated galaxies, groups and clusters (the \emph{intergalactic medium}) \cite{Pettini01,Tumlinson17,Werner08,Shull14,Cavaliere16,Zhang2018Galax}. 
Supermassive black holes at the centers of galactic nuclei are inextricably linked to the evolution of the hosting galaxy, as evidenced by the well-known relation between stellar velocity dispersion and supermassive black hole mass, which can range from millions to billions of $M_{\odot}$ \cite{Mo10,Schneider06,Sparke07}. %M-sigma relation %times the solar mass, $M_{\odot}$
An \emph{active galaxy} is thought to be a phase of galactic evolution where copious amounts of gas and stellar material are accreting onto the central supermassive black hole while moreover, material outflows associated with the accretion also populate the circumgalactic medium, the intragroup or intracluster medium, and the intergalactic medium with metals \cite{Ishibashi16,Moll07,Zhang2018Galax}. 
In any case, material expelled from galaxies can exist for long time periods as an extremely diffuse plasma while portions of it eventually coalesce with interstellar media, where it ultimately contributes to star formation and galactic evolution \cite{Zhang2018Galax,angles-alcazar17,Tumlinson17}. 
\begin{figure}
 \centering
 \includegraphics[scale=2.23]{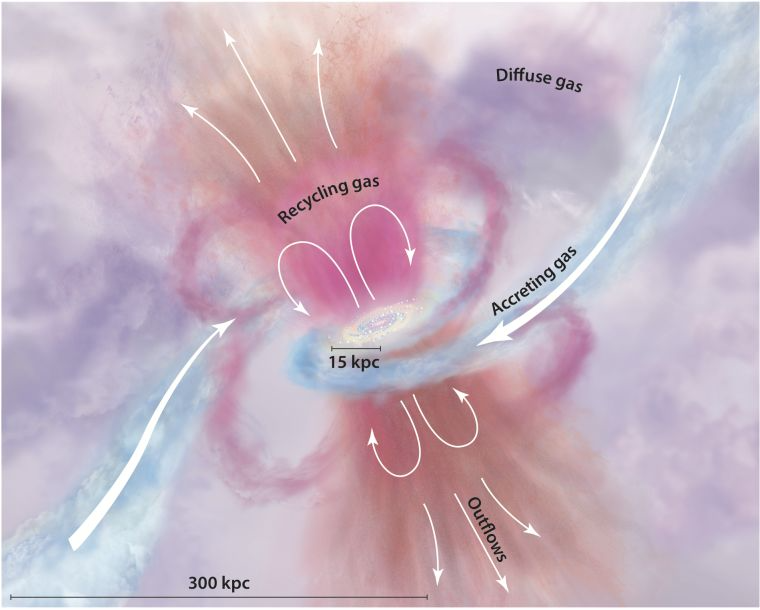}
 \caption[Cartoon illustrating how material cycles through a star-forming galaxy and the circumgalactic medium (credit: J.\ Tumlinson)]{Cartoon illustrating how material cycles through a star-forming galaxy and the circumgalactic medium. Material from the intergalactic medium accretes onto the galaxy (shown in blue) while outflows (pink and orange) eject material out of the galaxy. Nearby bound material (pink) is largely recycled back to the galaxy while expelled material (orange) either contributes to large-scale accretion (blue) or exists as diffuse gas (purple). Image credit: Tumlinson, et al.~(2017) \cite{Tumlinson17}.}\label{fig:CGM_Tumlinson}
 \end{figure}
The scenario of galactic feedback just described is illustrated\footnote{Image credit is owed to J.\ Tumlinson \cite{Tumlinson17}.} in \cref{fig:CGM_Tumlinson}, which depicts a galaxy with a diameter of about \num{15} \emph{kiloparsecs} (kpc)\footnote{Defined as the distance associated with a parallax measurement of one arcsecond ($\SI{1}{\arcsecond} \equiv \SI{1}{\degree}/3600$), one \emph{parsec} (\num{1}~pc) is about \SI{3.086e16}{\metre} or equivalently, \num{3.262} light years. For comparison, the distance to the nearest star, Proxima Centauri, is $\sim \num{1.3}$~pc while the diameter of the Milky Way is $\gtrapprox 30$~kpc.} and the circumgalactic medium with a size scale of \num{300}~kpc. 
As shown in the figure, the galaxy expels material through outflows while circumgalactic and intergalactic media feed accretion back to the galactic disk. 

Soft x-ray spectroscopy fits into this picture of the \emph{cosmic baryon cycle} with its exclusive ability to diagnose highly-ionized portions of diffuse plasmas that are expected to constitute a significant fraction of the total baryon content in the Universe \cite{Persic92,Bregman07,McGaugh10}. 
That is, while measurements of elemental abundance ratios from primordial nucleosynthesis \cite{Kirkman03} and acoustic oscillations observed in the \emph{cosmic microwave background} \cite{Komatsu09,Planck_collab} indicate that baryons should constitute $\gtrapprox \SI{4}{\percent}$ of the total mass-energy contained in the low-redshift Universe, stellar and gaseous material contained within galaxies, groups and clusters accounts for only $\gtrapprox \SI{10}{\percent}$ \cite{Persic92,Fukugita_2004,Bregman07}. 
The remaining $\lessapprox \SI{90}{\percent}$ is thought to reside in the extended halos of isolated galaxies, groups or clusters,\footnote{Stated differently, this refers to the outer portions of circumgalactic, intragroup or intracluster media, where plasma density is sufficiently low so as to prevent direct detection. This is in contrast to the diffuse x-ray emission easily observed throughout the bulk of rich clusters \cite{Sarazin86}.} and additionally, the intergalactic medium, where extremely low plasma density (down to $\sim 1$ particle \si{\per\metre\cubed}) makes spectroscopic detection difficult and often times prohibitive due to instrumental limitations \cite{Fukugita_2004,Bregman07,Shull12}. 
Approximately half of this material that exists outside of isolated galaxies, groups and clusters is expected to comprise the \emph{warm-hot intergalactic medium}, where filamentary structures of the \emph{cosmic web} are heated to \SIrange{e5}{e7}{\kelvin} by collisionless shocks induced by galactic feedback and gravitational effects of large-scale structure formation \cite{Cen99,Dave01,Nicastro05b,Nicastro08,Bykov08,deGraaff19}. 
Measuring this diffuse, highly-ionized baryonic content through soft x-ray absorption spectroscopy along the line-of-sight of active galactic nuclei is a main scientific objective for the \emph{Lynx X-ray Observatory} \cite{Lynx_web} that can only be carried out using state-of-the-art grating spectroscopy. 
One of four flagship mission concepts considered for the 2020 Astrophysics Decadal Survey \cite{Gaskin19,Gaskin15}, \emph{Lynx}, if selected, will replace the spacecraft x-ray observatories \emph{Chandra} \cite{Weisskopf00} and \emph{XMM-Newton} \cite{Jansen01,Lumb12} that have provided large amounts of scientific return since their deployment to Earth's orbit in the late 1990s. 
The \emph{X-ray Grating Spectrometer (XGS)} currently being planned for \emph{Lynx} baselines next-generation instrumentation that provides substantial improvements over the transmission grating spectrometers on board \emph{Chandra} \cite{Canizares05,Brinkman00} and the \emph{Reflection Grating Spectrometer (RGS)} of \emph{XMM-Newton} \cite{Kahn96,denHerder01} in terms of both spectral resolving power and spectral sensitivity \cite{McEntaffer19,Gunther19}. 

Leveraging from the \emph{RGS}, one of two main designs considered for the \emph{XGS} uses arrays of specialized reflection gratings integrated into the \emph{Lynx} telescope assembly to produce a soft x-ray diffraction pattern that is imaged and order-separated by detectors at the focal plane so that spectra can be extracted \cite{McEntaffer19}. 
Achieving high spectral resolving power and high spectral sensitivity with this spectrometer requires thousands of custom, high-precision gratings that each perform with high diffraction efficiency to be carefully manufactured and aligned in modular arrays. %\footnote{Discussed in \cref{sec:measure_efficiency,ap:grating_basics}, this quantity can be understood as the relative brightness of each diffracted beam produced by an illuminated grating.}
Bearing this in mind, the focus of this dissertation is on a series of laboratory experiments that demonstrate the ability of two recently-developed nanofabrication techniques in the areas of electron-beam lithography and nanoimprint lithography, namely \emph{thermally-activated selective topography equilibration} and \emph{substrate-conformal imprint lithography}, to pattern gratings with sawtooth surface reliefs (\emph{i.e.}, \emph{blazed gratings}) that are appropriate for such an application, with an emphasis on characterizing their diffraction-efficiency performance through beamline testing \cite{McCoy16,McCoy17,McCoy18,McCoy20,McCoy20b}. 
To provide background and motivation for this work, which was largely carried out at the Penn State Materials Research Institute \cite{PSU_MRI} and the Advanced Light Source at Lawrence-Berkeley National Laboratory \cite{ALS_web,LBNL_web}, this chapter first describes why sensitive, high-resolution grating spectroscopy is indispensable for measuring weak absorption lines in the soft x-ray that are produced as continuum emission from an active galactic nucleus passes through diffuse, intergalactic plasmas.  %circumgalactic and intergalactic media. 
Specifications for x-ray reflection gratings and previously-established methodology for their fabrication are then presented before the chapter concludes with an outline of this thesis document. %measuring weak absorption lines at soft x-ray wavelengths
This chapter builds on physics background outlined in \cref{ap:x-ray_intro,ap:quantum_spectral,app:x-ray_materials,ap:grating_basics} while SI units and the fundamental constants listed in \cref{tab:units,tab:fundamental_constants} are used throughout this dissertation.\footnote{I reserve credit for all micrographs, images and figures unless noted otherwise.} 

\section{Soft X-ray Spectroscopy of Highly-Charged Ions in Extended Galactic Halos and the Intergalactic Medium}\label{sec:astro_plasmas}
%%%%%%%%%%%%%%%%%%%%%%%%%%%%%%%%%%%%%%%%%-------------------------------------------------- 
In principle, highly-charged ions in extended halos of isolated galaxies, groups, or clusters, and the warm-hot intergalactic medium are observable spectroscopically as a given ionic species undergoes a bound-bound electronic transition that results in the emission or absorption of photons with energy centered around the difference in electronic binding energies, $\Delta \mathcal{E}_e$ [\emph{cf.\@} \cref{ap:quantum_spectral}]. 
A naturally-broadened spectral line takes the form of a \emph{Lorentzian distribution} [\emph{cf.\@} \cref{eq:normalized_lorentzian_QM}], which is written as a function of photon energy, $\mathcal{E}_{\gamma}$, as 
\begin{equation}\label{eq:Lorentzian_1} 
 \phi_{\text{nat}} \left( \mathcal{E}_{\gamma} , \Gamma \right) = \frac{1}{2 \pi} \left( \frac{\hbar^2 \Gamma}{\left(\mathcal{E}_{\gamma} - \Delta \mathcal{E}_e \right)^2 + \frac{\hbar^2}{4} \Gamma^2} \right) ,
 \end{equation}
where $\hbar$ is the reduced Planck's constant and $\Gamma$ is the quantum-mechanical transition rate derived in \cref{sec:transition_rate_sub}, which is equivalent to the relevant \emph{Einstein coefficient} for the process \cite{Rybicki86}. 
Spectral lines associated with bound-bound transitions occur at soft x-ray photon energies only in highly-charged ions that have ionization potentials, $\chi$, on the order of hundreds of \si{\electronvolt}, which is characteristic of an electron temperature, $T_e$, on the order of $\chi / k_{\mathcal{B}} \sim \SI{e6}{\kelvin}$, where $k_{\mathcal{B}}$ is the Boltzmann constant. 
With electrons and ions equilibrated in a plasma, $T_e$ is equal to the temperature of a particular species of ion, $T_{\text{ion}}$, and hence the effect of thermal broadening is expected to be significant for these spectral lines. 
\begin{figure}
 \centering
 \includegraphics[scale=1]{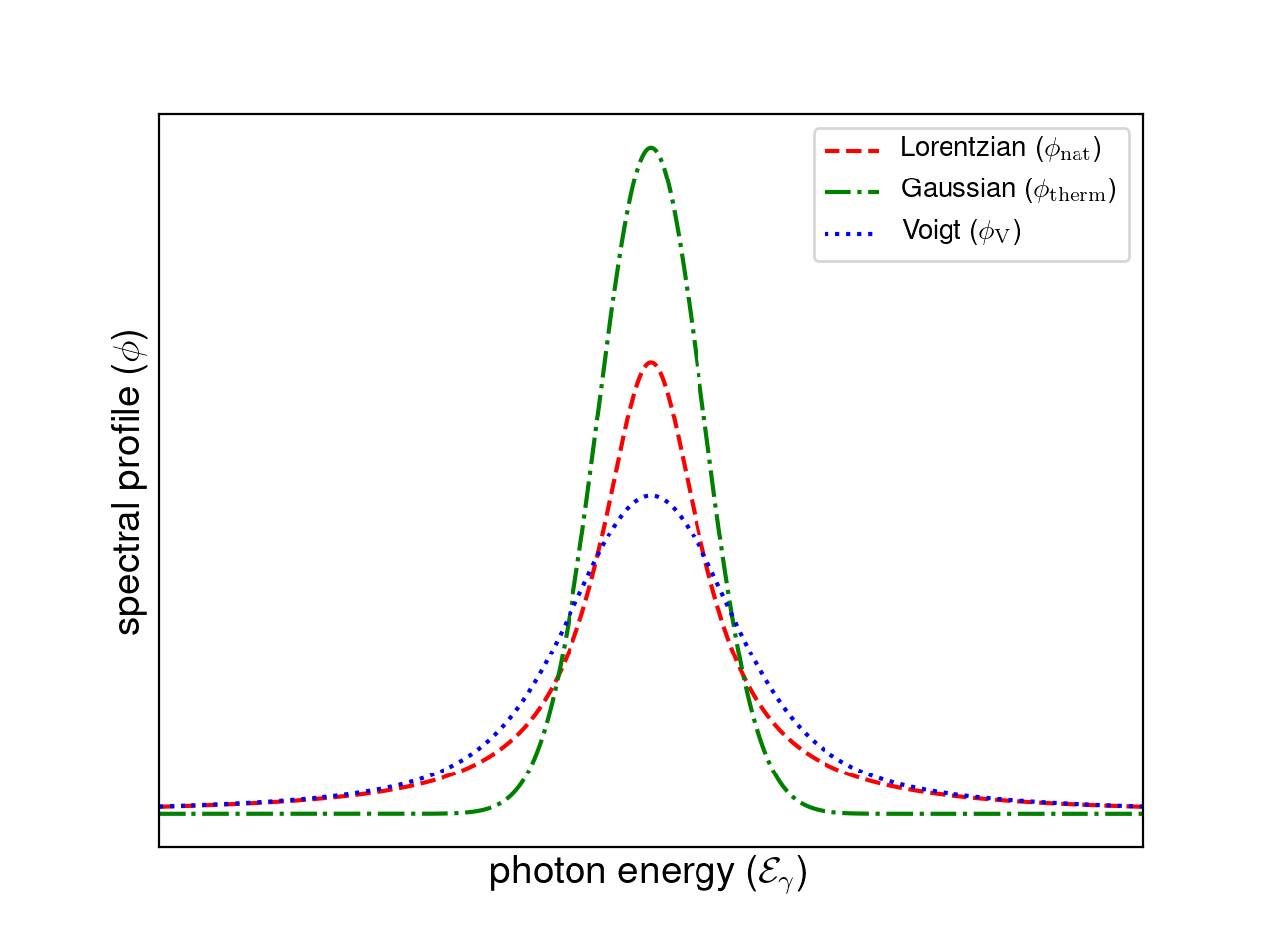}
 \caption[Spectral lines as a function of photon energy for Lorentzian, Gaussian and Voigt profiles.]{Spectral lines as a function of photon energy, $\mathcal{E}_{\gamma}$, for Lorentzian, Gaussian and Voigt profiles [\emph{cf.\@} \cref{eq:Lorentzian_1,eq:normalized_thermal_QM,eq:voigt}] with line centroids corresponding to $\mathcal{E}_{\gamma} = \Delta \mathcal{E}_e$, the electronic binding energy difference associated with the relevant bound-bound transition. The width of the Lorentzian profile describes natural broadening and depends on the quantum-mechanical transition rate, $\Gamma$ [\emph{cf.\@} \cref{sec:transition_rate_sub}], while the width of the Gaussian profile depends on the degree of thermal motion in the plasma, which is characterized by the quantity $\Delta \mathcal{E}_D$ [\emph{cf.\@} \cref{eq:Doppler_width}]. Neglecting collisional broadening, the Voigt profile is the convolution of these two functions, which then depends on both $\Gamma$ and $\Delta \mathcal{E}_D$.}\label{fig:spectral_lines}
 \end{figure}
This contribution to a spectral line profile takes the form of a \emph{Gaussian distribution}:
\begin{subequations}
\begin{equation}\label{eq:normalized_thermal_QM}
 \phi_{\text{therm}} \left( \mathcal{E}_{\gamma} , \Delta \mathcal{E}_D \right) \equiv \frac{1}{\Delta \mathcal{E}_D \sqrt{2 \pi}} \mathrm{e}^{-\left( \mathcal{E}_{\gamma} - \Delta \mathcal{E}_e \right)^2 / 2 \Delta \mathcal{E}_D^2} ,
 \end{equation}
where 
\begin{equation}\label{eq:Doppler_width}
 \Delta \mathcal{E}_D \equiv \Delta \mathcal{E}_e \sqrt{\frac{2 k_{\mathcal{B}} T_{\text{ion}}}{m_{\text{ion}} c_0^2}} %see pg 288 of R&L for derivation
 \end{equation} 
is the \emph{Doppler width} of the ions with $m_{\text{ion}}$ as the ion mass [\emph{cf.\@} \cref{tab:astro_element_nuclei_mass}] and $c_0$ as the speed of light \cite{Rybicki86}. 
\end{subequations}
Due to the extremely low densities characteristic of the plasmas considered, the effect of collisional broadening can be neglected and thus an overall \emph{Voigt profile} that combines \cref{eq:Lorentzian_1,eq:normalized_thermal_QM} can be written as the convolution of the functions $\phi_{\text{nat}} \left( \mathcal{E}_{\gamma} , \Gamma \right)$ and $\phi_{\text{therm}} \left( \mathcal{E}_{\gamma} , \Delta \mathcal{E}_D \right)$:
\begin{equation}\label{eq:voigt}
 \phi_{\text{V}} \left( \mathcal{E}_{\gamma}, \Gamma, \Delta \mathcal{E}_D \right) \equiv \int_{-\infty}^{\infty} \phi_{\text{therm}} \left( \mathcal{E}_{\gamma} , \Delta \mathcal{E}_D \right) \, \phi_{\text{nat}} \left( \mathcal{E}_{\gamma} - \mathcal{E}'_{\gamma} , \Gamma \right) \dd{\mathcal{E}'_{\gamma}},
 \end{equation}
which describes the shape of a general spectral line from a hot, diffuse plasma. %the theoretical shape
These three spectral profiles are compared in \cref{fig:spectral_lines}, where it is seen that \cref{eq:Lorentzian_1,eq:normalized_thermal_QM} convolve to yield a Voigt profile with a Gaussian-like core and Lorentzian-like wings. 

Discussed in \cref{ap:quantum_spectral}, the most abundant species of highly-charged ions that exist at $\sim \SI{e6}{\kelvin}$ temperatures are \emph{hydrogen-like} and \emph{helium-like} ions of low-to-mid $\mathcal{Z}$, where just one or two bound electrons remain, respectively. 
The strongest x-ray spectral lines from such ions are expected to be those associated with \emph{electric-dipole transitions}, also known as \emph{resonance lines}, that occur between the K-shell and L-shell [\emph{cf.\@} \cref{sec:multipole}; \cref{tab:astro_element_lyman_approx,tab:astro_element_he_res}]. 
Other x-ray spectral lines, such as those arising from \emph{intercombination} and \emph{forbidden} transitions in helium-like ions [\emph{cf.\@} \cref{tab:astro_element_he_interx,tab:astro_element_he_intery,tab:astro_element_he_forb}] meanwhile are relatively dim but play an important role in plasma diagnostics of emitting sources \cite{Porquet2010,Paerels03,Kahn02}. 
In extended galactic halos and the intergalactic medium, radiative transitions in highly-charged ions are driven by collisional excitement due to there being a lack of a strong x-ray source illuminating the plasma. 
This can be described using the \emph{principle of detailed balance} for a hypothetical two-level atom in a collisional plasma with an electronic binding energy difference $\Delta \mathcal{E}_e$ between two bound states:
\begin{equation}
 n_1 n_e C_{12} = n_2 n_e C_{21} + n_2 A_{21} ,
 \end{equation}
which relates the Einstein coefficients that describe transition rates for collisional excitement, $C_{12}$, to those for collisional de-excitement, $C_{21}$, and \emph{spontaneous emission},\footnote{The principle of spontaneous emission, where a photon is emitted as an excited atomic electron transitions to its ground state, is described using quantum mechanics of photons and non-relativistic bound electrons in \cref{sec:single_absorption_emission}.} $A_{21}$, where $n_e$ is the number density of energetic free electrons in the plasma while $n_1$ and $n_2$ are the number densities of bound electrons in the ground state and the excited state, respectively \cite{Mo10}. 

The average time in between collisions for a very diffuse plasma is much longer than the timescales associated with radiative decay in highly-charged ions, which are remarkably short due to the transition rate of such a transition, $\Gamma = A_{21}$, depending on $\Delta \mathcal{E}_e^2$ as described in \cref{sec:transition_rate_sub}. 
Thus, with collisional de-excitement neglected and excitations depending on two-body collisions between electrons and ions with volume densities $n_e$ and $n_{\text{ion}}$, respectively, the overall intensity of emission lines is then expected to be proportional to the \emph{emission measure} of the plasma, defined as 
\begin{equation}\label{eq:emission_measure}
 \text{EM} = \int n_e n_{\text{ion}} \dd[3]{\mathbold{r}} ,
 \end{equation}
where the integration is carried out over the volume of the emitting plasma \cite{Paerels03}. 
While x-ray grating spectroscopy is best suited for point sources as argued in \cref{sec:grating_tech_dev}, energy-dispersive detectors that are able to gather spectra of extended sources can measure emission lines from these diffuse plasmas in areas where $\text{EM}$ [\emph{cf.\@} \cref{eq:emission_measure}] is sufficiently large, such as in the bulk of intracluster media \cite{Bohringer10}. 
However, emission spectroscopy of highly-charged ions in extremely low-$n_e$ plasmas, such as those present in extended galactic halos and the intergalactic medium, demands levels of sensitivity that far surpass the capabilities the capabilities of state-of-the-art instruments of this type, such as the currently-planned \emph{Lynx Microcalorimeter (LXM)} \cite{Bandler19}. %that far surpass the capabilities
To circumvent this issue, x-ray absorption spectroscopy using active galactic nuclei as background sources must be employed, and as motivated in the following discussion, highly-efficient grating spectrometers capable of high spectral resolving power are required to detect faint absorption lines that characterize plasma density among other physical parameters. 

\subsection{Weak Absorption Lines From Active Galactic Nuclei}\label{sec:weak_absorb}
%%%%%%%%%%%%%%%%%%%%%%%%%%%%%%%%%%%%%%%%%--------------------------------------------------
An \emph{active galactic nucleus} describes a system where there exists an accretion disk around the galaxy's supermassive black hole, along with a hot corona and a surrounding torus of neutral gas and dust \cite{Schneider06,Carroll07,Mo10,Netzer13}. 
Associated with such a system are jets aligned with the axis of rotation and ionized winds emanating from the accretion disk; this scenario is illustrated\footnote{Image credit is owed to A.\ Simonnet (\url{http://auroresimonnet.com/}).} in \cref{fig:agn_cartoon}. 
%that are thought to populate circumgalactic and intergalactic media with baryonic matter. 
\begin{figure}
 \centering
 \includegraphics[scale=0.15]{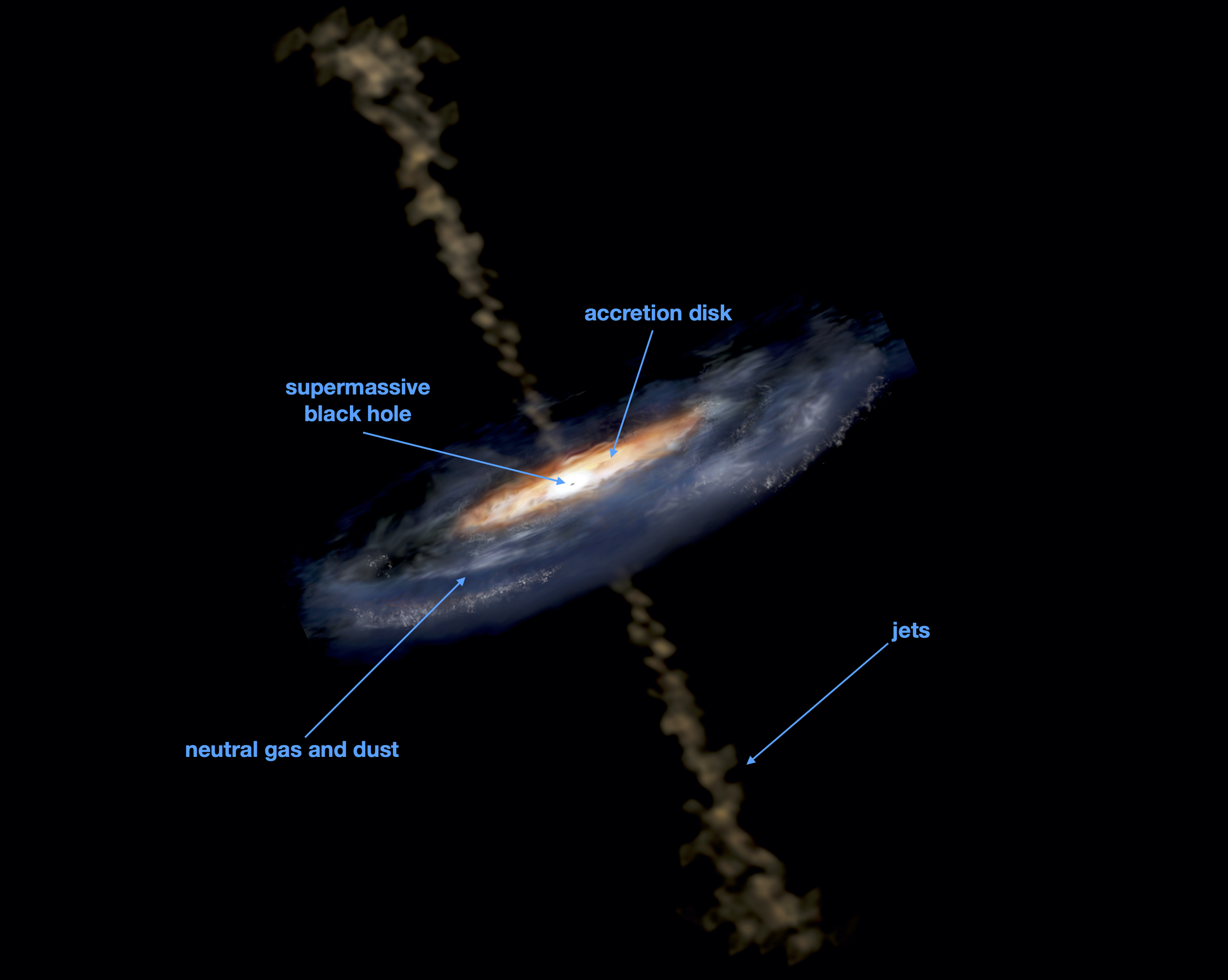}
 \caption[Cartoon of an active galactic nucleus (credit: A.\ Simmonnet)]{Cartoon of an active galactic nucleus. An accretion disk surrounds the immediate vicinity of the central supermassive black hole while further out radially, there exists a torus of neutral gas and dust. Moreover, a hot corona and ionizing winds (not shown) surround the inner accretion disk while jets of material are aligned with its axis of rotation. Image credit: A.\ Simonnet.}\label{fig:agn_cartoon}
 \end{figure}
In principle, the temperature of the accretion disk, $T$, is a decreasing function of radius, $r$, with $T \propto r^{-3/4}$ that typically maximizes around \SI{e5}{\kelvin} \cite{Schneider06}. 
The power radiated through a surface of unit area as a function of photon energy, $\mathcal{E}_{\gamma}$, produced by a single annulus of temperature $T$ is characteristic of a \emph{black-body radiator} described by \emph{Planck's law} \cite{Carroll07,Rybicki86}:
\begin{subequations}
\begin{equation}\label{eq:Planck_law}
 B (\mathcal{E}_{\gamma}, T) = \frac{\mathcal{E}_{\gamma}^3}{2 \pi^2 \hbar^2 c_0^2} \frac{1}{ \mathrm{e}^{\mathcal{E}_{\gamma} / k_{\mathcal{B}} T} - 1} ,
 \end{equation}
while according to \emph{Wein's displacement law}: 
\begin{equation}\label{eq:Wein_law}
 \lambda_{\text{max}} = \frac{b}{T} \quad \text{with } b = \SI{2.897771955e-3}{\meter\kelvin} ,
 \end{equation} 
\end{subequations}
the wavelength of radiation associated with a typical maximum temperature of $T \sim \SI{e5}{\kelvin}$, $\lambda_{\text{max}}$, falls in the far ultraviolet (UV). 
Taking into account the entire accretion disk with $T \propto r^{-3/4}$, the resulting UV continuum spectrum is a summation of blackbody spectra that manifests as a power law spectrum \cite{Schneider06}. 

X-rays are produced indirectly by such an accretion disk predominantly through the \emph{inverse Compton effect}, where highly-energetic electrons transfer momentum to the optical and UV photons generated by blackbody radiation \cite{Haardt91,Schneider06,Netzer13}. 
%Although the accretion disk itself is not a strong source of x-rays by the above argument, active galactic nuclei are indeed bright x-ray sources. 
%Active galactic nuclei typically are bright x-ray sources notwithstanding that the accretion disk itself is not a strong source of x-rays by the above argument. 
%This is thought to be because the corona is responsible for up-scattering UV photons produced by the accretion disk into x-rays through the \emph{inverse Compton effect}, where highly-energetic electrons transfer momentum to photons; 
While the result is also a continuum spectrum that appears as a power law, soft x-ray emission measured from active galactic nuclei typically exceeds extrapolations from higher $\mathcal{E}_{\gamma}$; this is the so-called \emph{soft x-ray excess}, which has several proposed emission mechanisms \cite{Walter93,Schneider06,Crummy06,Gliozzi20}. %, often taken as $\mathcal{E}_{\gamma}^{-0.7}$,
In cases where the jets of an active galactic nucleus are oriented toward Earth, the object is viewed as a \emph{blazar} with bright, beamed emission produced from \emph{synchrotron radiation}, which also manifests as a power law for a distribution of electron energies \cite{Rybicki86,Schneider06,Netzer13,Pal20}. %\cite{Pal20}
Active galactic nuclei thus are suitable background sources for absorption spectroscopy of highly-charged ions in extended galactic halos and the intergalactic medium, with blazars being desirable due to their particularly high x-ray brightness.\footnote{Another class of x-ray sources are \emph{x-ray binaries}: binary star systems where one of the stars exists as a compact object such as a neutron star or a black hole, which accretes material from the other. However, these x-ray sources are typically a few orders of magnitude dimmer than active galactic nuclei \cite{Schneider06}.}  %, and moreover, their intensity is often unstable over relatively short timescales

As described in \cref{sec:form_spectral}, an absorption line is formed as a particular species of ion along the line-of-site of a continuum source absorbs a narrow spectral distribution of soft x-rays centered around $\mathcal{E}_{\gamma} = \Delta \mathcal{E}_e$, the electronic binding energy difference associated with a bound-bound transition from the ground state to some excited bound state. 
The \emph{spectral flux}, $\mathcal{F}$,\footnote{This quantity is defined as radiative intensity per unit area, per unit time, per photon energy \cite{Rybicki86}.} absorbed by such a transition, in principle, takes the form of $\phi_{\text{V}} \left( \mathcal{E}_{\gamma}, \Gamma, \Delta \mathcal{E}_D \right)$ [\emph{cf.\@} \cref{eq:voigt}] with fixed values for $\Gamma$ and $\Delta \mathcal{E}_D$. 
The spectral flux of the absorption line, $\mathcal{F}_{\text{line}}$, then is the difference between the background continuum spectral flux, here taken as $\mathcal{F}_c \propto \mathcal{E}_{\gamma}^{-0.7}$, and this Voigt profile [\emph{cf.\@} \cref{fig:absorption_line}]. 
\begin{figure}
 \centering 
 \includegraphics[scale=0.245]{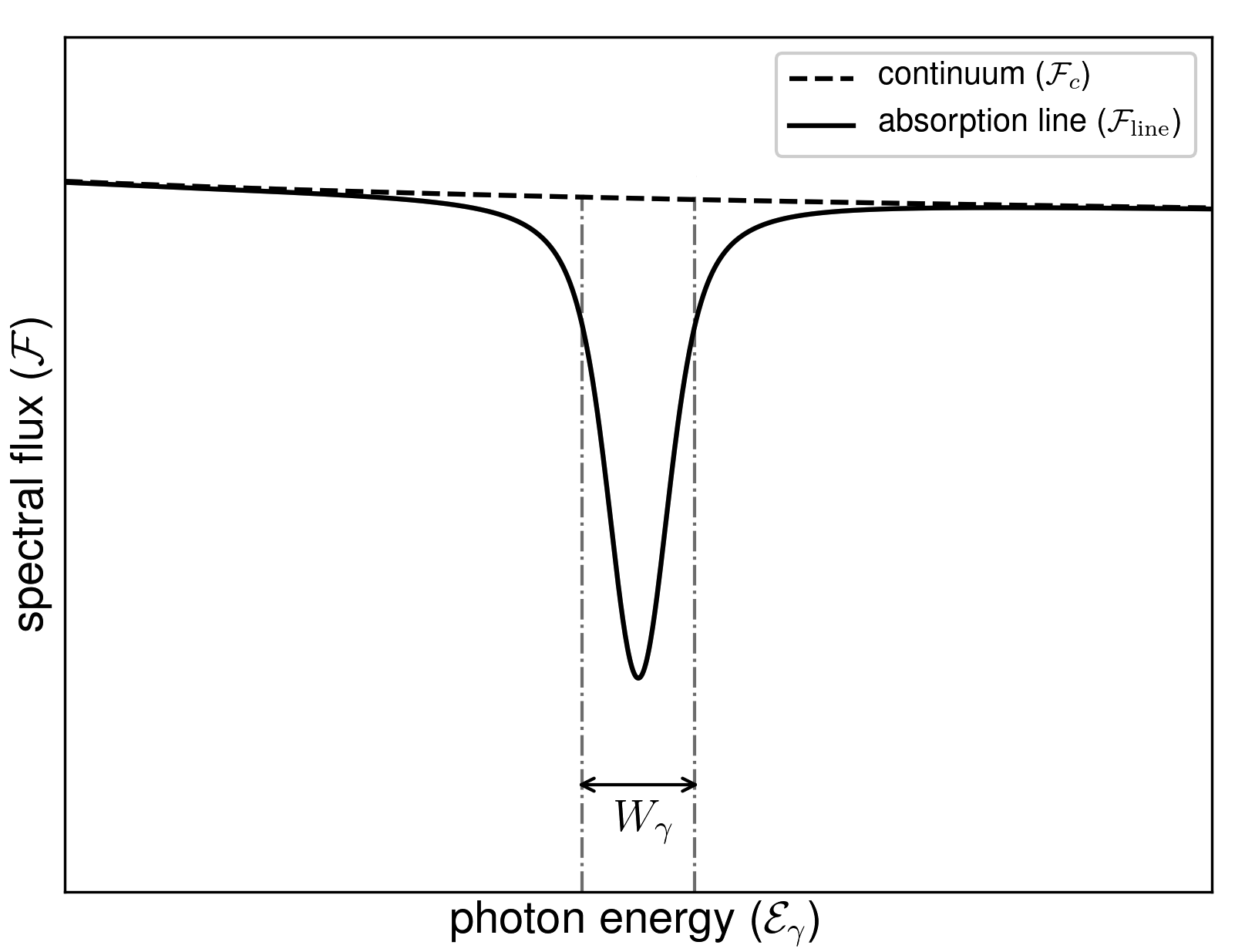}%absorption_line.png}
 \caption[Spectral flux of an absorption line as a function of photon energy]{Spectral flux of an absorption line as a function of photon energy, $\mathcal{F}_{\text{line}}$, where the line centroid corresponds to $\mathcal{E}_{\gamma} = \Delta \mathcal{E}_e$. By construction, $\mathcal{F}_{\text{line}}$ is the difference between the underlying continuum spectral flux, $\mathcal{F}_c \propto \mathcal{E}_{\gamma}^{-0.7}$, and a spectral profile associated with the absorption line, which is assumed to take the form of a Voigt profile [\emph{cf.\@} \cref{eq:voigt}].}\label{fig:absorption_line}
 \end{figure}
Depending on the \emph{optical depth} of a given species of highly-charged ion in the absorbing plasma, the strength of such an absorption line can be quantified using an \emph{equivalent width} \cite{Carroll07}, defined in terms of photon energy as 
%The strength of such an absorption line, which depends on the \emph{optical depth} of a given species of highly-charged ion in the absorbing plasma, can be quantified using an \emph{equivalent width} defined in terms of photon energy as 
\begin{equation}\label{eq:EW_gen}
 W_{\gamma} \equiv \int \left(  \frac{\mathcal{F}_c - \mathcal{F}_{\text{line}}}{\mathcal{F}_c} \right) \dd{\mathcal{E}_{\gamma}} = \int \left(  1 - \mathrm{e}^{- \tau^{\text{ion}}_{\gamma}} \right) \dd{\mathcal{E}_{\gamma}}  ,
 \end{equation} 
where, for a given absorbing ion, $\tau^{\text{ion}}_{\gamma} \equiv \sigma^{\text{ion}}_{\gamma} \mathcal{N}^{\text{ion}}_{\gamma}$ is the optical depth per photon energy, $\sigma^{\text{ion}}_{\gamma}$ is the cross-section per photon energy for absorption, which depends on $\Gamma$ for the relevant bound-bound absorption process [\emph{cf.\@} \cref{sec:transition_rate_sub}] and $\mathcal{N}^{\text{ion}}_{\gamma}$ is the column density per photon energy, defined as 
\begin{equation}
 \mathcal{N}^{\text{ion}}_{\gamma} = \int n^{\text{ion}}_{\gamma} \dd{\ell} , 
 \end{equation}
where $n^{\text{ion}}_{\gamma}$ is the volume density of ions absorbing at a photon energy $\mathcal{E}_{\gamma}$ and the integral over $\ell$ represents the line-of-sight of the observation \cite{Carroll07}.  
In diffuse plasmas of extended galactic halos and the intergalactic medium, $\mathcal{N}^{\text{ion}}_{\gamma}$ is very small and hence it is justified to make the approximation $\mathrm{e}^{-\tau^{\text{ion}}_{\gamma}} \approx 1 - \tau^{\text{ion}}_{\gamma}$ so that %the quantity $\mathcal{N}^{\text{ion}}_{\gamma}$ is very small
\begin{equation}\label{eq:EW_od_thin}
 W_{\gamma} \approx \int \tau^{\text{ion}}_{\gamma} \dd{\mathcal{E}_{\gamma}}
 \end{equation} 
and from comparing \cref{eq:EW_gen} and \cref{eq:EW_od_thin}, 
\begin{equation}
 \tau^{\text{ion}}_{\gamma} \approx \frac{\mathcal{F}_c - \mathcal{F}_{\text{line}}}{\mathcal{F}_c} .
 \end{equation}
Therefore, a measurement for $W_{\gamma}$ of a weak absorption line directly yields the optical depth for a specific highly-charged ion, and through $\mathcal{N}^{\text{ion}}_{\gamma} = \tau^{\text{ion}}_{\gamma} / \sigma^{\text{ion}}_{\gamma}$, the density of a given ionic species present in the plasma can be inferred. 

The overall function of an energy-dispersive spectrometer such as the \emph{LXM} is to bin collected photons by their energy, $\mathcal{E}_{\gamma}$, into spectral intervals of width $\Delta \mathcal{E}_{\gamma}$ so that the instrument's spectral resolving power, $\mathscr{R} = \mathcal{E}_{\gamma} / \Delta \mathcal{E}_{\gamma}$, by definition indicates its ability to distinguish between photons with $\mathcal{E}_{\gamma}$ and $\mathcal{E}_{\gamma} + \Delta \mathcal{E}_{\gamma}$ in the process of measuring a spectrum. 
Unresolved, weak absorption lines can be measured within a single resolution bin, $\Delta \mathcal{E}_{\gamma}$, and in this case \cref{eq:EW_gen} becomes\footnote{The following analysis regarding equivalent widths and a figure of merit for spectral line detection is based on an unpublished document written by J.\ Kaastra for the \emph{International X-ray Observatory} mission concept in 2008 \cite{McEntaffer09,McEntaffer10}.} 
\begin{equation}\label{eq:EW_int}
 W_{\gamma} = \left( \frac{\mathcal{F}_c - \mathcal{F}_{\text{line}}}{\mathcal{F}_c} \right) \Delta \mathcal{E}_{\gamma} \equiv \frac{F_{\text{line}}}{\mathcal{F}_c} \approx \tau^{\text{ion}}_{\gamma} \Delta \mathcal{E}_{\gamma} ,
 \end{equation}
where $F_{\text{line}} \equiv \left[ \mathcal{F}_c - \mathcal{F}_{\text{line}} \right] \Delta \mathcal{E}_{\gamma} = \mathcal{F}_c W_{\gamma}$ is the flux missing from the continuum due to the absorption line within the spectral interval $\Delta \mathcal{E}_{\gamma}$. 
Owing to their high $\mathcal{E}_{\gamma}$ and low count rates in astronomy, x-rays are practically measured as individual particles that obey \emph{Poisson statistics}, which dictate that the probability for the number of photons detected during an observation is given by 
\begin{equation}\label{eq:poisson}
 \mathscr{P} \left( N=N_{\text{tot}} \right) = \mathrm{e}^{-N_{\text{tot}}} \frac{N^N_{\text{tot}}}{N!} ,
 \end{equation}
where $N_{\text{tot}}$ is the average number of photons collected during a given observation time, $t_{\text{obs}}$, with $\sqrt{N_{\text{tot}}}$ as the standard deviation describing \emph{shot noise} \cite{siemiginowska_2011,Sturge03}. 
In the case of absorption spectroscopy, $N_{\text{tot}}$ is essentially the number of photons absorbed by the plasma subtracted from the number of continuum photons. 
The number of photons associated with the absorption line that are detected with an instrumental collecting area $A_{\text{col}}$ and an exposure time $t_{\text{obs}}$ is 
\begin{subequations}
\begin{equation}\label{eq:line_count}
 N_{\text{line}} \equiv F_{\text{line}} A_{\text{col}} t_{\text{obs}} 
 \end{equation}
while, using $A_{\text{col}} t_{\text{obs}} = N_{\text{line}} / F_{\text{line}}$ [\emph{cf.\@} \cref{eq:line_count}] and $W_{\gamma} = F_{\text{line}} / \mathcal{F}_c$ [\emph{cf.\@} \cref{eq:EW_int}], the number of photons from the continuum detected across the spectral interval $\Delta \mathcal{E}_{\gamma}$ is 
\begin{equation}\label{eq:cont_count}
 N_c \equiv \mathcal{F}_c \Delta \mathcal{E}_{\gamma} A_{\text{col}} t_{\text{obs}} = N_{\text{line}} \frac{\Delta \mathcal{E}_{\gamma}}{W_{\gamma}} .
 \end{equation}
 The total number of counts detected from a spectral line within an interval $\Delta \mathcal{E}_{\gamma}$, neglecting instrumental background,\footnote{In principle, such a background due to dark current, for example, contributes to this total through $N_b \equiv \mathcal{F}_b \Delta \mathcal{E}_{\gamma} A_{\text{col}} t_{\text{obs}}$, where $\mathcal{F}_b$ is the spectral flux associated with the signal.} then is
\begin{equation}\label{eq:total_count}
 N_{\text{tot}} \equiv N_c - N_{\text{line}} = N_{\text{line}} \left( \frac{\Delta \mathcal{E}_{\gamma}}{W_{\gamma}} - 1 \right) = N_c \left( 1 - \frac{W_{\gamma}}{\Delta \mathcal{E}_{\gamma}} \right) , 
 \end{equation}
\end{subequations}
with the corresponding counting error being approximately $\sqrt{N_c}$ for weak spectral lines with $\abs{W_{\gamma}} \ll \Delta \mathcal{E}_{\gamma}$. 

Detecting an absorption line with statistical significance hinges on having a sufficiently large \emph{signal-to-noise ratio}: 
\begin{equation}\label{eq:signal_noise}
 \mathcal{S} \equiv \frac{N_{\text{line}}}{\sqrt{N_{\text{tot}}}} = \sqrt{\frac{N_{\text{line}}}{\frac{\Delta \mathcal{E}_{\gamma}}{W_{\gamma}} - 1}} , 
 \end{equation}
where $N_{\text{line}}$, the number of photons absorbed by the spectral line, is the signal and $\sqrt{N_{\text{tot}}}$ is the shot noise [\emph{cf.\@} \cref{eq:poisson}]. 
With $N_{\text{line}} = N_c \left( W_{\gamma} / \Delta \mathcal{E}_{\gamma} \right)$ [\emph{cf.\@} \cref{eq:cont_count}] inserted into \cref{eq:signal_noise}, the minimum number of photons associated with the continuum that must be detected to achieve a signal-to-noise ratio $\mathcal{S}_0$ is described by the following inequality: 
\begin{equation}
 N_c > \mathcal{S}_0^2 \left( \frac{\Delta \mathcal{E}_{\gamma}}{W_{\gamma}} - 1 \right) \left( \frac{\Delta \mathcal{E}_{\gamma}}{W_{\gamma}} \right) \approx \left( \frac{\Delta \mathcal{E}_{\gamma}}{W_{\gamma}} \right)^2 \mathcal{S}_0^2 , 
 \end{equation}
which indicates that having a small resolution bin, $\Delta \mathcal{E}_{\gamma}$, or high $\mathscr{R} = \mathcal{E}_{\gamma} / \Delta \mathcal{E}_{\gamma}$, is needed for distinguishing weak absorption lines from dim continua. 
As $N_{\text{line}}$ [\emph{cf.\@} \cref{eq:line_count}] depends on the flux absorbed by the spectral line, $F_{\text{line}}$, the observation exposure time, $t_{\text{obs}}$, and the collecting area of the instrument, $A_{\text{col}}$, the only quantity that is inherent to a specific instrument is the latter, $A_{\text{col}}$. 
An appropriate figure of merit (FOM) for an instrument's ability to detect a spectral line then should be proportional to $\mathcal{S}$ [\emph{cf.\@} \cref{eq:signal_noise}], but without the dependence on $F_{\text{line}}$ and $t_{\text{obs}}$ for a specific observation: 
\begin{subequations}
\begin{equation}\label{eq:FOM}
 \text{FOM} \propto \frac{\mathcal{S}}{\sqrt{F_{\text{line}} t_{\text{obs}}}} = \frac{\sqrt{A_{\text{col}}}}{\sqrt{\frac{\Delta \mathcal{E}_{\gamma}}{W_{\gamma}} - 1}} .
 \end{equation}
Provided that they can be resolved, relatively strong spectral lines with $\abs{W_{\gamma}} \gg \Delta \mathcal{E}_{\gamma}$ have $\text{FOM} \propto \sqrt{A_{\text{col}}}$ and hence line detection depends on collecting area alone. 
On the other hand, for weak absorption lines with $\abs{W_{\gamma}} \ll \Delta \mathcal{E}_{\gamma}$, \cref{eq:FOM} reduces to 
\begin{equation}\label{eq:FOM_weak}
 \text{FOM}_{\text{weak}} \propto \frac{\mathcal{S}}{\sqrt{F_{\text{line}} t_{\text{obs}}}} \approx \sqrt{\frac{A_{\text{col}} W_{\gamma}}{\Delta \mathcal{E}_{\gamma}}} \propto \sqrt{A_{\text{col}} \mathscr{R}} ,
 \end{equation}
\end{subequations}
which shows that the detection of such a line has the same dependence on both instrument collecting area and spectral resolving power. %, $\mathscr{R} = \mathcal{E}_{\gamma} / \Delta \mathcal{E}_{\gamma}$. 
This latter relation can be rearranged to give 
\begin{equation}
 t_{\text{obs}} \propto \frac{\mathcal{S}^2}{\text{FOM}_{\text{weak}}^2 F_{\text{line}}} \propto \frac{1}{A_{\text{col}} \mathscr{R}} ,
 \end{equation}
which indicates that making improvements in instrument collecting area and spectral resolving power both serve to reduce the observation time required to achieve a signal-to-noise ratio $\mathcal{S}$ in a weak absorption line. 

\subsection{The Need for Next-Generation X-ray Gratings}\label{sec:need_gratings}
%%%%%%%%%%%%%%%%%%%%%%%%%%%%%%%%%%%%%%%%%--------------------------------------------------
The performance of a grating spectrometer differs fundamentally from an energy-dispersive instrument in that radiation is binned according to wavelength, $\lambda$, so that its wavelength-dispersive spectral resolving power, $\mathscr{R} = \lambda / \Delta \lambda$, indicates its ability to distinguish between radiation with $\lambda$ and $\lambda + \Delta \lambda$ in the process of measuring a spectrum [\emph{cf.\@} \cref{ap:grating_basics}]. 
This can be seen to be beneficial for soft x-ray absorption spectroscopy by noting that the spectral resolving power of an energy-dispersive instrument, $\mathscr{R} = \mathcal{E}_{\gamma} / \Delta \mathcal{E}_{\gamma}$, improves as $\mathcal{E}_{\gamma}$ increases and degrades as $\mathcal{E}_{\gamma}$ decreases.\footnote{That is, provided the resolution bin size, $\Delta \mathcal{E}_{\gamma}$, stays roughly constant across a given bandpass.} 
With the opposite being true for wavelength-dispersive instruments, where $\mathscr{R} = \lambda / \Delta \lambda$, a grating spectrometer with sufficiently small $\Delta \lambda$ lends itself to the detection of weak absorption lines with $\mathcal{E}_{\gamma} \lessapprox \SI{1}{\kilo\electronvolt}$, where the equivalent width in terms of wavelength, $W_{\lambda}$, is related to $W_{\gamma}$ [\emph{cf.\@} \cref{eq:EW_gen}] through 
\begin{equation}
 \frac{W_{\lambda}}{\lambda} = \frac{W_{\gamma}}{\mathcal{E}_{\gamma}} \implies W_{\lambda} = \frac{\lambda^2}{h c_0} W_{\gamma} = \frac{2 \pi h c_0}{\mathcal{E}^2_{\gamma}} W_{\gamma} 
 \end{equation}
with $h \equiv 2 \pi \hbar$ and $h c_0 \approx \SI{1240}{\electronvolt\nm}$. 
On the other hand, energy-dispersive instruments are beneficial for higher-energy spectral lines such as those associated with highly-charged ions of iron [\emph{cf.\@} \cref{tab:astro_element_lyman_approx,tab:astro_element_he_res}] and additionally, relatively bright emission lines from extended sources such as supernova remnants \cite{Dewey10,Ballet02}. 

As oxygen is the most abundant element beyond hydrogen and helium, hydrogen-like and helium-like charge states of this atom (\emph{i.e.}, \ion{O}{viii} and \ion{O}{vii}, respectively), with prominent absorption lines centered at $\mathcal{E}_{\gamma}^{\text{\ion{O}{viii}}} \approx \SI{653.4}{\electronvolt}$ and $\mathcal{E}_{\gamma}^{\text{\ion{O}{vii}}} \approx \SI{574.0}{\electronvolt}$, are the most important contributors for x-ray absorption studies of diffuse plasmas in extended galactic halos and the intergalactic medium. %circumgalactic and intergalactic media. 
Along with analogous spectral lines from highly-charged ions of other relatively abundant elements such as carbon, neon and nitrogen [\emph{cf.\@} \cref{tab:astro_element_lyman_approx,tab:astro_element_he_res}] these absorption lines moreover are generally Doppler shifted, predominantly by \emph{cosmological redshift}, $z$, which serves to decrease the measured photon energy, $\mathcal{E}'_{\gamma}$ \cite{Liddle03,Carroll07}: 
\begin{equation}\label{eq:energy_shift}
 \frac{\mathcal{E}_{\gamma}}{\mathcal{E}'_{\gamma}} \equiv 1 + z \implies \mathcal{E}'_{\gamma} = \frac{\mathcal{E}_{\gamma}}{1+z}.
 \end{equation}
Although a small number of absorbing \ion{O}{vii} systems associated with the warm-hot intergalactic medium have been claimed through statistical analyses of data gathered by the grating spectrometers of \emph{XMM-Newton} and \emph{Chandra} with stacked observing times, $t_{\text{obs}}$, on the order of \si{\mega\second} \cite{Nicastro18,Kovacs19}, a spectrometer with a large $\text{FOM}_{\text{weak}} \propto \sqrt{A_{\text{col}} \mathscr{R}}$ for $\mathcal{E}_{\gamma} \lessapprox \SI{1}{\kilo\electronvolt}$ is needed for detecting weak absorption lines in these diffuse plasmas with higher statistical confidence and reducing the required $t_{\text{obs}}$. 
Explicitly, with $\Delta \mathcal{E}_{\gamma} = \SI{3}{\electronvolt}$ and $A_{\text{col}} \approx \SI{1000}{\cm\squared}$ baselined for the main component of the \emph{LXM}, $\mathscr{R}$ is on the order of a few hundred for $\mathcal{E}_{\gamma} \lessapprox \SI{1}{\kilo\electronvolt}$. 
While this yields an improvement over the \emph{RGS} on board \emph{XMM-Newton} in terms of $\text{FOM}_{\text{weak}}$, which also exhibits $\mathscr{R}$ of a few hundred but with $A_{\text{col}} \approx \SI{150}{\cm\squared}$, both $\mathscr{R}$ and $A_{\text{col}}$ still must be increased to make significant improvements to $\text{FOM}_{\text{weak}}$ and hence \textbf{a next-generation grating spectrometer is required for the detection of weak absorption lines in extended galactic halos and the intergalactic medium}. 

A point of reference for $\mathscr{R}$ is the spectral resolving power associated with a thermally-broadened oxygen line: 
\begin{equation}
 \mathscr{R}_D \equiv \frac{\mathcal{E}_{\gamma}}{\Delta \mathcal{E}_D} = \frac{\mathcal{E}_{\gamma}}{\Delta \mathcal{E}_e} \sqrt{\frac{m_{\text{ion}} c_0^2}{2 k_{\mathcal{B}} T_{\text{ion}}}} \gtrapprox 5000 , 
 \end{equation}
where the Doppler width, $\Delta \mathcal{E}_D$, is given by \cref{eq:Doppler_width} using $T_{\text{ion}} \lessapprox \SI{e7}{\kelvin}$ as a typical ion temperature and $m_{\text{ion}} = \SI{2.656e-26}{\kilogram}$ as the mass of an oxygen-16 nucleus [\emph{cf.\@} \cref{tab:astro_element_nuclei_mass}]. 
The \emph{XGS} for \emph{Lynx} aims to achieve this spectral resolving power by maintaining $\Delta \lambda = \abs{n} \lambda / \mathscr{R}_D \lessapprox \SI{0.4}{\pico\metre} = \SI{4}{\milli\angstrom}$ across the soft x-ray bandpass using order numbers $n =$ \numrange{5}{8} for blazed reflection gratings \cite{McEntaffer19} or higher orders using \emph{critical-angle transmission gratings} \cite{Gunther19}. 
Additionally, the instrument baselines $A_{\text{col}} \gtrapprox \SI{4000}{\cm\squared}$ for spectral sensitivity, which requires gratings that perform with $\gtrapprox \SI{40}{\percent}$ \emph{total absolute diffraction efficiency}\footnote{This refers to the sum of absolute diffraction efficiency in all propagating orders with order number $n \neq 0$ [\emph{cf.\@} \cref{sec:grating_tech_intro,sec:measure_efficiency}].} at soft x-ray wavelengths. 
In principle, these specifications are achievable using x-ray reflection grating technology, which is the subject of \cref{sec:grating_tech_dev}. 

With a large $\text{FOM}_{\text{weak}}$ baselined for the \emph{XGS}, this instrument planned for \emph{Lynx} will be able to measure equivalent widths of soft x-ray absorption lines much smaller than what is possible with the \emph{LXM} as well as the grating spectrometers on board \emph{XMM-Newton Newton} or \emph{Chandra}. 
Additionally, a spectral resolving power of $\mathscr{R}_D \gtrapprox 5000$ enables the velocities of ionized outflows that emanate from active galactic nuclei to be measured with precision on the order of tens of \si{\kilo\metre\per\second}. 
This can be seen by noting that the definition of redshift [\emph{cf.\@} \cref{eq:energy_shift}] gives
\begin{subequations}
\begin{equation}\label{eq:wave_shift}
 \frac{\lambda'}{\lambda} \equiv 1 + z \implies z = \frac{\overbrace{\lambda'-\lambda}^{\Delta \lambda}}{\lambda} = \frac{1}{\mathscr{R}}
 \end{equation}
while for a velocity $v \ll c_0$, 
\begin{equation}
 \frac{v}{c_0} = \frac{\left( 1 + z \right)^2 - 1}{\left( 1 + z \right)^2 + 1} \approx z 
 \end{equation}
\end{subequations}
so that the smallest velocity resolved is $v = c_0 / \mathscr{R}_D \lessapprox c_0 / 5000$. 
By design, the \emph{XGS} is expected to be capable of measuring $W_{\lambda}$ for absorption lines on the level of $\SI{0.1}{\pico\metre} = \SI{1}{\milli\angstrom}$ while the \emph{RGS} of \emph{XMM-Newton} is typically used to measure $W_{\lambda}$ on the order of several \si{\pico\metre}, or tens of \si{\milli\angstrom}. 
\emph{Lynx} observation strategies for studies of intergalactic soft x-ray absorption sample plasmas along the line-of-sight of a large number of active galactic nuclei at various cosmological redshifts, $z$. 

As the most prominent soft x-ray absorption lines, the resonance lines from hydrogen-like and helium-like oxygen with rest-frame wavelength centroids $\lambda_{\text{\ion{O}{viii}}} \approx \SI{1.90}{\nm}$ and $\lambda_{\text{\ion{O}{vii}}} \approx \SI{2.16}{\nm}$, respectively, practically place a limit on these values for $z$. 
\begin{figure}
 \centering
 \includegraphics[scale=0.45]{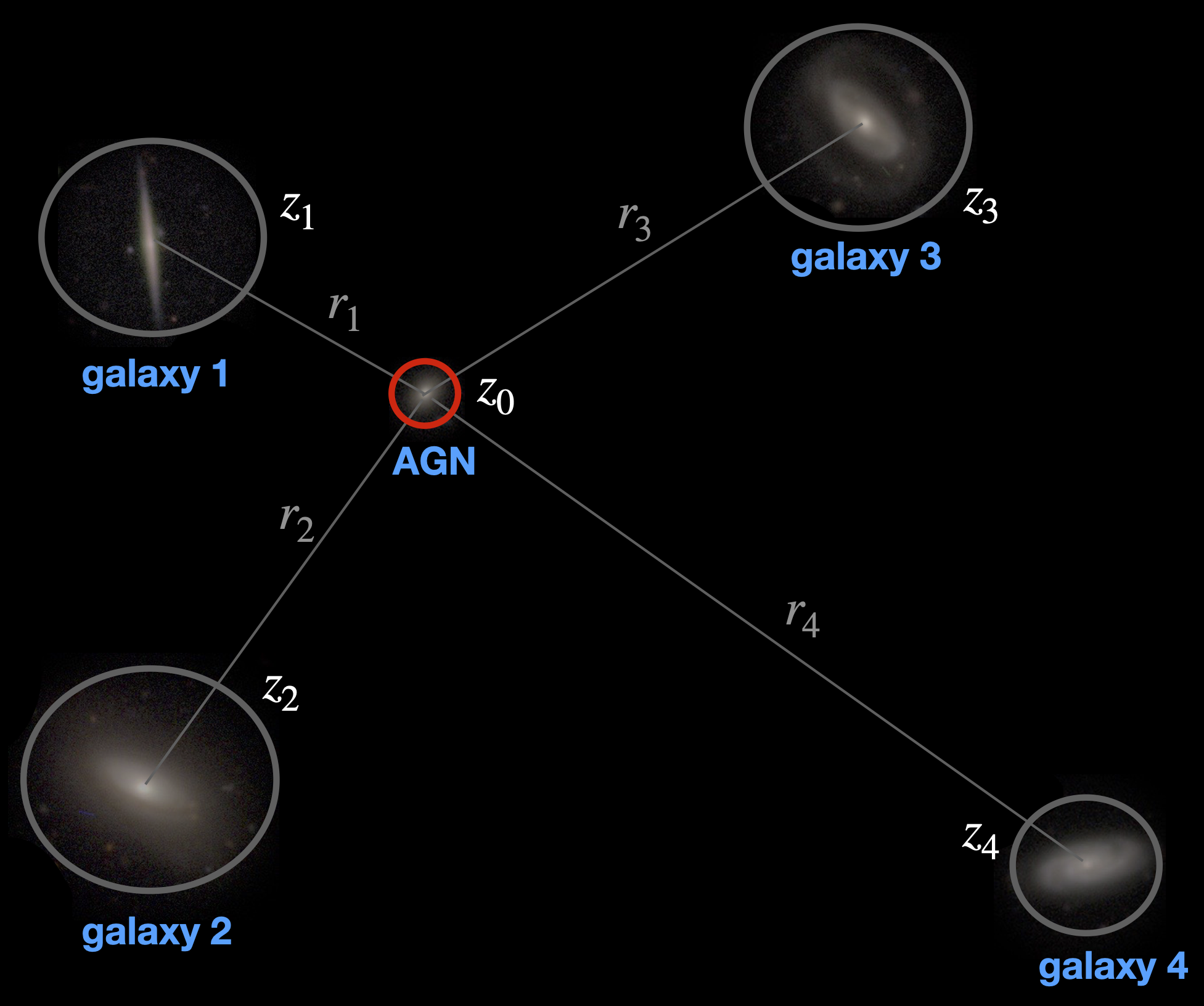} 
 \caption[Cartoon of an active galactic nucleus with four galaxies in the foreground for absorption spectroscopy]{Cartoon of an active galactic nucleus (AGN) at a redshift $z_0$ with four isolated galaxies at smaller redshifts $z_i$ (for $i=$ \numrange{1}{4}) in the foreground. In principle, a spectroscopic observation of the AGN with sufficiently large $\text{FOM}_{\text{weak}} \propto \sqrt{A_{\text{col}} \mathscr{R}}$ [\emph{cf.\@} \cref{eq:FOM_weak}] probes the density of highly-charged ions present in extended galactic halos and intergalactic medium at radial distances $r_i$ (for $i=$ \numrange{1}{4}) from the nuclear region of each galaxy. The result is a spectrum featuring prominent absorption lines from each intervening galaxy with centroids that depend on $z_i$ and equivalent widths that give an indication of plasma density, which in turn depends on $r_i$. The depicted galaxies are optical images obtained from \textsc{Galaxy Zoo}.}\label{fig:galaxy_field}
 \end{figure}
That is, using $\lambda' = (1 + z) \lambda$ [\emph{cf.\@} \cref{eq:wave_shift}] and noting that the red end of the soft x-ray exists at $\lambda \approx \SI{6}{\nm}$, \ion{O}{vii} and \ion{O}{viii} resonance lines are observable out to $z \approx 1.6$ and $z \approx 2$, respectively, where the Universe is observed \numrange{5}{10} billion years in the past. 
Illustrated\footnote{The galaxies depicted in \cref{fig:galaxy_field} are images obtained from \textsc{Galaxy Zoo} \cite{galaxy_zoo,Lintott08} but in this context, they are not intended to represent specific objects on the celestial sphere.} in \cref{fig:galaxy_field}, x-ray continuum emission from a distant active galactic nucleus with relatively high $z$ will generally be absorbed by highly-charged ions in the intergalactic medium and, if the line-of-sight passes near galaxies of lower $z$, circumgalactic, intragroup or intraculster media. 
Thus, by measuring $W_{\lambda}$ for many absorbing systems and correlating their measured redshift with that of intercepting galaxies, the density of extragalactic baryons can effectively be probed as a function of distance from a galactic nucleus and additionally, across cosmic time. 
Overall, these observations serve to track the highly-ionized portion of baryons that cycle through galaxies and are a prime example of a scientific objective that can only be accomplished with a next-generation grating spectrometer.

\section{Development of X-ray Reflection Gratings}\label{sec:grating_tech_dev}
%%%%%%%%%%%%%%%%%%%%%%%%%%%%%%%%%%%%%%%%%--------------------------------------------------
Making spectral observations in the soft x-ray, like in any spectral bandpass, requires channeling radiation according to its \emph{color} and then measuring the relative intensity of each component present to extract a spectrum. 
For UV and other electromagnetic radiation with $\lambda \gtrapprox \SI{100}{\nm}$, this is commonly carried out with wavelength-dispersive instruments in telescopes that utilize diffraction gratings \cite{Loewen97,Ryden10}. 
While x-rays are often measured spectroscopically as energetic particles, the soft x-ray bandpass has $\lambda$ ranging from several \si{\nm} down to about \SI{0.5}{\nm}, where diffraction gratings can be used to obtain sensitive, high-resolution spectra if implemented appropriately. 
Although this has been achieved with the grating spectrometers on board spacecraft observatories \emph{XMM-Newton} \cite{Jansen01,denHerder01,Lumb12} and \emph{Chandra} \cite{Canizares00,Weisskopf00}, these instruments face throughput limitations that result in poor signal-to-noise for distinguishing faint absorption lines from background continua [\emph{cf.\@} \cref{sec:astro_plasmas}]. 
With future x-ray observatories such as \emph{Lynx} \cite{Gaskin19,Lynx_web} calling for improvements in spectral sensitivity and spectral resolving power, $\mathscr{R} = \lambda / \Delta \lambda$, across the soft x-ray bandpass, this section motivates the investigation of new ways to produce highly-efficient reflection gratings with custom groove layouts that are capable of addressing these technological challenges. 

\subsection{General Grating-Design Considerations}\label{sec:grating_tech_intro}
%%%%%%%%%%%%%%%%%%%%%%%%%%%%%%%%%%%%%%%%%--------------------------------------------------
To introduce specifications for reflection gratings that push the state of the art for x-ray spectroscopy, it is useful to consider how an x-ray reflection grating spectrometer compares and contrasts with more conventional grating spectrometers used for astronomical spectroscopy at longer wavelengths. 
For example, in a common reflecting telescope designed for visible light, focusing mirrors are used at near-normal incidence and typically several reflections can occur without significant loss in throughput. 
This enables the implementation of an \emph{echelle spectrometer}, where light passes through an entrance pinhole or slit before being re-focused by various secondary optics so that collimated light is incident on a planar reflection grating with steep, triangular groove facets ruled uniformly over a relatively coarse groove spacing, $d$ \cite{Ryden10,Loewen97}. 
As described in \cref{sec:slit_array}, a low $\lambda / d$ ratio yields a large number of propagating orders while according to the scalar treatment of diffraction efficiency outlined in \cref{sec:sawtooth_principle}, far-field intensity is maximized in the direction of the diffracted angle $\beta = 2 \delta - \alpha$, where $\alpha$ is the incidence angle and $\delta$ is the \emph{blaze angle}, which characterizes the slope of the sawtooth facets. 
Thus, for such an \emph{echelle grating} with large $\delta$, diffraction efficiency is concentrated into high orders on one side of $0^{\text{th}}$ order, which in turn leads to high $\mathscr{R}$, with throughput maximized in a specified bandpass that depends on $\delta$ and $d$ \cite{Loewen97}. 
The result is a set of overlapping spectra, one for each propagating order, that are separated through the use of a second diffraction grating or other dispersive instrument (\emph{e.g.}, a prism) before they are imaged by a detector as an \emph{echellogram} \cite{Ryden10}. 
The extracted spectra in such a situation are often diffraction limited and hence require an echelle grating with a large grooved area to attain high $\mathscr{R}$ [\emph{cf.\@} \cref{sec:resolving_power}]. %  as described in \cref{sec:resolving_power}. 

In a somewhat similar vein to an echelle spectrometer, a traditional x-ray reflection grating spectrometer utilizes focusing mirrors and blazed gratings to disperse radiation collected by the telescope into spectra that are ultimately imaged by a detector at the focal plane. 
However, there are substantial differences that have their roots in the fact that significant x-ray reflection from an optical surface can only be achieved at grazing-incidence angles in the regime of \emph{total external reflection} [\emph{cf.\@} \cref{sec:planar_interface}]. %, which is described from first principles in \cref{sec:planar_interface}. 
With this restriction of grazing incidence, any single mirror or reflection grating appears small in projection to incident radiation; therefore, many nested mirrors and gratings, each with a substantial size, must be stacked and aligned into modules to achieve a sufficient collecting area for spectroscopy, $A_{\text{col}}$ \cite{Cash91,Kahn96,Allured13,Allured15,Donovan18,Shipley08,McEntaffer11}. 
Grazing-incidence focusing in an x-ray telescope is commonly achieved using the \emph{Wolter-I} design \cite{Wolter52}, where nested parabolic mirror segments begin to focus collimated light from an astronomical source before the focal length is reduced significantly by a following set of hyperbolic mirror segments that bring radiation to a focus several meters away \cite{Aschenbach85,schwartz_2011}. 
The fundamental difficulties of x-ray reflection [\emph{cf.\@} \cref{app:x-ray_materials}] combined with the fact that astrophysical spectral signatures of interest in the soft x-ray are quite faint\footnote{\emph{e.g.}, weak absorption lines from extended galactic halos and the intergalactic medium [\emph{cf.\@} \cref{sec:astro_plasmas}]} lead to a practical inability to re-collimate radiation collected by the telescope before it is incident on a grating. 
As an alternative strategy, modular grating arrays can be positioned to intercept the radiation coming to a focus in the telescope while an imaging detector placed at the focal plane is used to record the dispersed spectrum. %spectra?
In this case, order separation is carried out not by a secondary grating or prism as in a typical echelle spectrometer, but rather by the energy-dispersive response of the imaging detector, so that further reflections are not necessary \cite{Cash91}. 
Moreover, without an entrance slit or pinhole, the entire image collected by the Wolter-I optics passes through the spectrometer and as a result, high-resolution spectroscopy with such an instrument is functionally limited to point sources \cite{schwartz_2011}. 
The \emph{RGS} on board \emph{XMM-Newton} is the first space-borne instrument to realize this concept, where \num{182} identical gratings aligned into modular arrays intercept the radiation being focused by Wolter-I optics consisting of \num{58} nested mirror pairs while \emph{charge-coupled device} detectors are used to image the diffraction pattern and to carry out order separation \cite{Kahn96,denHerder01}. 

There exist two classes of grazing-incidence grating geometries that have been pursued for astronomical x-ray spectroscopy. 
This can be gleaned from noting the location of the $n^{\text{th}}$ order diffracting from a grating with a groove spacing $d$ at an arbitrary, oblique incidence angle, which is described by the \emph{generalized grating equation} \cite[\emph{cf.\@} \cref{eq:off-plane_incidence_orders}]{Loewen97}: 
\begin{equation}\label{eq:off-plane_orders}
 \sin \left( \alpha \right) + \sin \left( \beta \right) = \frac{n \lambda}{d \sin \left( \gamma \right)} \quad \text{for } n = 0, \pm 1, \pm 2, \pm 3 \dotsc 
 \end{equation}
As illustrated in the left panel of \cref{fig:conical_reflection_edit} and explained further in \cref{sec:off-plane_geo}, $\gamma$ is the half-opening angle of the cone, $\alpha$ is the azimuthal incidence angle and $\beta$ is the azimuthal diffracted angle of the $n^{\text{th}}$ diffracted order. 
\begin{figure}
 \centering
 \includegraphics[scale=0.165]{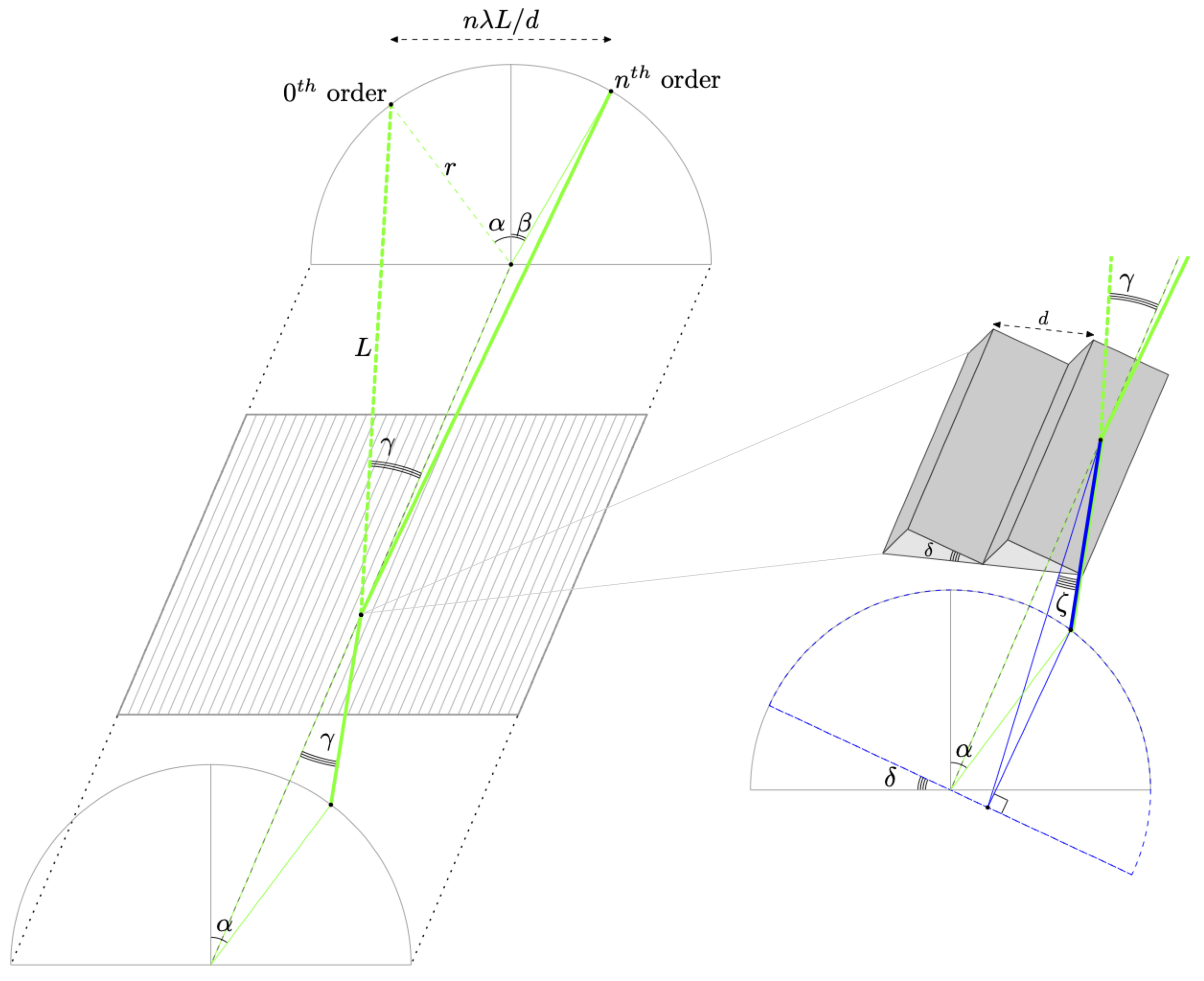}
 \caption[Reflection grating geometry for conical diffraction.]{Reflection grating geometry for conical diffraction. In an extreme off-plane mount, the incoming radiation is nearly parallel to the groove direction such that the half-angle of the cone opening, $\gamma$, is on the order of a degree. Because of this, the angle on the groove facets, $\zeta$, is a grazing-incidence angle for any value of $\alpha$ that illuminates the shallow side of the sawtooth profile with a blaze angle $\delta$. At a distance $L$ away from the point of incidence on the grating, diffracted orders are each separated by a distance $\lambda L / d$ along the direction of grating periodicity (\emph{i.e.}, the grating-dispersion direction), where $d$ is the groove spacing \cite{McCoy18,McCoy20}.}\label{fig:conical_reflection_edit}
 \end{figure} 
At a distance $L$ from the point of incidence on the grating (\emph{i.e.}, the \emph{throw} of the system), each propagating order with $n \neq 0$ is dispersed along the cross-groove direction by a distance from $0^{\text{th}}$ order given by
\begin{equation}\label{eq:linear_dispersion}
 x_n = \frac{n \lambda L}{d} 
 \end{equation}
and confined along the arc of a circle with a radius given by
\begin{equation}\label{eq:arc_radius}
 r = L \sin \left( \gamma \right) .
 \end{equation}
Radiation is incident on the sawtooth facets of a blazed grating at an angle $\zeta$ as illustrated in the right panel of \cref{fig:conical_reflection_edit}, which is related to $\alpha$ and $\gamma$ through the following relation:
\begin{equation}\label{eq:angle_on_groove}
 \sin \left( \zeta \right) = \sin \left( \gamma \right) \cos \left( \delta - \alpha \right) .
 \end{equation} 
Assuming that only the shallow side of the asymmetric, sawtooth-shaped grooves is illuminated, the result, in principle, is that radiation is preferentially diffracted to an angle $\beta = 2 \delta - \alpha$, so that for each propagating order coinciding with this angle, diffraction efficiency is maximized at the \emph{blaze wavelength}, which is given for the $n^{\text{th}}$ diffracted order by \cref{eq:blaze_wavelength_general}:
\begin{subequations}
\begin{equation}\label{eq:blaze_wavelength}
 \lambda_b = \frac{d \sin \left( \gamma \right)}{n} \left[ \sin \left( \alpha \right) + \sin \left( 2 \delta - \alpha \right) \right] .
 \end{equation}
Therefore, to take into account the grazing-incidence requirement that $\zeta$ be a small angle on the groove facets, commonly one of two grating geometries is employed:
\begin{enumerate}[noitemsep]
  \item A grazing-incidence, \emph{in-plane} geometry, where $\sin \left( \gamma \right) = 1$ such that incoming radiation is perpendicular to the groove direction and \cref{eq:angle_on_groove} yields $\zeta = \pi/2 - \alpha + \delta$, necessitating $\alpha$ approaching $90^{\circ}$ and a very shallow $\delta$ 
  \item An extreme \emph{off-plane} geometry, where $\gamma \lessapprox 2^{\circ}$ so that by \cref{eq:angle_on_groove}, $\zeta$ is always small and hence $\alpha$ is free to match $\delta$ in a projected \emph{Littrow configuration} \cite{Loewen97}, where $\alpha = \beta = \delta$ and $\zeta = \gamma$ 
\end{enumerate}
In the former case, a shallow blaze angle leads to diffraction efficiency being maximized in low order and moreover, higher orders with smaller $\beta$ are often vignetted in tightly-packed grating arrays. 
The latter case, on the other hand, allows for high efficiency in high order with appropriate choice of $\delta$, which in turn leads to high $\mathscr{R}$ in a bandpass of interest \cite{Werner77,Cash82,Cash91}. 
This is commonly achieved in a Littrow configuration, where $\alpha = \beta = \delta$ and \cref{eq:blaze_wavelength} reduces to \cref{eq:blaze_wavelength_Litt} \cite{Neviere78b}:
\begin{equation}\label{eq:blaze_wavelength_Littrow}
 \lambda_b = \frac{2 d \sin \left( \gamma \right) \sin \left( \delta \right)}{n} .
 \end{equation}
\end{subequations}
Additionally, an extreme off-plane geometry enables tight packing geometries for modular arrays without vignetting losses thanks to a small cone-opening angle, $2 \gamma$ \cite{Cash91}. 
While the \emph{RGS} on board \emph{XMM-Newton} was designed using a grazing-incidence, in-plane grating geometry, it is for these reasons that much of the recent and current instrument development for future observatories considers the extreme off-plane geometry to make advances in $\mathscr{R}$ and spectral sensitivity \cite{McEntaffer03,McEntaffer04,McEntaffer08b,McEntaffer09,McEntaffer10,McEntaffer11,McEntaffer13,DeRoo16thesis,DeRoo16,Tutt16,Miles18,Donovan18,Miles19,DeRoo20a,Donovan20}. 

Regardless of the geometry employed, reflection gratings in a Wolter-I telescope are used in a converging beam of radiation and this necessitates a \emph{variable-line-space} groove layout that matches the focal length of the telescope to reduce, or potentially eliminate, grating-induced aberrations in the resulting spectrum \cite{Hettrick83}. 
In such a scenario, $\mathscr{R}$ is limited not by the number of grating grooves as is often the case for an echelle spectrometer [\emph{cf.\@} \cref{sec:resolving_power}], but rather by the preservation of the \emph{point spread function} generated by the primary mirrors of the Wolter-I telescope as the converging radiation is diffracted by non-uniform grating grooves. 
With an in-plane geometry employed for the \emph{RGS}, the variable-line-space profile on each grating features parallel grooves that have a gradually decreasing spacing toward the telescope focus with an average of $d \approx \SI{1.5}{\um}$ \cite{Kahn96,denHerder01}. 
This gradient in $d$ serves to counteract the non-negligible change in $L$ that rays pick up as they strike different parts of the grating at grazing incidence in order for a constant linear dispersion [\emph{cf.\@} \cref{eq:linear_dispersion}] to be maintained. 
Meanwhile, each grating is positioned such that radiation is incident on the grating perpendicular to the groove direction and as a result, there is no need for non-parallel grooves. 

On the other hand, the variable-line-space equivalent for an extreme off-plane geometry manifests as a \emph{radial groove profile}, where each groove points toward the telescope focus in a fanned groove pattern \cite{Cash83}. 
This groove layout creates a gradient in $d$ across the grating similar to the in-plane case, but additionally, the tilted grooves in principle ensure that $\alpha$ is constant across each grating in a converging beam of radiation. 
However, the extreme off-plane geometries utilized for soft x-ray spectroscopy typically require $d$ on the order of a few hundred \si{\nm} or less in a Littrow configuration for a moderate blaze angle (\emph{e.g.}, $\delta \approx 30^{\circ}$) while in most cases, $d$ can be larger for grazing-incidence, in-plane geometries with very shallow blaze angles. 
Discussed further in \cref{sec:off-plane_geo}, this can be gleaned from \cref{eq:off-plane_orders}, which reduces to \cref{eq:normal_incidence_orders} for an in-plane geometry: 
\begin{subequations}
\begin{equation}\label{eq:in-plane_orders_ch1}
 \sin \left( \alpha \right) + \sin \left( \beta \right) = \frac{n \lambda}{d} \quad \text{for } n = 0, \pm 1, \pm 2, \pm 3 \dotsc ,
 \end{equation}
where both $\alpha$ and $\beta = 2 \delta - \alpha$ are large for grazing-incidence diffraction, but with opposite signs, so that for small $\delta$ and $\alpha \lessapprox 90^{\circ}$, \cref{eq:in-plane_orders_ch1} can be written approximately as 
\begin{equation}\label{eq:inplane_limit}
 2 \delta \left( \frac{\pi}{2} - \alpha \right) \approx \frac{n \lambda}{d} \quad \text{for } n = 0, \pm 1, \pm 2, \pm 3 \dotsc ,
 \end{equation}
\end{subequations}
which indicates $d \gg \lambda$ for small $\abs{n}$. 
In contrast, a grazing-incidence, off-plane geometry with small $\gamma$ and $\alpha = \beta = \delta$ has \cref{eq:off-plane_orders} approximated as 
\begin{equation}\label{eq:offplane_limit}
 2 \gamma \sin \left( \delta \right) \approx \frac{n \lambda}{d} \quad \text{for } n = 0, \pm 1, \pm 2, \pm 3 \dotsc ,
 \end{equation}
which shows that $d$ is still much larger than $\lambda$, but by a smaller margin for moderate values of $\delta$. 

The demand for variable-line-space profiles that match specific telescope focal lengths (to enable high $\mathscr{R}$) combined with the need to control $d$ and $\delta$ (to maximize diffraction efficiency near $\lambda_b$ for each order) has the consequence that \textbf{custom diffraction gratings are needed for x-ray spectroscopy}. 
The gratings used for the \emph{RGS} are gold-coated replicas of a master grating fabricated by \emph{mechanical ruling engine}, where a sculpted diamond tip burnishes one groove at a time into a layer of soft metal coated on a substrate \cite{Kahn96,denHerder01}. 
With heritage dating back to the early $19^{\text{th}}$ century \cite{Fraunhofer1823,Compton25}, the mechanical ruling process gives rise to a sawtooth-like topography, where the effective blaze angle, $\delta$, depends directly on the shape of ruling tip. 
Owing to the development of mechanical ruling engines with interferometric feedback control \cite{Harrison49,Harrison55}, nanoscale groove placement precision is possible \cite{Loewen97}. 
However, while many gratings used for near-visible wavelengths are also commonly fabricated by mechanical ruling, gratings generally are not custom-made for specific instruments due to the high cost and difficulties\footnote{In particular, environmental stabilities and tip wear must be kept under control over the course of the days, weeks or months of ruling time that may be required for gratings with relatively large areas.} associated with the ruling process; instead, spectrometers are often designed around standardized gratings provided by manufacturers. 
Additionally, surface-topography constraints are imposed by the shape of the ruling tip and hence commercial gratings are commonly ruled only with certain blaze angles, groove spacings and sizes \cite{Loewen97}. 
With specifications for radially-ruled, off-plane gratings departing significantly from standard gratings fabricated by mechanical ruling engine, realizing such an idealized groove profile for gratings with relatively large areas and controllable blaze angles requires identifying appropriate fabrication processes suitable for generating custom groove architectures that can be replicated to produce a large number of gratings. 
This motivates the investigation of techniques in the area of nanofabrication to produce new reflection gratings that each perform with high diffraction efficiency while maintaining an ability to realize a custom, variable-line-space profile that matches the focal length of a specific Wolter-I telescope. 

\subsection{Grating Fabrication}\label{sec:grating_fab_intro}
%%%%%%%%%%%%%%%%%%%%%%%%%%%%%%%%%%%%%%%%%--------------------------------------------------
Fabrication methods for x-ray reflection gratings have evolved significantly since the 1990s, when the gratings for the \emph{RGS} were manufactured by mechanical ruling engine and a replication process that involves casting replicas in a synthetic resin \cite{Kahn96,denHerder01}. 
This subsection outlines the main efforts for fabricating x-ray reflection gratings and motivates the investigation of new techniques that push the state of the art. 
The focus here is on reviewing techniques in maskless lithography coupled with supporting etching processes that are able to produce surface reliefs for blazed gratings. %surface-relief profiles
In particular, \cref{sec:holographic} discusses \emph{holographic gratings} while \cref{sec:binary_ebeam} introduces \emph{electron-beam lithography} as a method of grating manufacture.  
Moreover, \cref{sec:crystal_etching} describes how \emph{wet anisotropic etching} can be used to generate a sawtooth in mono-crystalline silicon while \cref{sec:nanoimprint} discusses \emph{nanoimprint lithography} for the replication of gratings. 
Considerations for metallic overcoats on gratings that enable high, broadband reflectivity at soft x-ray wavelengths are addressed in \cref{app:x-ray_materials}. 

\subsubsection{Holographic Recording}\label{sec:holographic}
%%%%%%%%%%%%%%%%%%%%%%%%%%%%%%%%%%%%%%%%%--------------------------------------------------
One of the fabrication techniques most pursued for x-ray reflection gratings \cite{McEntaffer03,McEntaffer04,Osterman04,Miles15,Tutt16} and other grating technology throughout the years is \emph{interference lithography}, also known as \emph{holographic recording}, where the interference pattern produced by two or more coherent light sources is recorded in a photo-sensitive material known as a \emph{photoresist} \cite{Loewen97,Hennessy11,Beesley70,Lu10}. 
In a classical holographic recording process, a layer of photoresist material is deposited on a substrate while layers of anti-reflective coatings are commonly required to prevent reflections at interfaces from occurring. 
\begin{figure} 
\centering
\includegraphics[scale=1]{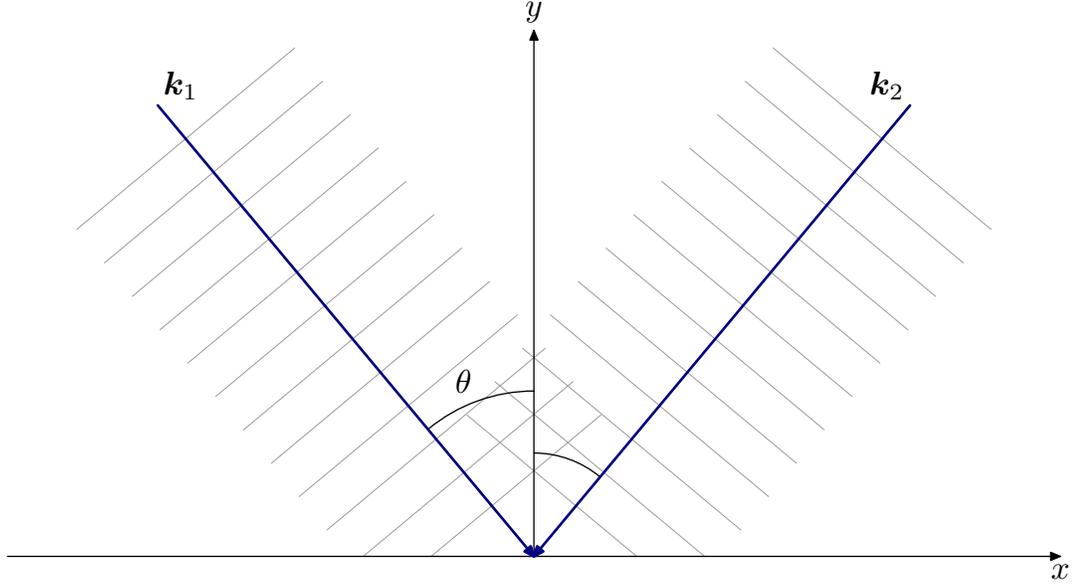}
\caption[Geometry for two-beam holographic recording]{Geometry for two-beam holographic recording with wave vectors $\mathbold{k}_1$ and $\mathbold{k}_2$ defined by \cref{eq:holographic_vectors}.}\label{fig:holographic} 
\end{figure}
The photoresist is then the recording medium for two offset laser sources, which can be approximated as monochromatic plane waves of wavelength $\lambda_0$ with wave vectors given by [\emph{cf.\@} \cref{fig:holographic}] 
\begin{subequations}
\begin{equation}\label{eq:holographic_vectors}
 \mathbold{k}_1 = \frac{2 \pi}{\lambda_0} \left[ \sin (\theta) \mathbold{\hat{x}} - \cos (\theta) \mathbold{\hat{y}} \right] \quad \text{and} \quad \mathbold{k}_2 = - \frac{2 \pi}{\lambda_0} \left[ \sin (\theta) \mathbold{\hat{x}} + \cos (\theta) \mathbold{\hat{y}} \right] .
 \end{equation}
The resulting interference pattern projected onto a surface parallel to the $x$-axis and orthogonal to the $y$-axis is sinusoidal, which can be seen by considering the superposition of the time-harmonic, scalar fields associated with each laser source as a function of $x$ for $y=0$. 
Assuming identical, ideal sources and defining $k_0 \equiv 2 \pi / \lambda_0$, the total scalar field representative of the interference pattern, $u(x)$, is proportional to 
\begin{equation}\label{eq:holographic_interfere}
 \left( \mathrm{e}^{i \mathbold{k}_1 \cdot \mathbold{r}} + \mathrm{e}^{i \mathbold{k}_2 \cdot \mathbold{r}} \right)\Big|_{y=0} = 
 \mathrm{e}^{i k_0 \sin \left( \theta \right) x} + \mathrm{e}^{-i k_0 \sin \left( \theta \right) x} = 2 \cos \left[ k_0 \sin \left( \theta \right) x \right] .
 \end{equation}
\end{subequations}

Using \cref{eq:holographic_interfere}, the intensity of radiation exposing the photoresist depends on $\norm{u(x)}^2$ and is proportional to 
\begin{subequations}
\begin{equation}
 \cos^2 \left[ k_0 \sin \left( \theta \right) x \right] = \frac{1}{2} \left( 1 + \cos \underbrace{\left[ 2 k_0 \sin \left( \theta \right) x \right]}_{K x} \right) ,
 \end{equation}
where $K \equiv 2 \pi / d$ is the \emph{grating wave number} [\emph{cf.\@} \cref{eq:grating_number}] of the projected interference pattern. 
Therefore, the nominal groove spacing of a grating fabricated by a two-beam holographic recording process is \cite{Loewen97}
\begin{equation}\label{holographic_groove_spacing}
 d = \frac{\lambda_0}{2 \sin (\theta)} .
 \end{equation} 
%is the groove spacing of a grating fabricated by a two-beam holographic recording process. 
\end{subequations}
In a manner similar in concept to film photography, this intensity pattern is recorded as the radiation affects the chemistry of the photoresist (\emph{e.g.}, via photogenerated acids) such that following wet development, the projected pattern is left in the exposed film \cite{Hennessy11,Beesley70}. 
More complex patterns, for example those with hexagonal and rectangular symmetry, can be realized with the inclusion of more than two recording beams \cite{deBoor09,Lu10,Burrow11,Zhurminsky19}. 
While interference lithography comes with its own set of challenges relating to environmental stabilities and optical control, it allows for better groove placement precision and potentially shorter manufacturing times as compared to mechanical ruling owing to the parallel, rather than serial, nature of the patterning method \cite{Loewen97}. 

Due to the sinusoidal topography generated by two-beam recording, a reflection grating fabricated by such a process, in principle, functions as a sinusoidal phase grating, where diffraction efficiency is expected to be distributed among low orders on both sides of $0^{\text{th}}$ order [\emph{cf.\@} \cref{sec:sinusoid_principle}]. 
Sinusoidal reflection gratings replicated from a holographic grating master fabricated by \textsc{HORIBA Jobin Yvon Inc.\@} \cite{Horiba_web} were used in an extreme off-plane mount for a series of sounding-rocket experiments for extended source spectroscopy of supernova remnants \cite{McEntaffer04b,McEntaffer05,McEntaffer06,McEntaffer07,McEntaffer08,Oakley09,Oakley10,Oakley11a,Oakley11b,Zeiger11,Oakley13,Zeiger13,Rogers13,Rogers15,Rogers16}. %\cite{McEntaffer03,McEntaffer04,Osterman04} 
Developed at the Center for Astrophysics \& Space Astronomy at the University of Colorado \cite{colorado_casa} and in part at the University of Iowa, Department of Physics \& Astronomy \cite{iowa_physics,McEntaffer09b}, each of these payload-integrated grating spectrometers featured an array of these sinusoidal gratings that intercepted radiation brought to a focus by a wire-grid collimator so that spectra could be recorded by \emph{gaseous electron-multiplier} detectors and telemetered to a ground station \cite{Rogers15,Rogers16,McCoy_telemetry,Rogers17,Rogers20}. 
\begin{figure}
 \centering
 \includegraphics[scale=0.7]{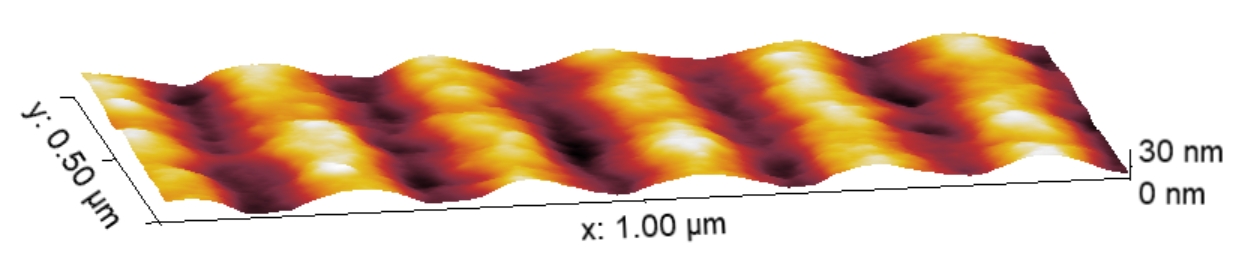}
 \caption[Atomic force micrograph (AFM) of a sinusoidal grating used for sounding-rocket experiments]{AFM of a sinusoidal grating used for sounding-rocket experiments.}\label{fig:OGRESS_AFM}
 \end{figure}
An \emph{atomic force micrograph (AFM)} of one these gratings, which has a groove density of \num{5670} grooves per \si{\mm} ($d \approx \SI{176.4}{\nm}$) and a nickel coat for soft x-ray reflectivity, is shown in \cref{fig:OGRESS_AFM}.\footnote{This image was taken at the Penn State Materials Characterization Laboratory (MCL) \cite{PSU_MRI_CL} using a \textsc{Bruker Dimension Icon}$^{\text{TM}}$ AFM under \textsc{PeakForce Tapping}$^{\text{TM}}$ mode \cite{Xu18}.} 

Holographic gratings ideally would be blazed to sawtooth profile to maximize diffraction efficiency for a particular diffracted angle [\emph{cf.\@} \cref{eq:blaze_wavelength}]. 
A topography that approximates a blazed profile can often be fabricated through holographic recording by tilting the substrate during exposure \cite{Loewen97}. 
Alternatively, a typical holographic grating can be blazed to a sawtooth-like profile by ablating one side of the sinusoidal grooves through directional ion etching to achieve a quasi-blaze profile in the photoresist or the underlying substrate \cite{Loewen97,Aoyagi76,Palmer95}. 
Although this type of grating has been shown experimentally to exhibit a blaze effect in an extreme off-plane mount, they have also been found to perform with lower overall diffraction efficiency in the soft x-ray spectrum than their parent sinusoidal gratings \cite{McEntaffer04,Miles15,Tutt16}. 
This is likely a result of there being extra surface roughness induced by the ion milling process that causes absorption and non-specular scatter [\emph{cf.\@} \cref{sec:rough_surface}]. 

Interference lithography can also be used to define a groove layout in a photoresist for crystallographic etching in silicon [\emph{cf.\@} \cref{sec:crystal_etching}] produces smooth, triangular groove facets \cite{Franke97,Chang03}. 
There are, however, fundamental aspects of this lithographic process that prevent an idealized x-ray reflection grating from being obtained. 
While gratings with $d \lessapprox \SI{100}{\nm}$ can be fabricated using interference lithography in some cases \cite[\emph{cf.\@} \cref{sec:summary_ch5}]{Chang08}, the wavelength of the recording laser, $\lambda_0$, often practically limits $d$ to a few hundred \si{\nm} [\emph{cf.\@} \cref{holographic_groove_spacing}]. 
More importantly, however, a true radial profile for off-plane gratings cannot be produced using this technique because curved grooves are produced when the recording lasers are offset rather than the targeted design, where grooves are straight and continuous but fan outward like spokes on a bicycle wheel \cite{Cash83}. 
With all of this considered, it is worthwhile to explore the capabilities of other techniques in the realm of nanofabrication to find methodology better suited for the task. 

\subsubsection{Electron-Beam Lithography}\label{sec:binary_ebeam}
%%%%%%%%%%%%%%%%%%%%%%%%%%%%%%%%%%%%%%%%%--------------------------------------------------
Direct-write patterning utilizing finely-focused beams of energetic electrons, ions or photons is commonly used in a variety of applications for fabricating photomasks or defining custom layouts in resist that serve as a template for subsequent etching processes. 
Among these techniques, \emph{electron-beam lithography (EBL)} \cite{Chen15,Mohammad2012,Wu10} has been pursued the most for x-ray reflection grating technology due to its ability to pattern custom groove layouts in a resist film with sub-\si{nm} precision combined with its widespread use in nanofabrication facilities and mature stages of technological development. 
Much of the recent work in this area \cite{McCoy18,Miles18,McCoy20,McCurdy20,DeRoo20a,DeRoo20b} has taken place at the Nanofabrication Laboratory of the Penn State Materials Research Institute using an \textsc{EBPG5200} tool for EBL built by \textsc{Raith Nanofabrication} \cite[\emph{cf.\@} \cref{fig:ebeam_tool_pic}]{raith_ebpg,PSU_MRI_nanofab,PSU_MRI_EBL}. 
\begin{figure}
 \centering
 \includegraphics[scale=0.057]{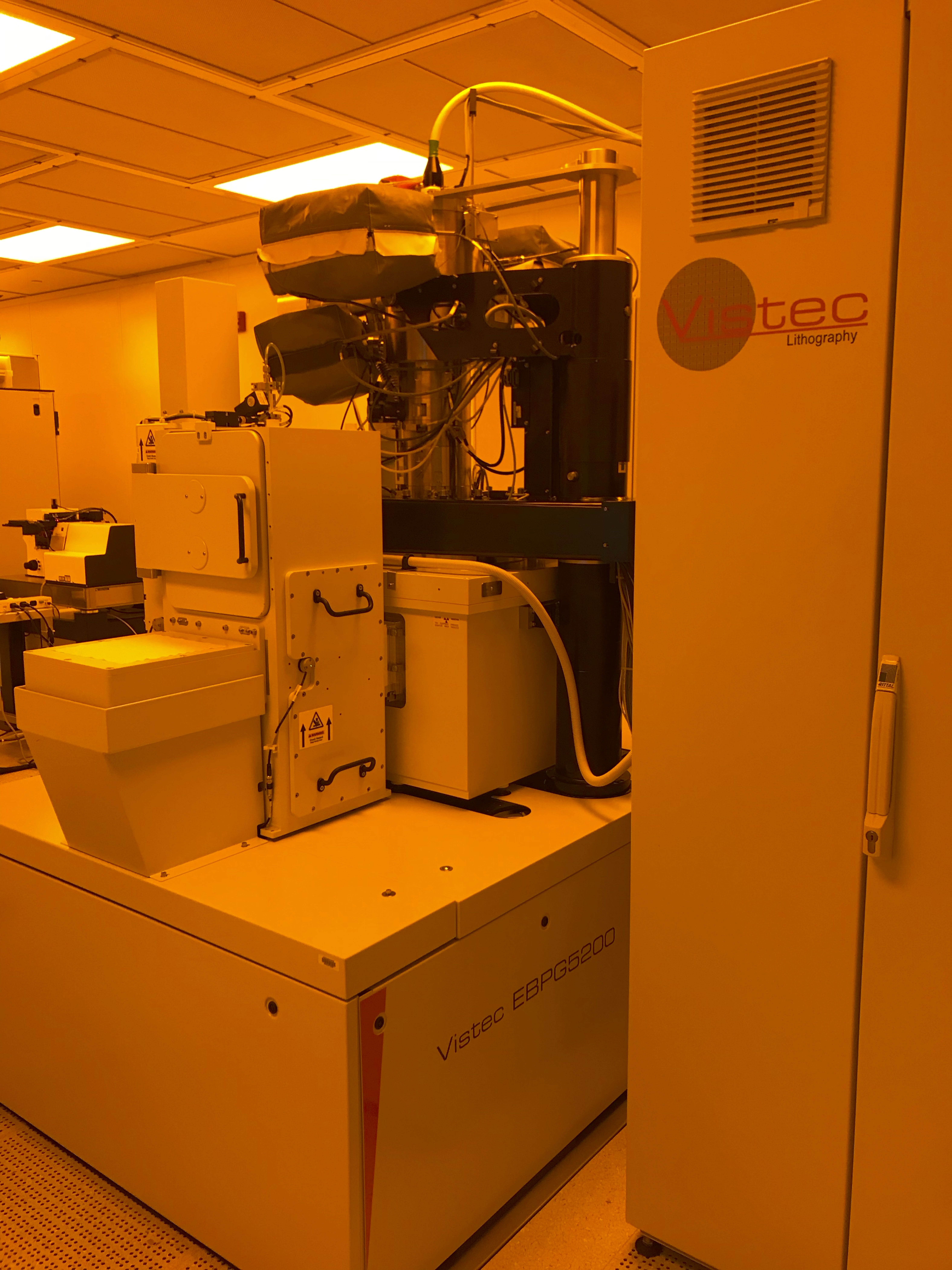}
 \caption[The \textsc{EBPG5200} electron-beam lithography (EBL) system installed at the Penn State Nanofabrication Laboratory]{The \textsc{EBPG5200} electron-beam lithography system installed at the Penn State Nanofabrication Laboratory \cite{PSU_MRI_EBL}.}\label{fig:ebeam_tool_pic}
 \end{figure}
Similar to other common EBL tools, the \textsc{EBPG5200} integrates a thermally-assisted, field-emission electron source with a system of electromagnetic lenses to focus a high-quality beam of electrons to a very small spot size \cite{Nakazawa88,Wu10}.  
This occurs under high vacuum to reduce the blurring effect of electrons scattering from gas molecules and moreover, small electric currents and large accelerating voltages generally are required to overcome the mutual electrostatic repulsion of the electrons to achieve a small spot size. 

The \textsc{EBPG5200} in particular uses a \SI{100}{\kilo\volt} accelerating voltage to enable an electron-beam with current measured in nanoamps (\si{\nano\ampere}) to be focused with a diameter less than \SI{10}{\nm} under optimized conditions \cite{PSU_MRI_EBL}. 
Meanwhile, computer-controlled electrostatic deflectors and beam-blanking electrodes in the tool serve to provide lithographic control of the beam so that a groove layout can be patterned serially in a somewhat similar manner to the process of mechanical ruling, but without the physical burnishing process \cite{Chen15,Mohammad2012,Wu10}. 
Direct-write patterning by beam deflection, however, is accessible only up to areas of \SI{1}{\mm} by \SI{1}{\mm} or less in most EBL tools, including the \textsc{EBPG5200}. 
A consequence of this is that for x-ray reflection gratings, which generally have patterned areas greatly exceeding \SI{1}{\mm\squared}, stage motion is required for a large number of these \emph{write fields} to be stitched together. 
This must be carried accurately because misalignments between these patterned areas lead to groove spacing error, and in principle, this can contribute to a degradation in $\mathscr{R}$ \cite{Dougherty01,DeRoo20b}. 
While this process enables the creation of custom, high-precision layouts for x-ray reflection gratings, it also leads to long manufacturing times and hence high laboratory costs for gratings of substantial size. 
It is for this primary reason that EBL is practically only used for the fabrication of a master grating while nanoimprint lithography [\emph{cf.\@} \cref{sec:nanoimprint}] is used to produce replicas for a spectrometer. 
%It is for this primary reason that EBL is practically only used for the fabrication of a master grating while techniques in nanoimprint lithography are used to produce replicas for a grating spectrometer [\emph{cf.\@} \cref{sec:nanoimprint}]. 

As alluded to in \cref{sec:holographic}, pattern resolution in holographic recording is limited fundamentally by the wavelength of the radiation used for lithographic exposure, $\lambda_0$. 
In contrast, pattern feature sizes in EBL are limited not by diffraction but rather by other factors such as the accuracy of the tool, the stability and size of the focused beam and additionally, the nature of the resist material \cite{Vieu00,Broers96,Manfrinato14}. 
This can be understood by considering the effective wavelength of a hypothetical mono-energetic beam of electrons, which is calculated by equating the relativistic kinetic energy of an electron to the change in electrostatic energy provided by a potential difference, $V_0$, that represents the accelerating voltage in an EBL tool \cite{Landau75}: 
\begin{subequations}
\begin{equation}\label{eq:rel_KE_voltage_balance}
 m_e c_0^2 \left( \frac{1}{\sqrt{1-\frac{v^2}{c_0^2}}} - 1 \right) = q_e V_0 ,
 \end{equation}
where $m_e$ is the electron rest mass, $c_0$ is the speed of light, and $q_e$ is the elementary charge, while $v$ is the speed of the electron. 
Solving \cref{eq:rel_KE_voltage_balance} for $v$ shows that an electron accelerated by a voltage of $\SI{100}{\kilo\volt}$ in the \textsc{EBPG5200} is relativistic: 
\begin{equation}
 v = c_0 \sqrt{1 - \left( \frac{1}{1 + \frac{q_e V_0}{m_e c_0^2}} \right)^2} \approx \num{0.548} c_0 \quad \text{for } V_0 = \SI{100}{\kilo\volt} .
 \end{equation}
Moreover, its relativistic momentum is given by 
\begin{equation}
 p = \frac{m_e v}{\sqrt{1 - \frac{v^2}{c_0^2}}} = m_e c_0 \sqrt{\left( 1 + \frac{q_e V_0}{m_e c_0^2} \right)^2 - 1} 
 \end{equation}
and then the \emph{de Broglie wavelength} for an electron, determined from 
\begin{equation}\label{eq:electron_wavelength}
 \lambda_e = \frac{h}{p} = \frac{\lambda_C}{\sqrt{\left( 1 + \frac{q_e V_0}{m_e c_0^2} \right)^2 - 1}} \quad \text{with } \lambda_C \equiv h / m_e c_0 \approx \SI{2.43}{\pm} ,
 \end{equation}
shows that an electron accelerated to \SI{100}{\kilo\electronvolt} has an effective wavelength of about \SI{3.7}{\pm}, indicating that from the standpoint of scalar diffraction, a mono-energetic beam of electrons can be thought of as being analogous to a mono-chromatic beam of electromagnetic radiation but with a much smaller wavelength. 
\end{subequations}
In practice, field-emission electron sources in modern EBL tools are characteristic of energy dispersion on a sub-\si{\electronvolt} level; this low-energy dispersion enables an electron-beam with electric current measured in \si{\nano\ampere} to be very finely focused by electromagnetic lenses \cite{Pala16}.

In a common EBL process for grating manufacture, a silicon wafer or a specialized optical flat is coated with a film of resist that acts as the recording medium for the electron beam \cite{Chen15,Mohammad2012,Wu10}. 
This film is typically a layer of polymers or copolymers that constitute a solid, predominantly amorphous, material network with a thickness of a few hundred \si{\nm} or more. 
A commonly used material is \emph{poly(methyl methacrylate)} (PMMA), which is a thermoplastic polymer composed of many repeating units of the monomer \emph{methyl methacrylate} (MMA; \ce{C5H8O2}) \cite{Hatzakis69,Greeneich75,Broers78,Ali15}. 
\begin{figure}
 \centering
 \includegraphics[scale=0.4]{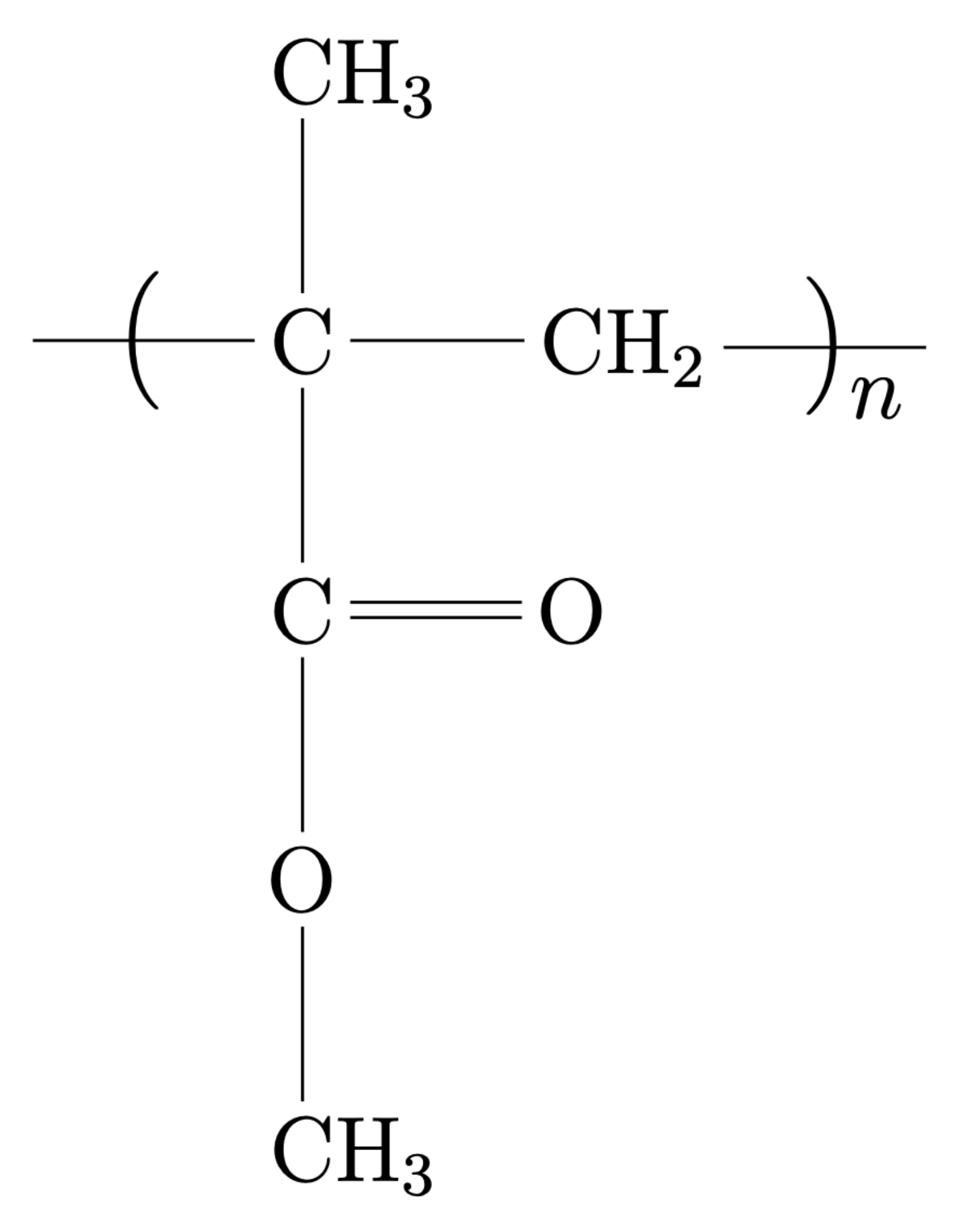}
 \caption[Structural formula of poly(methyl methacrylate) (PMMA).]{Structural formula of poly(methyl methacrylate) (PMMA). The average molecular weight, $M_w$, of PMMA resist depends directly on the length of these polymer chains, which is indicated by the degree of polymerization, $n$, that represents the number of monomers bonded together.}\label{fig:PMMA}
 \end{figure}
The structural formula for this macromolecule is drawn in \cref{fig:PMMA}, where MMA monomers with molar mass $M_0 \approx \SI{100}{\gram\per\mole}$ are covalently bonded together at the sites marked by brackets to form a linear polymer chain. 
In practice, synthesized PMMA contains a spread of polymer chain lengths; the number-averaged molecular mass is defined as 
\begin{subequations}
\begin{equation}
 M_n \equiv \frac{\sum_j N_j M_j}{\sum_j N_j} ,
 \end{equation}
where $N_j$ is the number of PMMA molecules of mass $M_j$ so that the average number of monomers bonded together in polymer chains is $M_n / M_0$, which is known as the \emph{degree of polymerization}. 
However, resist molecular weight provided by chemical manufacturers is usually a mass-averaged molar mass: 
\begin{equation}\label{eq:weight_avg_molecular_weight}
 M_w \equiv \frac{\sum_j N_j M_j^2}{\sum_j N_j M_j} ,
 \end{equation}
\end{subequations}
and this quantity is related to $M_n$ through the \emph{dispersity} of the material defined as $\DJ \equiv M_w / M_n \geq 1$, which is a measure of the polymer chain length distribution in the resist with $\DJ = 1$ corresponding to a uniform distribution \cite{Stepto09,Kirchner19}. %change $\text{PDI} = 1$ to $\DJ = 1$
A thin film of glass-state PMMA with a known value of $M_w$ is typically achieved by spin-coating a substrate with a solution of the polymer cast in a solvent that is removed with a hotplate bake [\emph{cf.\@} \cref{sec:TASTE}]. 

The electron-beam recording mechanism in PMMA and other positive-tone resists such as ZEP520A\footnote{This material is a copolymer of \emph{$\alpha$-chloromethacrylate} and \emph{$\alpha$-methylstyrene} with the chemical formula \ce{C13H6O2Cl} \cite{Czaplewski11}. Available from \textsc{Zeon Chemicals L.P.}, it is a thermoplastic resist similar to PMMA, but with improved dry etch selectivity and other differing parameters.} starts with polymer chain scission by high-energy electrons, which serves to lower $M_w$ in the resist locally \cite{Dobisz00,Greeneich75}. 
Lithographic electrons, such as those accelerated to \SI{100}{\kilo\electronvolt} in the \textsc{EBPG5200}, are energetic enough to forward-scatter through relatively thin resist layers with negligible intensity loss so that $M_w$ can be considered uniform throughout the depth of the exposed film \cite{Schleunitz10,Kirchner19}. 
The degree to which $M_w$ is reduced essentially depends on the number of electrons available for polymer chain scission in an exposed area, which is quantified by an electron dose, $D$ (usually given in units of \si{\micro\coulomb\per\cm\squared}). 
However, in addition to breaking polymer chains into fragments as they forward-scatter through a layer of resist, lithographic electrons back-scatter through the substrate, causing the resist to be inadvertently dosed over distances on the order of \SI{10}{\um} through the phenomenon known as the \emph{proximity effect} in EBL \cite{Pavkovich86,Wu10}. 
Algorithms for proximity effect correction, such as those provided by the \textsc{Layout BEAMER} software package (\textsc{GenISys GmbH}) \cite{beamer}, can be used to determine how $D$ should be adjusted across a layout to achieve a desired pattern in the resist. 
In any case, exposed PMMA that is fragmented to a sufficiently low $M_w$ is soluble in a suitable wet developer, such as a mixture of \emph{methyl isobutyl ketone} (MIBK; \ce{C6H12O}) and \emph{isopropyl alcohol} (IPA; \ce{C3H8O}), while unexposed resist with relatively large $M_w$ is virtually unaffected \cite{miller-chou03,Pratt75}. 
This enables PMMA with $M_w$ on the order of hundreds of \si{\kilogram\per\mole} to be used as a positive-tone resist in EBL, where, with a appropriate choice for $D$ and wet development parameters, exposed resist can be etched down to the substrate during wet development while the unexposed resist remains intact \cite{Greeneich75,Vutova01,Schift10,Kirchner19}. 

The end result of a standard EBL process for grating manufacture is a bi-level topography featuring lines and spaces of remaining resist and cleared substrate that define a groove spacing, $d$. 
For such a pattern to function as reflection grating, however, the layout is usually transferred into a more robust material such as an underlying substrate, which is illustrated schematically for a simplified fabrication process in \cref{fig:laminar_fab} \cite{Zeitner12}. 
\begin{figure}
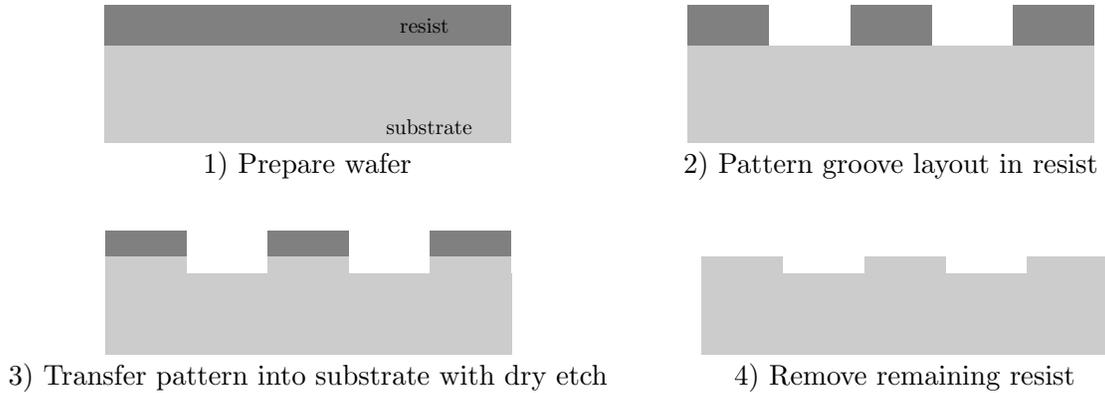
 
\centering
\includegraphics[scale=1.08]{Chapter-1/Figures/laminar_fab1.mps}
\includegraphics[scale=1.08]{Chapter-1/Figures/laminar_fab2.mps}
\includegraphics[scale=1.08]{Chapter-1/Figures/laminar_fab3.mps}
\includegraphics[scale=1.08]{Chapter-1/Figures/laminar_fab4.mps}
\caption[Outline of a simplified laminar grating fabrication recipe that uses EBL]{Outline of a simplified laminar grating fabrication recipe that uses EBL to define a groove layout.}\label{fig:laminar_fab}
\end{figure}
An AFM\footnote{As in \cref{fig:OGRESS_AFM}, this image was taken at the Penn State MCL \cite{PSU_MRI_CL} using a \textsc{Bruker Dimension Icon}$^{\text{TM}}$ AFM under \textsc{PeakForce Tapping}$^{\text{TM}}$ mode \cite{Xu18}.} of such a grating pattern is shown in \cref{fig:laminar_AFM}. 
\begin{figure} 
 \centering
 \includegraphics[scale=0.5]{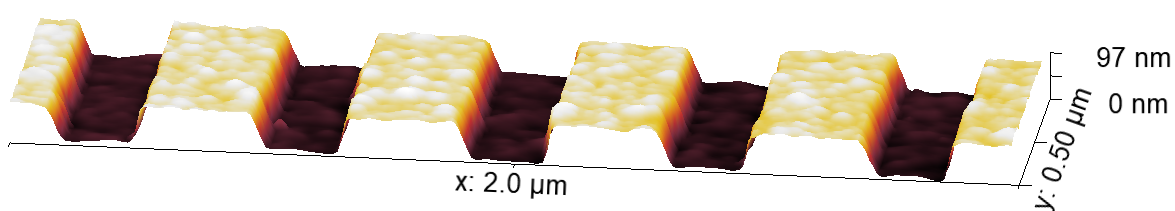}
 \caption[AFM of a laminar grating dry etched in amorphous silicon]{AFM of a laminar grating dry etched in amorphous silicon.}\label{fig:laminar_AFM}
 \end{figure}
This grating with $d = \SI{400}{\nm}$ was fabricated by staff at the Penn State Nanofabrication Laboratory \cite{PSU_MRI_nanofab} by using a dry-etch process (\emph{i.e.}, reactive ion etching)\footnote{Briefly, a \emph{dry etch} process uses a plasma driven by a radio-frequency electromagnetic field to etch a surface through the bombardment of ions. Due to the directional nature of the ions, such a process is anisotropic in the sense material is etched downward. However, \emph{reactive ion etching (RIE)} has the added component that ions react chemically to some degree with the surface and this introduces an isotropic component to the etch \cite{Franssila10}.\label{footnote:dry_etch_RIE}} to transfer the resist into a layer of amorphous silicon (not shown in \cref{fig:laminar_fab}) coated on a silicon wafer in a manner similar to the grating tested for soft x-ray spectral resolving power by DeRoo, et~al.~\cite{DeRoo20a}. 

The diffraction-efficiency behavior of a laminar grating depends on the angles of incidence, but in any case, the result is that certain propagating orders in principle are suppressed and hence a true blaze response is not achieved. %\footnote{Depending on the dry-etch selectivity of the resist used for EBL, an underlying layer of a more robust material (\emph{e.g.}, a silicon nitride footnote~\ref{footnote:nitride_mask}]) may be required to serve as a mask for the silicon etch.}
For angles of incidence close to $\alpha = 0$ in \cref{fig:conical_reflection_edit}, the scenario can be described approximately as a binary phase grating [\emph{cf.\@} \cref{eq:square_principle}], where radiation picks up a phase shift as it traverses the depth of the grating grooves; the result, according to this scalar treatment of diffraction, is a \emph{sinc-squared} modulation of the far-field intensity that depends on both the groove depth and the grating duty cycle. 
On the other hand, a large $\alpha$ in an extreme off-plane mount preferentially illuminates the sidewall of the laminar groove facets, which can lead to high efficiency in high order similar to a blazed grating, but still not without the suppression of other propagating orders. 
A sawtooth topography with a sharp apex and a well-defined blaze angle, $\delta$, is instead required to achieve high efficiency over a range of orders and to better customize the grating for a bandpass of interest. 
However, a laminar grating etched in silicon can be used as a template for ion milling, which ablates these facets at an angle to create a sawtooth-like topography \cite{Johnson79,Liu13}. 
While this is the subject of ongoing research at Penn State \cite{Miles2019_ion_mill}, an alternative, well-established strategy to achieving a grating blaze for a groove layout defined by EBL is crystallographic etching, which is described next. 

\subsubsection{Crystallographic Etching in Silicon}\label{sec:crystal_etching}
%%%%%%%%%%%%%%%%%%%%%%%%%%%%%%%%%%%%%%%%%--------------------------------------------------
A groove layout in a resist film patterned by virtually any lithography can be used to define a mask for anisotropic wet etching in mono-crystalline silicon, typically with a solution of \emph{potassium hydroxide} (\ce{KOH}), to achieve high-fidelity, sawtooth-shaped groove facets. 
This technique has been pursued previously to fabricate blazed reflection gratings for x-ray telescope applications \cite{Franke97,Chang03,Chang04,McEntaffer13,Peterson15,DeRoo16thesis,DeRoo16,Miles18} and multilayer-coated blazed gratings for extreme UV and x-ray monochromator applications \cite{Voronov10,Voronov11,Voronov16}, where in each case, either interference lithography, EBL or nanoimprint lithography [\emph{cf.\@} \cref{sec:holographic,sec:binary_ebeam,sec:nanoimprint}] was used to pattern a groove layout that was transferred by dry etch into an underlying hardmask layer\footnote{In many cases this etch mask is composed of stoichiometric silicon nitride (\ce{Si3N4}) or some other composition of the material (\ce{Si_{x}N_{y}}). A silicon-nitride film can be coated on a silicon wafer using an optimized, \emph{low-pressure chemical vapor deposition (LPCVD)} process, where gaseous chemical precursors are injected into a vacuum chamber while the wafer is heated to promote chemical reactions to occur at its surface to grow a film of a specified thickness on the wafer (typically a few tens of \si{\nm}) \cite{Franssila10}.\label{footnote:nitride_mask}} and then, following the removal of native oxide on the silicon surface (\ce{SiO2}), the crystal structure of a silicon substrate through a wet \ce{KOH} etch.\footnote{This can be carried out through a wet \emph{buffered oxide etch (BOE)} that consists of \emph{hydrofluoric acid} (\ce{HF}) diluted in \emph{ammonium fluoride} (\ce{NH4F}) \cite{Franssila10}.\label{footnote:buffered_oxide_etch}}
Such a process is described as anisotropic in the sense that \ce{KOH} etch rates in crystalline silicon depend strongly on crystallographic direction. 
If the groove layout is aligned with the crystal structure of the substrate appropriately, this manifests as a collection of atomically-smooth troughs that appear similar to a triangular sawtooth when viewed edge-on \cite{Tsang75,Sato99,Franssila10,Gosalvez15}.
The effective blaze angle for a reflection grating then depends on the surface-normal orientation of the wafer used for etching as well as the groove direction relative to crystallographic directions on the surface of the wafer \cite{McEntaffer13,DeRoo16,Miles18}. 

The ideal result of a \ce{KOH}-etching process can be envisioned by considering the \emph{face-centered cubic} crystal structure of silicon, which is illustrated\footnote{These diagrams were generated by T.\ Wood using \textsc{CrystalMaker X} \cite{crystal_maker,Palmer_CM}.} in \cref{fig:crystal_structure}. 
To start, the unit cell of this cubic crystal structure can be defined using an orthogonal set of principal-axis vectors, $\mathbold{a}$, $\mathbold{b}$ and $\mathbold{c}$, that each have identical magnitudes given by the lattice constant of silicon (\emph{i.e.}, $\abs{\mathbold{a}} = \abs{\mathbold{b}} = \abs{\mathbold{c}} = a_{\text{Si}} \approx \SI{0.54}{\nm}$) and directions given by $[ 1 0 0 ]$, $[ 0 1 0 ]$ and $[ 0 0 1 ]$, respectively [\emph{cf.\@} \cref{fig:crystal_structure}]. %\footnote{More generally, the directional distance between any two points on the crystal lattice can be written in this basis as $\mathbold{R} = u \mathbold{a} + v \mathbold{b} + w \mathbold{c}$, where $u$, $v$ and $w$ are integers \cite{Trolier-McKinstry17}.}
\begin{figure}
 \centering
 \includegraphics[scale=0.235]{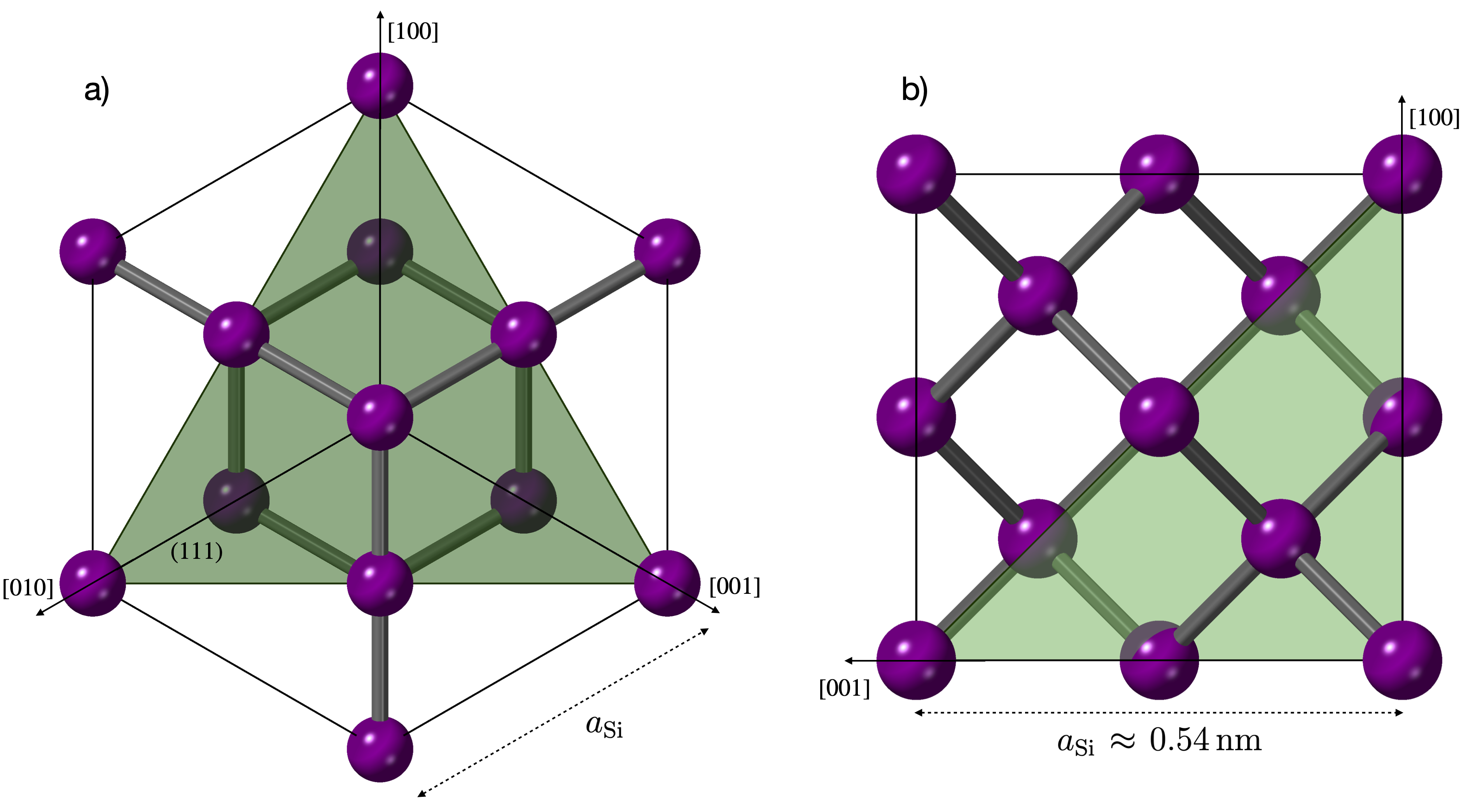}
 \includegraphics[scale=0.235]{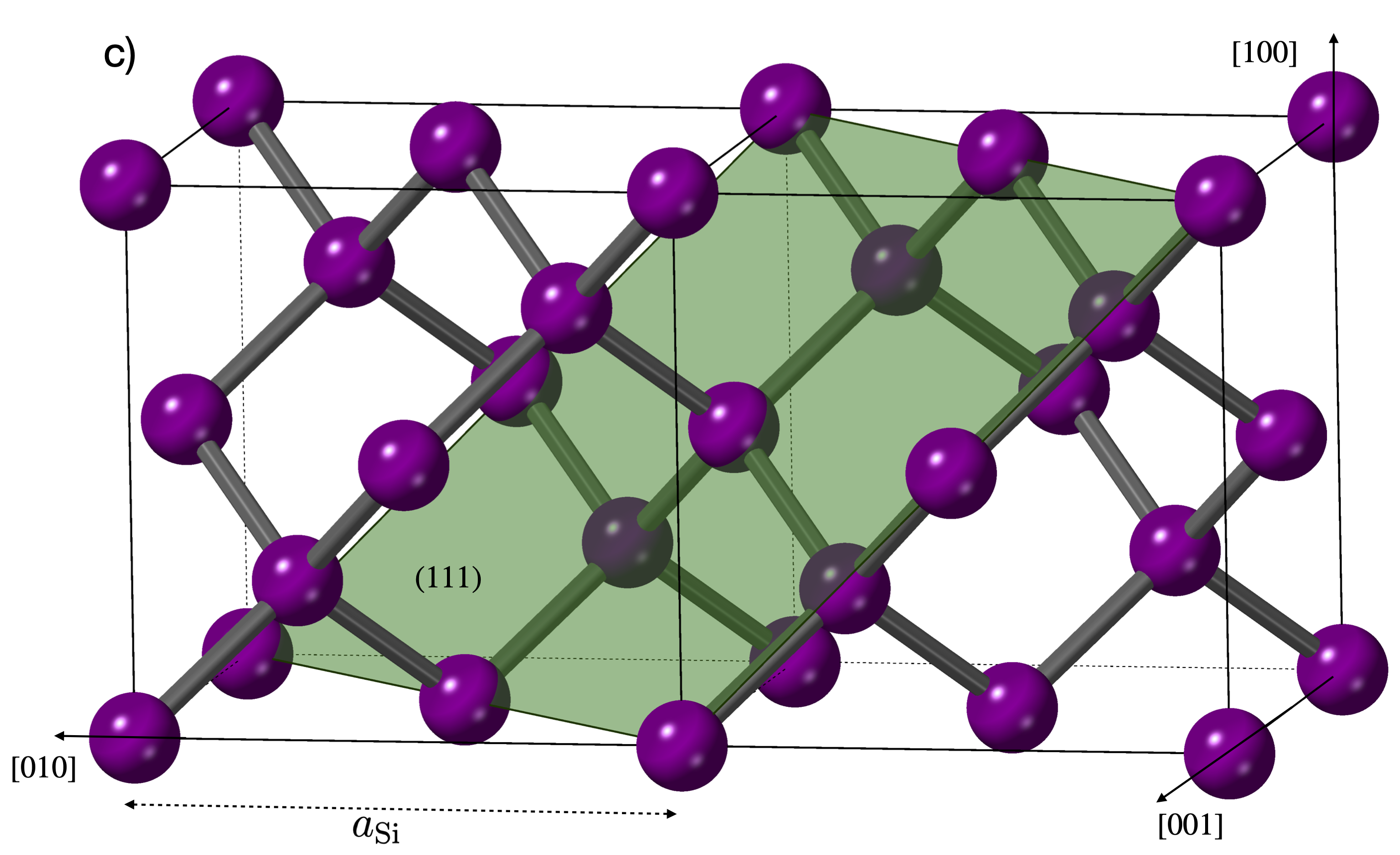}
 \caption[Face-centered cubic structure of silicon (credit: T.\ Wood using \textsc{CrystalMaker})]{Three views of the face-centered cubic structure of silicon, where $a_{\text{Si}} \approx \SI{0.54}{\nm}$ is the lattice constant, cube faces are $\{ 100 \}$ planes, purple spheres represent silicon atoms and gray bars indicate covalent bonds while a $(111)$ plane is shown in green: a) down the $[ 111 ]$ axis of a single unit cell, b) down the $[ 010 ]$ axis of a single unit cell and c) two adjacent cells. Image credit: T.\ Wood using \textsc{CrystalMaker X} \cite{crystal_maker}.}\label{fig:crystal_structure} 
 \end{figure}
Along with $[ \overline{1} 0 0 ]$, $[ 0 \overline{1} 0 ]$ and $[ 0 0 \overline{1} ]$ (where an overbar indicates a negative integer), $[ 1 0 0 ]$, $[ 0 1 0 ]$ and $[ 0 0 1 ]$ form a family of crystallographic directions that are equivalent by symmetry; this is denoted by $\langle 1 0 0 \rangle$ \cite{Franssila10,Trolier-McKinstry17}. %\footnote{In regard to crystallographic directions and planes, an overbar indicates a negative integer. Moreover, this notation holds for any $\langle u v w \rangle$ but the number of symmetry-related directions may vary.} 
A crystallographic plane that intersects the principal-axis vectors at $\mathbold{a} / h$, $\mathbold{b} / k$ and $\mathbold{c} / \ell$ is represented as $(h k \ell)$, where $h$, $k$ and $\ell$ are integers known as \emph{Miller indices}, and in the special case of a cubic crystal, $[ h k \ell ]$ and $(h k \ell)$ are always orthogonal such that the spacing between successive $(h k \ell)$ planes is given by \cite{Trolier-McKinstry17} 
\begin{equation}\label{eq:d_hkl}
 d_{h k \ell} = \frac{a_{\text{Si}}}{\sqrt{h^2 + k^2 + \ell^2}} . 
 \end{equation} 
Using $\{ h k \ell \}$ to denote a family of crystallographic planes that are equivalent by symmetry, $\{ 1 1 1 \}$ planes include $( 1 1 1 )$, $( \overline{1} 1 1 )$, $( 1 \overline{1} 1 )$, $( 1 1 \overline{1} )$, $( 1 \overline{1} \overline{1} )$, $( \overline{1} 1 \overline{1} )$, $( \overline{1} \overline{1} 1 )$ and $( \overline{1} \overline{1} \overline{1} )$ with a $( 1 1 1 )$ plane shown in \cref{fig:crystal_structure} as an example. 

\ce{KOH} is known to etch crystalline silicon along $\langle 1 1 1 \rangle$ directions at a rate much lower than all other families of crystallographic directions, and as a result, the $\{ 1 1 1 \}$ planes that intersect with the edges of the etch-mask layout are left exposed by the etch \cite{Franssila10,Gosalvez15}. 
To discuss this, the most common wafer geometry used for \ce{KOH} etching is one with a $\langle 100 \rangle$ surface normal and grooves aligned along any $\langle 011 \rangle$ direction, which is illustrated in \cref{fig:KOH_geo}. 
\begin{figure}
 \centering
 \includegraphics[scale=0.9]{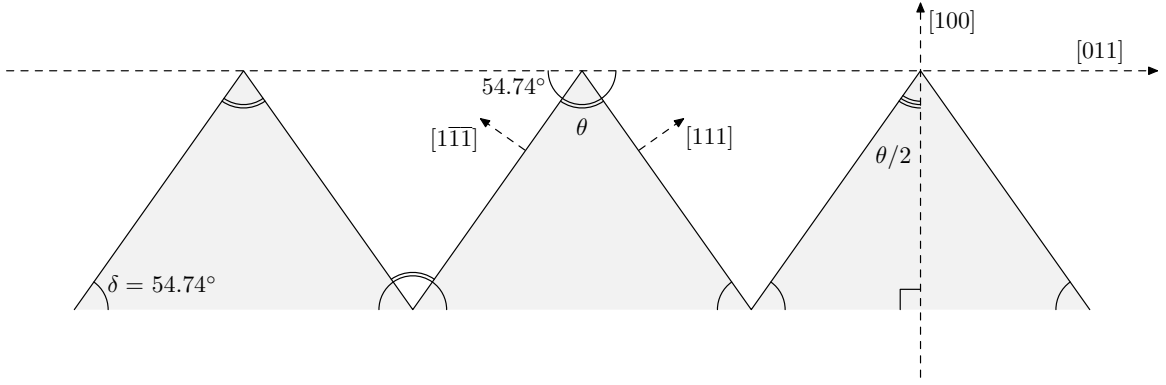}
 \caption[Geometry for \ce{KOH} etching in a $\langle 100 \rangle$-oriented silicon wafer]{Geometry for \ce{KOH} etching in a $\langle 100 \rangle$-oriented silicon wafer. The surface of the wafer is a $\{ 100 \}$ equivalent plane with a normal direction taken to be $[100]$. In this projection, the $[01\overline{1}]$ direction points out of the page while $(111)$ and $(1\overline{1} \overline{1})$ are the exposed $\{ 111 \}$ planes, defining an angle $\theta = 70.53^{\circ}$. The result is a symmetric sawtooth with a nominal blaze angle $\delta = 54.74^{\circ}$.}\label{fig:KOH_geo}
 \end{figure}
Taking the wafer surface normal to be $[100]$ with $[01\overline{1}]$ as the groove direction, the $\{ 111 \}$ planes exposed by a \ce{KOH} etch in this case are $(111)$ and $(1\overline{1} \overline{1})$, which intersect at an angle of $\theta = 70.53^{\circ}$. 
This can be seen by carrying out the dot product between two normalized vectors representing the $[111]$ and $[1 \overline{1} \overline{1}]$ directions, which are normal to the $(111)$ and $(1\overline{1} \overline{1})$ planes, and then taking the supplementary angle: 
\begin{equation}\label{eq:Si_angle}
 \frac{1}{\sqrt{3}} \begin{pmatrix} 1 & -1 & -1 \end{pmatrix}
 \frac{1}{\sqrt{3}} \begin{pmatrix} 1 \\ 1 \\ 1 \end{pmatrix} = -\frac{1}{3} = \cos \left( \pi - \theta \right) \implies \theta = 70.53^{\circ} . 
\end{equation}
Meanwhile, the angle between the wafer surface and both exposed $\{ 111 \}$ planes in the crystal is $\left( 180^{\circ} - \theta \right) / 2$ [\emph{cf.\@} \cref{fig:KOH_geo}]:
\begin{equation}\label{eq:100_angle}
 \frac{1}{\sqrt{3}} \begin{pmatrix} 1 & 1 & 1 \end{pmatrix}
 \begin{pmatrix} 1 \\ 0 \\ 0 \end{pmatrix} = \frac{1}{\sqrt{3}} = \cos \left( \delta \right) \implies \delta = 54.74^{\circ} ,
 \end{equation}
which defines $\delta$ of the symmetric sawtooth produced by such an etch and indicates that the distance between $\{ 111 \}$ planes is $d_{111} \equiv a_{\text{Si}}/\sqrt{3} \approx \SI{0.31}{\nm}$ [\emph{cf.\@} \cref{eq:d_hkl}]. 

Asymmetric sawtooth profiles with blaze angles other than $\delta = 54.74^{\circ}$ can be obtained by choosing an appropriate off-axis-cut silicon wafer with a surface-normal orientation rotated from $[100]$ about the axis coincident with the groove direction, $[01\overline{1}]$, so that $(111)$ and $(1\overline{1} \overline{1})$ are still the exposed $\{ 111 \}$ planes that define the angle $\theta$. 
\begin{figure}
 \centering
 \includegraphics[scale=1.4]{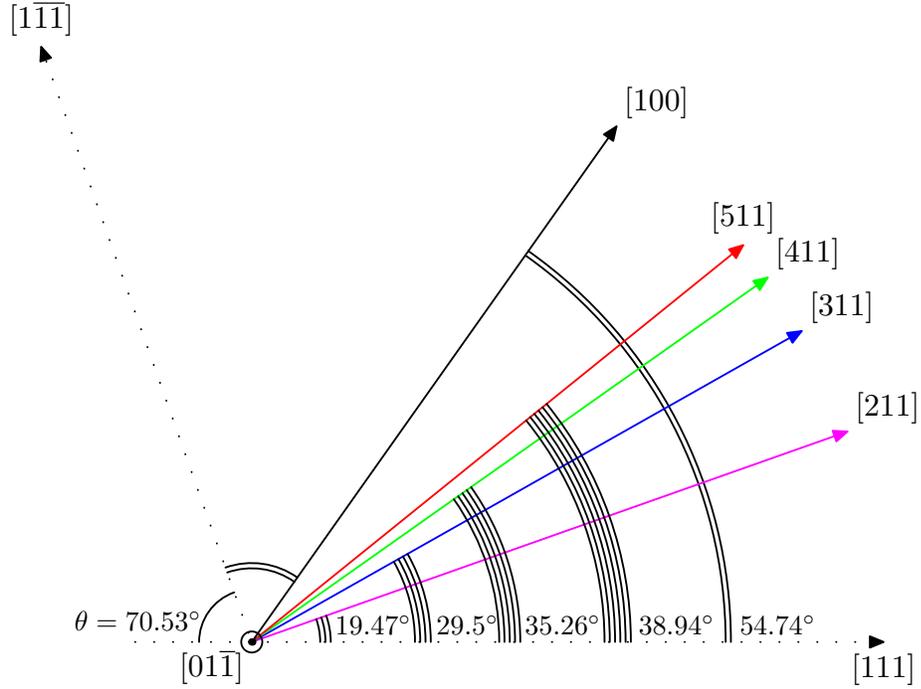}
 \caption[Various wafer orientations that are confined to a plane orthogonal to $\langle 011 \rangle$.]{Various wafer orientations that are confined to a plane orthogonal to $\langle 011 \rangle$. Each of these can be used to etch asymmetric sawtooth topographies into crystalline silicon using \ce{KOH} provided that the grating grooves are aligned with a $\langle 011 \rangle$ direction on the surface of the wafer.}\label{fig:KOH_angles}
 \end{figure}
This is demonstrated in \cref{fig:KOH_angles}, where as in the $\langle 100 \rangle$ case shown in \cref{fig:KOH_geo}, the $\langle 011 \rangle$ direction pointing out of the page is taken to be $[01 \overline{1}]$ while the relevant $\langle 111 \rangle$ directions are still $[111]$ and $[1 \overline{1} \overline{1}]$, which intersect at an angle $180^{\circ} - \theta = 109.47^{\circ}$ [\emph{cf.\@} \cref{eq:Si_angle}]. 
As indicated by \cref{fig:KOH_angles}, a series of vectors representing other wafer orientations also lay in this plane defined by $[111]$ and $[1\overline{1} \overline{1}]$. %, along with $[100]$.  
While a $[100]$ surface normal bisects these two $\langle 111 \rangle$ directions at an angle $\delta = 54.74^{\circ}$ [\emph{cf.\@} \cref{eq:100_angle}], those with crystallographic directions such as $[211]$ and $[311]$ intersect with $[111]$ at a smaller blaze angle, $\delta$, and $[1\overline{1} \overline{1}]$ at a different, steeper angle, $\bar{\delta} = 180^{\circ} - \theta - \delta$, to yield asymmetric sawtooth profiles that are rotated versions of the $\langle 100 \rangle$ case. 

With the angle $\theta$ defined by $\{ 111 \}$ plane intersections as in \cref{fig:KOH_geo}, the blaze angles produced by some of these off-axis wafer orientations are listed in \cref{tab:off_axis_si}. 
\begin{table}[]
 \centering
 \caption{Blaze angles ($\delta$) and corresponding opposite angles ($\bar{\delta} = 180^{\circ} - \theta - \delta$) on asymmetric sawtooth patterns generated by various wafer orientations with grooves aligned along $\langle 011 \rangle$ on the wafer surface.}\label{tab:off_axis_si}
 \begin{tabular}{ccc}
 \toprule
 wafer orientation & blaze angle ($\delta$) & opposite angle ($\bar{\delta}$)  \\ \midrule
 $\langle 211 \rangle$      & $19.47^{\circ}$        & $90^{\circ}$    \\
 $\langle 311 \rangle$      & $29.5^{\circ}$         & $79.97^{\circ}$ \\
 $\langle 411 \rangle$      & $35.26^{\circ}$        & $74.21^{\circ}$ \\
 $\langle 511 \rangle$      & $38.94^{\circ}$        & $70.53^{\circ}$ \\
 $\langle 100 \rangle$      & $54.74^{\circ}$        & $54.74^{\circ}$ \\ \bottomrule
\end{tabular}
\end{table}
A blaze angle of $\delta \approx 30^{\circ}$, for example, can be obtained using a $\langle 311 \rangle$-oriented wafer, which by definition, can be described using a vector that is rotated $25.24^{\circ}$ from $\langle 100 \rangle$, toward $\langle 111 \rangle$ in the plane orthogonal to $\langle 011 \rangle$ (\emph{i.e.}, the groove direction). 
\begin{figure}
 \centering
 \includegraphics[scale=0.9]{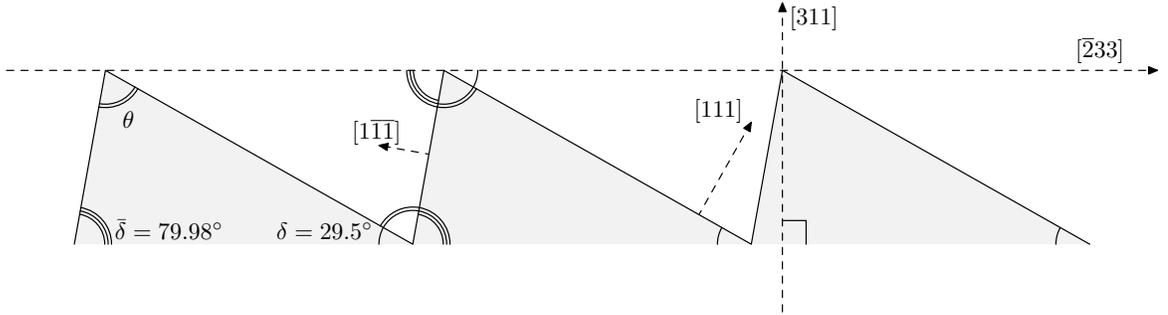}
 \caption[Geometry for \ce{KOH} etching in a $\langle 311 \rangle$-oriented silicon wafer]{Geometry for \ce{KOH} etching in a $\langle 311 \rangle$-oriented silicon wafer. The surface of the wafer is a $\{ 311 \}$ equivalent plane with a normal direction taken to be $[311]$. In this projection, the $[01\overline{1}]$ direction points out of the page while $(1\overline{1} \overline{1})$ and $(111)$ are the exposed $\{ 111 \}$ planes. Here, the $(111)$ surface acts as the blazed groove facet with a nominal blaze angle of $\delta = 29.5^{\circ}$ while the $(1\overline{1} \overline{1})$ plane defines a $\bar{\delta} = 79.98^{\circ}$ angle.}\label{fig:KOH_geo_311}
 \end{figure}
This is illustrated in \cref{fig:KOH_geo_311}, where the shallow facet defined by the $(111)$ plane serves as the blaze angle: 
\begin{subequations}
\begin{equation}\label{eq:311_angle}
 \frac{1}{\sqrt{11}} \begin{pmatrix} 3 & 1 & 1 \end{pmatrix}
 \frac{1}{\sqrt{3}} \begin{pmatrix} 1 \\ 1 \\ 1 \end{pmatrix} = \frac{5}{\sqrt{33}} = \cos \left( \delta \right) \implies \delta = 29.5^{\circ}
\end{equation}
and the steep side of the facet has an opening angle of $\bar{\delta} = 180^{\circ} - \theta - \delta$ defined by the $(1\overline{1} \overline{1})$ plane:
\begin{equation}\label{eq:311_angle2}
  \frac{1}{\sqrt{11}} \begin{pmatrix} 3 & 1 & 1 \end{pmatrix}
 \frac{1}{\sqrt{3}} \begin{pmatrix} 1 \\ -1 \\ -1 \end{pmatrix} = \frac{1}{\sqrt{33}} = \cos \left( \bar{\delta} \right) \implies \bar{\delta} =  79.98^{\circ} .
 \end{equation}
\end{subequations} 
While similar sawtooth profiles can be obtained using other wafer orientations [\emph{cf.\@} \cref{fig:KOH_angles}], it should be emphasized that not all off-axis orientations yield sawtooth profiles with two exposed $\{ 111 \}$ planes that intersect at the angle $\theta$. 
For example, a $\langle 110 \rangle$-oriented wafer can be used to achieve a symmetric sawtooth with $\delta \approx 35^{\circ}$ or a laminar-like grating with $\delta \approx 90^{\circ}$, depending on the alignment of the groove direction \cite{Kendall75,Rao17}.
Moreover, wafers with surface normals intermediate between $\langle 211 \rangle$ and $\langle 111 \rangle$ yield $19.47^{\circ} > \delta > 0^{\circ}$ for the exposed $(111)$ plane but accordingly, $\bar{\delta} > 90^{\circ}$, which may prevent the $(1\overline{1} \overline{1})$ plane from being exposed by a \ce{KOH} etch. 

%Moreover, this leads to the creation of a flat portion atop each groove instead of the pointed apex illustrated in \cref{fig:KOH_geo,fig:KOH_geo_311}. 

An example of an x-ray reflection grating fabricated via \ce{KOH} etching is shown though \emph{field-emission scanning electron microscopy (FESEM)} \cite{Zhou07} in \cref{fig:master_grating_SEM}, where the groove spacing is $d \lessapprox \SI{160}{\nm}$ and a nominal blaze angle of $\delta = 29.5^{\circ}$ was achieved through use of a $\langle 311 \rangle$-oriented silicon wafer [\emph{cf.\@} \cref{eq:311_angle}]. 
\begin{figure}
 \centering
 \includegraphics[scale=0.4]{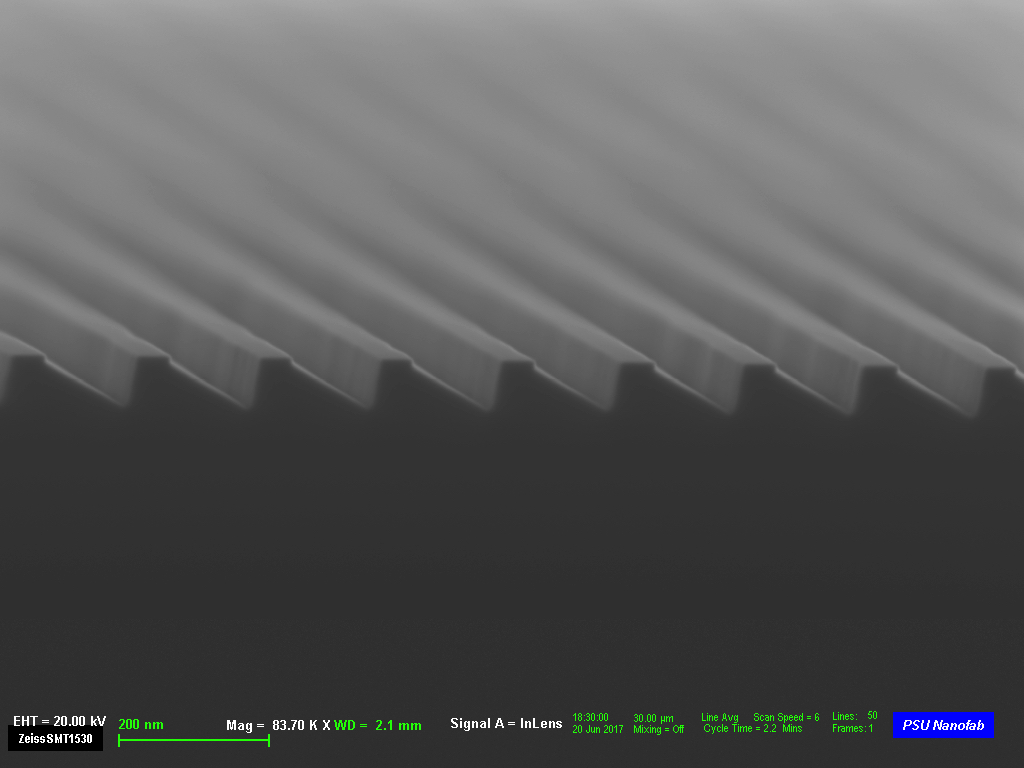}
 \caption[Field-emission scanning electron micrograph (FESEM) of a \ce{KOH}-etched grating viewed edge-on.]{Field-emission scanning electron micrograph (FESEM) of a \ce{KOH}-etched grating with $d \lessapprox \SI{160}{\nm}$ viewed edge-on, taken with a \textsc{Zeiss Leo 1530} instrument at the Penn State Nanofabrication Laboratory. Due to the $\langle 311 \rangle$ surface orientation of the silicon wafer used for processing, the etched sawtooth structure nominally features an active blaze angle of $\delta = 29.5^{\circ}$ and a steep-facet angle of $\bar{\delta} =  79.98^{\circ}$ [\emph{cf.\@} \cref{eq:311_angle,eq:311_angle2}]. Flat tops that protrude above the sawtooth pattern exist as a result of the \ce{Si_{x}N_{y}} etch mask [\emph{cf.\@} \cref{fig:master_fab}] while the groove depth follows from \cref{eq:groove_depth} \cite{McCoy17,Miles18,McCoy20b}.}\label{fig:master_grating_SEM}
 \end{figure}
This grating was fabricated by staff at the Penn State Nanofabrication Laboratory \cite{PSU_MRI_nanofab} in a process that used EBL to pattern a groove layout in resist, RIE to transfer the pattern into a \SI{30}{nm}-thick LPCVD layer of \ce{Si_{x}N_{y}} [\emph{cf.\@} footnotes~\ref{footnote:dry_etch_RIE} and \ref{footnote:nitride_mask}] and a timed, room-temperature \ce{KOH} etch to produce a sawtooth-like surface relief \cite{Miles18}. 
Following the procurement of a \SI{500}{\um}-thick, \SI{150}{\mm}-diameter silicon wafer from \textsc{Virginia Semiconductor} \cite{Virgsemi_web} with a $\langle 311 \rangle$ surface orientation and a pre-deposited, low-stress layer of \ce{Si_{x}N_{y}} \cite{Zheng13}, the groove layout was defined by EBL in a \SI{140}{\nm}-thick layer of ZEP520A resist, which was obtained by spin-coating a 1:1 mixture of the resist in Anisole (\emph{i.e.}, \emph{methoxybenzene}; \ce{C7H8O})\footnote{This was done at \num{3000} rotations per \si{\minute} for \SI{45}{\second} following a \SI{3}{\minute} dehydration hotplate bake at \SI{180}{\celsius}; the Anisole was then removed from the resist with an identical hotplate bake.} (\cref{fig:master_fab}, step 1). 
\begin{figure}
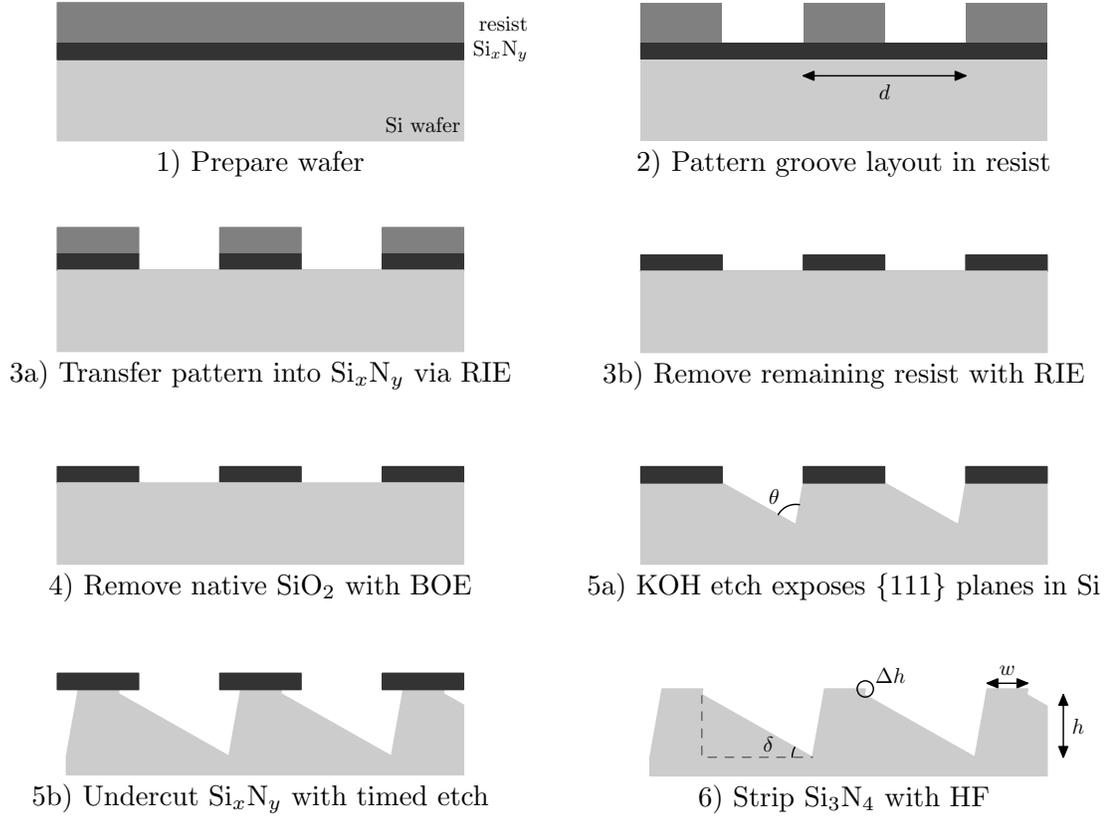
 
 \centering
 \includegraphics[scale=1.08]{Chapter-1/Figures/master_fab1.mps}
 \includegraphics[scale=1.08]{Chapter-1/Figures/master_fab2.mps}
 \includegraphics[scale=1.08]{Chapter-1/Figures/master_fab3.mps}
 \includegraphics[scale=1.08]{Chapter-1/Figures/master_fab4.mps}
 \includegraphics[scale=1.08]{Chapter-1/Figures/master_fab5.mps}
 \includegraphics[scale=1.08]{Chapter-1/Figures/master_fab6.mps}
 \includegraphics[scale=1.08]{Chapter-1/Figures/master_fab7.mps}
 \includegraphics[scale=1.08]{Chapter-1/Figures/master_fab8.mps}
 \caption[Outline of fabrication recipe for crystallographic etching in a $\langle 311 \rangle$-oriented silicon wafer using EBL]{Outline of fabrication recipe for crystallographic etching in a $\langle 311 \rangle$-oriented silicon wafer using EBL. A wafer coated with \ce{Si_{x}N_{y}} is spin-coated with resist, where a groove layout is defined by EBL. This pattern is transferred into the \ce{Si_{x}N_{y}} and then the resist is removed through reactive ion etching (RIE). Native \ce{SiO2} on the wafer is removed with a buffered oxide etch (BOE) before a timed crystallographic etch using \ce{KOH} is carried out to expose $\{ 111 \}$ planes that intersect at an angle of $\theta = 70.53^{\circ}$. Finally, the \ce{Si_{x}N_{y}} is removed using \ce{HF}, leaving groove structure with a depth $h$ and flat-top width $w$ that both depend on $d$ and the degree of \ce{KOH}-etch undercut by \cref{eq:groove_depth} \cite{McCoy17,Miles18}.}\label{fig:master_fab} 
 \end{figure}
Using the \textsc{EBPG5200} tool at Penn State described in \cref{sec:binary_ebeam}, the resist was patterned over a variable-line-space profile \SI{75}{\mm} by \SI{96}{\mm} in area that approximates a radial profile [\emph{cf.\@} \cref{sec:grating_tech_intro}] with groove spacing $d \lessapprox \SI{160}{\nm}$ to match a \SI{12}{\metre} focal length.\footnote{This was baselined for a preliminary design of \emph{Arcus}, a proposed soft x-ray spectrometer \cite{Smith19_arcus,Arcus_web}.} 
The groove layout was designed as six sections of parallel lines, each with a different value of $d$, and then fractured from computer-aided design into data understandable to the \textsc{EBPG5200} using the \textsc{Layout BEAMER} software package.\footnote{With a \SI{0.25}{\nm} resolution for the \textsc{EBPG5200} and a \SI{40}{\nm} beam step size in \textsc{Layout BEAMER}, $d$ for each section of parallel grooves ranges nominally from \SIrange{160}{158.25}{\nm}, in steps of \SI{0.25}{\nm}.} 
These lines and spaces were exposed using a nominal electron dose of $D = \SI{170}{\micro\coulomb\per\cm\squared}$, a \SI{60}{\nano\ampere} beam current, a \SI{200}{\um} aperture size, and a \SI{40}{\nm} beam step size, to a duty cycle of $\sim \SI{50}{\percent}$ and then developed at room temperature in \emph{n-amyl acetate} (\ce{C7H14O2}) for \SI{3}{\minute} and IPA for \SI{30}{\second} followed by a high-purity nitrogen blow dry (\cref{fig:master_fab}, step 2). 
\begin{figure}
 \centering
 \includegraphics[scale=0.15]{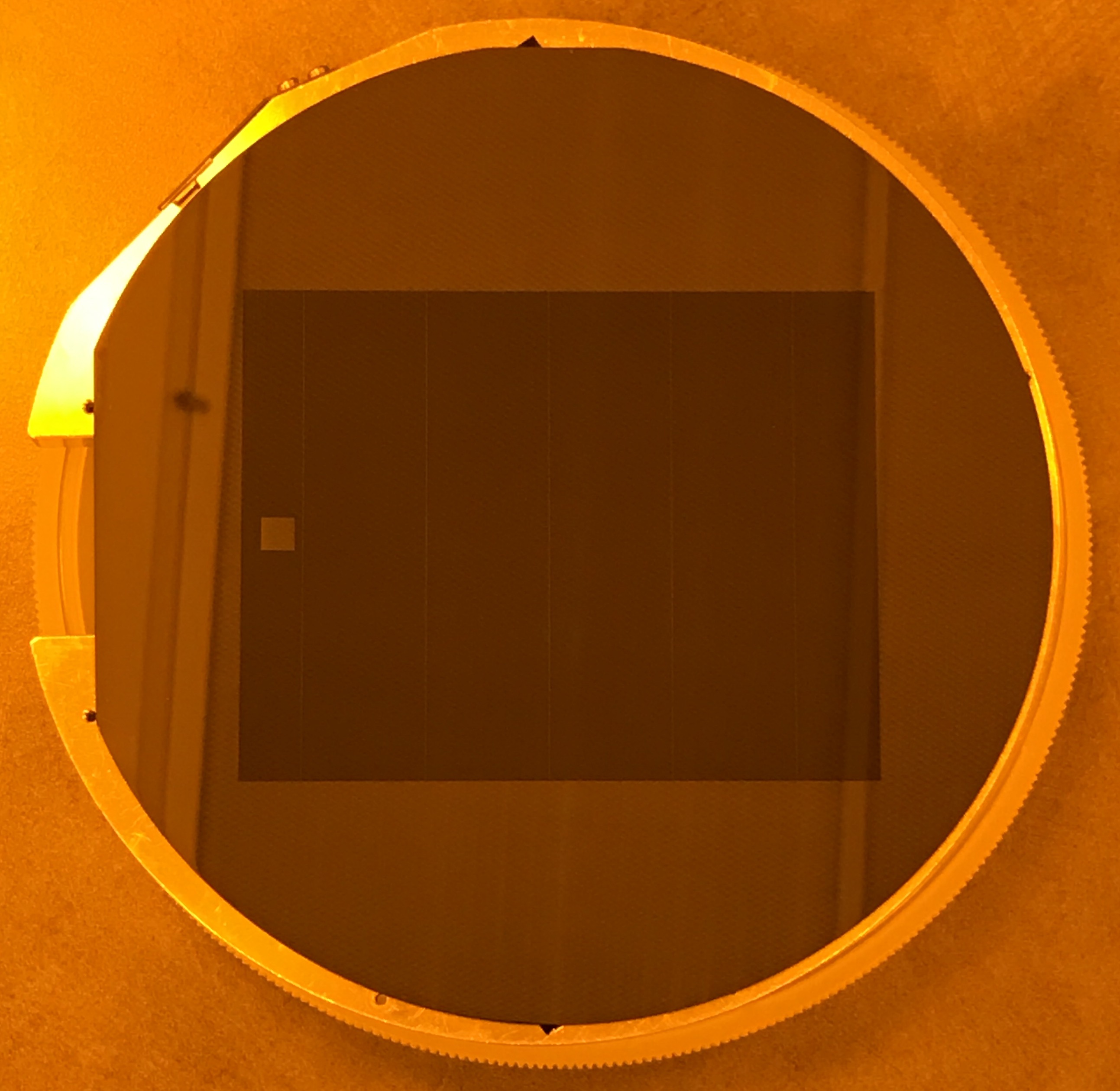}
 \caption[\ce{KOH}-etched grating patterned over a \SI{75}{\mm} by \SI{96}{\mm} area on a \SI{150}{\mm}-diameter, $\langle 311 \rangle$-oriented wafer.]{Photograph of the \SI{75}{\mm} by \SI{96}{\mm} grating depicted in \cref{fig:master_grating_SEM}, shown patterned on a \SI{150}{\mm}-dimater, $\langle 311 \rangle$-oriented wafer with grooves oriented horizontally. Also seen are thin lines of unpatterned area in between each section of parallel grooves and a small, square optical grating used for aligning off-plane gratings in a module \cite{McCoy17}.}\label{fig:master_grating} %\cite{Miles18}
 \end{figure}

The EBL-patterned resist was transferred into the underlying \ce{Si_{x}N_{y}} film through RIE processing carried out with a \textsc{Plasma-Therm Versalock} tool, which consisted of a short descum with an oxygen (\ce{O2}) plasma to remove residual exposed resist followed by plasma etch using a mixture of \emph{fluoroform} (\ce{CHF3}) and \ce{O2} (\cref{fig:master_fab}, step 3a). 
To prepare for the crystallographic etch, remaining resist was removed by oxidation with another \ce{O2} RIE (\cref{fig:master_fab}, step 3b) before a BOE [\emph{cf.\@} footnote~\ref{footnote:buffered_oxide_etch}] was carried out to remove native \ce{SiO2} on the wafer surface (\cref{fig:master_fab}, step 4). 
The wet anisotropic etch was performed in \SI{45}{\percent}-diluted \ce{KOH} at room temperature for \SI{24}{\min} to produce an asymmetric sawtooth with a narrow flat top beneath the \ce{Si_{x}N_{y}} mask and a sharp point at the bottom of each groove (\cref{fig:master_fab}, steps 5a and 5b). 
Finally, the \ce{Si_{x}N_{y}} mask was stripped with a \SI{10}{\min} etch in \SI{49}{\percent}-diluted \ce{HF} (\cref{fig:master_fab}, step 6), leaving the structure of a surface-relief grating with a nominal blaze angle of $\delta = 29.5^\circ$ and a top plateau of width $w \gtrapprox \SI{30}{\nm}$ so that the groove depth is given by 
\begin{equation}\label{eq:groove_depth}
 h \approx \frac{d - w}{\cot \left( \delta \right) - \cot \left( \theta + \delta \right)} + \Delta h ,
 \end{equation}
where $\Delta h$ is the vertical protrusion of the top plateau \cite{McCoy20b}. 
While these sharp grooves prevent $h$ to be measured by AFM without a high-aspect-ratio scanning probe tip, it is estimated that this quantity falls within the range \SIrange{65}{70}{\nm} with $\Delta h$ of a few \si{\nm}. 
The \SI{72}{\cm\squared} patterned area following all fabrication steps is shown on a \SI{150}{\mm}-diameter wafer in \cref{fig:master_grating}. 

Grating fabrication processes centering on \ce{KOH} etching are beneficial especially for producing sawtooth-like structures with exceptionally smooth facets at a specified blaze angle that depends on the crystallographic orientation of the wafer used for processing \cite{Franke97,McEntaffer13,DeRoo16thesis}. 
While these high-fidelity structures lend themselves to high overall diffraction efficiency and an effective blaze response, ideally the topography would be inverted to achieve a sharp groove apex with the flat tops produced by the \ce{Si_{x}N_{y}} etch mask laying at the groove base, which is usually shadowed by incident radiation. 
This can be achieved through grating replication via nanoimprint lithography \cite[\emph{cf.\@} \cref{sec:nanoimprint}]{Chang03} but there are still important disadvantages that come along with fabricating a master grating in this way. %through \ce{KOH} etching. 
First, \ce{KOH} etching demands precise alignment between the groove direction and the appropriate crystallographic axis (\emph{i.e.}, $\langle 011 \rangle$ in most cases) to achieve smooth and continuous groove facets. 
Moreover, the face-centered cubic structure of silicon [\emph{cf.\@} \cref{fig:crystal_structure}] ultimately prevents the formation of a true radial profile even with a hypothetically perfect alignment. 
In a \ce{KOH}-etched $\langle 311 \rangle$-oriented wafer, for example, the distance between $\{ 111 \}$ planes, $d_{111} \equiv a_{\text{Si}}/\sqrt{3}$, projects to a lateral spacing on the surface of the wafer of 
\begin{equation}
 \Delta d \equiv d_{111} \sin \left( \delta \right) \approx \SI{0.15}{\nm} \quad \text{for } \delta = 29.5^{\circ} . 
 \end{equation}
A radial profile patterned in a resist film by EBL, therefore, would still result in the formation of groove facets that are confined to this quantized groove spacing imposed by the crystal structure and, in principle, slight misalignments with the crystallographic planes combined with this latter issue lead to $\mathscr{R}$ being degraded in a Wolter-I grating spectrometer. 
Although this has yet to be confirmed by published testing results, it is worthwhile to investigate alternative nanofabrication techniques that enable a high-precision radial profile to be preserved while also achieving blazed groove facets to maximize spectral sensitivity in a given bandpass of interest. 

\subsubsection{Nanoimprint Lithography for Grating Replication}\label{sec:nanoimprint}
%%%%%%%%%%%%%%%%%%%%%%%%%%%%%%%%%%%%%%%%%--------------------------------------------------
The nanofabrication methods described in \cref{sec:holographic,sec:binary_ebeam,sec:crystal_etching} are suitable for a variety of custom reflection gratings but due to their complexity and potentially long manufacturing times, it is typically not practical to produce a large number of gratings for a Wolter-I spectrometer by using these processes to fabricate each grating directly. 
Reflection gratings fabricated by mechanical ruling engine are commonly replicated using processes where a suitable synthetic resin, such as epoxy, takes on the inverse mold of the master grating \cite{Loewen97,Franke97}.  
This resin is dispensed on the master grating and then a blank substrate is brought into contact while the material cures; a replica is produced on the blank substrate following separation of the mold and the master grating. 
Such processes, however, usually produce relatively low-fidelity replicas and become increasingly difficult for gratings with sub-\si{\um} groove period. 
Nonetheless, the surface-relief molds of the \num{182} gratings used for the \emph{RGS} on board \emph{XMM-Newton} were manufactured in this way before each was coated in gold for soft x-ray reflectivity \cite{Kahn96,denHerder01}. 

\emph{Nanoimprint lithography (NIL)} describes a class of nanofabrication techniques first developed in the mid 1990s as a method for high-throughput manufacturing of devices featuring nanoscale structures \cite{Chou96,Haisma96,Schift08,Schift10}. 
These techniques are resist-based similar to EBL [\emph{cf.\@} \cref{sec:binary_ebeam}] but rather than using radiation or particle exposure followed by wet development, the method of patterning is by molding through direct contact between a stamp (\emph{e.g.}, a master grating) and a layer of resist coated on a blank substrate. 
Moreover, the resist must be in a state where it is able to flow like a liquid with relatively low viscosity for the material to be able to mold to the inverse of the stamp. 
This typically occurs in polymers $\gtrapprox \SI{50}{\celsius}$ above their \emph{glass transition temperature}, $T_g$, which marks the smooth transition from a solid to a liquid in an amorphous material, which is in contrast to the thermodynamics of melting in a crystalline structure that involves latent heat \cite{Schift10,Kirchner19,Trolier-McKinstry17}. 

In the original variant of NIL, known as \emph{thermal NIL} or a type of \emph{hot embossing}, the resist can be heated to a temperature well above $T_g$ so that it is able to fill the stamp features with applied pressure on a specialized imprint tool \cite{Chou96,Schift10}. 
Then, by bringing the temperature back down below $T_g$, the stamp can be de-molded from the resist, leaving the imprinted pattern. 
\begin{figure}
 \centering
 \includegraphics[scale=0.65]{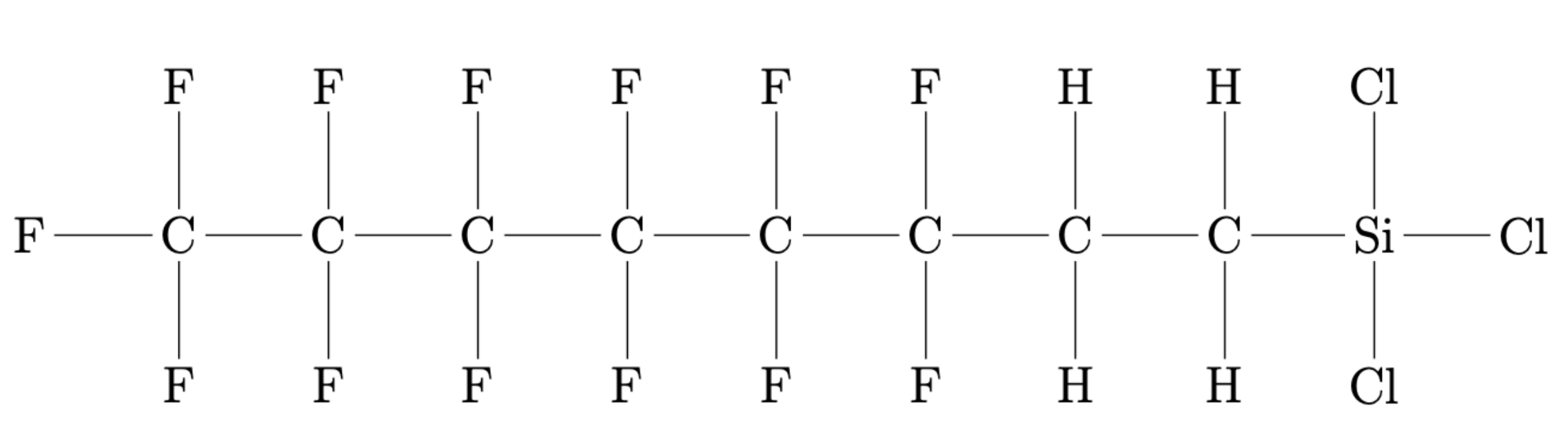}
 \caption[Structural formula of perfluorooctyltrichlorosilane (FOTS).]{Structural formula of FOTS with a \ce{-SiCl3} active head group and a \ce{-CF3} surface terminal group. Along with other similar materials, a self-assembled monolayer of FOTS can be used as an anti-adhesive layer in NIL for stamps with silicon surfaces.}\label{fig:FOTS}
 \end{figure}
However, this process must be carried out carefully to avoid degradation of the stamp and additionally, an anti-adhesive coating on the stamp is often required to prevent the imprint resist from sticking to the stamp. 
For a silicon surface with native oxide, this can be accomplished with a fluorinated trichlorosilane such as \emph{perfluorooctyltrichlorosilane} (FOTS; \ce{C8H4Cl3F13Si}) or \emph{perfluorodecyltrichlorosilane} (FDTS; \ce{C10H4Cl3F17Si}) \cite{Schift10,Zhuang07}.\footnote{Note, however, that different types of chemicals are required for other surfaces (\emph{e.g.}, gold \cite{Kim_2002}).} 
Such a material can be coated by \emph{molecular vapor deposition} on a hydroxylated silicon surface\footnote{\emph{i.e.}, such that hydroxyl groups (\ce{-OH}) are bonded to the native \ce{SiO2}} \cite{Kobrin05,Zhuang07,Verschuuren10}.
This process works by injecting the chosen chemical along with water vapor (\ce{H2O}) into a vacuum chamber so that the \ce{-SiCl3} active head groups on the molecule [\emph{cf.\@} \cref{fig:FOTS}] react with \ce{H2O} as well as \ce{SiO2} on the surface of the stamp. 
The sites on the molecule normally occupied by chlorine are in this case replaced by hydroxyl groups from the vapor as well as oxygen present in the native oxide. 
With \emph{hydrochloric acid} (\ce{HCl}) and \ce{H2O} as by-products, this bonds the fluorinated trichlorosilane to the stamp surface as a \emph{self-assembled monolayer} so that it becomes chemically inert and hydrophobic, thereby preventing adhesion with the imprint resist \cite{Zhuang05,Zhuang07,Schift10}. 

The first resist to be used for NIL was PMMA \cite{Chou96}, which was introduced in the context of EBL in \cref{sec:binary_ebeam} with its structural chemical composition shown in \cref{fig:PMMA}. 
While EBL typically calls for PMMA with relatively high $M_w$ (\emph{e.g.}, \SI{950}{\kilogram\per\mole}), PMMA used for NIL commonly has $M_w$ on the order of tens of \si{\kilogram\per\mole} so that with its thermoplastic properties, the material has decreased viscosity at temperatures above $T_g$, that correlate with $M_w$ below a critical value, $M_c$, which depends on various properties of the synthesized polymer \cite{Fuchs96,Geng16,Schift10,Kirchner19}. 
Although a low $M_w$ allows the molecules to fill small stamp features more easily, the resist is in a relatively soft state at room temperature due to $M_w$ being comparable to $M_c \approx \SI{10}{\kilogram\per\mole}$ for PMMA \cite{Schift10,Schleunitz10,Kirchner19}. 
This practically prevents the imprinted pattern from being used as a functional material and instead limits it to be used as an etch mask under appropriate conditions. 

NIL can also be used with UV-curable resists in a variant of the process known as \emph{UV-NIL} \cite{Haisma96}, where the resist is in a liquid state at room temperature so that the material can flow into the features of the applied stamp. 
This liquid-state material then is cross-linked by UV radiation, which serves to increase $M_w$ as well as $T_g$ so that the imprinted pattern becomes solidified at room temperature \cite{Schift08,Schift10}. 
With the resist being cured by UV radiation, its mechanical stability is improved so that the imprinted pattern can be used as a more functional structure for a reflection grating. 
However, the curing of the resist also changes the polymer structure of the material through cross-linking and hence some degree of volumetric shrinkage is expected in the imprinted pattern \cite{Horiba_2012}. 
While this effect may be small depending on the nature of the material, this shrinkage can, in principle, lead to imprinted gratings with a blaze angle $\delta'$ that is reduced relative to that of the master grating, $\delta$. 
Therefore, this effect of $\delta' < \delta$ generally must be accounted for to ensure that diffraction efficiency is maximized in the intended spectral bandpass [\emph{cf.\@} \cref{sec:discussion_scil}]. 

UV-NIL has been pursued previously to replicate gratings from master templates fabricated using the crystallographic etching technique described in \cref{sec:crystal_etching} \cite{Chang03,Chang04,McEntaffer13,DeRoo16thesis,DeRoo16,Miles18}. 
\begin{figure} 
 \centering
 \includegraphics[scale=0.55]{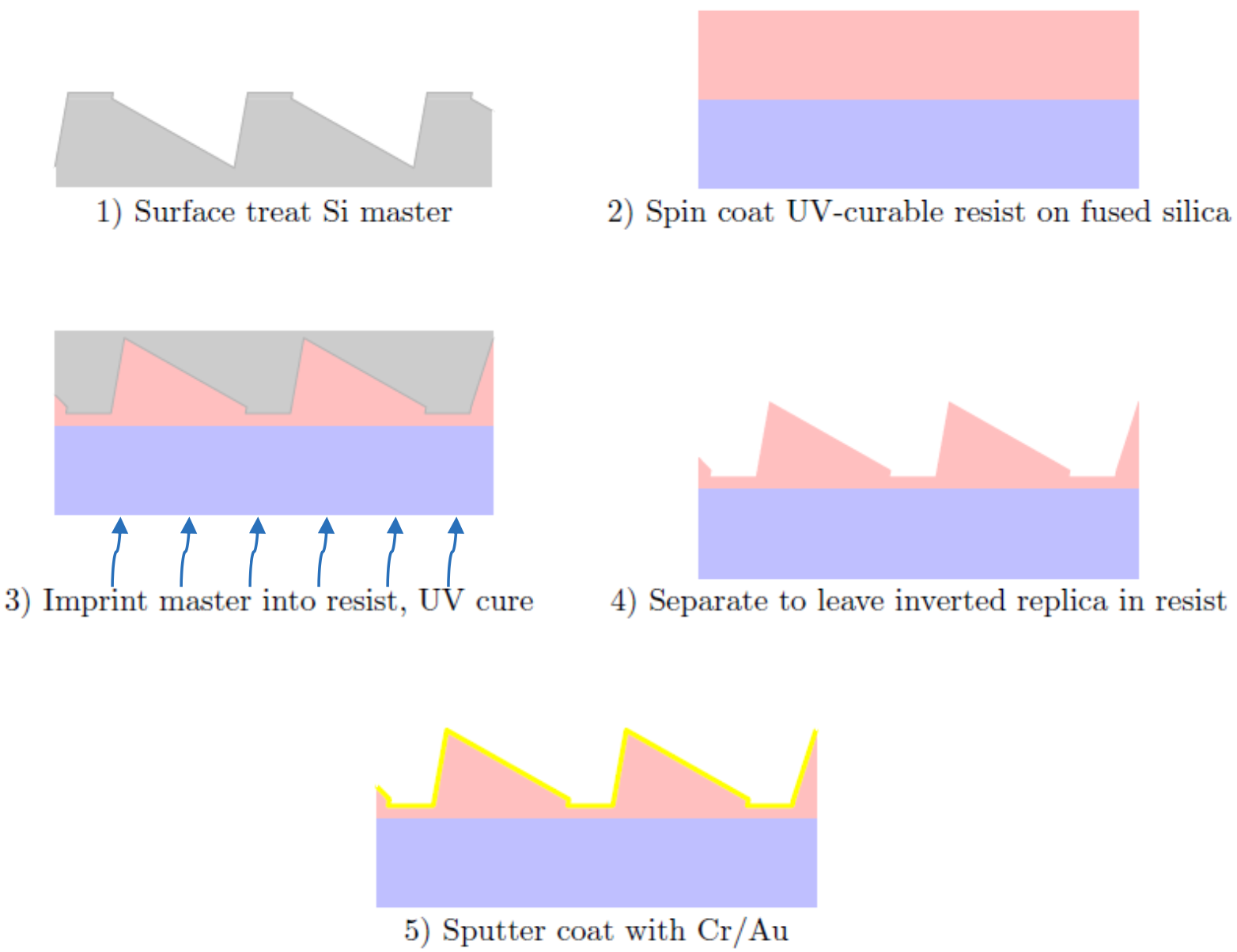}
 \caption[Outline of a grating replication process that uses a \ce{KOH}-etched master grating to imprint to ultraviolet-curable resist by UV-NIL.]{Outline of a grating replication process that used UV-NIL to imprint a \ce{KOH}-etched master grating into UV-curable resist \cite{Miles18}.}\label{fig:uv_nil} 
 \end{figure}
This process is illustrated in \cref{fig:uv_nil} for a grating fabrication scheme that used the \ce{KOH}-etched, $\langle 311 \rangle$-oriented wafer depicted in \cref{fig:master_grating_SEM,fig:master_fab,fig:master_grating} as a \SI{72}{\cm\squared} master grating; imprinting was carried out by \textsc{Nanonex Corporation} \cite{nanonex} using their \textsc{NX-2000} imprint tool and proprietary materials for the anti-adhesive layer (\textsc{NXT-130}), the imprint resist (\textsc{NXR-2050}), as well as an adhesion promoter (\textsc{NXT-404}) for the resist and a blank substrate \cite{Miles18}. 
While UV-NIL is commonly carried out using a stamp that is transparent to UV radiation (\emph{e.g.}, quartz) and a blank (opaque) silicon substrate \cite{Schift10}, the master grating in this case is made of silicon, which led to the need for an UV-transparent imprint wafer. 
Therefore, following the surface treatment for anti-stiction using \textsc{NXT-130} (\cref{fig:uv_nil}, step 1), a layer of \textsc{NXR-2050} resist was spin-coated on a \SI{150}{\mm}-diameter, fused-silica substrate, using \textsc{NXT-404} as an adhesion promoter (\cref{fig:uv_nil}, step 2). 
Based on the \SIrange{65}{70}{\nm} groove depth of the master grating [\emph{cf.\@} \cref{sec:crystal_etching}] this layer of resist was coated as a \SI{200}{\nm}-thick film so that the liquid resist could fill the features of the stamp adequately. 

Using an \textsc{NX-2000} imprinting tool, the master grating was then brought into contact with the coated substrate so that the \textsc{NXR-2050} resist could mold to the inverse of the stamp. 
The resist was then cured by exposing it to UV radiation through the fused-silica substrate (\cref{fig:uv_nil}, step 3) before the stamp was separated from the imprinted pattern (\cref{fig:uv_nil}, step 4). 
\begin{figure} 
 \centering
 \includegraphics[scale=0.37]{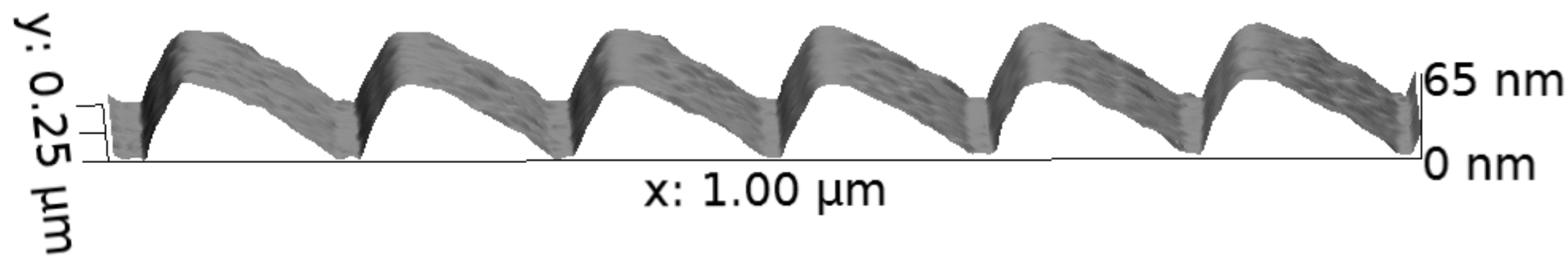}
 \caption[AFM of a blazed grating fabricated by UV-NIL from a \ce{KOH}-etched master (credit: T.\ Tighe)]{AFM of a blazed grating fabricated by UV-NIL from a \ce{KOH}-etched master. Image credit: T.\ Tighe at the Penn State MCL \cite{Miles18}.}\label{fig:uv_nil_AFM} 
 \end{figure}
Finally, the patterned resist featuring the inverse topography of the master was coated with a $\sim \SI{15}{\nm}$-thick layer of gold for soft x-ray reflectivity using a $\sim \SI{5}{\nm}$-thick layer of chromium for adhesion,\footnote{The reasoning for the use of the materials is discussed in \cref{sec:total_refl}.} which was accomplished via \emph{plasma sputter deposition} by staff at the Penn State Nanofabrication Laboratory \cite{PSU_MRI_nanofab} using a \textsc{Kurt J.\ Lesker CMS-18} tool \cite{Miles18}. 
This final, coated imprint is shown under AFM in \cref{fig:uv_nil_AFM}.\footnote{Credit for AFM is owed to T.\ Tighe at the Penn State MCL \cite{PSU_MRI_CL,Miles18}.}. 

Although NIL is a proven technology for replicating nanoscale patterns in a variety of applications, there are aspects of this process that lead to difficulties for x-ray reflection gratings. 
First, the rigidity of a silicon stamp requires a relatively high pressure to be applied for imprints of substantial area (\emph{e.g.}, the \SI{72}{\cm\squared} silicon master described above) so that conformal contact between the stamp and the resist-coated blank substrate can be achieved and air pockets can be avoided \cite{Schift08,Schift10}. 
These conditions also lead to imprinting imperfections arising from particulate contaminants that may be present in the laboratory environment and, potentially, damage to the stamp surface. 
Additionally, a rigid stamp is gradually degraded as it makes repeated imprints. 
For example, in the case of the \textsc{Nanonex} UV-NIL process just described \cite{Miles18}, a single stamp typically can produce tens of quality replicas \cite{Schift08,Schift10}. 
With the \emph{XGS} for \emph{Lynx} calling for thousands of replicated gratings \cite{McEntaffer19}, however, this becomes impractical especially for large, expensive silicon masters that are fabricated using EBL. 
Therefore, an alternative, high-throughput grating replication technique is needed to identify a process capable of meeting this manufacturing requirement. 

\section{Conclusions and Outline of This Thesis}\label{sec:ch1_conclusions}
%%%%%%%%%%%%%%%%%%%%%%%%%%%%%%%%%%%%%%%%%--------------------------------------------------
Measuring the diffuse, highly-ionized baryonic content in the extended halos of isolated galaxies, groups or clusters, and the warm-hot intergalactic medium through soft x-ray absorption spectroscopy of active galactic nuclei is a main scientific objective for the currently-planned \emph{Lynx X-ray Observatory} \cite[\emph{cf.\@} \cref{sec:astro_plasmas}]{Lynx_web,Gaskin19}. 
These observations demand high spectral sensitivity coupled with high spectral resolving power, $\mathscr{R}$, to enable faint absorption lines to be distinguished from continua at a level that offers a significant improvement over the capabilities of current instruments on board the spacecraft observatories \emph{Chandra} and \emph{XMM-Newton}. 
While microcalorimeters satisfy this need for many other areas of x-ray spectroscopy, a state-of-the-art grating spectrometer with a large effective collecting area is better suited for measuring faint absorption lines from abundant highly-charged ions that fall in the soft x-ray bandpass. 
The specialized x-ray reflection gratings described in \cref{sec:grating_tech_dev} are one of two technologies considered for the \emph{Lynx XGS} \cite{McEntaffer19}, with the other being critical-angle transmission gratings \cite{Heilmann11,Heilmann16,Gunther19}.
From the standpoint of grating fabrication, the main technological challenges are: 
\begin{enumerate}[noitemsep]
    \item producing a master grating with a sawtooth surface-relief profile that enables 
    \begin{enumerate}[noitemsep]
        \item total absolute diffraction efficiency in the soft x-ray exceeding \SI{40}{\percent}, and 
        \item $\mathscr{R} \gtrapprox 5000$ in a Wolter-I telescope
    \end{enumerate}
    and additionally,  
    \item mass-replicating blazed gratings to populate modular grating arrays for a soft x-ray spectrometer such as the \emph{XGS}. 
\end{enumerate}
This dissertation contributes to these two areas of reflection grating research with an emphasis on applied nanofabrication and beamline diffraction-efficiency testing. 
Mentioned at the start of this chapter, the focus in particular is on the implementation of two recently-developed techniques in nanofabrication that are hypothesized to be capable of meeting these goals and their subsequent testing for diffraction efficiency at beamline 6.3.2 of the Advanced Light Source \cite{ALS_web,ALS_632,Underwood96,Gullikson01}; beamline testing for spectral resolving power is beyond the scope of this dissertation. 

First, \cref{ch:master_fab} introduces the process of \emph{thermally-activated selective topography equilibration (TASTE)} \cite{Schleunitz14}, which utilizes \emph{grayscale electron-beam lithography} and \emph{polymer reflow} [\emph{cf.\@} \cref{sec:TASTE}] to generate smooth and continuous reliefs in a thermoplastic resist, as a means for fabricating a master grating. %sec:TASTE
This technique has the key advantage that blazed groove facets can, in principle, be patterned over a radial groove layout without dependence on crystallographic structure that can lead to a degradation in $\mathscr{R}$ as alluded to in \cref{sec:crystal_etching}. 
While \si{\um}-scale sawtooth structures fabricated by TASTE have been reported on in the literature \cite{Schleunitz10,Schleunitz14,Kirchner16,Kirchner19}, applying this process to x-ray reflection grating technology requires the realization of a blazed-grating surface relief with a groove spacing of a few hundred \si{\nm} that performs with high diffraction efficiency in an extreme off-plane mount when coated with an appropriate metal for reflectivity. 
As described in \cref{ch:master_fab}, this is carried out through the development of a new fabrication recipe for TASTE at the Penn State Nanofabrication Laboratory \cite{PSU_MRI_nanofab} and the implementation of this process for fabricating a blazed-grating prototype with parallel grooves that is suitable for diffraction-efficiency testing \cite{McCoy18,McCoy20}. 
These results serve as proof of concept for x-ray reflection grating fabrication by TASTE and a basis for future experiments that will examine the ability of this process to pattern a radial groove profile that enables high $\mathscr{R}$ in a Wolter-I spectrometer. 

The second subject of this dissertation, which is addressed in \cref{ch:grating_replication}, is on \emph{substrate-conformal imprint lithography (SCIL)} \cite{Verschuuren10,Verschuuren17} as a method for high-throughput grating replication. 
Developed by \textsc{Philips SCIL Nanoimprint Solutions} \cite{philis_scil}, this variant of NIL uses a flexible composite stamp formed from a rigid master template to pattern nanoscale features in an inorganic resist that cures thermodynamically through a silica \emph{sol-gel process}, which differs fundamentally from the behavior of the organic resists described in \cref{sec:nanoimprint}. %
Although SCIL enables the production of several hundred imprints before stamp degradation and avoids many of the detriments associated with large-area imprinting, the sol-gel resist suffers shrinkage dependent on the post-imprint cure temperature.  
To characterize this effect, \cref{ch:grating_replication} describes processing carried out by \textsc{Philips SCIL Nanoimprint Solutions} that uses the \ce{KOH}-etched silicon grating from \cref{sec:crystal_etching} as a master template for producing replicas similar to those fabricated by UV-NIL. 
The diffraction-efficiency performance of such a replica is then compared to that of the silicon master to show that sol-gel resist shrinkage induced by a low-temperature cure is responsible for an effective decrease in blaze wavelength, $\lambda_b$, that results from a facet angle reduction of a few degrees \cite{McCoy20b}. 
Leveraging from the first application of SCIL to x-ray reflection grating technology for the \emph{Water Recovery X-ray Rocket (WRXR)} \cite{Miles17,Miles18b,Miles19,Verschuuren18,Tutt19}, this research serves to inform grating production for future instruments such as \emph{The Rockets for Extended-source X-ray Spectroscopy (tREXS)} \cite{Miles19b,Tutt19b} and \emph{The Off-plane Grating Rocket Experiment (OGRE)} \cite{DeRoo13,McEntaffer14,DeRoo16thesis,Donovan18b,Tutt18,Donovan19a,Donovan19b}, in addition to the \emph{XGS} planned for \emph{Lynx} \cite{McEntaffer19} and other similar instruments, such as \emph{Arcus} \cite{Smith19_arcus,Arcus_web}. 
Conclusions and future outlook for this thesis are provided in \cref{ch:conclusions}.

%\emph{Arcus}, a proposed soft x-ray spectrometer \cite{Smith19_arcus}

% !TEX root = ../YourName-Dissertation.tex
\chapter[Beamline Characterization of Diffraction Efficiency]{Beamline Characterization of \\Diffraction Efficiency}\label{ch:diff_eff} 
%\chapter[Characterization of Diffraction Efficiency for X-ray Reflection Gratings]{Characterization of Diffraction \\Efficiency for X-ray Reflection \\Gratings}\label{ch:diff_eff} 
%%%%%%%%%%%%%%%%%%%%%%%%%%%%%%%%%%%%%%%%%--------------------------------------------------
Blazed reflection gratings are motivated in \cref{ch:introduction} as a technology suitable for the \emph{XGS} of the \emph{Lynx X-ray Observatory} \cite{Lynx_web,Gaskin19}, which has a main scientific objective of measuring the diffuse, highly-ionized baryonic content in extended galactic halos and the intergalactic medium through absorption spectroscopy of active galactic nuclei [\emph{cf.\@} \cref{sec:astro_plasmas}]. 
Described in \cref{sec:grating_tech_intro}, such a spectrometer uses many identical gratings aligned into modular arrays that each intercept radiation in an \emph{extreme off-plane mount}, where the cone half-opening angle, $\gamma$, is on the order of a degree while the azimuthal incidence angle, $\alpha$, is free to match the blaze angle, $\delta$, in a \emph{Littrow configuration}, where the azimuthal diffracted angle associated with the grating blaze is $\beta = \alpha = \delta$ [\emph{cf.\@} \cref{fig:conical_reflection_edit}]. 
With the research presented in \cref{ch:master_fab,ch:grating_replication} utilizing diffraction-efficiency testing as a means for characterizing the performance of fabricated x-ray reflection gratings [\emph{cf.\@} \cref{sec:ch1_conclusions}], this chapter outlines the relevant beamline methodology in \cref{sec:als_testing} following an established test procedure intended for the extreme off-plane geometry \cite{Miles18}. 
Additionally, \cref{sec:integral_method} describes how, with the aid of the \textsc{PCGrate-SX} software package (\textsc{I.\ I.\ G., Inc.\@}) \cite{PCGrate_web}, measured diffraction-efficiency data can be modeled using the \emph{integral method} to solve the Helmholtz equation for a periodic surface-relief boundary that represents the interface between vacuum and a reflection grating. 
A summary of these methods for characterizing soft x-ray diffraction efficiency is then provided in \cref{sec:ch2_summary}.\footnote{The beamline experimentation described in \cref{sec:als_testing} and utilized in \cref{ch:master_fab,ch:grating_replication} was carried out under an ALS general user proposal agreement lasting through 2017 and 2018.} 

\section[Reflection Grating Testing at the Advanced Light Source]{Reflection Grating Testing at the ALS}\label{sec:als_testing}
%%%%%%%%%%%%%%%%%%%%%%%%%%%%%%%%%%%%%%%%%--------------------------------------------------
Beamline 6.3.2 of the Advanced Light Source (ALS) \cite{ALS_web,ALS_632} at Lawrence-Berkeley National Laboratory \cite{LBNL_web} provides a test station for extreme ultraviolet (EUV) and soft x-ray reflectometry where, under high vacuum, highly-coherent, monochromatic radiation tunable across $\SI{40}{\nm} \gtrapprox \lambda \gtrapprox \SI{1}{\nm}$ strikes a stage-mounted optic while a movable photodiode detector is used to measure the intensity of ingoing and outgoing radiation \cite{Underwood96,Gullikson01}. 
For the physical reasons outlined in \cref{sec:SXR_med}, soft x-rays and EUV radiation are easily absorbed by carbon, nitrogen and oxygen atoms found in air molecules, and as a result, optical testing occurs under high vacuum with the aide of a graphical user interface that enables control of the beam wavelength, $\lambda$, in addition to optic-mount and detector motion. 
Using this facility, the specular reflectivity of a mirror [\emph{cf.\@} \cref{eq:fresnel_refl_def}]: 
\begin{equation}
 \mathcal{R} \left( \lambda \right) \equiv \frac{\mathcal{I}''\!\left( \lambda \right)}{\mathcal{I}_{\text{inc}} \!\left( \lambda \right)} ,
 \end{equation} 
can be experimentally determined for a given incidence angle by measuring the intensity of the reflected beam, $\mathcal{I}''\left( \lambda \right)$, relative to that of the incident beam, $\mathcal{I}_{\text{inc}} \!\left( \lambda \right)$. 
Similarly, the \emph{absolute diffraction efficiency} of a reflection grating, defined as the intensity ratio between the $n^{\text{th}}$ diffracted order and the unobstructed beam \cite{Loewen97}: %\cref{eq:scalar_efficiency}:
\begin{equation}\label{eq:diffraction_efficiency}
 \mathscr{E}_n \! \left( \lambda \right)  \equiv \frac{\mathcal{I}_n \! \left( \lambda \right)}{\mathcal{I}_{\text{inc}} \! \left( \lambda \right)} \quad \text{for } n = 0, \pm 1, \pm 2, \pm 3 \dotsc , 
 \end{equation} 
can also be measured using this beamline facility by comparing the intensity of each propagating order, $\mathcal{I}_n \! \left( \lambda \right)$, to $\mathcal{I}_{\text{inc}} \!\left( \lambda \right)$ for a particular grating geometry established using stage rotations inside the test chamber. 

A photograph of the chamber interior, which features an aperture for the monochromatic beam, detectors attached to goniometric staging and a central optic mount with linear and rotational degrees of freedom is shown in \cref{fig:beamline_photoA}, where the distance from the point of incidence on the optic mount to the detector staging (\emph{i.e.}, the \emph{throw}) is $L \approx \SI{235}{\mm}$.  
\begin{figure}
 \centering
 \includegraphics[scale=0.172]{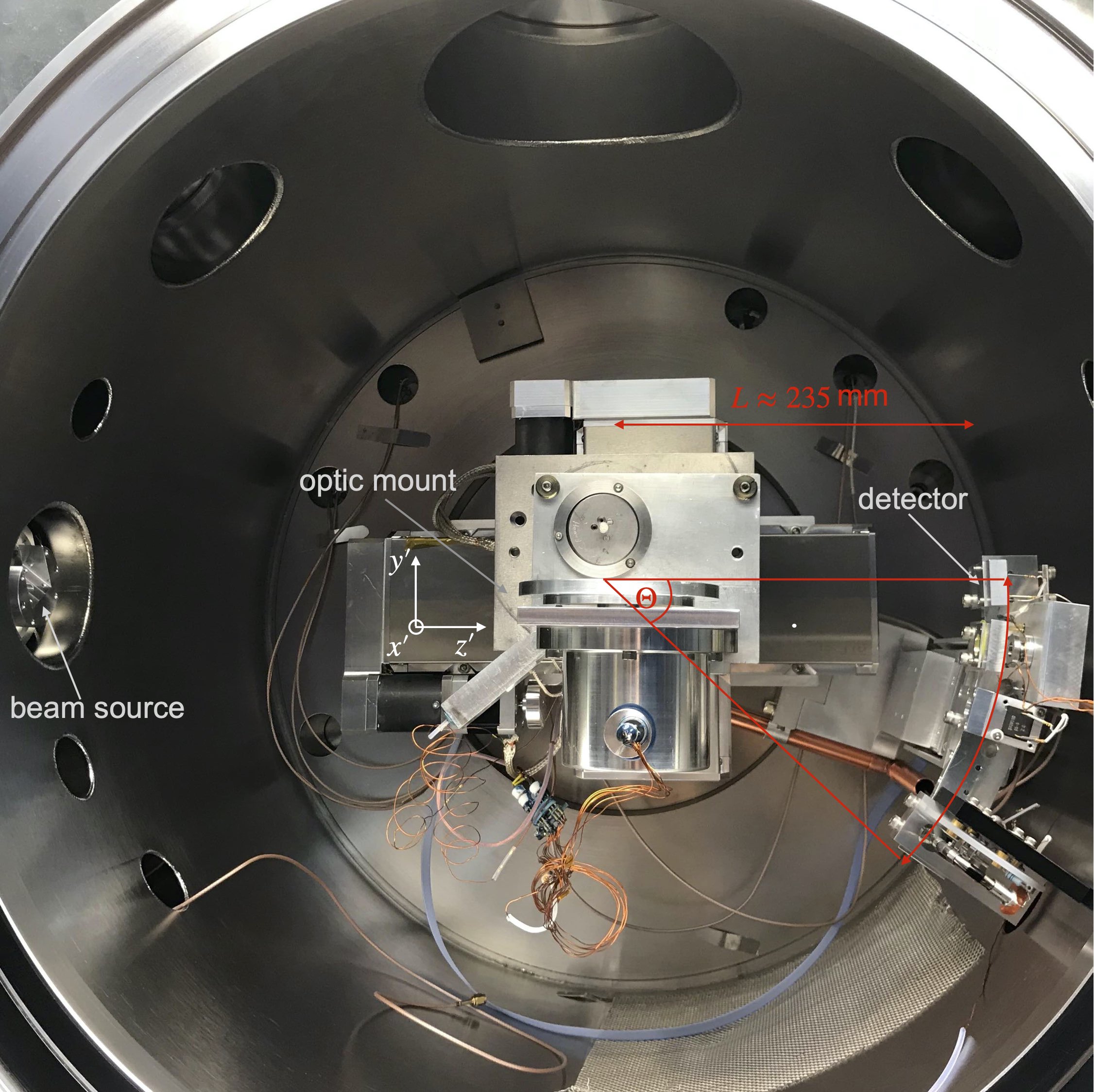}
 \caption[Overview of the station for soft x-ray reflectometry at beamline 6.3.2 of the Advanced Light Source (ALS)]{Beamline 6.3.2 of the ALS provides a station for EUV and soft x-ray reflectometry, where a highly-coherent beam produced by a monochromator travels along the $z'$-direction to a photodiode attached to controllable staging that, owing to the cylindrical shape of the vacuum chamber, can move linearly in the $x'$-direction and goniometrically in the $\Theta$-direction. A reflection grating is mounted on staging with translational and rotational degrees of freedom that can be moved to intercept the beam. This optic mount has rotational degrees of freedom that enable grating geometry to be set to a grazing-incidence, near-Littrow configuration.}\label{fig:beamline_photoA}
 \end{figure}
The detector staging can move linearly along the $x'$-direction and goniometrically about the $x'$-axis such that an angular position $\Theta$ maps to a vertical distance $y' \approx L \sin \left( \Theta \right)$, where any reference point can be used for $\Theta = 0$. 
As described in \cref{sec:grating_tech_intro} and illustrated in \cref{fig:conical_reflection_edit}, an extreme off-plane geometry causes a far-field diffraction pattern where propagating orders are confined to the surface of a cone with a half-opening angle, $\gamma$, of a few degrees; the locations of these orders are described by the \emph{generalized grating equation} \cite[\emph{cf.\@} \cref{eq:off-plane_orders,eq:off-plane_incidence_orders}]{Loewen97}: 
\begin{equation}\label{eq:off-plane_orders_ch2}
 \sin \left( \alpha \right) + \sin \left( \beta \right) = \frac{n \lambda}{d \sin \left( \gamma \right)} \quad \text{for } n = 0, \pm 1, \pm 2, \pm 3 \dotsc ,
 \end{equation}
with $d$ as the groove spacing, $\alpha$ as the azimuthal incidence angle and $\beta$ as the azimuthal diffracted angle of the $n^{\text{th}}$ order. 
Therefore, $\mathscr{E}_n \! \left( \lambda \right)$ can be determined according to \cref{eq:diffraction_efficiency} by sampling the diffracted arc as a function of $x'$ and $y'$ to measure $\mathcal{I}_n \! \left( \lambda \right)$ for each propagating order. 

Before $\mathscr{E}_n \! \left( \lambda \right)$ can be measured across a range wavelengths, a series of steps must be carried out to establish a desired grating geometry with specified values for $\alpha$ and $\gamma$. 
The condition for \emph{total external reflection (TER)} [\emph{cf.\@} \cref{sec:planar_interface}] requires that the groove-facet incidence angle for a grating with a blaze angle $\delta$, given by \cref{eq:angle_on_groove}, be smaller than the \emph{critical angle}, $\zeta_c (\omega)$ [\emph{cf.\@} \cref{eq:critical_angle}]: 
\begin{equation}\label{eq:TER_condition_grat}
 \zeta \equiv \arcsin \left[ \sin \left( \gamma \right) \cos \left( \delta - \alpha \right) \right] < \zeta_c (\omega) \approx \sqrt{2 \delta_{\nu} (\omega)} 
 \end{equation} 
with $\tilde{\nu}(\omega) \equiv 1 - \delta_{\nu} (\omega) + i \xi(\omega)$ as the complex index of refraction for given material [\emph{cf.\@} \cref{sec:x-ray_index}]. 
Moreover, \cref{sec:grating_tech_intro} motivates the use of the Littrow configuration with $\alpha = \beta = \delta$ and $\zeta = \gamma$. 
In practice, however, a \emph{near-Littrow} configuration with $\alpha \approx \delta$ and $\zeta \lessapprox \gamma$ is realized for beamline testing so that the critical-angle requirement is $\gamma \lessapprox \zeta_c (\omega)$. 
The methodology used for establishing such an extreme off-plane geometry is described in \cref{sec:geo_constrain} while \cref{sec:measure_efficiency} outlines how diffraction efficiency can be characterized by measuring $\mathscr{E}_n \! \left( \lambda \right)$ for a range of monochromator wavelengths. 
First, relevant details of the monochromatic beam used for testing are provided in \cref{sec:als_beam}. 

\subsection{The Monochromatic Beam}\label{sec:als_beam}
%%%%%%%%%%%%%%%%%%%%%%%%%%%%%%%%%%%%%%%%%--------------------------------------------------
The central component of the ALS is a particle accelerator that functions as an electron storage ring for each of the $\gtrapprox 40$ beamlines at the facility \cite{ALS_about}. 
This machine maintains highly-relativistic electrons that circulate a large vacuum system consisting of \num{12} straight sections that connect with one another so as to approximate a circle with a \SI{197}{\metre} circumference \cite{Attwood17}. 
Constituting an electrical current of about \SI{400}{\milli\ampere}, these electrons travel in pulsed bunches spaced by \si{\nano\second}-timescales such that the typical electron energy is
\begin{equation}\label{eq:ALS_energy}
 \mathcal{E}_e^{\text{ALS}} = \frac{m_e c_0^2}{\sqrt{1-\frac{v^2}{c_0^2}}} \approx \num{3720} m_e c_0^2 \approx \SI{1.9}{\giga\electronvolt} ,
 \end{equation}
where $v$ is an electron speed approaching the speed of light, $c_0$ \cite{Attwood17,Underwood96}.
At each intersection between straight sections in the storage ring, a \emph{bending magnet} provides a uniform magnetic field, $\mathbold{B}_0$, that deflects electrons in a curved path with a \emph{Lorentz force} given by \cref{eq:Lorentz_force} in the absence of an electric field:
\begin{equation}\label{eq:Lorentz_force_3}
 \mathbold{F} = m_e \mathbold{a} = - q_e \mathbold{v} \times \mathbold{B}_0 ,
 \end{equation}
where $\mathbold{v}$ is the velocity vector for a single electron and $\mathbold{a} \equiv \dv*{\mathbold{v}}{t}$ is its acceleration vector. 
Although many experiments at the ALS utilize magnetic devices with periodic structure (\emph{i.e.}, \emph{wigglers} and \emph{undulators}) to produce various forms of synchrotron radiation \cite{Attwood17}, beamline 6.3.2 uses bending-magnet radiation as a source for its grating monochromator system \cite{ALS_632,Underwood96,Gullikson01}. 
Therefore, basic properties of the monochromatic beam can be inferred from considering the nature of synchrotron radiation from a bending magnet and the way in which it is filtered by the grating monochromator system; this is outlined in the following paragraphs. 

As illustrated in \cref{fig:bending_magnet}, the vector $\mathbold{v}$ for electron velocity has a direction tangential to the curved path of an electron and orthogonal to $\mathbold{B}_0$ so that the vector $\mathbold{a}$ for acceleration points radially inward. 
\begin{figure}
 \centering
 \includegraphics[scale=0.3]{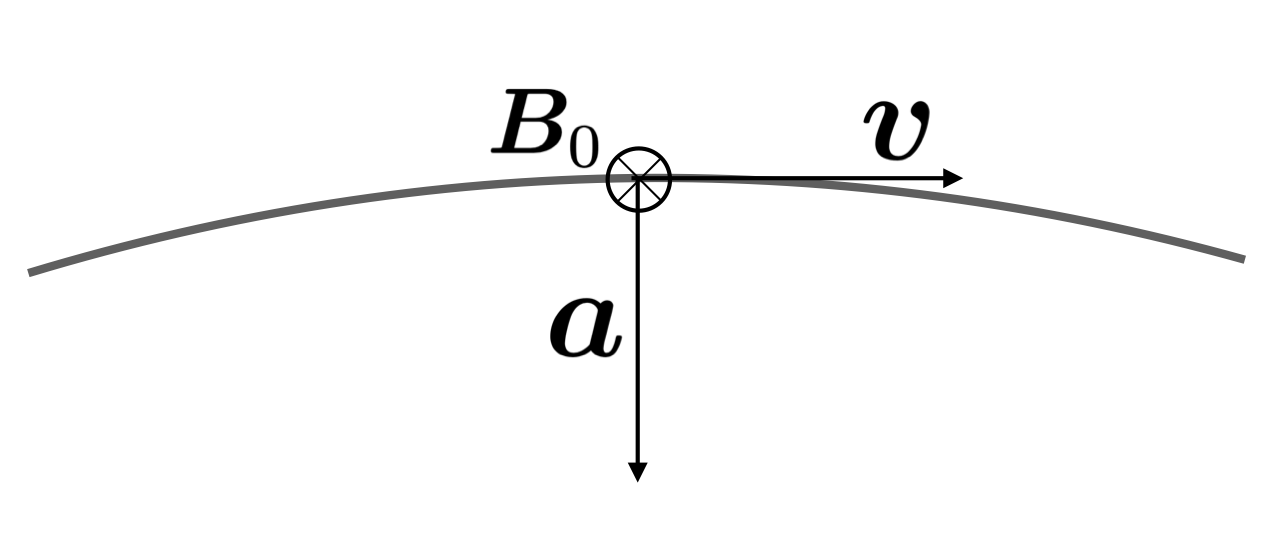}
 \caption[Orientation of velocity, acceleration and applied magnetic field vectors for bending-magnet synchrotron radiation.]{Orientation of velocity, acceleration and applied magnetic field vectors ($\mathbold{v}$, $\mathbold{a}$ and $\mathbold{B}_0$, respectively) for bending-magnet synchrotron radiation. These vectors are each orthogonal to one another, with $\mathbold{B}_0$, which is produced by the bending magnet, pointing into the page. The gray curve represents an approximately-circular electron trajectory with a radius of curvature, $R$, such that $\mathbold{v}$ points tangentially and $\mathbold{a}$ points radially inward.}\label{fig:bending_magnet}. 
 \end{figure}
In principle, an electron in its rest frame radiates as a result of this centripetal acceleration according to the $\sin^2 \left( \vartheta \right)$ torus shape for \emph{Larmor radiation} shown in \cref{fig:torus_radition}. 
In the laboratory frame on the other hand, this radiation pattern is highly beamed along the direction of $\mathbold{v}$ (\emph{i.e.}, $\vartheta \approx 0$), where the opening angle scales with $\sqrt{1- \left( v / c_0 \right)^2}$, the inverse of the \emph{Lorentz factor} \cite{Attwood17}. 
Because $v \equiv \abs{\mathbold{v}}$ is constant in time for purely centripetal motion, the relativistic acceleration vector in \cref{eq:Lorentz_force_3} can be approximated as
\begin{subequations}
\begin{equation}\label{eq:Lorentz_accel_approx}
 \mathbold{a} = \frac{\mathbold{F}}{m_e} = \dv{t} \left( \frac{\mathbold{v}}{\sqrt{1-\frac{v^2}{c_0^2}}} \right) \approx \frac{1}{\sqrt{1-\frac{v^2}{c_0^2}}} \dv{\mathbold{v}}{t} .
 \end{equation}
The magnitude of the Lorentz force, $\abs{\mathbold{F}} = m_e \abs{\mathbold{a}}$, then can be written as 
\begin{equation}\label{eq:Lorentz_force_3_mag_approx}
 \frac{m_e}{\sqrt{1-\frac{v^2}{c_0^2}}} \underbrace{\abs{\dv{\mathbold{v}}{t}}}_{v^2/R_e} = \underbrace{\frac{m_e c_0^2}{\sqrt{1-\frac{v^2}{c_0^2}}}}_{\mathcal{E}_e^{\text{ALS}} \text{ by \cref{eq:ALS_energy}}} \underbrace{\frac{v^2}{c_0^2}}_{\approx 1}  R_e^{-1} = q_e v \abs{\mathbold{B}_0},
 \end{equation}
where $R_e$ is the radius of curvature associated with the electron trajectory, which is given by approximately by 
\begin{equation}
 R_e \approx \frac{\mathcal{E}_e^{\text{ALS}}}{q_e c_0 \abs{\mathbold{B}_0}} \approx \SI{5}{\metre}
 \end{equation}
with $v \approx c_0$ and $\abs{\mathbold{B}_0} \approx \SI{1.27}{\tesla}$ at the ALS \cite{Attwood17}. 
\end{subequations}
In addition to this beaming effect, bending-magnet radiation can be shown to have a broad spectrum due to the pulsed nature of electrons traveling in the storage ring.\footnote{Stated differently, \emph{Heisenberg's uncertainty principle} for energy and time states that a short, pulsed time interval leads to an appreciable spread of possible photon energies \cite[\emph{cf.\@} \cref{sec:gen_first_order}]{Attwood17}.} 
From derivations found in textbooks \cite{Attwood17,Jackson75}, the central wavelength in terms of radiated power across the emitted spectrum is given by 
\begin{equation}
 \lambda_c = \frac{4 \pi m_e}{3 q_e \abs{\mathbold{B}_0}} \approx \SI{0.407}{\nm} \quad \text{with } \abs{\mathbold{B}_0} \approx \SI{1.27}{\tesla}
 \end{equation}
for an ALS bending magnet, which corresponds to a \emph{tender x-ray} [\emph{cf.\@} \cref{sec:histroical}] with a photon energy of $\mathcal{E}_{\gamma} \approx \SI{3}{\kilo\electronvolt}$. 

Bending-magnet radiation that enters the grating monochromator system at beamline 6.3.2 is spectrally purified to produce a nearly monochromatic beam centered at a specified wavelength $\SI{40}{\nm} \gtrapprox \lambda \gtrapprox \SI{1}{\nm}$. 
Briefly, this system consists of a set of optics that focuses the synchrotron radiation before it is incident on an appropriately-ruled reflection grating that directs radiation at the blazed diffracted angle, $\beta = 2 \delta - \alpha$ [\emph{cf.\@} \cref{sec:grating_tech_intro}], through an exit slit, which is then refocused into a collimated beam \cite{Underwood96}. 
Radiation that passes through this system, however, must be spectrally filtered further with the use of material slabs that have various spectral windows for transmission. 
Prominent absorption edges at EUV and soft x-ray wavelengths caused by these low-to-mid $\mathcal{Z}$ materials are also used for wavelength calibration procedures at the beamline. 
Moreover, the implementation of a triple-mirror order sorter is required to ensure spectral purity for radiation with $\lambda \gtrapprox \SI{2.8}{\nm}$ \cite{Gullikson01}. 
This beam of radiation ultimately is highly polarized along the vertical $y'$-direction labeled in \cref{fig:beamline_photoA}, which manifests as \emph{s-polarization} for an intercepting mirror flat\footnote{This refers to the electric field of the incident electromagnetic wave being perpendicular to plane of incidence and parallel to the surface. However, reflectivity is virtually polarization insensitive for grazing-incidence soft x-rays [\emph{cf.\@} \cref{sec:reflectivity_polarization}].} \cite{Muhlestein19}. 

As with any realistic source of radiation, the beam produced by the grating monochromator is associated with a finite coherence length, $\ell_{\text{coh}}$, that depends on $\lambda$ and a corresponding bandwidth, $\Delta \lambda$ [\emph{cf.\@} \cref{sec:coherent_states,fig:coherence_length}]. 
The grating monochromator at beamline 6.3.2 is reported to exhibit a relative spectral bandwidth of $\lambda / \Delta \lambda \lessapprox \num{7000}$ \cite{ALS_632,Underwood96,Gullikson01}, resulting in $\ell_{\text{coh}}$ on the order of \SI{10}{\um}. 
While the beam is mildly diverging with a cross-sectional diameter of $\lessapprox \SI{0.5}{\mm}$ at the optic mount, this level of coherence is sufficient for producing clearly-defined diffracted orders of wavelength $\lambda$, with locations given by \cref{eq:off-plane_orders} for a geometry parameterized by the incidence angles $\alpha$ and $\gamma$. 
\begin{figure}
 \centering
 \includegraphics[scale=0.5]{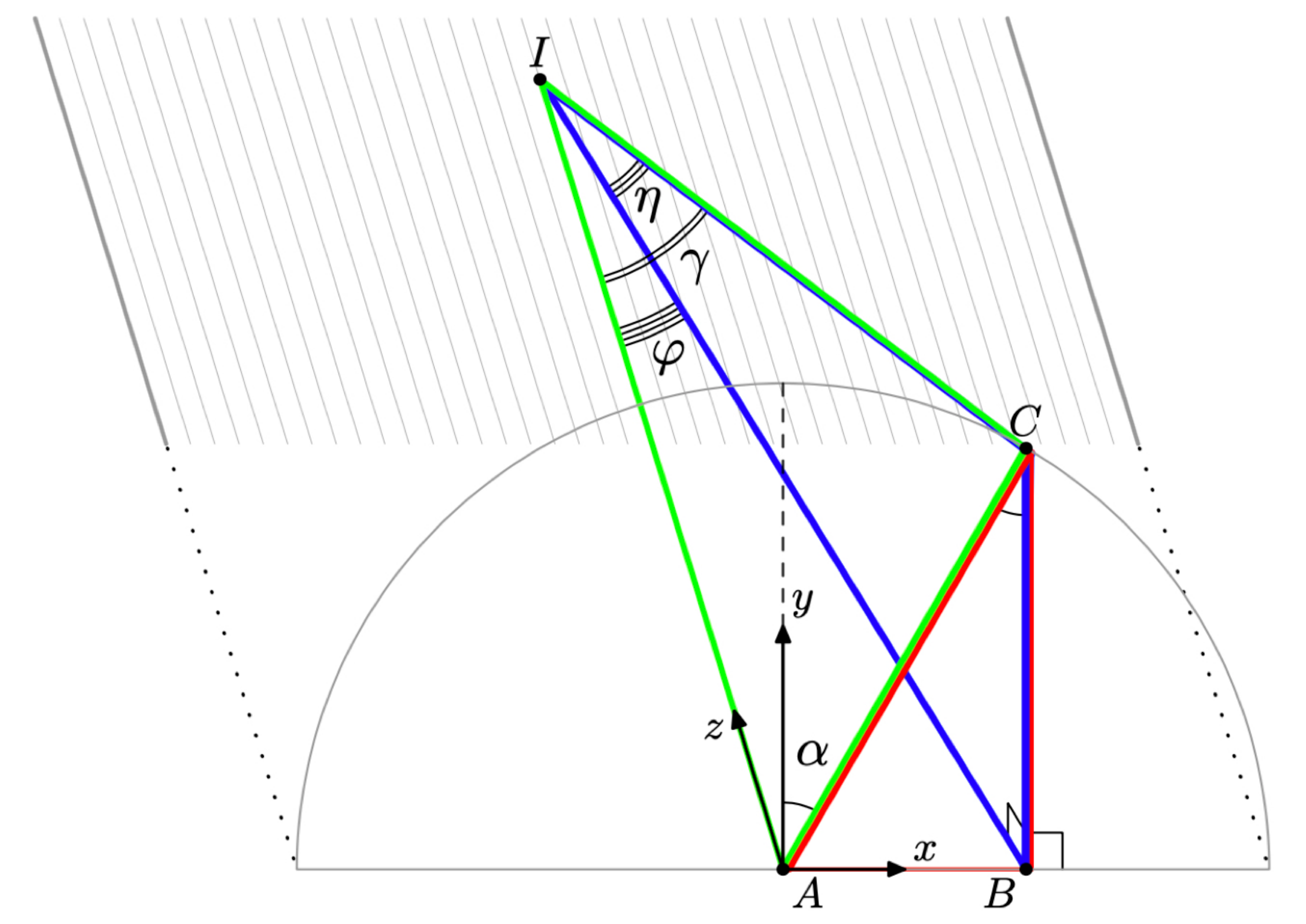}
 \caption[Angles relevant for beamline diffraction-efficiency testing]{Angles relevant for beamline diffraction-efficiency testing. The grating incidence angles $\alpha \equiv \angle ACB$ and $\gamma \equiv \angle AIC$ introduced in \cref{fig:conical_reflection_edit} are controlled through rotations about principal axes, which are parametrized by the angles $\eta \equiv \angle CIB$ and $\varphi \equiv \angle AIB$ given by \cref{eq:graze_angle,eq:yaw_angle} \cite{McCoy20}.}\label{fig:grating_angles} 
 \end{figure} 
Using the grating coordinate system defined in \cref{fig:grating_angles}, the central wave mode of the monochromatic beam with wave number $k_0 \equiv 2 \pi / \lambda$ has a wave vector written as 
\begin{subequations}
\begin{equation}\label{eq:beam_vec_grat}
 \mathbold{k} = k_x \mathbold{\hat{x}} + k_y \mathbold{\hat{y}} + k_z \mathbold{\hat{z}} = - k_0 \left[ \sin \left( \alpha \right) \sin \left( \gamma \right) \mathbold{\hat{x}} + \cos \left( \alpha \right) \sin \left( \gamma \right) \mathbold{\hat{y}} - \cos \left( \gamma \right) \mathbold{\hat{z}} \right] ,
 \end{equation}
whereas in the frame of the test chamber, this vector is taken to be oriented along the horizontal $z'$-direction labeled in \cref{fig:beamline_photoA} with 
\begin{equation}\label{eq:beam_vec_lab}
 \mathbold{k} = k_0 \mathbold{\hat{z}'} .
 \end{equation}
\end{subequations}
However, the implementation of the order sorter mentioned above shifts the beam direction so that $\mathbold{k}$ departs slightly from \cref{eq:beam_vec_grat,eq:beam_vec_lab}. 
Nonetheless, once a grating geometry has been established using the methodology outlined in \cref{sec:geo_constrain}, this beam can be used to measure $\mathscr{E}_n \! \left( \lambda \right)$ across a range of wavelengths [\emph{cf.\@} \cref{sec:measure_efficiency}]. 

\subsection{Constraining Grating Geometry}\label{sec:geo_constrain}
%%%%%%%%%%%%%%%%%%%%%%%%%%%%%%%%%%%%%%%%%--------------------------------------------------
Illustrated in \cref{fig:conical_reflection_edit}, angles $\alpha$ and $\gamma$ parameterize the direction of the incident radiation relative to the grating coordinate system defined in \cref{fig:grating_angles}, where $\mathbold{\hat{x}}$ is the grating-dispersion direction, $\mathbold{\hat{y}}$ is the cross-dispersion direction, and $\mathbold{\hat{z}}$ is the groove direction. 
At beamline 6.3.2, however, these angles are controlled indirectly through rotations about principal axes on the optic-mount stage \cite{Gullikson01,Miles18}. 
Two of these degrees of freedom are stage-controllable: the graze angle relative to the optic-mount surface, $\eta$, and another angle, $\varphi$, that describes the rotation of the grating about the axis normal to the optic-mount surface (\emph{i.e.}, grating yaw). 
As illustrated in \cref{fig:grating_angles}, the former angle is related to $\alpha$ and $\gamma$ through
\begin{equation}\label{eq:graze_angle}
 \sin \left( \eta \right) = \sin \left( \gamma \right) \cos \left( \alpha \right)
 \end{equation}
while, taking $\varphi = 90^{\circ}$ to correspond to an exact in-plane mount with $\sin \left( \gamma \right) = 1$, a relation for the latter can be written as 
\begin{equation}\label{eq:yaw_angle}
 \sin \left( \varphi \right) = \tan \left( \alpha \right) \tan \left( \eta \right) ,
 \end{equation}
or alternatively, $\cos \left( \varphi \right) = \cos \left( \gamma \right) / \cos \left( \eta \right)$ \cite{Miles18}.

The third degree of freedom (not shown in \cref{fig:grating_angles}) is an angle for grating roll, $\phi$, that nominally remains fixed at $\phi = 0$ so that the optic mount is leveled with $\eta = 0$. 
Although it is not stage-controllable, $\phi$ must be constrained experimentally along with $\eta$ and $\varphi$ to establish a near-Littrow configuration with $\alpha \approx \delta$ and $\gamma \approx \zeta < \zeta_c (\omega)$ at the beamline [\emph{cf.\@} \cref{eq:TER_condition_grat}]. 
This is done by first installing the grating on the optic mount, as shown in \cref{fig:beamline_photoB}, with $\eta \approx 0$ and $\phi \approx 0$ verified by the tilt of the optic mount using a spirit level. 
\begin{figure}
 \centering
 \includegraphics[scale=0.105]{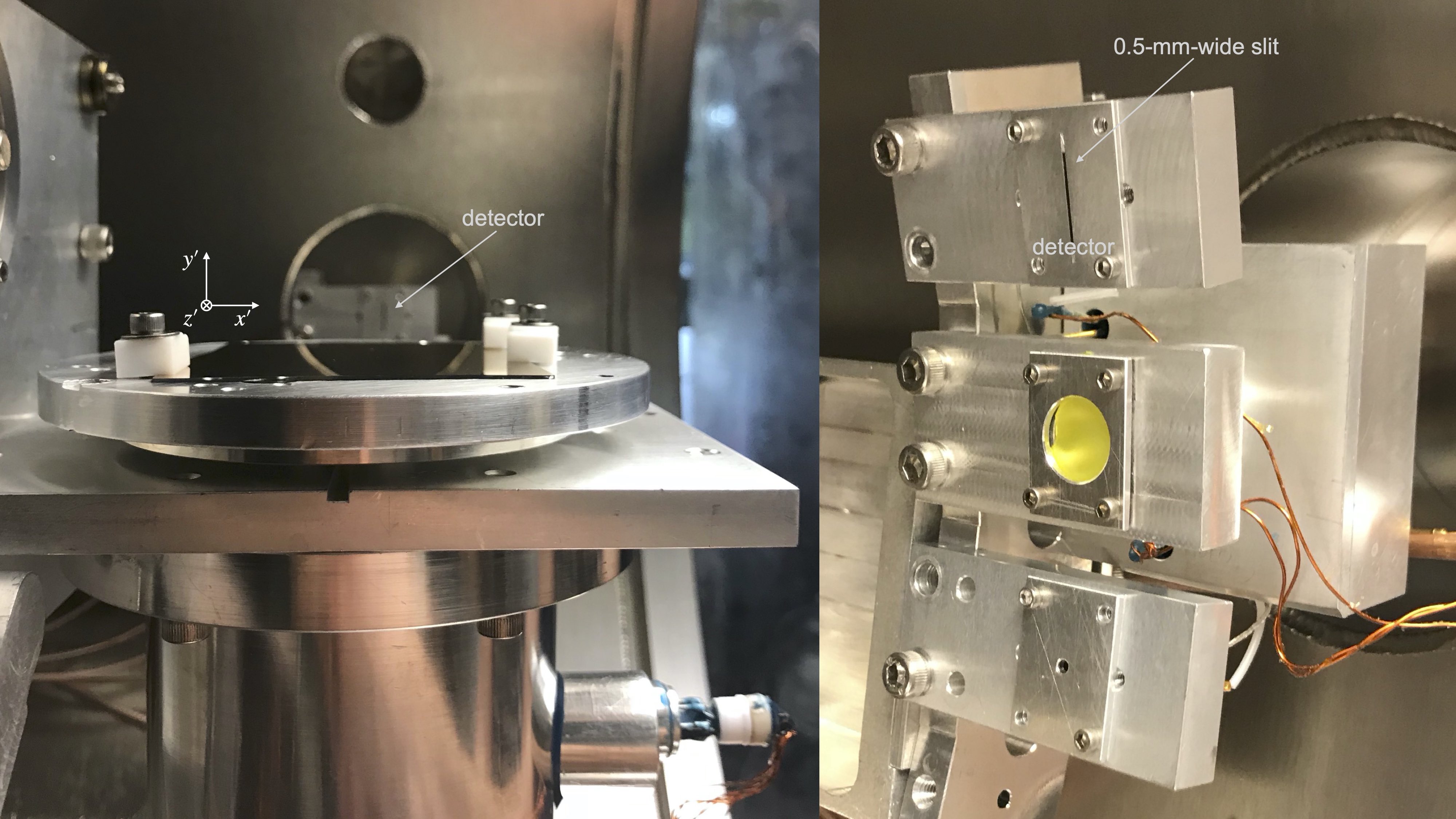}
 \caption[View of the optic mount and masked photodiode detector at beamline 6.3.2 of the ALS]{Inside the test chamber shown in \cref{fig:beamline_photoA}, a \SI{0.5}{\mm}-wide, vertical slit is installed to mask the photodiode detector so as to allow the intensity of each propagating order, $\mathcal{I}_n \! \left( \lambda \right)$, to be measured in isolation. With the grating groove direction approximately aligned with the $z'$-axis for an extreme off-plane mount, the grating dispersion and cross-dispersion directions are approximately aligned with the linear $x'$-direction and the goniometric $y'$-direction motion of the detector staging, respectively.}\label{fig:beamline_photoB}
 \end{figure}
Meanwhile, the groove direction is approximately aligned with the beam direction with $\varphi \approx 0$ so that the $(x',y',z')$ chamber coordinate system [\emph{cf.\@} \cref{fig:beamline_photoA,fig:beamline_photoB}] is approximately coincident with the $(x,y,z)$ grating coordinate system [\emph{cf.\@} \cref{fig:grating_angles}]. 
The grating is then carefully adjusted to occult the beam so as to position the point of interception close to the hub of stage-rotation axes. 

With the leveled grating intercepting the beam, the optic mount can be rotated about the $x'$-axis at the point of incidence to ensure that the angle between the direct beam and the $0^{\text{th}}$-order beam is roughly $2 \eta$, where $\eta$ is the targeted graze angle [\emph{cf.\@} \cref{eq:graze_angle}]. 
The grating-dispersion direction, $\mathbold{\hat{x}}$, then is approximately parallel with the linear stage movement of the detector along the $x'$-direction so that the $x$-position of each propagating order [\emph{cf.\@} \cref{eq:linear_dispersion}] can be measured in a configuration with $\varphi \approx 0$ and $\phi \approx 0$ as
\begin{equation}\label{eq:disp_approx}
 x_n \approx x_0 + \frac{n \lambda L}{d} ,
 \end{equation}
where $x_0$ is the $x$-position of $0^{\text{th}}$ order. 
The ability to separate each component of $\mathcal{I}_n \! \left( \lambda \right)$, however, depends on the collecting-area dimensions of the detector used for data collection. 
That is, with $\lambda L / d$ typically on the order of \si{\mm} while the diffracted-arc radius [\emph{cf.\@} \cref{eq:blaze_wavelength}] is $r = L \sin \left( \gamma \right) \lessapprox \SI{8}{\mm}$ for $\gamma \lessapprox 2^{\circ}$, %eq:arc_radius
a detector with a narrow collecting area is required to allow each order to be measured in isolation. 
Moreover, if the detector length, $\ell_{\text{det}}$, is larger than $r$, $\mathcal{I}_n \! \left( \lambda \right)$ for each propagating order can be measured using purely horizontal detector motion with an appropriate choice for $\Theta$. 
Although this condition $\ell_{\text{det}} > r$ is not satisfied in both \cref{ch:master_fab,ch:grating_replication} due the use of detectors with different dimensions, a \SI{0.5}{\mm}-wide, vertical slit was in each case installed for the purposes of order separation [\emph{cf.\@} \cref{fig:beamline_photoB}]. %; an example of this is shown in \cref{fig:beamline_photoB}. 

For a grating geometry with zero yaw (\emph{i.e.}, $\varphi \approx 0$) and $\eta$ of a few degrees, \cref{eq:yaw_angle} requires that $\alpha \approx 0$ while the arc radius is given by $r \approx L \sin \left( \eta \right)$ [\emph{cf.\@} \cref{eq:arc_radius,eq:graze_angle}]. 
Increasing grating yaw by azimuthal stage rotation for $\varphi$ while $\eta$ remains fixed causes $\alpha$ to increase by \cref{eq:yaw_angle} and $r$ to increase as 
\begin{equation}
 r = L \sqrt{\sin^2 \left( \eta \right) + \sin^2 \left( \varphi \right) \cos^2 \left( \eta \right)} = L \sqrt{1 - \cos^2 \left( \varphi \right) \cos^2 \left( \eta \right)} .
 \end{equation}
A grating geometry with $\alpha \approx \delta$ hence can be established by setting $\eta$ and $\varphi$ appropriately according to \cref{eq:graze_angle,eq:yaw_angle} with $\phi \approx 0$. 
In principle, $L$ changes as the the detector moves along the $x'$-direction with focal corrections on the order of tens of \si{\um} within \SI{10}{mm} of detector travel, but for the analysis considered in this dissertation, these corrections are ignored so that $L$ is fixed for all detector positions considered. 
Order locations given by the generalized grating equation [\emph{cf.\@} \cref{eq:off-plane_orders,eq:off-plane_orders_ch2}] then can be mapped using $(x',y')$ positions associated with the detector staging that are taken to coincide directly with the $(x,y)$ coordinates for grating dispersion such that aberrations arising from the misalignment between these two planes are neglected. 
While the $x$-positions of propagating orders can be determined by sampling the diffracted arc along the $x'$-direction, their $y$-positions require knowledge of system throw, $L \approx \SI{235}{\mm}$ [\emph{cf.\@} \cref{fig:conical_reflection_edit}], to map the goniometric angle associated with the stage motion, $\Theta$, to a $y'$-coordinate using $y' = L \sin \left( \Theta \right)$ [\emph{cf.\@} \cref{fig:beamline_photoA}]. 

A measured value for $L$ can be measured experimentally by comparing the known detector length, $\ell_{\text{det}}$, 
\begin{figure}
 \centering
 \includegraphics[scale=0.8]{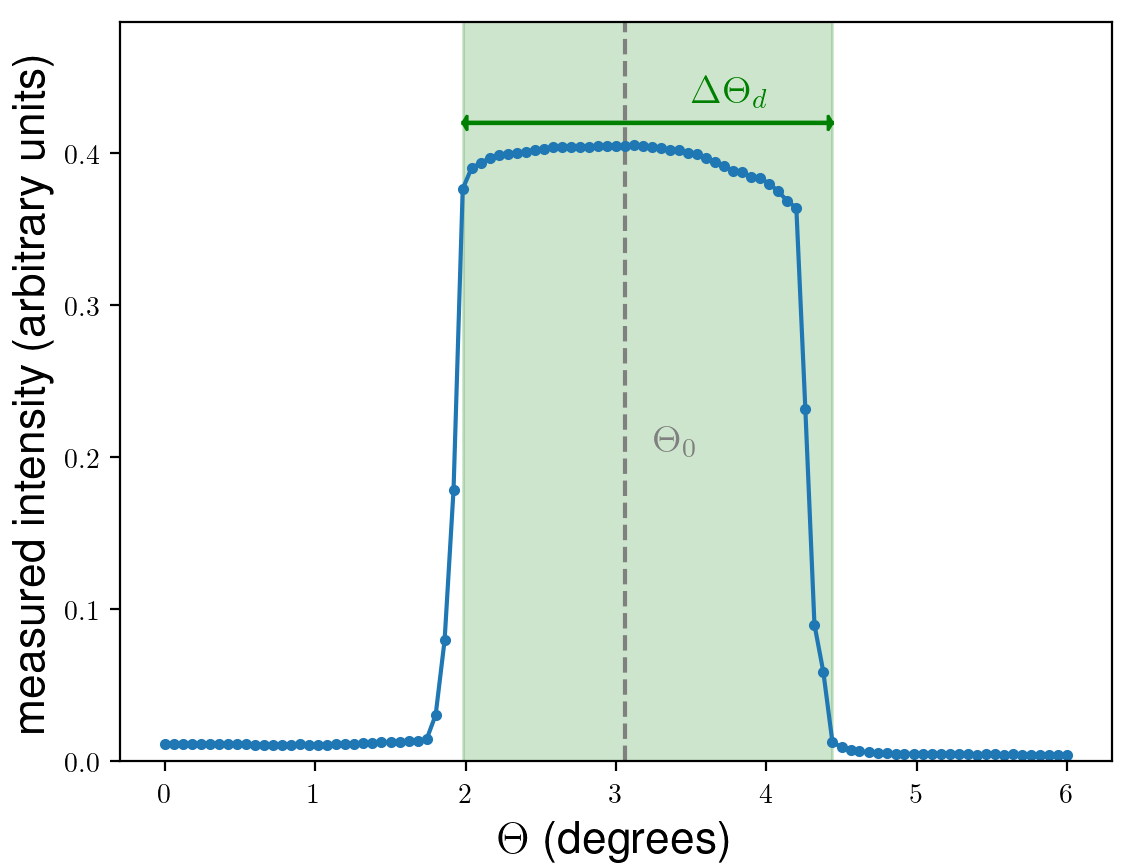}
 \caption[Example of measured intensity data for a $0^{\text{th}}$-order diffracted beam as a function of goniometric angle]{Example of measured intensity data for a $0^{\text{th}}$-order diffracted beam as a function of goniometric angle, $\Theta$, in steps of $\sim$0.06$^{\circ}$. The green shaded region highlights the measured angular size of the detector, $\Delta \Theta_d$, which corresponds to one full translation of the beam in the goniometric direction, along the long direction of the \SI{0.5}{\mm}-wide slit. Additionally, the goniometric centroid angle associated with the measured beam, $\Theta_0$, can be determined through a weighted mean calculation of the data. This process is repeated for each propagating order to derive $\Theta_n$ for the $n^{\text{th}}$ order and ultimately, the corresponding $y$-component, $y_n = L \sin \left( \Theta_n \right)$.}\label{fig:ALS_cent_throw}
 \end{figure}
to the angular size of the detector as measured by a goniometric scan of the $0^{\text{th}}$-order beam, $\Delta \Theta_d$:
\begin{equation}
 \sin \left( \Delta \Theta_d \right) = \frac{\ell_{\text{det}}}{L} \implies L \approx \frac{\ell_{\text{det}}}{\Delta \Theta_d}
 \end{equation}
An example of this is shown in \cref{fig:ALS_cent_throw}, which depicts measured intensity data as a function of $\Theta$, in steps of $\sim$0.06$^{\circ}$. 
Using these data, a measurement for $\Delta \Theta_d$ can be extracted by identifying start and end points for $\Theta$ that correspond to one beam translation across the detector. 
This $\Theta$ range includes the quasi-flat region of the curve, where the full length of the beam falls on the detector, as well as one of its wings, which represents the beam either entering or exiting the active detector area. % [\emph{cf.\@} \cref{fig:ALS_cent_throw}]. 

With a measured value for $L$, a $y$-coordinate for any given diffracted beam can be determined by performing a goniometric scan at an appropriate $x'$-position and then extracting a centroid angle, $\Theta_n$, so that $y_n = L \sin \left( \Theta_n \right)$. 
This is shown for the case of $n=0$ in \cref{fig:ALS_cent_throw}, where a weighted-mean treatment of data is used to extract $\Theta_0$. 
The $x$-positions of propagating orders, on the other hand, require horizontal scans of the diffracted arc at a fixed goniometric angle so that centroids can be extracted. 
 %with an example shown in \cref{fig:ALS_xdata}. 
\begin{figure}
 \centering
 \includegraphics[scale=0.78]{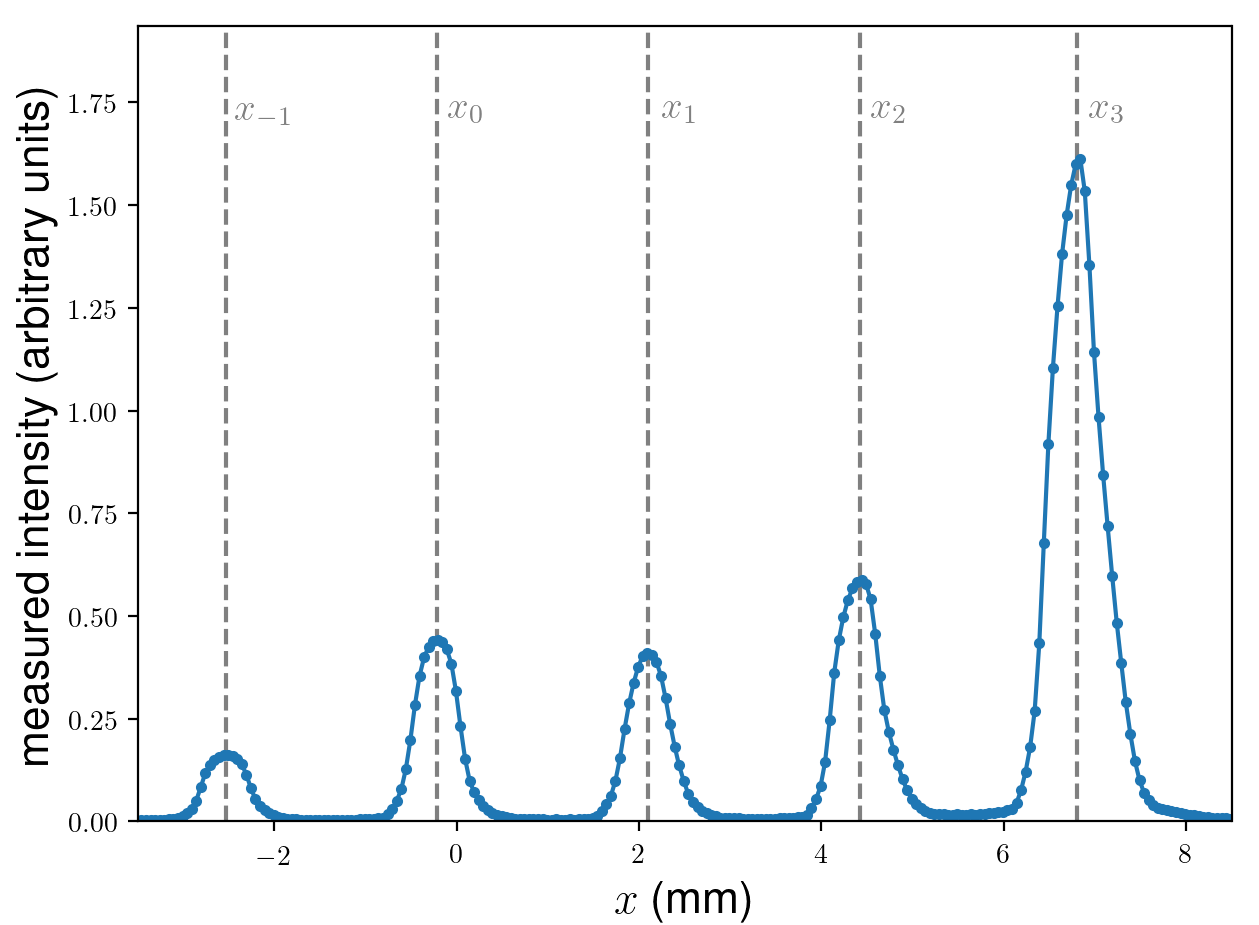}
 \caption[Example of measured intensity data for the diffracted arc as a function of horizontal detector position]{Example of measured intensity data for the diffracted arc as a function of horizontal detector position, $x'$, in steps of \SI{50}{\um}. The $x$-component of each propagating order, $x_n$, is determined using the weighted-mean centroid of each local maximum spaced approximately by $n \lambda L /d$.}\label{fig:ALS_xdata}
 \end{figure}
Example data that were gathered from a diffracted arc scanned horizontally in \SI{50}{\um} steps are shown in \cref{fig:ALS_xdata}, where a clear peak is seen for propagating orders with $n$ ranging from \numrange{-1}{3}. 
The weighted-mean centroid for each peak then yields an $x$-coordinate for the corresponding order [\emph{cf.\@} \cref{eq:disp_approx}] so that with $(x_n,y_n)$ measurements gathered for the diffracted arc, the data can be fit to a circle described by 
\begin{equation}\label{eq:arc_circle} 
 \left( x - x_{\text{cen}} \right)^2 + \left( y - y_{\text{cen}} \right)^2 = r^2,
 \end{equation}
where $(x_{\text{cen}},y_{\text{cen}})$ is the location of the arc center and $r = L \sin (\gamma)$ is its radius. 
\begin{figure}
 \centering
 \includegraphics[scale=1.05]{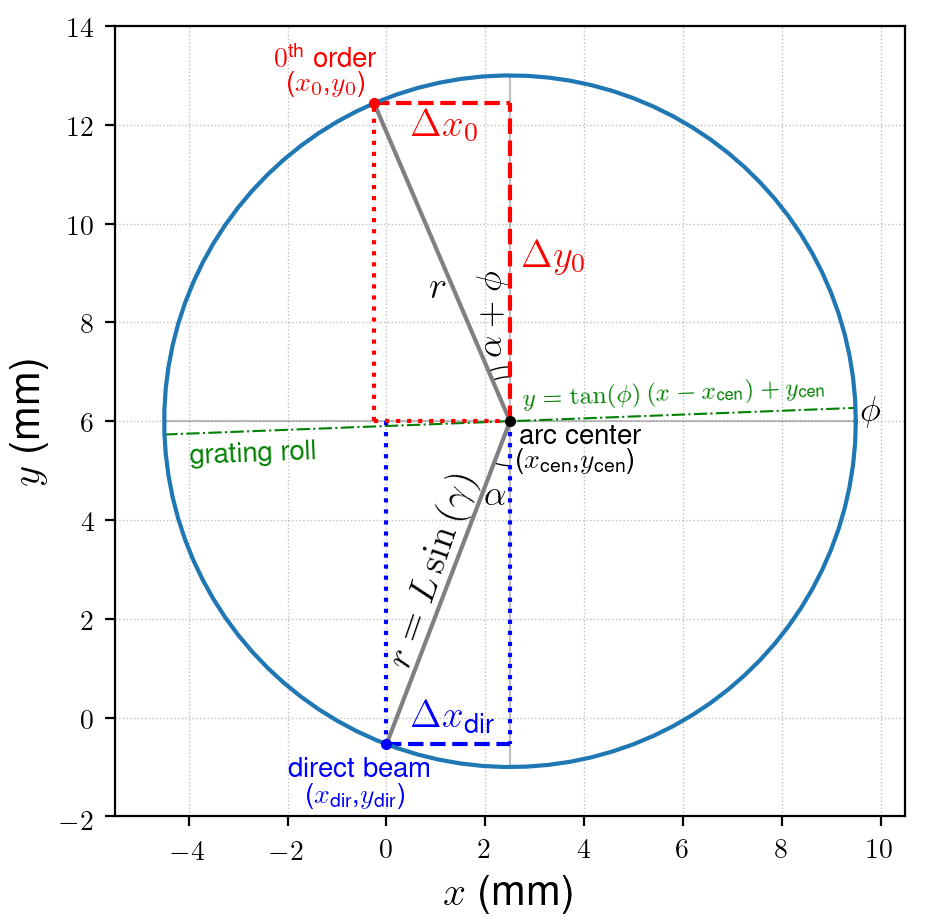}
 \caption[Example of a diffracted arc fit to a circle]{Example of a diffracted arc fit to a circle by \cref{eq:arc_circle} with radius $r$ and arc center $(x_{\text{cen}},y_{\text{cen}})$ from gathered $(x_n,y_n)$ data. The cone opening half-angle, $\gamma$, is related to the system throw, $L$, through $r = L \sin \left( \gamma \right)$ while the azimuthal incidence angle, $\alpha$, follows from $\sin \left( \alpha \right) = \Delta x_{\text{dir}}/r$, where $\Delta x_{\text{dir}}$ is the $x$-distance between $x_{\text{cen}}$ and the direct beam. Grating roll, $\phi$, can be constrained using $\sin \left( \alpha + \phi \right) = \Delta x_0/r$, where $\Delta x_0$ is the $x$-distance between $x_{\text{cen}}$ and $0^{\text{th}}$ order. By \cref{eq:measure_graze}, the substrate graze angle, $\eta$, can be determined using the $y$-distance between $y_{\text{cen}}$ and $0^{\text{th}}$ order, $\Delta y_0$, while the angle for grating yaw, $\varphi$, follows from \cref{eq:measure_yaw}.}\label{fig:ALS_arc}
 \end{figure}
This is illustrated in \cref{fig:ALS_arc}, which depicts a circle of radius $r = \SI{7}{\mm}$ that represents an example diffracted arc with $\gamma \approx 1.7^{\circ}$. 

The positions of $0^{\text{th}}$ order, $(x_0,y_0)$, and the direct beam, $(x_{\text{dir}},y_{\text{dir}})$, are required along with $r$ and $(x_{\text{cen}},y_{\text{cen}})$ to constrain the angular degrees of freedom for the optic mount. 
By \cref{eq:arc_radius}, measured values for $L$ and $r$ yield the half-angle of the cone opening for the diffracted arc:
\begin{equation}\label{eq:measure_gamma}
 \gamma = \arcsin \left( \frac{r}{L} \right) \approx \frac{r}{L} , % + \frac{1}{6} \left( \frac{r}{L} \right)^3 ,
 \end{equation}
where the approximation holds for $r \ll L$. 
The azimuthal incidence angle, $\alpha$, then follows from \cref{eq:graze_angle} with a measured value for the graze angle relative to the grating substrate, $\eta$, which can be determined by comparing goniometric centroid angles for the direct beam and $0^{\text{th}}$ order with knowledge of grating roll \cite{Miles18}. 
Alternatively, this angle can be measured independently of $\gamma$ and $\eta$ using the $x$-distance between the direct beam and the arc center together with the arc radius [\emph{cf.\@} \cref{fig:ALS_arc}]:
\begin{equation}\label{eq:measure_alpha}
 \alpha = \arcsin \left( \frac{\Delta x_{\text{dir}}}{r} \right) ,
 \end{equation}
where $\Delta x_{\text{dir}} \equiv x_{\text{cen}} - x_{\text{dir}}$. 
While the angles $\alpha$ and $\gamma$ fully describe a conical grating geometry for the purposes of modeling diffraction efficiency in \cref{sec:integral_method}, the principle-axis angles defined in \cref{fig:grating_angles} should also be constrained to ensure consistency with \cref{eq:graze_angle,eq:yaw_angle}. 

By definition, a grating with zero roll is perfectly leveled with the $x'$-axis of the laboratory coordinate system so that $x_0 = x_{\text{dir}}$ and as a result, $\eta$ can be determined from $\sin \left( \eta \right) = \Delta y_0 / L$, where $\Delta y_0 \equiv y_0 - y_{\text{cen}}$ is the $y$-distance between $0^{\text{th}}$ order and the arc center. 
With a roll angle $\phi \lessapprox 1^{\circ}$, however, the diffracted arc is rotated slightly with $\Delta x_0 \equiv x_{\text{cen}} - x_0 > \Delta x_{\text{dir}}$. 
This angle can be determined using [\emph{cf.\@} \cref{fig:ALS_arc}] %As shown geometrically in \cref{fig:ALS_arc}, 
\begin{equation}\label{eq:measure_roll}
 \phi = \arcsin \left( \frac{\Delta x_0}{r} \right) - \alpha \approx \frac{\Delta x_0}{r} - \sin \left( \alpha \right) ,
 \end{equation}
where the approximation holds for $\abs{\Delta x_0 - \Delta x_{\text{dir}}} \ll r$. 
A nonzero roll also causes a measurement of $\Delta y_0$ to be underestimated by a factor of $\cos \left( \phi \right)$ such that 
a relation for $\eta$ is given by 
\begin{equation}\label{eq:measure_graze}
 \sin \left( \eta \right) = \frac{\Delta y_0 \sec \left( \phi \right)}{L} \implies \eta \approx \frac{\Delta y_0}{L} \left(1 + \frac{\phi^2}{2} \right)  .
 \end{equation}
Finally, $\varphi$ can be determined geometrically using the right triangle defined by $\Delta x_{\text{dir}}$ and the projection of the $0^{\text{th}}$-order ray into the plane of the grating, $L \cos \left( \eta \right)$, with $L$ as a hypotenuse:\footnote{This can be gleaned from \cref{fig:grating_angles}, where the line segment $\overline{IG}$ represents $L$ so that the line segment $\overline{IB}$ shown in solid blue is equivalent to $L \cos \left( \eta \right)$.}
\begin{equation}\label{eq:measure_yaw}
 \sin \left( \varphi \right) = \frac{\Delta x_{\text{dir}}}{L \cos \left( \eta \right)} \implies \varphi \approx \frac{\Delta x_{\text{dir}}}{L} \left(1 + \frac{\eta^2}{2} \right) ,
 \end{equation}
where the approximations in \cref{eq:measure_graze,eq:measure_yaw} hold for small angles. 

\subsection{Measuring Diffraction Efficiency}\label{sec:measure_efficiency}
%%%%%%%%%%%%%%%%%%%%%%%%%%%%%%%%%%%%%%%%%--------------------------------------------------
Once a desired grating geometry with $\alpha \approx \delta$ and $\gamma \approx \zeta < \zeta_c (\omega)$ has been established using the methodology described in \cref{sec:geo_constrain}, absolute diffraction efficiency, $\mathscr{E}_n$, can be experimentally determined according to \cref{eq:diffraction_efficiency} as function of wavelength, $\lambda$, or photon energy, $\mathcal{E}_{\gamma} \equiv h c_0 / \lambda$ with $h c_0 \approx \SI{1240}{\electronvolt\nm}$. 
For a given monochromatic-beam configuration, the intensity of each propagating order, $\mathcal{I}_n$, can be calculated by identifying the appropriate maxima from a horizontal detector scan [\emph{cf.\@} \cref{fig:ALS_xdata}] and then taking three measurements around the centroid after subtracting out the appropriate noise floors. 
Contributions to this noise include dark current from the photodiode detector, which can be measured using the detector readout in the absence of incident radiation, and diffuse scatter arising from surface roughness on the groove facets [\emph{cf.\@} \cref{sec:rough_surface}], which, in principle, can be estimated using the continuum level in between order maxima. 
The intensity of the direct beam, $\mathcal{I}_{\text{inc}}$, can be measured in a similar way while the optic mount is moved out of the beam path so that the only expected source of noise is dark current. 
These two processes for measuring $\mathcal{I}_n$ and $\mathcal{I}_{\text{inc}}$ can then be repeated as function of $\lambda$ (or $\mathcal{E}_{\gamma}$) using appropriate procedures for adjusting the monochromatic beam. 

Other useful quantities that characterize the performance of a reflection grating can be derived using measured data for $\mathscr{E}_n$ [\emph{cf.\@} \cref{eq:diffraction_efficiency}]. 
First, the \emph{total absolute diffraction efficiency} of a grating is defined as $\mathscr{E}_{\text{tot}} \equiv \sum_n \mathscr{E}_n$ for all propagating orders with $n \neq 0$. 
This quantity describes the fraction of radiation from a grating that is available for spectroscopy whereas the \emph{total response} of the grating, defined as $\mathscr{E}_{\text{tot}} + \mathscr{E}_0$, is a measure of all radiation that is \emph{specularly diffracted}\footnote{In analogy to a specularly-reflected beam from a mirror flat, this definition here is generalized to include propagating orders with diffracted angles described by the grating equation.} from the grating relative to the incident beam. 
In principle, $\mathscr{E}_{\text{tot}} + \mathscr{E}_0$, is equivalent to a measure of specular reflection from a mirror flat with a grazing-incidence angle equal to the groove facet incidence angle for a blazed grating, $\zeta$ [\emph{cf.\@} \cref{sec:refl_account}]. 
Moreover, the \emph{relative diffraction efficiency} of a reflection grating is defined as $\mathscr{E}_n / \mathcal{R}_F$, where $\mathcal{R}_F$ is the Fresnel reflectivity of an equivalent mirror flat at a grazing-incidence angle $\zeta$, which is virtually polarization insensitive at EUV and soft x-ray spectral wavelengths [\emph{cf.\@} \cref{sec:reflectivity_polarization}]. 
Related definitions are the \emph{total relative diffraction efficiency}, $\mathscr{E}_{\text{tot}} / \mathcal{R}_F$, and the \emph{total relative response}, $\left( \mathscr{E}_{\text{tot}} + \mathscr{E}_0 \right) / \mathcal{R}_F$. 
This latter quantity is a measure of all radiation lost to phenomena other than $\lambda$-dependent reflectivity, which includes, but is not limited to, the surface roughness effects described in \cref{sec:rough_surface}. 

\section{Modeling Diffraction Efficiency}\label{sec:integral_method}
%%%%%%%%%%%%%%%%%%%%%%%%%%%%%%%%%%%%%%%%%--------------------------------------------------
Basic physics of gratings are formulated according to a scalar theory of diffraction in \cref{sec:amplitude_phase}, where a reflection grating is treated as a surface of point-source emitters with periodically-varying phase (\emph{i.e.}, a \emph{phase grating}). 
Through performing a spatial Fourier transform of this periodic phase function, it is found that a sawtooth profile gives rise to an intensity pattern where diffraction efficiency is maximized near a certain \emph{blaze wavelength}, $\lambda_b$, for each propagating order [\emph{cf.\@} \cref{eq:blaze_wavelength,eq:blaze_wavelength_general}]. 
Although such a scalar treatment roughly describes the impact of groove shape on grating response, a vector theory of diffraction that takes into account rigorous boundary conditions at the surface of a grating is instead needed for predicting this behavior more accurately. 
With \cref{ch:master_fab,ch:grating_replication} utilizing the \textsc{PCGrate-SX} software package \cite{PCGrate_web} for this purpose, \cref{sec:grating_boundary} describes the general grating value problem while \cref{sec:grat_bound_consider} outlines how the integral method can be used to calculate relative diffraction efficiency under the simplifying assumption of a \emph{perfectly conducting} boundary. 
Finally, \cref{sec:refl_account} describes how the reflectivity phenomena outlined in \cref{sec:planar_interface} can be folded into these results to determine absolute diffraction efficiency. 

\subsection{The Grating Boundary Value Problem}\label{sec:grating_boundary}
%%%%%%%%%%%%%%%%%%%%%%%%%%%%%%%%%%%%%%%%%-------------------------------------------------
A time-harmonic boundary value problem that describes electromagnetic waves in the presence of a reflection grating can be formulated using framework similar to what is utilized in \cref{sec:planar_interface} for the ideal mirror flat and the pseudo-random rough surface. 
Using the Cartesian coordinate system from \cref{sec:geo_constrain} with $\mathbold{\hat{x}}$, $\mathbold{\hat{y}}$ and $\mathbold{\hat{z}}$ as unit vectors for the dispersion, cross-dispersion and groove directions, respectively,\footnote{Note that as in the case of the mirror flat described in \cref{sec:planar_interface}, the $y$-axis here is normal to the grating substrate. In this case, however, the $(x,z)$ coordinates are rotated relative to what is defined in \cref{fig:2Dmirror_angles,fig:in_plane_scatter,fig:off_plane_scatter} for the mirror flat, where $k_z = 0$ by convention.} the surface-relief profile of a grating can be understood as a special case of the framework for a rough surface presented in \cref{sec:rough_surface}, where the surface-profile function, $Y(x,z)$, is replaced by a specified \emph{groove function}, $g(x)$, which defines the cross-sectional shape of the grating groove profile. 
This function is assumed to be periodic over the groove spacing, $d$, so that $g(x) = g(x+d)$ and the function can be be expressed as a Fourier series with $K \equiv 2 \pi / d$ as the \emph{grating wave number} [\emph{cf.\@} \cref{eq:grating_number}]:
\begin{subequations}%http://mathworld.wolfram.com/FourierSeries.html
\begin{equation} \label{eq:groove_fourier_series}
 g(x) = \sum_{n=-\infty}^{\infty} g_n \mathrm{e}^{i n K x} \equiv \sum_{n=-\infty}^{\infty} g_n (x), 
 \end{equation} 
where the complex Fourier coefficients are given by
\begin{equation} \label{eq:groove_fourier_coeff}
 g_n = \frac{1}{d} \int_{0}^{d} g \left( x \right) \mathrm{e}^{- i n K x} \dd{x} \quad \text{for } n = 0, \pm 1, \pm 2, \pm 3 \dotsc 
 \end{equation} 
\end{subequations}
Meanwhile, the incident wave mode is taken to be the central wavelength, $\lambda$, of the monochromatic beam [\emph{cf.\@} \cref{sec:als_beam}] with a wave vector given by \cref{eq:beam_vec_grat}: 
\begin{subequations}
\begin{equation}\label{eq:beam_vec_grat_2}
 \mathbold{k} = k_x \mathbold{\hat{x}} + k_y \mathbold{\hat{y}} + k_z \mathbold{\hat{z}} = - k_0 \left[ \sin \left( \alpha \right) \sin \left( \gamma \right) \mathbold{\hat{x}} + \cos \left( \alpha \right) \sin \left( \gamma \right) \mathbold{\hat{y}} - \cos \left( \gamma \right) \mathbold{\hat{z}} \right] ,
 \end{equation}
where the angles $\alpha$ and $\gamma$ are defined geometrically in \cref{fig:grating_angles} and $k_0 \equiv 2 \pi / \lambda$.  
The accompanying wavefront, which is composed of an electric field, $\mathbold{E} \left( \mathbold{r} \right)$, and a magnetic field, $\mathbold{H} \left( \mathbold{r} \right)$, can be written in a form similar to \cref{eq:incident_fields}: 
\begin{equation} \label{eq:incident_fields_grat} 
 \mathbold{E} \left( x, y, z \right) = \mathbold{E}_0 \mathrm{e}^{i \left( k_x x + k_y y + k_z z \right)} \text{ and } \mathbold{H} \left( x, y, z \right) = \frac{1}{Z_0 k_0} \mathbold{k} \times \mathbold{E} \left( x, y, z \right) ,
 \end{equation}
\end{subequations}
where $Z_0$ is the impedance of free space and $\mathbold{E}_0 = E_{0,x} \mathbold{\hat{x}} + E_{0,y} \mathbold{\hat{y}} + E_{0,z} \mathbold{\hat{z}}$ has components  
\begin{subequations}
\begin{align}\label{eq:inc_E_comp}
 %\begin{split}
 E_{0,x} &= \mathcal{A}_0 \left[ \sin \left( \theta_p \right) \cos \left( \alpha \right) + \cos \left( \theta_p \right) \sin \left( \alpha \right) \cos \left( \gamma \right) \right] \\
 E_{0,y} &= \mathcal{A}_0 \left[ \cos \left( \theta_p \right) \cos \left( \alpha \right) \cos \left( \gamma \right) - \sin \left( \theta_p \right) \sin \left( \alpha \right) \right] \\
 E_{0,z} &= \mathcal{A}_0 \cos \left( \theta_p \right) \sin \left( \gamma \right) ,
 %\end{split}
 \end{align}
\end{subequations}
where $\mathcal{A}_0 \equiv \abs{\mathbold{E}_0}$ describes the amplitude of the electric field and the angle $\theta_p$ parameterizes the orientation of $\mathbold{E}_0 - E_{0,z} \mathbold{\hat{z}}$ relative to $\mathbold{k} - k_z \mathbold{\hat{z}}$ \cite[\emph{cf.\@} \cref{fig:polarization_angle}]{Petit80}. %projected onto the $(x,y)$ plane \cite{Petit80} [\emph{cf.\@} \cref{fig:polarization_angle}]. 

Similar to the treatment of specular reflectivity from a mirror flat [\emph{cf.\@} \cref{sec:reflectivity_polarization}], evaluating polarization sensitivity for a reflection grating can be handled by considering two special cases of \cref{eq:inc_E_comp} along with the corresponding amplitude vector for the magnetic field given by $\mathbold{H}_0 \equiv \left( Z_0 k_0 \right)^{-1} \mathbold{k} \times \mathbold{E}_0$. 
\begin{figure}
 \centering
 \includegraphics[scale=1.25]{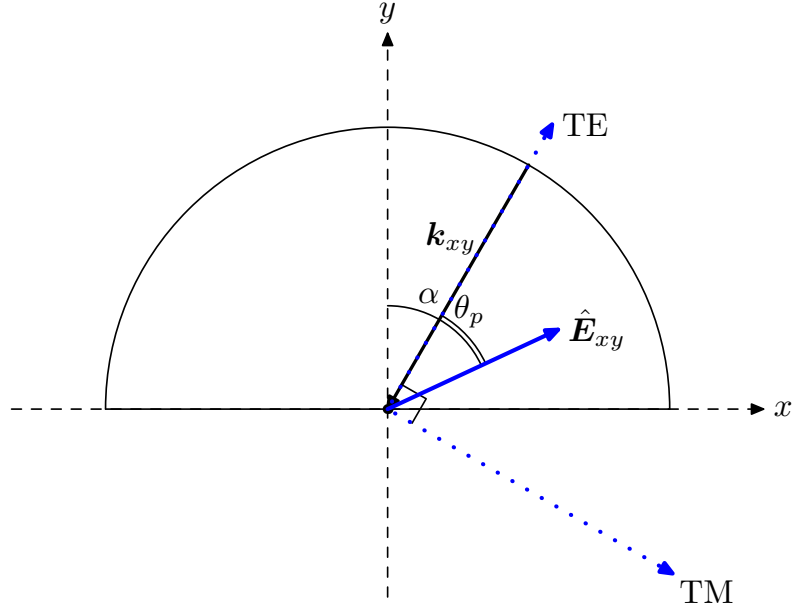}
 \caption[Definition of the polarization angle, $\theta_p$]{Definition of the polarization angle, $\theta_p$. The projection of the incident electric field amplitude, $\mathbold{E}_0$, onto the $(x,y)$ plane is represented by the unit vector $\hat{\mathbold{E}}_{xy}$, which is aligned with $\mathbold{E}_0 - E_{0,z} \mathbold{\hat{z}}$. This vector, along with the projection of $\mathbold{k}$ onto the $(x,y)$ plane given by $\mathbold{k}_{xy} \equiv \mathbold{k} - k_z \mathbold{\hat{z}} = - k_0 \sin \left( \gamma \right) \left[ \sin \left( \alpha \right) \mathbold{\hat{x}} + \cos \left( \alpha \right) \mathbold{\hat{y}} \right]$, define the angle $\theta_p$ with $\theta_p = 0$ and $\theta_p = \pi/2$ corresponding to transverse-electric (TE) and transverse-magnetic (TM) polarization, respectively.}\label{fig:polarization_angle}
 \end{figure} 
These are: $\theta_p  = 0$ such that the field amplitudes become
\begin{subequations}
\begin{align}
 \mathbold{E}_0 &= \mathcal{A}_0 \left[ \sin \left( \alpha \right) \cos \left( \gamma \right) \mathbold{\hat{x}} + \cos \left( \alpha \right) \cos \left( \gamma \right) \mathbold{\hat{y}} + \sin \left( \gamma \right) \mathbold{\hat{z}} \right] \\
 \mathbold{H}_0 &= \left( \frac{\mathcal{A}_0}{Z_0} \right) \left[ \cos \left( \alpha \right) \mathbold{\hat{x}} -  \sin \left( \alpha \right) \mathbold{\hat{y}} \right]
 \end{align}
\end{subequations}
and $\theta_p  = \pi/2$ such that
\begin{subequations}
\begin{align}
 \mathbold{E}_0 &= \mathcal{A}_0 \left[ \cos \left( \alpha \right) \mathbold{\hat{x}} - \sin \left( \alpha \right) \mathbold{\hat{y}} \right] \\
 \mathbold{H}_0 &= \left( \frac{\mathcal{A}_0}{Z_0} \right) \left[ \sin \left( \alpha \right) \cos \left( \gamma \right) \mathbold{\hat{x}} + \cos \left( \alpha \right) \cos \left( \gamma \right) \mathbold{\hat{y}} + \sin \left( \gamma \right) \mathbold{\hat{z}} \right] .
 \end{align}
\end{subequations}
The distinguishing feature between these two sets of equations is that $H_{0,z} = 0$ for $\theta_p  = 0$ while $E_{0,z} = 0$ for $\theta_p  = \pi/2$ and as a result, these two orthogonal polarizations are referred to as \emph{transverse-electric (TE) polarization} and \emph{transverse-magnetic (TM) polarization}, respectively \cite{Petit80,Loewen97}. 
Until \cref{sec:grat_bound_consider}, discussion centers on a generic case with $E_{0,z} \neq 0$ and $H_{0,z} \neq 0$. 

\subsubsection{Helmholtz Equation and Boundary Conditions}\label{sec:grating_Helmholtz}
%%%%%%%%%%%%%%%%%%%%%%%%%%%%%%%%%%%%%%%%%-------------------------------------------------
As generalizations of the reflection and refraction phenomena that occur from an incident wave striking an ideal mirror flat [\emph{cf.\@} \cref{sec:reflectivity_polarization}], \emph{reflected-diffracted} and \emph{refracted-diffracted} fields exist in the presence of a surface-relief grating along with the incident fields [\emph{cf.\@} \cref{eq:incident_fields_grat}]. % given by \cref{eq:incident_fields_grat}. 
While it will be shown that both sets of fields are split into propagating orders that depend on the order number $n$, they are first described generically using \cref{eq:3D_reflected_wave_vector} for the reflected-diffracted wave vector:
\begin{subequations}
\begin{equation}\label{eq:refl_diff_wave0}
 \mathbold{k''} = k_x'' \mathbold{\hat{x}} + k_y'' \mathbold{\hat{y}} + k_z'' \mathbold{\hat{z}} \quad \text{where} \quad k_0 = \sqrt{k_x''^2 + k_y''^2 + k_z''^2} , 
 \end{equation}
with \cref{eq:reflected_fields} for the reflected-diffracted fields:
\begin{equation}
\mathbold{E''} \left( x, y, z \right) = \mathbold{E}_0'' \mathrm{e}^{i \left( k_x'' x + k_y'' y + k_z'' z \right)} \text{ and } \mathbold{H''} \left( x, y, z \right) = \frac{1}{Z_0 k_0} \mathbold{k''} \times \mathbold{E''} \left( x, y, z \right) ,
 \end{equation} 
\end{subequations}
where $\mathbold{E}_0''$ describes polarization and amplitude in analogy to $\mathbold{E}_0$ for the incident wave. 
Similarly, \cref{eq:3D_refracted_wave_vector} is used for the complex refracted-diffracted wave vector that depends on the complex index of refraction of the grating material, $\tilde{\nu} (\omega)$: 
\begin{subequations}
\begin{equation}
 \mathbold{\tilde{k}'} = \tilde{k}_x' \mathbold{\hat{x}} + \tilde{k}_y' \mathbold{\hat{y}} + \tilde{k}_z' \mathbold{\hat{z}} \quad \text{where} \quad \tilde{\nu} (\omega) k_0 = \sqrt{\tilde{k}_x'^2 + \tilde{k}_y'^2 + \tilde{k}_z'^2} ,
 \end{equation}
with \cref{eq:refracted_fields} for the refracted-diffracted fields that decay as they propagate according to the imaginary parts of $\tilde{k}_x'$, $\tilde{k}_y'$ and $\tilde{k}_z'$:
\begin{equation}
 \mathbold{E'} \left( x, y, z \right) = \mathbold{E}_0' \mathrm{e}^{i \left( \tilde{k_x}' x + \tilde{k_y}' y + \tilde{k_z}' z \right)} \text{ and } \mathbold{H'} \left( x, y, z \right) = \frac{1}{Z_0 k_0} \mathbold{\tilde{k}'} \times \mathbold{E'} \left( x, y, z \right) ,
 \end{equation}
\end{subequations}
where $\mathbold{E}_0'$ describes polarization and the initial amplitude. 
Together, these fields must be solutions of the Helmholtz equation in vacuum [\emph{cf.\@} \cref{eq:Helmholtz_above_mirror}]: 
\begin{subequations}
\begin{equation}\label{eq:Helmholtz_above_mirror_ch2} 
 \left( \laplacian + k_0^2 \right) \left\{
 \begin{array}{lr}
 \mathbold{E} \left( x, y, z \right) + \mathbold{E''} \left( x, y, z \right) \\
 \mathbold{H} \left( x, y, z \right) + \mathbold{H''} \left( x, y, z \right) 
 \end{array}
 \right\} = \mathbf{0} \quad \text{for } y > g(x) ,
 \end{equation}
and inside the material of the grating [\emph{cf.\@} \cref{eq:Helmholtz_below_mirror}]:
\begin{equation}\label{eq:Helmholtz_below_mirror_ch2} 
 \left( \laplacian + \tilde{k}'^2 \right) \left\{
 \begin{array}{lr}
 \mathbold{E'} \left( x, y, z \right) \\
 \mathbold{H'} \left( x, y, z \right) 
 \end{array}
 \right\} = \mathbf{0} \quad \text{for } y < g(x) 
 \end{equation}
\end{subequations}
with $\tilde{k}' \equiv \tilde{\nu} (\omega) k_0$ and $\mathbf{0}$ as the null vector. 
\begin{figure}
 \centering
 \includegraphics[scale=1.25]{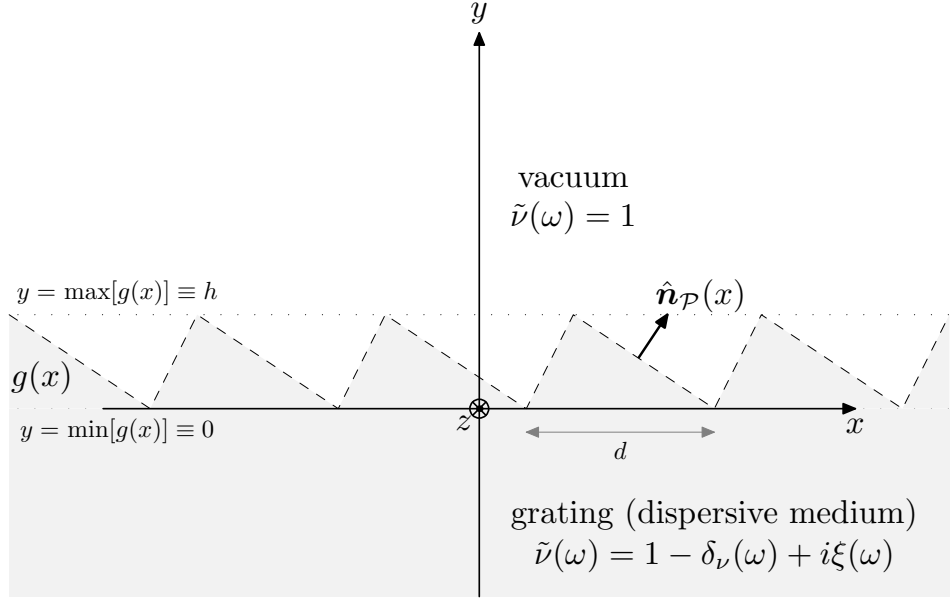} 
 \caption[Surface-relief boundary between vacuum and a dispersive medium representing the functional material of a reflection grating]{Surface-relief boundary between vacuum and a dispersive medium representing the functional material of a reflection grating. With grooves aligned with the $z$-direction as in \cref{fig:grating_angles}, the cross-sectional profile, $\mathcal{P}$ (shown as a dashed line), of a grating with a groove spacing $d$ is described by the periodic function $g(x)$ confined to $0 \leq y \leq h$, where $h$ is the groove depth. The unit vector $\mathbold{\hat{n}}_{\mathcal{P}}(x)$ describes the direction normal to grating surface at a position $x$.}\label{fig:grating_boundary_def} 
 \end{figure}
They must also satisfy the appropriate boundary conditions at the surface of the grating profile, $\mathcal{P}$, described by $y = g(x)$, which can be written as special cases of \cref{eq:rough_boundary_1,eq:rough_boundary_2,eq:rough_boundary_3,eq:rough_boundary_4} for a rough surface with the pseudo-random function $Y(x,z)$ replaced by the periodic function $g(x)$:\footnote{These boundary conditions are valid in the absence of surface charge and surface current. This treatment is justified in the soft x-ray regime, where materials exhibit low electrical conductivity due to $\omega$ approaching their plasma frequency [\emph{cf.\@} \cref{sec:x-ray_index}]. For the scenario of a perfectly conducting grating described in \cref{sec:grat_bound_consider}, however, the refracted-diffracted fields are null and a term for surface current must be included.} 
\begin{subequations}
\begin{align}
 \mathbold{\hat{n}}_{\mathcal{P}}(x) \times \left[ \mathbold{E} \left( x, g(x), z \right) + \mathbold{E''} \left( x, g(x), z \right) \right] &= \mathbold{\hat{n}}_{\mathcal{P}}(x) \times \mathbold{E'} \left( x, g(x), z \right) \label{eq:grating_boundary_1} \\
 \mathbold{\hat{n}}_{\mathcal{P}}(x) \times \left[\mathbold{H} \left( x, g(x), z \right)  + \mathbold{H''} \left( x, g(x), z \right) \right] &= \mathbold{\hat{n}}_{\mathcal{P}}(x) \times \mathbold{H'} \left( x, g(x), z \right) \label{eq:grating_boundary_2} \\
 \mathbold{\hat{n}}_{\mathcal{P}}(x) \cdot \left[ \mathbold{E} \left( x, g(x), z \right) + \mathbold{E''} \left( x, g(x), z \right) \right] &= \tilde{\nu}^2 (\omega) \, \mathbold{\hat{n}}_{\mathcal{P}}(x) \cdot \mathbold{E'} \left( x, g(x), z \right) \label{eq:grating_boundary_3} \\
 \mathbold{\hat{n}}_{\mathcal{P}}(x) \cdot \left[ \mathbold{H} \left( x, g(x), z \right) + \mathbold{H''} \left( x, g(x), z \right) \right] &= \mathbold{\hat{n}}_{\mathcal{P}}(x) \cdot \mathbold{H'} \left( x, g(x), z \right)  \label{eq:grating_boundary_4} , 
 \end{align}
\end{subequations}
where $\mathbold{\hat{n}}_{\mathcal{P}}(x)$ 
is a unit vector normal to the grating surface at a position $x$ with $\mathbold{\hat{n}}_{\mathcal{P}}(x) \cdot \mathbold{\hat{z}} = 0$ [\emph{cf.\@} \cref{fig:grating_boundary_def}]. 

In analogy to the space-harmonic quality of an ideal mirror flat that leads to $k_{\parallel} = k_{\parallel}'' = \tilde{k_{\parallel}}'$ at a planar boundary [\emph{cf.\@} \cref{sec:reflectivity_polarization}], the fields on $\mathcal{P}$ have the property that a translation in $z$ introduces only a phase shift $\mathrm{e}^{i k_z z}$ to the incident fields: 
\begin{subequations}
\begin{equation}
 \left\{
 \begin{array}{lr}
 \mathbold{E} \left( x, y, z \right) \\
 \mathbold{H} \left( x, y, z \right)
 \end{array} 
 \right\} =  \left\{
 \begin{array}{lr}
 \mathbold{E} \left( x, y \right) \\
 \mathbold{H} \left( x, y \right)
 \end{array} 
 \right\} \mathrm{e}^{i k_z z} .
 \end{equation}
The boundary conditions given by \cref{eq:grating_boundary_1,eq:grating_boundary_2,eq:grating_boundary_3,eq:grating_boundary_4} then require $k_z = k_z'' = \tilde{k_z}'$ so that the reflected-diffracted and refracted-diffracted fields are given by 
\begin{equation}
 \left\{
 \begin{array}{lr}
 \mathbold{E}'' \left( x, y \right) \\
 \mathbold{H}'' \left( x, y \right)
 \end{array} 
 \right\} \mathrm{e}^{i k_z z} \quad \text{and} \quad   \left\{
 \begin{array}{lr}
 \mathbold{E}' \left( x, y \right) \\
 \mathbold{H}' \left( x, y \right)
 \end{array} 
 \right\} \mathrm{e}^{i k_z z} ,
 \end{equation}
\end{subequations}
respectively. 
These $\mathrm{e}^{i k_z z}$ terms the drop out of the Helmholtz equation so that, using the following shorthand notation: 
\begin{subequations}
\begin{align}
 \begin{split}
 \nabla_{xy}^2 \equiv \nabla^2 - \pdv[2]{z} &= \pdv[2]{x} + \pdv[2]{y} \quad \text{with} \\
 k_{xy}^2 \equiv k_0^2  - k_z^2 = k_x^2 + k_y^2 \quad &\text{and} \quad \tilde{k}_{xy}^2 \equiv \tilde{k}'^2 - \tilde{k_z}'^2 = \tilde{k_x}'^2 + \tilde{k_y}'^2 ,
 \end{split}
 \end{align} 
\end{subequations}
\cref{eq:Helmholtz_above_mirror_ch2} for vacuum reduces to 
\begin{subequations}
\begin{equation}\label{eq:Helmholtz_above_grating}
 \left( \nabla_{xy}^2 +  k_{xy}^2 \right) \left\{
 \begin{array}{lr}
 \mathbold{E} \left( x, y \right) + \mathbold{E''} \left( x, y \right) \\
 \mathbold{H} \left( x, y \right) + \mathbold{H''} \left( x, y \right) 
 \end{array}
 \right\} = \mathbf{0} \quad \text{for } y > g(x) 
 \end{equation}
and \cref{eq:Helmholtz_below_mirror_ch2} for the grating material reduces to 
\begin{equation}\label{eq:Helmholtz_below_grating}
 \left( \nabla_{xy}^2 + \tilde{k}_{xy}^2 \right) \left\{
 \begin{array}{lr}
 \mathbold{E'} \left( x, y \right) \\
 \mathbold{H'} \left( x, y \right) 
 \end{array}
 \right\} = \mathbf{0} \quad \text{for } y < g(x)  .
 \end{equation}
\end{subequations}

The framework just presented can be simplified for TE and TM polarization states by representing Faraday's law in time-harmonic form [\emph{cf.\@} \cref{eq:Maxwell_vac_TH_3}], $\curl \mathbold{E} \left( \mathbold{r} \right) = i \omega \mu_0 \mathbold{H} \left( \mathbold{r} \right)$, as 
\begin{subequations}
\begin{align}
 \pdv{E_z (x,y)}{y} - i k_z E_y (x,y) &= i \omega \mu_0 H_x (x,y) \label{eq:maxgrat1a} \\
 i k_z E_x (x,y) - \pdv{E_z (x,y)}{x} &= i \omega \mu_0 H_y (x,y) \label{eq:maxgrat1b} \\
 \pdv{E_y (x,y)}{x} - \pdv{E_x (x,y)}{y} &= i \omega \mu_0 H_z (x,y) \label{eq:maxgrat1bc} 
 \end{align}
\end{subequations}
and similarly, expressing the time-harmonic version of the Amp\`ere-Maxwell equation [\emph{cf.\@} \cref{eq:Maxwell_vac_TH_4}], $\curl \mathbold{H} \left( \mathbold{r} \right) = -i \omega \epsilon_0 \mathbold{E} \left( \mathbold{r} \right)$, as %without electric current,
\begin{subequations}
\begin{align}
 \pdv{H_z (x,y)}{y} - i k_z H_y (x,y) &= - i \omega \epsilon_0 E_x (x,y) \label{eq:maxgrat2a} \\
 i k_z H_x (x,y) - \pdv{H_z (x,y)}{x} &= - i \omega \epsilon_0 E_y (x,y) \label{eq:maxgrat2b} \\
 \pdv{H_y (x,y)}{x} - \pdv{H_x (x,y)}{y} &= - i \omega \epsilon_0 E_z (x,y) \label{eq:maxgrat2c} ,
 \end{align}
\end{subequations}
where, in both sets of equations, the factors of $k_z$ appear due to the the assumed $\mathrm{e}^{i k_z z}$ dependence along the groove direction. 
In the case of TE polarization with $H_z (x,y) = 0$, \cref{eq:maxgrat1a,eq:maxgrat1b,eq:maxgrat2a,eq:maxgrat2b} can be combined to give 
\begin{subequations}
\begin{align}
 \begin{split}\label{eq:TE_comps}
 \curl \left[ E_z (x,y) \mathbold{\hat{z}} \right] &= \frac{-i}{\omega \epsilon_0} k^2_{xy} \left[ H_x (x,y) \mathbold{\hat{x}} + H_y (x,y) \mathbold{\hat{y}} \right] \\
 &= \frac{-i}{k_z} k^2_{xy} \left[ E_x (x,y) \mathbold{\hat{x}} + E_y (x,y) \mathbold{\hat{y}} \right]
 \end{split}
 \end{align}
while for the case of TM polarization with $E_z (x,y) = 0$, these equations yield 
\begin{align}
 \begin{split}\label{eq:TM_comps}
 \curl \left[ H_z (x,y) \mathbold{\hat{z}} \right] &= \frac{-i}{\omega \mu_0} k^2_{xy} \left[ E_x (x,y) \mathbold{\hat{x}} + E_y (x,y) \mathbold{\hat{y}} \right] \\
 &= \frac{-i}{k_z} k^2_{xy} \left[ H_x (x,y) \mathbold{\hat{x}} + H_y (x,y) \mathbold{\hat{y}} \right] ,
 \end{split}
 \end{align}
\end{subequations}
where identical expressions hold for diffracted fields. 
Together, \cref{eq:TE_comps,eq:TM_comps} show that for the two orthogonal polarization states, the transverse component of the relevant field can be used to determine the remaining field components. 
With the incident wave being a solution to the Helmholtz equation by default, \cref{eq:Helmholtz_above_grating} reduces to a scalar equation for the relevant $z$-component of the fields for each polarization:
\begin{subequations}
\begin{equation}\label{eq:Helmholtz_above_grat_scal}
 \left( \nabla_{xy}^2 +  k_{xy}^2 \right) u'' (x,y) = 0 \quad \text{for } y > g(x) 
 \end{equation}
with 
\begin{equation}
 u'' (x,y) \equiv
   \begin{cases}
     E_z'' (x,y) & \text{for }\ \text{TE polarization} \\
     H_z'' (x,y) & \text{for }\ \text{TM polarization} 
   \end{cases} 
\end{equation}
\end{subequations}
and \cref{eq:Helmholtz_below_grating} for the grating material reduces to
\begin{subequations}
\begin{equation}\label{eq:Helmholtz_below_grat_scal}
 \left( \nabla_{xy}^2 + \tilde{k}_{xy}^2 \right) u' (x,y) = 0 \quad \text{for } y < g(x) 
 \end{equation}
with 
\begin{equation}
 u' (x,y) \equiv
   \begin{cases}
     E_z' (x,y) & \text{for }\ \text{TE polarization} \\
     H_z' (x,y) & \text{for }\ \text{TM polarization.} 
   \end{cases} 
\end{equation}
\end{subequations}
These two cases for the reflected-diffracted and refracted-diffracted fields are considered separately in \cref{sec:reflected_diffracted,sec:refracted_diffracted}. 

\subsubsection{Reflected-Diffracted Fields}\label{sec:reflected_diffracted}
%%%%%%%%%%%%%%%%%%%%%%%%%%%%%%%%%%%%%%%%%-------------------------------------------------
The condition $k_z = k_z''$ implies a $z$-direction law of reflection analogous to \cref{eq:reflection}, which ensures the cone opening half-angle to be equal to the angle $\gamma$ defined for the incident wave vector. 
This can be seen by writing the generic reflected-diffracted wave vector [\emph{cf.\@} \cref{eq:refl_diff_wave0,eq:3D_reflected_wave_vector}] as 
\begin{align}\label{eq:refl_diff_wave1}
 \begin{split}
 \mathbold{k''} &= k_x'' \mathbold{\hat{x}} + k_y'' \mathbold{\hat{y}} + k_z'' \mathbold{\hat{z}} \\
 &= k_0 \left[ \sin \left( \beta \right) \sin \left( \gamma'' \right) \mathbold{\hat{x}} + \cos \left( \beta \right) \sin \left( \gamma'' \right) \mathbold{\hat{y}} + \cos \left( \gamma'' \right) \mathbold{\hat{z}} \right] ,
 \end{split}
 \end{align}
where $\beta$ is the azimuthal diffracted angle [\emph{cf.\@} \cref{fig:conical_reflection_edit}] and $\gamma''$ is the cone opening half-angle: with $k_z = k_0 \cos \left( \gamma \right)$ by \cref{eq:beam_vec_grat}, $k_z = k_z''$ verifies $\gamma'' = \gamma$. 
The fields along the dispersion direction, on the other hand, experience a phase shift $\mathrm{e}^{i k_x x}$ due to incident wave and additionally, they are modulated by $\mathcal{P}$ defined by $g(x)$. 
Thus, with $\mathrm{e}^{i k_x x}$ accounted for, the relevant field component of the diffracted fields given by $u'' (x,y)$ can be expressed as a Fourier series similar to \cref{eq:groove_fourier_series} \cite{Petit80,antonakakis14}: 
\begin{subequations}
\begin{equation}\label{eq:refl_plane_spec_1}
  u'' \left[ x, g(x) \right] \mathrm{e}^{- i k_x x} = \sum_{n=-\infty}^{\infty} u_n'' \left[ g(x) \right] \mathrm{e}^{i n K x} ,
 \end{equation}
with $u_n'' (y)$ as an unknown function that describes the $n^{\text{th}}$ spatial harmonic of the diffracted field. 
Rewriting \cref{eq:refl_plane_spec_1} as
\begin{equation}\label{eq:refl_plane_spec_2}
 u'' \left[ x, g(x) \right] = \sum_{n=-\infty}^{\infty} u_n'' \left[ g(x) \right] \mathrm{e}^{i \left( k_x + n K \right) x} \equiv \sum_{n=-\infty}^{\infty} u_n'' \left[ g(x) \right] \mathrm{e}^{i k_{x,n}'' x} 
 \end{equation}
\end{subequations}
indicates that $k_x'' = k_x + n K \equiv k_{x,n}''$, which is recognized as the generalized grating equation [\emph{cf.\@} \cref{eq:off-plane_orders}].\footnote{This can be seen using $k_x'' = k_0 \sin \left( \beta \right) \sin \left( \gamma \right)$ from \cref{eq:refl_diff_wave1} and $k_x = - k_0 \sin \left( \alpha \right) \sin \left( \gamma \right)$ from \cref{eq:beam_vec_grat} with $k_0 \equiv 2 \pi / \lambda$ and $K \equiv 2 \pi / d$ such that from $k_x'' - k_x = nK$,
\begin{equation*}
 k_0 \sin \left( \gamma \right) \left[ \sin \left( \beta \right) + \sin \left( \alpha \right) \right] = nK \implies \sin \left( \alpha \right) + \sin \left( \beta \right) = \frac{n K}{k_0 \sin \left( \gamma \right)} = \frac{n \lambda}{d \sin \left( \gamma \right)} .
 \end{equation*}} 
The diffracted wave vector [\emph{cf.\@} \cref{eq:refl_diff_wave1}] then becomes 
\begin{align}
 \begin{split}\label{eq:refl_diff_wave2}
 \mathbold{k}_n'' &= k_{x,n}'' \mathbold{\hat{x}} + k_{y,n}'' \mathbold{\hat{y}} + k_z \mathbold{\hat{z}} \\
 \text{with} \quad k_{x,n}'' \equiv k_x &+ nK \quad \text{and} \quad k_{y,n}'' \equiv \sqrt{k_0^2 - \left( k_{x,n}''\right)^2 - k_z^2} , 
 \end{split}
 \end{align}
which shows that for all $n$ such that $k_0^2 > \left( k_x + nK \right)^2 + k_z^2$, there is a propagating order confined to the surface of a cone with opening angle $2 \gamma$ [\emph{cf.\@} \cref{sec:off-plane_geo}]. 

From the above considerations, the diffracted fields are expected to be of the form of a \emph{Rayleigh expansion}:
\begin{equation}\label{eq:refl_plane_spec_3}
 u'' (x,y) = \sum_{n=-\infty}^{\infty} u_n'' (y) \mathrm{e}^{i \left( k_x + n K \right) x} ,
 \end{equation}
which describes a \emph{spectrum of plane waves}, where $k_0$ remains constant but the direction of $\mathbold{k''}$ varies with $n$ so that the boundary conditions given by \cref{eq:grating_boundary_1,eq:grating_boundary_2,eq:grating_boundary_3,eq:grating_boundary_4} are satisfied \cite{Petit80,antonakakis14}. 
Inserting \cref{eq:refl_plane_spec_3} into \cref{eq:Helmholtz_above_grat_scal} and noting that the fields of the incident wave satisfy the Helmholtz equation by default gives 
\begin{align}\label{eq:Helmholtz_above_grating_expand}
 \begin{split}
 \left( \nabla_{xy}^2 +  k_{xy}^2 \right) \sum_{n=-\infty}^{\infty} u_n'' (y) \mathrm{e}^{i \left( k_x + n K \right) x}  &= 0 \\
 \implies \sum_{n=-\infty}^{\infty} \left[ \dv[2]{u_n'' (y)}{y} + \left( k_{y,n}'' \right)^2 u_n'' (y) \right] \mathrm{e}^{i n K x} &= 0 ,
 \end{split}
 \end{align}
where the phase factor $\mathrm{e}^{i k_x x}$ drops out of the second line. For $y > h \equiv \text{max} \! \left[g(x)\right]$ such that $\tilde{\nu} (\omega) = 1$ holds for all $x$, \cref{eq:Helmholtz_above_grating_expand} reduces to a collection of differential equations for each value of $n$ \cite{Petit80,antonakakis14}:\footnote{Fields in the grooved region with $0 \leq y \leq h$ are considered in \cref{sec:grat_bound_consider}.}
\begin{equation}\label{eq:Helmholtz_above_grating_order}
 \dv[2]{u_n'' (y)}{y} + \left( k_{y,n}'' \right)^2 u_n'' (y) = 0 .
 \end{equation}
A physically-realistic solution to \cref{eq:Helmholtz_above_grating_order} is one which diffracted waves propagate away from the grating surface, in vacuum:
\begin{subequations}
\begin{equation}\label{eq:Helmholtz_above_grat_order_sol}
  u_n'' (y) = \mathcal{U}_n'' \mathrm{e}^{i k_{y,n}'' y}
 \end{equation}
with
\begin{equation}\label{eq:generic_amp}
 \mathcal{U}_n'' \equiv
   \begin{cases}
     \mathcal{A}_{\text{TE},n}'' \sin \left( \gamma \right) & \text{for } \text{TE polarization} \\
     Z_0^{-1} \mathcal{A}_{\text{TM},n}'' \sin \left( \gamma \right) & \text{for } \text{TM polarization,} 
   \end{cases} 
\end{equation}
\end{subequations}
where $\mathcal{A}_{\text{TE},n}''$ and $\mathcal{A}_{\text{TM},n}''$ are the electric-field amplitudes of the $n^{\text{th}}$ order in each orthogonal polarization while $Z_0$ is the impedance of free space. 
The total diffracted field then can be written as a sum of diffracted orders:
\begin{equation}\label{eq:diff_field_exp}
u'' (x,y) = \sum_{n=-\infty}^{\infty} \mathcal{U}_n'' \mathrm{e}^{i \left( k_{x,n}'' x + k_{y,n}'' y \right)} ,
 \end{equation}
which results from combining \cref{eq:refl_plane_spec_3,eq:Helmholtz_above_grat_order_sol}. 

In analogy to the definition of specular reflectivity [\emph{cf.\@} \cref{eq:fresnel_refl_def}], the diffraction efficiency in each orthogonal polarization is defined as the outgoing flux of propagating orders through an imagined surface above the grating substrate relative to the incoming flux associated with the incident wave [\emph{cf.\@} \cref{eq:diffraction_efficiency}]. 
By defining a scalar function for the $z$-component of the incident field as
\begin{subequations}
\begin{equation}
 u (x,y) \equiv \left\{
 \begin{array}{lr}
  E_z (x,y) \\
  H_z (x,y)
 \end{array}
 \right\} \equiv \mathcal{U}_0 \mathrm{e}^{i \left( k_x x + k_y y \right)}  
\end{equation}
with 
\begin{equation}
 \mathcal{U}_0 \equiv
   \begin{cases}
     \mathcal{A}_0 \sin \left( \gamma \right) & \text{for } \text{TE polarization} \\
     Z_0^{-1} \mathcal{A}_0 \sin \left( \gamma \right) & \text{for } \text{TM polarization,} 
   \end{cases} 
\end{equation}
\end{subequations}
these quantities can be expressed as 
\begin{subequations}
\begin{align}
 \mathscr{E}_{\text{TE},n} &= \frac{k_{y,n}''}{k_y} \norm{\frac{\mathcal{A}_{\text{TE},n}''}{\mathcal{A}_0}}^2 = \frac{\cos \left( \beta \right)}{\cos \left( \alpha \right)} \norm{\tilde{r}_{\text{TE}}^{(n)}}^2\label{eq:TE_efficiency} \\ 
 \mathscr{E}_{\text{TM},n} &= \frac{k_{y,n}''}{k_y} \norm{\frac{\mathcal{A}_{\text{TM},n}''}{\mathcal{A}_0}}^2 = \frac{\cos \left( \beta \right)}{\cos \left( \alpha \right)} \norm{\tilde{r}_{\text{TM}}^{(n)}}^2 ,
 \end{align} 
where $\mathcal{A}_0$ is the amplitude of the incident electric field [\emph{cf.\@} \cref{eq:inc_E_comp,fig:polarization_angle}] while $\tilde{r}_{\text{TE}}^{(n)} \equiv \mathcal{A}_{\text{TE},n}'' / \mathcal{A}_0$ and $\tilde{r}_{\text{TM}}^{(n)} \equiv \mathcal{A}_{\text{TE},n}'' / \mathcal{A}_0$ are unitless amplitude coefficients that are generally complex. 
Diffraction efficiency for an arbitrary polarization state can then be written as 
\begin{equation}\label{eq:diffraction_efficiency_theo}
 \mathscr{E}_n = \mathscr{E}_{\text{TE},n} \cos^2 \left( \theta_p \right) + \mathscr{E}_{\text{TM},n} \sin^2 \left( \theta_p \right) ,
 \end{equation}
\end{subequations}
where $\theta_p$ is the polarization angle \cite[\emph{cf.\@} \cref{eq:inc_E_comp,fig:polarization_angle}]{Petit80}. 
Determining amplitude coefficients $\mathcal{A}_{\text{TE},n}''$ and $\mathcal{A}_{\text{TM},n}''$ follows from enforcing boundary conditions for a given polarization state [\emph{cf.\@} \cref{sec:grat_bound_consider}]. 

\subsubsection{Refracted-Diffracted Fields}\label{sec:refracted_diffracted}
%%%%%%%%%%%%%%%%%%%%%%%%%%%%%%%%%%%%%%%%%-------------------------------------------------
In analogy to Snell's law of refraction for a planar interface [\emph{cf.\@} \cref{eq:refraction}], the requirement $k_z = \tilde{k_z}'$ yields a relationship between $\gamma$ and the cone opening half-angle of refracted-diffracted radiation. 
The complex, refracted-diffracted wave vector can be written as
\begin{align}
 \begin{split}\label{eq:refr_diff_wave1}
 \mathbold{\tilde{k}'} &= \tilde{k}_x' \mathbold{\hat{x}} + \tilde{k}_y' \mathbold{\hat{y}} + \tilde{k}_z' \mathbold{\hat{z}} \\
 &=  \tilde{k}' \left[ \sin \left( \tilde{\beta}' \right) \sin \left( \tilde{\gamma}' \right) \mathbold{\hat{x}} - \cos \left( \tilde{\beta}' \right) \sin \left( \tilde{\gamma}' \right) \mathbold{\hat{y}} + \cos \left( \tilde{\gamma}' \right) \mathbold{\hat{z}} \right] ,
 \end{split}
 \end{align}
where $\tilde{\beta}'$ and $\tilde{\gamma}'$ are complex angles that, for propagating orders, have real components that represent the diffracted angle and cone opening half-angle, respectively, and imaginary components that are related to wave attenuation in the material, as described in \cref{sec:total_refl} for the case of a mirror flat.\footnote{The geometry for conical diffraction from a transmission grating is illustrated in \cref{fig:conical_transmission}. However, this figure neglects effects related to refraction and hence the diffracted angle and the cone opening half-angle are labeled as $\beta$ and $\gamma$, respectively.} 
From $k_z = k_0 \cos (\gamma)$ and $k_z = \tilde{k}' \cos \left( \tilde{\gamma}' \right)$, the relevant form of Snell's law along the $z$-direction is 
\begin{subequations}
\begin{equation}\label{eq:comp_snell_grat}
\cos \left( \gamma \right) = \tilde{\nu}(\omega) \cos \left( \tilde{\gamma}' \right) ,
 \end{equation}
which can be decomposed into two expressions as in \cref{eq:refractionA,eq:refractionB}:
\begin{align}
 \nu (\omega) \cos \left( \gamma'_{\Re} \right) \cosh \left( \gamma'_{\Im} \right) + \xi (\omega) \sin \left( \gamma'_{\Re} \right) \sinh \left( \gamma'_{\Im} \right) &= \cos \left( \zeta \right) \label{eq:refractionA_grat} \\
 \xi (\omega) \cos \left( \gamma'_{\Re} \right) \cosh \left( \gamma'_{\Im} \right) - \nu (\omega) \sin \left( \gamma'_{\Re} \right) \sinh \left( \gamma'_{\Im} \right) &=0 \label{eq:refractionB_grat} ,
 \end{align} 
\end{subequations} 
where $\tilde{\gamma}' \equiv \gamma'_{\Re} + i \gamma'_{\Im}$ and $\tilde{\nu} (\omega) \equiv \nu (\omega) + i \xi (\omega)$. 
These equations imply that if $\xi (\omega)$ can be neglected, $\gamma'_{\Im} = 0$ with \cref{eq:comp_snell_grat} reducing to a purely real equation similar to \cref{eq:Snell_real}:\footnote{As described in \cref{sec:total_refl}, the effect of $\xi (\omega) \neq 0$ causes the fields to penetrate slightly into the material with $\gamma'_{\Re}$ being a small angle given approximately by \cref{eq:zetaprime_real} with $\zeta \to \gamma$.}
\begin{equation}\label{eq:Snell_real_grat}
 \cos \left( \gamma \right) = \nu (\omega) \cos \left( \gamma'_{\Re} \right) .
 \end{equation}
TER in an extreme off-plane geometry is enabled for values of $\gamma$ that yield $\gamma'_{\Re} = 0$ by \cref{eq:Snell_real_grat}. 
However, the topography of the grating grooves must also be taken into account; for a blazed grating with an active facet angle $\delta$, the angle on the groove facet, $\zeta$, must be smaller than the critical angle for TER [\emph{cf.\@} \cref{eq:TER_condition_grat}]. 

In any case where radiation is considered to penetrate into the grating material, the refracted fields on $\mathcal{P}$ take on the periodicity of the grating as in \cref{sec:reflected_diffracted} for the reflected fields. 
This can be written analogously to \cref{eq:refl_plane_spec_1} as 
\begin{subequations}
\begin{equation}\label{eq:refr_plane_spec_1}
  u' \left[ x, g(x) \right] \mathrm{e}^{- i k_x x} = \sum_{n=-\infty}^{\infty} u_n' \left[ g(x) \right] \mathrm{e}^{i n K x} ,
 \end{equation}
$u' (x,y)$ is either $E_z' (x,y)$ or $H_z' (x,y)$ depending on polarization and $u_n' (y)$ is an unknown function that describes the $n^{\text{th}}$ spatial harmonic of the diffracted field. 
As in \cref{sec:reflected_diffracted}, \cref{eq:refr_plane_spec_1} indicates a spectrum of planes waves for $y <0$:
\begin{equation}\label{eq:refr_plane_spec_2} %for the case of reflected-diffracted fields
 u' (x,y) = \sum_{n=-\infty}^{\infty} u_n' (y) \mathrm{e}^{i \left( k_x + n K \right) x} \equiv \sum_{n=-\infty}^{\infty} u_n' (y) \mathrm{e}^{i \tilde{k}_{x,n}' x} ,
 \end{equation}
\end{subequations}
where the expression $\tilde{k}_x' = k_x + n K \equiv \tilde{k}_{x,n}'$ is recognized as the \emph{grating equation in a dispersive medium}:\footnote{This can be seen using $\tilde{k_x}' = \tilde{k}' \sin \left( \tilde{\beta}' \right) \sin \left( \tilde{\gamma}' \right)$ from \cref{eq:refr_diff_wave1} and $k_x = - k_0 \sin \left( \alpha \right) \sin \left( \gamma \right)$ from \cref{eq:beam_vec_grat} along with \cref{eq:comp_snell_grat} to eliminate $\tilde{\gamma}'$.}  
\begin{equation}\label{eq:grat_disp_med}
 \sin \left( \alpha \right) \sin \left( \gamma \right) + \sin \left( \tilde{\beta}' \right) \sqrt{\tilde{\nu}^2 \left( \omega \right) - \cos^2 \left( \gamma \right)} = \frac{n \lambda}{d} . 
 \end{equation}
The diffracted wave vector given by \cref{eq:refr_diff_wave1} then becomes 
\begin{align}
 \begin{split}
 \mathbold{\tilde{k}}_n' &= \tilde{k}_{x,n}' \mathbold{\hat{x}} + \tilde{k}_{y,n}' \mathbold{\hat{y}} + k_z \mathbold{\hat{z}} \\
 \text{with} \quad \tilde{k}_{x,n}'  \equiv k_x &+ nK \quad \text{and} \quad \tilde{k}_{y,n}'  \equiv \sqrt{\tilde{k}'^2 - \left( \tilde{k}_{x,n}'  \right)^2 - k_z^2},
 \end{split}
 \end{align} 
which describes a decaying, propagating order for all $n$ at which yield $\abs{\Re \left[ \tilde{\beta}' \right]} < 90^{\circ}$ according to \cref{eq:grat_disp_med}. 
Meanwhile, $\Im \left[ \tilde{\beta}' \right]$ is related to the attenuation of a diffracted ray that, in analogy to the case of a mirror flat [\emph{cf.\@} \cref{sec:total_refl}], causes electromagnetic waves to be highly distorted as they propagate and decay in nearly orthogonal directions. 
The \emph{penetration depth}, $\mathcal{D}_{\perp}$, defined by \cref{eq:penetration_depth} as the vertical distance into a material over which the intensity of radiation drops to $1/\mathrm{e}$ its initial value, informs thickness requirements for reflective overcoats on gratings. 
That is, with a thickness of \numrange{4}{5} $\mathcal{D}_{\perp}$, reflections at underlying material interfaces can be neglected and the layer can be considered virtually infinitely thick. 
For a high-$\mathcal{Z}$ material such as gold, this thickness requirement is $\sim \SI{15}{\nm}$ for grazing-incidence soft x-rays [\emph{cf.\@} \cref{fig:X-ray_atten}]. 

\subsection{The Integral Method for X-ray Reflection Gratings}\label{sec:grat_bound_consider}
%%%%%%%%%%%%%%%%%%%%%%%%%%%%%%%%%%%%%%%%%-------------------------------------------------
The grating boundary value problem presented in \cref{sec:grating_boundary} can be simplified with two assumptions that are justified in the following discussion:
\begin{enumerate}[noitemsep]
	\item The grating medium is considered to be \emph{perfectly conducting} such that the refracted fields are null with $\mathbold{E'} \left( x, y \right) = \mathbold{H'} \left( x, y \right) = \mathbf{0}$
	\item Diffraction efficiency is polarization insensitive for x-ray reflection gratings but with the assumption of perfect conductivity, gratings are modeled most accurately using TE polarization \cite{Marlowe16}
 \end{enumerate}
Although real materials are poorly conducting at soft x-ray frequencies for the reasons outlined in \cref{sec:SXR_med}, the electromagnetic fields involved in lossless TER behave similarly to those striking a medium with perfect conductivity, where the \emph{skin depth}\footnote{In a conducting medium with permittivity, permeability and conductivity given by $\epsilon$, $\mu$ and $\sigma$, respectively, solving Maxwell's equations using Ohm's Law given by $\mathbfcal{J} = \sigma \mathbold{E}$, with $\mathbfcal{J}$ as current density, yields $\tilde{k}^2 = k^2 + i \mu \sigma \omega$ as dispersion relation with $k \equiv \omega \sqrt{\epsilon \mu}$ and $\tilde{k}$ as a complex wave number. For very high conductivity or very low frequencies, the skin depth, which describes the distance over which the fields to $1/\mathrm{e}$ their initial amplitude, is given by 
\begin{equation*}
 \mathcal{D}_{\sigma} \approx \sqrt{\frac{2}{\omega \mu \sigma}} ,
 \end{equation*}
which approaches zero as $\sigma \to \infty$ \cite{Landau60}.\label{footnote:skin_depth}}
approaches zero \cite{Griffiths17,Jackson75}. 
This can be understood by considering that the refracted angle at an interface between vacuum and a dispersive medium with index of refraction $\tilde{\nu}(\omega) = \nu (\omega) + i \xi (\omega)$ is null for TER in the limit that $\xi (\omega) \to 0$ [\emph{cf.\@} \cref{sec:total_refl}]. 
Under such conditions, the fields do not penetrate into the dispersive medium so that the penetration depth, $\mathcal{D}_{\perp}$, approaches zero while Fresnel reflectivity, $\mathcal{R}_F$, is unity. 
While this assumption is not strictly justified for TER in the soft x-ray, where $\mathcal{D}_{\perp}$ is on the order of a few \si{\nm} and $\mathcal{R}_F<1$ [\emph{cf.\@} \cref{sec:planar_interface}], it has been demonstrated experimentally that x-ray reflection gratings used in the extreme off-plane mount exhibit polarization-insensitive diffraction efficiency that can be modeled with a fair degree of accuracy using the assumption of perfect conductivity \cite{Marlowe16}. 
However, this approach requires the use of TE polarization to achieve physical results and additionally, the reflectivity of the grating material must be taken into account separately to predict absolute diffraction efficiency. 

With the assumption of a perfectly conducting grating, the physical picture of soft x-rays interacting with atomic electrons [\emph{cf.\@} \cref{sec:SXR_med}] is abandoned and instead, electric current confined to an infinitesimal skip depth [\emph{cf.\@} footnote~\ref{footnote:skin_depth}] at the surface of the grating is considered as the source for the reflected-diffracted fields \cite{Loewen97,Petit80}. 
This mathematical problem then can be posed using an inhomogeneous version of the Helmholtz equation defined by \cref{eq:Helmholtz_above_grating}, where $\mathbold{E}$ and $\mathbold{E}''$ are the incident and reflected-diffracted electric fields, respectively:\footnote{This expression can be obtained by combining \cref{eq:Maxwell_vac_start_1,eq:Maxwell_vac_start_2,eq:Maxwell_vac_start_3,eq:Maxwell_vac_start_4} with $\rho(\mathbold{r},t)= 0$ and then assuming time-harmonic fields as in \cref{sec:time_harmonic}.}
\begin{subequations}
\begin{equation}\label{eq:Helmholtz_inhomog}
 \left( \nabla^2 + k_0^2 \right) \left[ \mathbold{E} \left( \mathbold{r} \right) + \mathbold{E}'' \left( \mathbold{r} \right) \right] = - i \mu_0 \omega \mathbfcal{J}(\mathbold{r})
 \end{equation}
with $\mathbfcal{J}(\mathbold{r})$ as electric current density, which appears in the time-harmonic version of Amp\`ere's law given by $\curl \mathbold{H} \left( \mathbold{r} \right) = \mathbfcal{J}(\mathbold{r}) - i \omega \epsilon_0 \mathbold{E} \left( \mathbold{r} \right)$ [\emph{cf.\@} \cref{eq:Maxwell_vac_TH_4}]. 
By the same argument for symmetry along the groove direction described in \cref{sec:reflected_diffracted}, \cref{eq:Helmholtz_inhomog} reduces to a scalar equation, where, as a generalization of \cref{eq:Helmholtz_above_grat_scal} for the case of TE polarization, the relevant component of the field, $u'' (x,y) = E_z'' (x,y)$, must satisfy 
\begin{align}\label{eq:Helmholtz_inhomog_scalar}
 \begin{split}
 \left( \nabla_{xy}^2 + k_{xy}^2 \right) E_z'' (x,y) = - i \mu_0 \omega \mathcal{J}_z (x,y) \\
 \text{with } \nabla_{xy}^2 \equiv \nabla^2 - \pdv[2]{z} \quad \text{and} \quad k_{xy}^2 \equiv k_0^2 - k_z^2 , 
 \end{split}
 \end{align}
\end{subequations}
where $\mathcal{J}_z \equiv \mathbfcal{J} \cdot \mathbold{\hat{z}}$ is the $z$-component of the current density. 
A current density confined to the grating surface, $\mathcal{P}$, with infinitesimal thickness can be written in terms of a surface-current density vector, $\pmb{\mathscr{J}}$:
\begin{subequations}
\begin{equation}\label{eq:surf_curr_term}
 \mathcal{J}_z (x,y) = \mathscr{J}_z (x) \delta_D \left[y - g(x) \right] ,
 \end{equation}
where $\mathscr{J}_z \equiv \pmb{\mathscr{J}} \cdot \mathbold{\hat{z}}$ and $\delta_D \left[y - g(x) \right]$ is a \emph{Dirac delta function} that ensures $\mathscr{J}_z = 0$ unless $y=g(x)$. 
This surface-current density is assumed to be confined to $\mathcal{P}$ such that 
\begin{equation}\label{eq:surf_curr_expan}
 \mathscr{J}_z (x) \mathrm{e}^{-i k_x x} = \sum_{n=-\infty}^{\infty} \mathscr{J}_{z,n} \mathrm{e}^{i n K x} ,
 \end{equation} 
where $\mathrm{e}^{-i k_x x}$ accounts for the phase shift from the incident wave and the Fourier coefficients are defined by 
\begin{equation}
 \mathscr{J}_{z,n}  = \frac{1}{d} \int_0^d \mathscr{J}_z (x) \mathrm{e}^{- i k_{x,n}'' x} \dd{x} \quad \text{for } n = 0, \pm 1, \pm 2, \pm 3 \dotsc,
 \end{equation}
with $k_{x,n}'' \equiv k_x + n K$. 
\end{subequations}
Calculating theoretical diffraction efficiency by the integral method involves the following mathematical steps:
\begin{enumerate}[noitemsep]
 \item The inhomogeneous Helmholtz equation given by \cref{eq:Helmholtz_inhomog_scalar} is solved for the case of a point-source grating using \emph{Green's functions} \cite{Jackson75,Petit80}.
 \item An expression for $E_z'' (x,y)$ is formulated as the convolution of the appropriate Green's function and $\mathscr{J}_z (x)$ [\emph{cf.\@} \cref{eq:surf_curr_expan}].
 \item A \emph{Dirichlet boundary condition} \cite{Jackson75,Petit80} for TE polarization is enforced on $\mathcal{P}$ to yield a system of coupled integrals that can be solved numerically to determine $\mathscr{J}_{z,n}$ for a truncated range of $n$.
 \item The expression for $E_z'' (x,y)$ is truncated accordingly and the amplitude of each propagating order is found using calculated values for $\mathscr{J}_{z,n}$. 
 \item The amplitudes for each propagating order are used to calculate diffraction efficiency, $\mathscr{E}_n$, in TE polarization [\emph{cf.\@} \cref{eq:TE_efficiency}]. Due to the assumption of perfect conductivity, however, this quantity is equivalent to relative diffraction efficiency with $\sum_n \mathscr{E}_n = 1$ over propagating orders. 
 \item Absolute diffraction efficiency is determined by modulating the results for relative diffraction efficiency by an appropriate expression for soft x-ray reflectivity from a mirror flat using the framework presented in \cref{sec:planar_interface}. 
 \end{enumerate}
Following textbooks on electromagnetic grating theory \cite{Petit80,antonakakis14}, these items are described in \cref{sec:imhomog_green,sec:dirch_bound,sec:rel_eff_perf,sec:refl_account}. 

\subsubsection[Solving the Inhomogeneous Helmholtz Equation]{Solving the Inhomogeneous Helmholtz Equation}\label{sec:imhomog_green}
%%%%%%%%%%%%%%%%%%%%%%%%%%%%%%%%%%%%%%%%%-------------------------------------------------
Following the method of \emph{Green's functions} for differential equations \cite{Jackson75,Petit80}, the inhomogeneous Helmholtz equation [\emph{cf.\@} \cref{eq:Helmholtz_inhomog_scalar}] with \cref{eq:surf_curr_term} can be solved by first considering a solution for diffracted fields in the presence of a periodic array of point sources for surface current. 
Using notation similar to \cref{eq:Helmholtz_inhomog_scalar,eq:surf_curr_term}, this can be written as 
\begin{equation}\label{eq:Green_Helmholtz}
 \left( \nabla^2_{xy} + k^2_{xy} \right) \mathcal{G} (x,y) = \mathrm{e}^{i k_x x} \delta_D \left( y \right) \Sha_d \left( x \right) = \mathrm{e}^{i k_x x} \delta_D \left( y \right) \sum_{n=-\infty}^{\infty} \delta_D \left( x - n d \right) ,
 \end{equation}
where $\mathrm{e}^{i k_x x}$ accounts for the phase shift of the incident wave, $\Sha_d \left( x \right)$ is a \emph{Dirac comb} of period $d$ [\emph{cf.\@} \cref{sec:resolving_power}] and $\mathcal{G} (x,y)$ is a unitless Green's function of the form 
\begin{subequations}
\begin{equation}\label{eq:Green_func1}
 \mathcal{G} (x,y) = \mathrm{e}^{i k_x x} \sum_{n=-\infty}^{\infty} \mathcal{G}_n (y) \mathrm{e}^{i n K x} = \sum_{n=-\infty}^{\infty} \mathcal{G}_n (y) \mathrm{e}^{i k_{x,n}'' x}   
 \end{equation}
with Fourier coefficients given by 
\begin{equation}
 \mathcal{G}_n (y) = \frac{1}{d} \int_0^d \mathcal{G} (x,y) \mathrm{e}^{- i k_{x,n}'' x} \dd{x} \quad \text{for } n = 0, \pm 1, \pm 2, \pm 3 \dotsc
 \end{equation}
\end{subequations}
Writing $\Sha_d \left( x \right)$ as a Fourier series\footnote{This is given by 
\begin{equation*}
 \Sha_d \left( x \right) \equiv \sum_{n=-\infty}^{\infty} \delta_D \left( x - n d \right) = \frac{1}{d} \sum_{n=-\infty}^{\infty} \mathrm{e}^{i n K x} .
 \end{equation*}} and then inserting \cref{eq:Green_func1} into \cref{eq:Green_Helmholtz} yields 
\begin{subequations}
\begin{equation}
 \sum_{n=-\infty}^{\infty} \left[ \dv[2]{\mathcal{G}_n (y)}{y} + \left( k_{xy}^2 - \left[ k_{x,n}'' \right]^2 \right) \mathcal{G}_n (y) \right] \mathrm{e}^{i k_{x,n}'' x} = \frac{\delta_D \left( y \right)}{d} \sum_{n=-\infty}^{\infty} \mathrm{e}^{i k_{x,n}'' x} ,
 \end{equation}
which reduces to  
\begin{equation}
 \dv[2]{\mathcal{G}_n (y)}{y} + \left( k_{y,n}''\right)^2 \mathcal{G}_n (y) = \frac{\delta_D \left( y \right)}{d}
 \end{equation}
with $\left( k_{y,n}'' \right)^2 \equiv k_0^2 - \left( k_{x,n}'' \right)^2 - k_z^2$. 
This differential equation has outgoing-wave solutions given by 
\begin{equation}\label{eq:Green_y_sol}
 \mathcal{G}_n (y) = \frac{1}{2 i d k_{y,n}''} \mathrm{e}^{i k_{y,n}'' \abs{y}} ,
 \end{equation}
where the Fourier coefficients are determined from the condition that the jump in $\dv*{\mathcal{G}_n (y)}{y}$ across $y=0$ is equal to $d^{-1}$ \cite{Petit80,Goray10}. 
\end{subequations}
The diffracted field from \cref{eq:Helmholtz_inhomog_scalar}, $E_z'' (x,y)$, then can be expressed a convolution of the full Green's function, determined from \cref{eq:Green_func1} with \cref{eq:Green_y_sol} and $(x,y) \to (x-x',y-y')$:
\begin{equation}\label{eq:full_green}
 \mathcal{G} (x-x',y-y') = \frac{1}{2id} \sum_{n=-\infty}^{\infty} \frac{1}{k_{y,n}''} \mathrm{e}^{i \left( k_{x,n}'' \left( x - x' \right) + k_{y,n}'' \abs{\left( y - y' \right)} \right)} ,
 \end{equation}
with the source term for current given by \cref{eq:surf_curr_term}: 
\begin{subequations}
\begin{align}\label{eq:E_z_integral1}
 \begin{split}
 E_z'' (x,y) &= - i \omega \mu_0 \left[ \mathcal{G} (x,y) * \mathscr{J}_z (x) \delta_D \left(y - g(x) \right) \right] \\
 &= - i \omega \mu_0 \iint_{-\infty}^{\infty} \mathcal{G} (x-x',y-y') \mathscr{J}_z (x') \delta_D \left[y' - g(x') \right] \dd{x'} \dd{y'} . 
 \end{split}
 \end{align}
Because both $\mathcal{G} (x-x',y-y')$ and $\mathscr{J}_z (x')$ have periodicity in $d$ while the Dirac delta ensures $y' = g(x')$, \cref{eq:E_z_integral1} can be written as a finite integral on $\mathcal{P}$, over the groove period, $d$:
\begin{align}
 \begin{split}\label{eq:E_z_integral2}
 E_z'' (x,y) &= - \frac{\omega \mu_0}{2 d} \sum_{n=-\infty}^{\infty} \frac{1}{k_{y,n}''} \int_{\mathcal{P}} \mathrm{e}^{i \left( k_{x,n}'' \left( x - x' \right) + k_{y,n}'' \abs{\left( y - g(x') \right)} \right)} \mathscr{J}_z (x') \dd{\ell'} \\
 &= - \frac{\omega \mu_0}{2 d} \sum_{n=-\infty}^{\infty} \frac{1}{k_{y,n}''} \int_0^d \mathrm{e}^{i \left( k_{x,n}'' \left( x - x' \right) + k_{y,n}'' \abs{\left( y - g(x') \right)} \right)} \mathscr{J}_z (x') \ell (x') \dd{x'} \\ &\quad \text{with } \ell (x') \equiv \sqrt{1 + \left( \dv{g(x')}{x'} \right)^2} ,
 \end{split}
 \end{align}
\end{subequations}
where $\dd{\ell'} = \sqrt{\dd{x'}^2 + \dd{y'}^2}$ is a line element on $\mathcal{P}$ with $\dd{y'} = \left( \dv*{g(x')}{x'} \right) \dd{x'}$.  

\subsubsection{Enforcing a Dirichlet Boundary Condition}\label{sec:dirch_bound} 
%%%%%%%%%%%%%%%%%%%%%%%%%%%%%%%%%%%%%%%%%-------------------------------------------------
The integral for $E_z'' (x,y)$ [\emph{cf.\@} \cref{eq:E_z_integral2}] must be consistent with boundary conditions that follow from \cref{eq:grating_boundary_1,eq:grating_boundary_2,eq:grating_boundary_3,eq:grating_boundary_4} with null refracted fields and the inclusion of surface-current density, $\pmb{\mathscr{J}} (x)$:\footnote{While the continuity equation [\emph{cf.\@} \cref{eq:continuity}] implies dynamic charge density in the presence of current density, the characteristic timescale for decay in a medium with permittivity $\epsilon$ and conductivity $\sigma$ is $\epsilon / \sigma$, which approaches zero for $\sigma \to \infty$ \cite[\emph{cf.\@} footnote~\ref{footnote:skin_depth}]{Griffiths17}. Therefore, surface-charge density is neglected in \cref{eq:perf_grat_boundary_3}.}
\begin{subequations}
\begin{align}
 \left[ n_x \mathbold{\hat{x}} + n_y \mathbold{\hat{y}}  \right] \times \left[ \mathbold{E} \left( x, g(x) \right) + \mathbold{E''} \left( x, g(x) \right) \right] &= \mathbf{0} \label{eq:perf_grat_boundary_1} \\
 \left[ n_x \mathbold{\hat{x}} + n_y \mathbold{\hat{y}}  \right] \times \left[\mathbold{H} \left( x, g(x) \right)  + \mathbold{H''} \left( x, g(x) \right) \right] &= \pmb{\mathscr{J}} (x) \label{eq:perf_grat_boundary_2} \\
 \left[ n_x \mathbold{\hat{x}} + n_y \mathbold{\hat{y}}  \right] \cdot \left[ \mathbold{E} \left( x, g(x) \right) + \mathbold{E''} \left( x, g(x) \right) \right] &= 0 \label{eq:perf_grat_boundary_3} \\
 \left[ n_x \mathbold{\hat{x}} + n_y \mathbold{\hat{y}}  \right] \cdot \left[ \mathbold{H} \left( x, g(x) \right) + \mathbold{H''} \left( x, g(x) \right) \right] &= 0 \label{eq:perf_grat_boundary_4} , 
 \end{align}
\end{subequations} 
where $\mathbold{\hat{n}}_{\mathcal{P}}(x) \equiv n_x \mathbold{\hat{x}} + n_y \mathbold{\hat{y}}$ is the surface-normal vector [\emph{cf.\@} \cref{fig:grating_boundary_def}]. 
By decomposing the field vectors into Cartesian components, it is found that the only condition relevant for $E_z'' (x,y)$ is a \emph{Dirichlet boundary condition} for the $z$-component of the electric field \cite{Petit80,Jackson75}:
\begin{subequations} 
\begin{equation}\label{eq:Dirch_a}
 E_z'' \left[ x, g(x) \right] = - E_z \left[ x, g(x) \right] = - \mathcal{A}_0 \sin \left( \gamma \right) \mathrm{e}^{i \left( k_x x + k_y g(x) \right)} 
 \end{equation}
with $k_x = - k_0 \sin \left( \alpha \right) \sin \left( \gamma \right)$ and $k_y = - k_0 \cos \left( \alpha \right) \sin \left( \gamma \right)$ as components of the incident wave vector [\emph{cf.\@} \cref{eq:beam_vec_grat_2}]. 
Using \cref{eq:E_z_integral2,eq:surf_curr_expan}, \cref{eq:Dirch_a} becomes
\begin{equation}\label{eq:Dirch_b}
 \int_0^d \mathcal{K} (x,x') \sum_{n=-\infty}^{\infty} \mathscr{J}_{z,n} \mathrm{e}^{i n K x'} \dd{x'} = \mathcal{A}_0 \sin \left( \gamma \right) \mathrm{e}^{i k_y g(x)} \equiv I (x) ,
 \end{equation}
where a kernel function has been defined as
\begin{align}\label{eq:kern_def}
 \begin{split}
 \mathcal{K} (x,x') &\equiv - i \omega \mu_0 \ell (x') \mathrm{e}^{i k_x (x-x')} \mathcal{G}(x-x',g(x)-g(x')) \\
  &= \frac{\omega \mu_0}{2 d} \sqrt{1 + \left( \dv{g(x')}{x'} \right)^2} \sum_{n=-\infty}^{\infty} \frac{1}{k_{y,n}''} \mathrm{e}^{i \left( n K \left( x - x' \right) + k_{y,n}'' \abs{\left( g(x) - g(x') \right)} \right)} , 
 \end{split}
 \end{align}
\end{subequations} 
and the phase factor $\mathrm{e}^{i k_x x}$ drops out from both sides of the equation \cite{Petit80}. 

To determine $\mathscr{J}_{z,n}$ from the integral expression given by \cref{eq:Dirch_b}, $\mathcal{K} (x,x')$ is Fourier-expanded in both $x$ and $x'$:
\begin{subequations} 
\begin{equation}\label{eq:kern_exp}
 \mathcal{K} (x,x') = \sum_{m=-\infty}^{\infty} \sum_{n=-\infty}^{\infty} A_{mn} \mathrm{e}^{i K \left( m x - n x' \right)}
 \end{equation}
with coefficients defined by
\begin{equation}\label{eq:kern_coef}
 A_{mn} = \frac{1}{d^2} \iint_0^d \mathcal{K} (x,x') \mathrm{e}^{-i K \left( m x - n x' \right)} \dd{x} \dd{x'} ,
 \end{equation}
\end{subequations} 
and similarly, $I (x)$ [\emph{cf.\@} \cref{eq:Dirch_b}] is expanded in $x$: 
\begin{subequations} 
\begin{equation}\label{eq:inc_exp}
 I (x) \equiv \mathcal{A}_0 \sin \left( \gamma \right) \mathrm{e}^{i k_y g(x)} = \sum_{n=-\infty}^{\infty} B_n \mathrm{e}^{i n K x}
 \end{equation}
with coefficients defined by 
\begin{equation}\label{eq:inci_coef}
 B_n = \frac{\mathcal{A}_0 \sin \left( \gamma \right)}{d} \int_0^d \mathrm{e}^{i \left( k_y g(x) - n K x \right)} \dd{x} . 
 \end{equation}
\end{subequations} 
By inserting \cref{eq:kern_exp,eq:inc_exp} into \cref{eq:Dirch_b} and noting that
\begin{equation}
 \int_0^d \sum_{\ell=-\infty}^{\infty} \sum_{m=-\infty}^{\infty} A_{\ell m} \mathrm{e}^{i K \left( \ell x - m x' \right)} \sum_{n=-\infty}^{\infty} \mathscr{J}_{z,n} \mathrm{e}^{i n K x'} \dd{x'} = d \sum_{\ell=-\infty}^{\infty} \sum_{m=-\infty}^{\infty} A_{\ell m} \mathrm{e}^{i \ell K x} \mathscr{J}_{z,m} ,
 \end{equation}
where the only terms that survive the integral are those with $m=n$, the expression reduces to
\begin{subequations} 
\begin{equation}\label{eq:Dirch_c}
 d \sum_{m=-\infty}^{\infty} \sum_{n=-\infty}^{\infty} A_{m n} \mathrm{e}^{i m K x} \mathscr{J}_{z,n} = \sum_{m=-\infty}^{\infty} B_m \mathrm{e}^{i m K x} ,
 \end{equation}
which implies the equality of each Fourier component:
\begin{equation}\label{eq:Dirch_d}
 d \sum_{n=-\infty}^{\infty} A_{m n} \mathscr{J}_{z,n} = B_m .
 \end{equation}
 \end{subequations}
This is a linear system of equations for $\mathscr{J}_{z,n}$ that can be solved with the aid of a computer using \cref{eq:kern_coef,eq:inci_coef} for $A_{m n}$ and $B_m$, respectively \cite{Petit80,antonakakis14}.
In order to determine $\mathscr{J}_{z,n}$ numerically, \cref{eq:Dirch_d} is to be truncated into a sum from $-\mathcal{N}$ to $\mathcal{N}$:
\begin{equation}
 d \sum_{n=-\mathcal{N}}^{\mathcal{N}} A_{m n} \mathscr{J}_{z,n} = B_m ,
 \end{equation}
where $A_{m n}$ is calculated using \cref{eq:kern_coef} with a truncated version of \cref{eq:kern_def}:
\begin{equation}\label{eq:kern_def_trunc}
 \mathcal{K} (x,x') \equiv \frac{\omega \mu_0}{2 d} \sqrt{1 + \left( \dv{g(x')}{x'} \right)^2} \sum_{n=-\mathcal{N}}^{\mathcal{N}} \frac{1}{k_{y,n}''} \mathrm{e}^{i \left( n K \left( x - x' \right) + k_{y,n}'' \abs{\left( g(x) - g(x') \right)} \right)} 
 \end{equation}
and $B_m$ is calculated using \cref{eq:inci_coef}. 
The relevant component of diffracted field, $E_z'' (x,y)$, can then be calculated through \cref{eq:E_z_integral2} using these data for $\mathscr{J}_{z,n}$:
\begin{align}
 \begin{split}
 E_z'' (x,y) = i \omega \mu_0 \int_0^d \mathcal{G} (x-x',y-g(x')) \mathscr{J}_z (x') \mathrm{e}^{i k_{x,n}'' x} \ell (x') \dd{x'} \\
 \text{with} \quad \mathscr{J}_z (x') =  \sum_{n=-\mathcal{N}}^{\mathcal{N}} \mathscr{J}_{z,n} \mathrm{e}^{i k_{x,n}'' x} \quad \text{and} \quad \ell (x') \equiv \sqrt{1 + \left( \dv{g(x')}{x'} \right)^2} ,
 \end{split}
 \end{align}
where $\mathcal{G} (x-x',y-y')$ is the Green's function given by \cref{eq:full_green}. 
By expanding the result in the form of \cref{eq:diff_field_exp}:
\begin{subequations}
\begin{equation}\label{eq:known_field_exp}
 E_z'' (x,y) = \sin \left( \gamma \right) \sum_{n=-\infty}^{\infty} \mathcal{A}_n'' \mathrm{e}^{i \left( k_{x,n}'' x + k_{y,n}'' y \right)} ,
 \end{equation}
the amplitude for each diffracted order, $\mathcal{A}_n''$, can be determined from
\begin{equation}\label{eq:known_field_coef}
 \mathcal{A}_n'' = \frac{\mathrm{e}^{- i k_{y,n}'' y}}{d} \int_0^d E_z'' (x,y) \mathrm{e}^{- i k_{x,n}'' x} \dd{x} ,
 \end{equation}
where $n$ that satisfy $k^2_0 > \left( k_x + n K \right)^2 + k_z^2$ correspond to propagating orders with 
\begin{equation}\label{eq:known_field_coef_norm}
 \norm{\mathcal{A}_n''}^2 = \frac{1}{d^2} \norm{\int_0^d E_z'' (x,y) \mathrm{e}^{- i k_{x,n}'' x} \dd{x}}^2 
 \end{equation}
\end{subequations}
as a result of $k_{y,n}''$ being purely real. 

\subsubsection{On Perfectly Conducting Gratings}\label{sec:rel_eff_perf} 
%%%%%%%%%%%%%%%%%%%%%%%%%%%%%%%%%%%%%%%%%-------------------------------------------------
With $\norm{\mathcal{A}_n''}^2$ determined from \cref{eq:known_field_coef_norm} for propagating orders, diffraction efficiency can be calculated using \cref{eq:TE_efficiency}: 
\begin{equation}\label{eq:rel_efficiency}
 \mathscr{E}_n = \frac{k_{y,n}''}{k_y} \norm{\frac{\mathcal{A}_n''}{\mathcal{A}_0}}^2 \equiv \frac{\cos \left( \beta \right)}{\cos \left( \alpha \right)} \norm{\tilde{r}^{(n)}}^2 ,
\end{equation}
where $\mathcal{A}_0$ is the amplitude of the incident wave and $\tilde{r}^{(n)} \equiv \mathcal{A}_n'' / \mathcal{A}_0$ is a unitless amplitude coefficient. 
In analogy to a perfectly conducting mirror flat with $\mathcal{R}_F = 1$, however, the sum of $\mathscr{E}_n$ over all propagating orders is equal to unity. 
To show that $\mathscr{E}_n$ does in fact represent relative efficiency, \emph{Green's theorem} is invoked for two arbitrary scalar fields, $U(\mathbold{r})$ and $V(\mathbold{r})$, that are considered to be enclosed within a volume $\mathcal{V}$ such that \cite{Petit80}
\begin{align}\label{eq:green_theorem}
 \begin{split}
 &\int_{\mathcal{V}} \left[ U(\mathbold{r}) \nabla^2 V(\mathbold{r}) - V(\mathbold{r}) \nabla^2 U(\mathbold{r}) \right] \dd[3]{\mathbold{r}} \\
 &\quad \quad = \oint_{\mathcal{S}} \left[ U(\mathbold{r}) \, \mathbold{\hat{n}} \cdot \grad{V(\mathbold{r})} - V(\mathbold{r}) \, \mathbold{\hat{n}} \cdot \grad{U(\mathbold{r})} \right] \dd{a}, 
 \end{split}
 \end{align}
where $\dd{a}$ is an infinitesimal area element on an enclosing surface $\mathcal{S}$ with an outward-normal unit vector $\mathbold{\hat{n}}$ \cite{Jackson75}. 
In this case, $U(\mathbold{r})$ and $V(\mathbold{r})$ are taken to be solutions to the Helmholtz equation in vacuum [\emph{cf.\@} \cref{eq:Helmholtz_above_mirror_ch2,eq:Helmholtz_above_mirror}]:
\begin{equation} 
 \left( \laplacian + k_0^2 \right) \left\{
 \begin{array}{lr}
 U(\mathbold{r})\\
 V(\mathbold{r})
 \end{array}
 \right\} = \mathbf{0} \quad \text{for } y > g(x) ,
 \end{equation}
so that $U(\mathbold{r}) \nabla^2 V(\mathbold{r}) = V(\mathbold{r}) \nabla^2 U(\mathbold{r}) = - k_0^2 U(\mathbold{r}) V(\mathbold{r})$ and \cref{eq:green_theorem} becomes
\begin{equation}\label{eq:green_theorem_mod1}
 \oint_{\mathcal{S}} \left[ U(\mathbold{r}) \, \mathbold{\hat{n}} \cdot \grad{V(\mathbold{r})} - V(\mathbold{r}) \, \mathbold{\hat{n}} \cdot \grad{U(\mathbold{r})} \right] \dd{a} = 0 .
 \end{equation}

By the \emph{theorem of invariance} for perfectly conducting gratings \cite{Petit80,Loewen97}, any off-plane geometry parameterized by a cone opening half-angle, $\gamma$, is equivalent in terms of diffraction efficiency to an in-plane geometry with a reduced wavelength $\bar{\lambda} = \lambda \csc (\gamma)$ for a fixed azimuthal incidence angle, $\alpha$ [\emph{cf.\@} \cref{sec:off-plane_geo}]. 
The fields $U(\mathbold{r})$ and $V(\mathbold{r})$ hence can be taken to be solutions to the Helmholtz equation for an in-plane geometry without loss of generality for examining $\mathscr{E}_n$. 
With $k_z = 0$ in such a scenario, the fields depend only on $x$ and $y$ so that the Helmholtz equation can be written as
\begin{equation} 
 \left( \pdv[2]{x} + \pdv[2]{y} + k_0^2 \right) \left\{
 \begin{array}{lr}
 U(x,y) \\
 V(x,y) 
 \end{array}
 \right\} = 0 \quad \text{for } y > g(x) 
 \end{equation}
and the surface-area integral given by \cref{eq:green_theorem_mod1} becomes a line integral over a path $\mathcal{L}$ that defines a cross-section in the $(x,y)$ plane:
\begin{equation}\label{eq:green_theorem_mod2}
 \int_{\mathcal{L}} \left[ U(x,y) \, \mathbold{\hat{n}} \cdot \grad{V(x,y)} - V(x,y) \, \mathbold{\hat{n}} \cdot \grad{U(x,y)} \right] \dd{\ell} = 0 ,
 \end{equation}
where $\dd{\ell}$ is an infinitesimal line element and $\mathbold{\hat{n}}$ is also confined to the $(x,y)$ plane. 
\begin{figure}
 \centering
 \includegraphics[scale=1.25]{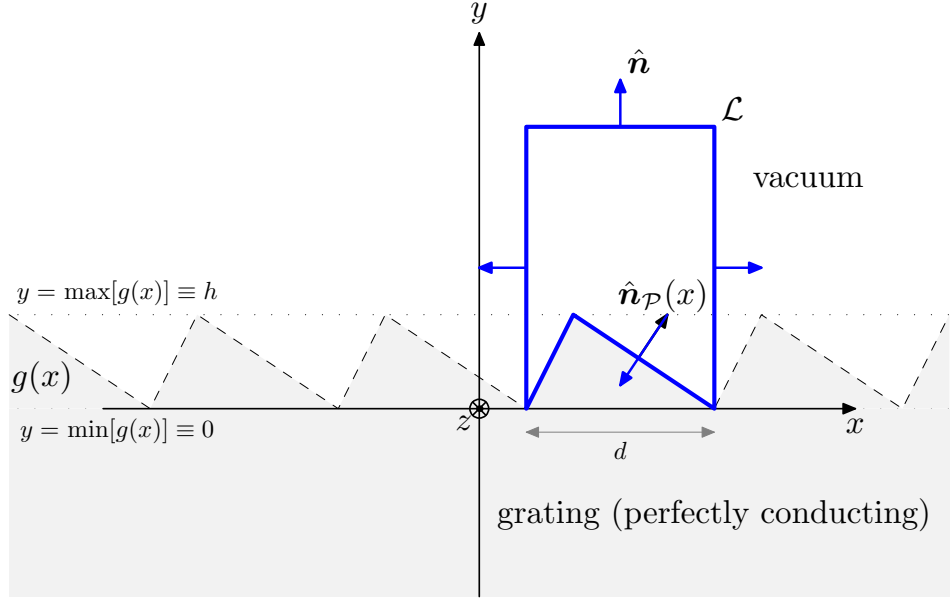} 
 \caption[Surface-relief boundary between vacuum and a perfectly conducting medium]{Surface-relief boundary between vacuum and a perfectly conducting medium. As in \cref{fig:grating_boundary_def}, the grating grooves are aligned with the $z$-direction so that the cross-sectional profile, $\mathcal{P}$ (shown as a dashed line), is described by the periodic function $g(x)$ confined to $0 \leq y \leq h$ while the unit vector $\mathbold{\hat{n}}_{\mathcal{P}} (x)$ describes the direction normal to grating surface at a position $x$. The blue line, $\mathcal{L}$, represents the cross-section of an areal surface $\mathcal{S}$ that encloses a volume $\mathcal{V}$ where two scalar fields, $U(\mathbold{r})$ and $V(\mathbold{r})$, satisfy Green's theorem given by \cref{eq:green_theorem} using the outward-normal unit vector $\mathbold{\hat{n}}$.}\label{fig:grating_boundary_perf} 
 \end{figure}
This is illustrated in \cref{fig:grating_boundary_perf}, where $\mathcal{L}$ (shown in blue) is taken to be a path that outlines one groove period and extends upward to an unspecified $y$ value. 
Assuming the fields obey a Dirichlet boundary condition on $\mathcal{P}$ with 
\begin{equation}
\left\{
 \begin{array}{lr}
 U \left[ x, g(x) \right] \\
 V \left[ x, g(x) \right] 
 \end{array}
 \right\} = 0 ,
\end{equation}
the portion of the line integral along $\mathcal{P}$ vanishes and \cref{eq:green_theorem_mod2} reduces to\footnote{By symmetry, the vertical portions of the path $\mathcal{L}$ cancel each other out in \cref{eq:green_theorem_mod2}. The piece of $\mathcal{L}$ that survives the line integral is the top portion, where $\mathbold{\hat{n}} = \mathbold{\hat{y}}$.} 
\begin{equation}\label{eq:grat_lemma1}
 \int_0^d \left( U (x,y) \pdv{V (x,y)}{y} - V (x,y) \pdv{U (x,y)}{y} \right) \dd{x} = 0 ,
 \end{equation}
which is a lemma valid for $y > \text{max} [g(x)] \equiv h$ \cite{Petit80}. 

Conservation of energy for diffraction efficiency from a perfectly conducting grating can be examined by taking $U (x,y)$ to be the sum of the incident field and the reflected-diffracted field, $u (x,y) + u'' (x,y)$, and $V(x,y)$ to be its complex conjugate, $U^{*} (x,y)$ \cite{Petit80,antonakakis14}. 
To do this, it is useful to write the scalar diffracted field as two separate sums of propagating and evanescent orders
\begin{equation}
 u'' (x,y) = \sum_{n}^{\text{prop.}} \mathcal{A}_n'' \mathrm{e}^{i \left( k_{x,n}'' x + k_{y,n}'' y \right)} + \sum_{n}^{\text{evan.}} \mathcal{A}_n'' \mathrm{e}^{i \left( k_{x,n}'' x + k_{y,n}'' y \right)} , 
 \end{equation}
where $k_{y,n}''$ is either real or imaginary, respectively. 
With $k_{x,n}'' \equiv k_x + n K$, the two functions become
\begin{subequations}
\begin{equation}
  U (x,y) = \underbrace{\mathcal{A}_0 \mathrm{e}^{i \left( k_x x + k_y y \right)}}_{u (x,y)} + \mathrm{e}^{i k_x x} \left[ \sum_{n}^{\text{prop.}} \mathcal{A}_n'' \mathrm{e}^{i \left( n K x + k_{y,n}'' y \right)} + \sum_{n}^{\text{evan.}} \mathcal{A}_n'' \mathrm{e}^{i \left( n K x + k_{y,n}'' y \right)} \right] 
  \end{equation}
and
\begin{align}
\begin{split}
  V (x,y) &= u^* (x,y) \\
  & + \mathrm{e}^{-i k_x x} \left[ \sum_{n}^{\text{prop.}} \left(\mathcal{A}_n''\right)^* \mathrm{e}^{-i \left( n K x + k_{y,n}'' y \right)} + \sum_{n}^{\text{evan.}} \left(\mathcal{A}_n''\right)^* \mathrm{e}^{-i \left( n K x - k_{y,n}'' y \right)} \right]
  \end{split}
 \end{align}
\end{subequations}
while their partial derivatives are given by 
\begin{subequations}
\begin{align}
\begin{split}
 \pdv{U (x,y)}{y} &= i k_y u (x,y) \\
 & + i \mathrm{e}^{i k_x x} \left[ \sum_{n}^{\text{prop.}} k_{y,n}'' \mathcal{A}_n'' \mathrm{e}^{i \left( n K x + k_{y,n}'' y \right)} + \sum_{n}^{\text{evan.}} k_{y,n}'' \mathcal{A}_n'' \mathrm{e}^{i \left( n K x + k_{y,n}'' y \right)} \right] 
 \end{split}
 \end{align}
and
\begin{align}
\begin{split}
 &\pdv{V (x,y)}{y} = - i k_y u^* (x,y) \\
 & \quad - i \mathrm{e}^{-i k_x x} \left[ \sum_{n}^{\text{prop.}} k_{y,n}'' \left(\mathcal{A}_n''\right)^* \mathrm{e}^{-i \left( n K x + k_{y,n}'' y \right)} - \sum_{n}^{\text{evan.}} k_{y,n}'' \left(\mathcal{A}_n''\right)^* \mathrm{e}^{-i \left( n K x - k_{y,n}'' y \right)} \right]. 
 \end{split}
 \end{align}
\end{subequations}
When these expressions are inserted into \cref{eq:grat_lemma1}, the only terms that survive the integral are those that are not periodic over the groove spacing\footnote{This includes cross-terms from the summations, where $\mathrm{e}^{i n K}$ phase factors vanish, and additionally, the incident fields by definition.} such that the two pieces of the integral become
\begin{subequations}
\begin{align}
 \int_0^d U (x,y) \pdv{V (x,y)}{y} \dd{x} &= - i d \left[ k_y \abs{\mathcal{A}_0}^2 + \sum_{n}^{\text{prop.}} k_{y,n}'' \norm{\mathcal{A}_n'}^2 + \sum_{n}^{\text{evan.}} k_{y,n}'' \norm{\mathcal{A}_n''}^2 \mathrm{e}^{2 i k_{y,n}''} \right] \\
 \int_0^d V (x,y) \pdv{U (x,y)}{y} \dd{x} &= i d \left[ k_y \abs{\mathcal{A}_0}^2 + \sum_{n}^{\text{prop.}} k_{y,n}'' \norm{\mathcal{A}_n'}^2 + \sum_{n}^{\text{evan.}} k_{y,n}'' \norm{\mathcal{A}_n''}^2 \mathrm{e}^{2 i k_{y,n}''} \right] .
 \end{align}
\end{subequations}
The end result yields
\begin{equation}\label{eq:prop_order_unity}
 1 = \sum_{n}^{\text{prop.}} \frac{k_{y,n}''}{k_y} \frac{\norm{\mathcal{A}_n''}^2}{\abs{\mathcal{A}_0}^2} = \sum_{n}^{\text{prop.}} \mathscr{E}_n ,
 \end{equation}
which indicates that diffraction efficiency summed over propagating orders is unity \cite{Petit80}. 

The relation just derived can be verified using \textsc{PCGrate-SX} under \emph{perfect conductivity mode} by defining a grating profile with a very large, complex of refraction\footnote{As described in \cref{sec:refl_account}, \textsc{PCGrate-SX} modulates $\mathscr{E}_n$ by Fresnel reflectivity, $\mathcal{R}_F$, to predict absolute diffraction efficiency under perfect conductivity mode. By using the maximum allowed value of $\tilde{\nu} (\omega) = \num{10000} (1+ i)$ for index of refraction, $\mathcal{R}_F$ is within a fraction of a percent of unity.} that approximates a perfectly-conducting material. 
\begin{figure}
 \centering
 \includegraphics[scale=0.23]{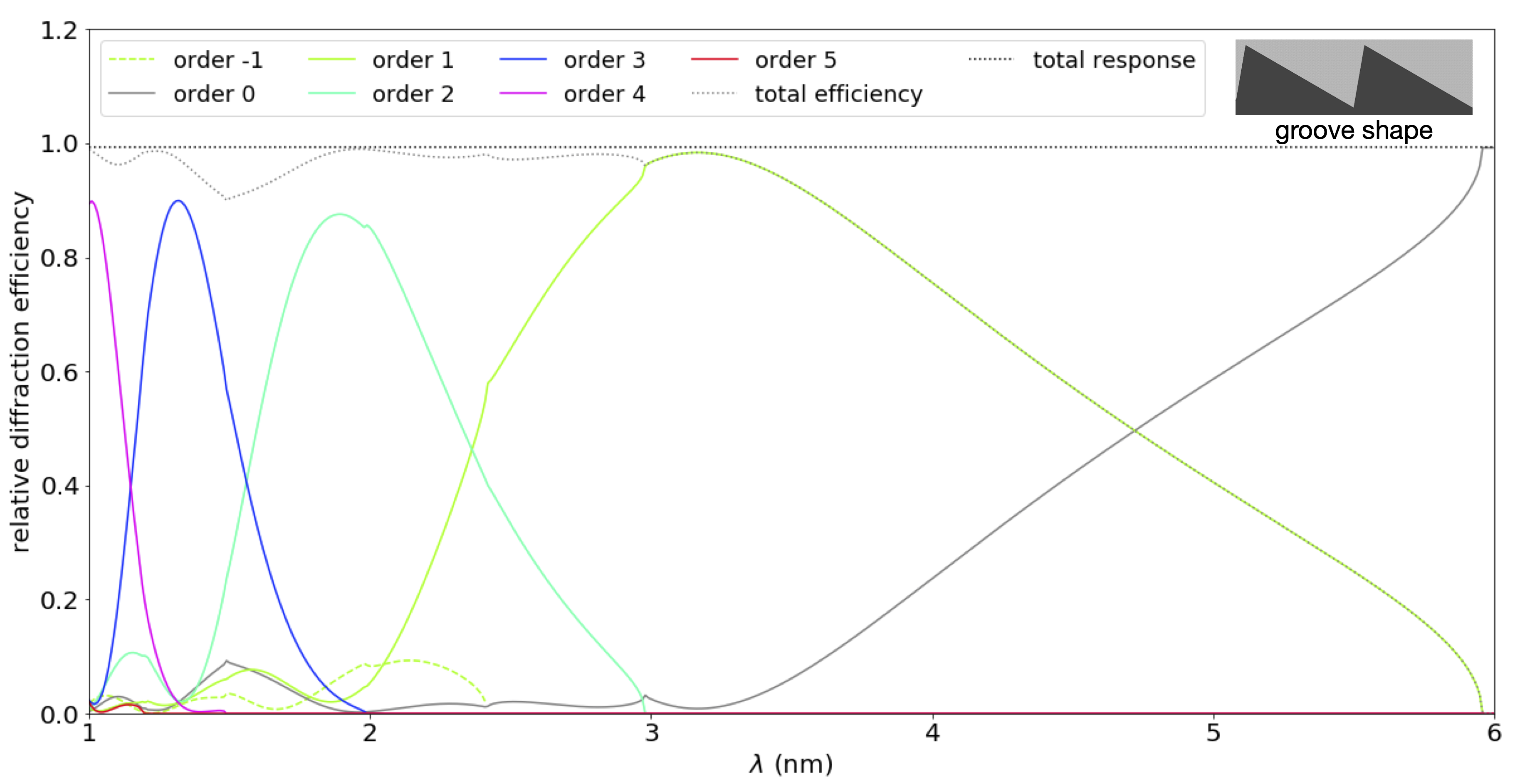}
 \caption[Example of predicted diffraction efficiency from a perfectly conducting grating using \textsc{PCGrate-SX}]{Example of predicted diffraction efficiency from a perfectly conducting grating emulated using $\tilde{\nu} (\omega) = \num{10000} (1+ i)$ for a border index of refraction in \textsc{PCGrate-SX} (v.\ 6.1) under \emph{perfect conductivity mode}. With the total response of the grating, $\mathscr{E}_{\text{tot}} + \mathscr{E}_0$, approaching unity, $\mathscr{E}_n$ is here interpreted as relative efficiency such that reflectivity must be taken into account separately [\emph{cf.\@} \cref{fig:pcgrate_example_au}].}\label{fig:pcgrate_example_perf}
 \end{figure}
An example of this is shown in \cref{fig:pcgrate_example_perf}, where, as indicated in the figure inset, the grating profile is assumed to be triangular with a $\delta = 30^{\circ}$ active blaze angle and an $80^{\circ}$ opposite angle to yield a groove depth of $h \lessapprox \SI{85}{\nm}$ at $d = \SI{160}{\nm}$. 
Meanwhile, a plane wave of radiation with $\SI{1}{\nm} \leq \lambda \leq \SI{5}{\nm}$ is taken to be incident with a polar angle of $\alpha = 25^{\circ}$ and a half-cone opening angle of $\gamma = 1.5^{\circ}$ to yield propagating orders with $n$ ranging from \numrange{-1}{5}. 
The diffraction efficiency, $\mathscr{E}_n$, for each order is shown as a colored curve, where each maximum corresponds to the blaze wavelength in each propagating order.\footnote{Note that the locations of these maxima may differ from what is predicted by \cref{eq:blaze_wavelength,eq:blaze_wavelength_general} for the blaze wavelength. This is a symptom of the scalar treatment of gratings [\emph{cf.\@} \cref{ap:grating_basics}] being insufficient for modeling the efficiency of x-ray reflection gratings accurately.} 
Moreover, the total diffraction efficiency, $\mathscr{E}_{\text{tot}} \equiv \sum_{n \neq 0} \mathscr{E}_n$, and the total response, $\mathscr{E}_{\text{tot}} + \mathscr{E}_0$, are shown as gray and black dotted lines, respectively. 
Because $\mathscr{E}_{\text{tot}}$ here approaches unity, $\mathscr{E}_n$ can be taken to represent relative efficiency. 

\subsection{Accounting for Soft X-ray Reflectivity}\label{sec:refl_account} 
%%%%%%%%%%%%%%%%%%%%%%%%%%%%%%%%%%%%%%%%%-------------------------------------------------
As a result of the integral method for a perfectly-conducting grating interface effectively giving relative efficiency [\emph{cf.\@} \cref{sec:grat_bound_consider}], reflectivity must be taken into account separately to predict absolute efficiency. 
For an ideal blazed grating, this can be treated by multiplying the predicted relative efficiency by an expression for specular reflectivity, $\mathcal{R}$, from the surface of the groove facets \cite{antonakakis14}. 
If it can be assumed that only one side of the groove facets is illuminated\footnote{This is often the case for a near-Littrow configuration defined by $\alpha \approx \delta$, where $\alpha$ is the azimuthal incidence angle and $\delta$ is the blaze angle. However, this approximation may break down for values of $\alpha$ considerably smaller than $\delta$ or for topographies with a relatively shallow opposite angle. Nonetheless, \textsc{PCGrate-SX} takes into account reflectivity for any grating profile.} and that surface roughness is negligible [\emph{cf.\@} \cref{sec:rough_surface}], this can be written as 
\begin{equation}\label{eq:absolute_eff_approx}
 \mathscr{E}_n \to \mathscr{E}_n \mathcal{R}_F (\zeta) = \underbrace{\frac{\cos \left( \beta \right)}{\cos \left( \alpha \right)} \norm{\tilde{r}^{(n)}}^2}_{\text{\cref{eq:rel_efficiency}}} \mathcal{R}_F (\zeta) ,
 \end{equation}
where $\tilde{r}^{(n)} \equiv \mathcal{A}_n'' / \mathcal{A}_0$ and $\mathcal{R}_F (\zeta)$ is Fresnel reflectivity at a grazing-incidence angle $\zeta$ [\emph{cf.\@} \cref{eq:fresnel_refl_def}].\footnote{Relative to the groove-facet plane defined by a blaze angle $\delta$, this incidence angle is given by \cref{eq:angle_on_groove} [\emph{cf.\@} \cref{fig:conical_reflection_edit}].} 
Because reflectivity is virtually polarization insensitive for grazing-incidence soft x-rays [\emph{cf.\@} \cref{sec:reflectivity_polarization}], $\mathcal{R}_F (\zeta)$ can be taken to represent Fresnel reflectivity in s-polarization, given for a thick slab by \cref{eq:s_reflectivity}. 

To illustrate the validity of \cref{eq:absolute_eff_approx}, the \textsc{PCGrate-SX} model from \cref{fig:pcgrate_example_perf} is reproduced in \cref{fig:pcgrate_example_au} for a gold interface. 
\begin{figure}
 \centering
 \includegraphics[scale=0.23]{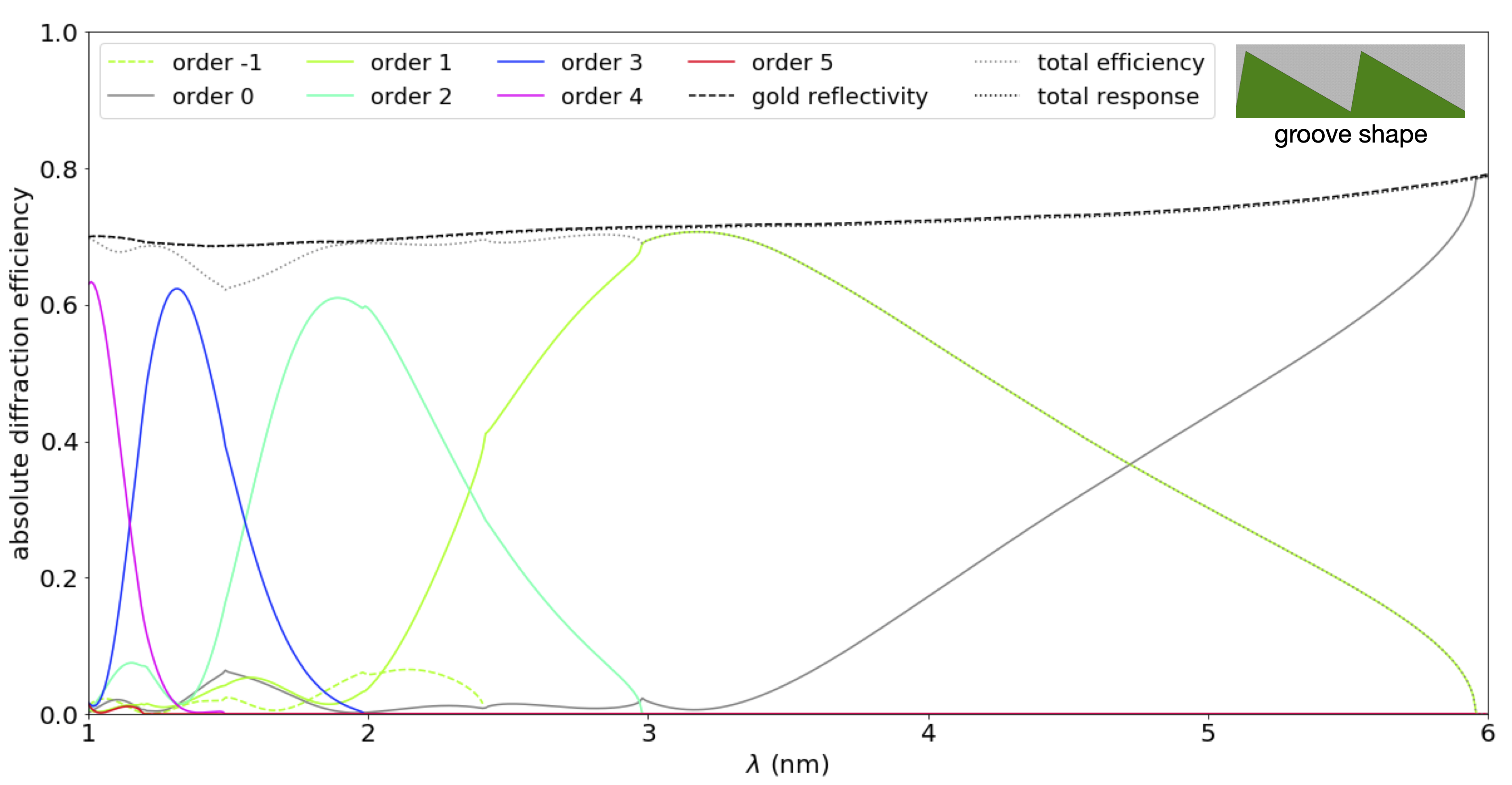}
 \caption[Example of predicted diffraction efficiency for a gold grating using \textsc{PCGrate-SX} under perfect conductivity mode]{Example of predicted diffraction efficiency for a gold grating using \textsc{PCGrate-SX} (v.\ 6.1) under perfect conductivity mode using the same parameters as \cref{fig:pcgrate_example_perf}. The total response of the grating predicted by \textsc{PCGrate-SX} very closely matches Fresnel reflectivity of an equivalent surface at a grazing incidence angle $\zeta \equiv \arcsin \left[ \sin \left( \gamma \right) \cos \left( \delta - \alpha \right) \right]$ [\emph{cf.\@} \cref{eq:angle_on_groove}].}\label{fig:pcgrate_example_au}
 \end{figure}
That is, using data for $\tilde{\nu} (\omega)$ from \cref{fig:X-ray_index}, which were obtained from the CXRO online database \cite{CXRO_database} assuming standard density, \textsc{PCGrate-SX} modulates $\mathscr{E}_n$ predicted for a perfectly conducting grating by the Fresnel reflectivity of a gold slab, which can be taken to represent an overcoat with a thickness several times larger than the penetration depth, $\mathcal{D}_{\perp}$ [\emph{cf.\@} \cref{eq:penetration_depth}].  
In \cref{fig:pcgrate_example_au}, the total response predicted by \textsc{PCGrate-SX} matches $\mathcal{R}_F (\zeta)$ for $\zeta \approx 1.49^{\circ}$, which is determined from $\zeta \equiv \arcsin \left[ \sin \left( \gamma \right) \cos \left( \delta - \alpha \right) \right]$ using $\gamma = 1.5^{\circ}$ and $\abs{\delta - \alpha} = 5^{\circ}$ [\emph{cf.\@} \cref{eq:angle_on_groove}]. 
This indicates that the sum of propagating orders is equal to the Fresnel reflectivity of an equivalent surface:
\begin{equation}
 \sum_{n}^{\text{prop.}} \mathscr{E}_n = \mathcal{R}_F , 
 \end{equation}
which is expected from \cref{eq:absolute_eff_approx}. 
While this relation can be shown more rigorously by solving the Helmholtz equation for a grating border with finite conductivity \cite{Goray10}, this method of predicting absolute diffraction is often sufficient for modeling the behavior of x-ray reflection gratings \cite{Marlowe16}. 

The effect of surface roughness on groove facets can be treated in a similar manner to the scenario of rough mirror flat [\emph{cf.\@} \cref{sec:rough_surface}]. % described in \cref{sec:rough_surface}. 
For the special case of an isotropic, normally-distributed rough surface with a small \emph{root mean square (RMS) roughness} $\sigma$ and a \emph{correlation length} $\ell_{\text{corr}}$ [\emph{cf.\@} \cref{eq:surface_variance,eq:auto_correlation_normal}], $\mathcal{R}_F$ is damped by an exponential factor in the limit of either small or large $\ell_{\text{corr}}$. 
That is, for $\abs{k_{\perp}} \sigma \ll 1$, where $k_{\perp} \equiv - k_0 \sin \left( \zeta \right)$ is the component of $\mathbold{k}$ perpendicular to the groove facet, the \emph{Nevot-Croce factor} derived in appendix~\ref{sec:NC_factor} is valid for small $\ell_{\text{corr}}$ such that $\ell_{\text{corr}} k_{\perp}^2 \ll k_0$ \cite{deBoer95}. 
With the typical size of rough features being too small to produce a diffraction pattern, this quantity describes the fraction of specularly reflected radiation lost due to absorption. % as described in appendix~\ref{sec:NC_factor}. 
For a reflective overcoat that can be treated as a thick slab, the Nevot-Croce factor can be used to modify Fresnel reflectivity [\emph{cf.\@} \cref{eq:NC_refl_ratio}]:
\begin{equation}
 \mathcal{R}_F \to \mathcal{R}_F \norm{\mathrm{e}^{- 2 k_{\perp} \tilde{k_{\perp}}' \sigma^2}}^2 = \mathcal{R}_F \mathrm{e}^{-4 k_0^2 \sin \left( \zeta \right) \Re \left[ \sqrt{\tilde{\nu}^2 (\omega) - \cos^2 \left( \zeta \right) } \right] \sigma^2} ,
 \end{equation}
where $\tilde{k_{\perp}}' = - k_0 \sqrt{\tilde{\nu}^2 (\omega) - \cos^2 \left( \zeta \right)}$ is the component of $\mathbold{\tilde{k}}'$ perpendicular to the groove facet. 
Depending on $\zeta$ and $\lambda$, this Nevot-Croce factor serves an an approximate treatment for groove facets that are dominated by very small lateral surface features. 

Diffuse scattering from facet surface roughness is expected to occur in situations where $\ell_{\text{corr}}$ takes on a value too large to justify use of the Nevot-Croce factor. 
While treating such a scenario rigorously requires using an integral method similar to that described in \cref{sec:grat_bound_consider} \cite{Daillant2009}, a single mode of surface roughness can be treated as a sinusoidal diffraction grating with the effective \emph{surface wavelength}, $\Lambda$, replacing $d$, [\emph{cf.\@} appendix~\ref{sec:intermediate_rough}]. 
Relative to the groove-facet plane, the resulting diffraction pattern has in-plane scattering dominating over off-plane scattering by a factor of $\zeta^{-1}$ [\emph{cf.\@} \cref{fig:in_plane_scatter,fig:off_plane_scatter,eq:in-plane_scatter_angle,eq:off-plane_scatter_angle}]. 
In the limit of very large $\ell_{\text{corr}}$, the \emph{Debye-Waller factor} derived in appendix~\ref{sec:DW_factor} [\emph{cf.\@} \cref{eq:DW_refl_s,eq:DW_factor}] is valid for modeling a reduction in $\mathcal{R}_F$ due to surface roughness: 
\begin{equation}
 \mathcal{R}_F \to \mathcal{R}_F \abs{\mathrm{e}^{-2 k_{\perp}^2 \sigma^2}}^2 = \mathcal{R}_F \mathrm{e}^{-4 k_0^2 \sin^2 \left( \zeta \right) \sigma^2} .
 \end{equation}
A similar factor can be formulated for line-edge and roughness in laminar gratings but its applicability to blazed gratings is limited and not used in this dissertation \cite{Fernandez19,McCurdy20}. 

\section{Summary}\label{sec:ch2_summary} 
%%%%%%%%%%%%%%%%%%%%%%%%%%%%%%%%%%%%%%%%%-------------------------------------------------
The spectral sensitivity of an x-ray grating spectrometer such as the \emph{XGS}, in addition to spectral resolving power, hinges on the instrumental collecting area, which depends directly on the absolute diffraction efficiency of individual reflection gratings used in an extreme off-plane, near-Littrow configuration with $\alpha \approx \delta$ and $\zeta \lessapprox \gamma$ [\emph{cf.\@} \cref{ch:introduction}]. 
As described in \cref{sec:als_testing}, this quantity can be measured in the EUV and soft x-ray at beamline 6.3.2 of the ALS \cite{ALS_632,Underwood96,Gullikson01} by measuring the intensity of the diffracted beam for each propagating order relative to the incident beam after establishing a desired grating geometry using stage rotations and in-situ analysis of the diffracted arc \cite{Miles18}. 
Measured results can be modeled using the \textsc{PCGrate-SX} software package, which employs the integral method for Green's functions to solve the Helmholtz equation for a user-defined grating boundary [\emph{cf.\@} \cref{sec:integral_method}].  
This boundary value problem can be simplified with the assumption of a perfectly-conducting grating boundary and polarization insensitivity, which has been justified experimentally \cite{Marlowe16}. 
When taking into account Fresnel reflectivity for blazed groove facets, the effect of surface roughness can be treated in certain limiting cases. 
The methodology described in this chapter for taking measurements at the beamline and modeling the gathered data using \textsc{PCGrate-SX} is utilized for the experiments described in \cref{ch:master_fab,ch:grating_replication}.
% !TEX root = ../McCoy-Dissertation.tex
\chapter[Thermally-Activated Selective Topography Equilibration (TASTE) for Custom X-ray Reflection Gratings]{Thermally-Activated Selective \\Topography Equilibration for \\Custom X-ray Reflection Gratings}\label{ch:master_fab} 
%%%%%%%%%%%%%%%%%%%%%%%%%%%%%%%%%%%%%%%%%--------------------------------------------------
The process of \emph{thermally-activated selective topography equilibration (TASTE)} \cite{Schleunitz14} is hypothesized in \cref{sec:ch1_conclusions} to be capable of producing blazed-grating surface reliefs that enable both high spectral sensitivity and high spectral resolving power, $\mathscr{R} = \lambda / \Delta \lambda$, in a grazing-incidence spectrometer that employs reflection gratings. 
With a key scientific objective of measuring the diffuse, highly-ionized baryonic content in extended galactic halos and the intergalactic medium through soft x-ray absorption spectroscopy of active galactic nuclei [\emph{cf.\@} \cref{sec:astro_plasmas}], the currently-planned \emph{XGS} for the \emph{Lynx X-ray Observatory} requires several thousand blazed gratings with fanned groove layouts stacked and aligned into modular arrays to intercept soft x-rays coming to a focus in a Wolter-I telescope \cite{Lynx_web,Gaskin19,McEntaffer19}. 
A main challenge from the standpoint of grating fabrication is the realization of a lithographic process that can generate non-parallel groove layouts with high fidelity while also maintaining blazed grooves that enable high diffraction efficiency. 
In particular, sensitivity requirements for \emph{Lynx} require that \emph{total absolute diffraction efficiency}, $\mathscr{E}_{\text{tot}}$ [\emph{cf.\@} \cref{sec:measure_efficiency}], exceeds \SI{40}{\percent} across the soft x-ray bandpass \cite{McEntaffer19}. 

The state of the art for blazed reflection gratings that perform with high $\mathscr{E}_{\text{tot}}$ at soft x-ray wavelengths are those fabricated by crystallographic etching in silicon [\emph{cf.\@} \cref{sec:crystal_etching}], where typically either interference lithography [\emph{cf.\@} \cref{sec:holographic}] or electron-beam lithography (EBL) [\emph{cf.\@} \cref{sec:binary_ebeam}] is used to define a groove layout in resist before the pattern is transferred into the crystalline substrate to produce atomically-smooth sawtooth facets \cite{Franke97,Chang03,Voronov11,McEntaffer13,DeRoo16,Miles18}. 
However, interference lithography faces severe limitations in its ability to pattern non-parallel groove layouts, and even with the direct-write capabilities of EBL, the cubic structure of mono-crystalline silicon [\emph{cf.\@} \cref{fig:crystal_structure}] prevents the formation of fanned or curved grooves with smooth and continuous triangular facets. 
Additionally, these anisotropic etching processes demand precise alignment between the groove layout in resist and the crystallographic planes of silicon to produce a high-fidelity grating [\emph{cf.\@} \cref{sec:crystal_etching}]. 
An alternative to these methods of grating manufacture is TASTE, which combines \emph{grayscale lithography} and polymer \emph{thermal reflow} to produce smooth and continuous surface reliefs in PMMA [\emph{cf.\@} \cref{fig:PMMA}] or other thermoplastic resists such as ZEP520A and mr-PosEBR \cite{Schleunitz14,Kirchner16,Pfirrmann16}. 
With no dependence on the crystal structure of the substrate, TASTE has the potential for realizing reflection gratings that feature both a blazed surface topography and a non-parallel groove layout, thereby enabling high sensitivity and high $\mathscr{R}$ in a soft x-ray spectrometer. 

This chapter describes the first application of TASTE to x-ray reflection grating technology\footnote{Supported by a NASA Space Technology Research Fellowship lasting from 2015 to 2019, much of this research is published in two peer-reviewed articles \cite{McCoy18,McCoy20}.} through the fabrication and subsequent diffraction-efficiency testing of a blazed, \SI{400}{\nm}-period grating prototype patterned in \SI{130}{\nm}-thick, 950k PMMA resist\footnote{950k refers to $M_w = \SI{950}{\kilogram\per\mole}$, the weight-averaged molecular mass [\emph{cf.\@} \cref{eq:weight_avg_molecular_weight}].} and coated with gold for reflectivity, using titanium for adhesion. 
All EBL described in this chapter was carried out using the \textsc{EBPG5200} tool installed at the Penn State Nanofabrication Laboratory \cite[\emph{cf.\@} \cref{fig:ebeam_tool_pic}]{PSU_MRI_nanofab} with process development for TASTE detailed in \cref{sec:TASTE}. 
Fabrication of the grating prototype then is described in \cref{sec:TASTE_prototype} before measurements of its diffraction efficiency in an extreme off-plane mount are presented and analyzed in \cref{sec:taste_eff_results} using the experimental procedures and theoretical modeling discussed in \cref{ch:diff_eff}. 
Conclusions and a summary of this chapter are provided in \cref{sec:ch3summary}. 

\section[Process Development for TASTE]{Process Development for TASTE}\label{sec:TASTE} 
%%%%%%%%%%%%%%%%%%%%%%%%%%%%%%%%%%%%%%%%%-------------------------------------------------
The TASTE process can be understood as a variant of polymer thermal reflow, where a thermoplastic resist heated above its \emph{glass transition temperature}, $T_g$ [\emph{cf.\@} \cref{sec:nanoimprint}], flows with a viscosity that decreases with increasing temperature, $T$, as polymer chains gain mobility to move past one another \cite{Schift10,Kirchner19}. %expected T_g for PMMA? %Petersen02, %, for a given molecular weight, $M_w$
%For PMMA with unaltered molecular weight, $M_{w,0}$, on the order of hundreds of \si{\kilogram\per\mole}, 
This quantity $T_g$ depends on the commercial composition of the resist and typically falls in the range $\SI{100}{\celsius} \lessapprox T_g \lessapprox \SI{130}{\celsius}$ for PMMA with unaltered molecular weight, $M_{w,0}$, on the order of hundreds of \si{\kilogram\per\mole} \cite{Schleunitz10,Ali15}. 
Traditional thermal reflow is often carried out at a temperature $T \gtrapprox T_g + \SI{50}{\celsius}$, where, for example, a laminar structure fabricated by EBL evolves into an energetically-favorable, convex topography [\emph{cf.\@} \cref{fig:reflow_diagram}] as the molten resist slowly equilibrates as surface tension drives the liquid to minimize its \emph{surface free energy}\footnote{This quantity is defined as $\mathscr{F} \equiv \alpha_{\text{i}} \mathscr{A}$, where $\alpha_{\text{i}}>0$ is the surface-tension coefficient for an interface with surface area $\mathscr{A}$ such that the infinitesimal work required to increase $\mathscr{A}$ by $\dd{\mathscr{A}}$ is $\alpha_{\text{i}} \dd{\mathscr{A}}$. For a given $\alpha_{\text{i}}$, a minimization of $\mathscr{F}$ manifests as minimization of $\mathscr{A}$ \cite{landau2013statistical}.} for the both the resist-substrate interface and the resist-air interface while the resist volume remains constant \cite{Kirchner14,Kirchner14a,Kirchner19}. 
\begin{figure}
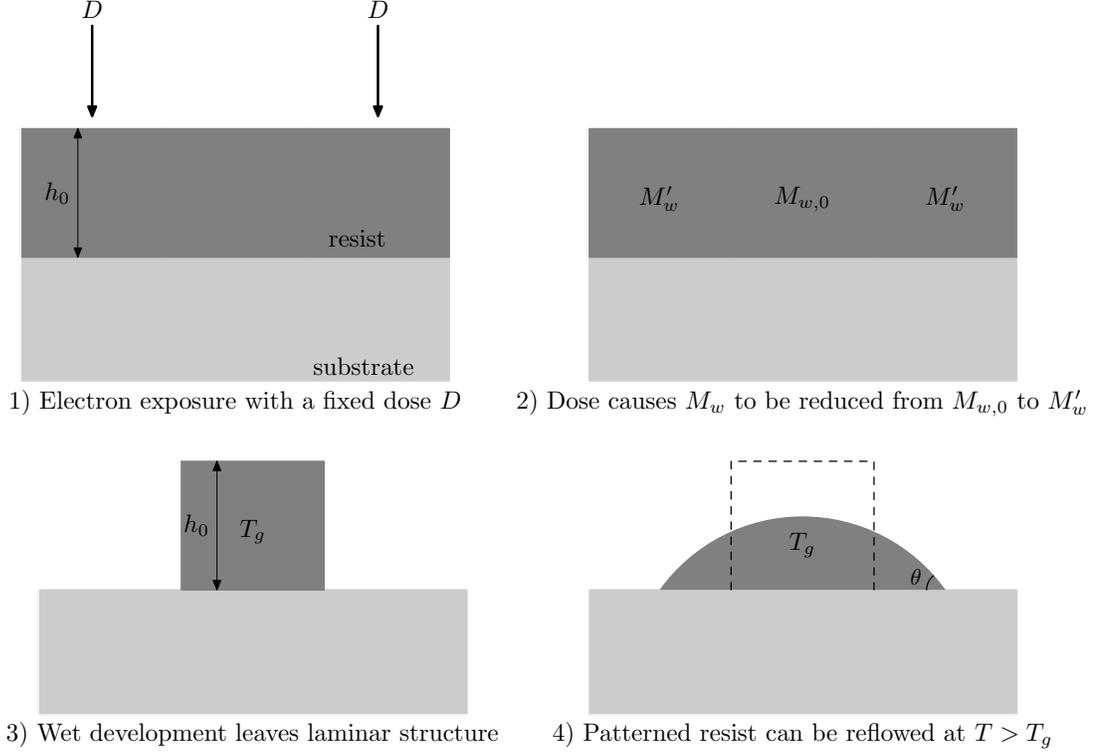

 \centering
 \includegraphics[scale=0.95]{Chapter-3/Figures/reflow_fab1.mps}
 \includegraphics[scale=0.95]{Chapter-3/Figures/reflow_fab2.mps}
 \includegraphics[scale=0.95]{Chapter-3/Figures/reflow_fab3.mps}
 \includegraphics[scale=0.95]{Chapter-3/Figures/reflow_fab4.mps}
 \caption[Outline of a traditional thermal reflow process]{Outline of a traditional thermal reflow process. EBL can be used to pattern a laminar structure in resist such that an electron dose $D$ causes exposed resist with reduced molecular weight, $M_w'$, to be etched away during wet development. With the remaining, unexposed resist having a uniform average molecular weight, $M_{w,0}$, and a glass transition temperature, $T_g$, thermal reflow can be induced for $T \gtrapprox T_g + \SI{50}{\celsius}$, which causes the laminar structure to become molten so that it evolves toward a convex topography with an energy-optimal contact angle, $\theta$, according to a time-dependent, viscoelastic creep process.}\label{fig:reflow_diagram}
 \end{figure} 
With surface pressure and inter-facial wetting being the dominant drivers in reflow of PMMA \cite{Kirchner19}, this equilibrated topography ideally is a surface of constant curvature with minimized area and an energy-optimized contact angle at the resist-substrate interface, $\theta$, which in principle, is determined from \emph{Young's equation} \cite{landau2013statistical}
\begin{equation}\label{eq:contact_angle}
 \cos \left( \theta \right) = \frac{\alpha_{\text{s-a}} - \alpha_{\text{s-r}}}{\alpha_{\text{r-a}}} ,
 \end{equation}
where $\alpha_{\text{s-a}}$, $\alpha_{\text{s-r}}$ and $\alpha_{\text{r-a}}$ are the surface-tension coefficients for the substrate-air, substrate-resist and resist-air interfaces, respectively.\footnote{For a molten polymer at a given temperature $T > T_g$, these quantities are often unknown in practice. A rough estimation for PMMA on silicon with native oxide, however, gives $\theta \approx 22^{\circ}$ \cite{Kirchner14}.} 
Due to the relatively high viscosity of molten resist, however, a sufficiently long reflow time, $t$, is required for the material to reach such an energetically-favorable state with an optimized contact angle. 
Thus, through this time-dependent, \emph{viscoelastic creep process}, resist can be reshaped to the equilibrated state, or any intermediate state with careful control of $t$, by cooling the material back to its glass state with $T < T_g$ \cite{Kirchner14,Kirchner14a,Kirchner19}. 

TASTE differs from the process just described in that it relies on a lateral contrast in $T_g$ imparted in resist so that selective equilibration via poylmer reflow can be achieved. 
This can be established indirectly using \emph{grayscale electron-beam lithography (GEBL)}, where dose-modulated electron exposure locally reduces $M_w$ from $M_{w,0}$ to a value that depends on the electron dose, $D$ (units of \si{\micro\coulomb\per\cm\squared}) \cite{Dobisz00,Stauffer92}. 
For electron-exposed resist with $M_w$ reduced to $\lessapprox \SI{10}{\kilogram\per\mole}$ throughout the depth of the film [\emph{cf.\@} \cref{sec:binary_ebeam}], the developer solubility and $T_g$ both depend strongly on local $M_w$ such that following an appropriately-chosen, timed, wet development process, a multi-level topography is produced \cite{Schleunitz10,Schleunitz14}. 
\begin{figure}
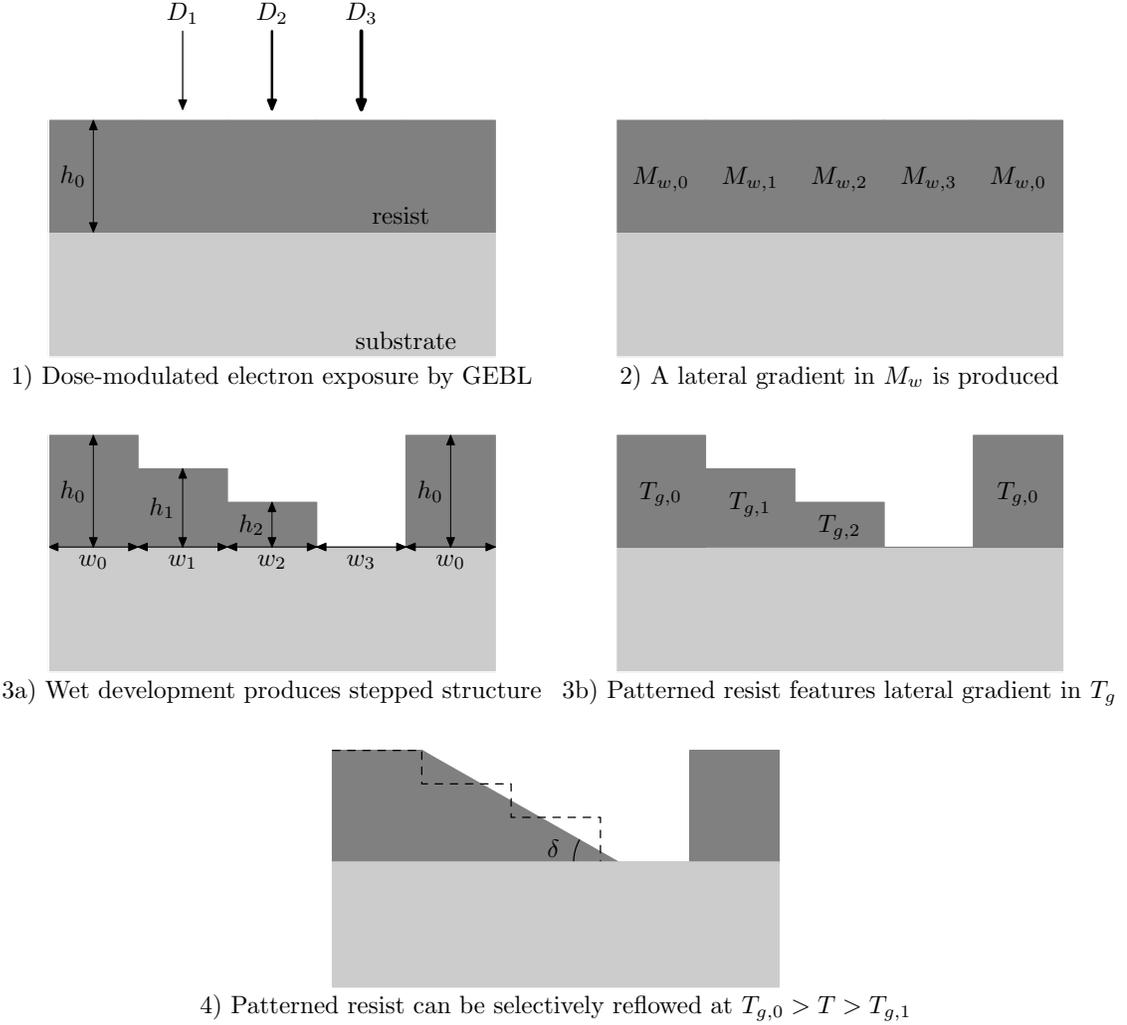

 \centering
 \includegraphics[scale=0.95]{Chapter-3/Figures/taste_fab1.mps}
 \includegraphics[scale=0.95]{Chapter-3/Figures/taste_fab2.mps}
 \includegraphics[scale=0.95]{Chapter-3/Figures/taste_fab3.mps}
 \includegraphics[scale=0.95]{Chapter-3/Figures/taste_fab4.mps}
 \includegraphics[scale=0.95]{Chapter-3/Figures/taste_fab5.mps}
 \caption[Physical properties of PMMA resist processed by grayscale EBL (GEBL) that enable thermally-activated selective topography equilibration (TASTE)]{Physical properties of PMMA resist processed by grayscale electron-beam lithography (GEBL) that enable TASTE. Dose-modulated electron exposure gives rise to a lateral gradient in $M_w$ so that varying resist thicknesses result from $M_w$-dependent etch rates that occur during wet development. The GEBL-fabricated structure also exhibits a lateral gradient in $T_g$, thereby enabling selective thermal reflow at a temperature $T_{g,0} > T > T_{g,1}$ that smooths the exposed staircase steps into a quasi-linear slope. The end result emulates a blazed grating with a groove spacing $d$ and a blaze angle $\delta$ \cite{McCoy20}.}\label{fig:TASTE_diagram}
 \end{figure}
Neglecting the effect of lateral development, which is expected to cause tilted surfaces and rounded corners in a GEBL structure \cite{Kirchner16}, this is illustrated in \cref{fig:TASTE_diagram} (steps \numrange{1}{3}), where electron doses $D_1 < D_2 < D_3$ give rise to local molecular weights $M_{w,1} > M_{w,2} > M_{w,3}$, and, if $D_3$ is large enough to clear the resist, local resist thicknesses $h_1 > h_2 > h_3 = 0$ following wet development. 
The local molecular weights $M_{w,0} > M_{w,1} > M_{w,2}$ left in the patterned resist correspond to glass transition temperatures $T_{g,0} > T_{g,1} > T_{g,2}$, with $T_{g,0}$ as $T_g$ of unexposed resist. 
In principle, this lateral contrast in $T_g$ enables selective equilibration such that with an an appropriate temperature for thermal reflow, $T_{g,0} > T_{\text{reflow}} > T_{g,1}$, electron-exposed resist becomes molten while unexposed resist remains in its glass state \cite{Schleunitz10,Schleunitz14,Kirchner14a,Kirchner19}. 

In the example illustrated in \cref{fig:TASTE_diagram}, heating the resist to a temperature $T_{\text{reflow}}$ causes the intermediate staircase steps to tend toward a surface of least energy while this molten portion of the resist is pinned between the top staircase step and the substrate surface. 
As this occurs at a rate that depends on $T_{\text{reflow}}$, the molten material creeps along the substrate as it wets the surface while its stepped topography smooths into a sloped structure to minimize its surface area \cite{Schleunitz10,Schleunitz14,Kirchner19}. 
Slight levels of curvature near the base of this slope, however, are expected as a result of the contact angle formed at the resist-substrate interface [\emph{cf.\@} \cref{eq:contact_angle}], which is unknown in practice if not determined experimentally \cite{Kirchner14,Kirchner14a}.\footnote{This is the case for the TASTE research described in this chapter.} 
Moreover, rounding in top staircase steps can occur during thermal reflow if unexposed resist is granted viscosity that allows it to soften sufficiently at a given temperature; this can result from dosing errors in GEBL that yield poor contrast in $T_g$, or if the applied temperature for reflow is too high. 
The final product expected from an ideal TASTE process, neglecting possible influence of the flowed contact angle and top-step rounding, is illustrated in \cref{fig:TASTE_diagram} (step 4), where the sawtooth-like topography serves as a grating mold with a blaze angle, $\delta$, that depends essentially on the geometry of the initial GEBL pattern. 
Previous publications from the last decade \cite{Schleunitz10,Schleunitz11a,Schleunitz11b,Schleunitz14} show that TASTE is capable of generating a \si{\um}-scale, sawtooth-like topography in PMMA in this way: by first patterning multilevel staircase structures through GEBL that repeat over a period $d$ as described above, and then heating the resist by hotplate to an appropriate-chosen temperature, $T_{\text{reflow}}$, in order to induce selective equilibration. 

This section describes a series of experiments that establish a TASTE process for sub-\si{\um} blazed gratings \cite{McCoy18} with \cref{sec:resist_contrast,sec:GEBL} devoted to GEBL and \cref{sec:reflow_results} to thermal reflow. 
For the processing described throughout this chapter, \SI{130}{\nm}-thick resist films for test samples were attained by spin-coating PMMA with $M_w = \SI{950}{\kilogram\per\mole}$ diluted \SI{3}{\percent} in Anisole (\ce{C7H8O})\footnote{950k PMMA resist prepared in this way is available from \textsc{Kayaku Advanced Materials, Inc.}, formerly known as \textsc{MicroChem Corp.}} onto clean, dehydrated silicon wafers purchased from \textsc{Virginia Semiconductor} \cite{Virgsemi_web}, all \SI{100}{\mm} in diameter and \SI{0.5}{\mm} thick. %\emph{i.e.}, methoxybenzene; 
In each case, the wafer was baked for dehydration before being spin-coated at \num{3000} rotation per \si{\minute} using a dynamic dispense and then baked again to remove the Anisole.\footnote{All dehydration and solvent bakes were carried out at $\SI{180}{\celsius}$ for \SI{3}{\minute} using a hotplate.}  
Electron exposure was performed at \SI{100}{\kilo\volt} using the \textsc{EBPG5200} system at the Penn State Nanofabrication Laboratory \cite[\emph{cf.\@} \cref{fig:ebeam_tool_pic}]{PSU_MRI_nanofab,PSU_MRI_EBL,raith_ebpg} with data preparation facilitated using \textsc{GenISys}'s \textsc{Layout BEAMER} software package \cite{beamer}.  
Samples were developed at room temperature for \SI{2}{\minute} in a 1:1 mixture of methyl isobutyl ketone (MIBK) and isopropyl alcohol (IPA) [\emph{cf.\@} \cref{sec:binary_ebeam}], followed by a \SI{30}{\second} IPA bath and a nitrogen blow-dry. 
To allow solvent-induced gel and resist swelling to subside, test samples were not characterized until $\sim \SI{24}{\hour}$ hours post development. 
All \emph{atomic force microscopy (AFM)} was carried out at the Penn State Materials Characterization Laboratory (MCL) \cite{PSU_MRI_CL} using a \textsc{Bruker Dimension Icon}$^{\text{TM}}$ under \textsc{PeakForce Tapping}$^{\text{TM}}$ mode \cite{Xu18}, with a \textsc{SCANASYST-AIR} tip as in \cref{fig:OGRESS_AFM,fig:laminar_AFM,fig:uv_nil_AFM}. 

\subsection[Resist Contrast in 130~nm-thick PMMA]{Resist Contrast in 130~nm-thick PMMA}\label{sec:resist_contrast}
%%%%%%%%%%%%%%%%%%%%%%%%%%%%%%%%%%%%%%%%%--------------------------------------------------
Practicing GEBL requires knowledge of \emph{resist contrast} to map electron dose, $D$, to remaining film thickness for a given wet development recipe. 
Data for resist contrast in \SI{130}{\nm}-thick PMMA using the MIBK/IPA development process described above were gathered through \emph{spectroscopic ellipsometry (SE)} by patterning a \num{5} by \num{5} array of \SI{250}{\um}-wide squares dosed to a range of values using the \textsc{Feature Dose Assignment} module in \textsc{Layout BEAMER} \cite{beamer}. 
Fundamentally, SE is rooted in measuring the change in polarization that electromagnetic radiation experiences as it is reflected from a thin film or a stratified medium over a range of visible and near-visible wavelengths \cite{Losurdo09,Tompkins05}. 
This change in polarization for a wavelength $\lambda$ can be parameterized through $\Psi$ and $\Delta$ defined by 
\begin{equation}
 \tan \left( \Psi \right) \mathrm{e}^{i \Delta} = \frac{\tilde{r}_p}{\tilde{r}_s} , 
 \end{equation}
where $\tilde{r}_s$ and $\tilde{r}_p$ are the complex reflection coefficients in orthogonal, \emph{s- and p-polarizations} [\emph{cf.\@} \cref{sec:planar_interface}] that can be determined for a stratified medium using the \emph{transfer-matrix method} \cite{Born80,Gibaud2009}. 
\begin{figure}
 \centering
 \includegraphics[scale=1.2]{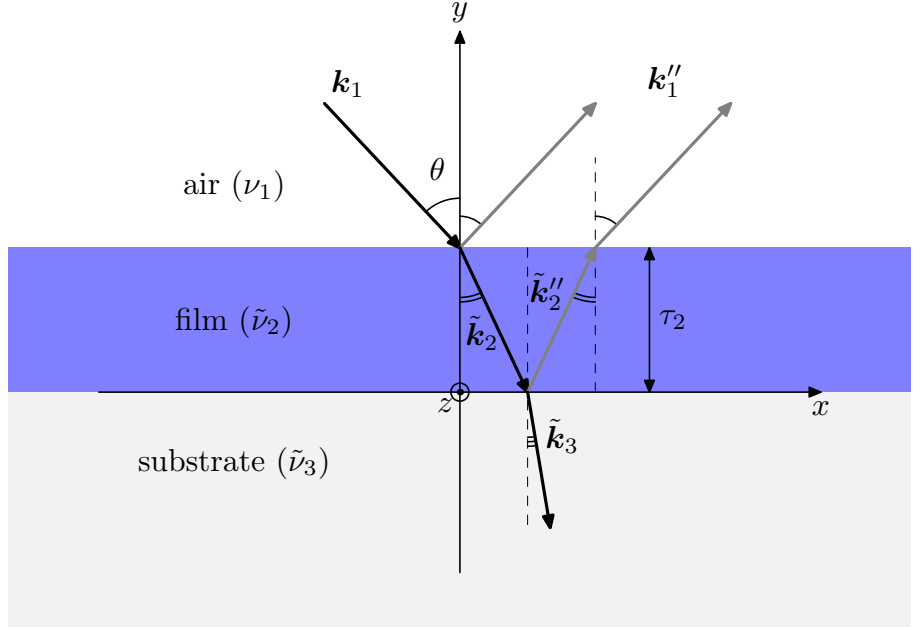}
 \caption[Geometry for spectroscopic ellipsometry (SE) of a film coated on a substrate, in air]{Geometry for spectroscopic ellipsometry (SE) of a film of thickness $\tau_2$ coated on a substrate, in air. The indexes of refraction in air, the film, and the substrate are given by $\nu_1$, $\tilde{\nu}_2$ and $\tilde{\nu}_3$, respectively, so that the corresponding, incident wave vectors in each region are $\mathbold{k}_1$, $\tilde{\mathbold{k}}_2$ and $\tilde{\mathbold{k}}_3$, where $\nu_1$ and $\mathbold{k}_1$ are assumed to be real. The component of $\mathbold{k}_1$ perpendicular to the surface is related to the angle $\theta$: $k_{y,1} = -k_0 \nu_1 \cos \left( \theta \right)$ with $k_0 \equiv 2 \pi / \lambda$ as the wave number in vacuum. By Snell's law, the corresponding components of $\tilde{\mathbold{k}}_2$ and $\tilde{\mathbold{k}}_3$ are given by $\tilde{k}_{y,2} = -k_0 \sqrt{\tilde{\nu}_2^2 - \nu_1^2 \sin^2 \left( \theta \right)}$ and $\tilde{k}_{y,3} = -k_0 \sqrt{\tilde{\nu}_3^2 - \tilde{\nu}_2^2 \sin^2 \left( \theta \right)}$, respectively. The reflected rays in the film and in air, $\tilde{\mathbold{k}}_2''$ and $\mathbold{k}_1''$, have $x$-components identical to their respective incident rays and $y$-components given by $\tilde{k}_{y,2}'' = - \tilde{k}_{y,2}$ and $k_{y,1}'' = - k_{y,1}$.}\label{fig:2DSE} 
 \end{figure}
For example, the reflection coefficient for a thin film of thickness $\tau_2$ on a substrate with an incidence angle $\theta$ [\emph{cf.\@} \cref{fig:2DSE}] is given by \emph{Airy's formula}:
\begin{equation}\label{eq:refl_coeff_example}
 \tilde{r} = \frac{\tilde{r}_{1,2} + \tilde{r}_{2,3} \mathrm{e}^{-2 i \tilde{k}_{y,2} \tau_2}}{1 + \tilde{r}_{1,2} \tilde{r}_{2,3} \mathrm{e}^{-2 i \tilde{k}_{y,2} \tau_2}} ,
 \end{equation} 
where $\tilde{k}_{y,2}$ is the component of the wave vector perpendicular to surface, in the film, with complex index of refraction $\tilde{\nu}_2$, while $\tilde{r}_{1,2}$ and $\tilde{r}_{2,3}$ are the reflection coefficients for the two interfaces, in either s- or p-polarization \cite{Tompkins05,Born80,Gibaud2009}. 
When $\Psi$ and $\Delta$ are measured at $\theta$ near the \emph{Brewster's angle} for the interface between air and the top layer of the sample,\footnote{\emph{i.e.}, $\theta_B \equiv \arctan \left( \nu_2 / \nu_1 \right)$ with $\nu_2 \equiv \Re \left[ \tilde{\nu}_2 \right]$ \cite{Born80}} 
the data are sensitive to the film thickness, $\tau_2$, so that with knowledge of the index of refraction for each layer, $\tau_2$ can be determined \cite{Tompkins05}. 

Shown as computer-aided design (CAD) in \cref{fig:dose_layout}, the layout of the pattern used for measuring resist contrast features a \num{5} by \num{5} array of \SI{250}{\um}-wide squares along with alignment markers for automated data collection (large crosses). 
\begin{figure}
 \centering
 \includegraphics[scale=0.9]{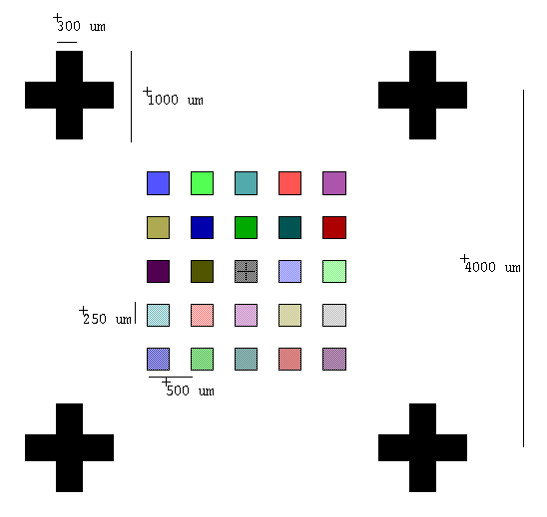}
 \caption[Computer-aided design for resist-contrast data collection]{CAD layout for resist-contrast data collection prepared using the \textsc{Tanner L-Edit} software package, where an array of \num{25} squares, each \SI{250}{\um} in width, is assigned to a range of values using \textsc{Layout BEAMER} \cite{McCoy18}.}\label{fig:dose_layout}
 \end{figure}
This pattern was exposed using a \SI{30}{\nano\ampere} beam current with a \SI{200}{\um} aperture in the \textsc{EBPG5200} to dose each square in \SI{7}{\percent} increments starting with $D = \SI{50}{\micro\coulomb\per\cm\squared}$ to impart a range of $M_w$ in the resist that correlates inversely with $D$. 
During wet development in MIBK/IPA, the resist in each square is etched at a rate that increases with decreasing $M_w$ to produce a different thickness of remaining resist \cite{miller-chou03}. 
These developed thicknesses were measured at the center of each square \SI{24}{\hour} post development using a focused \textsc{M-2000} ellipsometer (\textsc{J.A.\ Woollam}) installed at the Penn State Nanofabrication Laboratory \cite{PSU_MRI_nanofab}. 
To model the gathered data using \textsc{Woollam}'s \textsc{CompleteEASE} software package \cite{complete_ease}, SE analysis was restricted to $\SI{450}{\nm} < \lambda < \SI{1000}{\nm}$, where the resist is approximately transparent. 
\begin{figure}
 \centering
 \includegraphics[scale=0.12]{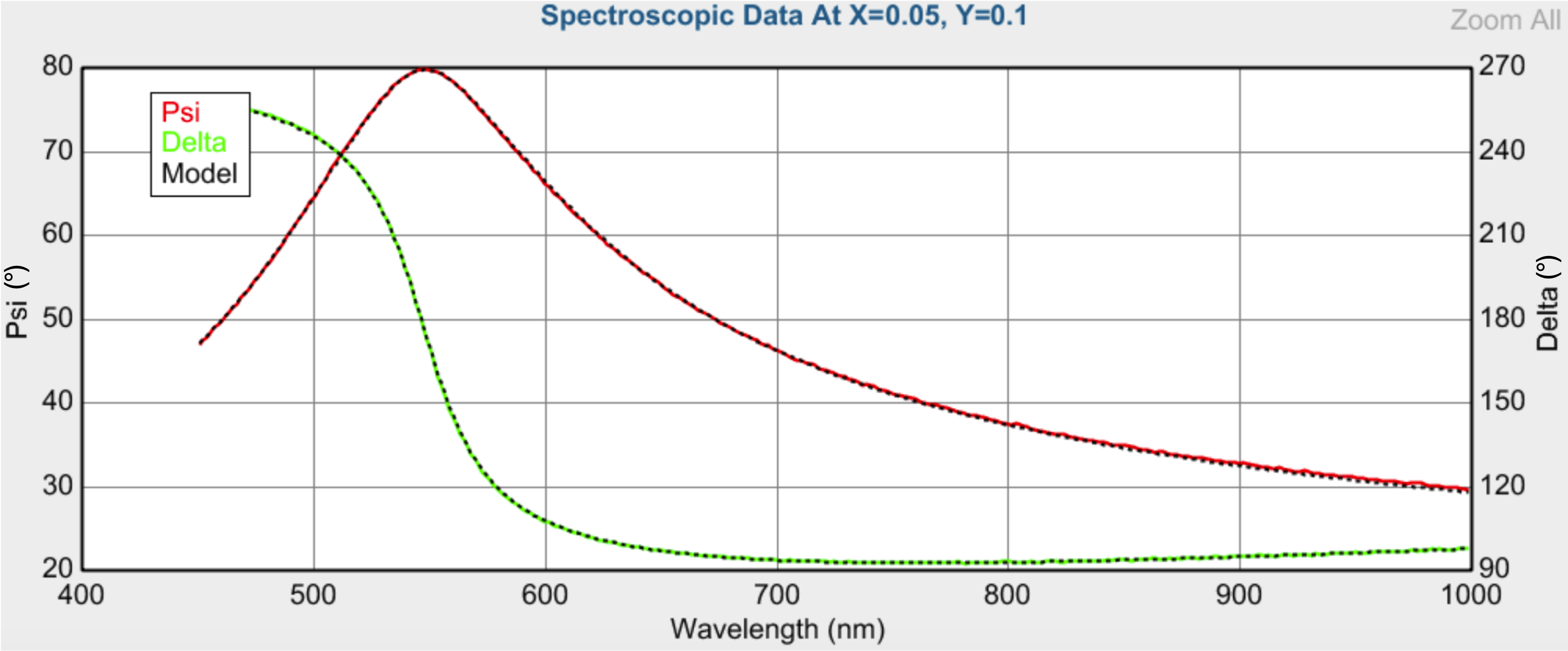}
 \includegraphics[scale=1.2]{Chapter-3/Figures/diffraction_arc28.mps}
 \caption[Example of a resist thickness measurement using SE]{Example of a resist thickness measurement using SE with \textsc{Woollam}'s \textsc{CompleteEASE} software package, where $\Psi$ and $\Delta$ are shown as red and green curves with units given in degrees (\emph{top}). The sample is treated as a bilayer on a substrate, where a film of native oxide is included between the resist and the silicon substrate, where both films are assumed to be transparent (\emph{bottom}; layers re not drawn to scale.). Using \cref{eq:Cauchy_index} for $\nu_2$ and tabulated data in \textsc{CompleteEASE} for $\nu_3$ and $\tilde{\nu}_4$, the thickness of resist in each dose square ($\tau_2$) could be measured with a fixed value of $\tau_3 = \SI{1.5}{\nm}$ for the native-oxide layer.}\label{fig:SE_example}
 \end{figure}
The sample was treated as a bilayer on a substrate, to account for the native \ce{SiO2} on the silicon wafer, with a reflectivity coefficient being a generalization of \cref{eq:refl_coeff_example} \cite{Gibaud2009}: 
\begin{equation}\label{eq:refl_coeff_bilayer}
 \tilde{r} = \frac{\tilde{r}_{1,2} + \tilde{r}_{2,3} \mathrm{e}^{-2 i \tilde{k}_{y,2} \tau_2} + \tilde{r}_{3,4} \mathrm{e}^{-2 i \left( \tilde{k}_{y,2} \tau_2 + \tilde{k}_{y,3} \tau_3 \right)} + \tilde{r}_{1,2} \tilde{r}_{2,3} \tilde{r}_{3,4} \mathrm{e}^{-2 i \tilde{k}_{y,3} \tau_3} }{1 + \tilde{r}_{1,2} \tilde{r}_{2,3} \mathrm{e}^{-2 i \tilde{k}_{y,2} \tau_2} + \tilde{r}_{2,3} \tilde{r}_{3,4} \mathrm{e}^{-2 i \tilde{k}_{y,3} \tau_3} + \tilde{r}_{1,2} \tilde{r}_{3,4} \mathrm{e}^{-2 i \left( \tilde{k}_{y,2} \tau_2 + \tilde{k}_{y,3} \tau_3 \right)}} ,
 \end{equation}
where the subscripts \numlist{1;2;3;4} correspond to air, PMMA, \ce{SiO2} and the silicon substrate as in \cref{fig:SE_example} with $\tau_2$ and $\tau_3$ representing the thicknesses of the resist and native-oxide layers, respectively. 

As described in \cref{sec:x-ray_index}, the complex index of refraction, $\tilde{\nu} (\omega) \equiv \nu (\omega) + i \xi (\omega)$, is related to the \emph{atomic scattering factor} defined by \cref{eq:atomic_scattering_factor_reduced2}, $f^0 \left( \omega \right)$, which is valid for forward-scattering in the soft x-ray, and more generally, for radiation with $\lambda$ much larger than the \emph{Bohr radius}, $a_0 \approx \SI{0.05}{\nm}$ [\emph{cf.\@} \cref{sec:SXR_med}]: 
\begin{subequations}
\begin{equation}\label{eq:index_atomic1}
 \tilde{\nu}^2 (\omega) = 1 - 4 \pi c_0 r_e \mathcal{N}_a \frac{f^0 \left( \omega \right)}{\omega^2} , 
 \end{equation}
where $c_0$ is the speed of light, $r_e$ is the classical electron radius and $\mathcal{N}_a$ is the atomic density of the material. 
A material can be considered transparent to radiation if $\omega$ is far from atomic resonances so that, neglecting effects related to absorption described by $\xi (\omega)$ and the damping terms of $f^0 \left( \omega \right)$, \cref{eq:index_atomic1} can be written in terms of the real index, $\nu (\omega)$, as 
\begin{equation}\label{eq:index_atomic2}
 \nu^2 (\omega) - 1 =  4 \pi c_0 r_e \mathcal{N}_a \sum_s \frac{g_s}{\left( \omega_s^2 - \omega^2 \right)} ,
 \end{equation}
\end{subequations}
where $g_s$ is the oscillator strength for an atomic resonance frequency $\omega_s$ [\emph{cf.\@} \cref{sec:scattering_atoms}]. 
For small $\nu$ and $\omega \ll \omega_s$, \cref{eq:index_atomic2} can be expanded as a Taylor series, rearranged and expressed in terms of $\lambda$ to arrive at \emph{Cauchy's equation} for a transparent film \cite{Born80}:
\begin{equation}\label{eq:Cauchy_index}
 \nu \left( \lambda \right) = A + \frac{B}{\lambda^2} + \frac{C}{\lambda^4} + \dotsc ,
 \end{equation}
where $A$, $B$ and $C$ are coefficients that are to be determined. 
\begin{figure}
 \centering
 \includegraphics[scale=0.82]{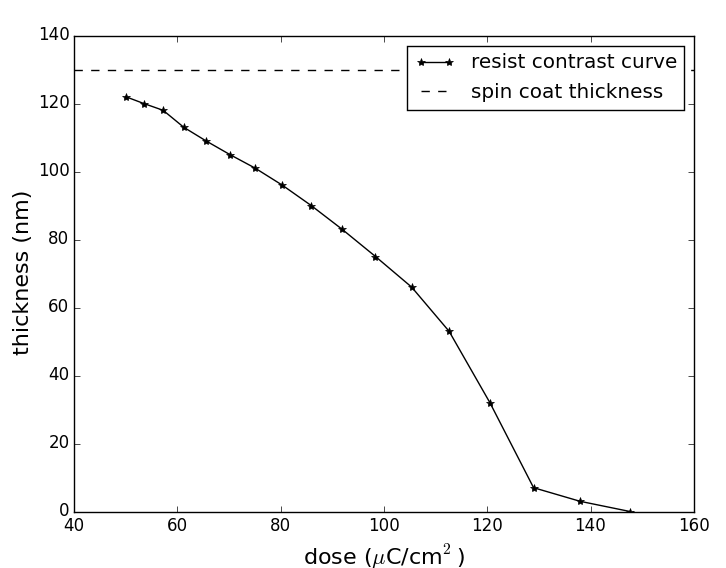}
 \caption[Resist contrast for \SI{130}{\nm}-thick PMMA with $M_w = \SI{950}{\kilogram\per\mole}$]{Resist contrast for \SI{130}{\nm}-thick PMMA with $M_w = \SI{950}{\kilogram\per\mole}$ developed at room temperature using 1:1 MIBK/IPA for \SI{2}{\minute} and IPA for \SI{30}{\second}: the center of each square of the pattern shown in \cref{fig:dose_layout} is measured using spectroscopic ellipsometry and a film thickness is extracted. Resist contrast is plotted as remaining resist thickness as a function of electron dose, $D$ \cite{McCoy18}.}\label{fig:resist_contrast}
 \end{figure}
As $\nu$ is expected to change in each square of \cref{fig:dose_layout} due to electron exposure, $A$ and $B$ were fit using \textsc{Woollam}'s \textsc{CompleteEASE} software package \cite{complete_ease} for each measurement while keeping $C$ and all higher-order coefficients equal to zero. 
The native-oxide layer was accounted for as a \SI{1.5}{\nm}-thick film using tabulated data for $\nu$ provided in \textsc{CompleteEASE}; an example of such a measurement with both films is shown in \cref{fig:SE_example}. 
By repeating SE measurements for each of the \num{25} squares of \cref{fig:dose_layout}, resist thickness was extracted and plotted as a function of $D$ as a \emph{resist-contrast curve}, shown in \cref{fig:resist_contrast}, which served as the basis for GEBL in \SI{130}{\nm}-thick PMMA. 

\subsection[Test Patterns for Grayscale Electron-Beam Lithography]{Test Patterns for GEBL}\label{sec:GEBL}
%%%%%%%%%%%%%%%%%%%%%%%%%%%%%%%%%%%%%%%%%--------------------------------------------------
The standard approach employed for GEBL process development was to draw CAD for multi-level layouts that map design layer to intended resist thickness. 
Although a contrast curve gives remaining resist thickness as a function of electron dose by definition, \emph{proximity effect correction (PEC)} must be carried out to attain accurate GEBL structures as alluded to in \cref{sec:binary_ebeam} \cite{Pavkovich86}. 
This was handled using the \textsc{3DPEC} algorithm of \textsc{Layout BEAMER}, which uses input resist-contrast data and a point spread function for electron backscattering in silicon to calculate how dose is distributed for PEC \cite{beamer,Unal10}. %, determined through \emph{Monte Carlo simulation},\footnote{(very briefly explain this)}
The profile of energy distribution from electron scattering from an infinitesimally narrow source of current is described as a normalized sum of Gaussian distributions for forward-scattering in the resist and backscattering in the substrate with characteristic widths $\beta_f$ and $\beta_b$, respectively \cite{Pavkovich86}: 
\begin{align}
 \begin{split}
 f(\varrho) &= \frac{1}{1 + \eta_{\beta}} \left( \frac{1}{\pi \beta_f^2} \right) \mathrm{e}^{- \frac{\varrho^2}{\beta_f^2}} + \frac{\eta_{\beta}}{1 + \eta_{\beta}} \left( \frac{1}{\pi \beta_b^2} \right) \mathrm{e}^{- \frac{\varrho^2}{\beta_b^2}} \\
 &\approx \frac{1}{1 + \eta_{\beta}}  \left[ \delta_D (\varrho) + \frac{\eta_{\beta}}{\pi \beta_b^2} \mathrm{e}^{- \frac{\varrho^2}{\beta_b^2} } \right] ,
 \end{split}
 \end{align}
where $\varrho$ is a radial distance in the plane of the resist surface, $\eta_{\beta}$ is the ratio of deposited energy over $\beta_b$ to that of $\beta_f$. 
The approximation holds for small $\beta_f$ such that forward-scattering is treated as a Dirac delta function and backscattering in the substrate is entirely responsible for the proximity effect with $\beta_b \approx \SI{30}{\um}$ in a silicon wafer \cite{Pavkovich86,Czaplewski12}. 
By convolving this function with the incident current density of the electron beam, the energy deposited in the resist from backscattering can be inferred with the aid of \textsc{Layout BEAMER} to produce a dose-corrected layout fractured into \num{100} layers; this is exported to \textsc{EBPG5200} machine format as \num{100} \emph{pattern-generator (PG) shapes}. 
Patterns exposed in this way used a \SI{1}{\nano\ampere} beam current, a \SI{200}{\um} aperture and a \SI{10}{\nm} beam step size. 

Due to the added overhead associated with switching between PG shapes, an alternative approach utilizing \textsc{EBPG5200} \emph{sequencing} was also pursued. 
Unique to \textsc{EBPG} systems, sequencing is a set of commands that defines a series of lines and beam jumps to be executed as a custom, subfield-sized PG shape. 
This shape can be patterned over an arbitrary area using an array of subfield-sized rectangles defined in CAD. 
\begin{figure}
 \centering
 \includegraphics[scale=0.7]{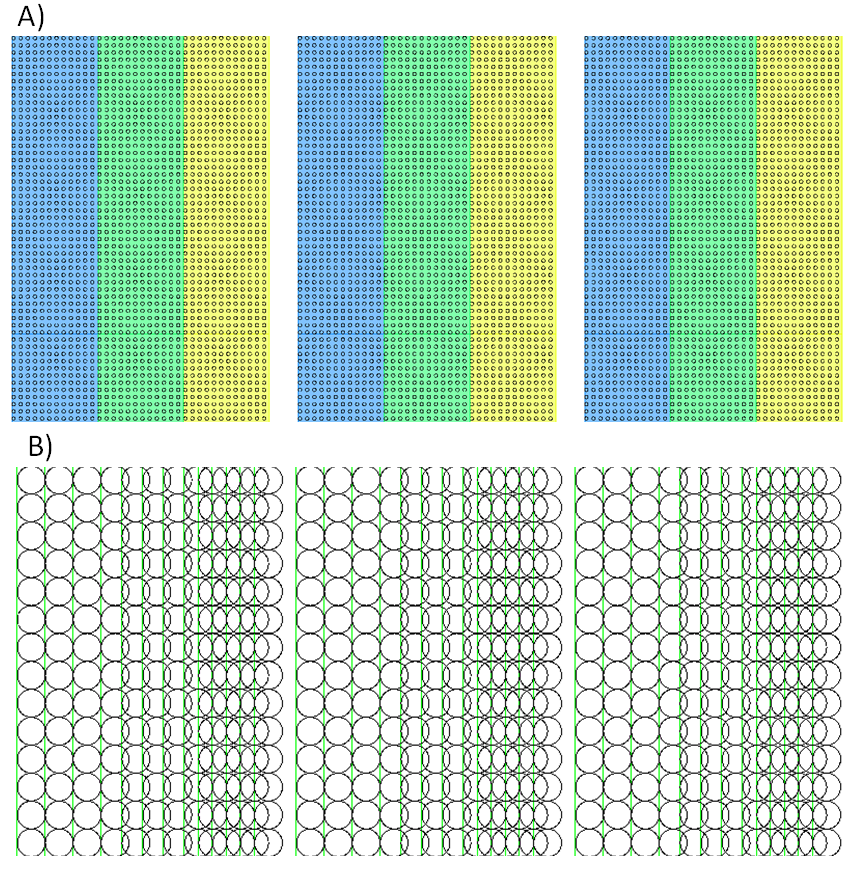}
 \caption[\textsc{Layout BEAMER 3DPEC} and \textsc{EBPG5200} sequencing approaches to GEBL]{Two approaches to pattern generation for GEBL: a) Experimentally-determined data for resist contrast [\emph{cf.\@} \cref{fig:resist_contrast}] are input into \textsc{Layout BEAMER}, where the 3DPEC algorithm is used to generate a dose-corrected layout. Each dose is a different pattern generator (PG) shape, represented here with colorscale. b) Code for \textsc{EBPG5200} sequencing defines a custom subfield-sized pattern generator shape consisting of a series of lines and beam jumps using a fixed dose, where black circles represent \SI{40}{\nm}-diameter beam shots. Lines are programmed to overlap to varying degrees to emulate dose modulation achieved using the 3DPEC approach \cite{McCoy18}.}\label{fig:GEBL_approaches} 
 \end{figure}
Although sequencing is limited to lines and jumps at a single user-input value for $D$, dose modulation can be emulated using overlapping beams [\emph{cf.\@} \cref{fig:GEBL_approaches}]. 
In this way, GEBL patterns were produced by generating sequencing code for a \SI{4}{\um}-large custom PG shape to approximate the dose-corrected layout from 3DPEC. 
These patterns used a \SI{25}{\nano\ampere} beam current, a \SI{400}{\um} aperture and a \SI{40}{\nm} beam step size with the electron beam defocused to a \SI{40}{\nm} diameter. 
The input dose was determined through analysis of dose test arrays and the highest possible beam current\footnote{This beam current was limited by the \SI{25}{\mega\hertz} \textsc{EBPG5200} clock frequency at the time of the experiment (2017) \cite{McCoy18}.} was used to write each pattern. 

Using these two GEBL approaches, multi-level staircase structures with \SI{840}{\nm} and \SI{400}{\nm} periodicities were patterned in PMMA resist. 
The width of staircase steps generally were designed to be comparable to the nominal spin-coat thickness of \SI{130}{\nm}. 
A series of GEBL test samples were written over \SI{0.5}{\mm} by \SI{2}{\mm} areas featuring the following patterns:
\begin{itemize}[noitemsep] 
 \item \emph{pattern A}: 6-level staircase where each step is \SI{140}{\nm} wide to give $d = \SI{840}{\nm}$
 \item \emph{pattern B}: 4-level staircase where each step is \SI{100}{\nm} wide to give $d = \SI{400}{\nm}$
 \item \emph{pattern C}: 4-level staircase where the width of the top step is reduced to \SI{40}{\nm}; remaining levels are \SI{120}{\nm} wide to give $d = \SI{400}{\nm}$
 \item \emph{pattern D}: Emulation of pattern C attained through \textsc{EBPG5200} sequencing\footnote{This was carried out according to the pattern shown in \cref{fig:GEBL_approaches}(B), where the overlapping beams emulate doses $D$, $1.5D$ and $2D$ with $D = \SI{75}{\micro\coulomb\per\cm\squared}$ as the user-input dose.}
 \end{itemize}%
As described at the start of \cref{sec:TASTE}, the staircase steps intermediate between the top step and the exposed substrate will, in principle, equilibrate into sloped surfaces with application of an optimized thermal treatment \cite{Schleunitz11a,Schleunitz14}. 
The width of the top step in patterns C and D was reduced in an effort to minimize flat area atop groove structures in the final product. 
Ideally, grooves are sharp, triangular sawtooth facets to provide an effective blazed grating response. 

Untreated GEBL structures in PMMA for patterns A-D are shown under AFM in \cref{fig:GEBL_ABCD}.  
\begin{figure}
 \centering
 \includegraphics[scale=1.2]{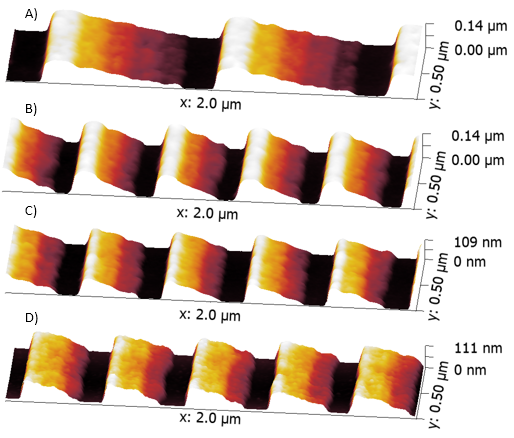}
 \caption[GEBL test patterns in \SI{130}{\nm}-thick PMMA examined under AFM]{GEBL in \SI{130}{\nm}-thick PMMA examined under AFM. Patterns A-C were attained through the standard 3DPEC approach while pattern D is a result from the \textsc{EBPG5200} sequencing approach: a) Pattern A: GEBL with 6 levels where each are \SI{140}{\nm} wide to give a \SI{840}{\nm} periodicity, b) Pattern B: GEBL with 4 levels where each are \SI{100}{\nm} wide to give a \SI{400}{\nm} periodicity, c) Pattern C: GEBL with 4 levels where the width of the top level is reduced to \SI{40}{\nm}; remaining levels are \SI{120}{\nm} wide to give a periodicity of \SI{400}{\nm}, d) Pattern D: same as pattern C but attained using EBPG sequencing \cite{McCoy18}.}\label{fig:GEBL_ABCD}
 \end{figure}
Using \textsc{Bruker}'s \textsc{PeakForce Tapping}$^{\text{TM}}$ mode \cite{Xu18}, patterns were scanned over \SI{2}{\um} at \num{512} samples per line to give a \SI{3.9}{\nm} AFM pixel size. 
In patterns A and B, staircase steps are equal in width and the overall structure height is comparable to the initial spin-coat thickness as measured by AFM. 
However, due to the narrowed width of the top step in patterns C and D, which is the same size as the defocused electron-beam diameter, the overall structure height is reduced to about \SI{75}{\percent} of the original film thickness achieved by spin coating. 
Patterns A-C were fabricated using the \textsc{Layout BEAMER 3DPEC} approach, where a \SI{1}{\mm\squared} area is exposed in $\sim \SI{20}{\minute}$ using the EBPG5200. 
In contrast, pattern D over the same area can be written using \textsc{EBPG5200} sequencing in $\sim \SI{1}{\minute}$. 
Although pattern D exhibits higher noise than pattern C under AFM, the end result should be comparable owing to the smoothing effect of thermal reflow. 
%after thermal reflow should be comparable. 
For this reason, the \textsc{EBPG5200} sequencing approach is expected to be valuable for patterning by GEBL over large areas. 

\subsection{Selective Thermal Reflow}\label{sec:reflow_results} 
%%%%%%%%%%%%%%%%%%%%%%%%%%%%%%%%%%%%%%%%%--------------------------------------------------
To induce selective thermal reflow in the GEBL test patterns [\emph{cf.\@} \cref{fig:GEBL_ABCD}] such that the top, unexposed staircase step remains largely unaffected, the resist must be heated at an appropriate temperature, $T = T_{\text{reflow}}$, for a duration, $t$, that is long enough to allow the exposed staircase steps to equilibrate into a wedge-like, smooth surface as has been shown previously for \si{\um}-scale GEBL patterns \cite{Schleunitz10,Schleunitz11a,Schleunitz14,Kirchner19}. 
All thermal reflow experimentation to probe this parameter space for the smaller-scale patterns considered was carried out using an automated hotplate on a resist stabilization system built by \textsc{Fusion Semiconductor} at the Penn State Nanofabrication Laboratory \cite{PSU_MRI_nanofab}. 
The tool, which accepts wafers with \SI{100}{\mm} and \SI{150}{\mm} diameters, was employed for heating test samples in a controllable and reproducible way. 
A series of identical GEBL samples were fabricated and thermally treated under different conditions to probe the $T$-$t$ parameter space for TASTE. 
Using a short heating duration ($t = \SI{20}{\second}$), samples were treated using a range of $T$ around the expected $T_{\text{reflow}}$ for PMMA (\SIrange{110}{130}{\celsius}). 
Conversely, samples were heated for different values of $t$, (\SIlist{20;60;120}{\second}) while holding $T$ constant at \SI{120}{\celsius}. 
Each sample was allowed to cool on the cassette rack of the automated hotplate tool after heating.

Thermal-reflow test results for pattern A as a function of $T$ and $t$ are shown in \cref{fig:A_temp_scan,fig:A_time_scan}, respectively. 
\begin{figure}
 \centering
 \includegraphics[scale=0.1505]{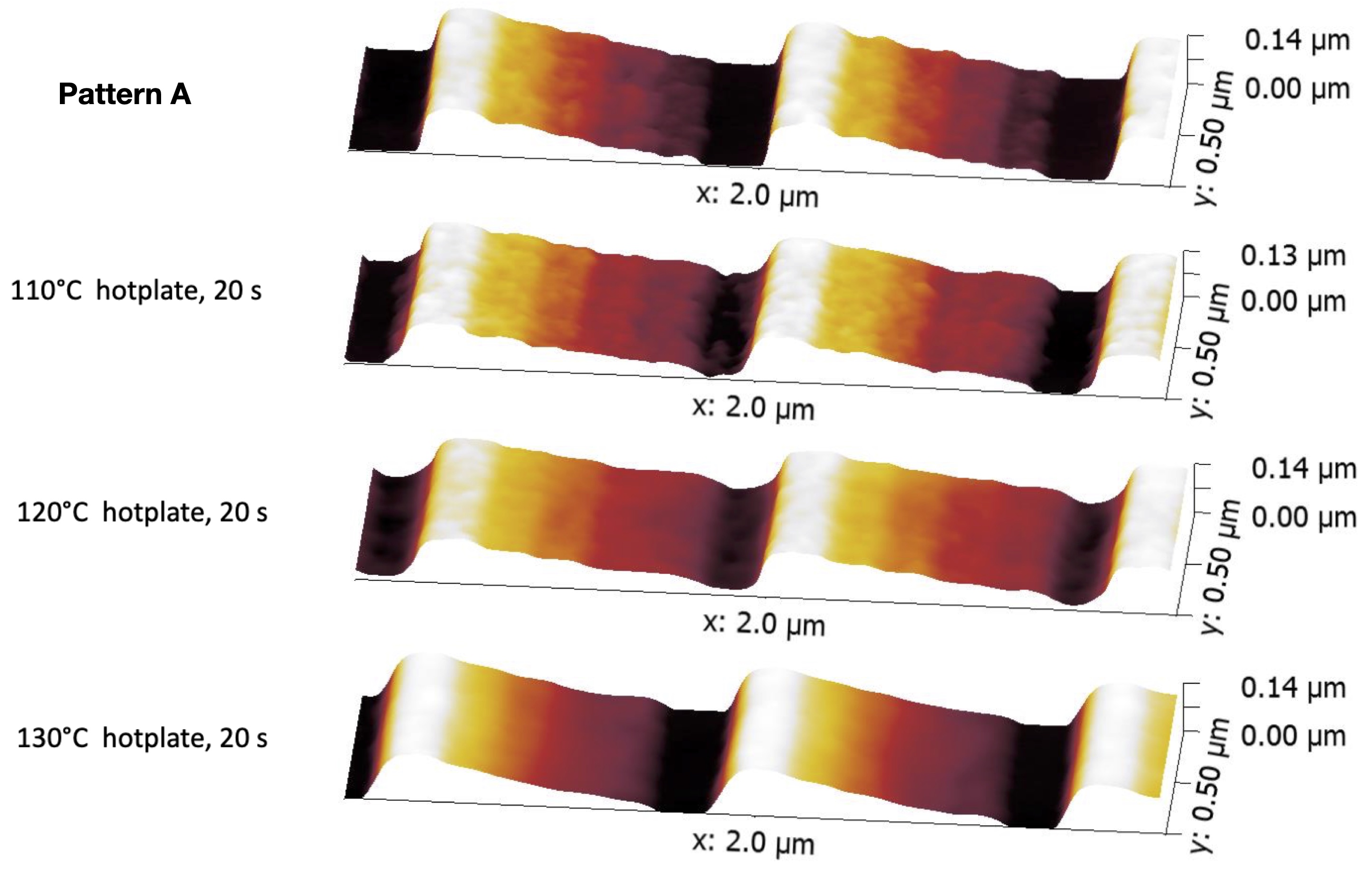}
 \caption[Thermal reflow results as a function of heating temperature for grayscale pattern A]{Thermal reflow results as a function of heating temperature, $T$, for GEBL in \SI{130}{\nm}-thick PMMA with \num{6} levels: each are \SI{140}{\nm} wide to give $d = \SI{840}{\nm}$ \cite{McCoy18}.}\label{fig:A_temp_scan}
 \end{figure}
\begin{figure}
 \centering
 \includegraphics[scale=0.1505]{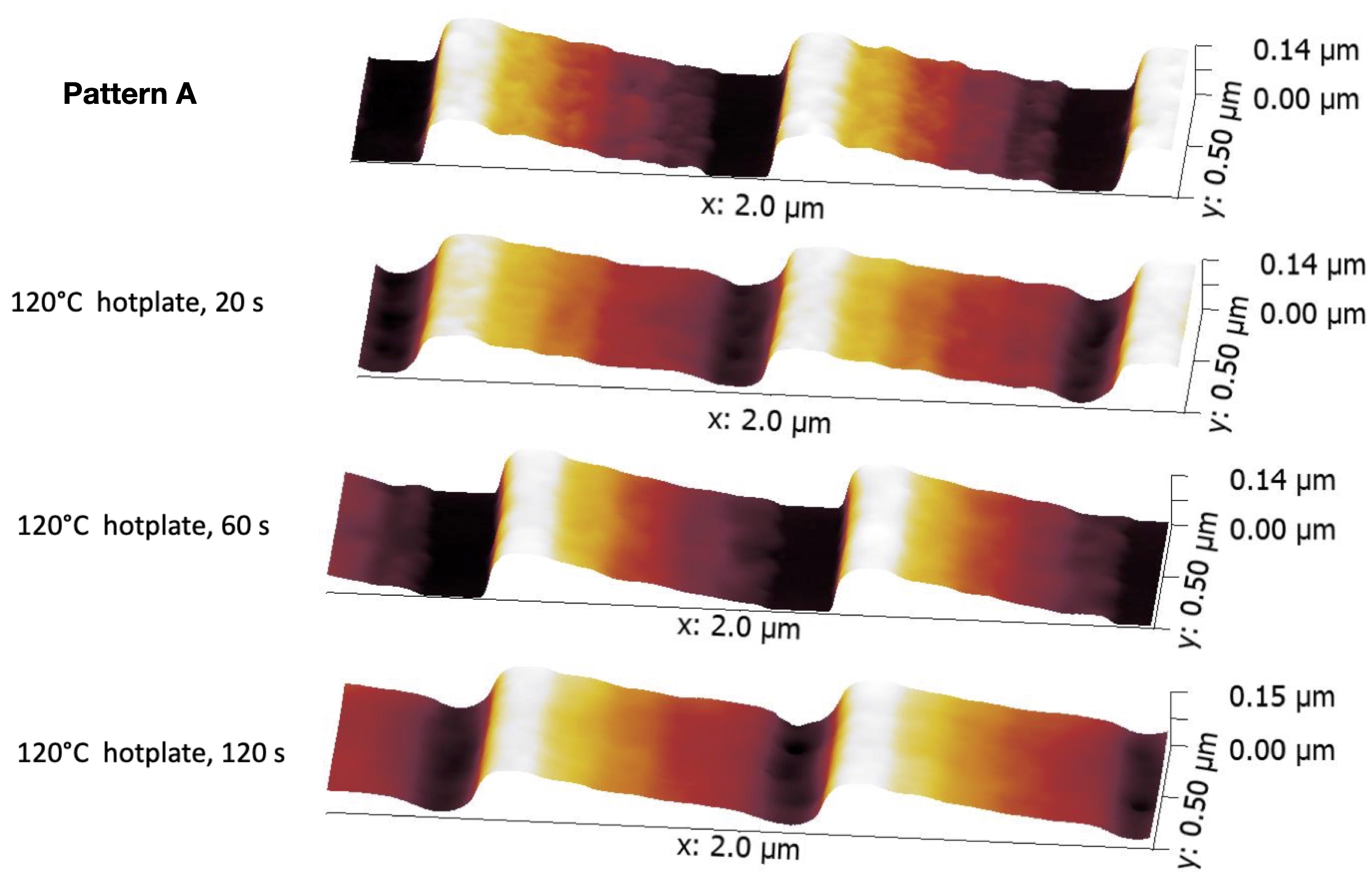}
 \caption[Thermal reflow results as a function of heating duration for grayscale pattern A]{Thermal reflow results as a function of heating duration, $t$, for GEBL in \SI{130}{\nm}-thick PMMA with \num{6} levels as in \cref{fig:A_temp_scan} \cite{McCoy18}.}\label{fig:A_time_scan}
 \end{figure}
Together, these include six AFM scans: three samples treated at \SIlist{110;120;130}{\celsius} keeping heating time constant at \SI{20}{\second}, and, three samples heated with durations of \SIlist{20;60;120}{\second} holding temperature at \SI{120}{\celsius}. 
Analogous AFM scans of thermal reflow results are shown in \cref{fig:B_temp_scan,fig:B_time_scan} for pattern B and in \cref{fig:C_temp_scan,fig:C_time_scan} for pattern C \cite{McCoy18}.
\begin{figure}
 \centering
 \includegraphics[scale=0.1505]{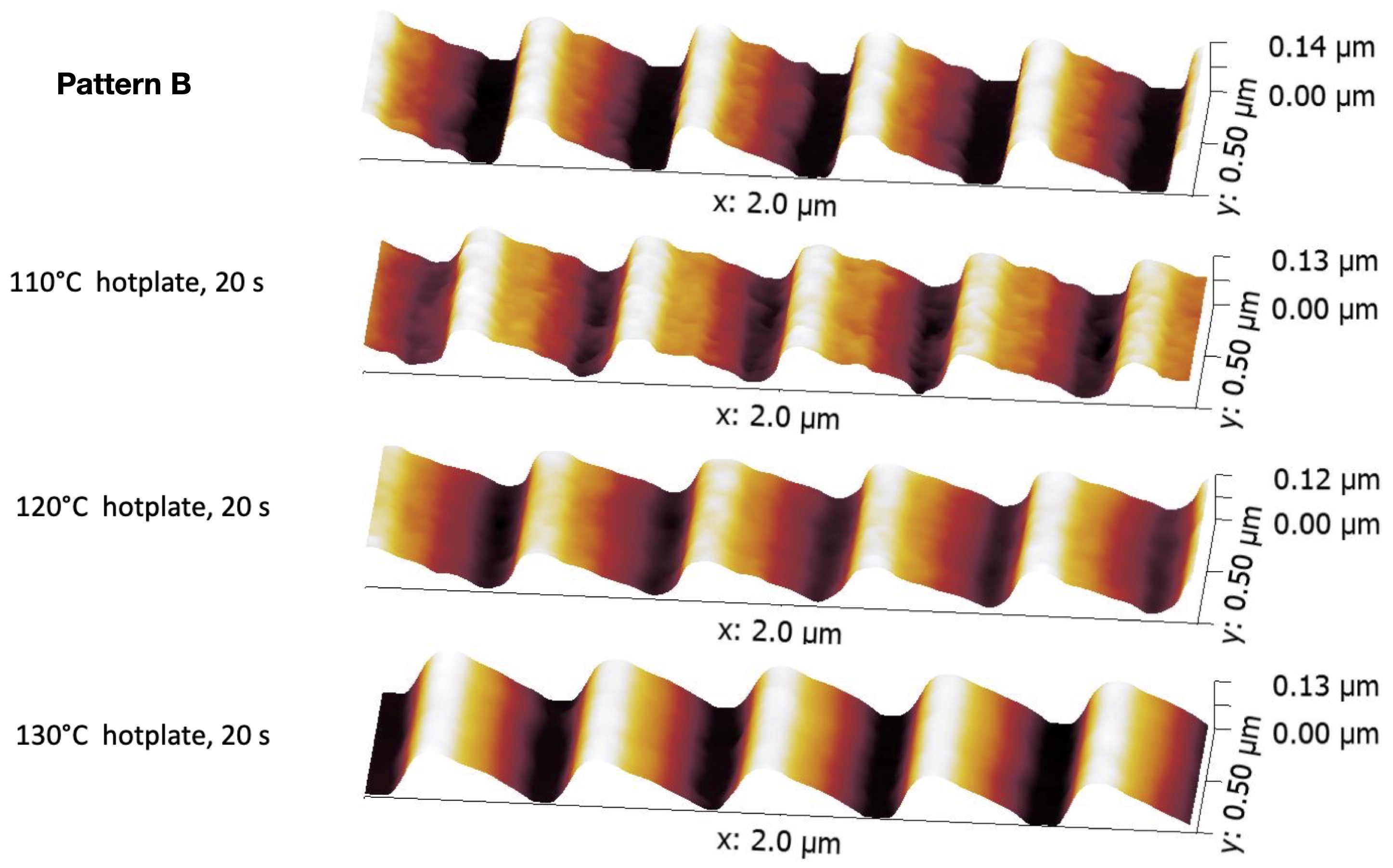}
 \caption[Thermal reflow results as a function of heating temperature for grayscale pattern B]{Thermal reflow results as a function of $T$ for GEBL in \SI{130}{\nm}-thick PMMA with \num{4} levels: each are \SI{100}{\nm} wide to give $d = \SI{400}{\nm}$ \cite{McCoy18}.}\label{fig:B_temp_scan}
 \end{figure}
\begin{figure}
 \centering
 \includegraphics[scale=0.1505]{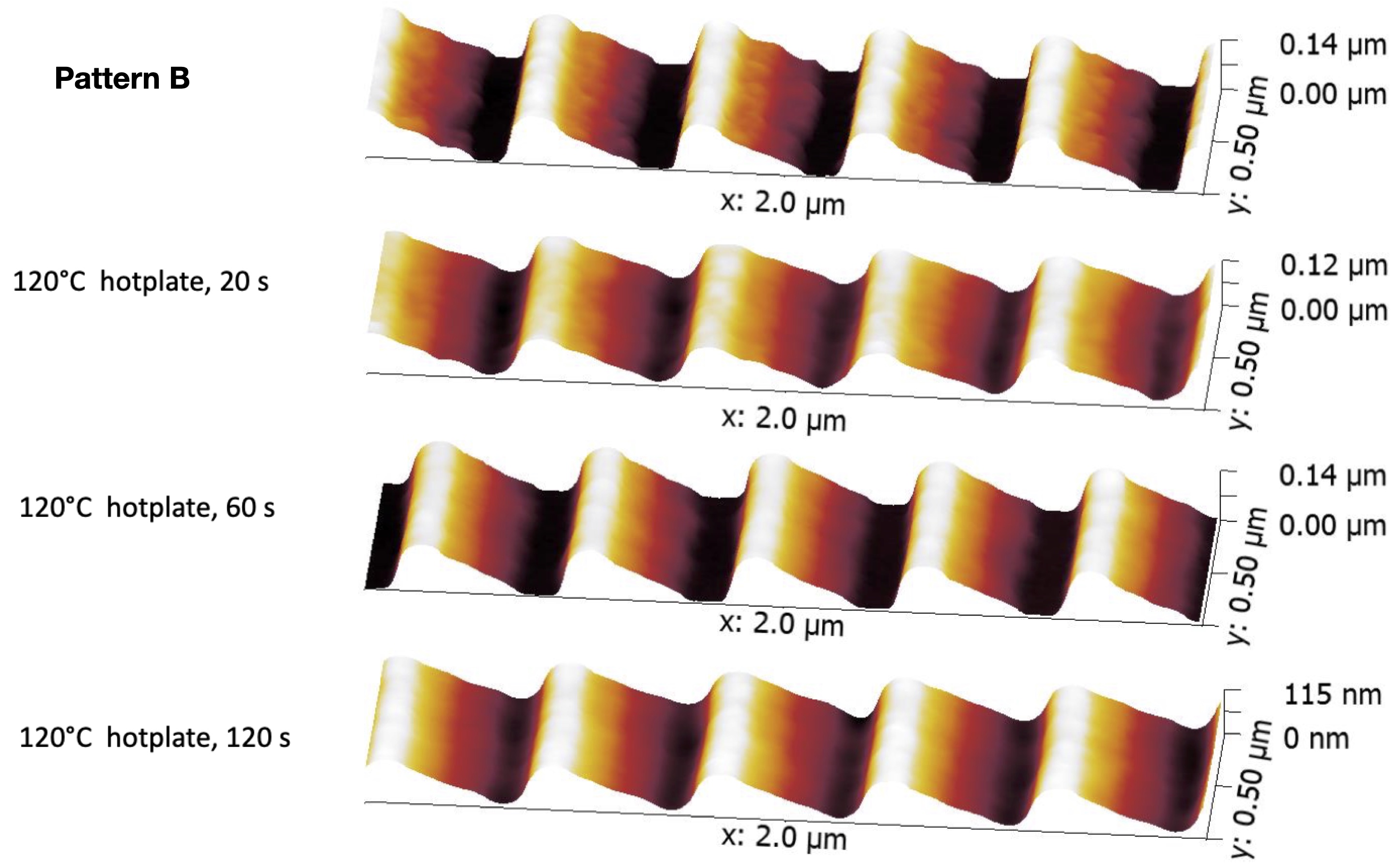}
 \caption[Thermal reflow results as a function of heating duration for grayscale pattern B]{Thermal reflow results as a function of $t$ for GEBL in \SI{130}{\nm}-thick PMMA with \num{4} levels as in \cref{fig:B_temp_scan} \cite{McCoy18}.}\label{fig:B_time_scan}
 \end{figure}
\begin{figure}
 \centering
 \includegraphics[scale=0.1505]{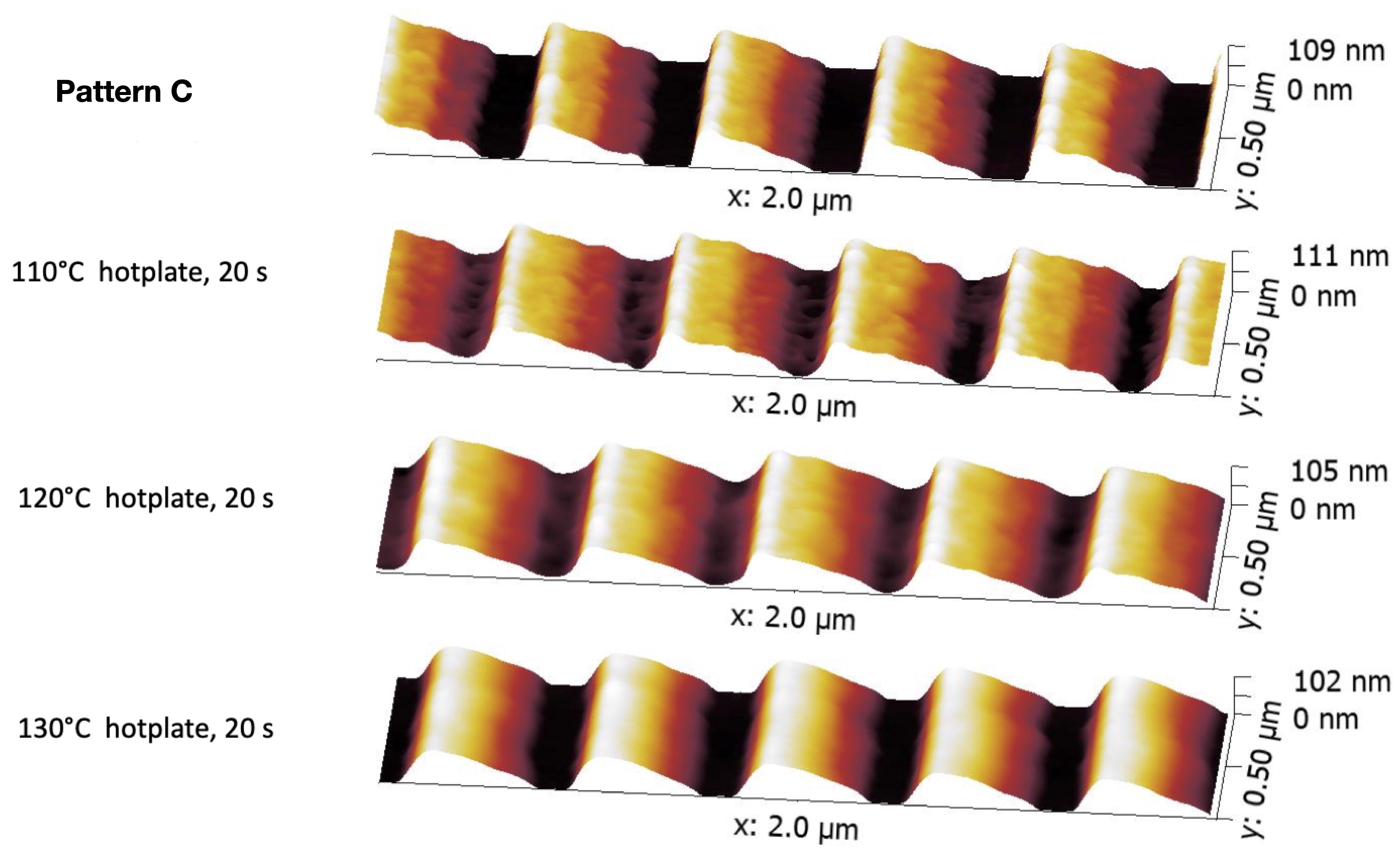}
 \caption[Thermal reflow results as a function of heating temperature for grayscale pattern C]{Thermal reflow results as a function of $T$ for GEBL in \SI{130}{\nm}-thick PMMA with \num{4} levels: the width of the top level is reduced to \SI{40}{\nm} while remaining levels are \SI{120}{\nm} wide to give $d = \SI{400}{\nm}$ \cite{McCoy18}.}\label{fig:C_temp_scan}
 \end{figure}
\begin{figure}
 \centering
 \includegraphics[scale=0.1505]{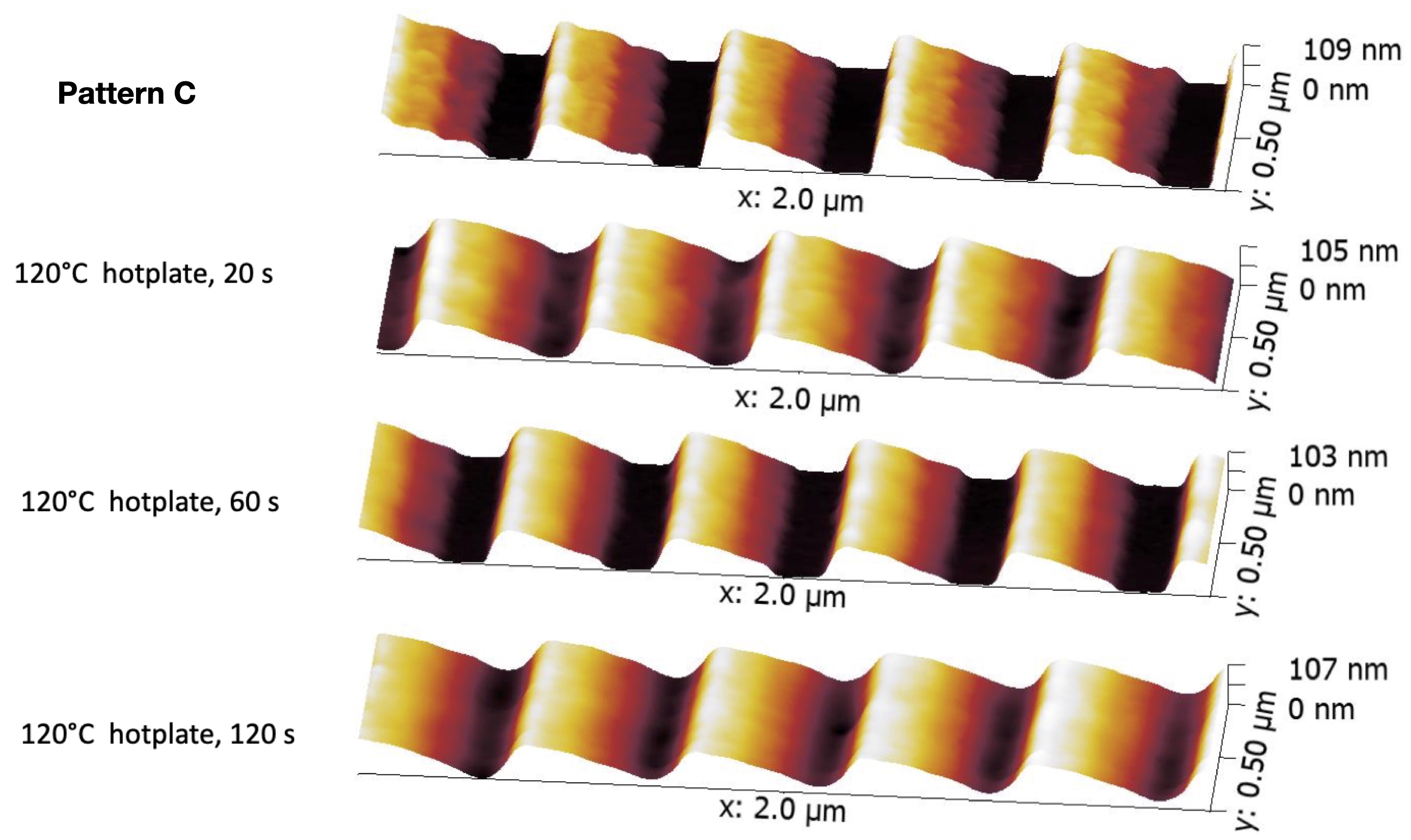}
 \caption[Thermal reflow results as a function of heating duration for grayscale pattern C]{Thermal reflow results as a function of $t$ for GEBL in \SI{130}{\nm}-thick PMMA with \num{4} levels as in \cref{fig:C_temp_scan} \cite{McCoy18}.}\label{fig:C_time_scan}
 \end{figure}
These initial results demonstrate the principle of TASTE: stepped surfaces of resist begin to smooth into inclines with an appropriate thermal treatment. 
However, only a small sample of the $T$-$t$ parameter space for TASTE has been explored. 
The reflow temperature should be optimized such that intermediate steps are able to flow while the top step is unaffected; this is to achieve a sharp sawtooth by avoiding groove rounding effects. 
Meanwhile, heating duration should also be optimized to allow flowing resist to equilibrate into smooth, sloped surfaces. 
Preferred temperature-time parameters will depend on the initial GEBL structure. 

Shown under AFM in \cref{fig:A_temp_scan}, a small reflow effect at $T = \SI{110}{\celsius}$ is seen in the intermediate steps of pattern A. 
The effect becomes more pronounced at $T = \SI{120}{\celsius}$ while at $T = \SI{130}{\celsius}$, the entire structure, including the top step, is noticeably affected. 
Holding $T = \SI{120}{\celsius}$ constant as in \cref{fig:A_time_scan}, the intermediate steps become increasingly smooth with increasing $t$. 
At $t = \SI{120}{\second}$, however, it appears that equilibration has not yet been reached as evidenced by the slight waviness on the facets left over from the GEBL staircase steps. 
This suggests that the optimum temperature for TASTE of pattern A is close to $T_{\text{reflow}} = \SI{120}{\celsius}$ while the reflow duration required for equilibration is $t > \SI{120}{\second}$. 
Shown in \cref{fig:B_temp_scan,fig:B_time_scan,fig:C_temp_scan,fig:C_time_scan}, a similar effect is observed for patterns B and C. 
However, the optimum reflow temperature for patterns B and C may be slightly lower as the narrowed top step in each case is expected to have a reduced $M_w$ relative to that of pattern A. 
This is evidenced by the $\sim \SI{25}{\percent}$ reduction in structure height relative to the original film thickness, as mentioned previously, and the slight rounding observed in the top step at $T = \SI{120}{\celsius}$ for $t = \SI{120}{\second}$, which both reflect a slightly lowered $T_{\text{reflow}}$ as compared to pattern A. 
In these cases, selectivity for equilibration is degraded such that the top step experiences a small degree of viscous flow along with other steps, and as a result, the overall structure exhibits a convex-like quality at the groove apex along with a quasi-linear slope that emulates a blazed facet. 

\section[Grating Prototype Fabrication]{Grating Prototype Fabrication}\label{sec:TASTE_prototype} 
%%%%%%%%%%%%%%%%%%%%%%%%%%%%%%%%%%%%%%%%%-------------------------------------------------
Although not exhaustive, the TASTE results presented in \cref{sec:reflow_results} show that a process window exists to transform a repeating staircase structure fabricated by GEBL into a sawtooth-like surface relief using thermal reflow. 
The experimentation described in this section leverages from the process development outlined in \cref{sec:TASTE} to fabricate a blazed grating prototype that is suitable for soft x-ray diffraction-efficiency testing in an extreme off-plane mount. 
Based on the layout for \emph{pattern B} introduced in \cref{sec:GEBL}, this grating prototype was designed to have $d = \SI{400}{\nm}$, and from the AFM-measured angle of the sawtooth facets in $\SI{130}{\nm}$-thick PMMA [\emph{cf.\@} \cref{fig:B_temp_scan,fig:B_time_scan}], the effective blaze angle is expected to be $\delta \approx 27^{\circ}$. %a groove spacing of
In a near-Littrow configuration for an extreme off-plane geometry with order locations given by the generalized grating equation \cite[\emph{cf.\@} \cref{eq:off-plane_orders,eq:off-plane_orders_ch2,eq:off-plane_incidence_orders}]{Loewen97}, the azimuthal incidence angle [\emph{cf.\@} \cref{fig:conical_reflection_edit,fig:grating_angles}] is $\alpha \approx \delta$ while the groove-facet incidence angle is $\zeta \lessapprox \gamma$ by \cref{eq:angle_on_groove}, where $\gamma$ is the cone opening half-angle of the diffraction pattern. 
This grazing-incidence angle $\gamma$ must be smaller than the critical angle for \emph{total external reflection (TER)} of the grating material, $\zeta_c (\omega)$ [\emph{cf.\@} \cref{sec:als_testing}]. 
To satisfy these conditions, the grating prototype was designed for use at a nominal graze angle of $\eta = 1.5^{\circ}$ in a Littrow configuration, where $\alpha = \beta = \delta \approx 27^{\circ}$ and $\zeta = \gamma \approx 1.7^{\circ}$ by \cref{eq:angle_on_groove,eq:graze_angle} with \cref{eq:blaze_wavelength} for the blaze wavelength becoming  
\begin{equation}\label{eq:blaze_wavelength_TASTE}
 \lambda_b = \frac{2 d \sin \left( \gamma \right) \sin \left( \delta \right) }{n} \approx \frac{\SI{11}{\nm}}{n} ,
 \end{equation}
which provides an estimate for the spectral locations of peak orders [\emph{cf.\@} \cref{fig:pcgrate_example_au}]. 

As described in \cref{sec:als_testing}, a grating geometry defined by $\alpha$ and $\gamma$ is parameterized by the principal-axis angles $\eta$, $\varphi$ and $\phi$ at beamline 6.3.2 of the ALS [\emph{cf.\@} \cref{fig:grating_angles,fig:ALS_arc}]. 
With $\eta = 1.5^{\circ}$, \cref{eq:yaw_angle} yields $\varphi \approx 0.8^{\circ}$ so that for $\phi \approx 0^{\circ}$, the grating-dispersion direction is virtually parallel to the direction of the horizontal linear stage motion [\emph{cf.\@} \cref{sec:geo_constrain}]. 
Because the incident beam is then nearly parallel with both the groove direction and the surface of the grating substrate, the grooves of the grating prototype must be long enough to encompass the incident beam in projection at the chosen value of $\eta \approx 1.5^{\circ}$. 
With knowledge that the cross-sectional size of the beam at the ALS is $\lessapprox \SI{0.5}{\mm}$ as it is incident on an optic, the grating prototype was designed to be \SI{50}{\mm} along the groove direction and \SI{7.5}{\mm} along the grating-dispersion direction so as to allow the beam to be positioned on the grooved area with relative ease. 
Considering the wavelengths of radiation at which there exist propagating orders in this geometry and the separation of these orders defined by \cref{eq:linear_dispersion} with $d = \SI{400}{\nm}$ and $L \approx \SI{235}{\mm}$ relative to the \SI{0.5}{\mm} slit width, diffraction-efficiency testing was carried out over $\SI{15.5}{\nm} \gtrapprox \lambda \gtrapprox \SI{1.55}{\nm}$, or equivalently, extreme ultraviolet (EUV) and soft x-ray photon energies, $\mathcal{E}_{\gamma}$, ranging from \SIrange{80}{800}{\electronvolt}. 

\subsection{TASTE Processing}\label{sec:prototype_TASTE_process}
%%%%%%%%%%%%%%%%%%%%%%%%%%%%%%%%%%%%%%%%%--------------------------------------------------
GEBL processing for fabrication of the grating-prototype surface relief is outlined in \cref{fig:TASTE_diagram}: the staircase topography features two electron-exposed steps, a cleared area and an unexposed step, all of equal width consistent with a periodicity of $d = \SI{400}{\nm}$.\footnote{\emph{i.e.}, $w_0 = w_1 = w_2 = w_3 = \SI{100}{\nm}$} 
Electron dosing for GEBL was performed according to the resist-contrast curve provided in \cref{sec:resist_contrast}, which is based on a room-temperature development recipe consisting of \SI{2}{\minute} in 1:1 MIBK/IPA followed by a \SI{30}{\second} rinse in IPA and a high-purity nitrogen blow-dry. 
This contrast curve is shown in \cref{fig:resist_contrast}, where post-development PMMA thickness as measured by SE is plotted as a function of assigned electron dose, $D$. 
These data were processed using the \textsc{Layout BEAMER 3DPEC} algorithm [\emph{cf.\@} \cref{sec:GEBL}] to generate a dose-corrected layout appropriate for achieving exposed staircase steps with $h_1 \approx 0.66 h_0$ and $h_2 \approx 0.33 h_0$, where $h_0 \approx \SI{130}{\nm}$ is the spin-coat thickness. 
Electron exposure for GEBL was carried out using an \SI{8}{\nano\ampere} beam current and a \SI{400}{\um} aperture with a beam step size and a writing-grid resolution of \SI{10}{\nm}, which is comparable to the beam spot size realized by the \textsc{EBPG5200} under these conditions. 
Due to a \SI{100}{\mega\hertz} clock frequency upgrade to the \textsc{EBPG5200} at Penn State in 2018, these beam conditions differ from the recipe described in \cref{sec:GEBL}, which was limited by the \SI{25}{\mega\hertz} frequency at the time of the experiment \cite{McCoy18,McCoy20}. %

Using the GEBL process outlined above, test patterns were exposed, wet-developed and characterized by AFM to verify that the previously-reported staircase topography could be readily reproduced using the increased value for beam current enabled by a \SI{100}{\mega\hertz} clock frequency. 
\begin{figure}
 \centering
 \includegraphics[scale=0.7]{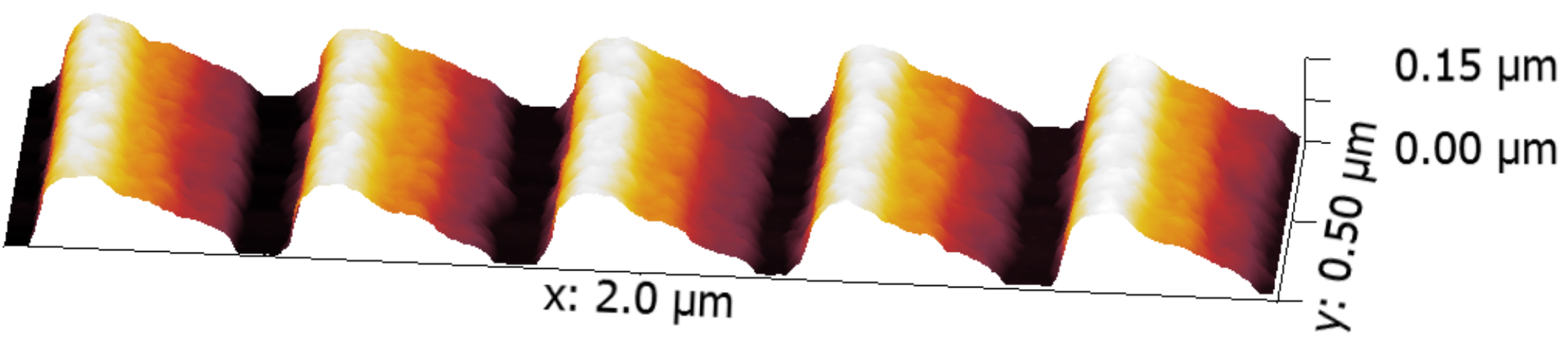}
 \includegraphics[scale=0.7]{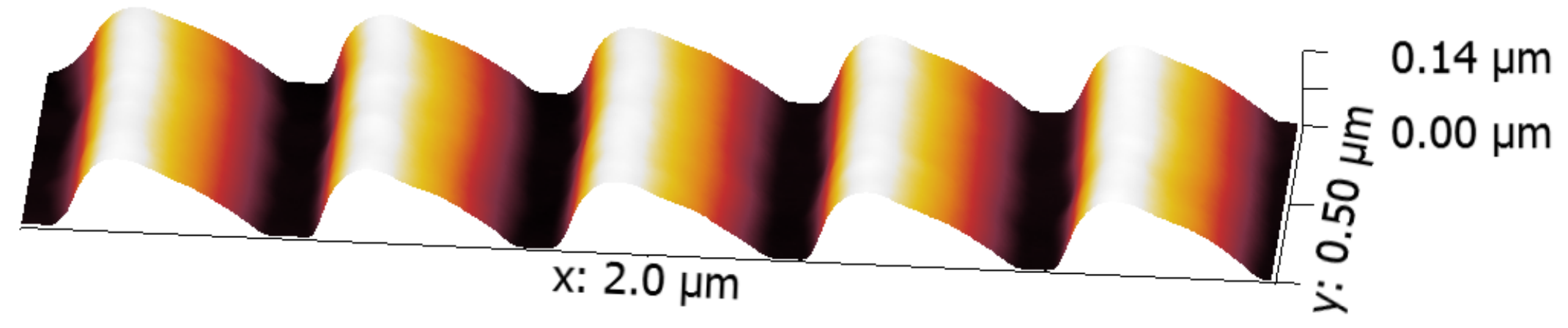}
 \caption[AFMs of the GEBL-processed resist and the resist following thermal reflow.]{AFMs of the GEBL-processed resist (\emph{top}) and the resist following thermal reflow (\emph{bottom}) \cite{McCoy20}.}\label{fig:TASTE_AFMs}
 \end{figure}
As in \cref{sec:TASTE}, AFM was carried out over \SI{2}{\um} in the grating-dispersion direction at \num{512} samples per line to yield a \SI{3.9}{\nm} pixel size.  
A scan of the GEBL pattern exposed using an \SI{8}{\nano\ampere} beam current and a \SI{400}{\um} aperture is shown in the top panel of \cref{fig:TASTE_AFMs}, where it is verified that the topography appears virtually indistinguishable from the previous result from \cref{sec:TASTE} obtained using a \SI{1}{\nano\ampere} beam current and a \SI{200}{\um} aperture. 
Next, thermal-reflow experimentation on test samples was carried out using an automated hotplate tool built by \textsc{Fusion Semiconductor} [\emph{cf.\@} \cref{sec:reflow_results}], where from the results reported in \cref{sec:TASTE}, it is expected that the optimum value for $T_{\text{reflow}}$ is near \SI{120}{\celsius}. 
Through a series of reflow tests, it was found that $T_{\text{reflow}} = \SI{116}{\celsius}$ applied for a duration of \SI{30}{\minute}\footnote{Due to the \SI{999}{\second} time-out of the \textsc{Fusion Semiconductor} automated hotplate tool, thermal reflow was carried out in two consecutive \SI{15}{\minute} intervals.} produced a topography that most closely resembled a sawtooth. 
An AFM of a test pattern hotplate-treated in this way is shown in the bottom panel of \cref{fig:TASTE_AFMs}, where a slight convex quality is observed, which is likely a result of degraded reflow selectivity [\emph{cf.\@} \cref{sec:TASTE_ch5}]. % as discussed further in \cref{sec:TASTE_ch5}.  
Nonetheless, the overall structure resembles a sawtooth and based on these results, a \SI{7.5}{\mm} by \SI{50}{\mm} area was exposed for GEBL using a \SI{300}{\um} by \SI{300}{\um} mainfield with \SI{10}{\nm} resolution, a \SI{4}{\um} by \SI{4}{\um} subfield with \SI{5}{\nm} resolution and \textsc{Large Rectangle Fine Trapezoid} fracturing in \textsc{Layout Beamer} \cite{beamer}. 
\begin{figure}
 \centering
 \includegraphics[scale=0.15]{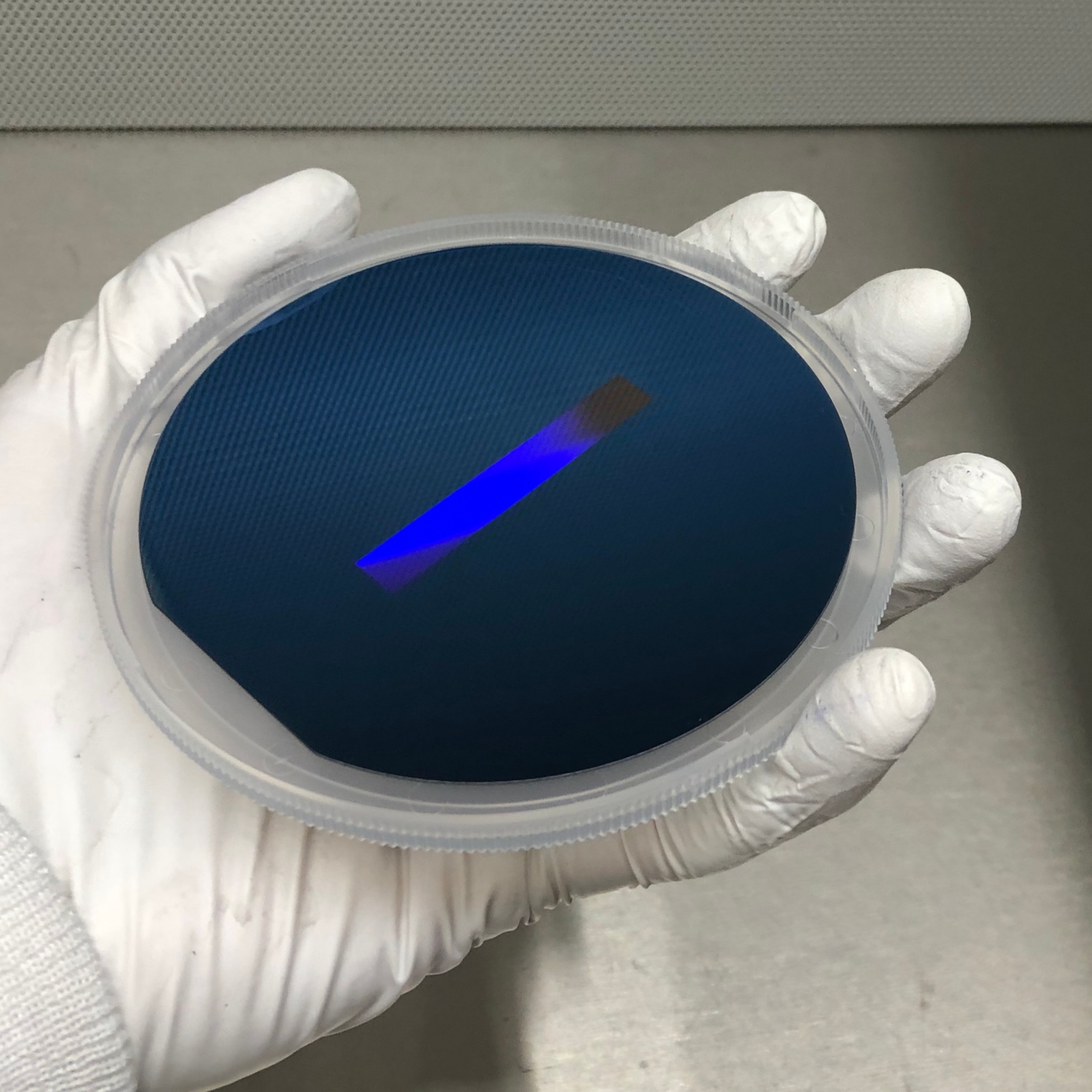}
 \caption[Photograph of the (uncoated) grating prototype patterned in \SI{130}{\nm}-thick PMMA on a \SI{100}{\mm}-diameter silicon wafer using TASTE.]{Surface-relief mold for the grating prototype patterned in \SI{130}{\nm}-thick PMMA on a \SI{100}{\mm}-diameter silicon wafer using TASTE. The grating measures \SI{50}{\mm} in the groove direction and \SI{7.5}{\mm} in the grating-dispersion direction \cite{McCoy20}.}\label{fig:uncoated_grating_pic}
 \end{figure}
Under these conditions, the \textsc{EBPG5200} exposure duration (including tool overhead) was $\lessapprox \SI{20}{\hour}$. 
The grating mold resulting from the entire TASTE process patterned on a \SI{100}{\mm}-diameter wafer is pictured in \cref{fig:uncoated_grating_pic}. 

\subsection{Coating for Reflectivity}\label{sec:prototype_coating}
%%%%%%%%%%%%%%%%%%%%%%%%%%%%%%%%%%%%%%%%%--------------------------------------------------
Although achieving an effective blaze response from the grating prototype hinges on the shape of the sawtooth facets produced by TASTE, \emph{absolute diffraction efficiency}, $\mathscr{E}_n$, is also dependent on the reflectivity of the sawtooth facets at a nominal incidence angle $\zeta \approx 1.7^{\circ}$ [\emph{cf.\@} \cref{ch:diff_eff}]. 
Having an overcoating on the grating surface relief described in \cref{sec:prototype_TASTE_process} is important primarily for avoiding prominent absorption edges of carbon and oxygen in PMMA and for achieving high overall reflectivity at this value of $\zeta$ [\emph{cf.\@} \cref{fig:X-ray_index,fig:X-ray_refl}]. 
As described in \cref{sec:reflectivity_polarization}, Fresnel reflectivity, $\mathcal{R}_F$, is virtually insensitive to polarization and hence the quantity can be expressed in s-polarization with a high degree of accuracy.  
For a thick slab of material realized by a film thickness several times larger than the $1 / \mathrm{e}$ \emph{penetration depth} for radiation under TER [\emph{cf.\@} \cref{sec:total_refl,eq:penetration_depth}]: 
\begin{equation}\label{eq:ch3_pene_dep}
 \mathcal{D}_{\perp} = \frac{\lambda}{4 \pi \, \text{Im} \! \left[ \sqrt{\tilde{\nu}^2 (\omega) - \cos^2 \left( \zeta \right)} \right]} ,
 \end{equation}
$\mathcal{R}_F$ in s-polarization is given by \cref{eq:s_reflectivity}: %[\emph{cf.\@} \cref{eq:s_reflectivity}]
\begin{equation}\label{eq:ch3_fres_refl}
 \mathcal{R}_F = \norm{ \frac{\sin \left( \zeta \right) - \sqrt{\tilde{\nu}^2 (\omega) - \cos^2 \left( \zeta \right)}}{\sin \left( \zeta \right) + \sqrt{\tilde{\nu}^2 (\omega) - \cos^2 \left( \zeta \right)}} }^2 , 
 \end{equation}
where $\tilde{\nu} (\omega) = 1 - \delta_{\nu} (\omega) + i \xi (\omega)$ is the complex index of the material tabulated by the CXRO online database \cite[\emph{cf.\@} \cref{tab:optical_constants}]{CXRO_database}. 
\begin{figure}
 \centering
 \includegraphics[scale=0.75]{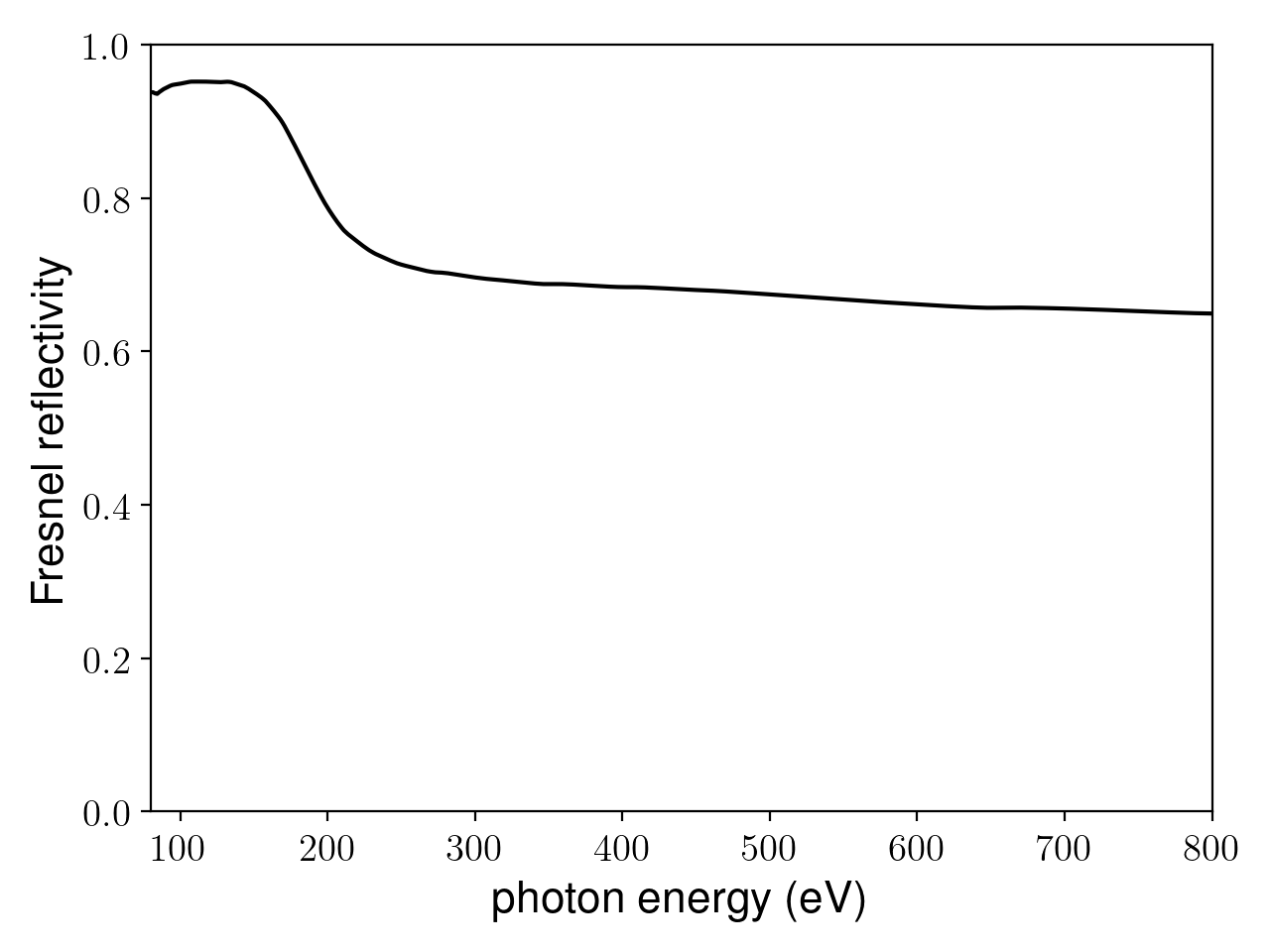}
 \includegraphics[scale=0.75]{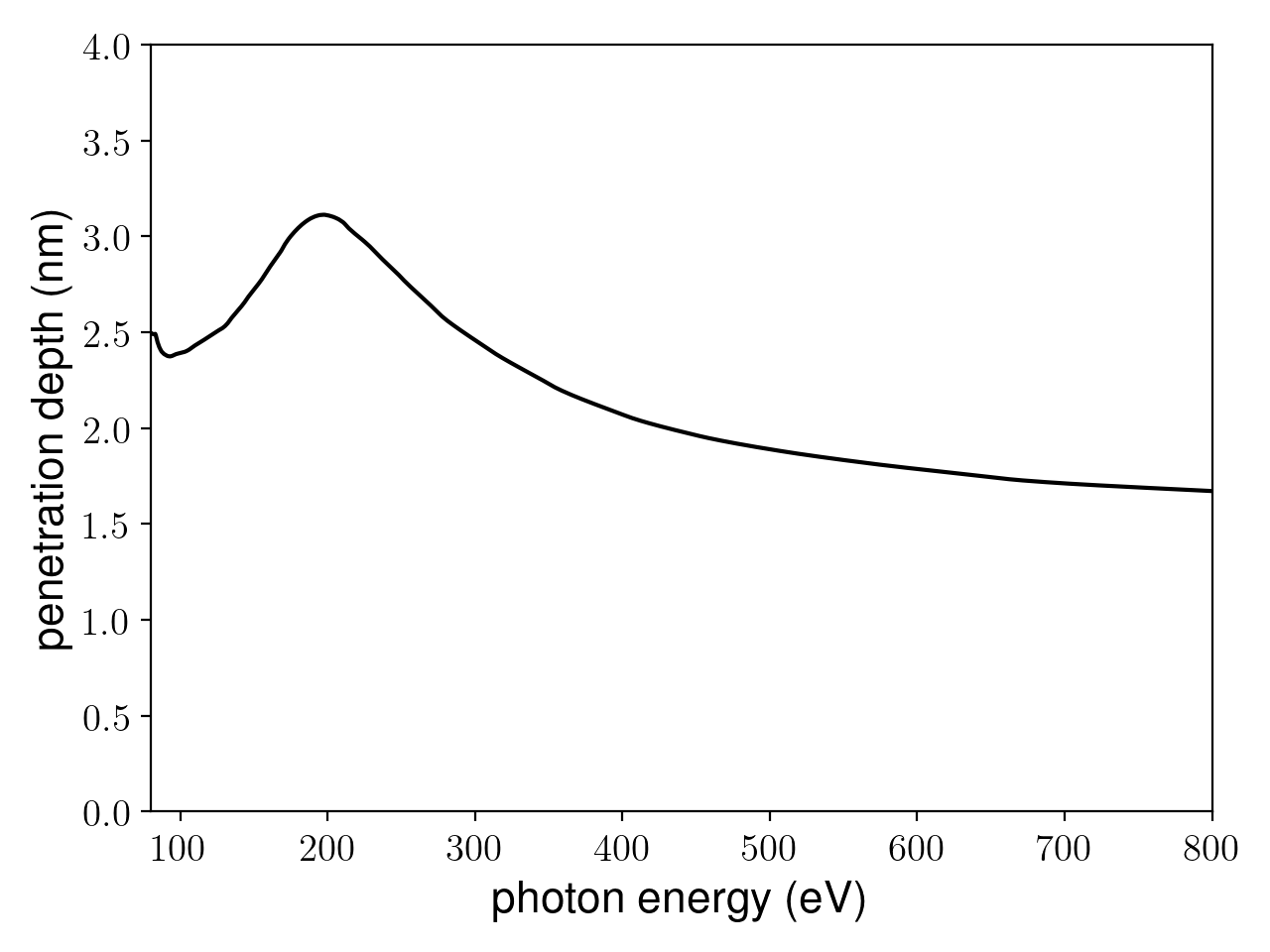}
 \caption[Fresnel reflectivity and penetration depth in a thick slab of gold for nominal test geometry]{Fresnel reflectivity, $\mathcal{R}_F$, and penetration depth, $\mathcal{D}_{\perp}$, in a thick slab of gold at $\zeta = 1.7^{\circ}$ over $\SI{80}{\electronvolt} \leq  \mathcal{E}_{\gamma} \leq \SI{800}{\electronvolt}$ (data obtained from CXRO \cite{CXRO_database}) \cite{McCoy20}.}\label{fig:Au_refl_attn} %($\mathcal{R}_F$ given by \cref{eq:ch3_fres_refl}) %($\mathcal{D}_{\perp}$ given by \cref{eq:ch3_pene_dep}) 
 \end{figure}
Based on the x-ray properties of various materials examined in \cref{app:x-ray_materials}, gold is a suitable choice for this overcoat due to its high $\mathcal{R}_F$ and nanoscale $\mathcal{D}_{\perp}$ across the spectral bandpass considered [\emph{cf.\@} \cref{fig:Au_refl_attn}]. 
In principle, a $\sim \SI{15}{\nm}$-thick layer is sufficient to treat the material as a thick slab so that further reflections at underlaying material interfaces can be neglected. 

Due to its broadband response over the wavelength range for diffraction-efficiency testing, $\SI{15.5}{\nm} \gtrapprox \lambda \gtrapprox \SI{1.55}{\nm}$, gold was chosen as the reflective overcoat for the grating. 
However, because gold is non-reactive toward PMMA, a thin film of an oxidizing metal such as chromium or titanium must be first deposited on the patterned resist to promote wetting and adhesion for the top, reflective layer \cite[\emph{cf.\@} \cref{sec:reflectivity_polarization}]{Trolier-McKinstry17}. 
Ideally, the result is a gold coating that maintains the fidelity of the sawtooth topography while also realizing blazed groove facets with RMS surface roughness, $\sigma$, low enough to reduce absorption and non-specular scatter at the short $\lambda$ considered [\emph{cf.\@} \cref{sec:rough_surface}]. 
From the \emph{Fraunhofer criterion} for a smooth surface [\emph{cf.\@} \cref{eq:roughness_inequality_general}]: 
\begin{equation}\label{eq:roughness_inequality}
 \sigma < \frac{\lambda}{32 \sin \left( \zeta \right)} , 
 \end{equation}
$\sigma$ should be on the level of \SI{1}{\nm} RMS to satisfy this condition for $\SI{15.5}{\nm} \gtrapprox \lambda \gtrapprox \SI{1.55}{\nm}$. 
The impact of surface roughness, however, also depends on the the typical lateral size scale of the surface features [\emph{cf.\@} \cref{sec:discussion_taste}]. 

Deposition for the grating overcoating was performed by \emph{electron-beam physical vapor deposition (EBPVD)} \cite{Singh05} using a \textsc{Kurt J.\ Lesker Lab-18} system at the Penn State Nanofabrication Laboratory \cite{PSU_MRI_nanofab}. 
First, a \SI{5}{\nm}-thick film of titanium was deposited on the patterned TASTE wafer described in \cref{sec:prototype_TASTE_process} at a previously-determined rate of \SI{0.5}{\angstrom\per\second} under high vacuum. 
This allowed a thin oxide layer to form between the resist surface and the titanium coating, providing a wetted, metallic surface for the gold layer to adhere to. %titanium and some of the oxygen present in PMMA to form 
Without breaking vacuum, the gold was then deposited at a rate of \SI{1.0}{\angstrom\per\second} to achieve a layer $\sim \SI{15}{\nm}$ thick. 
\begin{figure}
 \centering
 \includegraphics[scale=0.7]{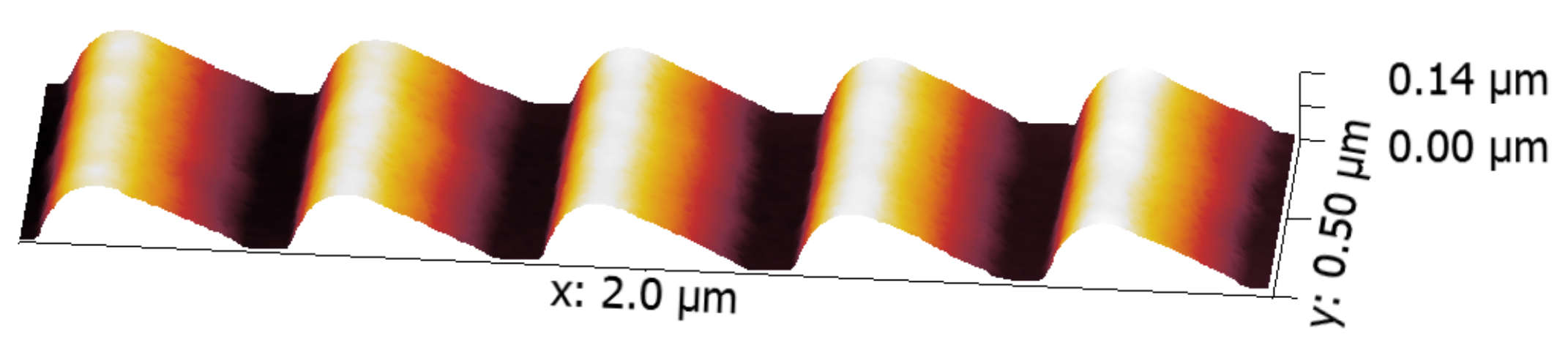}
 \caption[AFM of the grating prototype grooves following electron-beam physical vapor deposition of gold, using titanium as an adhesion layer on PMMA]{AFM of the grating prototype grooves following EBPVD of gold, using titanium as an adhesion layer on PMMA \cite{McCoy20}.}\label{fig:coating_AFM}
 \end{figure}
The final, coated grating prototype appears under AFM as a sawtooth-like topography very similar to the uncoated, TASTE-processed resist from \cref{fig:TASTE_AFMs}; this image of the coated grating grooves, taken using the AFM methodology described in \cref{sec:prototype_TASTE_process}, is shown in \cref{fig:coating_AFM}. 

The coated grooves were imaged over a larger area by \emph{field-emission scanning electron microscopy (FESEM)} using a \textsc{Zeiss Leo 1530} system at the Penn State Nanofabrication Laboratory \cite{PSU_MRI_nanofab,Zhou07}; this micrograph, taken using a \SI{0.5}{\kilo\volt} electron accelerating voltage, is shown in \cref{fig:coating_SEM}. 
\begin{figure}
 \centering
 \includegraphics[scale=0.5]{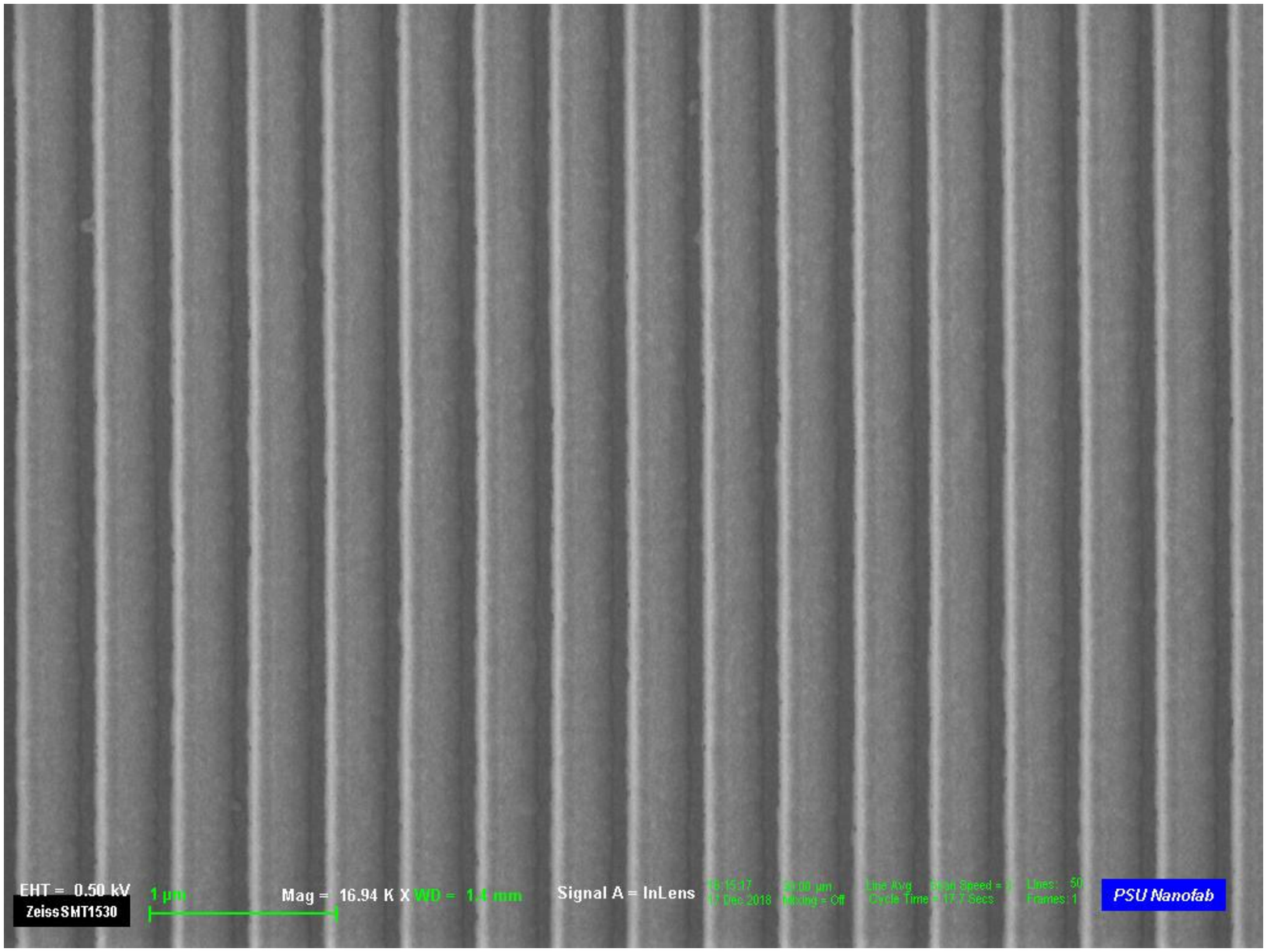}
 \caption[FESEM of the gold-coated grating prototype grooves viewed top-down]{Field-emission scanning electron micrograph (FESEM) of the gold-coated grating prototype grooves viewed top-down, taken with a \textsc{Zeiss Leo 1530} instrument at the Penn State Nanofabrication Laboratory \cite{McCoy20}.}\label{fig:coating_SEM}
 \end{figure}
From the gathered AFM data, $\sigma$ on the groove facets measures about \SI{1.5}{\nm} RMS using the \textsc{Bruker NanoScope Analysis} software package, whereas prior to the coating [\emph{cf.\@} \cref{fig:TASTE_AFMs}, bottom panel], $\sigma \approx \SI{1.25}{\nm}$ RMS on PMMA. 
While the blaze angle measures $\delta \approx 27^{\circ}$ as expected, the groove depth measures $\sim \SI{10}{\nm}$ less than the uncoated grooves shown in \cref{fig:TASTE_AFMs}. 
Moreover, the bottom plateau of the coated grooves appears slightly widened relative to the bottom plateau of the uncoated grooves, where the surface of the silicon substrate is exposed; this suggests that the EBPVD process utilized produces a non-uniform coating, which is likely due to the geometry of the EBPVD chamber coupled with the topography of the grating grooves. 
Because these regions are to a high degree shadowed to the incoming radiation in a near-Littrow configuration, however, this is not expected to have a large impact on diffraction efficiency. 

\section{Beamline Experiments}\label{sec:taste_eff_results} 
%%%%%%%%%%%%%%%%%%%%%%%%%%%%%%%%%%%%%%%%%--------------------------------------------------
Following the methodology outlined in \cref{sec:als_testing}, the grating prototype described in \cref{sec:TASTE_prototype} was tested for EUV and soft x-ray diffraction efficiency at beamline 6.3.2 of the ALS \cite{ALS_632,Underwood96,Gullikson01}. 
The gold-coated grating prototype installed inside the beamline test chamber in an extreme off-plane mount is shown in \cref{fig:test_chamber}, where the grating-dispersion direction, $x$, is roughly parallel with the direction of horizontal stage motion for the photodiode detector, which is seen masked with a \SI{0.5}{\mm}-wide, vertical slit. 
\begin{figure}
 \centering
 \includegraphics[scale=0.31]{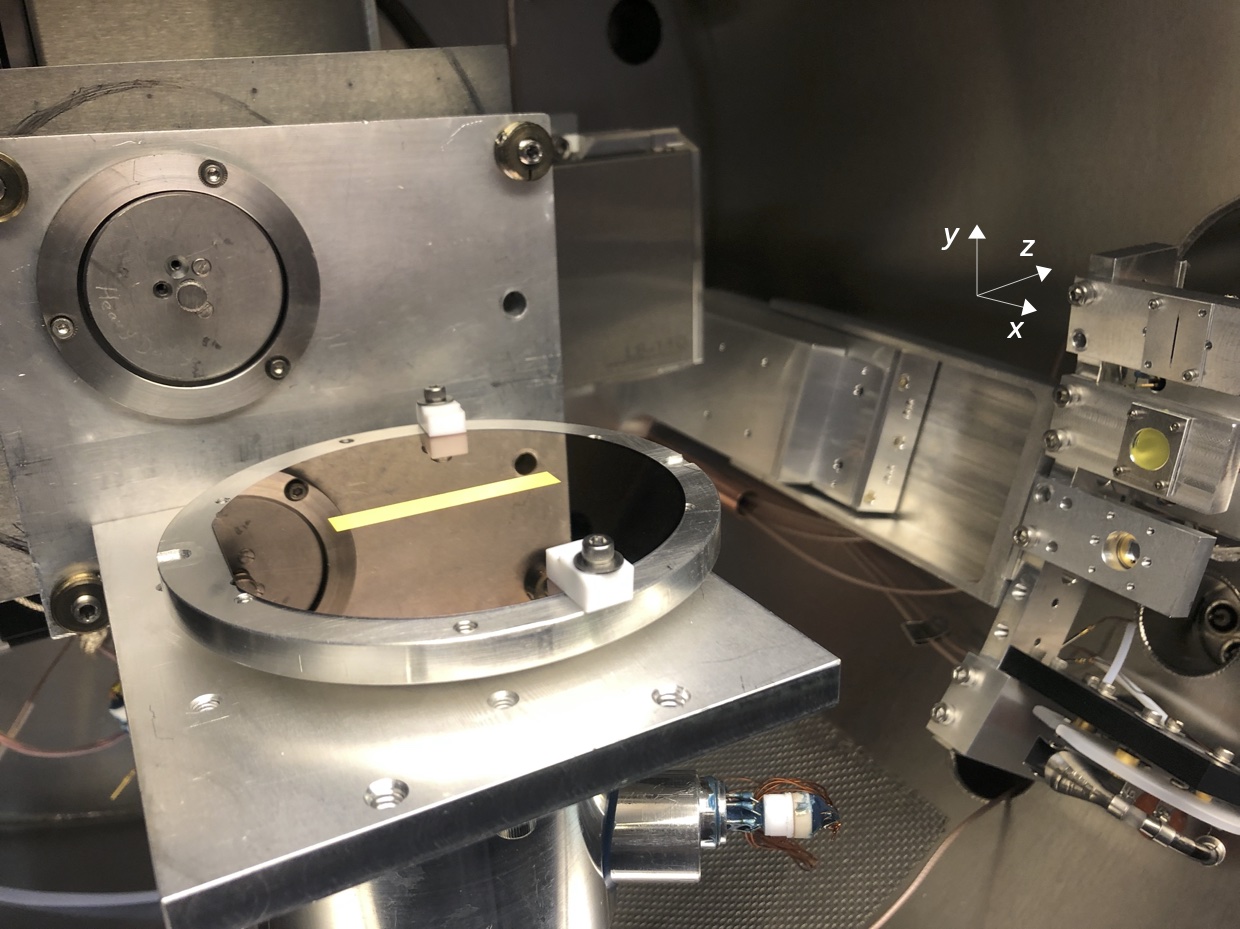}
 \caption[Photograph of the TASTE grating prototype installed inside the test chamber of ALS beamline 6.3.2.]{The TASTE grating prototype [\emph{cf.\@} \cref{sec:prototype_TASTE_process,sec:prototype_coating}] installed inside the test chamber of ALS beamline 6.3.2 \cite{McCoy20}.}\label{fig:test_chamber} 
 \end{figure}
The grating was first oriented at a yaw angle of $\varphi \approx 0^{\circ}$ with graze and roll angles, $\eta$ and $\phi$ respectively, being approximately zero as measured by the tilt of the optic mount using a spirit level. 
The angle $\eta$ was then adjusted to the nominal test value of $1.5^{\circ}$ by using the goniometric stage motion of the photodiode to ensure that the angle between the direct beam and the reflected beam is roughly $2 \eta \approx 3^{\circ}$. 
Next, all grating geometric angles introduced in \cref{sec:geo_constrain} were determined experimentally through analyzing the arc of diffraction as sampled by the photodiode. 
From these measurements, the grating was set to a near-Littrow configuration by adjusting $\varphi$ to ensure that $\alpha \approx 27^{\circ}$ and $\gamma \approx 1.7^{\circ}$ at $\eta \approx 1.5^{\circ}$. 

\subsection{Constraining Grating Geometry}\label{sec:grat_geo_taste}
%%%%%%%%%%%%%%%%%%%%%%%%%%%%%%%%%%%%%%%%%--------------------------------------------------
The throw of the system at the location of $0^{\text{th}}$ order was measured to be $L = 233.0 \pm \SI{1.4}{\mm}$ by comparing the known detector length of $\ell_{\text{det}} = \SI{10}{\mm}$ to the angular size of the detector as measured by a goniometric scan of the beam [\emph{cf.\@} \cref{fig:ALS_cent_throw}]. 
Described in \cref{sec:geo_constrain}, $L$ changes as the the detector moves along the direction $x$ with focal corrections on the order of tens of \si{\um} within \SI{10}{\mm} of travel; however, these corrections are ignored so that order locations are mapped using $x$ and $y = L \sin \left( \Theta \right)$ with $L$ fixed at the measured value. 
In the final test geometry, the diffracted arc was mapped using data gathered at \SIlist{450;500}{\electronvolt} in steps of \SI{50}{\um} along the $x$-direction of the photodiode staging. 
\begin{figure}
 \centering
 \includegraphics[scale=0.4]{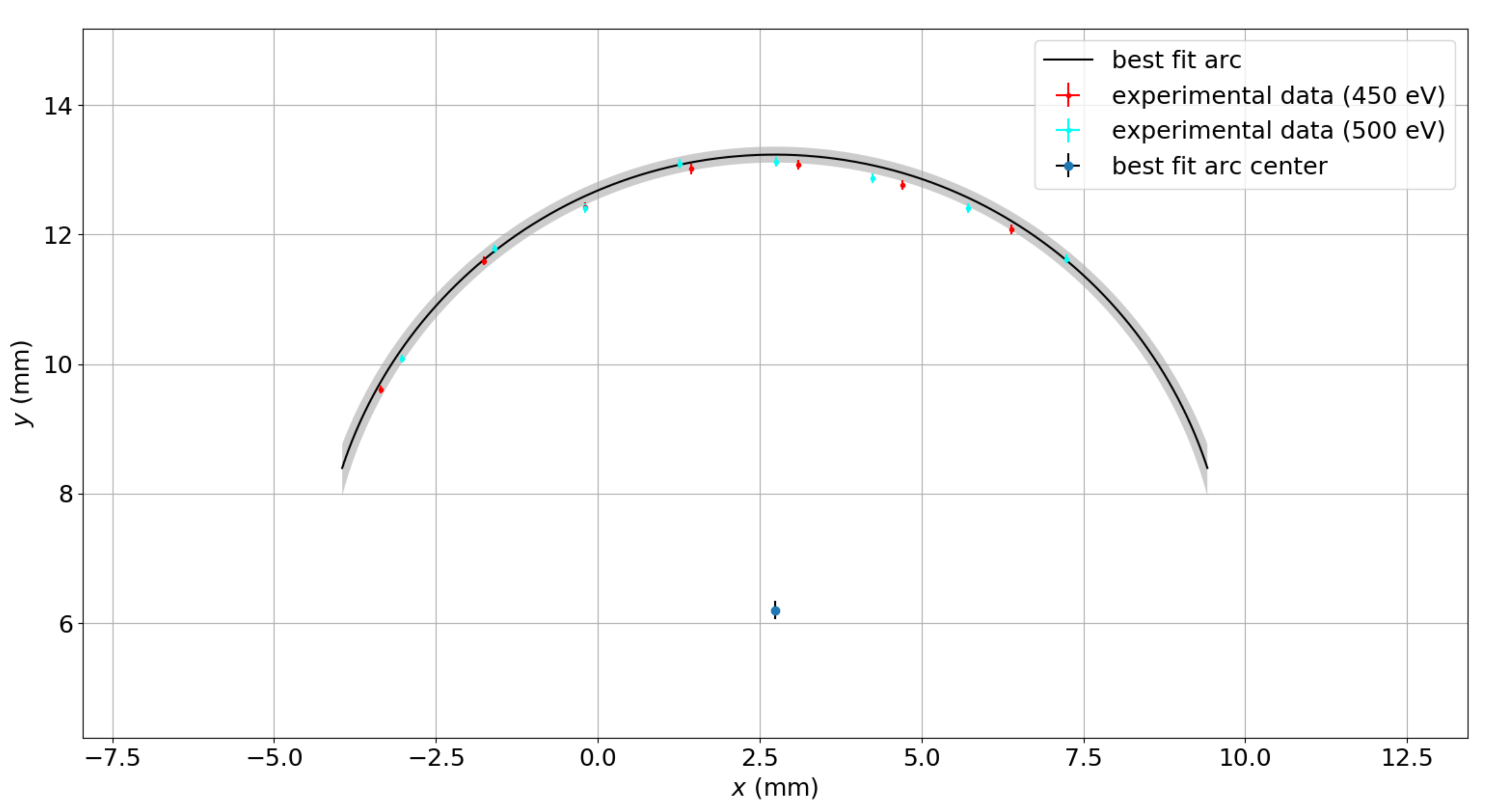}
 \caption[Test-configuration diffracted arc for the TASTE prototype mapped using data gathered at \SI{450}{\electronvolt} and \SI{500}{\electronvolt} fit to a circle]{Test-configuration diffracted arc mapped using data gathered at \SI{450}{\electronvolt} and \SI{500}{\electronvolt} and fit to a circle. Grayed regions represent one standard deviation uncertainty \cite{McCoy20}.}\label{fig:arc_fit_taste}
 \end{figure}
By fitting these data to a half-circle as shown in \cref{fig:arc_fit_taste}, the arc radius was measured as $r = 7.03 \pm \SI{0.12}{\mm}$, and from $r = L \sin \left( \gamma \right)$ [\emph{cf.\@} \cref{eq:arc_radius}], the cone opening half-angle for the diffraction pattern was determined to be $\gamma = 1.73 \pm 0.03^{\circ}$. 

The azimuthal incidence angle, $\alpha$, was measured independently of the roll angle, $\phi$, by \cref{eq:measure_alpha} using $\Delta x_{\text{dir}}$ as the $x$-distance between the direct beam (not shown in \cref{fig:arc_fit_taste}) and the center of the diffracted arc determined from the fit [\emph{cf.\@} \cref{fig:ALS_arc}]. 
With a measured value of $\alpha = 23.4 \pm 0.6^{\circ}$, the roll angle was constrained as $\phi = 1.14 \pm 0.04^{\circ}$ by \cref{eq:measure_roll} using $\Delta x_0$ as the $x$-distance between $0^{\text{th}}$ order and the center of the diffracted arc. 
Using this result and $\Delta y_0$, the $y$-distance between $0^{\text{th}}$ order and the center of the diffracted arc, a graze angle of $\eta \approx 1.5^{\circ}$ was verified through \cref{eq:measure_graze} to give $\eta = 1.56 \pm 0.04^{\circ}$. 
\begin{table}[]
\centering
\caption[Measured parameters for the test-configuration diffracted arc of the TASTE grating prototype at the ALS]{Measured parameters for the test-configuration diffracted arc of the TASTE grating prototype at the ALS \cite{McCoy20}.}\label{tab:arc_params_taste}
\begin{tabular}{@{}ll@{}}
\toprule
parameter                                                                 & measured value             \\ \midrule
system throw ($L$)                                                        & $232.0 \pm 1.4$ mm     \\
arc radius ($r$)                                                          & $7.03 \pm 0.12$ mm       \\
$x$-distance between direct beam and arc center ($\Delta x_{\text{dir}}$) & $2.80 \pm 0.05$ mm       \\
$x$-distance between 0$^{\text{th}}$ order and arc center ($\Delta x_0$)  & $2.92 \pm 0.05$ mm       \\
$y$-distance between 0$^{\text{th}}$ order and arc center ($\Delta y_0$)  & $6.33 \pm 0.14$ mm       \\ \midrule
cone opening half-angle ($\gamma$) by \cref{eq:arc_radius}        	& $1.73 \pm 0.03^{\circ}$  \\
azimuthal incidence angle ($\alpha$) by \cref{eq:measure_alpha}       & $23.4 \pm 0.6^{\circ}$ \\ \midrule
roll (rotation about $z$-axis; $\phi$) by \cref{eq:measure_roll}      & $1.14 \pm 0.04^{\circ}$  \\
graze (rotation about $x$-axis; $\eta$) by \cref{eq:measure_graze}      & $1.56 \pm 0.04^{\circ}$  \\
yaw (rotation about $y$-axis; $\varphi$) by \cref{eq:measure_yaw}     & $0.69 \pm 0.01^{\circ}$  \\ \bottomrule
\end{tabular}
\end{table}
Finally, grating yaw was measured using \cref{eq:measure_yaw} to yield $\varphi = 0.69 \pm 0.01^{\circ}$. 
Summarized in \cref{tab:arc_params_taste}, these measurements indicate a near-Littrow test configuration at $\eta \approx 1.5^{\circ}$ for a blaze angle of $\delta \approx 27^{\circ}$. 

\subsection{Test Results for Diffraction Efficiency}\label{sec:test_results_taste}
%%%%%%%%%%%%%%%%%%%%%%%%%%%%%%%%%%%%%%%%%--------------------------------------------------
In the test geometry established in \cref{sec:grat_geo_taste}, diffraction-efficiency data were gathered as a function of photon energy, $\mathcal{E}_{\gamma}$, where for each measurement, both the diffracted arc and the direct beam were scanned along the $x$-direction in \SI{50}{\um} increments. 
From these measurements, $\mathscr{E}_n$, defined as $\mathcal{I}_n / \mathcal{I}_{\text{inc}}$ [\emph{cf.\@} \cref{eq:diffraction_efficiency}], was calculated according to the methodology described in \cref{sec:measure_efficiency}, with dark current from the photodiode detector taken into account. 
Diffuse scatter arising from surface roughness on the groove facets, which in principle only affects $\mathcal{I}_n$, was estimated using the continuum level in between order maxima and was found to be significant only for $\mathcal{E}_{\gamma} \gtrapprox \SI{600}{\electronvolt}$, where it contributed to $\mathscr{E}_n$ on the level of \SI{1}{\percent} or less for each propagating order. 
These measurements were taken in \SI{20}{\electronvolt} steps, first from \SIrange{440}{800}{\electronvolt} and then from \SIrange{80}{420}{\electronvolt} with the triple-mirror order sorter described in \cref{sec:als_beam}.   

Although the implementation of the order sorter is expected to shift slightly the position of the beam on the grating, and hence the measured parameters listed in \cref{tab:arc_params_taste}, the effect is small and not apparent in the measured absolute efficiency data, which are shown in \cref{fig:abs_eff_taste} compared to $\mathcal{R}_{F}$ for gold at $\zeta = 1.73^{\circ} \approx \gamma$ using \cref{eq:ch3_fres_refl}. 
\begin{figure}
 \centering
 \includegraphics[scale=0.395]{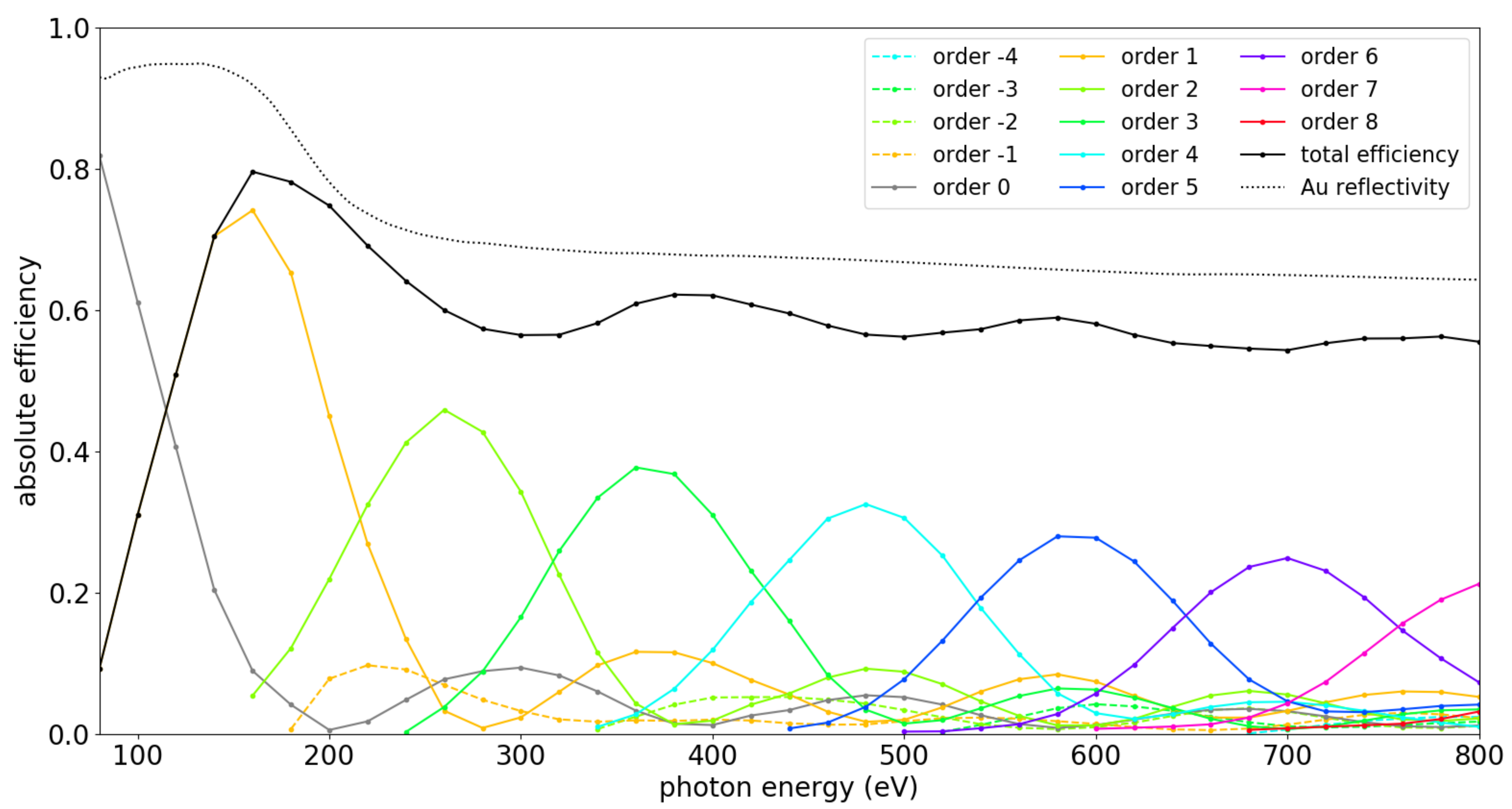}
 \caption[Measured absolute diffraction efficiency for the TASTE grating prototype compared to the Fresnel reflectivity of gold]{Data for absolute diffraction efficiency gathered at the ALS compared to the Fresnel reflectivity of gold, $\mathcal{R}_{F}$. Total diffraction efficiency below \SI{240}{\electronvolt} misses contributions from orders $n = \num{2}$ and $n = \num{3}$ on the order of a few percent \cite{McCoy20}.}\label{fig:abs_eff_taste}
 \end{figure} 
However, as indicated most clearly by the sharp cut-off in the measured $n=2$ curve at $\mathcal{E}_{\gamma} \approx \SI{160}{\electronvolt}$, the beam shift evidently caused measurements of propagating orders of $n=\num{2}$ and $n=\num{3}$ with large diffracted angle, $\beta$, to be missed by the photodiode during data collection. 
These data nonetheless show that peak-order efficiency ranges from about \SI{75}{\percent} down to \SI{25}{\percent} as $\mathcal{E}_{\gamma}$, and $n$, increase. 
$\mathscr{E}_{\text{tot}}$, defined as $\sum_n \mathscr{E}_n$ for all propagating orders with $n \neq 0$, is also plotted in \cref{fig:abs_eff_taste} but due to the missing $n=\num{2}$ and $n=\num{3}$ measurements in the EUV, this curve underestimates the true total diffraction efficiency for $\mathcal{E}_{\gamma} < \SI{240}{\electronvolt}$. 
Moreover, relative diffraction efficiency was calculated by dividing each $\mathscr{E}_n$ measurement from \cref{fig:abs_eff_taste} by $\mathcal{R}_{F}$ [\emph{cf.\@} \cref{sec:measure_efficiency}]. 
\begin{figure}
 \centering
 \includegraphics[scale=0.39]{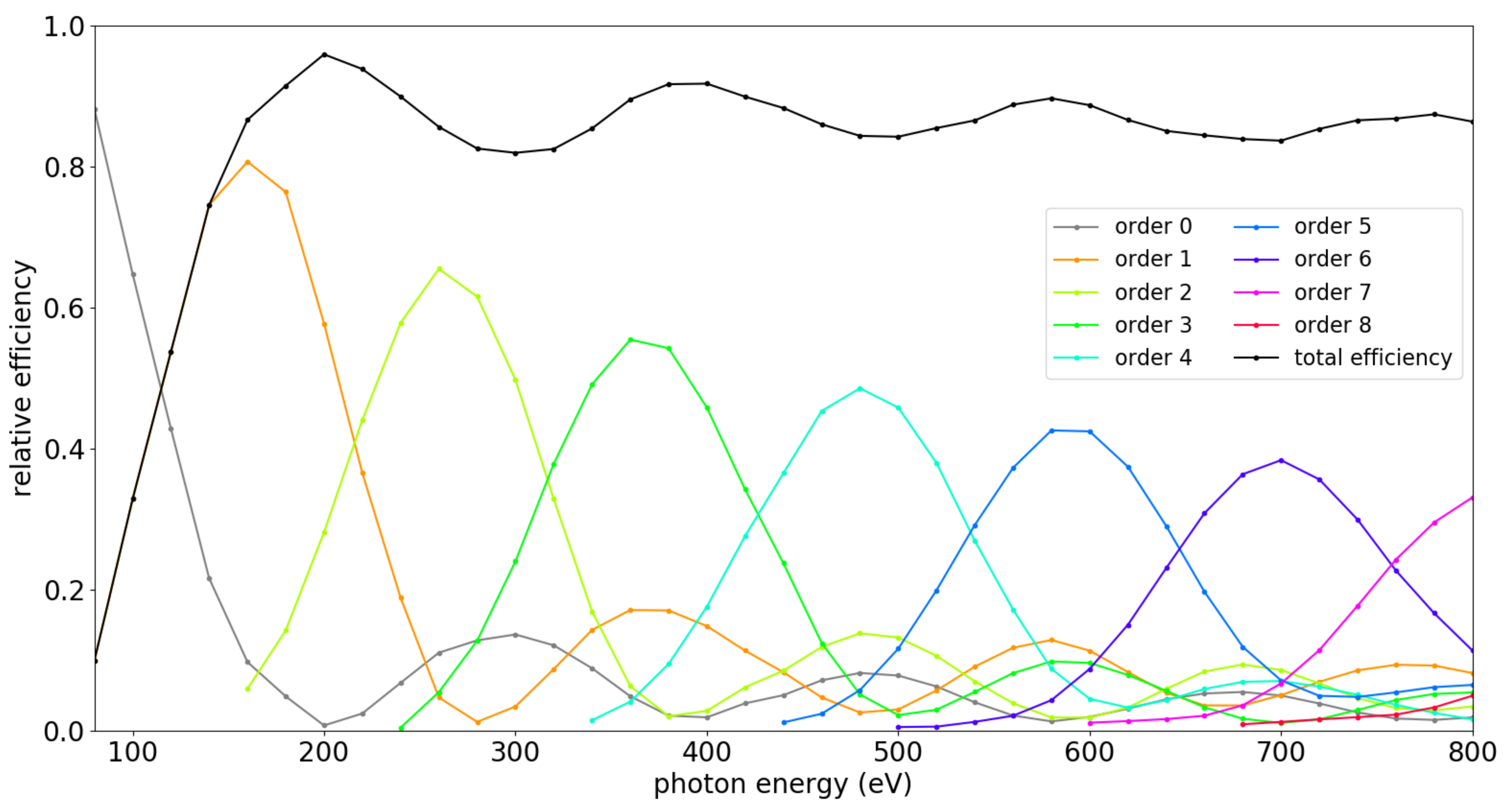}
 \caption[Relative diffraction efficiency calculated by dividing absolute diffraction efficiency by the Fresnel reflectivity of gold]{Relative diffraction efficiency calculated by dividing the absolute diffraction efficiency from \cref{fig:abs_eff_taste} by the Fresnel reflectivity of gold, $\mathcal{R}_{F}$. Total relative diffraction efficiency below \SI{240}{\electronvolt} misses contributions from orders $n = \num{2}$ and $n = \num{3}$ on the order of a few percent \cite{McCoy20}.}\label{fig:rel_eff_taste}
 \end{figure}
This result is plotted in \cref{fig:rel_eff_taste}, where total relative diffraction efficiency, $\mathscr{E}_{\text{tot}} / \mathcal{R}_{F}$, ranges from about \SIrange{95}{88}{\percent} as $\mathcal{E}_{\gamma}$ increases from \SIrange{240}{800}{\electronvolt}, where all propagating orders are accounted for [\emph{cf.\@} \cref{tab:eff_benchmark}]. 

\section{Analysis and Discussion}\label{sec:discussion_taste}
%%%%%%%%%%%%%%%%%%%%%%%%%%%%%%%%%%%%%%%%%--------------------------------------------------
The beamline measurements presented in \cref{sec:test_results_taste} indicate that the grating prototype yields an approximate blaze response at EUV and soft x-ray wavelengths in a near-Littrow configuration.  
This is evidenced by the total diffraction efficiency, $\mathscr{E}_{\text{tot}}$, being dominated by single orders with $n > 0$ and peak positions close to those predicted by \cref{eq:blaze_wavelength_TASTE} for the blaze wavelength [\emph{cf.\@} \cref{fig:abs_eff_taste}]. 
However, along with the peak orders that resemble a blaze response, propagating orders of lower $n$ each contribute to $\mathscr{E}_{\text{tot}}$ at a level of $\sim \SI{10}{\percent}$. 
Thus, toward the blue end of the measured bandpass, where a relatively large number of propagating orders exist, peak-order diffraction efficiency is comparatively low and comprises a smaller fraction of $\mathscr{E}_{\text{tot}}$. %exist by \cref{eq:off-plane_orders}
This suggests that diffracted orders gradually become suppressed with increasing $n$ due to an imperfect sawtooth topography generated by the TASTE process outlined in \cref{sec:prototype_TASTE_process}. 
That is, while an ideal blazed grating exhibits a sharp sawtooth topography, the grating prototype features a quasi-flat apex produced by the \SI{100}{\nm}-wide, top staircase step in the GEBL pattern that is nominally unexposed to high-energy electrons and hence largely unaffected by the thermal reflow process. %\footnote{As noted in \cref{sec:reflow_results,sec:prototype_TASTE_process}, however, slight rounding in this top step is thought to have occurred from a degraded reflow selectivity.} 

In addition to an imperfect sawtooth topography, peak-order diffraction efficiency, especially toward the blue end of the spectrum, is impacted by $\lambda$-dependent losses that arise from surface roughness on the groove facets [\emph{cf.\@} \cref{sec:refl_account}]. 
This can be gleaned from analyzing the total relative response from the grating, $\left( \mathscr{E}_{\text{tot}} + \mathscr{E}_0 \right) / \mathcal{R}_{F}$ [\emph{cf.\@} \cref{sec:measure_efficiency}], with $\mathcal{R}_{F}$ given by \cref{eq:ch3_fres_refl}. 
Due to the short, nanoscale penetration depth, $\mathcal{D}_{\perp}$, of gold at grazing incidence [\emph{cf.\@} \cref{fig:Au_refl_attn}], it is justified to treat the grating overcoat material as an infinitely-thick layer of gold using index of refraction data provided by CXRO \cite{CXRO_database}.
The grating's total relative response is plotted in \cref{fig:rough_eff_taste} over $\SI{240}{\electronvolt} \leq \mathcal{E}_{\gamma} \leq \SI{800}{\electronvolt}$, where the data show a monotonic decrease from about \SI{96}{\percent} down to \SI{88}{\percent} as wavelength decreases, suggesting that $\lambda$-dependent losses are occurring. %the range of measured photon energies that include all propagating orders,
\begin{figure}
 \centering
 \includegraphics[scale=0.39]{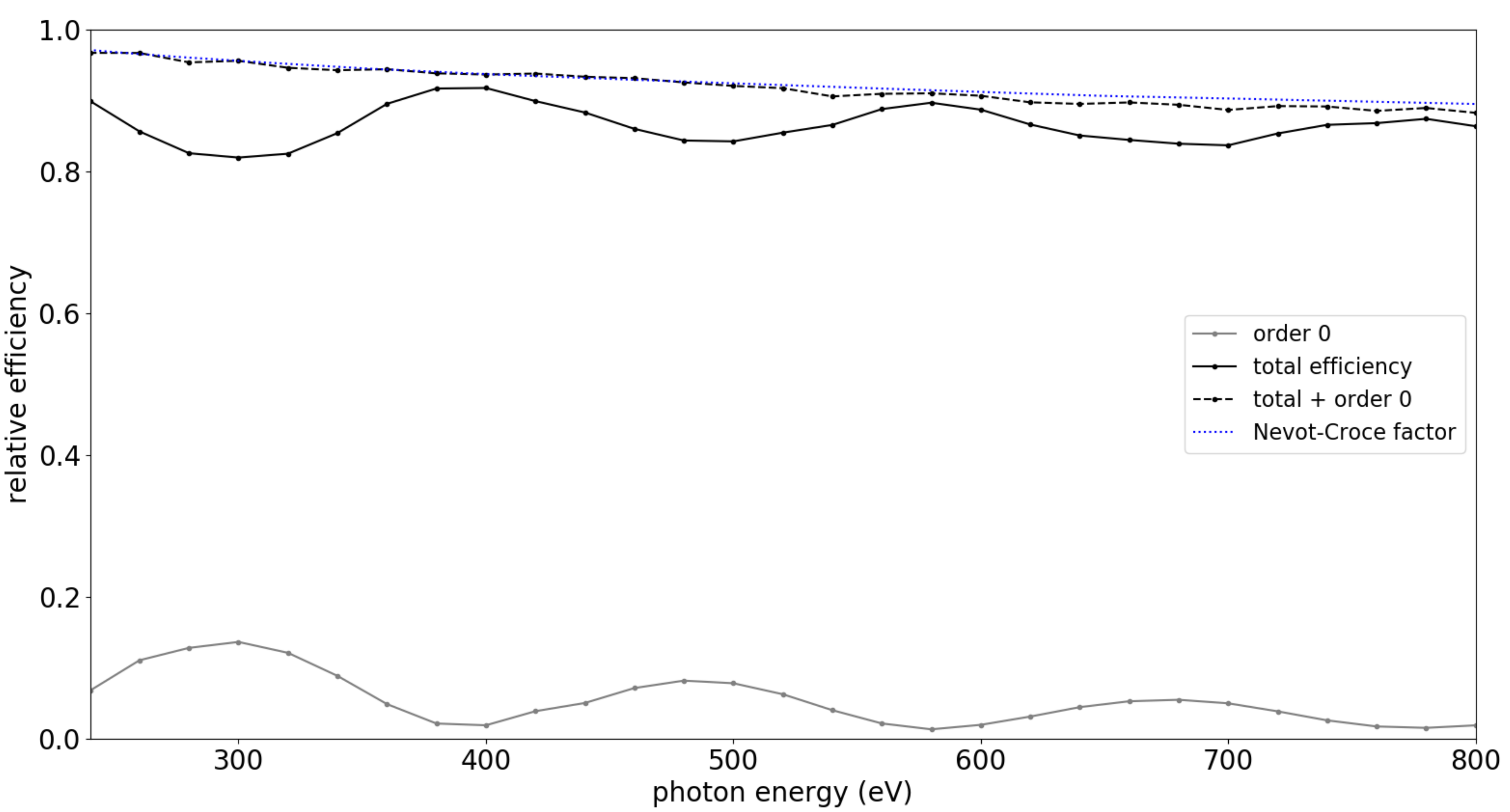}
 \caption[Total relative response of the grating prototype, compared to the Nevot-Croce factor]{Total relative response of the grating prototype, defined as the sum of total diffraction efficiency and zero order, relative to the reflectivity of gold. Overlaid is the Nevot-Croce factor [\emph{cf.\@} \cref{eq:nc_factor_taste}] for $\zeta = 1.73^{\circ}$ and $\sigma = \SI{1.5}{\nm}$ RMS, which indicates the theoretical specular reflectivity of a rough surface relative to Fresnel reflectivity, $\mathcal{R}_{F}$ \cite{McCoy20}.}\label{fig:rough_eff_taste}
 \end{figure}
This is to be compared with the specular reflectivity of a hypothetical mirror flat relative to $\mathcal{R}_{F}$ such that its total relative response is \SI{100}{\percent} in the absence of surface roughness. 
In the regime of TER, the reduced specular reflectivity from a rough surface, $\mathcal{R}_{\text{rough}}$, is described approximately by the \emph{Nevot-Croce factor} [\emph{cf.\@} \cref{sec:refl_account}]. 
For a thick slab of gold with a complex index of refraction $\tilde{\nu} (\omega)$, a grazing-incidence angle $\zeta$ and RMS surface roughness $\sigma$, the norm-squared of this factor is given by \cref{eq:NC_refl_ratio} with $\mathcal{R}_{NC} = \mathcal{R}_{\text{rough}}$:
\begin{equation}\label{eq:nc_factor_taste}
  \frac{\mathcal{R}_{NC}}{\mathcal{R}_F} = \norm{\mathrm{e}^{- 2 k_{\perp} \tilde{k}_{\perp} \sigma^2}}^2 = \mathrm{e}^{-4 k_0^2 \sin \left( \zeta \right) \Re \left[ \sqrt{\tilde{\nu}^2 (\omega) - \cos^2 \left( \zeta \right) } \right] \sigma^2} ,
 \end{equation}
where $k_{\perp} = - k_0 \sin \left( \zeta \right)$ and $\tilde{k}_{\perp} = -k_0 \sqrt{\tilde{\nu}^2 (\omega) - \cos^2 \left( \zeta \right)}$ are the components of the wave vector normal to the surface in vacuum and gold, respectively, with $k_0 \equiv 2 \pi / \lambda$. 
Moreover, $\Re \left[ \sqrt{\tilde{\nu}^2 (\omega) - \cos^2 \left( \zeta \right) } \right]$ represents the real part of $-\tilde{k}_{\perp} / k_0$. 

As described in \cref{sec:refl_account}, the Nevot-Croce factor defined by \cref{eq:nc_factor_taste} is valid for small roughness features taking on a Gaussian height distribution with $\abs{k_{\perp}} \sigma \ll 1$ so that using $\sigma \approx \SI{1.5}{\nm}$ RMS as measured by AFM and $\zeta = 1.73^{\circ}$, this condition is satisfied for  
\begin{equation}\label{eq:rough_condition}
 \lambda \gg 2 \pi \sigma \sin \left( \zeta \right) \approx \SI{0.3}{\nm}  .
 \end{equation}
Additionally, derivations of the Nevot-Croce factor assume a very small surface correlation length, $\ell_{\text{corr}}$ [\emph{cf.\@} \cref{eq:auto_correlation_normal}], that satisfies $\ell_{\text{corr}} k^2_{\perp} \ll k_0$ \cite{deBoer95}. 
Keeping $\ell_{\text{corr}}$, which represents the lateral size scale of roughness features, as an unknown, this yields 
\begin{equation}\label{eq:corr_condition}
 \lambda \gg 2 \pi \ell_{\text{corr}} \sin^2 \left( \zeta \right) \approx 0.006 \ell_{\text{corr}} . 
 \end{equation}
If \cref{eq:rough_condition,eq:corr_condition} are fulfilled, diffuse scatter in vacuum can in principle be neglected and $\lambda$-dependent losses attributed to absorption as radiation scatters into the medium. 
Otherwise, radiation of wavelength $\lambda$ is able to diffract from roughness spatial frequencies on the order of $\ell_{\text{corr}}^{-1}$, producing diffuse scatter that can be detected by the photodiode, in which case details of the \emph{power spectral density (PSD) function} for surface roughness [\emph{cf.\@} \cref{eq:PSD_general}] are required to obtain a more accurate expression for $\mathcal{R}_{\text{rough}}$ \cite{deBoer95,Wen15}. 

Because the PSD function for surface roughness is not known for the groove facets on the grating prototype, \cref{eq:nc_factor_taste} was taken to approximate $\left( \mathscr{E}_{\text{tot}} + \mathscr{E}_0 \right) / \mathcal{R}_{F}$ for the grating prototype in the presence of surface roughness. 
This is plotted in \cref{fig:rough_eff_taste}, where it is seen that the data closely match the Nevot-Croce factor with the experimentally-determined values of $\zeta \approx \gamma = 1.73^{\circ}$ and $\sigma \approx \SI{1.5}{\nm}$ RMS. 
This supports the idea that absorption due to surface roughness on the groove facets is responsible for the losses in the grating's total response over $\SI{240}{\electronvolt} \leq \mathcal{E}_{\gamma} \leq \SI{800}{\electronvolt}$.  %the measured bandpass that encompasses all propagating orders. 
Although the detection of diffuse scatter for $\mathcal{E}_{\gamma} \gtrapprox \SI{600}{\electronvolt}$ ($\lambda \lessapprox \SI{2}{\nm}$) [\emph{cf.\@} \cref{sec:taste_eff_results}] suggests that the conditions for the Nevot-Croce factor to be valid are not strictly fulfilled at these relatively short wavelengths, \cref{fig:rough_eff_taste} indicates that \cref{eq:nc_factor_taste} is a decent approximation across the bandpass considered.
However, future diffraction efficiency test campaigns should better quantify diffuse scatter due to surface roughness in a similar manner to x-ray reflectivity experiments that aim to characterize surfaces, materials and inter-facial roughness \cite{Gay1999,Baumbach1999}. 

To investigate the impact that an imperfect sawtooth topography with an unpointed apex has on the measured diffraction efficiency, data for $\mathscr{E}_n$ were modeled according to the vector diffraction theory framework outlined in \cref{sec:integral_method}. 
This was handled using the software package \textsc{PCGrate-SX} (v.\ 6.1) \cite{PCGrate_web}, which solves the Helmholtz equation through the integral method for a custom grating boundary and incidence angles input by the user \cite[\emph{cf.\@} \cref{sec:grat_bound_consider}]{Goray10}. 
Previous beamline experiments have verified a lack of polarization sensitivity for x-ray reflection gratings used in extreme off-plane mounts \cite{Marlowe16} and as a result, \textsc{PCGrate-SX} calculations were carried out assuming a perfectly conducting grating boundary with perfectly smooth groove facets and an incident wavefront with \emph{TE polarization} [\emph{cf.\@} \cref{fig:polarization_angle,sec:dirch_bound}].  
Although perfect conductivity combined with the absence of surface roughness implies a lossless response from the grating grooves [\emph{cf.\@} \cref{sec:rel_eff_perf}], \textsc{PCGrate-SX} modulates the predicted diffraction efficiency by the reflectivity of a user-input, stratified medium defined by index of refraction data and custom layer thicknesses [\emph{cf.\@} \cref{sec:refl_account}]. 

Taking the grating material to be an infinitely-thick layer of gold as discussed above, the cross-sectional groove shape of the grating prototype was approximated as an acute trapezoid with a near-vertical lateral side opposite a slope that emulates the active blaze facet. 
Additionally, a flat bottom portion was included to represent the cleared portion of the resist described in \cref{sec:TASTE_prototype}. 
\begin{figure}
 \centering
 \includegraphics[scale=0.65]{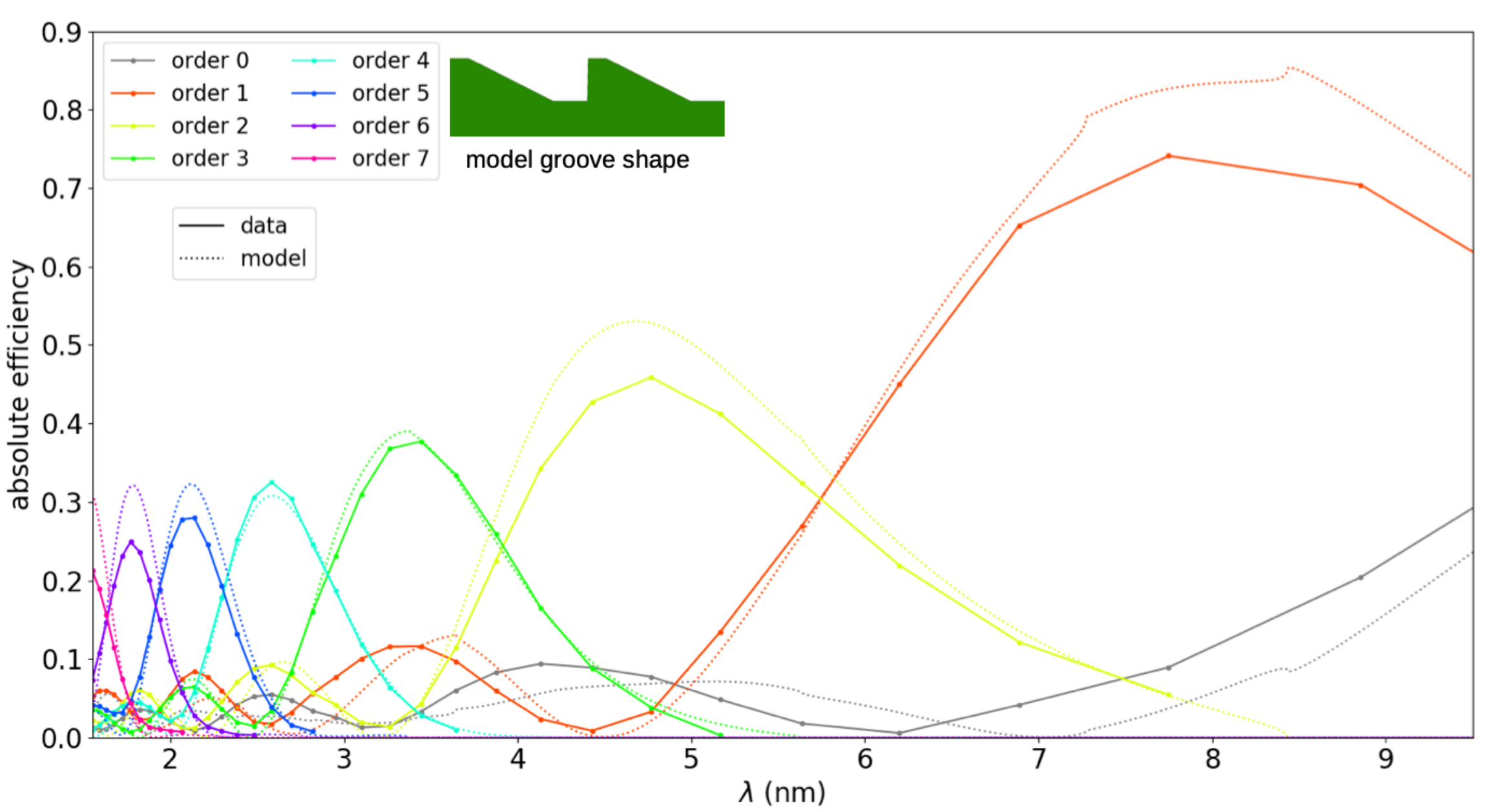} %85 nm land, 120 nm depth, 77 nm ridge, 27 blaze angle, 89 steep angle
 \caption[Measured absolute diffraction efficiency compared to theoretical diffraction efficiency modeled using \textsc{PCGrate-SX}]{Absolute diffraction efficiency from \cref{fig:abs_eff_taste} compared to theoretical diffraction efficiency modeled using \textsc{PCGrate-SX}. Modeled data are multiplied by the Nevot-Croce factor [\emph{cf.\@} \cref{eq:nc_factor_taste}] for $\zeta = 1.73^{\circ}$ and $\sigma = \SI{1.5}{\nm}$ RMS \cite{McCoy20}.}\label{fig:model_eff_taste}
 \end{figure}
Using the nominal values of $\alpha$ and $\gamma$ listed in \cref{tab:arc_params_taste} for grating incidence angles and $d = \SI{400}{\nm}$ for the groove spacing, a series of \textsc{PCGrate-SX} calculations were performed for a range of trapezoids with slightly-varying dimensions close to those measured by AFM in \cref{fig:coating_AFM}. 
The model matching the measured data most closely was one with a blaze angle of $\delta = 27^{\circ}$, a groove depth of \SI{120}{\nm}, a flat-top width of \SI{77}{\nm} and bottom-width of \SI{85}{\nm}. 
These predicted data, modulated by the Nevot-Croce factor from \cref{fig:rough_eff_taste}, are plotted as a function of $\lambda$ in \cref{fig:model_eff_taste} and compared to $\mathscr{E}_n$ measured for orders $n=0$ through $n=7$. 

It is seen in \cref{fig:model_eff_taste} that the measured peak-order positions match roughly those predicted by the model, demonstrating that the grating prototype has an efficiency response similar to that of a blazed grating with $d=\SI{400}{\nm}$~ and $\delta = 27^{\circ}$ at the experimentally-determined incidence angles of $\alpha = 23.4^{\circ}$ and $\gamma = 1.73^{\circ}$. 
However, the amplitudes of the peak orders generally fall short of the model with the apparent exceptions of $n=3$ and $n=4$. 
This phenomenon seems to be due in part to the mismatches that exist between the measured data and the model for secondary diffraction peaks, suggesting that the groove shape trapezoidal approximation is not sufficient to reproduce these results to a high degree of accuracy. 
The grating prototype grooves likely have an apex that is slightly rounded as a result of the thermal reflow process but this is difficult to verify through AFM because the shape of the \textsc{SCANASYST-AIR} tip is convolved with the true grating topography in the micrographs shown in \cref{fig:TASTE_AFMs,fig:coating_AFM}. 
Nonetheless, rounding of corners or other deviations from an ideal acute trapezoid are expected to have an impact on the distribution of diffraction efficiency among orders. 
It is also seen in \cref{fig:model_eff_taste} that measured peak-order efficiency becomes increasingly diminished relative to the model as order number increases beyond $n=4$, which is consistent with the observation already mentioned that the relatively large number of orders at short $\lambda$ each contribute substantially to $\mathscr{E}_{\text{tot}}$ while the peak order comprises a relatively smaller fraction. 
This is another indication of there being groove-shape imperfections that diminish the grating prototype's blaze response. 
A possible explanation beyond rounded corners at the apex is irregularity or non-flatness of the sloped surfaces of the grating grooves across the prototype. 
In principle, this could be caused in part by a non-uniform spin-coat thickness but it is expected that the imperfect blazed grating topography produced by TASTE process described in \cref{sec:prototype_TASTE_process} is the largest contributor to this issue.  

\section{Summary and Conclusions}\label{sec:ch3summary}
%%%%%%%%%%%%%%%%%%%%%%%%%%%%%%%%%%%%%%%%%--------------------------------------------------
A prototype for a reflection grating with a groove spacing of \SI{400}{\nm} was fabricated by generating an approximate sawtooth topography in \SI{130}{\nm}-thick PMMA resist coated on a silicon wafer using the TASTE process established in \cref{sec:TASTE} and then coating the grating grooves with a thin layer of gold via EBPVD for reflectivity, using titanium for adhesion. %at the Penn State Materials Research Institute
Diffraction-efficiency measurements gathered at beamline 6.3.2 of the ALS \cite{ALS_632,Underwood96,Gullikson01}, which span $\SI{15.5}{\nm} \gtrapprox \lambda \gtrapprox \SI{1.55}{\nm}$ in a grazing incidence, extreme off-plane mount, demonstrate that the prototype behaves approximately as a blazed grating with $\delta \approx 27^{\circ}$. % groove cross-sections shaped like an acute trapezoid and a blaze angle of $\delta \approx 27^{\circ}$. 
The total response from the grating relative to the reflectivity of the gold overcoat measures between \SIlist{96;88}{\percent} in the soft x-ray, with losses attributed to absorption and diffuse scatter from grating facets with $\sim \SI{1.5}{\nm}$ RMS surface roughness. 
However, even with losses accounted for, the blaze response is observed to diminish for peak orders with $n \geq 5$. 
While this phenomenon is a result of the TASTE process yielding an imperfect sawtooth topography, these results show that TASTE is a promising fabrication technique for the manufacture of custom reflection gratings for soft x-ray spectroscopy \cite{McCoy18,McCoy20}.

An especially important feature of the TASTE process is its ability to define a sawtooth-like topography over a groove layout defined by EBL while also avoiding the dependences on crystallographic structure that exist in processes that \ce{KOH} etching to provide a grating blaze \cite{Franke97,Chang03,McEntaffer13,Miles18}. 
This is particularly advantageous for realizing fanned, curved or other variable-line-space groove layouts that are required for achieving high $\mathscr{R} = \lambda / \Delta \lambda$ while also having blazed groove facets that enable high spectral sensitivity. %also, possibly curved substrates
With $\mathscr{E}_{\text{tot}}$ exceeding \SI{40}{\percent} in the soft x-ray bandpass, these results show that gratings fabricated by TASTE are capable of meeting \emph{Lynx} requirements in terms of spectral sensitivity [\emph{cf.\@} \cref{tab:eff_benchmark}]. 
Additionally, an absolute efficiency of \SI{75}{\percent} in $n=1$ at $\mathcal{E}_{\gamma} \approx \SI{160}{\electronvolt}$ gives an indication that TASTE can realize a highly-efficient grating for EUV spectroscopy with modification of grating parameters. 
However, further work in nanofabrication and beamline testing for $\mathscr{R}$ is required to determine to what degree TASTE is able to make improvements in these areas of technological development. 
In particular, producing gratings with groove spacing significantly smaller than \SI{400}{\nm} that maintain a satisfactory sawtooth topography is challenging from the standpoint of fabrication by TASTE \cite{McCurdy20}; this is discussed further in \cref{ch:conclusions}. 
% !TEX root = ../McCoy-Dissertation.tex
\chapter[Substrate-Conformal Imprint Lithography (SCIL) for Replication of X-ray Reflection Gratings]{Substrate-Conformal Imprint \\Lithography for Replication of \\X-ray Reflection Gratings}\label{ch:grating_replication} 
%%%%%%%%%%%%%%%%%%%%%%%%%%%%%%%%%%%%%%%%%-------------------------------------------------- 
The process of \emph{substrate-conformal imprint lithography (SCIL)} \cite{Verschuuren10} is introduced in \cref{sec:ch1_conclusions} as a method for high-throughput replication of x-ray reflection gratings, which is essential for achieving a sufficient collecting area for spectroscopy, $A_{\text{col}}$, in future instruments that call for large numbers of grating replicas such as the \emph{XGS} for the \emph{Lynx} mission concept \cite[\emph{cf.\@} \cref{sec:astro_plasmas,sec:grating_tech_dev}]{Gaskin19,McEntaffer19}. 
Unlike rigid-stamp nanoimprint processes such as UV-NIL \cite[\emph{cf.\@} \cref{sec:nanoimprint}]{Chou96,Haisma96,Schift08,Schift10}, SCIL centers on the use of a low-cost, flexible stamp that is molded from a master template, which in this context, defines a grating surface relief \cite{Verschuuren10,Verschuuren17,Verschuuren19}. 
This lithographic approach enables nanoscale patterns to be imprinted in resist over large areas with a stamp that conforms locally to particulate contaminants and globally to any slight bow of the replica substrate so as to avoid damage to the master template by eliminating the need for an applied high pressure. 
Additionally, wave-like sequential imprinting made possible by the flexibility of the stamp and specialized pneumatic tooling serves to eliminate large trapped air pockets. 
For these reasons, SCIL is an attractive technique for imprinting surface reliefs for x-ray reflection gratings, which are often patterned on \SI{150}{\mm}-diameter wafers [\emph{cf.\@} \cref{sec:grating_fab_intro}]. 

Although SCIL stamps are compatible with many UV-curable, organic resists similar to those used for UV-NIL \cite{Ji10,Schift10}, high-volume production that relies on long stamp lifetime is best suited for use with a brand of inorganic resist, synthesized by \textsc{Philips SCIL Nanoimprint Solutions} \cite{philis_scil} (referred to as \textsc{Philips} hereafter) and known commercially as \textsc{NanoGlass}, which cures through a thermodynamically-driven, silica \emph{sol-gel process} \cite{Verschuuren10,Verschuuren17}. 
Packaged equipment that automates sol-gel resist spin-coating and the pneumatic-based SCIL wafer-scale imprint method for high-volume replication has also been developed by \textsc{Philips} \cite{philis_scil}. 
Using this production platform, known as \textsc{AutoSCIL}, a single composite stamp is capable of producing $\gtrapprox \num{700}$ imprints in \textsc{NanoGlass} resist at a rate of \num{60}, \SI{150}{\mm}-diameter wafers per hour without pattern degradation \cite{Verschuuren17,Verschuuren18,Verschuuren19}. 
This technique was first applied to x-ray reflection grating technology for the development of \emph{WRXR} \cite{Miles17,Miles18b,Tutt19}, which utilized \textsc{AutoSCIL} to produce \num{26} replicas of a \SI{110}{\cm\squared} master grating fabricated via electron-beam lithography (EBL) and crystallographic etching in a manner similar to what is described in \cref{sec:crystal_etching} \cite{Miles19,Verschuuren18}. 
While this relatively low imprint throughput can, in principle, be achieved by other means, future instruments such as \emph{tREXS}, \emph{OGRE} and the \emph{XGS} \cite{Miles19b,Tutt18,McEntaffer19}, which each require many more grating replicas, are expected to benefit from the capabilities of \textsc{AutoSCIL}. 

The application of SCIL to grating replication technology was motivated by an initial set of experiments, carried out in collaboration with \textsc{Philips}, that characterize the quality of imprints produced from a master grating template wet-etched in silicon \cite{McCoy17}, which was fabricated by staff at Penn State Nanofabrication Laboratory \cite[\emph{cf.\@} \cref{sec:crystal_etching}]{Miles18,PSU_MRI_nanofab}. 
These experiments and related studies are the subject of this chapter, with a special focus on the impact that resist shrinkage has on blaze angle in imprinted gratings.\footnote{Supported by a NASA Space Technology Research Fellowship lasting from 2015 to 2019, much of this research is published in a peer-reviewed article \cite{McCoy20b}. In addition to the Penn State Materials Research Institute, resources of the Quattrone Nanofabrication Facility at the Singh Center for Nanotechnology \cite{Quattrone} (U.\ of Pennsylvania) were used for SCIL process development carried out in 2017 and 2018 \cite{McCoy17}.} 
First, \cref{sec:grating_fab} outlines a SCIL process for x-ray reflection grating manufacture along with the expected level of shrinkage that occurs in \textsc{NanoGlass} as the silica sol-gel network densifies with a post-imprint thermal treatment. 
Soft x-ray diffraction-efficiency measurements of an imprinted grating and its corresponding master template are then presented in \cref{sec:beamline_testing}. 
These results are analyzed in \cref{sec:discussion_scil} through a comparison of the experimental data to theoretical models for diffraction efficiency from \cref{sec:integral_method} in order to demonstrate a non-negligible level of blaze angle reduction due to resist shrinkage. 
Finally, \cref{sec:conclusion} provides conclusions and a summary for this chapter. 

\section{Grating Fabrication by SCIL}\label{sec:grating_fab}
%%%%%%%%%%%%%%%%%%%%%%%%%%%%%%%%%%%%%%%%%-------------------------------------------------- 
Regardless of the resist used for imprinting, the SCIL process relies on having stamp features carried in a material that is flexible enough for surface-conformal imprinting but stiff enough to maintain a high-fidelity stamp topography at sub-\si{\um} size scales \cite{Verschuuren10,Verschuuren17,Verschuuren19}. 
\begin{figure}
 \centering
 \includegraphics[scale=0.45]{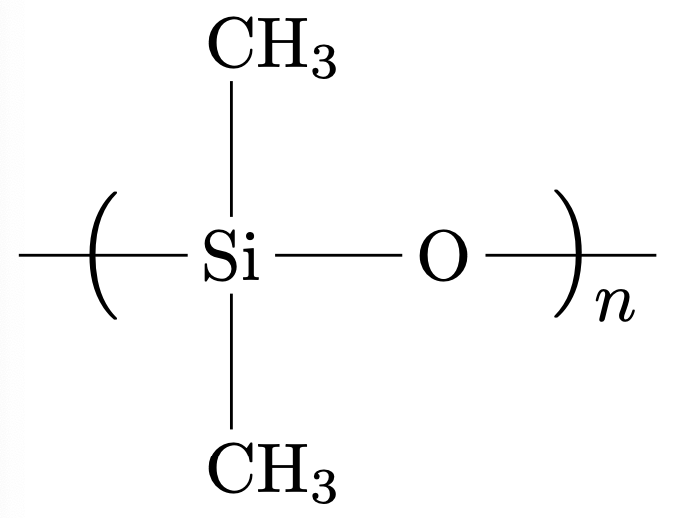}
 \caption[Structural formula of polydimethylsiloxane (PDMS).]{Structural formula of the elastomer polydimethylsiloxane (PDMS), where siloxanes connect at the sites marked by brackets to form a longer polymer chain with side methyl groups.}\label{fig:PDMS}
 \end{figure}
A common stamp material in \emph{soft lithography} \cite{Xia98,Qin10} is \emph{polydimethylsiloxane} (PDMS), a type of \emph{elastomer} composed of many repeating units of the monomer \emph{dimethylsiloxane} (\ce{C2H6OSi}), where linear siloxanes (\emph{i.e.}, \ce{-Si-O-}) connect at the sites marked by brackets while methyl groups (\emph{i.e.}, \ce{-CH3}) terminate the remaining bonding sites of silicon \cite[\emph{cf.\@} \cref{fig:PDMS}]{Zaman19}. %as shown in \cref{fig:PDMS}, %valence-electron
With a \emph{Young's modulus}\footnote{This quantity is defined as the ratio of tensile stress to tensile strain, which essentially characterizes the stiffness of a material \cite{landau1986theory}.} on the order of \SI{1}{\mega\pascal}, however, sub-\si{\um} features in a PDMS stamp are subject to collapse and additionally, deformation of sharp corners due to surface tension \cite{Bietsch00,Hui02,Verschuuren10}. 
To circumvent this issue, modified versions of PDMS have been synthesized to have a Young's modulus that is high enough for smaller-scale imprinting but still much smaller than \SI{1}{\giga\pascal} so as to enable conformal contact over large areas. 
An example of such a material is \emph{hard-PDMS (H-PDMS)}, which features silicon-ethyl bonds (\emph{i.e.}, \ce{-Si-CH2CH2-Si-}) between linear polymer chains\footnote{Briefly, these bonds are produced from an addition reaction between vinyl-modified PDMS and hydride-modified PDMS \cite{Schmid00,Verschuuren10}. That is, either vinyl groups (\emph{i.e.}, \ce{-CH=CH2}) or hydride groups (\emph{i.e.}, \ce{-H}) replace some of the methyl groups in PDMS and reactions between these two components produce silicon-ethyl bonds.} to yield a cured material with a Young's modulus of $\sim \SI{10}{\mega\pascal}$, which, in principle, allows features with size scales down to $\sim \SI{200}{\nm}$ to be imprinted using a flexible stamp \cite{Schmid00,Kim08,Verschuuren10}. 
\textsc{Philips} provides a similar material, known commercially as \emph{X-PDMS}, which features an increased cross-link density in the material network\footnote{The (proprietary) synthesis of this material is based on the chemistry of H-PDMS, but with added components that participate in the addition reaction such as vinyl-modified, quaternary siloxanes, which serve to increase cross-link density further \cite{Verschuuren10}.\label{footnote:X-PDMS}} to enable the construction of SCIL stamps that carry nanoscale features with high fidelity by achieving a Young's modulus on the order of several tens of \si{\mega\pascal} \cite{Verschuuren10,Verschuuren17,Verschuuren19}. 

A composite stamp for SCIL consists primarily of two components that are supported by a flexible sheet of glass with a thickness of about $\SI{200}{\um}$: a $\SI{50}{\um}$-thick layer of H-PDMS or X-PDMS that carries the inverse topography of a master template and an underlaying, $\gtrapprox \SI{0.5}{\mm}$-thick layer of standard, soft PDMS that attaches to the glass sheet by application of an adhesion promoter \cite{Verschuuren10,Verschuuren17}. 
When such a stamp is applied without high pressure to a wafer freshly spin-coated with a film of suitable imprint resist, its features are filled through capillary action as the liquid-state material cures to form a solid structure that takes on the inverse topography of the stamp. 
\textsc{NanoGlass} resist is stored as a $\SI{-20}{\celsius}$ sol that consists of chemical precursors \emph{tetramethylorthosilicate} (TMOS; \ce{Si(OCH3)4}) and \emph{methyltrimethoxysilane} (MTMS; \ce{CH3Si(OCH3)3}) suspended in a mixture of water and alcohols. 
\begin{figure}
 \centering
 \includegraphics[scale=0.45]{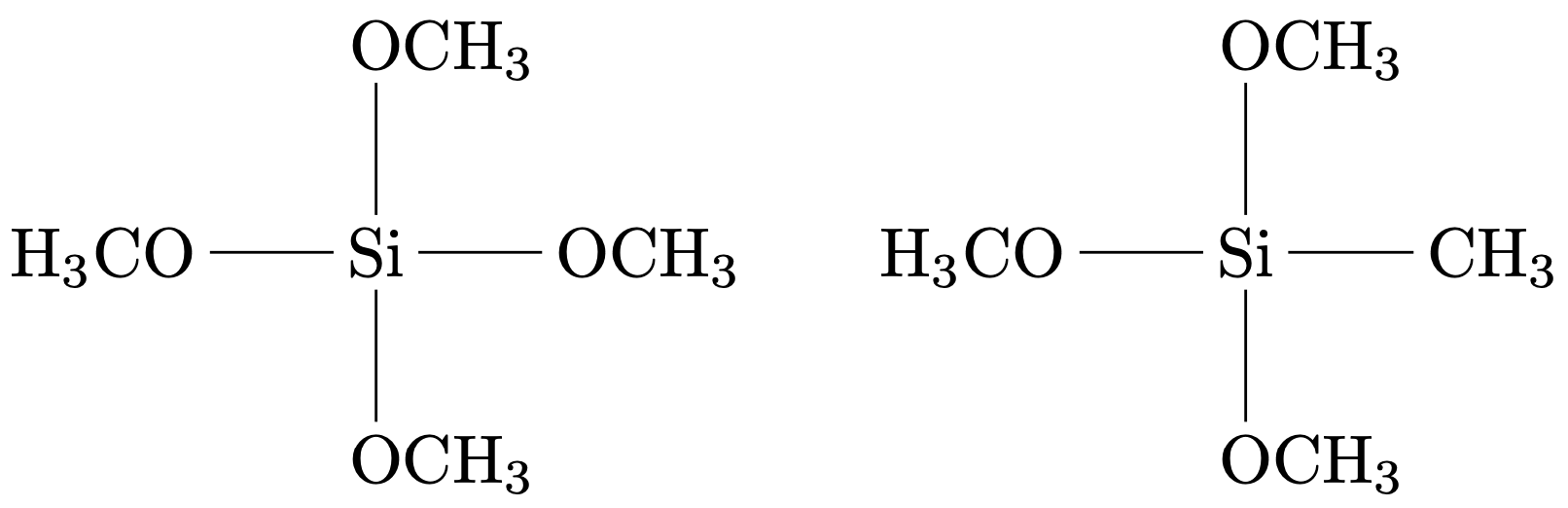}
 \caption[Structural formula of the sol-gel precursors used for \textsc{NanoGlass} imprint resist in substrate-conformal imprint lithography (SCIL).]{Structural formula of the sol-gel precursors used for \textsc{NanoGlass} imprint resist: TMOS (\emph{left}; fully inorganic with four methoxy groups) and MTMS (\emph{right}; organically modified with three methoxy groups and one methyl group).}\label{fig:precursors}
 \end{figure}
These precursors, with structural formulas shown in \cref{fig:precursors}, react to form a gel, and ultimately a solid silica-like network, along with alcohols and water left as reaction products \cite{Brinker90,Hench90,Verschuuren10,Verschuuren19}. 
Briefly, methoxy groups (\ce{-O-CH3}) bonded to silicon in TMOS and MTMS undergo hydrolysis so that they are replaced with hydroxyl groups (\ce{-OH}): 
\begin{equation}
 \ce{Si-(O-CH3)} + \ce{H2O} \longrightarrow \ce{Si-OH} + \ce{HO-CH3} , 
 \end{equation}
where methanol (\ce{HO-CH3}) is produced as a reaction product. 
Hydroxylated silicon sites can then react with one another to form a siloxane bond with the release of water: 
\begin{equation}
 \ce{Si-OH} + \ce{HO-Si} \longrightarrow \ce{Si-O-Si} + \ce{H2O} 
 \end{equation}
and as these condensation reactions continue, a silica-like material network is formed. 
This sol-gel process carries out over the course of \SI{15}{\minute} at room temperature while reaction products and trapped air diffuse into the stamp, leaving solidified resist molded to the inverse of the stamp topography after stamp separation. 

The imprinted resist initially has $\sim \SI{70}{\percent}$ the density of fused silica due to the presence of nanoscale pores that result from the organic component of the MTMS precursor (\emph{i.e.}, \ce{Si-CH3} [\emph{cf.\@} \cref{fig:precursors}]), which does not participate in the sol-gel reaction \cite{Verschuuren10,Verschuuren17,Verschuuren19}. 
However, the material can be densified for stability through a \SI{15}{\minute} bake at $T_{\text{cure}} \gtrapprox \SI{50}{\celsius}$ to induce further cross-linking in the material network, where $T_{\text{cure}} \gtrapprox \SI{450}{\celsius}$ breaks the organic bonds in MTMS and causes a moderate level of shrinkage while $T_{\text{cure}} \gtrapprox \SI{850}{\celsius}$ gives rise to the density of maximally cross-linked \ce{SiO2}. 
It has been previously reported that $T_{\text{cure}} \approx \SI{200}{\celsius}$ leads to $\sim \SI{15}{\percent}$ volumetric shrinkage in imprinted laminar gratings while $T_{\text{cure}}$ in excess of $\SI{1000}{\celsius}$ results in a maximal, $\sim \SI{30}{\percent}$ shrinkage \cite{Verschuuren10,Verschuuren19}.   
Based on these results, it is hypothesized that a low-$T_{\text{cure}}$ treatment should lead to $\sim \SI{10}{\percent}$ volumetric shrinkage in the resist, which is comparable to typical levels of imprint-resist shrinkage in UV-NIL \cite{Schift10}. 
To examine the impact on blaze angle in an x-ray reflection grating from this phenomenon, several test replicas of a master grating were produced by \textsc{Philips} and cured at $T_{\text{cure}} \approx \SI{90}{\celsius}$ in an effort to induce a $\sim \SI{10}{\percent}$ shrinkage in the silica sol-gel network. 

\subsection{Master Grating}\label{sec:master_for_SCIL}
%%%%%%%%%%%%%%%%%%%%%%%%%%%%%%%%%%%%%%%%%--------------------------------------------------
The master grating chosen for this study was originally used as a direct stamp for grating fabrication by UV-NIL \cite[\emph{cf.\@} \cref{fig:master_grating,fig:uv_nil}]{Miles18}. 
As summarized in \cref{fig:master_fab}, this \SI{75}{\mm} by \SI{96}{\mm} (\SI{72}{\cm\squared}) grating was fabricated through a multi-step process centering on crystallographic etching in a $\langle 311 \rangle$-oriented, \SI{500}{\um}-thick, \SI{150}{\mm}-diameter silicon wafer. 
The groove layout was a variable-line-space profile defined by EBL with groove spacing, $d$, ranging nominally from \SIrange{160}{158.25}{\nm} along the groove direction, which is aligned with the $\langle 110 \rangle$ direction in the $\{ 311 \}$ plane of the wafer surface [\emph{cf.\@} \cref{sec:crystal_etching}]. 
This layout was then transferred by reactive ion etch into a thin film of low-stress \ce{Si_{x}N_{y}} before the native \ce{SiO2} on the exposed wafer surface was removed with a buffered oxide etch. 
Next, a timed, room-temperature \ce{KOH} etch was carried out to generate the asymmetric sawtooth defined by exposed $\{ 111 \}$ planes that form sharp points at the bottom of each groove with an angle $\theta \equiv \arccos \left( 1/3 \right) \approx 70.5^{\circ}$ [\emph{cf.\@} \cref{eq:Si_angle}]. 
Due to the $\langle 311 \rangle$ surface orientation of the silicon wafer, the exposed $\{ 111 \}$ planes define nominal facet angles of $\delta = 29.5^{\circ}$ and $\bar{\delta} = 180^{\circ} - \theta - \delta \approx 80^{\circ}$ [\emph{cf.\@} \cref{tab:off_axis_si}]. 
A cross-section image of the grating, with the \ce{Si_{x}N_{y}} mask removed by an \ce{HF} soak, is shown under \emph{field-emission scanning electron microscopy (FESEM)} in \cref{fig:master_grating_SEM}. 

Illustrated in \cref{fig:master_grating_profile}, the groove topography resulting from the process just described resembles a series of acute trapezoids with flat tops of width $w$ that each protrude a distance $\Delta h$ of a few \si{\nm} so that the groove depth, $h$, is given approximately by \cref{eq:groove_depth}. 
Although the depth of these sharp grating grooves could not be verified by \emph{atomic force microscopy (AFM)}\footnote{As in \cref{ch:master_fab}, all AFM for this chapter was carried out using a \textsc{Bruker Icon} instrument equipped with a \textsc{SCANASYST-AIR} tip and \textsc{PeakForce Tapping}$^{\text{TM}}$ mode at the Penn State MCL \cite{PSU_MRI_CL,Xu18}.} due to the aspect ratio of the scanning probe tip used for imaging, it is estimated that this quantity falls in the range $\SI{65}{\nm} \lessapprox h \lessapprox \SI{70}{\nm}$ with $w \gtrapprox \SI{30}{\nm}$. 
\begin{figure}
 \centering
 \includegraphics[scale=0.25]{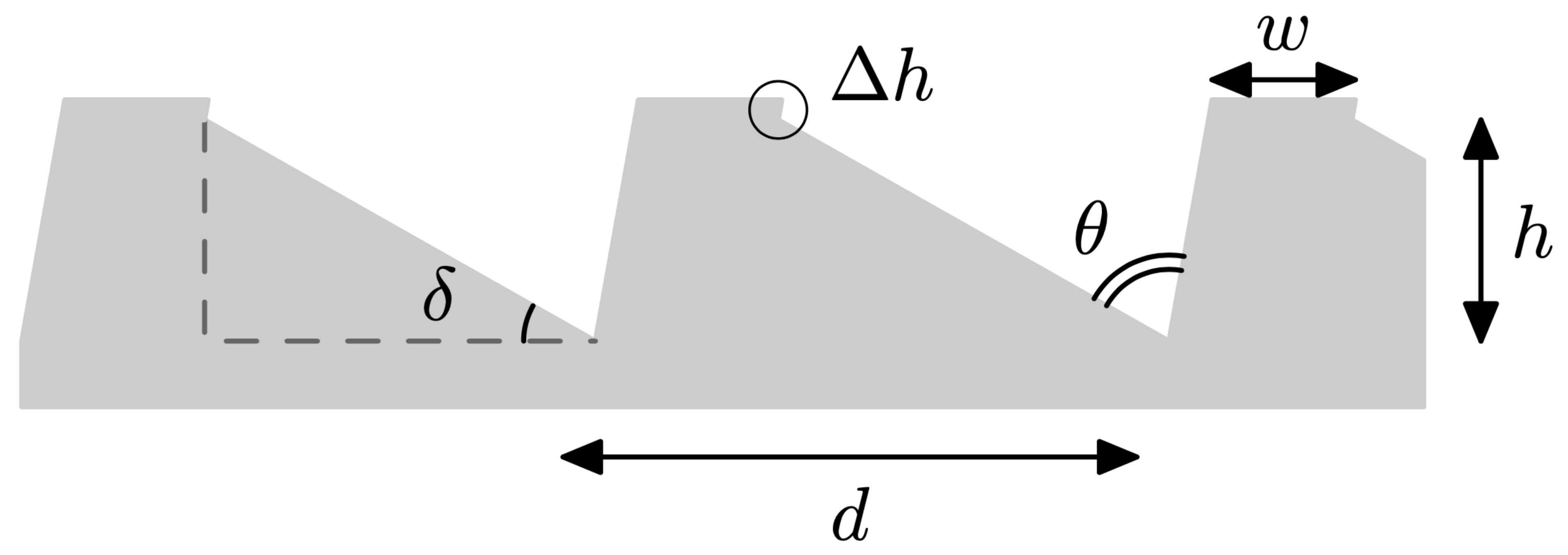}
 \caption[Illustration of the \ce{KOH}-etched silicon master surface profile]{Illustration of the \ce{KOH}-etched silicon master surface profile from \cref{fig:master_fab} with $\delta = 29.5^{\circ}$ as the nominal blaze angle and $\theta \approx 70.5^{\circ}$ defined by the intersection of exposed $\{ 111 \}$ planes. At a groove spacing of $d \lessapprox \SI{160}{\nm}$, the flat-top regions have widths $w \gtrapprox \SI{30}{\nm}$ as a result of the etch undercut while the groove depth is $\SI{65}{\nm} \lessapprox h \lessapprox \SI{70}{\nm}$ by \cref{eq:groove_depth}. Indicated by the circle, the indented portion of the etched topography cannot be described with a functional form for the diffraction-efficiency analysis in \cref{sec:discussion_scil} \cite{McCoy20b}.}\label{fig:master_grating_profile} 
 \end{figure}
Under AFM, facet surface roughness, $\sigma$ [\emph{cf.\@} \cref{sec:rough_surface}], measures $\lessapprox \SI{0.4}{\nm}$ RMS while the average of \num{30} blaze angle measurements over a \SI{0.5}{\um} by \SI{1}{\um} area yields $\delta = 30.0 \pm 0.8^{\circ}$, where the uncertainty is one standard deviation. 
Although these AFM data were gathered with vertical measurements calibrated to a \SI{180}{\nm} standard, this measurement for $\delta$ is limited in its accuracy due to a relatively poor lateral resolution on the order of a few \si{\nm}. %, at the Penn State Materials Characterization Laboratory,
The measurement is, however, consistent with the nominal value of $\delta = 29.5^\circ$ and is considered a reasonable estimation for the blaze angle of the silicon master, which is constrained through diffraction-efficiency testing in \cref{sec:discussion_scil}. 

\subsection{Stamp Construction}\label{sec:stamp_construction}
%%%%%%%%%%%%%%%%%%%%%%%%%%%%%%%%%%%%%%%%%--------------------------------------------------
Prior to construction of the composite stamp used for imprint production, the silicon master described in \cref{sec:master_for_SCIL} was cleaned in a heated bath of \textsc{Nano-Strip}$^{\text{TM}}$ (\textsc{VWR Int.}), which consists primarily of \emph{sulfuric acid} (\ce{H2SO4}), and then by an oxygen plasma treatment; a top-down FESEM of the cleaned grating is shown in \cref{fig:post_nanostrip}. 
\begin{figure}
 \centering
 \includegraphics[scale=0.38]{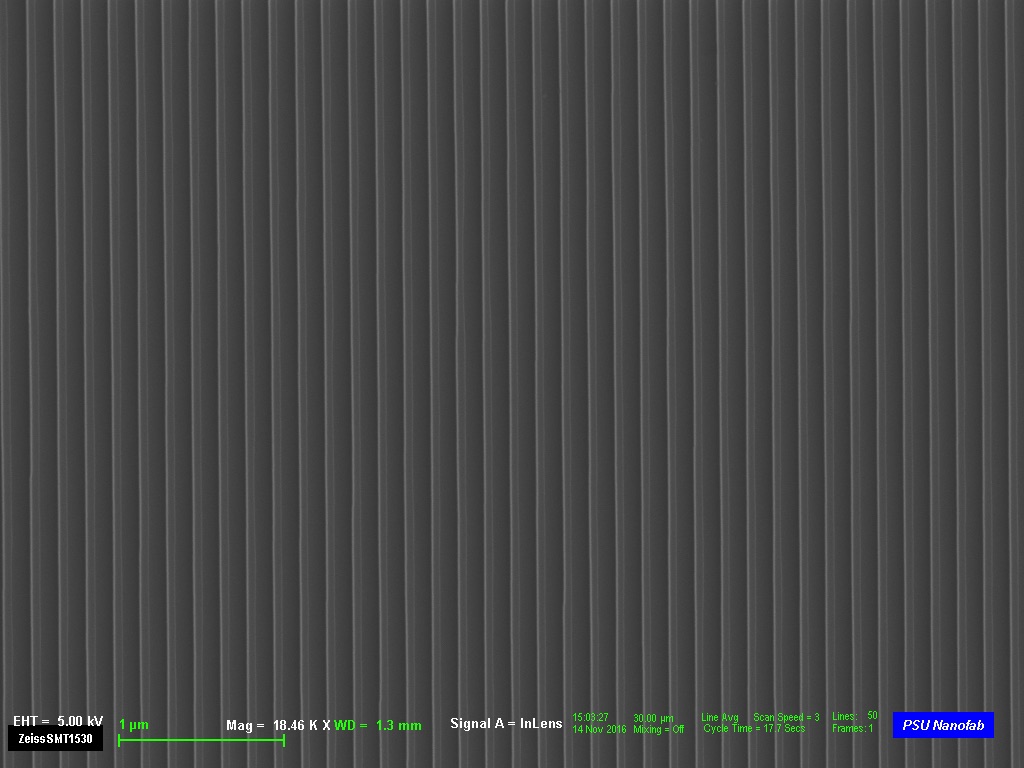}
 \caption[Top-down FESEM of the silicon master following wafer cleaning.]{Field-emission scanning electron micrograph (FESEM) of the silicon master viewed top-down, following wafer cleaning. Image was taken with a \textsc{Zeiss Leo 1530} instrument at the Penn State Nanofabrication Laboratory as in \cref{fig:master_grating_SEM}.}\label{fig:post_nanostrip}
 \end{figure}
The grating was then surface-treated for anti-stiction with a self-assembled monolayer of perfluorodecyltrichlorosilane (FDTS; \ce{C10H4Cl3F17Si}) \cite{Zhuang07}, which was achieved through a $\SI{50}{\celsius}$ molecular vapor deposition process similar to what is outlined in \cref{sec:nanoimprint}.\footnote{While this wafer cleaning was carried out at Penn State, the surface treatment was performed by \textsc{Philips}. Similar processes using FOTS (\ce{C8H4Cl3F13Si}) [\emph{cf.\@} \cref{fig:FOTS}] were performed by staff at the Cornell NanoScale Science and Technology Facility \cite{Cornell_nano} for SCIL process development carried out at the Quattrone Nanofabrication Facility \cite{Quattrone}.} 
As described by Verschuuren, et~al.~\cite{Verschuuren17} and illustrated in \cref{fig:scil_stamps}(a), a standard SCIL stamp consists of a $\SI{50}{\um}$-thick layer of modified PDMS that carries the inverse topography of the master template, and an underlaying, $\gtrapprox \SI{0.5}{\mm}$-thick layer of standard PDMS that attaches to a flexible, $\SI{200}{\um}$-thick glass sheet by application of an adhesion promoter. 
\begin{figure}
 \centering
 \includegraphics[scale=0.5]{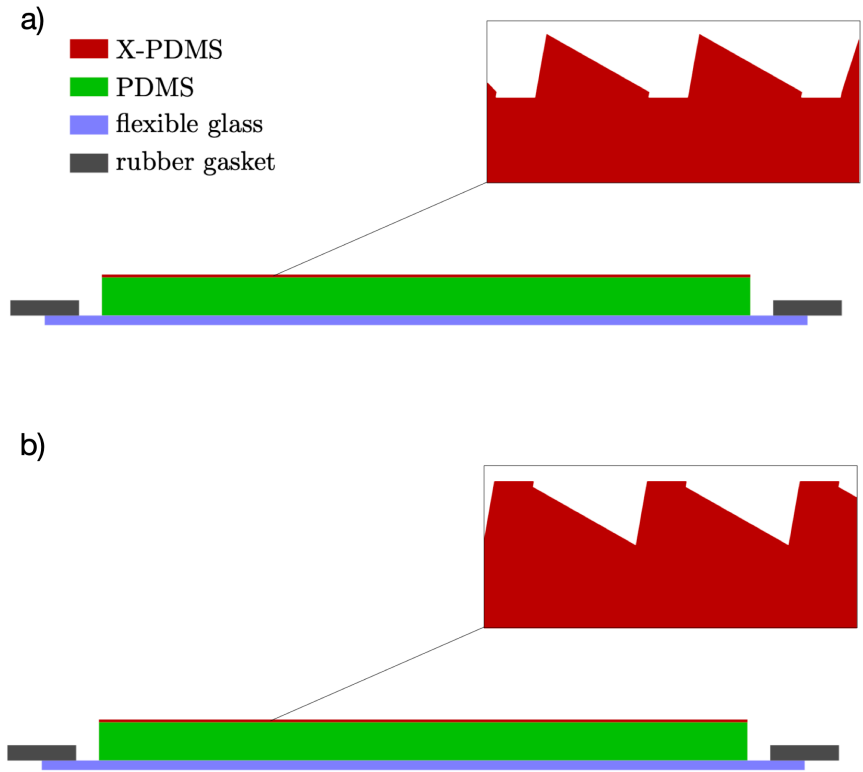}
 \caption[Schematic for SCIL composite stamps of two varieties]{Schematic for SCIL composite stamps of two varieties: a) an initial stamp featuring an inverted topography molded directly from the silicon master and b) a second stamp  featuring a topography similar to the master grating, which was molded using the first stamp as a master template. In either case, grating grooves are carried in a layer of X-PDMS tens of \si{\um} thick that sits on a \SI{200}{\mm}-diameter, flexible glass sheet buffered by a $\gtrapprox \SI{0.5}{\mm}$-thick layer of soft PDMS. A rubber gasket can be attached for use with the pneumatic-based SCIL wafer-scale imprint method. This illustration neglects slight rounding that can occur in sharp corners under the influence of surface tension in X-PDMS \cite{McCoy20b}.}\label{fig:scil_stamps}
 \end{figure}
A rubber gasket can then be glued to the outer perimeter of the square glass sheet for use with the pneumatic-based SCIL wafer-scale imprint method\footnote{This technique is possible with \textsc{AutoSCIL} (for high volume production) or with equipment provided by \textsc{S{\"U}SS MicroTec} that interfaces with a masker aligner (for low-volume production) \cite{philis_scil,suss_microtec}.} to produce imprints with topographies that resemble the silicon master. % shown in \cref{fig:master_grating_profile}.
However, in an effort to produce imprints that emulate the UV-NIL replica described by Miles, et~al.~\cite{Miles18}, which was fabricated using the silicon master as a direct stamp [\emph{cf.\@} \cref{fig:uv_nil_AFM}], this process was modified to construct a stamp with an inverted topography [\emph{cf.\@} \cref{fig:scil_stamps}(b)] so as to allow the production of imprints with sharp apexes and flat portions at the bottom of each groove \cite{Verschuuren18}. 

The variety of modified PDMS used for this study was X-PDMS (v.\ 3), a proprietary material available from \textsc{Philips} \cite{philis_scil}, which was dispensed over the surface of the surface-treated silicon master and then solidified through two rounds of spin-coating and baking steps using primary and accompanying components of the material (\emph{i.e.}, the X-PDMS that carries the groove shape and an additional \emph{intermediate layer}) \cite{Verschuuren17}. 
First, after the silicon master was cleaned again using deionized water (\ce{H2O}) and isopropyl alcohol (\ce{C3H8O}), $\sim \SI{3}{\gram}$ of the primary component was dispensed through a short spin-coat process at \num{2000} rotations per \si{\minute} using a low spin acceleration, leaving a layer tens of \si{\um} thick. 
This was followed immediately by a $\SI{50}{\celsius}$ hotplate bake for \SI{3}{\minute} and a room-temperature cool-down of \SI{10}{\minute}, leaving the material in a tacky state. 
Next, $\sim \SI{3}{\gram}$ of the accompanying component was spin-coated over this material in a similar way before the wafer was baked by hotplate at \SI{70}{\celsius} for \SI{10}{\minute} to form an intermediate layer also tens of \si{\um} thick. 
The doubly-coated silicon master was then oven-baked at $\SI{75}{\celsius}$ for \SI{20}{\hour}, forming a $\sim \SI{50}{\um}$-thick layer of cured X-PDMS with a Young's modulus on the order of several tens of \si{\mega\pascal}. 
In principle, this level of stiffness is sufficient for the stamp to carry $d \lessapprox \SI{160}{\nm}$ grating grooves without pattern distortion or feature collapse \cite{Verschuuren17,Verschuuren19}. 
\begin{figure}
 \centering
 \includegraphics[scale=0.15]{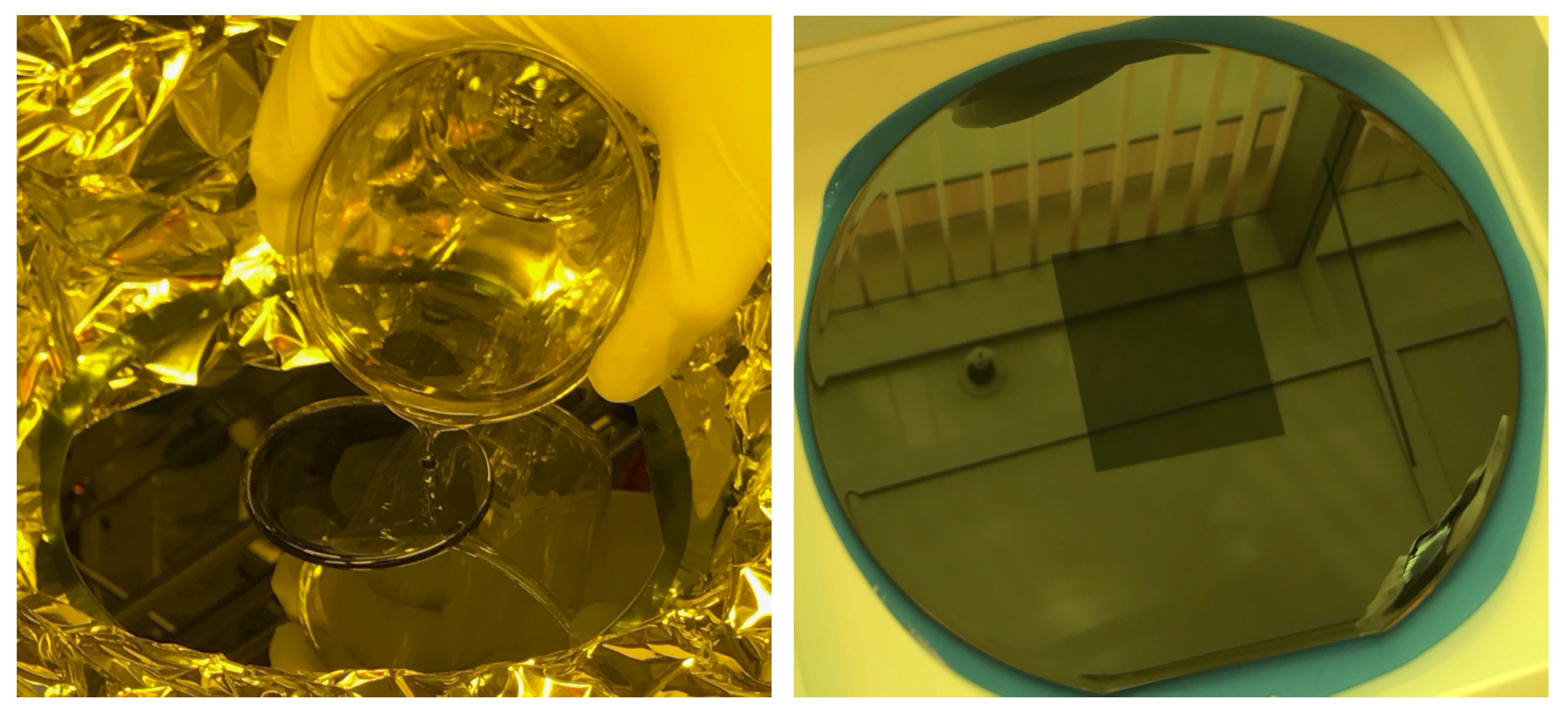}
 \caption[A \SI{150}{\mm}-diameter, \ce{KOH}-etched, $\langle 100 \rangle$ silicon grating spin-coated with H-PDMS at the Quattrone Nanofabrication Facility (QNF).]{A \SI{150}{\mm}-diameter silicon master spin-coated with H-PDMS at Quattrone Nanofabrication Facility (QNF). Due to the high viscosity of the material, a gradual spin acceleration must be used to achieve a uniform layer \cite{McCoy17}.}\label{fig:UPenn_spin}
 \end{figure}
A similar silicon master\footnote{The grating shown in \cref{fig:UPenn1} was fabricated by staff at the Penn State Nanofabrication Laboratory \cite{PSU_MRI_nanofab} in a similar fashion to the master grating described in \cref{sec:master_for_SCIL}: parallel lines and spaces with $d = \SI{160}{\nm}$ were patterned over a \SI{40}{\mm} by \SI{50}{\mm} (\SI{20}{\cm\squared}) area by EBL before the pattern was transferred into $\langle 100 \rangle$-oriented silicon through a \ce{KOH} etch to yield a symmetric, sawtooth-like topography with an effective blaze angle of $\delta = 54.74^{\circ}$ [\emph{cf.\@} \cref{fig:KOH_geo,tab:off_axis_si}].} coated with H-PDMS at the Quattrone Nanofabrication Facility (QNF) \cite{Quattrone} is shown in \cref{fig:UPenn_spin}. 

Using the SCIL \emph{Stamp Making Tool (SMT)} built by \textsc{Philips}, the initial, non-inverted stamp was formed by curing soft, \textsc{Sylgard 184} PDMS (\textsc{Dow, Inc.}) between the X-PDMS layer and a \SI{200}{\um}-thick sheet of \textsc{D 263} glass (\textsc{Schott AG}) cut into a circle with a \SI{200}{\mm} diameter. 
Consisting primarily of two opposite-facing vacuum chucks heated to $\SI{50}{\celsius}$ with surfaces flat to $\lessapprox \SI{10}{\um}$ peak-to-valley, this tool was used to spread $\sim \SI{12}{\gram}$ of degassed PDMS evenly over the X-PDMS layer. 
With the doubly-coated master secured to the bottom chuck, the \textsc{D 263} glass sheet secured to the top chuck was carefully brought into contact with the PDMS and then slowly clamped down using micrometer spindles to form a uniformly-thick layer while ensuring that the two chuck surfaces are parallel to within \SI{20}{\um}. 
These materials were baked in this configuration at \SI{50}{\celsius} until the PDMS was cured before the stamp was carefully separated from the silicon master. 
Using the initial, non-inverted stamp as a master template, the second, inverted stamp was then constructed on a square sheet of glass through steps identical to those outlined above. 
This processing was enabled by the first stamp being constructed on a round sheet of glass, which allowed it to be spin-coated with X-PDMS and subsequently cured like the silicon master. 

As an additional item for clarification, laboratory space for a similar SCIL stamp-making process carried out at QNF is pictured in the top panel of \cref{fig:UPenn1}. 
\begin{figure}
 \centering
 \includegraphics[scale=0.19]{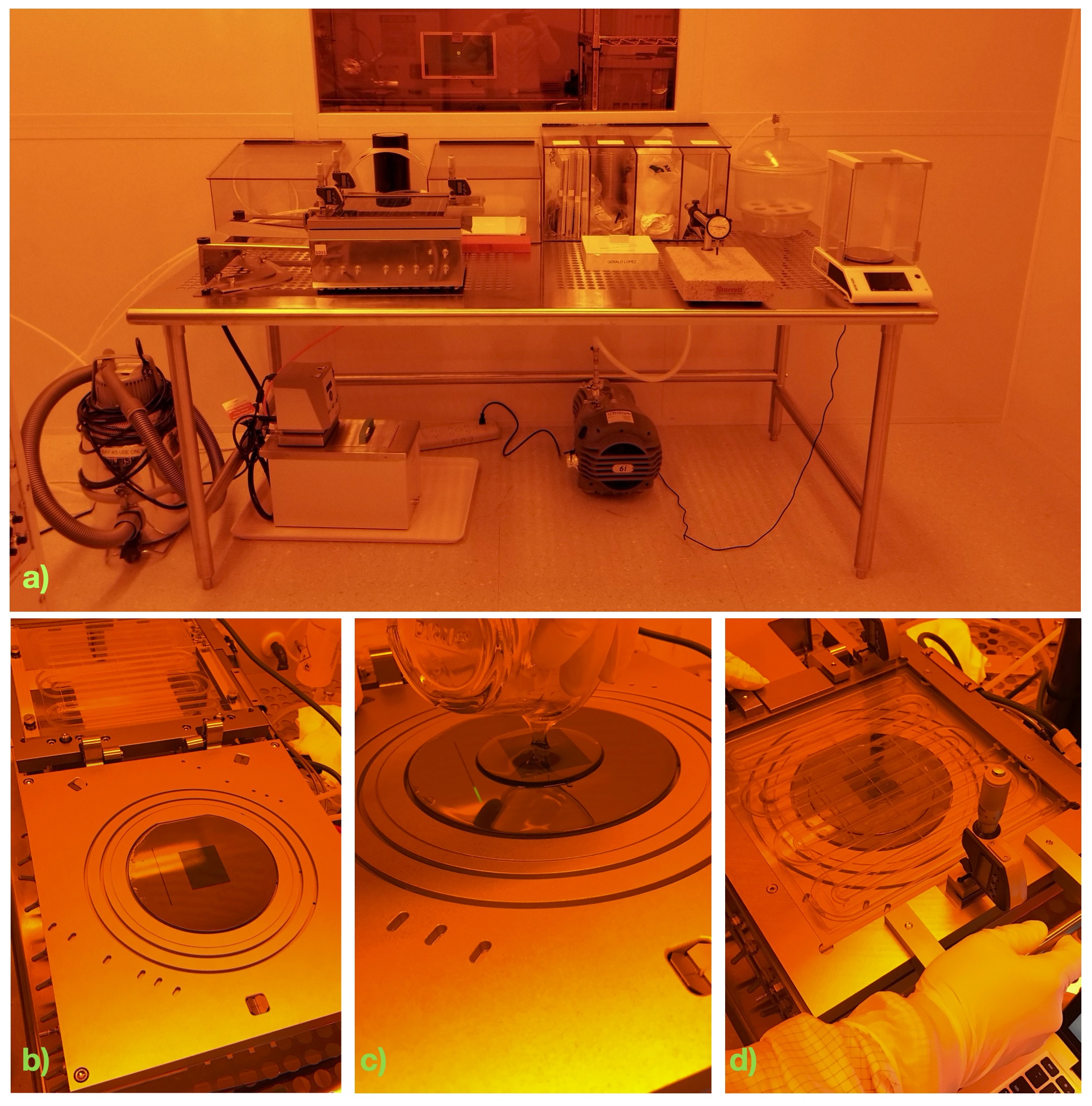}
 \caption[SCIL stamp construction at QNF]{SCIL stamp construction at QNF: a) laboratory space featuring an MRT (\textsc{S{\"U}SS MicroTec} \cite{suss_microtec}) and supporting equipment for surface heating and PDMS preparation, b) heated surfaces on the MRT for the silicon master and a flexible sheet of glass on the inside of the MRT lid, c) PDMS is poured over the surface of the silicon master, d) the MRT lid is closed and the PDMS is cured at \SI{50}{\celsius} \cite{McCoy17}.}\label{fig:UPenn1}
 \end{figure}
Making use of the \emph{Master Replication Tooling (MRT)} built by \textsc{S{\"U}SS MicroTec} \cite{suss_microtec}, which provides the functionality of the SMT, the bottom panels of \cref{fig:UPenn1} show steps involved with curing \textsc{Sylgard 184} PDMS in between the master grating template spin-coated with H-PDMS from \cref{fig:UPenn_spin}. 
First, the coated master is secured on a surface heated to \SI{50}{\celsius} with a vacuum chuck provided by the MRT; a clean sheet of \textsc{AF32} glass (\textsc{Schott AG}) is also secured to a heated vacuum chuck on the inside of the MRT lid.  
\textsc{Sylgard 184} PDMS is then carefully poured over the surface of the master before the top chuck is carefully brought in contact before micrometer spindles are used to spread the material evenly over the wafer surface. 
After the materials have been cured, the MRT is opened and the SCIL stamp is carefully separated from the silicon master. 

\subsection{Imprint Production}\label{sec:imprint_production}
%%%%%%%%%%%%%%%%%%%%%%%%%%%%%%%%%%%%%%%%%--------------------------------------------------
Performed by \textsc{Philips}, several blazed-grating surface reliefs were imprinted by hand into $\sim \SI{100}{\nm}$-thick films of \textsc{NanoGlass T1100} sol-gel resist spin-coated on \SI{1}{\mm}-thick, \SI{150}{\mm}-diameter silicon wafers using the inverted X-PDMS stamp described in \cref{sec:stamp_construction}.  
\begin{figure}
 \centering
 \includegraphics[scale=0.5]{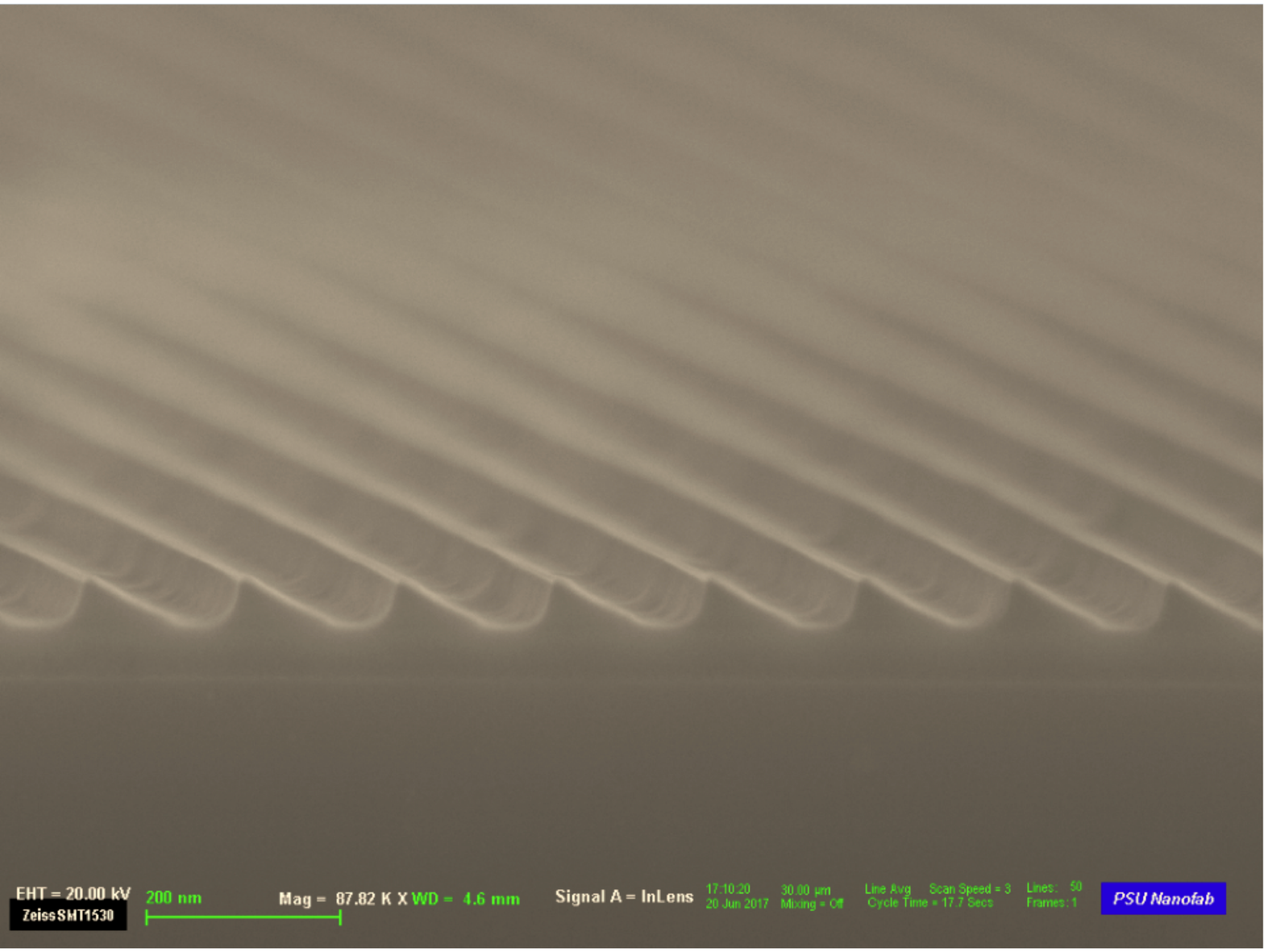}
 \caption[Cross-section FESEM of a grating imprint in \textsc{NanoGlass T1100}]{Cross-section FESEM of a grating imprint with $d \lessapprox \SI{160}{\nm}$ in $\sim \SI{100}{\nm}$-thick \textsc{NanoGlass T1100} sol-gel resist coated on a silicon wafer. This topography was produced from an inverted composite stamp [\emph{cf.\@} \cref{fig:scil_stamps}(b)]. Image was taken with a \textsc{Zeiss Leo 1530} instrument as in \cref{fig:master_grating_SEM,fig:post_nanostrip} \cite{McCoy17,McCoy20b}.}\label{fig:SCIL_SEMs} %and then baked to $\SI{90}{\celsius}$ following stamp separation
\end{figure} 
Although the pneumatic-based SCIL wafer-scale imprint method (\emph{e.g.}, via \textsc{AutoSCIL}) is best equipped for minimizing pattern distortion over \SI{150}{\mm} wafers, imprinting by hand is sufficient for producing a small number of grating molds suitable for the diffraction-efficiency testing described in \cref{sec:beamline_testing}, which depends primarily on groove facet shape over a local area defined by the projected size of the monochromatic beam at the ALS [\emph{cf.\@} \cref{sec:als_beam}]. 
\begin{figure}
 \centering
 \includegraphics[scale=0.52]{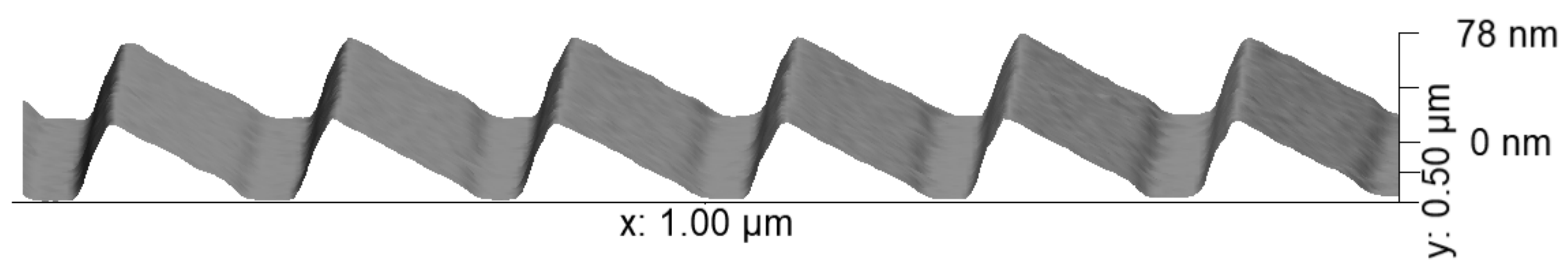}
 \caption[AFM of a grating imprint in \textsc{NanoGlass T1100}]{AFM of a grating imprint with $d \lessapprox \SI{160}{\nm}$ in $\sim \SI{100}{\nm}$-thick \textsc{NanoGlass T1100} sol-gel resist coated on a silicon wafer as in \cref{fig:SCIL_SEMs}. Facet roughness and blaze angle under AFM measure $\sigma \approx \SI{0.6}{\nm}$ RMS and $27.9 \pm 0.7^{\circ}$, respectively \cite{McCoy17,McCoy20b}.}\label{fig:SCIL_AFMs} 
 \end{figure}
With imprinting taking place at room temperature, \SI{15}{\minute} of stamp-resist contact was required for the sol-gel process to carry out. 
Each wafer following stamp separation was baked by hotplate to $\SI{90}{\celsius}$ for \SI{15}{\minute} to densify the imprinted material to a small degree, thereby inducing resist shrinkage. 
An imprinted replica produced in this way is shown under FESEM in \cref{fig:SCIL_SEMs}, where grating grooves are seen imprinted over a residual layer of resist several tens of \si{\nm} thick. 
An identical imprint is shown under AFM in \cref{fig:SCIL_AFMs}, where the average blaze angle measures $\delta' = 27.9 \pm 0.7^{\circ}$, giving a $\delta' / \delta \approx 0.93$ or $\delta - \delta' \approx 2^{\circ}$ reduction in blaze angle relative to $\delta = 30.0 \pm 0.8^{\circ}$ measured for the silicon master \cite[\emph{cf.\@} \cref{sec:master_for_SCIL}]{McCoy20b}. 

\section{Beamline Experiments}\label{sec:beamline_testing}
%%%%%%%%%%%%%%%%%%%%%%%%%%%%%%%%%%%%%%%%%-------------------------------------------------- 
This section describes how the reduced blaze angle of the SCIL replica described in \cref{sec:grating_fab} is verified through beamline testing by comparing its diffraction-efficiency response in the soft x-ray to that of the silicon master. 
These experiments for measuring \emph{absolute diffraction efficiency}, $\mathscr{E}_n$, took place at beamline 6.3.2 of the ALS \cite{ALS_632,Underwood96,Gullikson01} using methodology similar to \cref{sec:taste_eff_results} for TASTE prototype testing.  %similar to \cref{sec:taste_eff_results}.
Discussed in \cref{sec:als_testing}, this facility provides a monochromatic beam of EUV radiation and soft x-rays with photon energy, $\mathcal{E}_{\gamma} = h c_0 / \lambda$, up to \SI{1300}{\electronvolt}. 
Radiation with $\mathcal{E}_{\gamma} < \SI{440}{\electronvolt}$, however, requires the use of a triple-mirror order sorter to ensure a spectrally pure beam [\emph{cf.\@} \cref{sec:als_beam}]. 
%Although $\alpha$ and $\gamma$ are controlled 
With a specified grating geometry being established through the movement of stage rotations at the beamline [\emph{cf.\@} \cref{sec:geo_constrain}], the implementation of this order sorter can cause a slight beam shift that serves to perturb a precisely-set grating geometry; because of this, $\mathcal{E}_{\gamma} = \SI{440}{\electronvolt}$ was chosen as the lower limit for diffraction-efficiency testing. 

The silicon master was tested without a reflective overcoat to maintain its sharply defined surface-relief profile whereas the SCIL replica was coated with a thin layer of gold, using chromium as an adhesion layer, in an effort to emulate the UV-NIL replica described by Miles, et~al.~\cite[\emph{cf.\@} \cref{sec:nanoimprint}]{Miles18}. 
\cref{sec:reflectivity} describes how the upper photon-energy limit for testing was determined from considering Fresnel reflectivity for silicon and gold surfaces at a grazing-incidence angle $\zeta \approx 1.7^{\circ}$ [\emph{cf.\@} \cref{eq:angle_on_groove,eq:graze_angle}]: 
\begin{equation}
 \sin \left( \zeta \right) = \sin \left( \eta \right) \frac{\cos \left( \alpha - \delta \right)}{\cos \left( \alpha \right)} \implies \sin \left( \zeta \right) \approx \sin \left( \eta = 1.5^{\circ} \right) \sec \left( \delta = 29.5^{\circ} \right) ,
 \end{equation}
where $\eta = 1.5^{\circ}$ is the graze angle relative to the optic mount surface used by Miles, et~al.~\cite{Miles18} and $\alpha \approx \delta$ is the azimuthal incidence angle in a near-Littrow configuration with $\delta = 29.5^{\circ}$ as the nominal blaze angle of the silicon master described in \cref{sec:grating_fab}.\footnote{These angles are defined geometrically in \cref{fig:conical_reflection_edit,fig:grating_angles}.}
Following this, \cref{sec:grat_geo_scil,sec:test_results_scil}, describe how test geometries were constrained and present the gathered diffraction-efficiency data, respectively.  

\subsection{Reflectivity Considerations}\label{sec:reflectivity}
%%%%%%%%%%%%%%%%%%%%%%%%%%%%%%%%%%%%%%%%%-------------------------------------------------- 
To model the surface the silicon master, a thin layer of native \ce{SiO2} is taken into account so that an expression for specular reflectivity is determined from phase-matching the electromagnetic fields at a boundary between vacuum and \ce{SiO2}, as well as an accompanying boundary between \ce{SiO2} and the silicon substrate. 
Although the monochromatic beam at the ALS is \emph{s-polarized} to a high degree, Fresnel reflectivity in the soft x-ray is virtually polarization-independent at grazing-incidence angles [\emph{cf.\@} \cref{sec:als_beam,sec:reflectivity_polarization}]. 
Taking $\tilde{\nu}_1$ as the complex index of refraction for \ce{SiO2}, the complex reflection coefficient for the first interface in s-polarization, which is approximately equal to the corresponding term in \emph{p-polarization}, is [\emph{cf.\@} \cref{eq:refl_coeff01}]
\begin{subequations}
\begin{equation}
 \tilde{r}_{0,1} = \frac{k_{\perp,0} - \tilde{k}_{\perp,1}}{k_{\perp,0} + \tilde{k}_{\perp,1}} = \frac{\sin \left( \zeta \right) - \sqrt{\tilde{\nu}_1^2 - \cos^2 \left( \zeta \right)}}{\sin \left( \zeta \right) + \sqrt{\tilde{\nu}_1^2 - \cos^2 \left( \zeta \right)}} ,
 \end{equation}
where $k_{\perp,0} = -k_0 \sin \left( \zeta \right)$ and $\tilde{k}_{\perp,1} = -k_0 \sqrt{\tilde{\nu}_1^2 - \cos^2 \left( \zeta \right)}$ are the components of the wave vector normal to the boundary in vacuum and in \ce{SiO2}, respectively, with $k_0 \equiv 2 \pi / \lambda$ as the wave number in vacuum. 
Similarly, with $\tilde{\nu}_2$ as the refractive index for silicon, the reflection coefficient for the second interface in the same polarization is [\emph{cf.\@} \cref{eq:refl_coeff12}] 
\begin{equation}
 \tilde{r}_{1,2} = \frac{\tilde{k}_{\perp,1} - \tilde{k}_{\perp,2}}{\tilde{k}_{\perp,1} + \tilde{k}_{\perp,2}} = \frac{ \sqrt{\tilde{\nu}_1^2 - \cos^2 \left( \zeta \right)} - \sqrt{\tilde{\nu}_2^2 - \cos^2 \left( \zeta \right)} }{ \sqrt{\tilde{\nu}_1^2 - \cos^2 \left( \zeta \right)} + \sqrt{\tilde{\nu}_2^2 - \cos^2 \left( \zeta \right)}} , 
 \end{equation}
\end{subequations}
where $\tilde{k}_{\perp,2} = -k_0 \sqrt{\tilde{\nu}_2^2 - \cos^2 \left( \zeta \right)}$ is the component of the wave vector normal to the surface in silicon. 
Using $\tilde{r}_{0,1}$ and $\tilde{r}_{1,2}$, the overall specular reflectivity for a silicon substrate with a native-oxide layer of thickness $\tau$ can be written as the norm-squared of \cref{eq:refl_coeff_example} for a thin film:
\begin{equation}\label{eq:Si_layer_refl}
 \mathcal{R}_{\text{film}} = \norm{ \frac{\tilde{r}_{0,1} + \tilde{r}_{1,2} \mathrm{e}^{-2i \tilde{k}_{\perp,1} \tau}}{1 + \tilde{r}_{0,1} \tilde{r}_{1,2} \mathrm{e}^{-2i \tilde{k}_{\perp,1} \tau}} }^2 .
 \end{equation}
%with $i \equiv \sqrt{-1}$ as the imaginary unit. 

The absorbing effect of surface roughness in the limit of very small lateral size scales can be taken into account by modulating $\tilde{r}_{0,1}$ and $\tilde{r}_{1,2}$ by the following \emph{Nevot-Croce factors} [\emph{cf.\@} \cref{sec:refl_account}]:
\begin{equation}\label{eq:NC_factors}
 \tilde{r}_{0,1} \to \tilde{r}_{0,1} \mathrm{e}^{-2 k_{\perp,0} \tilde{k}_{\perp,1} \sigma_1^2} \quad \text{and} \quad \tilde{r}_{1,2} \to \tilde{r}_{1,2} \mathrm{e}^{-2 \tilde{k}_{\perp,1} \tilde{k}_{\perp,2} \sigma_2^2} ,
 \end{equation}
where $\sigma_1$ and $\sigma_2$ are the RMS roughness values for the surface of the \ce{SiO2} film and the underlaying boundary between \ce{SiO2} and the silicon substrate, respectively. 
 \begin{figure}
 \centering
 \includegraphics[scale=0.6]{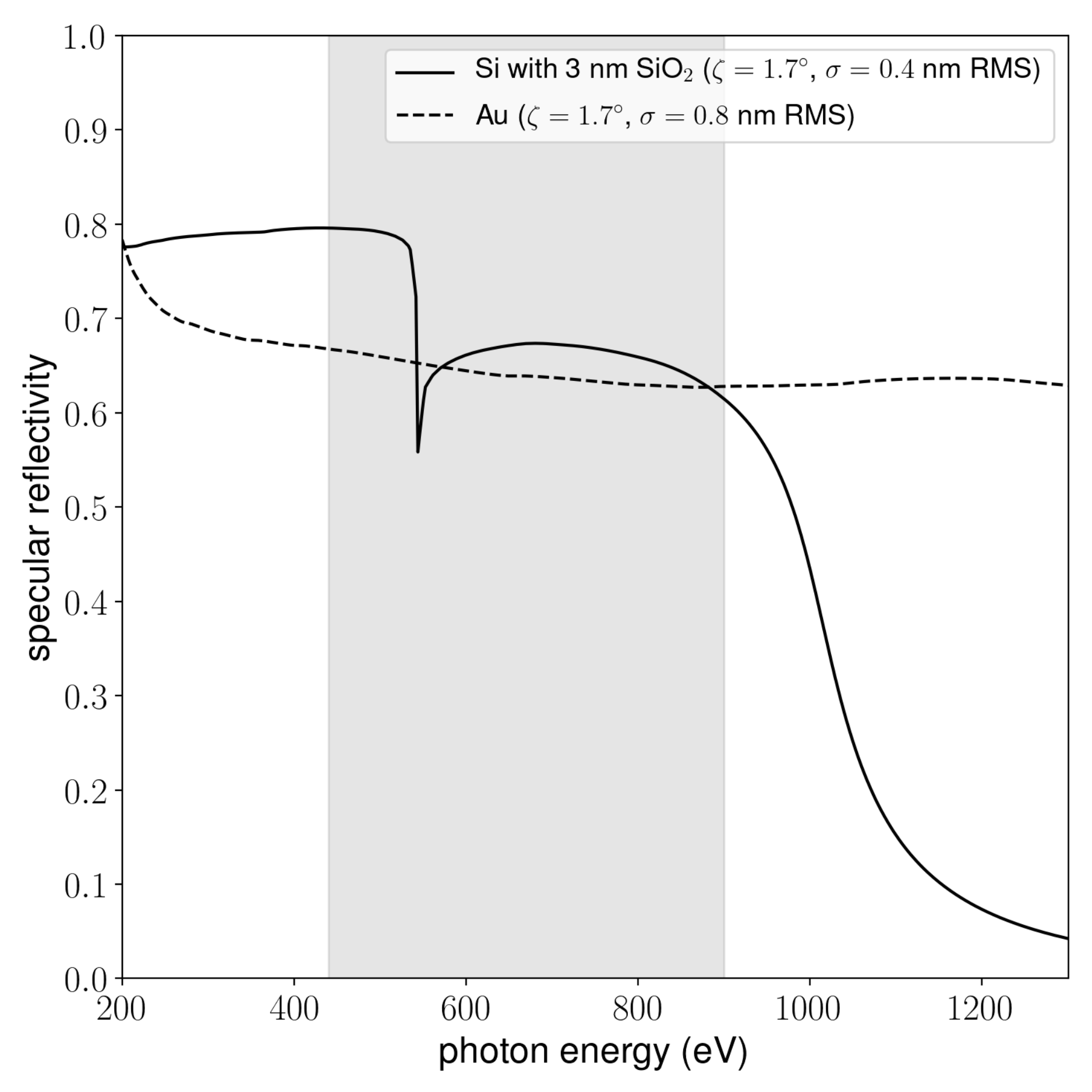}
 \caption[Specular reflectivity curves for a silicon substrate and a thick slab of gold at a $1.7^{\circ}$ graze angle]{Specular reflectivity curves for a silicon substrate with \SI{3}{\nm} of native oxide (solid line) and a thick slab of gold (dashed line) for a $1.7^{\circ}$ graze angle, which are indicative of the overall responses expected for the silicon master and the coated SCIL replica, respectively. Because the use of the order sorter at the beamline (required below \SI{440}{\electronvolt}) causes a slight beam shift relative to the grating grooves while the reflectivity of silicon drops above \SI{900}{\electronvolt}, diffraction-efficiency testing was restricted to the gray-shaded region (data obtained from CXRO \cite{CXRO_database}).}\label{fig:reflectivity} 
 \end{figure}
Using data for $\tilde{\nu}_1$ and $\tilde{\nu}_2$ obtained from the CXRO online database \cite{CXRO_database}, assuming standard material densities, and taking $\sigma_1 = \sigma_2 = \SI{0.4}{\nm}$ RMS, which is estimated from AFM measurements, $\mathcal{R}_{\text{film}}$ [\emph{cf.\@} \cref{eq:Si_layer_refl}] for $\tau = \SI{3}{\nm}$  at a $\zeta = 1.7^{\circ}$ graze angle is plotted as a function of $\mathcal{E}_{\gamma}$ in \cref{fig:reflectivity}. 
In addition to a prominent oxygen absorption edge near \SI{550}{\electronvolt}, it is seen that $\mathcal{R}_{\text{film}}$ decreases substantially for $\mathcal{E}_{\gamma} \gtrapprox \SI{900}{\electronvolt}$ as $\zeta \approx 1.7^{\circ}$ approaches the the critical angle for \emph{total external reflection} on silicon [\emph{cf.\@} \cref{sec:planar_interface}]. 
Because of this decrease in $\mathcal{R}_{\text{film}}$ the spectral range for diffraction-efficiency testing was further restricted to $\SI{440}{\electronvolt} \leq \mathcal{E}_{\gamma} \leq \SI{900}{\electronvolt}$, or equivalently, $\SI{2.82}{\nm} \gtrapprox \lambda \gtrapprox \SI{1.38}{\nm}$. 

In principle, \textsc{NanoGlass} sol-gel resist has an index of refraction, $\tilde{\nu}$, similar to that of \ce{SiO2} in the soft x-ray [\emph{cf.\@} \cref{fig:X-ray_index}] but with deviations arising from a lower density that results from nanoscale porosity in the material network caused by the MTMS sol-gel precursor, which has silicon-methyl bonds that do not participate in the sol-gel reaction \cite[\emph{cf.\@} \cref{sec:grating_fab}]{Verschuuren10}. 
Although these methyl groups present in the resist also should cause a carbon absorption edge near \SI{280}{\electronvolt} [\emph{cf.\@} \cref{fig:atomic_scattering_data}], the inverted SCIL replica was coated with a thin layer of gold, which was chosen for this study due to its broadband reflectivity across the $\SI{2.82}{\nm} \gtrapprox \lambda \gtrapprox \SI{1.38}{\nm}$ spectral range considered. %, with tabulated data for $\tilde{\nu}$. 
Because it is comprised primarily of an \ce{SiO2} network, this resist requires its surface to be primed with an oxidizing metal so as to promote wetting and adhesion for gold and other non-oxidizing metals [\emph{cf.\@} \cref{sec:prototype_coating}]. 
The distance normal to a gold surface at which radiation loses $1/\mathrm{e}$ of its original intensity is given by the \emph{penetration depth}, $\mathcal{D}_{\perp}$ [\emph{cf.\@} \cref{eq:ch3_pene_dep}]. 
Tabulated data from CXRO \cite{CXRO_database} indicate that $\mathcal{D}_{\perp} \lessapprox \SI{2}{\nm}$ at a graze angle of $\zeta \approx 1.7^{\circ}$ for $\SI{2.82}{\nm} \gtrapprox \lambda \gtrapprox \SI{1.38}{\nm}$ and as a result, the gold layer need only be \SI{15}{\nm} thick to reduce incident radiation at the resist interface by $\sim \SI{99.9}{\percent}$. 
This justifies the treatment of this layer as a thick slab, with specular reflectivity in the Nevot-Croce regime for surface roughness [\emph{cf.\@} \cref{sec:refl_account,eq:nc_factor_taste}] given by the norm squared of $\tilde{r} \, \mathrm{e}^{-2 k_{\perp} \tilde{k}_{\perp} \sigma^2}$ for s-polarization: %,sec:rough_surface
\begin{equation}\label{eq:Au_refl}
 \mathcal{R}_{\text{slab}} = \norm{ \frac{k_{\perp} - \tilde{k}_{\perp}}{k_{\perp} + \tilde{k}_{\perp}} }^2 \mathrm{e}^{-4 \sigma^2 k_{\perp} \, \text{Re} \left[ \tilde{k}_{\perp} \right]} ,
 \end{equation}
where $\text{Re} \left[ \tilde{k}_{\perp} \right]$ is the real component of $\tilde{k}_{\perp} = - k_0 \sqrt{\tilde{\nu}^2 - \cos^2 \left( \zeta \right)}$, the component normal to the surface of the wave vector in gold. 

Shown under AFM in \cref{fig:SCILcoat_AFM}, the gold film was sputter-coated on the replica in an identical fashion to Miles, et~al.~\cite{Miles18} using a \textsc{Kurt J.\ Lesker CMS-18} tool at the Penn State Nanofabrication Laboratory \cite{PSU_MRI_nanofab}: \SI{5}{\nm} of chromium was deposited for adhesion followed immediately by \SI{15}{\nm} of gold, without breaking vacuum.  
\begin{figure}
 \centering
 \includegraphics[scale=0.52]{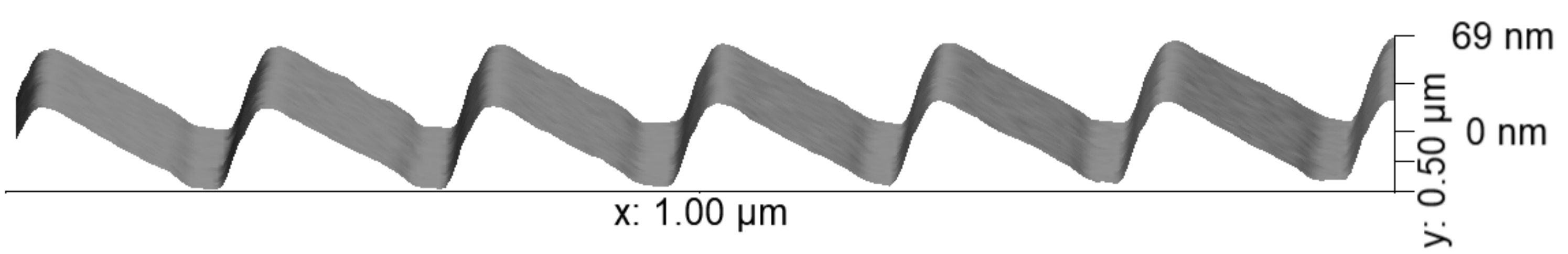}
 \caption[AFM of a SCIL imprint sputter-coated with \SI{15}{\nm} of gold, using \SI{5}{\nm} of chromium as an adhesion layer.]{AFM of the SCIL imprint shown in \cref{fig:SCIL_AFMs}, sputter-coated with \SI{15}{\nm} of gold, using \SI{5}{\nm} of chromium as an adhesion layer. Following this deposition, facet roughness and blaze angle under AFM measure $\sigma \approx \SI{0.8}{\nm}$ RMS and $28.4 \pm 0.8^{\circ}$, respectively \cite{McCoy17,McCoy20b}.}\label{fig:SCILcoat_AFM} 
 \end{figure}
Using $\zeta \approx 1.7^{\circ}$, a measured facet roughness of $\sigma \approx \SI{0.8}{\nm}$ RMS\footnote{This offers improvement over what is observed in the corresponding UV-NIL imprint, where facet roughness measures $\sigma \approx \SI{1.4}{\nm}$ RMS under AFM [\emph{cf.\@} \cref{fig:uv_nil_AFM}].} and data for $\tilde{\nu}$ obtained from CXRO, $\mathcal{R}_{\text{slab}}$ is also plotted as a function of $\mathcal{E}_{\gamma}$ in \cref{fig:reflectivity}, where it is seen that $\sim \SI{65}{\percent}$ reflectivity is maintained across the test bandpass of \SIrange{400}{900}{\electronvolt}. 
Following the deposition of this coating, the blaze angle under AFM measures $\delta ' = 28.4 \pm 0.8^{\circ}$ from \num{30} individual measurements, which is consistent with the pre-coating measurement of $\delta ' = 27.9 \pm 0.7^{\circ}$. 
The statistical consistency between these two measurements suggests that coating effects had a minimal impact on the blaze angle and that $\delta' / \delta$ constrained from diffraction-efficiency testing results is expected to be indicative of resist shrinkage alone. 

\subsection{Constraining Grating Geometry}\label{sec:grat_geo_scil}
%%%%%%%%%%%%%%%%%%%%%%%%%%%%%%%%%%%%%%%%%--------------------------------------------------
Following the test procedure outlined in \cref{sec:geo_constrain}, near-Littrow configurations for both the silicon master and the coated SCIL replica were established at the beamline using the principal-axis rotations illustrated in \cref{fig:grating_angles}. 
These two gratings were installed one at a time on the optic mount [\emph{cf.\@} \cref{fig:beamline_photoB}] with $\eta \approx 0$ and $\phi \approx 0$ verified by the tilt of the optic mount using a spirit level. 
The groove direction was in each case approximately aligned with the beam direction with $\varphi \approx 0$ so that the $(x',y',z')$ chamber coordinate system [\emph{cf.\@} \cref{fig:beamline_photoA}] is roughly coincident with the $(x,y,z)$ grating coordinate system [\emph{cf.\@} \cref{fig:grating_angles}]. 
In this configuration, the tested grating was carefully adjusted to occult the beam so as to position the point of interception close to the hub of rotation axes. 
With the $\lessapprox \SI{0.5}{\mm}$ cross-sectional diameter of the beam projecting to tens of \si{\mm} at grazing incidence and the point of incidence being the central grooved region of each grating, the effective groove spacing for both gratings was taken as $d = \SI{159.125}{\nm}$, which is the nominal average of the silicon master variable-line-space profile [\emph{cf.\@} \cref{sec:master_for_SCIL}]. 

The detector used for this test campaign was a gallium-arsenide-phosphide Schottky photodiode with a square, \SI{4.6}{\mm} by \SI{4.6}{\mm} collecting area (\textsc{Hamamatsu Photonics}). 
For a grating installed in such an extreme off-plane mount, the linear stage movement of the detector is roughly parallel with the grating-dispersion direction so that each propagating order is confined along an arc with a radius $r = L \sin \left( \gamma \right)$ while being dispersed from $0^{\text{th}}$ order over a horizontal distance given by \cref{eq:disp_approx}, which neglects focal plane corrections on the order of tens of \si{\um} for a fixed value of $L \approx \SI{235}{\mm}$ [\emph{cf.\@} \cref{sec:geo_constrain}]. 
Because the horizontal spacing between each order, $\lambda L / d$, is smaller than the length of this detecting area, $\ell_{\text{det}} = \SI{4.6}{\mm}$, for $d \lessapprox \SI{160}{\nm}$ and $\lambda$ characteristic of soft x-rays, a vertical, \SI{0.5}{\mm}-wide slit was installed to mask this detector so as to allow the intensity of each order, $\mathcal{I}_n \! \left( \lambda \right)$, to be measured in isolation with horizontal stage motion [\emph{cf.\@} \cref{sec:geo_constrain}].  
This is shown in \cref{fig:beamline_photoB}, which depicts a grating installed on the optic mount and the masked detector in the background, with the slit aligned with the goniometric $y'$-direction. 
Because $\ell_{\text{det}} = \SI{4.6}{\mm}$ is also smaller than the expected diffracted-arc radius of $r \approx \SI{7}{\mm}$ for $\gamma \approx 1.7^{\circ}$, however, it is not possible to measure all propagating orders consistently using purely horizontal detector stage motion. 
While measuring diffraction efficiency in this way prevents total diffraction efficiency, $\mathscr{E}_{\text{tot}}$, defined as $\sum_n \mathscr{E}_n$ for $n \neq 0$ [\emph{cf.\@} \cref{sec:measure_efficiency}], from being determined experimentally, it does allow for peak orders to be constrained and compared to theoretical models. 

Described at the start of \cref{sec:beamline_testing}, the nominal value for $\eta$ in a Littrow configuration consistent with $\gamma \approx 1.7^{\circ}$ and $\alpha = \delta = 29.5^{\circ}$ is $\eta \approx 1.5^{\circ}$ by \cref{eq:graze_angle}. 
Thus, with the leveled grating intercepting the beam, the optic mount was rotated about the $x$-axis at the point of incidence to ensure that the angle between the direct beam and the reflected beam was roughly $2 \eta \approx 3^{\circ}$ as measured by the detector as it scans goniometrically in the vertical direction at the location of $0^{\text{th}}$ order. 
Starting with this basic configuration for each tested grating, small adjustments to $\eta$ and $\varphi$ were made while the arc of diffraction was analyzed in-situ after being mapped as a function of $x$ and $y$ to confirm a near-Littrow configuration. 
Unlike the $x$-positions of propagating orders given by \cref{eq:disp_approx}, which can be measured directly by scanning the detector linearly along the horizontal direction and then calculating weighted-mean centroids [\emph{cf.\@} \cref{fig:ALS_xdata}], their corresponding $y$-positions require precise knowledge of the system throw, $L \approx \SI{235}{\mm}$, to map the goniometric angle associated with vertical stage motion, $\Theta_n$, to a linear $y$-coordinate using $y_n = L \sin \left( \Theta_n \right)$ [\emph{cf.\@} \cref{sec:geo_constrain}]. 

The quantity $L$ was experimentally determined separately for each installed grating by comparing $\ell_{\text{det}}$ to the angular size of the detector as measured by a goniometric scan of the beam at the location of $0^{\text{th}}$ order [\emph{cf.\@} \cref{fig:ALS_cent_throw}]. 
\begin{figure}
 \centering
 \includegraphics[scale=0.31]{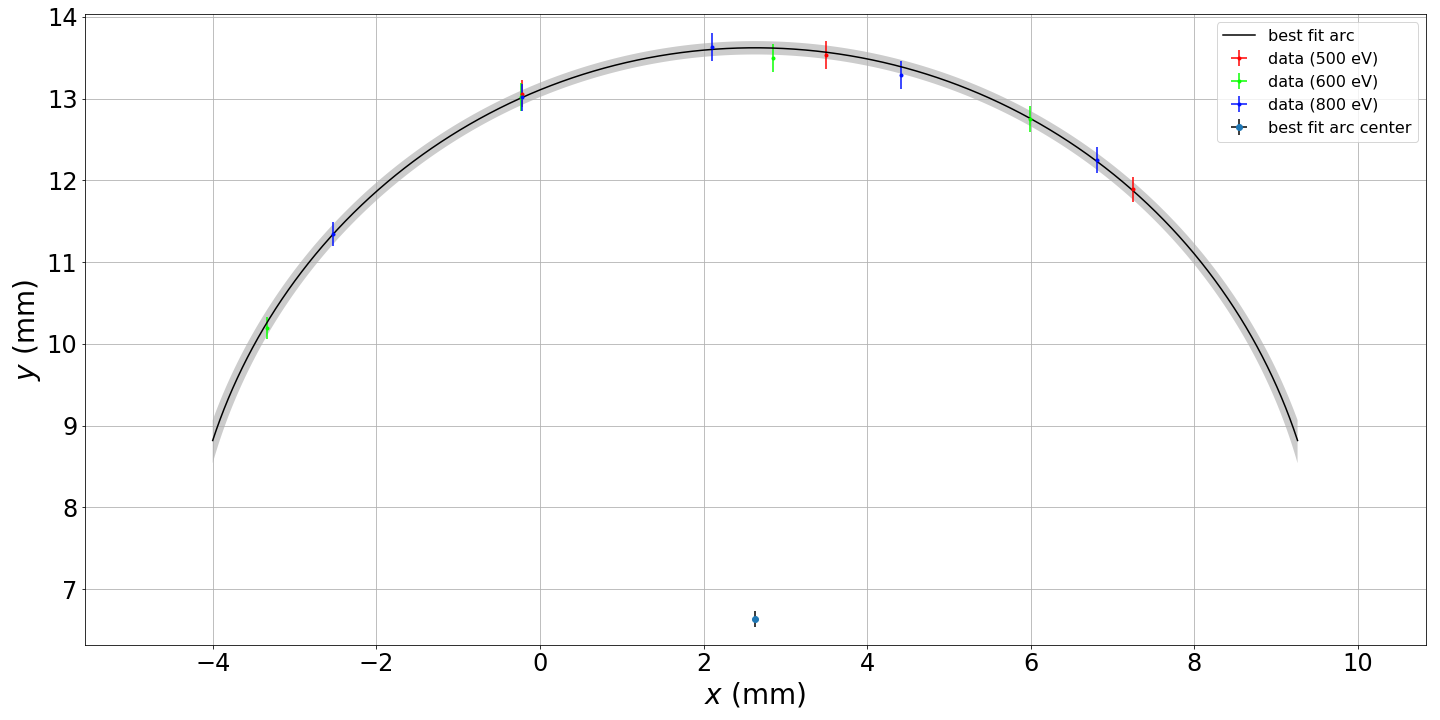}
 \includegraphics[scale=0.31]{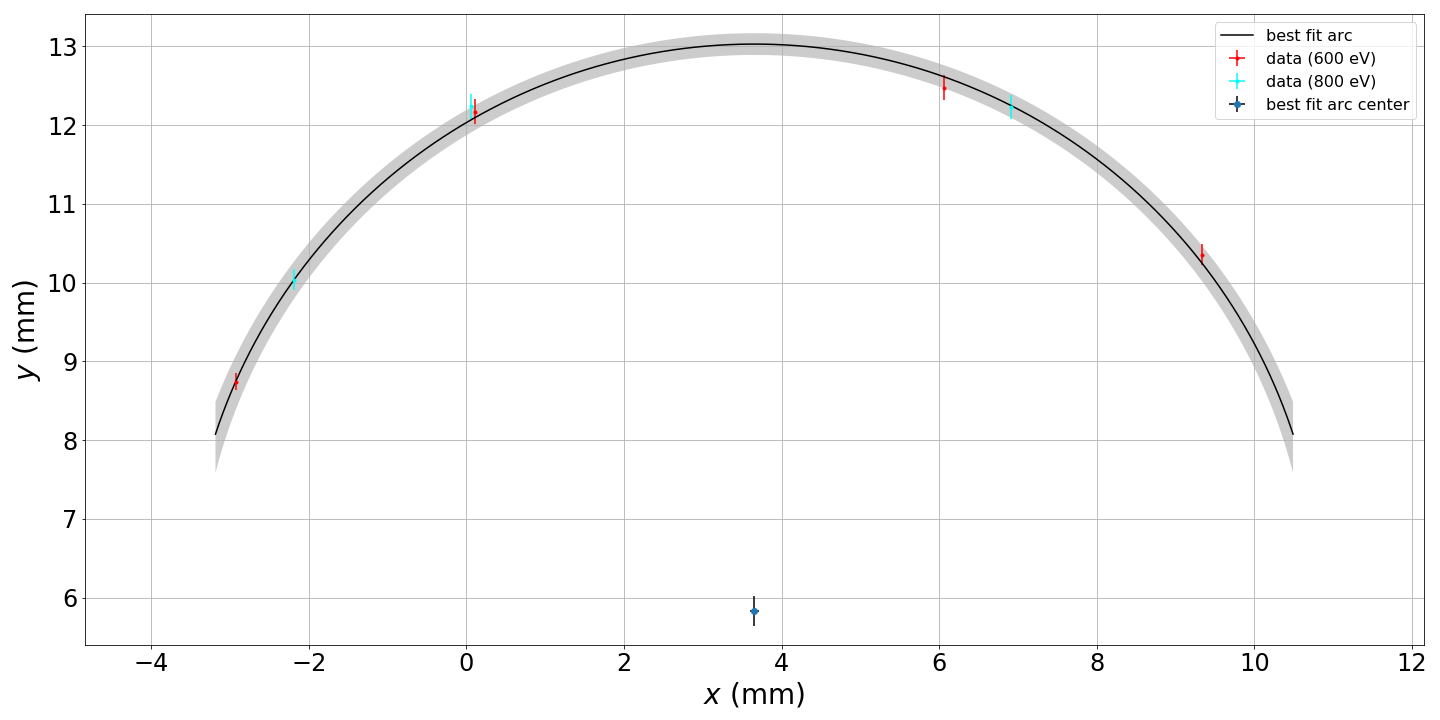}
 \caption[Measured diffracted arcs for the silicon master and the coated SCIL replica]{Measured diffracted arcs for the silicon master (\emph{top}) and the coated SCIL replica (\emph{bottom}) in their respective test configurations, fit to half-circles. Grayed regions represent one standard deviation uncertainty \cite{McCoy20b}.}\label{fig:arc_fit}
 \end{figure}
This allowed $r$ to be inferred from fitting measured $(x,y)$ data, gathered at a few photon energies, to a half circle, which is shown in \cref{fig:arc_fit} for both the silicon master and the coated SCIL replica. 
The angle $\gamma$ for each grating was then determined from $\sin \left( \gamma \right) = r/ L$ [\emph{cf.\@} \cref{eq:arc_radius}] before the remaining angles were calculated using the fit values for the arc center coordinates as they compare to the measured $(x,y)$ locations of $0^{\text{th}}$ order as well as the direct beam. 
First, the angle $\alpha$ was measured using \cref{eq:measure_alpha} and similarly, the grating roll characterized by the angle $\phi$ was constrained through \cref{eq:measure_roll} [\emph{cf.\@} \cref{fig:ALS_arc}]. 
With this value for $\phi$ and a measured $y$-distance between $0^{\text{th}}$ order and the center of the diffracted arc, $\Delta y_0$, the substrate graze angle, $\eta$, was measured for each grating using \cref{eq:measure_graze} before grating yaw, which is parameterized by the angle $\varphi$, was determined from \cref{eq:measure_yaw}. 
\begin{table}[]
 \centering
 \caption[Measured diffracted-arc parameters for the silicon master and the coated SCIL replica in their respective test configurations at the ALS]{Measured diffracted-arc parameters for the silicon master and the coated SCIL replica in their respective test configurations at the ALS \cite{McCoy20b}.}\label{tab:arc_params}
 \begin{tabular}{cccc}
 \cline{1-3}
 measured parameter                                                     & \ master         		  & \ replica     \\ \hline
 $L$                                    & \ $234.7 \pm \SI{3.0}{\mm}$   & \ $235.6 \pm \SI{3.0}{\mm}$ \\
 $r$                                    & \ $6.98 \pm \SI{0.08}{\mm}$   & \ $7.20 \pm \SI{0.14}{\mm}$ \\
 $\Delta x_{\text{dir}}$ 				& \ $2.80 \pm \SI{0.03}{\mm}$   & \ $3.68 \pm \SI{0.07}{\mm}$ \\
 $\Delta x_0$  							& \ $2.86 \pm \SI{0.03}{\mm}$   & \ $3.57 \pm \SI{0.06}{\mm}$ \\
 $\Delta y_0$  							& \ $6.41 \pm \SI{0.10}{\mm}$   & \ $6.34 \pm \SI{0.19}{\mm}$ \\ 
 $\gamma$ by \cref{eq:arc_radius}       & \ $1.71 \pm 0.03^{\circ}$ 	& \ $1.75 \pm 0.04^{\circ}$   \\
 $\alpha$ by \cref{eq:measure_alpha}    & \ $23.7 \pm 0.7^{\circ}$  	& \ $30.7 \pm 0.9^{\circ}$    \\ 
 $\phi$ by \cref{eq:measure_roll}       & \ $0.50 \pm 0.14^{\circ}$ 	& \ $0.98 \pm 0.39^{\circ}$   \\
 $\eta$ by \cref{eq:measure_graze}      & \ $1.57 \pm 0.03^{\circ}$ 	& \ $1.54 \pm 0.05^{\circ}$   \\
 $\varphi$ by \cref{eq:measure_yaw}     & \ $0.68 \pm 0.01^{\circ}$ 	& \ $0.89 \pm 0.02^{\circ}$   \\ \hline
 \end{tabular}
 \end{table}
These measured parameters are listed in \cref{tab:arc_params} for both gratings. %the silicon master and the coated SCIL replica. 

Although the measured values for $\eta$ and $\varphi$ depart slightly from the targeted values of $\eta \approx 1.5^{\circ}$ and $\varphi \approx 0.84^{\circ}$, the arc parameters listed in \cref{tab:arc_params} indicate near-Littrow configurations for both gratings. 
By \cref{eq:blaze_wavelength} for the blaze wavelength, the diffraction efficiency of propagating orders with $n=2$ and $n=3$ are expected to maximize in the spectral range $\SI{440}{\electronvolt} \leq \mathcal{E}_{\gamma} \leq  \SI{900}{\electronvolt}$ for a grating with $d \lessapprox \SI{160}{\nm}$ in a near-Littrow configuration with $\gamma \approx 1.7^{\circ}$. 
Moreover, rewriting \cref{eq:blaze_wavelength} for the blaze wavelength as [\emph{cf.\@} \cref{eq:angle_on_groove,eq:blaze_wavelength_general}]
\begin{subequations}
\begin{equation}\label{eq:blaze_wavelength_alt}
 \lambda_b = \frac{2 d \sin \left( \zeta \right) \sin \left( \delta \right)}{n} = \frac{2 d \sin \left( \gamma \right) \sin \left( \delta \right)}{n} \cos \left( \delta - \alpha \right) 
 \end{equation}
and then invoking small angle approximations for $\gamma$ as well as $\abs{\delta - \alpha}$ gives %in terms of radians
\begin{equation}\label{eq:blaze_wavelength_alt_approx}
 \lambda_b \approx \frac{2 d \gamma \sin \left( \delta \right)}{n} \left( 1 - \frac{\left( \delta - \alpha \right)^2}{2} \right) .
 \end{equation}
\end{subequations}
This approximate expression suggests that the locations of peak orders are most sensitive to $\delta$ and $\gamma$ in an extreme off-plane mount rather than $\alpha$ provided that $\abs{\delta - \alpha} \ll \SI{1}{\radian}$, which describes a near-Littrow configuration. 
With both gratings loosely satisfying this condition for $\alpha$, the grating geometries listed in \cref{tab:arc_params} were employed for testing.

\subsection{Test Results for Diffraction Efficiency}\label{sec:test_results_scil}
%%%%%%%%%%%%%%%%%%%%%%%%%%%%%%%%%%%%%%%%%--------------------------------------------------
Experimental data for $\mathscr{E}_n$ were gathered as a function of $\mathcal{E}_{\gamma}$ over the range \SIrange{440}{900}{\electronvolt} using test configurations defined by the parameters listed in \cref{tab:arc_params}. 
As described in \cref{sec:measure_efficiency} and carried out in \cref{sec:test_results_taste}, $\mathcal{I}_n$ for each propagating order, at each value of $\mathcal{E}_{\gamma}$, was measured using the masked photodiode by scanning the diffracted arc horizontally, in \SI{50}{\um} steps, and then determining the maximum of each diffracted order. 
Meanwhile, $\mathcal{I}_{\text{inc}}$ for the unobstructed beam was measured at each value of $\mathcal{E}_{\gamma}$ in an analogous way, with the grating moved out of the path of the beam. 
The quantity $\mathscr{E}_n$ was then determined through $\mathcal{I}_n / \mathcal{I}_{\text{inc}}$ [\emph{cf.\@} \cref{eq:diffraction_efficiency}] after subtracting dark current as in \cref{sec:test_results_taste}, but due to the use of a detector that is smaller than the radius of the diffracted arc [\emph{cf.\@} \cref{sec:grat_geo_scil}], $\mathcal{I}_n$ measurements for propagating orders near the bottom of the diffracted arc were missed systematically during data collection. 
\begin{figure}
 \centering
 \includegraphics[scale=0.4]{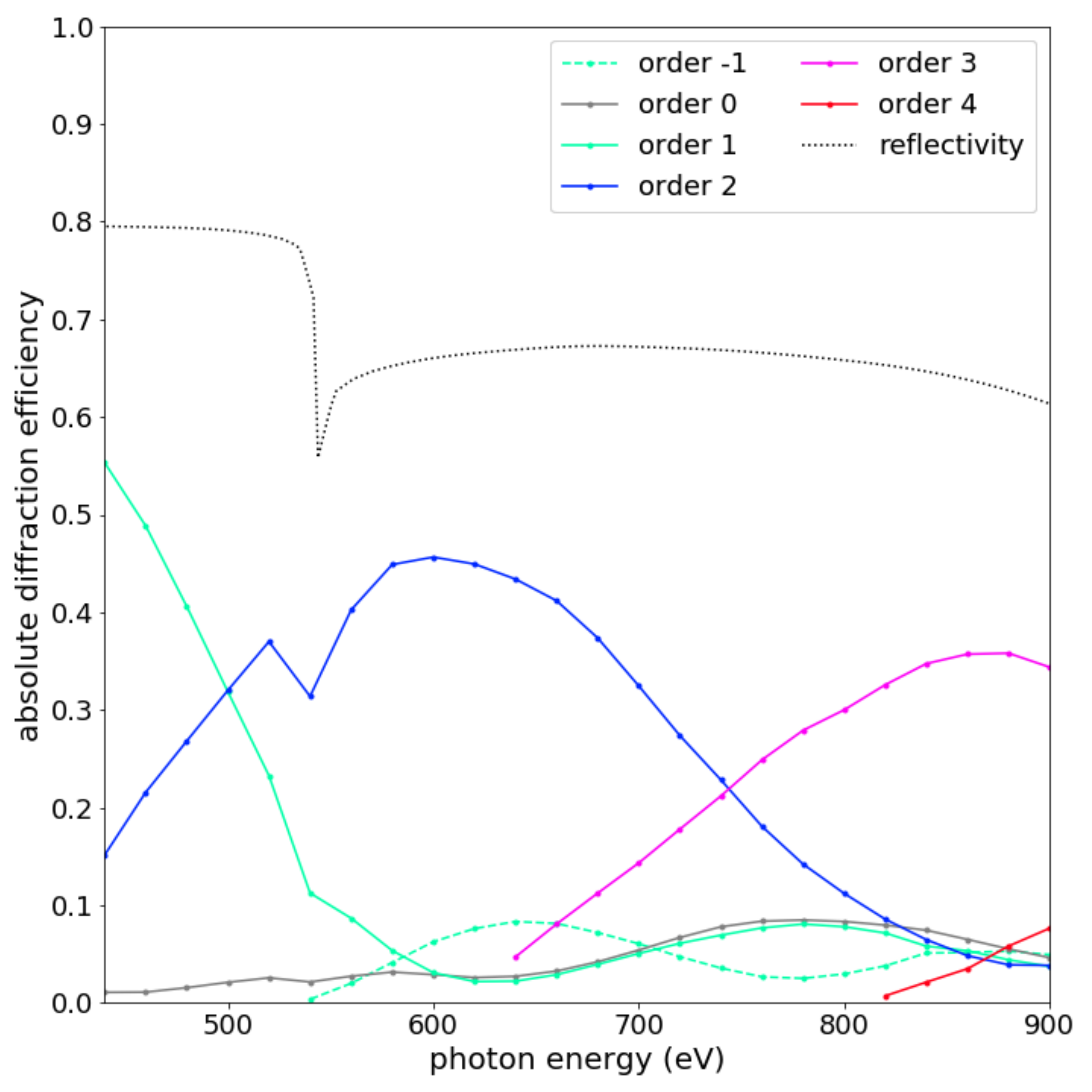}
 \includegraphics[scale=0.4]{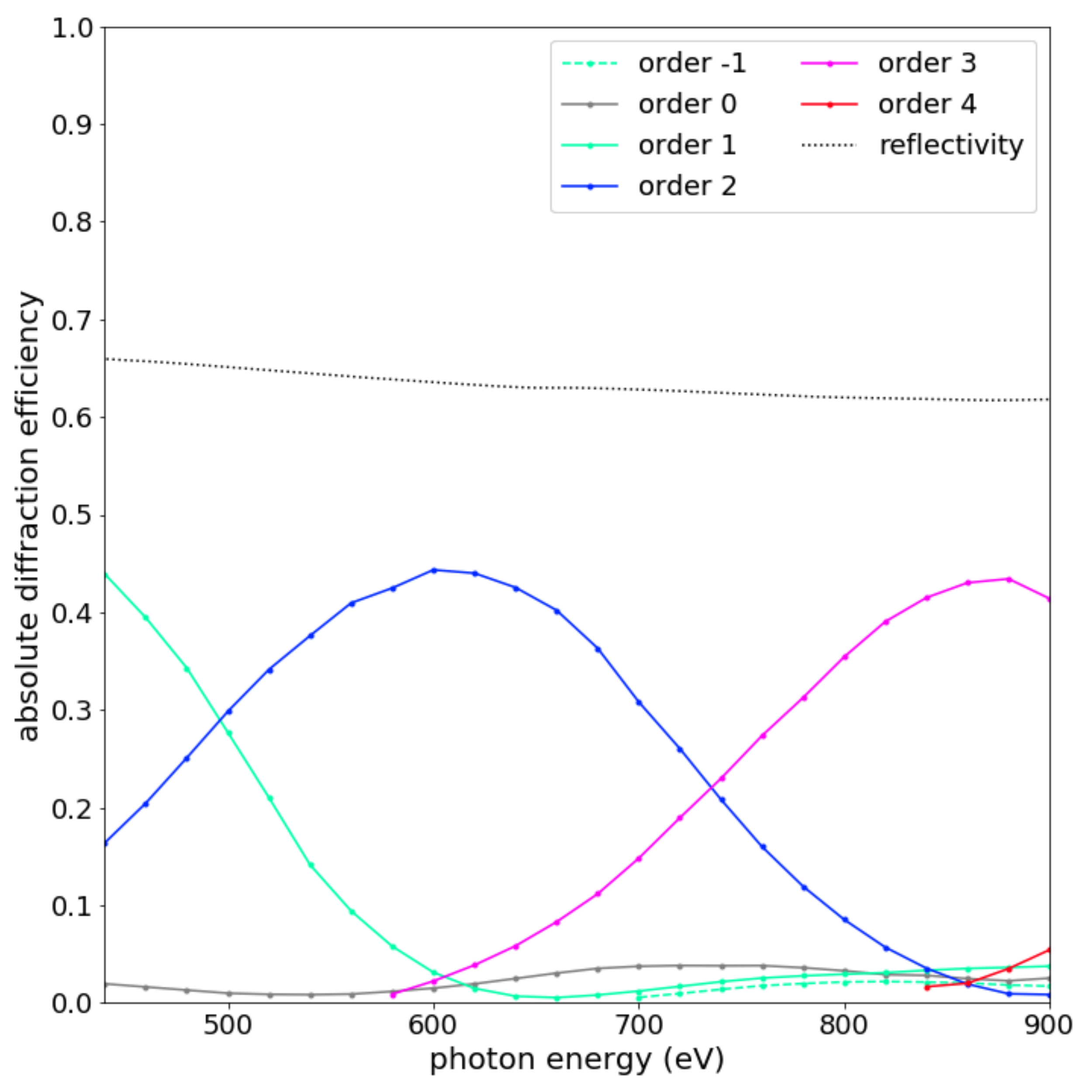}
 \caption[Measured absolute diffraction efficiency data for the silicon master and the coated SCIL replica]{Measured absolute diffraction efficiency data for the silicon master (\emph{top}) and the coated SCIL replica (\emph{bottom}) in geometrical configurations described by the parameters listed in \cref{tab:arc_params} compared to specular reflectivity calculated using the parameters listed in \cref{tab:nom_params} \cite{McCoy20b}.}\label{fig:abs_eff}
 \end{figure} 
Notwithstanding this, $\mathscr{E}_n$ was measured every \SI{20}{\electronvolt} between \SI{440}{\electronvolt} and \SI{900}{\electronvolt}; these results are plotted in \cref{fig:abs_eff} for both gratings. %the silicon master and the SCIL replica. 

The data plotted in \cref{fig:abs_eff} are compared to specular reflectivity curves relevant for each case, which are plotted as dotted lines. 
\begin{table}[]
 \centering
 \caption[Nominal parameters relevant for specular reflectivity analysis and diffraction-efficiency modeling of the silicon master and the coated SCIL replica ]{Nominal parameters relevant for specular reflectivity analysis and diffraction-efficiency modeling of the silicon master and the coated SCIL replica \cite{McCoy20b}.}\label{tab:nom_params} 
 \begin{tabular}{cccc}
 \cline{1-3}
 nominal parameter                                  & \ master         		  								  & \ replica            \\ \hline
 $\gamma$											& $1.71^{\circ}$								  		  & $1.75^{\circ}$	 	 \\
 $\alpha$											& $23.7^{\circ}$								          & $30.7^{\circ}$	 	 \\
 estimated blaze angle ($\delta$, $\delta'$) 		& $29.5^{\circ}$								  		  & $28^{\circ}$	 	 \\
 blaze angle by AFM ($\delta$, $\delta'$) 			& $30.0^{\circ}$								  		  & $28.4^{\circ}$	 	 \\
 facet incidence angle ($\zeta$) 					& $1.70^{\circ}$								          & $1.75^{\circ}$	 	 \\
 reflective material 								& \SI{3}{\nm} \ce{SiO2} on \ce{Si}						 & \ce{Au} (thick slab)		 \\
 eq.\ for specular reflectivity 					& \cref{eq:Si_layer_refl} with \cref{eq:NC_factors} & \cref{eq:Au_refl} \\
 facet roughness ($\sigma$)							& $\SI{0.4}{\nm}$ RMS						  			  & $\SI{0.8}{\nm}$ RMS	 	 \\ \hline
 \end{tabular}
 \end{table}
Summarized in \cref{tab:nom_params}, these curves were calculated according to the considerations outlined in \cref{sec:reflectivity} but with updated values for the facet-incidence angle, $\zeta$, which follow from the nominal values for $\gamma$ and $\alpha$ as well as the expected blaze angles, $\delta$ and $\delta'$. 
Peak-order values of $\mathscr{E}_n$ range from \SIrange{40}{45}{\percent} for both gratings or equivalently, \SIrange{65}{70}{\percent} measured relative to the reflectivity in each case, which is comparable to the results reported for the corresponding UV-NIL replica presented by Miles, et~al.~\cite{Miles18}. 
However, because not all diffracted orders are accounted for due to the condition $\ell_{\text{det}} < r$, $\mathscr{E}_{\text{tot}}$ is not plotted in \cref{fig:abs_eff} [\emph{cf.\@} \cref{sec:geo_constrain}]. 

\section{Analysis and Discussion}\label{sec:discussion_scil}
%%%%%%%%%%%%%%%%%%%%%%%%%%%%%%%%%%%%%%%%%--------------------------------------------------
The soft x-ray diffraction-efficiency measurements presented in \cref{sec:test_results_scil} demonstrate that both the silicon master and the corresponding SCIL replica exhibit a significant blaze response in a near-Littrow, grazing-incidence configuration. 
Using these data, the following analysis seeks to constrain the impact of resist shrinkage on the blaze angle of the SCIL replica by estimating $\delta' / \delta$ through comparing measured single-order efficiency curves to those predicted by theoretical models of diffraction efficiency. 
The models utilized for this study were produced with the aid of the software package \textsc{PCGrate-SX} (v.\ 6.1) \cite{PCGrate_web}, which solves the Helmholtz equation through the integral method for a custom grating boundary and incidence angles input by the user \cite[\emph{cf.\@} \cref{sec:grat_bound_consider}]{Goray10}. 
In each case, the grating boundary is taken to be perfectly conducting so that the electromagnetic fields inside the grating material are null. 
While this implies lossless diffraction efficiency for a perfectly smooth grating boundary as shown in \cref{sec:rel_eff_perf}, the overall response is modulated by the appropriate specular reflectivity in each case [\emph{cf.\@} \cref{sec:reflectivity,sec:refl_account}]. 
The incident radiation is treated as having a wave vector given by \cref{eq:beam_vec_grat} and \emph{TE polarization} [\emph{cf.\@} \cref{fig:polarization_angle}] for the reasons outlined in \cref{sec:grat_bound_consider}.  
Additionally, the absorbing effect surface roughness in the limit of small surface features is taken into account by using the appropriate Nevot-Croce factors [\emph{cf.\@} \cref{eq:NC_factors}] with the values for facet roughness listed in \cref{tab:nom_params}.  
The groove spacing in each case is assumed to be $d = \SI{159.125}{\nm}$ based on the considerations outlined in \cref{sec:grat_geo_scil}.
With \cref{sec:silicon_master_sub,sec:scil_replica_sub} treating the silicon master and the SCIL replica, respectively, \cref{sec:shrinkage_model} presents an approximate model for resist shrinkage based on simple geometric considerations to interpret this result. 

\subsection{Silicon Master}\label{sec:silicon_master_sub}
%%%%%%%%%%%%%%%%%%%%%%%%%%%%%%%%%%%%%%%%%--------------------------------------------------
As a point of reference for evaluating resist shrinkage in the SCIL replica, the diffraction-efficiency results for the silicon master [\emph{cf.\@} \cref{fig:abs_eff}, \emph{top panel}] are compared to \textsc{PCGrate-SX} models in \cref{fig:model_si1}. 
\begin{figure}
 \centering
 \includegraphics[scale=0.25]{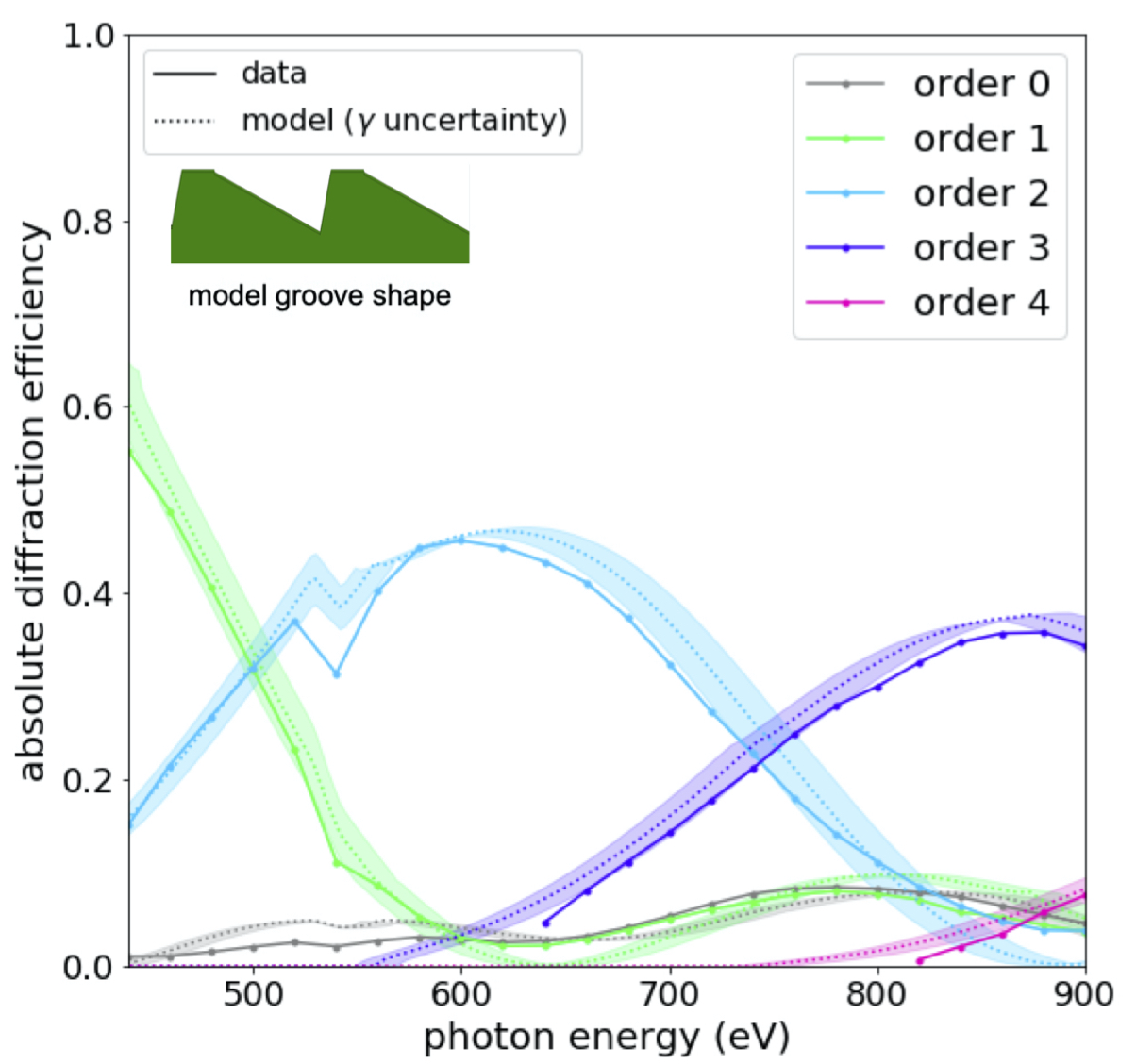}
 \includegraphics[scale=0.25]{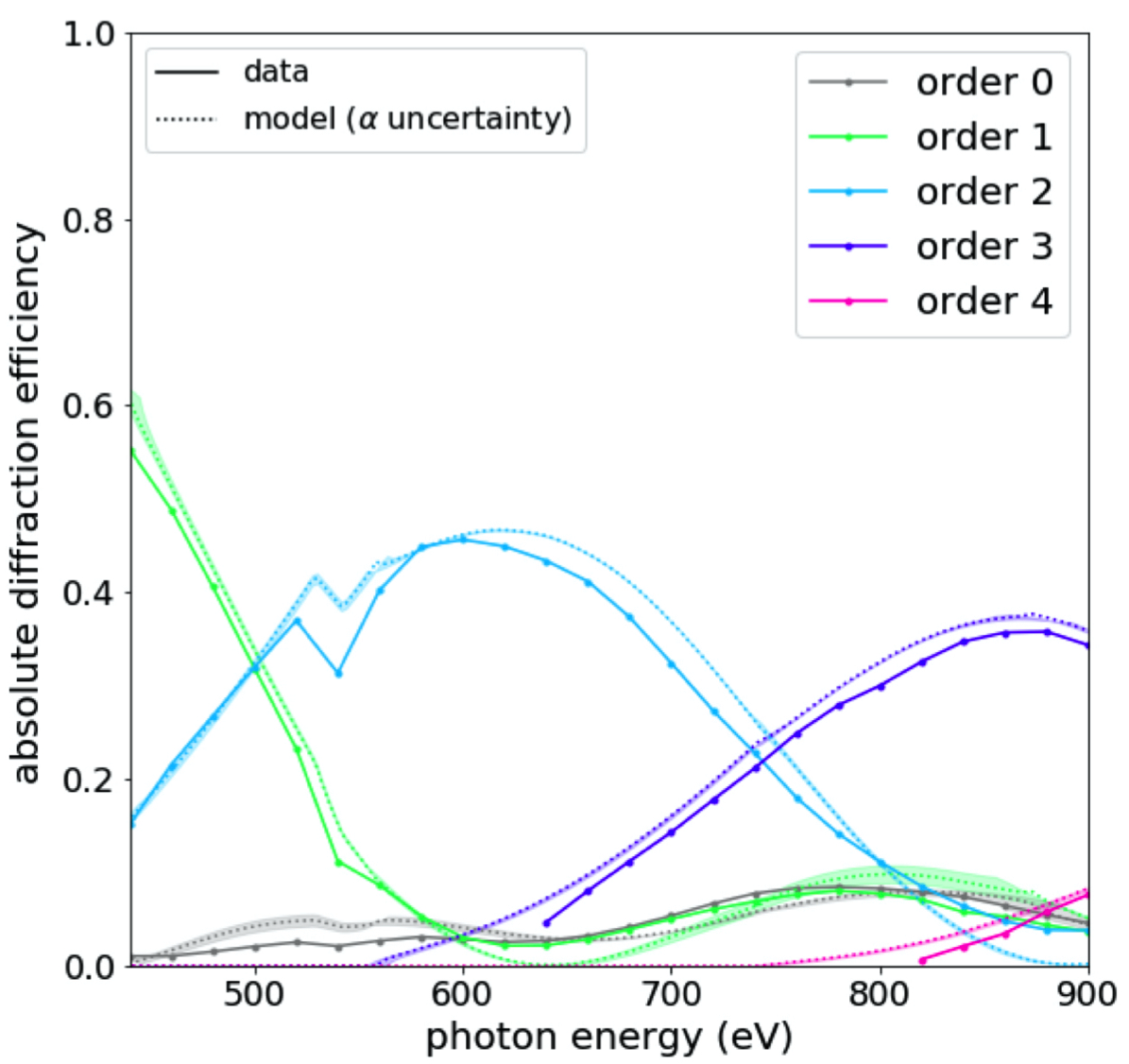}
 \caption[Measured absolute diffraction efficiency data for the silicon master compared to \textsc{PCGrate-SX} models that assume a groove profile similar to a \ce{KOH}-etched $\langle 311 \rangle$ silicon grating]{Measured absolute diffraction efficiency data for the silicon master [\emph{cf.\@} \cref{fig:abs_eff}, \emph{top panel}] compared to \textsc{PCGrate-SX} models that assume a groove profile similar to a \ce{KOH}-etched $\langle 311 \rangle$ silicon grating with $\delta = 29.5^{\circ}$, $\bar{\delta} \approx 80^{\circ}$, and a grove depth of $h \approx \SI{67}{\nm}$ with $w = \SI{35}{\nm}$ and $\Delta h = \SI{3}{\nm}$ [\emph{cf.\@} \cref{eq:groove_depth}]. In the top and bottom panels, respectively, $\gamma$ and $\alpha$ are allowed to vary at levels of $1.71 \pm 0.03^{\circ}$ and $23.7 \pm 0.7^{\circ}$ (\emph{shaded swaths}) \cite{McCoy20b}.}\label{fig:model_si1}
 \end{figure} 
These models are based on the wet-etched grating topography described in \cref{sec:master_for_SCIL} together with incident radiation parameterized by $\gamma$ and $\alpha$ for the test-configuration geometry established in \cref{sec:grat_geo_scil}. %in \cref{sec:crystal_etching} 
The grating boundary is defined using the trapezoid-like groove shape shown in the figure inset, with nominal sawtooth angles of $\delta = 29.5^{\circ}$ and $\bar{\delta} \approx 80^{\circ}$ for $\langle 311 \rangle$-oriented, \ce{KOH}-etched silicon [\emph{cf.\@} \cref{tab:off_axis_si}], a flat-top width of $w = \SI{35}{\nm}$, a nub-protrusion height of $\Delta h = \SI{3}{\nm}$ and a groove depth of $h \approx \SI{67}{\nm}$ [\emph{cf.\@} \cref{eq:groove_depth}]. %[\emph{cf.\@} \cref{fig:master_grating_profile}]
In both panels of \cref{fig:model_si1}, the model corresponding to the nominal values $\gamma = 1.71^{\circ}$ and $\alpha = 23.7^{\circ}$ [\emph{cf.\@} \cref{tab:nom_params}] is plotted as a series of dotted lines for each diffracted order shown. 
Models that take into account the uncertainties for $\gamma$ and $\alpha$ listed in \cref{tab:arc_params} are then plotted as shaded swaths in the top and bottom panels of \cref{fig:model_si1}, respectively. 

The results presented in \cref{fig:model_si1} show that the geometry constrained in \cref{sec:grat_geo_scil} leads to the production of models that roughly match the experimental data. 
Mismatches between the models and the data may be in part due to the detailed shape of nubs atop of each groove, which cannot be described with a functional form\footnote{This refers to the periodic groove function, $g(x)$, defined in \cref{sec:grating_boundary}.} due to the presence of a small, indented sidewall [\emph{cf.\@} \cref{fig:master_grating_profile,fig:master_grating_SEM}]. % as illustrated in \cref{fig:master_grating_profile} and shown under FESEM in \cref{fig:master_grating_SEM}. 
Although this limits the accuracy of the \textsc{PCGrate-SX} models utilized, the model uncertainty swaths indicate that $\gamma$ serves to shift the centroids of peak orders\footnote{\emph{i.e.}, the photon-energy equivalent to the blaze wavelength, $h c_0 / \lambda_b$} while $\alpha$ has a relatively small impact as expected from \cref{eq:blaze_wavelength_alt_approx}. 
\begin{figure}
 \centering
 \includegraphics[scale=0.25]{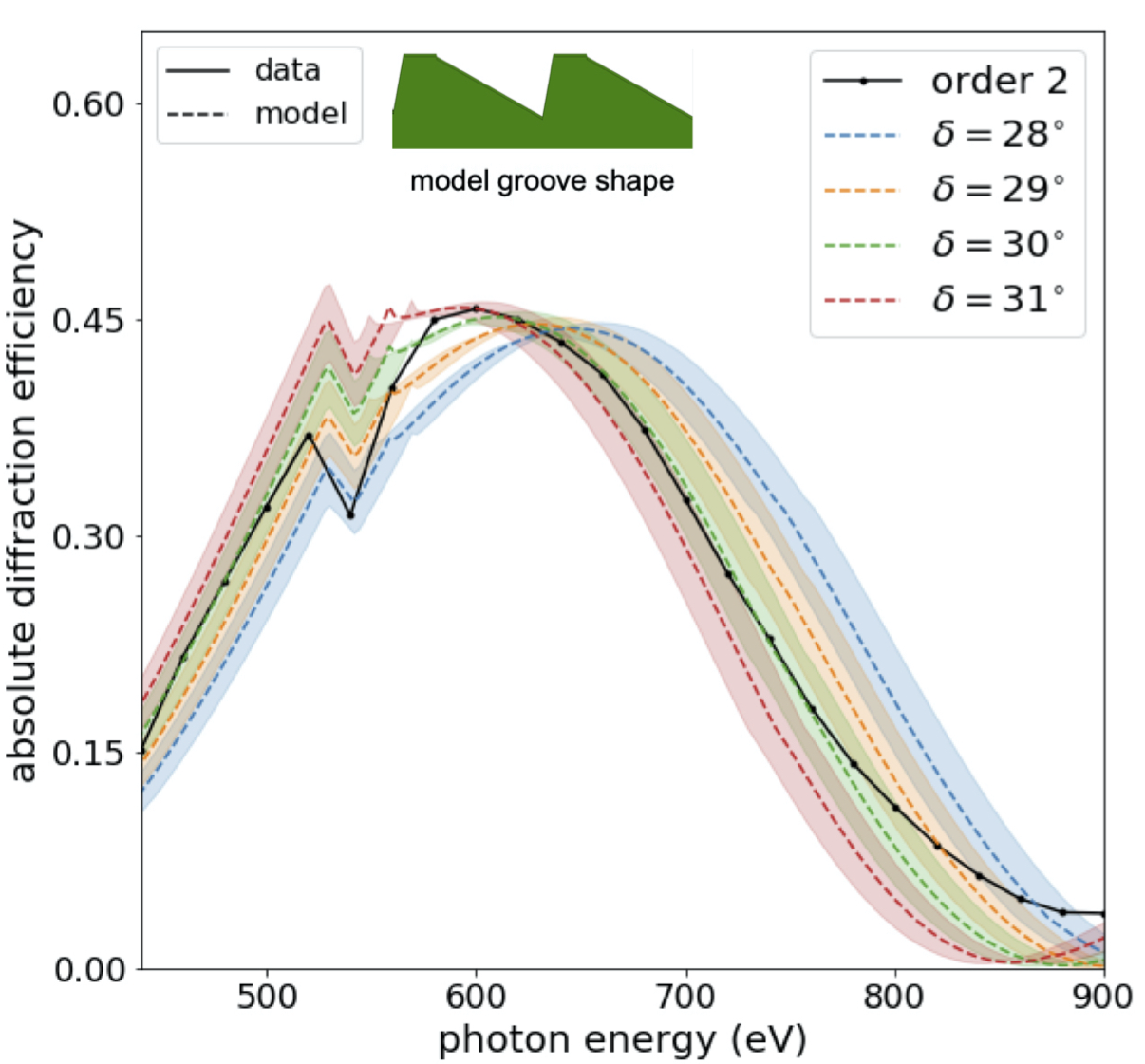}
 \includegraphics[scale=0.25]{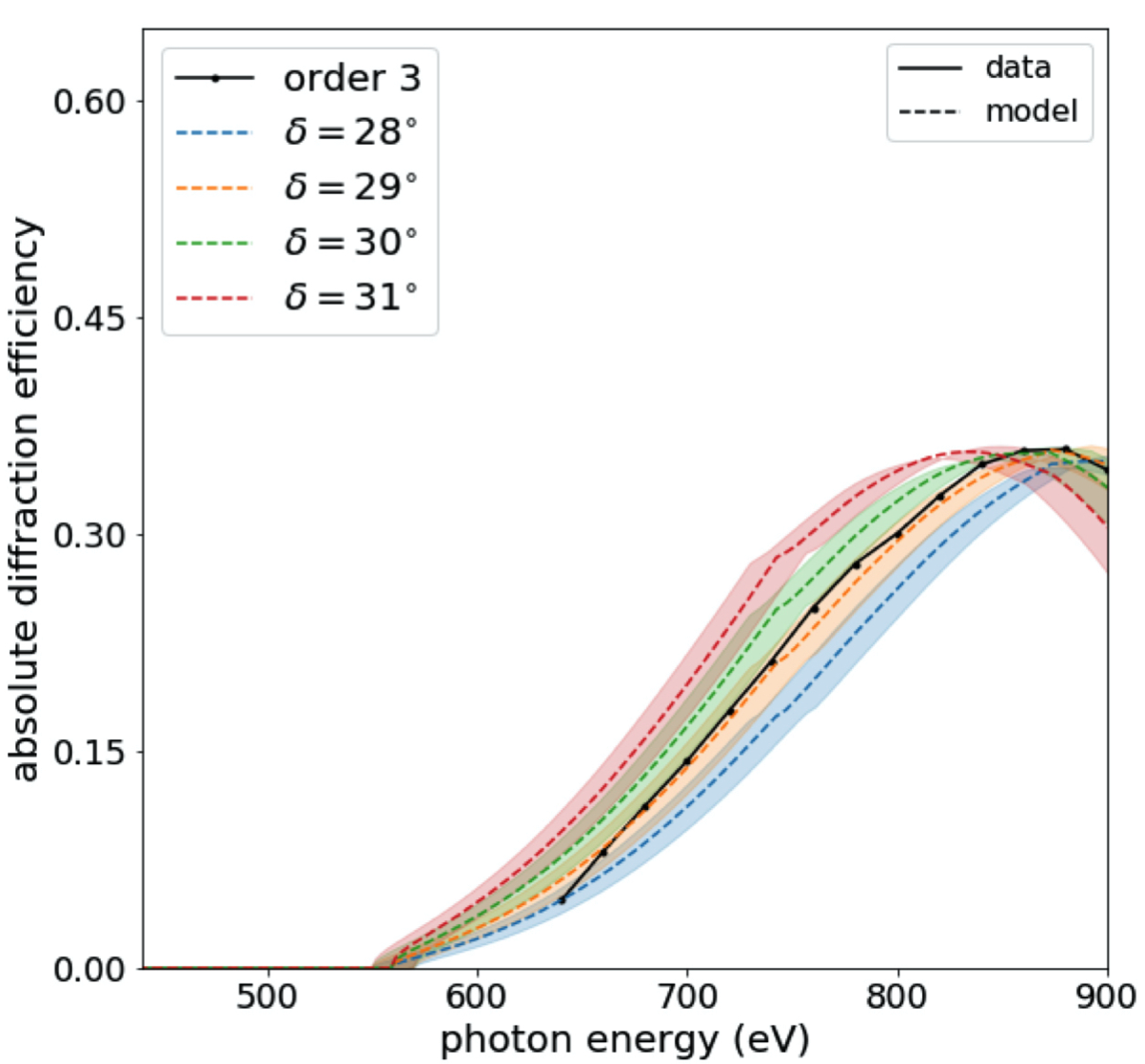}
 \caption[Measured absolute diffraction efficiency data in second and third orders for the silicon master compared to \textsc{PCGrate-SX} models of varying blaze angle]{Measured absolute diffraction efficiency data with $n=2$ and $n=3$ for the silicon master compared to \textsc{PCGrate-SX} models with $28^{\circ} \leq \delta \leq 31^{\circ}$, $\alpha = 23.7^{\circ}$, and $\gamma = 1.71 \pm 0.03^{\circ}$ (\emph{shaded swaths}) that are normalized to match the data in terms of peak efficiency. These results indicate that $\delta$ for the silicon master is close to the value of $\delta = 30^{\circ}$ measured by AFM \cite{McCoy20b}.}\label{fig:model_si2} %while the shaded swaths represent the $\pm 0.03^{\circ}$ uncertainty in $\gamma$ [\emph{cf.\@} \cref{fig:model_si1}, \emph{top panel}]
 \end{figure} 
With these centroids also depending directly on $\sin \left( \delta \right)$ by \cref{eq:blaze_wavelength_alt_approx}, a series of models with $28^{\circ} \leq \delta \leq 31^{\circ}$ in steps of $1^{\circ}$ are compared to absolute-efficiency data for $n=2$ and $n=3$ in \cref{fig:model_si2}. 

In all of the models used for \cref{fig:model_si2}, $w = \SI{35}{\nm}$ and $\Delta h = \SI{3}{\nm}$ are fixed while the sawtooth angles vary as $\delta$ and $\bar{\delta} = 180^{\circ} - \theta - \delta$ so that the overall groove depth, $h$, follows from \cref{eq:groove_depth}. 
Moreover, the measured data have been corrected for the absorbing effect of small-scale facet surface roughness using the ratio of Nevot-Croce reflectivity (\emph{i.e.}, \cref{eq:Si_layer_refl} with $\sigma_1 = \sigma_2 = \SI{0.4}{\nm}$~RMS) to Fresnel reflectivity (\emph{i.e.}, \cref{eq:Si_layer_refl} with $\sigma_1 = \sigma_2 \to 0$), at the nominal incidence angle $\zeta = 1.70^{\circ}$ [\emph{cf.\@} \cref{tab:nom_params}]. 
The modeled efficiency in each case, which assumes a perfectly smooth grating boundary, was then normalized to match the peak efficiency of the measured data. 
Additionally, dotted lines represent the nominal model with $\gamma = 1.71^{\circ}$ and $\alpha = 23.7^{\circ}$ while the shaded swaths show the $\pm 0.03^{\circ}$ uncertainty in $\gamma$. 
These results support the expectation that the blaze angle of the silicon master is in the neighborhood of the nominal $\langle 311 \rangle$ value of $\delta = 29.5^{\circ}$ as well as the AFM-measured value of $\delta = 30.0 \pm 0.8^{\circ}$ reported in \cref{sec:grating_fab}. 

\subsection{SCIL Replica}\label{sec:scil_replica_sub}
%%%%%%%%%%%%%%%%%%%%%%%%%%%%%%%%%%%%%%%%%--------------------------------------------------
Knowing that variance in $\alpha$ has a small effect on peak-order centroids for $\alpha \approx \delta$ [\emph{cf.\@} \cref{eq:blaze_wavelength_alt_approx}], two different \textsc{PCGrate-SX} models for the coated SCIL replica are plotted in \cref{fig:model_scil1} and compared to the experimental data [\emph{cf.\@} \cref{fig:abs_eff}, \emph{bottom panel}] using uncertainty swaths for $\gamma$ alone. 
\begin{figure}
 \centering
 \includegraphics[scale=0.25]{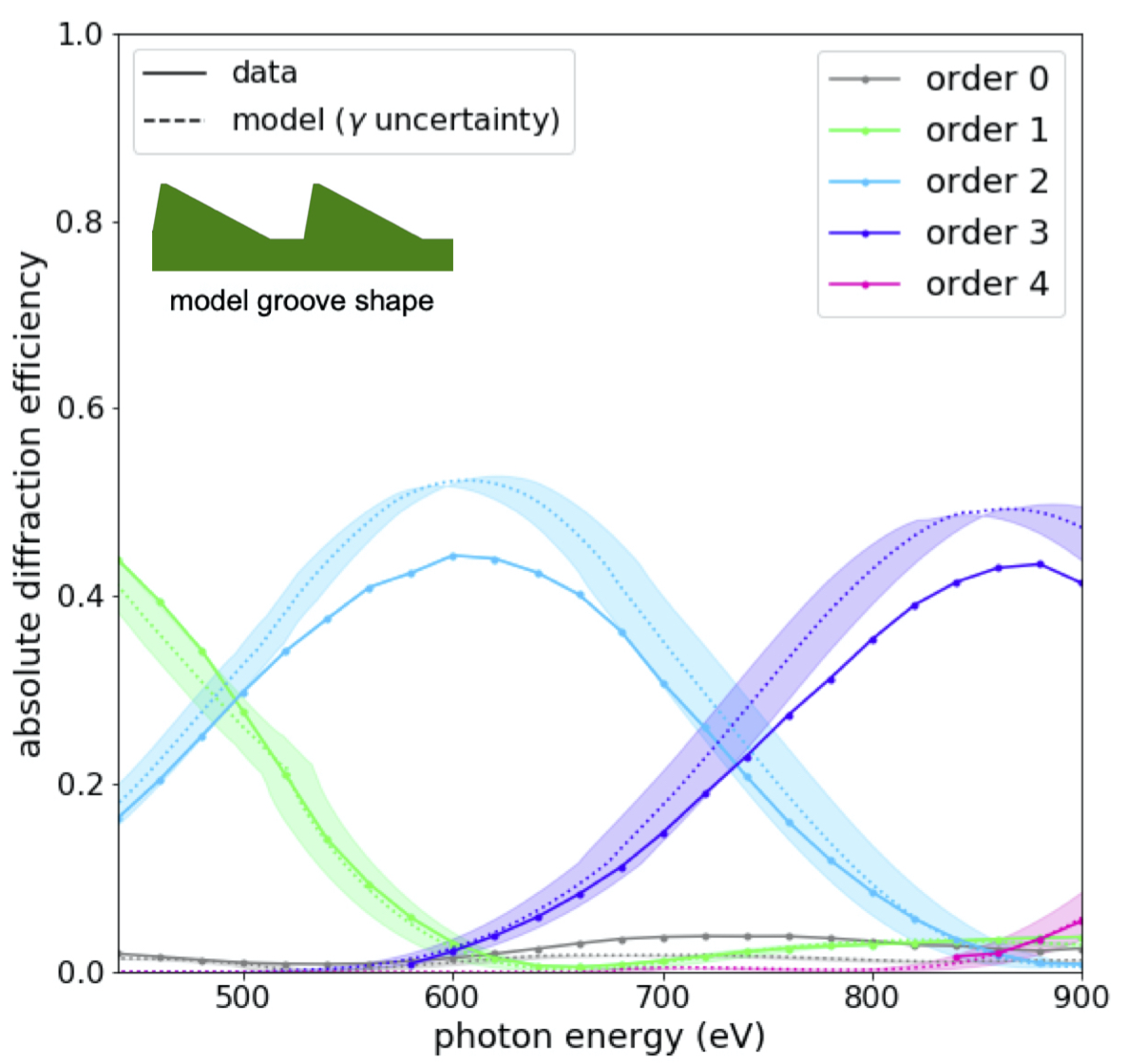}
 \includegraphics[scale=0.25]{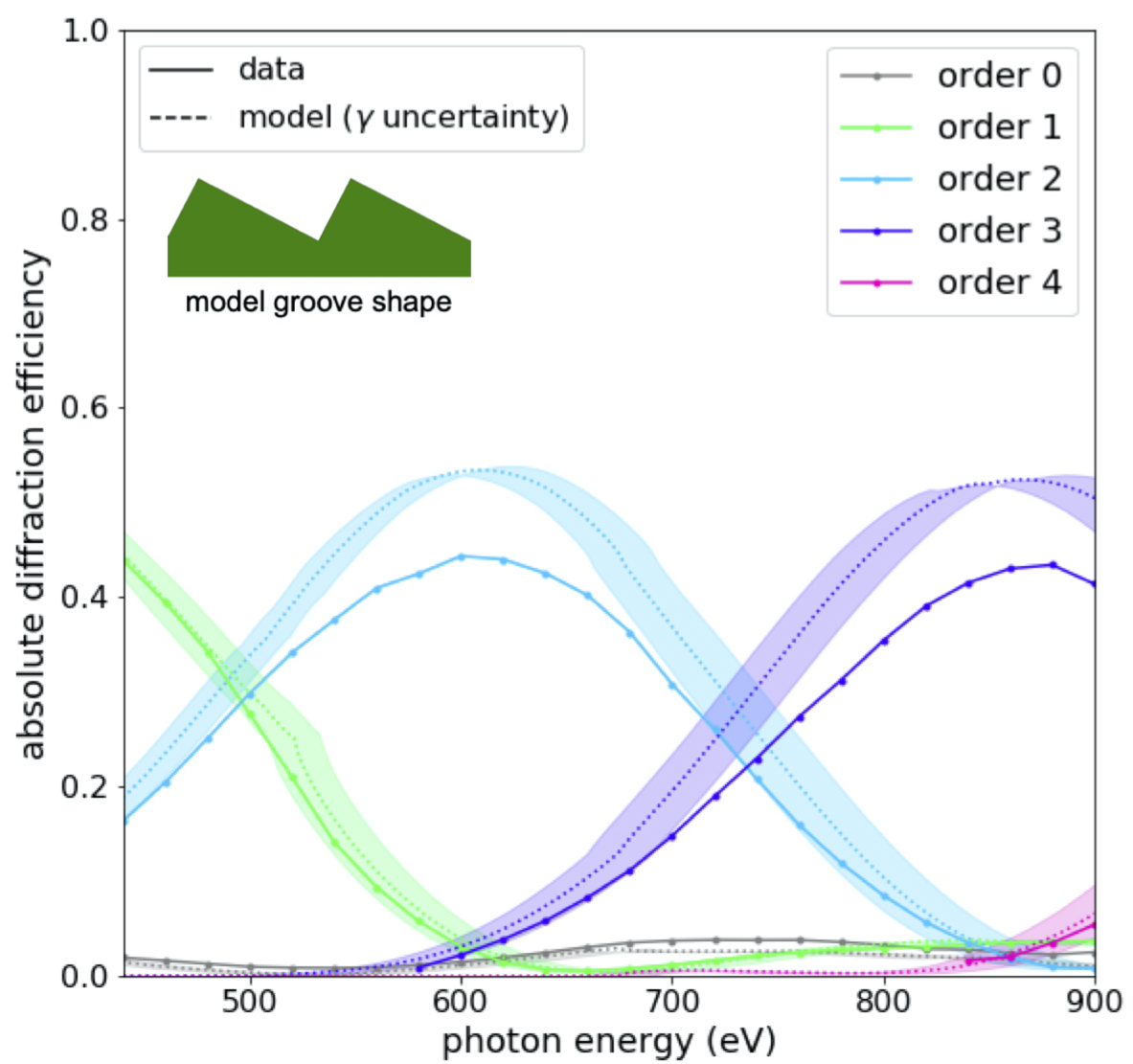}
 \caption[Measured absolute diffraction efficiency data for the coated SCIL replica compared to \textsc{PCGrate-SX} models featuring two different groove shapes, each with a blaze angle of $28^{\circ}$.]{Measured absolute diffraction efficiency data for the coated SCIL replica [\emph{cf.\@} \cref{fig:abs_eff}, \emph{bottom panel}] compared to \textsc{PCGrate-SX} models featuring two different groove shapes, each with $\delta' = 28^{\circ}$, $\alpha = 30.7^{\circ}$ and $\gamma = 1.75 \pm 0.04^{\circ}$ (\emph{shaded swaths}). While the top-panel groove shape more closely resembles the topography of the SCIL replica, it yields model results that are comparable to those produced from the simpler, ideal sawtooth (\emph{bottom panel}) \cite{McCoy20b}.}\label{fig:model_scil1} %The simpler, ideal sawtooth model thus can be used to match predicted peak-order centroids to the measured data.
 \end{figure} 
Both of these models use the nominal values $\gamma = 1.75^{\circ}$ and $\alpha = 30.7^{\circ}$ with $\pm 0.04^{\circ}$ uncertainty in $\gamma$ [\emph{cf.\@} \cref{tab:arc_params}] along with a shrunken blaze angle of $\delta' = 28^{\circ}$, which is an approximate value based on the AFM measurements from \cref{sec:reflectivity}, and specular reflectivity for a gold slab [\emph{cf.\@} \cref{eq:Au_refl}]. 
However, the two models differ in their groove shape details; the model in the top panel of \cref{fig:model_scil1} attempts to emulate the topography of the SCIL replica with a $\bar{\delta} \approx 80^{\circ}$ steep angle, a groove depth of $h' \approx \SI{58}{\nm}$, which was estimated from AFM measurements, a flat-bottom portion of width $w \approx \SI{35}{\nm}$ and a \SI{5}{\nm}-wide flat top so as to approximate a slightly rounded groove apex. 
On the other hand, the model in the bottom panel is taken to be the ideal case of a sawtooth with a sharp, $90^{\circ}$ apex angle and no flat-bottom portion, which is consistent with $h' \approx \SI{66}{\nm}$. 
As is apparent in \cref{fig:model_scil1}, the two models for the SCIL replica yield similar results, where in each case, the centroids of peak orders roughly match the data for $\delta' = 28^{\circ}$ while the overall efficiency in each peak order is over-estimated, even with facet roughness taken into account. 
The primary difference between the two models is that the more detailed groove facet in the top panel predicts slightly lower efficiency in peak orders so as to match the data more closely than does the model that assumes an ideal groove facet. 

With both model variants in \cref{fig:model_scil1} giving nearly-identical results in terms of peak-order centroids, the ideal sawtooth model in the right panel offers more simplicity for analyzing the impact of resist shrinkage due to $\delta'$ essentially being the only free parameter for a fixed grating geometry. 
This latter model therefore can be used to constrain the shrunken blaze angle by comparing the data to several models with varying values for $\delta'$, so long as the models are normalized to match the peaks of the measured data as in \cref{fig:model_si2}. 
\begin{figure}
 \centering
 \includegraphics[scale=0.25]{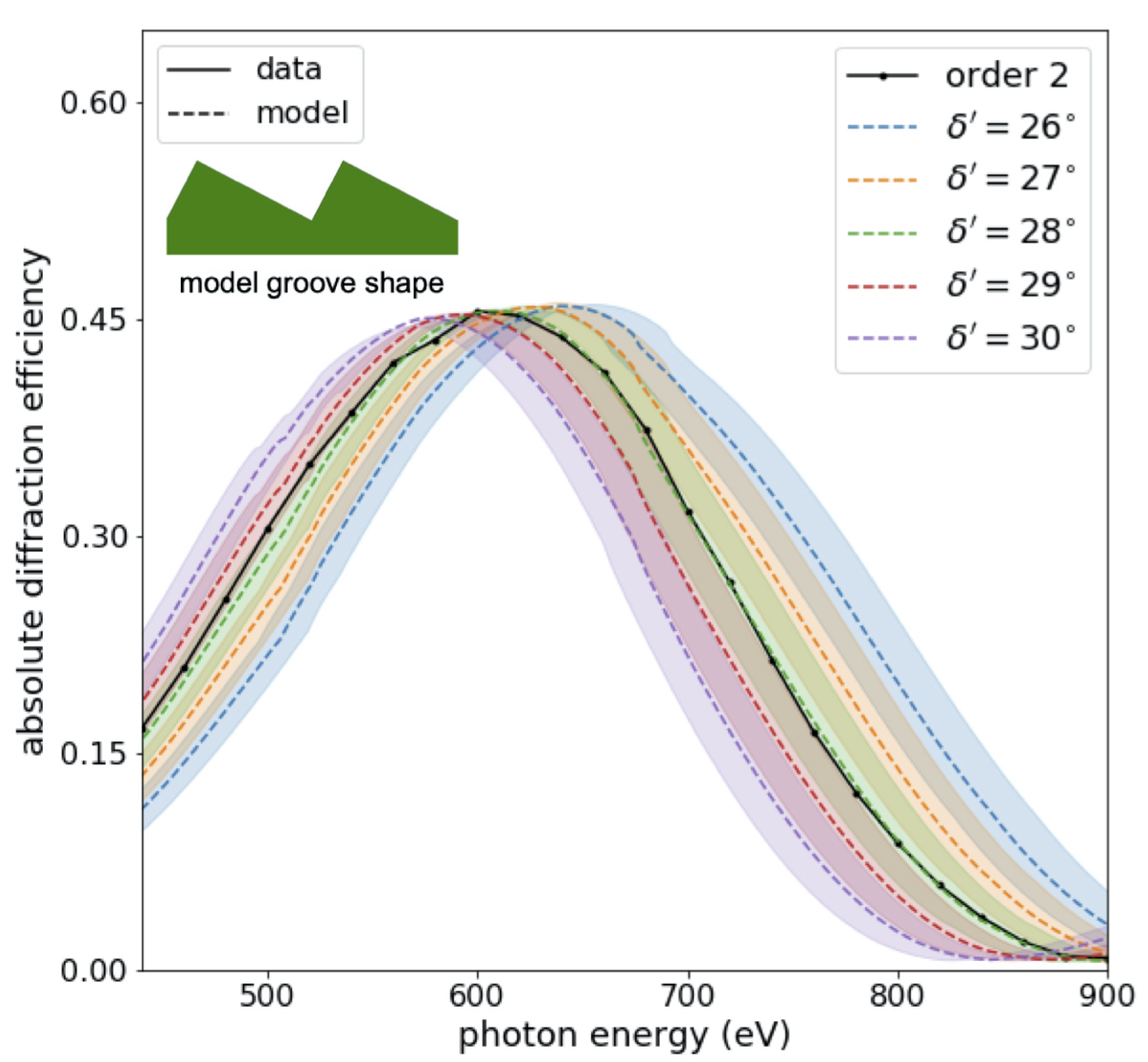}
 \includegraphics[scale=0.25]{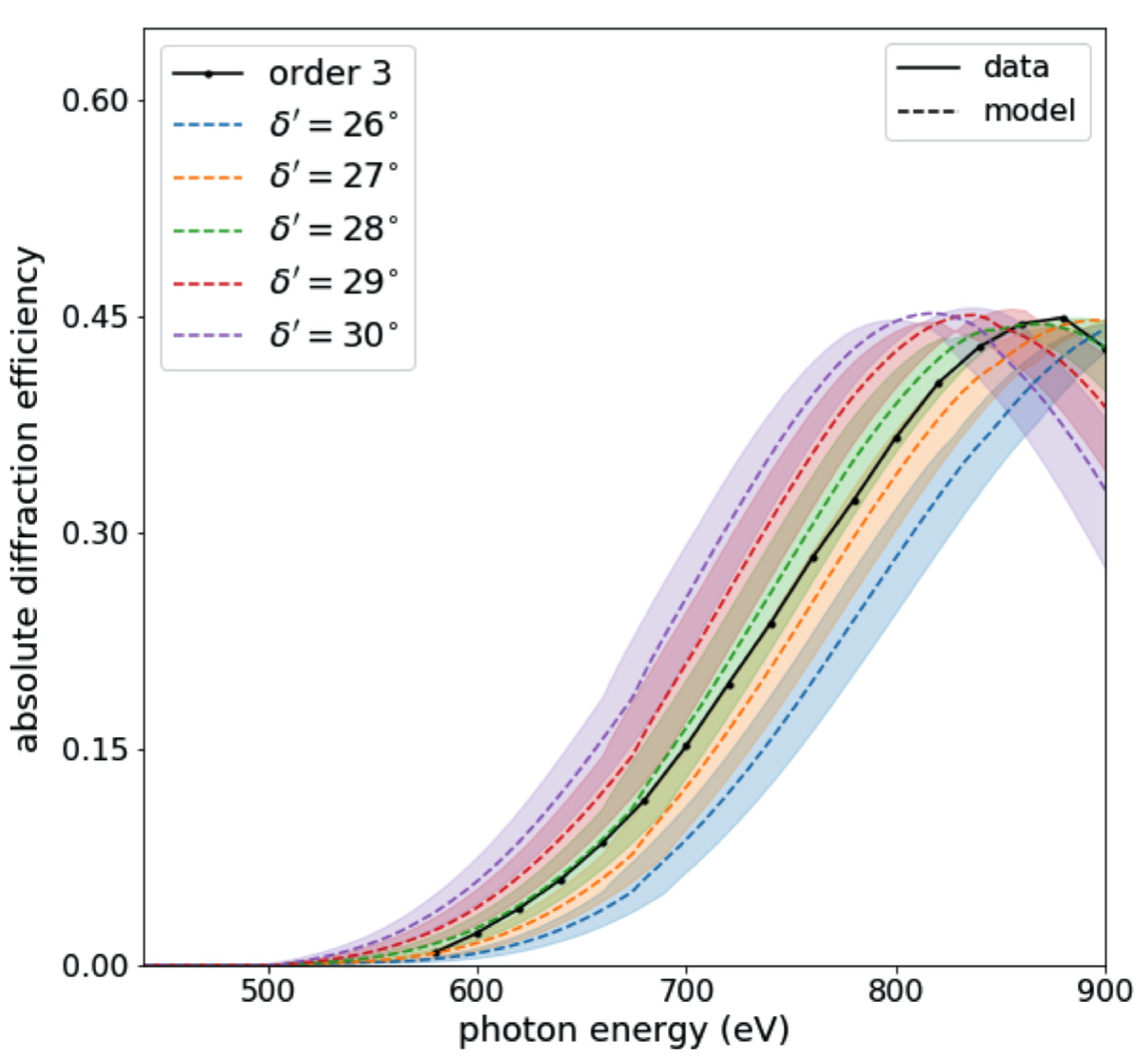}
 \caption[Measured absolute diffraction efficiency data in in second and third orders for the coated SCIL replica compared to \textsc{PCGrate-SX} models that assume an ideal sawtooth with varying blaze angles.]{Measured absolute diffraction efficiency data with $n=2$ and $n=3$ for the coated SCIL replica, corrected for surface roughness using \cref{eq:Au_refl}, and compared to \textsc{PCGrate-SX} models that assume an ideal sawtooth [\emph{cf.\@} \cref{fig:model_scil1}, \emph{bottom panel}] with $26^{\circ} \leq \delta' \leq 30^{\circ}$, $\alpha = 30.7^{\circ}$ and $\gamma = 1.75 \pm 0.04^{\circ}$ (\emph{shaded swaths}), which have been normalized to match the data to show that the measured data most closely match a grating with $\delta' = 28^{\circ}$ \cite{McCoy20b}.}\label{fig:model_scil2} 
 \end{figure} 
This is demonstrated in \cref{fig:model_scil2}, where the absolute diffraction data for the SCIL replica in orders $n=2$ and $n=3$ are each plotted against five \textsc{PCGrate-SX} models with $26^{\circ} \leq \delta' \leq 30^{\circ}$ in steps of $1^{\circ}$, all with $\alpha = 30.7^{\circ}$ and $\gamma = 1.75 \pm 0.04^{\circ}$, the latter of which is represented by uncertainty swaths. 
The data plotted in \cref{fig:model_scil2} have been corrected for surface-roughness losses through division of the Nevot-Croce exponential term in \cref{eq:Au_refl} while the models assume perfectly smooth surfaces and are normalized to the data in terms of peak efficiency. 

From the results presented in \cref{fig:model_scil2}, it is apparent that the measured data are most consistent with the $\delta' = 28^{\circ}$ model, as expected from AFM measurements.  
Although this analysis does not tightly constrain $\delta'$, it does demonstrate that the SCIL replica functions as a blazed grating with a facet angle reduced by $\sim 2^{\circ}$ relative to the silicon master that has been shown to exhibit $\delta \approx 30^{\circ}$. % a blaze angle close to $\delta = 30^{\circ}$. 
\begin{figure}
 \centering
 \includegraphics[scale=0.25]{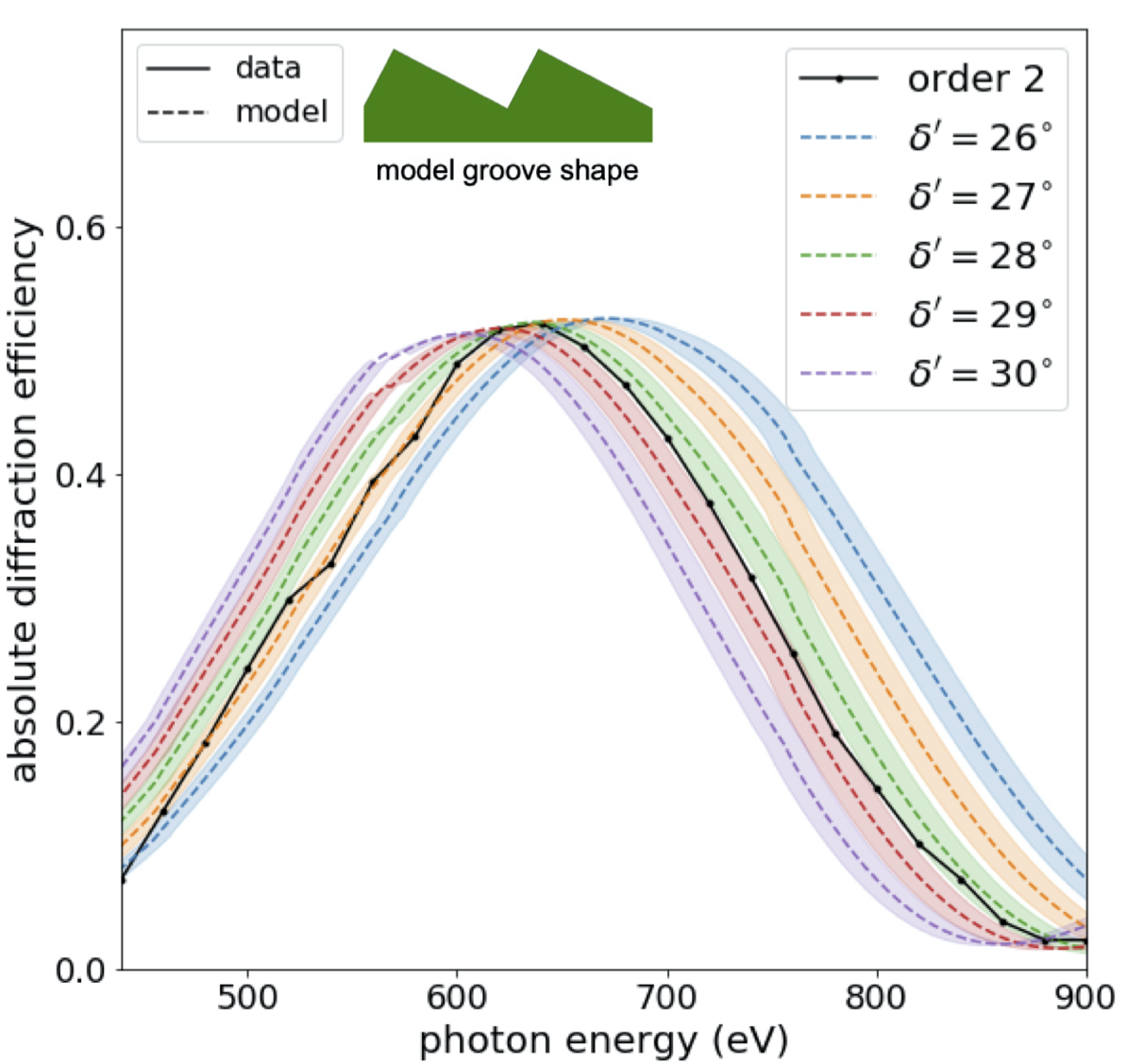}
 \includegraphics[scale=0.25]{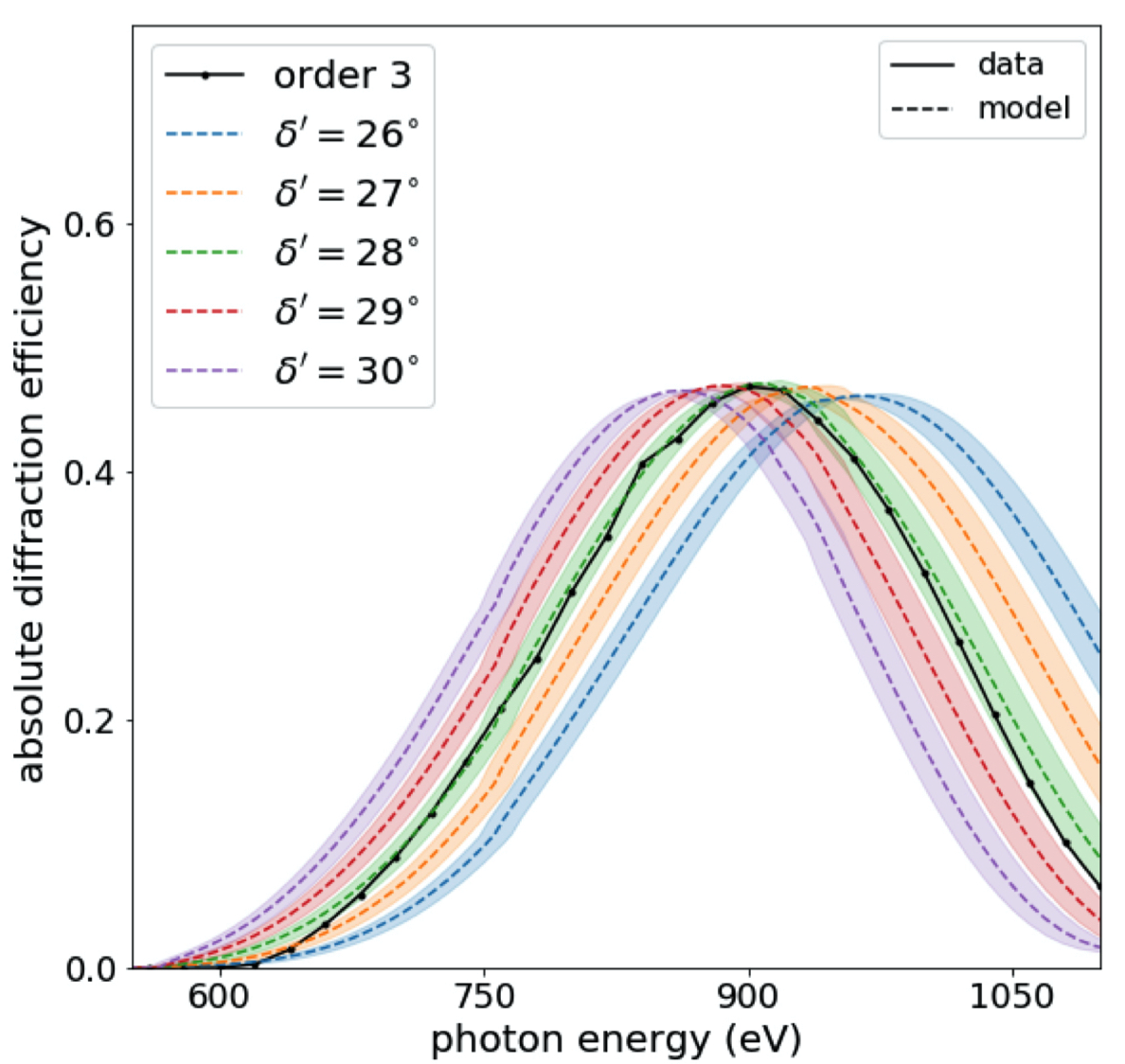}
 \caption[Measured absolute diffraction efficiency data in second and third orders for a coated UV-NIL replica]{Measured absolute diffraction efficiency data with $n=2$ and $n=3$ for the coated UV-NIL replica \cite{Miles18} compared to \textsc{PCGrate-SX} models that assume an ideal sawtooth [\emph{cf.\@} \cref{fig:model_scil1}, \emph{bottom panel}] with $26^{\circ} \leq \delta' \leq 30^{\circ}$, $\alpha = 24.5^{\circ}$ and $\gamma = 1.66 \pm 0.02^{\circ}$ (\emph{shaded swaths}). As in \cref{fig:model_scil2}, the measured data were corrected for surface roughness using \cref{eq:Au_refl} and normalized to match the data to show that the data most closely match $\delta' = 28^{\circ}$.}\label{fig:model_uvnil} 
 \end{figure} %These results also show that the measured data most closely match a grating with $\delta' = 28^{\circ}$.
This modeling scheme was repeated using the soft x-ray diffraction efficiency data presented by Miles, et~al.~\cite{Miles18}, which were gathered using a gold-coated UV-NIL replica produced from the same silicon master [\emph{cf.\@} \cref{fig:uv_nil_AFM}] in a geometry with $\alpha = 24.5 \pm 1.6^{\circ}$ and $\gamma = 1.66 \pm 0.02^{\circ}$. 
While the shrinkage mechanism for UV-curable resists differs fundamentally from thermodynamically-curable sol-gel resist, these results shown in \cref{fig:model_uvnil} indicate that the measured data also most closely match $\delta' = 28^{\circ}$. 
This suggests that the impact of resist shrinkage in \SI{90}{\celsius}-treated sol-gel resist is comparable to that of standard UV-NIL with $\sim \SI{10}{\percent}$ volumetric shrinkage, as expected from the considerations presented in \cref{sec:grating_fab}. 

\subsection{Geometric Resist-Shrinkage Model}\label{sec:shrinkage_model}
%%%%%%%%%%%%%%%%%%%%%%%%%%%%%%%%%%%%%%%%%--------------------------------------------------
To model the observed reduction in blaze angle between the silicon master and the SCIL replica from \cref{sec:silicon_master_sub,sec:scil_replica_sub} in terms of resist shrinkage, it is first assumed that shrinkage effects in the SCIL stamp can be neglected owing to the high intrinsic cross-link density of X-PDMS \cite[\emph{cf.\@} footnote~\ref{footnote:X-PDMS}]{Verschuuren19}. 
The surface-relief profile of the imprinted blazed grating, before resist shrinkage, then is considered to be composed of a series of groove facets with spacing $d \lessapprox \SI{160}{\nm}$ that resembles the inverse of the silicon master [\emph{cf.\@} \cref{fig:master_grating_profile}]. 
These facets are separated from one another approximately by the distance $w \gtrapprox \SI{30}{\nm}$ defined by the width of flat-tops in the silicon master [\emph{cf.\@} \cref{fig:master_grating_profile}] so that the base of each groove facet has a width $b \approx d - w \lessapprox \SI{130}{\nm}$, which is assumed to be a small enough size scale for material relaxation in sol-gel resist. 
\begin{figure}
 \centering
 \includegraphics[scale=0.55]{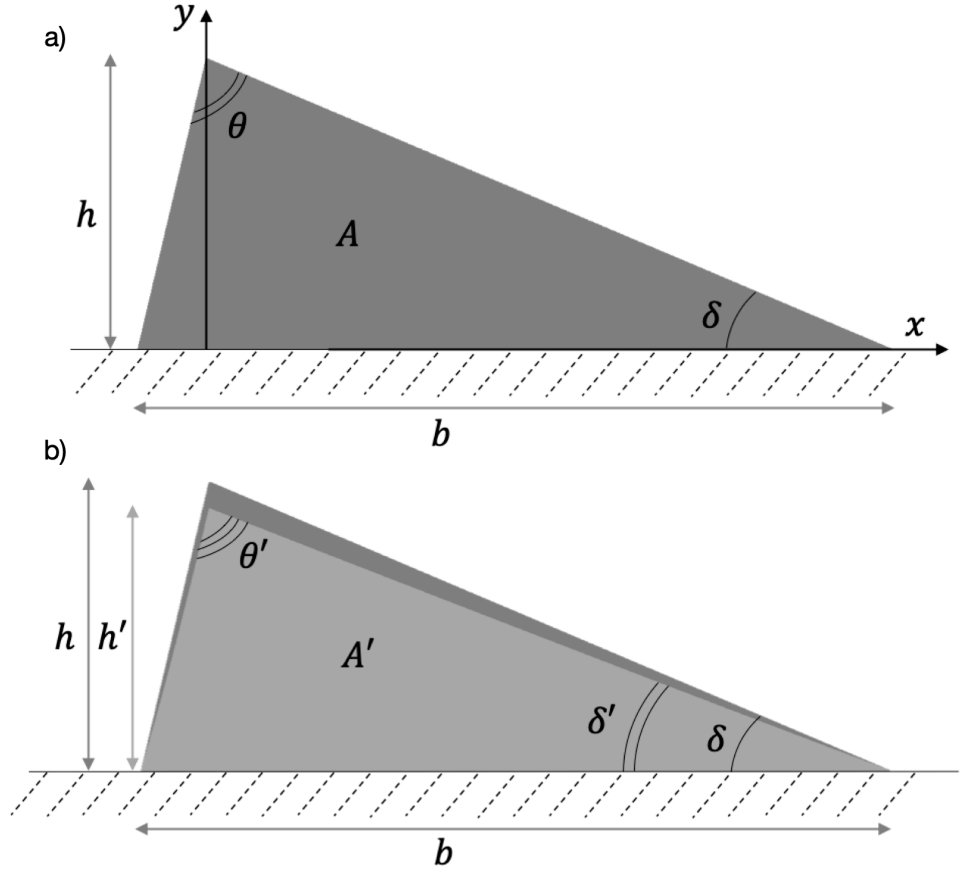}
 \caption[Approximate model for resist shrinkage]{Approximate model for resist shrinkage with $y=0$ representing a fixed boundary defined by the residual layer. a) The original facet shape has a blaze angle $\delta = 29.5^{\circ}$, an apex angle $\theta \approx 70.5^{\circ}$ and a cross-sectional area $A = bh/2$. b) A shrunken facet is generated by dividing the original facet shape into \num{1000} layers along the $y$-direction and then requiring that the area of each is reduced by \SI{10}{\percent} with $\chi = 0.1$ and $A'= 0.9 A$ while the ratio between lateral and vertical shrinkage varies as a function of $y$ according to \cref{eq:exp_shrink} for $S$ with $\ell_e / h = 0.05$. The result is reduction in groove depth with $h'/h \approx 0.91$, a steep sidewall featuring slight curvature, an increased apex angle with $\theta' / \theta \approx 1.05$ and crucially, a quasi-flat facet surface with a reduced blaze angle, $\delta' \approx 27.4^{\circ}$ with $\delta' / \delta \approx 0.93$ \cite{McCoy20b}.}\label{fig:shrinkage} 
 \end{figure} 
As illustrated in \cref{fig:shrinkage}(a), it is assumed that the shallow side of the facet takes on the nominal value of $\delta = 29.5^{\circ}$ while the protruding nubs of the master grating then are ignored for simplicity so that the groove depth with $\Delta h = 0$ is $h \lessapprox \SI{67}{\nm}$ by \cref{eq:groove_depth}, where $b/h \approx \cot \left( \delta \right) - \cot \left( \theta + \delta \right) \approx 1.944$ using $\theta \approx 70.5^{\circ}$. 
This angle, $\theta$, which is defined by the crystal structure of silicon [\emph{cf.\@} \cref{fig:master_grating_profile}], serves as the pre-shrinkage apex angle shown in \cref{fig:shrinkage}(a). 

Simulations of resist shrinkage in UV-NIL based on continuum mechanics of elastic media \cite{Shibata10,Horiba_2012} indicate that on average, local volumetric reduction is uniform such that a volume element, $V$, of a continuous medium shrinks to $V'$ given by 
\begin{equation}\label{eq:volume_shrinkage}
 V' = V \left( 1 - \chi \right) ,
 \end{equation}
with $\chi$ as the fractional loss in volume. 
In this regard, the residual layer of resist that exists beneath the groove facets [\emph{cf.\@} \cref{fig:SCIL_SEMs}] is expected to experience reduction in thickness alone such that the initial thickness, $\tau$, reduces to $\tau' = \tau \left( 1 - \chi \right)$ while stress-induced substrate deformation from this laterally-constrained shrinkage is considered to be negligible owing to the \SI{1}{\mm} thickness of the silicon wafer used for the grating replica. 
Each groove facet of the grating profile considered then has a volume given by $V = A \mathcal{L}$, where $A = b h / 2$ is its cross-sectional area and $\mathcal{L}$ is its length. 
Because $\mathcal{L}$ is orders of magnitude larger than the vertical and lateral extents of each groove facet over the \SI{72}{\cm\squared} grating area, however, shrinkage in an imprinted grating is considered to be completely restricted along the grove direction due to the inability of the material network to relax over such a macroscopic size scale. 
The shrunken volume then is given by $V' = A' \mathcal{L}$ so that \cref{eq:volume_shrinkage} reduces to 
\begin{equation}\label{eq:area_shrinkage}
 A' = A \left( 1 - \chi \right) , 
 \end{equation}
where $A'$ is cross-sectional area of a groove facet following resist shrinkage. 
Based on simulations for resist shrinkage in UV-NIL that verify a lack of dependence on residual-layer thickness \cite{Horiba_2012}, it is expected that residual-layer shrinkage does not impact the cross-sectional shape of the groove facets and hence the surface of the residual layer is treated as a fixed boundary. 

The simple resist-shrinkage model formulated here stems from the assumption that throughout each imprinted groove facet, the reduction in cross-sectional area is uniform in magnitude while the ratio of lateral shrinkage to vertical shrinkage varies spatially in a manner consistent with the boundary condition provided by the residual layer. 
This can be expressed by introducing $s_x$ and $s_y$ as functions of position that describe lateral and vertical shrinkage, respectively, so that \cref{eq:area_shrinkage} can be written as 
\begin{subequations}
\begin{equation}\label{eq:area_shrinkage2}
 \frac{A'}{A} = \left( 1 - \chi \right) = \left( 1 - s_x \right) \left( 1 - s_y \right) .
 \end{equation}
Then, defining $S \equiv s_x/s_y$ as the shrinkage aspect ratio, \cref{eq:area_shrinkage2} becomes
\begin{equation}
 \left( 1 - \chi \right) = \left( 1 - S f \right) \left( 1 - f \right) \implies S f^2 - \left( 1 + S \right) f + \chi = 0
 \end{equation}
with $s_x = S f$ and $s_y = f$. 
Considering only $1 \geq S > 0$, this quadratic equation has solutions for $f$ given by 
\begin{equation}\label{eq:quad_solution}
 f = \frac{1 + S - \sqrt{(1+S)^2 - 4 S \chi}}{2 S} ,
 \end{equation}
\end{subequations}
where $f \leq 1$ is ensured so that $A'$ is always positive.  
The fixed boundary provided by the residual layer requires $S = 0$ at $y=0$ so that $s_x = 0$ and $s_y = \chi$ near the base of each groove facet of width $b$. 
On the other hand, $s_x$ should increase from zero as the distance from the boundary increases until ultimately, free shrinkage occurs both laterally and vertically due to the previously-stated assumption that $b \lessapprox \SI{130}{\nm}$ is a small enough size scale for material relaxation in sol-gel resist. 
The shrinkage aspect ratio, $S$, therefore should be a function of $y$ that grows from zero at $y=0$ to a value approaching unity with $s_x = s_y = f = 1 - \sqrt{1 - \chi}$ as $y$ increases toward $h$. 
For the present discussion, this behavior for $0 \leq y \leq h$ is taken to be described approximately by 
\begin{equation}\label{eq:exp_shrink}
 S = 1 - \mathrm{e}^{- y / \ell_e} ,
 \end{equation}
where $\ell_e$, while formally an unknown quantity, is defined as the vertical distance from the boundary at which $S$ reaches $1 - \left( 1 / \mathrm{e} \right) \approx 0.63$. 

The spatially-varying function for $S$ given by \cref{eq:exp_shrink} was incorporated into the resist-shrinkage model by first considering the original groove facet shape shown in \cref{fig:shrinkage}(a) to be composed of \num{1000} rectangular layers, each with an identical, thin, vertical thickness and lateral widths that vary with $y$ so as to emulate a sawtooth with $\delta = 29.5^{\circ}$ and $\bar{\delta} = 180^{\circ} - \theta - \delta \approx 80^{\circ}$ slopes. 
Using \cref{eq:exp_shrink} discretized into \num{1000} $y$ steps for $S$ along with \cref{eq:quad_solution} for $f$, a shrunken facet profile could be produced by requiring the area of each of these layers to be reduced according to $s_x = S f$ and $s_y = f$ for specified values of $\chi$ and $\ell_e$. 
The result in any case is a reduction in groove depth with $h' < h$ while the base of the facet retains its width $b$, causing an increased apex angle, $\theta' > \theta$, and a small level of curvature where $y \lessapprox \ell_e$. 
The active groove facet then flattens to a linear slope as $y$ becomes larger than $\ell_e$ and because of this, the reduced blaze angle, $\delta'$, is extracted from the model by measuring slope only in the upper-half of the facet, where $S \lessapprox 1$ for relatively small values of $\ell_e / h$. 
While the model outputs $h'$ and $\delta'$ are independent from one another, they are related approximately through
\begin{equation}\label{eq:reduced_depth}
 h' \lessapprox \frac{\tan \left( \delta ' \right)}{\tan \left( \delta \right)} h ,
 \end{equation}
but without accurate depth measurements of the silicon master, or any AFM measurements of the composite stamp, it cannot be ruled out that groove-apex rounding during stamp construction and imprint production also contributes to this groove depth reduction. 
Consequently, a measured value for $h'/h$ in this case does not provide a meaningful constraint on resist shrinkage and is not considered further. 
\begin{figure}
 \centering
 \includegraphics[scale=0.55]{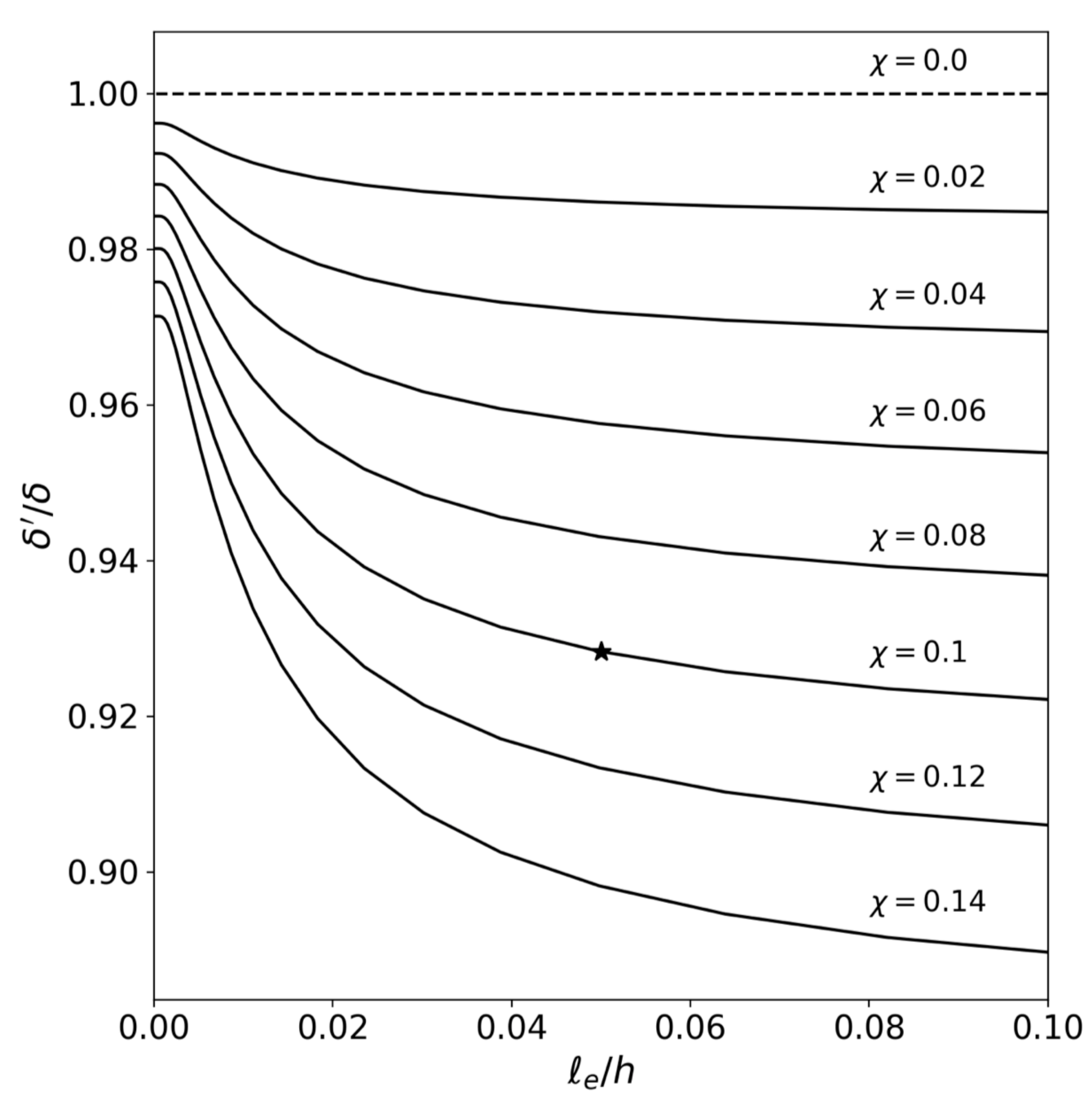}
 \caption[Reduced blaze angle predicted by the resist-shrinkage model relative to the initial blaze angle as a function of various parameters.]{Reduced blaze angle predicted by the resist-shrinkage model relative to the initial blaze angle, $\delta' / \delta$, as a function of $\ell_e / h$ for various values of $\chi$. With $\chi = 0.1$ and $\ell_e / h = 0.05$ marked by the star, the modeled shrunken facet depicted in \cref{fig:shrinkage} exhibits $\delta' / \delta \approx 0.93$ with $\delta' = 27.4^{\circ}$ and $\delta = 29.5^{\circ}$ \cite{McCoy20b}.}\label{fig:shrinkage_compare} 
 \end{figure}
Similarly, an approximate expression for $\theta'$ depends on $h'$ and $\delta'$:
\begin{equation}
 \cot \left( \theta' \right) \lessapprox \frac{h'}{b} \csc^2 \left( \delta' \right) - \cot \left( \delta' \right)
 \end{equation}
but this parameter is also not constrained experimentally in the present study. 

Predicted values for $\delta' / \delta$ are plotted as a function of $\ell_e / h$ for various values of $\chi$ in \cref{fig:shrinkage_compare}, where the marked star indicates the inputs used for the shrunken facet depicted in \cref{fig:shrinkage}(b). 
Despite $\ell_e / h$ remaining poorly constrained without measurements for $h'/h$ and $\theta' / \theta$, the comparison between the resist-shrinkage model just presented and $\delta' / \delta \approx 0.93$ determined from diffraction-efficiency analysis along with AFM measurements supports the hypothesis stated in \cref{sec:grating_fab} that the level of volumetric shrinkage for a $T_{\text{cure}} = \SI{90}{\celsius}$-treated sol-gel imprint is approximately \SI{10}{\percent}.  
Although this analysis does not tightly constrain $\delta'$, it does demonstrate that the SCIL replica functions as a blazed grating with a facet angle reduced by $\sim 2^{\circ}$ relative to the silicon master, which has been shown to exhibit a blaze angle of $\delta \approx 30^{\circ}$, giving a value for $\delta' / \delta$ that is consistent with a typical shrunken facet with $\chi \approx 0.1$. 
The outputs from this illustrated model using $\chi = 0.1$ and $\ell_e / h = 0.05$ as inputs are $h' \approx 0.91 h$, $\theta' \approx 74.1^{\circ}$ and $\delta' \approx 27.4^{\circ}$ with $\delta' / \delta \approx 0.93$ relative to the assumed initial blaze angle.\footnote{While these models were formulated assuming $\delta = 29.5^{\circ}$, the quantity $\delta' / \delta$ scales approximately with choice of $\delta$.} 

\section{Summary and Conclusions}\label{sec:conclusion}
%%%%%%%%%%%%%%%%%%%%%%%%%%%%%%%%%%%%%%%%%--------------------------------------------------
This chapter describes a SCIL fabrication process for a blazed-grating surface relief imprinted in \textsc{NanoGlass T1100}, a thermodynamically-curable silica sol-gel resist, and the subsequent soft x-ray diffraction-efficiency testing of the grating in an extreme off-plane mount after it was sputter-coated with a thin layer of gold for reflectivity using chromium as an adhesion layer. 
%Although a few diffracted orders were missed systemically during data collection at beamline 6.3.2 of the ALS, 
The collected peak-order efficiency measurements are comparable to previous results obtained for a UV-NIL replica coated with the same material and produced from the same master template, which was fabricated through a process centering on EBL and crystallographic etching in $\langle 311 \rangle$-oriented silicon \cite[\emph{cf.\@} \cref{sec:crystal_etching}]{Miles18}. 
Further analysis that consists of matching theoretical models to measured data shows that while this silicon master yields diffraction efficiency results that are close to a nominal $\langle 311 \rangle$ blaze angle with $\delta \approx 30^{\circ}$, the response of the coated SCIL replica is consistent with a reduced blaze angle of $\delta' \approx 28^{\circ}$, which is the same approximate result obtained for the corresponding UV-NIL replica \cite[\emph{cf.\@} \cref{sec:nanoimprint}]{Miles18}. 
This supports the hypothesis that the replicated grating, which was treated with a $T_{\text{cure}} = \SI{90}{\celsius}$ bake following stamp separation, experienced volumetric shrinkage in the sol-gel resist on the level of \SI{10}{\percent} to provide a blaze angle reduced by $\sim \SI{7}{\percent}$ to give $\delta - \delta' \approx 2^{\circ}$ for $\delta \approx 30^{\circ}$.  
While this result could be better constrained through further diffraction-efficiency testing and more rigorous modeling for resist shrinkage,
it serves as experimental evidence for resist shrinkage in SCIL impacting the performance of a reflection grating in terms of its ability to maximize diffraction efficiency for a specific diffracted angle, $\beta = 2 \delta - \alpha$. 
This is of particular relevance to instrument development for astrophysical soft x-ray spectroscopy that relies on the production of large numbers of identical gratings \cite{Miles19b,Tutt18,McEntaffer19}. 
The impact of resist shrinkage therefore should be compensated for in the fabrication of the master grating to ensure that grating replicas perform as expected. 
% !TEX root = ../McCoy-Dissertation.tex
\chapter{Conclusions and Outlook}\label{ch:conclusions}
%%%%%%%%%%%%%%%%%%%%%%%%%%%%%%%%%%%%%%%%%--------------------------------------------------
This dissertation contributes to the field of high-energy astrophysics through technological development of specialized reflection gratings that push the state of the art for x-ray spectroscopy. 
A main scientific goal for the currently-planned \emph{Lynx X-ray Observatory} \cite{Gaskin19} is to diagnose highly-ionized portions of extended galactic halos and intergalactic medium by measuring the equivalent widths of weak absorption lines produced from hydrogen-like and helium-like species of oxygen, and other $\mathcal{Z} \geq 6$ elements [\emph{cf.\@} \cref{tab:astro_element_nuclei_mass}], along the line-of-sight of active galactic nuclei [\emph{cf.\@} \cref{sec:astro_plasmas}].  
The detection of these weak spectral lines, which fall in the soft x-ray bandpass [\emph{cf.\@} \cref{tab:astro_element_lyman_approx,tab:astro_element_he_res}], relies not only on having a large instrument collecting area for spectral sensitivity, $A_{\text{col}}$, but also on achieving high spectral resolving power, $\mathscr{R}$, with a figure of merit depending on $\sqrt{A_{\text{col}} \mathscr{R}}$ [\emph{cf.\@} \cref{eq:FOM_weak}]. 
As discussed in \cref{sec:grating_tech_dev,sec:ch1_conclusions}, implementing reflection gratings for the \emph{X-ray Grating Spectrometer (XGS)} planned for \emph{Lynx} \cite{McEntaffer19} centers on
\begin{enumerate}[noitemsep]
	\item the production of a master grating with a sawtooth surface-relief profile that enables 
	\begin{enumerate}[noitemsep]
    	\item total absolute diffraction efficiency in the soft x-ray exceeding \SI{40}{\percent}, and 
        \item $\mathscr{R} \gtrapprox 5000$ in a Wolter-I telescope
    \end{enumerate}
	and additionally,  
	\item the mass manufacture of grating replicas to populate modular grating arrays, which involves 
	\begin{enumerate}[noitemsep]
    	\item imprinting grating surface-relief molds, and 
        \item coating each replica for soft x-ray reflectivity.  
    \end{enumerate}
\end{enumerate}
With these requirements motivated in \cref{ch:introduction}, \cref{ch:master_fab,ch:grating_replication} focus on applying the processes of \emph{thermally-activated selective topography equilibration (TASTE)} \cite{Schleunitz14} and \emph{substrate-conformal imprint lithography (SCIL)} \cite{Verschuuren17} to x-ray reflection grating technology, with an emphasis on characterizing diffraction efficiency using the beamline methodology described in \cref{ch:diff_eff}. 
The findings from these studies on TASTE and SCIL are summarized in \cref{sec:TASTE_ch5,sec:SCIL_ch5}, respectively, with future work described therein. 
Finally, outlook for potential future studies is discussed in \cref{sec:summary_ch5}.
%outlook for future work provided therein. 
%Finally, a summary and a discussion on future work is provided in \cref{sec:summary_ch5}.

\section{TASTE for Master Grating Fabrication}\label{sec:TASTE_ch5}
%%%%%%%%%%%%%%%%%%%%%%%%%%%%%%%%%%%%%%%%%--------------------------------------------------
Coupling \emph{grayscale electron-beam lithography (GEBL)} with selective \emph{thermal reflow} \cite{Schleunitz10,Schleunitz11a,Schleunitz14,Kirchner19}, TASTE offers an avenue for patterning sawtooth-like topographies in polymeric resist over a custom groove layout. 
%Utilizing this approach for 
This approach to grating fabrication eliminates dependence on substrate crystal structure, thereby providing a way for a radially-ruled, blazed grating layout to be generated with the precision of EBL, which, in principle, enables high $\mathscr{R}$ by reducing spectral aberrations from an imperfect radial profile [\emph{cf.\@} \cref{sec:crystal_etching}]. 
Through carrying out process development for TASTE using the \textsc{EBPG5200} tool at the Penn State Nanofabrication Laboratory \cite[\emph{cf.\@} \cref{fig:ebeam_tool_pic}]{PSU_MRI_EBL,raith_ebpg}, it was found that sub-\si{\um}, sawtooth-like topographies could be generated in \SI{130}{\nm}-thick poly(methyl methacrylate) (PMMA) [\emph{cf.\@} \cref{fig:PMMA}] by patterning repeating staircase patterns by GEBL and then performing an appropriate thermal annealing step. 
The results presented in \cref{sec:TASTE} show a path forward for fabricating x-ray reflection gratings using TASTE. 
Patterns with a groove spacing of $d = \SI{840}{\nm}$ yield a blaze angle $\delta \approx 10^{\circ}$ and are more suitable for in-plane geometries while patterns with $d = \SI{400}{\nm}$ provide a base for off-plane gratings with $\delta \approx 25^{\circ}$ \cite{McCoy18}. 

Starting with a $d = \SI{400}{\nm}$ test pattern from \cref{sec:TASTE}, a grating prototype was fabricated by coating TASTE-processed PMMA on a silicon wafer with a thin layer of gold for reflectivity by \emph{electron-beam physical vapor deposition (EBPVD)}, using titanium as an adhesion layer [\emph{cf.\@} \cref{sec:TASTE_prototype}]. 
This functional reflection grating was then tested for diffraction efficiency in an extreme off-plane mount over the photon energy range \SIrange{80}{800}{\electronvolt} at beamline 6.3.2 of the Advanced Light Source (ALS) at Lawrence-Berkeley National Laboratory \cite{ALS_632,Underwood96,Gullikson01,LBNL_web}. 
%As described in \cref{sec:taste_eff_results}, 
The results show absolute, peak-order efficiency ranging from \SIrange{75}{25}{\percent} as photon energy, $\mathcal{E}_{\gamma}$ increases while total diffraction efficiency, $\mathscr{E}_{\text{tot}}$, is maintained at $\lessapprox \SI{60}{\percent}$ across the tested bandpass, which demonstrates that x-ray reflection gratings fabricated by TASTE are capable of meeting \emph{Lynx} requirements in terms of spectral sensitivity \cite[\emph{cf.\@} \cref{sec:taste_eff_results}]{McCoy20}. 
\begin{table}[]
\centering
\caption[Benchmark table of blazed, off-plane x-ray reflection gratings in terms of total diffraction efficiency near the L-shell resonance transition in \ion{O}{vii}]{Benchmark table of blazed, off-plane x-ray reflection gratings in terms of \emph{total absolute diffraction efficiency} ($\mathscr{E}_{\text{tot}} \equiv \sum_n \mathscr{E}_n$ for all propagating orders with $n \neq 0$) near the L-shell resonance transition in \ion{O}{vii} at $\mathcal{E}_{\gamma} \approx \SI{574}{\electronvolt}$ [\emph{cf.\@} \cref{tab:astro_element_he_res}]. Refs~\cite{Werner77,Cash82,McEntaffer04} use data gathered at discrete energies using electron-impact sources; in these cases $\mathscr{E}_{\text{tot}}$ is referenced at the $\mathcal{E}_{\gamma} \approx \SI{525}{\electronvolt}$ K-shell fluorescence line in neutral oxygen. Refs~\cite{Tutt16,DeRoo16thesis,Miles18,Miles19,McCoy20,McCurdy20} use data gathered at synchrotron facilities; in these cases $\mathscr{E}_{\text{tot}}$ is referenced using measurements closest to $\mathcal{E}_{\gamma} \approx \SI{574}{\electronvolt}$ (typically within \SI{25}{\electronvolt}). For comparison, critical transmission gratings yield $\mathscr{E}_{\text{tot}} \approx \SI{30}{\percent}$ near  $\mathcal{E}_{\gamma} \approx \SI{574}{\electronvolt}$ \cite{Heilmann11,Heilmann16}. Items are listed in order of publication year (1977-2020).}\label{tab:eff_benchmark}
\begin{tabular}{cccc}
\cline{1-4}
fabrication           & grating grooves & geometry & $\mathscr{E}_{\text{tot}}$ \\ \hline
mechanical ruling \cite{Werner77} & $\delta \approx 5^{\circ}$, $d \approx \SI{278}{\nm}$ (\ce{Au}) & $\alpha \approx 0$, $\gamma \lessapprox 2^{\circ}$   & $\sim \SI{20}{\percent}$            \\
mechanical ruling \cite{Cash82}   & $\delta \approx 21^{\circ}$, $d \approx \SI{167}{\nm}$ (\ce{Au})  & $\alpha \approx \delta$, $\gamma \lessapprox 2^{\circ}$  & $\sim \SI{25}{\percent}$           \\
holography/ion etch \cite{McEntaffer04} & $\delta \approx 9^{\circ}$, $d \approx \SI{236}{\nm}$ (\ce{Pt})  & $\alpha \approx \delta$, $\gamma \approx 2.7^{\circ}$      & \SI{27}{\percent}            \\
holography/ion etch \cite{Tutt16} & $\delta \approx 9^{\circ}$, $d \approx \SI{236}{\nm}$ (\ce{Pt})  & $\alpha \approx 0$, $\gamma \approx 2^{\circ}$ & $\sim \SI{25}{\percent}$  \\
holography/ion etch \cite{Tutt16} & $\delta \approx 16^{\circ}$, $d \approx \SI{170}{\nm}$ (\ce{Pt})  & $\alpha \approx 0$, $\gamma \approx 2^{\circ}$ & $\sim \SI{30}{\percent}$  \\
\ce{KOH} etch (via EBL) \cite{DeRoo16thesis}  & $\delta \lessapprox 55^{\circ}$, $d = \SI{160}{\nm}$ (\ce{Au})  & $\alpha \approx  \delta$, $\gamma \approx 2.6^{\circ}$  & $\gtrapprox \SI{35}{\percent}$  \\
\ce{KOH} etch/UV-NIL \cite{Miles18}  & $\delta \lessapprox 30^{\circ}$, $d \lessapprox \SI{160}{\nm}$ (\ce{Au})  & $\alpha \approx 25^{\circ}$, $\gamma \approx 1.7^{\circ}$         & $\lessapprox \SI{60}{\percent}$  \\
\ce{KOH} etch/SCIL \cite{Miles19}  & $\delta \lessapprox 30^{\circ}$, $d \lessapprox \SI{174}{\nm}$ (\ce{Ni})  & $\alpha \approx 23^{\circ}$, $\gamma \approx 2.5^{\circ}$         & $\sim \SI{55}{\percent}$  \\
TASTE \cite[\emph{cf.\@} \cref{ch:master_fab}]{McCoy20}  & $\delta \approx 27^{\circ}$, $d = \SI{400}{\nm}$ (\ce{Au})  &  $\alpha \approx 23^{\circ}$, $\gamma \approx 1.7^{\circ}$        & $\lessapprox \SI{60}{\percent}$            \\
TASTE \cite{McCurdy20}  & $\delta \approx 35^{\circ}$, $d = \SI{160}{\nm}$ (\ce{Au}) & $\alpha \approx \delta$, $\gamma \approx 1.8^{\circ}$         & \SI{43}{\percent}  \\ \hline
\end{tabular}
\end{table}
This measurement for $\mathscr{E}_{\text{tot}}$ is compared to analogous results from other blazed x-ray reflection gratings in \cref{tab:eff_benchmark}, where it can be seen that \ce{KOH}-etched gratings and TASTE-fabricated gratings offer substantial improvement over gratings manufactured by mechanical ruling engine or by holographic recording coupled with directional ion etching [\emph{cf.\@} \cref{sec:grating_tech_dev}]. 

%Because this latter condition relies on the realization of a radial groove profile that matches the focal length of the telescope, beamline testing for $\mathscr{R}$ is required to make this determination. 
%Achieving a high-fidelity radial profile in part requires careful control of field stitching with the \textsc{EBPG5200} especially along the grating-dispersion direction so that irregularities in $d$, which can lead to a degradation in $\mathscr{R}$ as alluded to in \cref{sec:binary_ebeam}, can be avoided. 

While the prototype diffraction-efficiency results presented in \cref{sec:taste_eff_results} have been shown to be characteristic of a blazed grating with $\delta \approx 27^{\circ}$ through \textsc{PCGrate-SX} modeling \cite[\emph{cf.\@} \cref{sec:integral_method}]{PCGrate_web,Goray10}, there exists room for improvement in this fabrication process. 
In particular, the fidelity of the sawtooth-like groove facets hinges on the selectivity of the thermal reflow step, which in turn depends on the distribution of average molecular weight in the resist, $M_w$, imparted via high-energy electron exposure in GEBL [\emph{cf.\@} \cref{fig:TASTE_diagram}]. 
This material contrast in $M_w$ establishes a lateral gradient in the polymer \emph{glass transition temperature}, $T_g$, such that the viscosity of molten resist at a given temperature decreases with $M_w$. 
Selective thermal reflow is then induced by heating the resist by hotplate to a carefully chosen temperature, $T_{\text{reflow}}$, that is lower than $T_g$ for unexposed portions of the resist (\emph{i.e.}, top steps in a staircase pattern), but higher than the range of $T_g$ associated with electron-exposed resist (\emph{i.e.}, the remaining, lower steps in a staircase pattern) \cite{Schleunitz14,Kirchner19}. 
Because resist can be dosed inadvertently from electron scattering, especially when the widths of staircase steps are comparable to the size of the focused electron beam (typically $\lessapprox \SI{40}{\nm}$ in the \textsc{EBPG5200}), the size of this process window in practice depends on critical dimensions of the GEBL layout. 
Thermal reflow selectivity in the $d = \SI{400}{\nm}$ grating prototype described in \cref{sec:TASTE_prototype} appears to be diminished relative to \si{\um}-scale TASTE patterns previously reported in the literature \cite{Schleunitz10,Schleunitz14,Kirchner16,Pfirrmann16}, which is evidenced by the slight bulging effect seen in \cref{fig:B_time_scan,fig:TASTE_AFMs}.  
Moreover, McCurdy, et~al.~\cite{McCurdy20} show that this effect is exacerbated at $d = \SI{160}{\nm}$, where a decreased selectivity for thermal reflow causes significant rounding in the top staircase step as well as facet-surface irregularities that ultimately cause lower overall diffraction efficiency compared to what is reported in \cref{sec:taste_eff_results} [\emph{cf.\@} \cref{tab:eff_benchmark}]. %arising from the GEBL layout 

Other limitations of TASTE for grating fabrication stem from its dependence on resist thickness for a given GEBL recipe. 
Beyond the need for a uniformly-thick layer of resist, which can be difficult to achieve consistently through spin-coating, the groove depth (\emph{i.e.}, the height of top steps in a staircase pattern) is determined essentially by the resist thickness. 
This then places a restriction on $\delta$ for a fixed value of $d$, where $\delta$ decreases with increasing $d$; for example, the results from \cref{sec:TASTE} show that groove spacings of $d = \SI{400}{\nm}$ and $d = \SI{840}{\nm}$ at a resist thickness of \SI{130}{\nm} yield $\delta \approx 25^{\circ}$ and $\delta \approx 10^{\circ}$, respectively. 
Changing the resist thickness for GEBL requires the establishment of a new resist-contrast curve [\emph{cf.\@} \cref{sec:resist_contrast}]. 
For any given resist thickness, however, the blaze angle is expected to be limited practically to $\delta \lessapprox 45^{\circ}$ due to the difficulties involved with patterning high-aspect-ratio structures in GEBL. 
On the other hand, a TASTE pattern can be transferred into silicon through a dry etch, and depending on the etch selectivity of the resist relative to the substrate, $\delta$ can, in principle, be steepened relative to what is patterned directly in resist. 
A desired etch selectivity may be achieved by using a different resist (\emph{e.g.}, ZEP520A, mr-PosEBR \cite{Kirchner16,Pfirrmann16}) or by modifying PMMA  through a process such as \emph{sequential infiltration synthesis} \cite{Tseng11}. 

Overall, TASTE provides an alternative approach to fabricating blazed gratings that perform with high diffraction efficiency at soft x-ray wavelengths. 
While \ce{KOH} etching may be capable of producing gratings that perform with higher diffraction efficiency owing to the smooth facets and well-defined structures generated by the process [\emph{cf.\@} \cref{sec:crystal_etching}], TASTE has the key advantage for variable-line-space gratings that the precision of $d$ is not limited by the crystal structure of the substrate [\emph{cf.\@} \cref{fig:crystal_structure}]. 
This is expected to be beneficial for the radially-ruled gratings that the \emph{XGS} reflection grating design calls for \cite{McEntaffer19}, but due to the relatively low throughput of TASTE, like other EBL-based processes, the implementation of a grating replication process is required [\emph{cf.\@} \cref{sec:SCIL_ch5}]. 
Moreover, the \emph{Extreme-Ultraviolet Stellar Characterization for Atmospheric Physics and Evolution (ESCAPE)} mission concept, with a goal of characterizing high-energy radiation in habitable zones surrounding M-dwarfs and their impact on the atmospheres of exoplanets, baselines a spectrometer design that incorporates two large-area blazed gratings featuring curved groove layouts that play a similar role for extreme ultraviolet (EUV) radiation in a \emph{Hettrick-Bowyer-I} telescope \cite{France19,France20,Hettrick84}. 
With \emph{ESCAPE} baselining $\sim \SI{60}{\percent}$ single-order diffraction efficiency in the EUV, the beamline results presented in \cref{sec:taste_eff_results} give an indication that TASTE is capable of meeting this goal for spectral sensitivity. 

Further work involving beamline testing for $\mathscr{R}$ is required to determine whether TASTE is capable of producing gratings that meet performance requirements for instruments such as \emph{ESCAPE} and the \emph{XGS} for \emph{Lynx} \cite{France19,McEntaffer19}. 
Smaller-scale development for efficient, high-resolution, x-ray reflection gratings is taking place for the \emph{The Off-plane Grating Rocket Experiment (OGRE)} \cite{DeRoo13,McEntaffer14,Donovan18b,Tutt18,Donovan19a,Donovan19b}, which calls for $\delta \lessapprox 30^{\circ}$ blazed gratings with $d \lessapprox \SI{160}{\nm}$ radial-groove layouts that match a \SI{3.5}{\metre} focal length. 
Along with optical-system testing for $\mathscr{R}$, grating development for \emph{ESCAPE} and \emph{OGRE} warrants further diffraction-efficiency testing to determine which blazed-grating manufacture produces curved or fanned grooves that are efficient while also offering high $\mathscr{R}$. 
A new ALS general user proposal to carry out diffraction-efficiency experimentation for \emph{ESCAPE} and \emph{OGRE} grating prototypes has recently been submitted for beamline testing in late 2021 and 2022. 
These prototypes include four grating variants, all with the same groove layout: one fabricated via EBL coupled with \ce{KOH} etching \cite{Miles18}, one via TASTE \cite{McCoy20,McCurdy20}, one via EBL coupled with directional ion milling \cite{Miles2019_ion_mill} and one analogous, laminar grating for comparison. 
Together with beamline results for $\mathscr{R}$, these results will inform which fabrication technique is best suited for achieving high efficiency and $\mathscr{R}$ simultaneously. 

%which seeks to observe emission lines that emanate from coronal activity in the Capella star system

%Further work involving beamline testing for $\mathscr{R}$, however, is required to determine if TASTE can produce gratings that perform with $\mathscr{R} \gtrapprox 5000$ in the converging beam of a Wolter-I telescope. 

\section{SCIL for Grating Replication}\label{sec:SCIL_ch5}
%%%%%%%%%%%%%%%%%%%%%%%%%%%%%%%%%%%%%%%%%--------------------------------------------------
SCIL is a variant of \emph{nanoimprint lithography (NIL)} designed for high-throughput patterning of nanoscale structures over large areas \cite{Verschuuren10,Verschuuren17}. 
%As described in \cref{sec:grating_fab}, 
The process utilizes a flexible stamp that is molded from a rigid master template to imprint features in resist with the aid of specialized pneumatic tooling [\emph{cf.\@} \cref{sec:grating_fab}]. 
In contrast to rigid-stamp UV-NIL, which has been pursued previously for x-ray reflection grating replication \cite[\emph{cf.\@} \cref{sec:nanoimprint}]{Chang03,Chang04,McEntaffer13,DeRoo16,Miles18}, SCIL enables conformal imprinting over wafers up to \SI{200}{\mm} in diameter without the need for high applied pressure. 
This in turn avoids damage to an expensive master template while reducing pattern defects and trapped air pockets. % and, 
Moreover, packaged equipment developed by \textsc{Philips SCIL Nanoimprint Solutions} \cite{philis_scil}, known commercially as \textsc{AutoSCIL}, serves to automate the imprinting process for high-volume production, which is most compatible with \textsc{NanoGlass}, an inorganic resist that cures through a thermodynamically-driven, silica sol-gel process \cite{Verschuuren10,Verschuuren17,Verschuuren19}. 
Using \textsc{AutoSCIL}, a single stamp is capable of producing $\gtrapprox \num{700}$ imprints in \textsc{NanoGlass} without pattern degradation, at a rate of \num{60} wafers per hour \cite{Verschuuren17,Verschuuren18}. 

In collaboration with \textsc{Philips SCIL Nanoimprint Solutions}, the \textsc{AutoSCIL} technique was first applied to x-ray reflection grating technology for the \emph{Water Recovery X-ray Rocket (WRXR)}, which was a Penn State sounding-rocket payload that utilized \num{26} replicas of a \ce{KOH}-etched silicon master for a diffuse-object, soft x-ray spectrometer \cite{Miles17,Miles18b,Miles19,Verschuuren18,Tutt19}. 
This production run served as a trial, indicating that the process is well suited for producing many replicas of a master grating for the purpose of populating modular arrays in a Wolter-I spectrometer, which is of particular importance for \emph{XGS} as it calls for thousands of identical gratings \cite{McEntaffer19}, in addition to upcoming rocket payloads that each call for hundreds of replicas: \emph{The Rockets for Extended-source X-ray Spectroscopy (tREXS)} \cite{Miles19b,Tutt19b} and \emph{OGRE} \cite{Tutt18,Donovan19a,Donovan19b}. 
As in virtually any NIL process, however, imprinted surface-relief features are prone to topographic distortion induced by resist shrinkage \cite{Horiba_2012}. 
This phenomenon occurs in \textsc{NanoGlass} as the silica precursors tetramethylorthosilicate (TMOS) and methyltrimethoxysilane (MTMS) contained in the resist [\emph{cf.\@} \cref{fig:precursors}] react to form a silica-like network with a cross-link density that depends on the post-imprint cure temperature, $T_{\text{cure}}$ \cite{Verschuuren10}. 
In \cref{ch:grating_replication}, it is demonstrated through diffraction-efficiency testing at beamline 6.3.2 of the ALS \cite{ALS_632,McCoy20b} that $T_{\text{cure}} \approx \SI{90}{\celsius}$ gives rise to a $\sim \SI{10}{\percent}$ volumetric shrinkage in the resist, which causes a $\sim 2^{\circ}$ reduction in $\delta$ relative to a $\sim 30^{\circ}$ facet angle defined by \ce{KOH} etching in $\langle 311 \rangle$-oriented silicon [\emph{cf.\@} \cref{fig:KOH_geo_311}]. 
With support from AFM measurements, this was carried out by comparing the blaze response of a test imprint provided by \textsc{Philips SCIL Nanoimprint Solutions} [\emph{cf.\@} \cref{fig:SCIL_SEMs}] and the corresponding \ce{KOH}-etched silicon master [\emph{cf.\@} \cref{fig:master_grating_SEM}]. %, which was fabricated by staff at the Penn State Materials Research Institute [\emph{cf.\@} \cref{fig:master_grating_SEM}]. 

The result just described shows that the impact of resist shrinkage on blaze angle is non-negligible and hence should be compensated for in the fabrication of a master grating \cite{McCoy20b}. 
Further studies to examine how the reduced blaze angle, $\delta'$, evolves with $T_{\text{cure}}$ should aid in this process, where fabricating a master grating with a precise value for $\delta$ may prove to be difficult. 
This is expected to be useful for achieving a blaze angle that is intermediate between what is is produced from \ce{KOH} etching in silicon with standard wafer orientations [\emph{cf.\@} \cref{tab:off_axis_si}]. 
Although the \textsc{AutoSCIL} production platform provides an avenue for high-volume production of grating imprints, PVD techniques such as EBPVD and \emph{plasma sputter deposition} are limited in throughput, and moreover, the impact of ion bombardment (from the latter process) on the sol-gel network has not yet been investigated. 
This motivates the pursuit of deposition strategies that are both capable of high throughput and compatible with \textsc{NanoGlass} resist. 

In addition to studying how $\delta'$ depends on $T_{\text{cure}}$ for a given groove geometry imprinted in resist and the effects of PVD coatings on groove shape, film-stress studies are warranted to determine how depositions of \textsc{NanoGlass} resist and metallic, reflective coatings contribute to substrate deformation \cite{Scheuerman69}. 
Similar to studies that examine how film stress affects the performance of x-ray telescope optics in terms of angular resolution \cite{Chan13,Probst18,Molnar-Fenton19}, compensating for this effect is motivated by the drive to preserve the flat figure of each grating in a spectrometer so that $\mathscr{R}$ is not degraded by unintentional substrate curvature. 
With the grooves of grating replicas being carried in \textsc{NanoGlass}, the nanoscale porosity of the resist network caused by the MTMS precursor is prone to the entrapment of water vapor in a manner similar to other sol-gel systems \cite{Hench90,Brinker90,Vishnevskiy19}. 
While it has been found experimentally that water present in the resist network can lead to minor imprinting issues that can be overcome \cite{Verschuuren10}, the effect of water vapor on aged imprints coated for reflectivity should be further investigated. 
Alternatively, the sol-gel network can be fully densified with $T_{\text{cure}} \gtrapprox \SI{850}{\celsius}$ to eliminate porosity but this comes with increased film stress as well as a smaller value for $\delta'$. % and potential groove-shape deformation. 

%stress studies and effect of water vapor on sol-gel network
%[To be expanded]

The nanofabrication processes of TASTE and SCIL, together, offer an avenue for manufacturing next-generation x-ray reflection gratings. 
With established process development for TASTE and equipment for SCIL stamp construction as well as low-volume, pneumatic imprinting (built by \textsc{S{\"U}SS MicroTec} \cite{suss_microtec}) recently installed at the Penn State Nanofabrication Laboratory \cite{PSU_MRI_nanofab}, future work will include studying the compatibility of SCIL with a master grating fabricated by TASTE.  %(using a mask aligner add-on built by \textsc{S{\"U}SS MicroTec})
Because polydimethylsiloxane (PDMS) [\emph{cf.\@} \cref{fig:PDMS}] and PMMA do not bond well without the use of an adhesive layer \cite{Tan10}, it is hypothesized that an anti-stiction treatment is not needed to construct an X-PDMS SCIL stamp from TASTE-processed resist; this is supported by a study that shows how PDMS and H-PDMS stamps used in \emph{soft UV-NIL} processes can be molded directly from untreated PMMA patterned by EBL \cite{Huelsen14}. As oxidized silicon is exposed between groove facets in a TASTE grating, however, it is likely that a small level of stiction will occur and hence stamp separation should be performed carefully.
The processes of TASTE and SCIL are nevertheless expected to be compatible for the production of grating replicas. 
This will be pursued experimentally using a duplicated, uncoated version of the TASTE prototype described in \cref{sec:TASTE_prototype}. 

\section{Outlook for Future Studies}\label{sec:summary_ch5}
%%%%%%%%%%%%%%%%%%%%%%%%%%%%%%%%%%%%%%%%%--------------------------------------------------
With its ability to generate blazed groove facets in resist over a custom layout defined by EBL, TASTE is a promising technique for the manufacture of state-of-the-art x-ray reflection gratings \cite[\emph{cf.\@} \cref{sec:TASTE_ch5}]{McCoy18,McCoy20}. 
The process does, however, face limitations in its ability to produce sharply-defined groove facets as the groove periodicity, $d$, becomes significantly smaller than \SI{400}{\nm} such that the width of each GEBL staircase step is comparable to the diameter of the focused electron beam \cite{McCurdy20}. 
While pattern resolution has potential for improvement through the implementation of a cold-development process \cite{Ocola06}, a fundamental size-scale limitation likely exists for GEBL \cite{Fallica17}. 
This motivates the use of \emph{EUV lithography} \cite{Wu07,Tallents10} or \emph{EUV interference lithography (EUV-IL)} \cite{Auzelyte09} at $\lambda \approx \SI{13.5}{\nm}$, which are processes capable of patterning on the sub-\SI{10}{\nm} level. 
Of particular interest is grayscale EUV-IL, which has been demonstrated to be capable of generating a $d = \SI{100}{\nm}$ tri-level staircase pattern in PMMA with the use of a transmission-grating EUV exposure mask that provides a two-fold reduction in pattern periodicity and a mask-shifting technique that enables dose-modulated exposure \cite{Fallica17,Buitrago16}. 
Such a process can, in principle, be coupled with thermal reflow to realize an EUV-IL variant of the TASTE process to produce a surface relief for a blazed x-ray reflection \cite{Fallica17,Schleunitz14}. 
Further investigation, however, is required to determine if this fabrication approach can produce a large-area groove layout with a high-fidelity radial profile that also enables high diffraction efficiency. 
This would necessitate the manufacture of a transmission-grating mask that is designed appropriately so as to produce a radial-groove diffraction pattern and additionally, the use of projection lithography equipment with a high-brightness source of EUV radiation (\emph{e.g.}, beamline 12.0.1 of the ALS \cite{ALS_EUVlitho}). 

This dissertation has considered only the fabrication of planar gratings; \emph{concave gratings} \cite{Loewen97}, however, are also of interest for their ability to focus and disperse radiation simultaneously. 
The fabrication of gratings on curved surfaces has been pursued through a variety of techniques throughout the years, including mechanical ruling \cite{Harada80,Kita83,Kita92}, holographic recording \cite{Sokolova04}, directional ion etching \cite{Liu15,Guo18}, soft lithography \cite{Xia98,Paul03,Choi04}, projection lithography \cite{Klosner02}, as well as hybrid approaches that combine aspects of soft and projection lithography \cite{Kim09,Park12}. 
Additionally, EBL is capable of patterning on curved substrates \cite{Wilson03,PSU_MRI_EBL,Lee18,Arat19} and this is of special interest in x-ray spectroscopy primarily for the following reason. 
That is, if a blazed x-ray grating topography can be patterned on the surface of a hyperbolic mirror appropriately, a two-element Wolter-I grating spectrometer could be realized \cite{DeRoo19,DeRoo19b}. 
Such an optical system would reduce the number of reflections required to produce x-ray spectra while reducing instrument mass and eliminating the need for grating-array alignment. 
However, strategies for the mass production of hyperbolic gratings should be considered. 
This may involve using a SCIL stamp to pattern on curved surfaces in a manner similar to soft-lithographic approaches \cite{Paul03,Choi04} but further research and experimentation is needed to make this assessment. 

Beyond telescopic applications of concave gratings, the nanofabrication technology discussed in this thesis is also of interest for laboratory x-ray astrophysics experiments \cite{Beiersdorfer03,Hell20} that utilize an \emph{electron-beam ion trap (EBIT)} \cite{Levine88a,Levine88b} to produce spectra from quasi-stationary, highly-charged ions. 
Such an x-ray source effectively provides diverging rays from a central electron beam, where a particular species of ion is trapped; this radiation must be focused and dispersed to produce spectra with appreciable $\mathscr{R}$ and high diffraction efficiency. % is desirable for spectral sensitivity. 
While commercially-available, mechanically-ruled concave gratings with in-plane, variable-line-space layouts have been implemented for EUV and soft x-ray spectroscopy on EBIT systems previously \cite{Beiersdorfer99,Utter99,Beiersdorfer04,Ohashi11,Shi14}, these instruments achieve at most $\mathscr{R} \approx 1200$ for long-$\lambda$ soft x-rays. 
This performance can, in principle, be improved by constructing a new EBIT grating spectrometer that is centered around a custom, blazed grating patterned on an ellipsoidal mirror segment that brings diverging rays to a focus while also producing an off-plane diffraction pattern that enables high-$\mathscr{R}$ spectra to be extracted through the use of high diffracted orders [\emph{cf.\@} \cref{sec:grating_tech_intro}]. 
Owing to its ability to pattern a blazed grating topography with the precision of EBL, TASTE is hypothesized to be capable of fabricating such a custom, concave grating provided that an appropriate curved substrate that focuses radiation effectively can be obtained. 
The development of a next-generation grating spectrometer for EBIT spectroscopy and its installment at an EBIT laboratory, such as those at Lawrence-Livermore National Laboratory \cite{llnl_ebit} and the Harvard-Smithsonian Center for Astrophysics \cite{cfa_ebit}, is crucial for furthering the fields of astrophysical x-ray spectroscopy and theoretical atomic physics by enabling wavelength centroids, transition rates and cross-sections of faint, poorly-studied, soft x-ray spectral lines to be measured in the laboratory so that they can complement astrophysical observations \cite{Smith19_labastro}. 
At the time of this writing, preliminary research is being undertaken to move forward with this project. 
%rather than relying on photon-starved observations of astrophysical plasmas

%EBIT \cite{Levine88a,Levine88b,Beiersdorfer03,Smith19_labastro}

%[To be expanded]

%%%%%%%%%%%%%%%%%%%%%%%%%%%%%%%%%%%%%%%%%%%%%%%%%%%%%%%%%%%%%%%
% Appendices
%
% Because of a quirk in LaTeX (see p. 48 of The LaTeX
% Companion, 2e), you cannot use \include along with
% \addtocontents if you want things to appear the proper
% sequence.
%%%%%%%%%%%%%%%%%%%%%%%%%%%%%%%%%%%%%%%%%%%%%%%%%%%%%%%%%%%%%%%
\appendix
\titleformat{\chapter}[display]{\fontsize{30}{30}\selectfont\bfseries\sffamily}{Appendix \thechapter\textcolor{gray75}{\raisebox{3pt}{|}}}{0pt}{}{}
% If you have a single appendix, then to prevent LaTeX from
% calling it ``Appendix A'', you should uncomment the following two
% lines that redefine the \thechapter and \thesection:
%\renewcommand\thechapter{}
%\renewcommand\thesection{\arabic{section}}
% !TEX root = ../McCoy-Dissertation.tex
\Appendix{Physics Introduction}\label{ap:x-ray_intro}
%%%%%%%%%%%%%%%%%%%%%e%%%%%%%%%%%%%%%%%%%%--------------------------------------------------
Basic physics relevant to this dissertation are outlined in \cref{ap:x-ray_intro,ap:quantum_spectral,app:x-ray_materials,ap:grating_basics} using SI units and the fundamental constants listed in \cref{tab:units,tab:fundamental_constants} throughout. 
\begin{table}[]
 \centering
 \caption[SI units for physical quantities]{Units for physical quantities in the \emph{System of International Units} (SI units) and SI derived units} \label{tab:units}
 \begin{tabular}{@{}llll@{}} 
 \toprule
 quantity & unit name & symbol & base units \\ \midrule 
 distance & meter & \si{\metre} & - \\
 time & second & \si{\second} & - \\
 mass & kilogram & \si{\kilogram} & - \\
 electric current & ampere & \si{\ampere} & - \\
 amount of substance & mole & \si{\mole}  & \num{6.02214076e23} \\ 
 temperature & kelvin & \si{\kelvin} & - \\ \midrule 
 energy & joule & \si{\joule} & \si{\kilogram\square\metre\per\square\second} \\
 power & watt & \si{\watt} & \si{\kilogram\square\metre\per\second\tothe{3}} \\ 
 electric charge & coulomb & \si{\coulomb} & \si{\ampere\second} \\
 electric potential & volt & \si{\volt} & \si{\kilogram\square\metre\per\ampere\per\second\tothe{3}} \\
 electrical resistance & ohm & \si{\ohm} & \si{\kilogram\square\metre\per\ampere\tothe{2}\per\second\tothe{3}} \\
 electrical capacitance & farad & \si{\farad} & \si{\second\tothe{4}\ampere\tothe{2}\per\square\metre\per\kilogram} \\
 electrical inductance & henry & \si{\henry} & \si{\kilogram\square\metre\per\square\second\per\square\ampere} \\
 magnetic flux density & tesla & \si{\tesla} & \si{\kilogram\per\square\second\per\ampere} \\
 pressure & pascal & \si{\pascal} & \si{\kilogram\per\metre\per\square\second} \\ %\\
 %temperature & degree Celsius & \si{\celsius} & $\SI{0}{\celsius} = \SI{273.15}{\kelvin}$ \\ 
 \bottomrule
 \end{tabular}
 \end{table}
\begin{table}[]
 \centering
 \caption[Relevant physical constants in SI units]{Physical constants expressed in SI units and definition of the electronvolt (the amount of energy that an electron posses after passing through an electric potential difference of one volt).} \label{tab:fundamental_constants}
 \begin{tabular}{@{}lll@{}} 
 \toprule
 quantity & symbol & value in SI units \\ \midrule 
 \emph{Planck's constant} & $h$ & \SI{6.62607015e-34}{\joule\second} \\ 
 \emph{reduced Planck's constant} & $\hbar \equiv h / 2 \pi$ & \SI{1.05457800e-34}{\joule\second} \\ 
 \emph{vacuum permittivity} & $\epsilon_0$ & \SI{8.85418782e-12}{\farad\per\metre} \\ 
 \emph{vacuum permeability} & $\mu_0$ & \SI{4 \numpi e-7}{\henry\per\metre} \\ 
 \emph{speed of light (in vacuum)} & $c_0 \equiv \left( \epsilon_0 \mu_0 \right)^{-1/2}$ & \SI{299792458}{\metre\per\second} \\ 
 \emph{impedance of free space} & $Z_0 \equiv \mu_0 c_0$ & \SI{119.9169832 \numpi}{\ohm} \\
 \emph{elementary charge} & $q_e$ & \SI{1.60217662e-19}{\coulomb} \\ 
 \emph{electron rest mass} & $m_e$ & \SI{9.1093835e-31}{\kilogram} \\ 
 \emph{Bohr radius} & $a_0 \equiv 4 \pi \epsilon_0 \hbar^2 / m_e q_e^2$ & \SI{5.2917721067e-11}{\metre} \\ 
 \emph{classical electron radius} & $r_e \equiv q_e^2 / 4 \pi \epsilon_0 m_e c_0^2$ & \SI{2.81794032e-15}{\metre} \\
 \emph{Rydberg constant} & $R_{\infty} \equiv m_e q_e^4 / 8 \epsilon_0^2 h^3 c_0$ & \SI{10973731.568508}{\per\metre} \\ 
 \emph{Thomson cross-section} & $\sigma_e \equiv 8 \pi r_e^2 / 3$ & \SI{6.6524587e-29}{\metre\square} \\
 \emph{Boltzmann constant} & $k_{\mathcal{B}}$ & \SI{1.380648e-23}{\joule\per\kelvin} \\
 \emph{fine-structure constant} & $\alpha_f \equiv q^2_e / 4 \pi \epsilon_0 \hbar c_0$ & \num{0.0072973525693} $\approx 1/137$ \\ \midrule
 \emph{electronvolt} & \SI{1}{\electronvolt} & \SI{1.6021766209e-19}{\joule} \\ 
 %\emph{centigrade} & \si{\celsius} & $\SI{0}{\celsius} = \SI{273.15}{\kelvin}$ \\
 \bottomrule
 \end{tabular}
 \end{table}
To start, all types of electromagnetic radiation can be described physically as both classical electromagnetic waves and massless particles (\emph{i.e.}, \emph{photons}) according to \emph{wave-particle duality}. 
This principle was first proposed by Planck and Einstein in the early 1900s \cite{Planck1901,Einstein1905a} and later validated by Compton in the early 1920s through the discovery of \emph{Compton scattering}, where x-rays transfer momentum to electrons as they are ejected from atoms \cite{Compton23a,Als-Nielsen11}. 
While this process can usually be neglected for radiation with wavelength, $\lambda$, much longer than the \emph{Compton wavelength} of the electron ($\lambda_C \equiv h / m_e c_0 \approx$~\SI{2.43}{\pico\metre}) [\emph{cf.\@} \cref{eq:electron_wavelength}], their interaction with matter can still be explained in terms of particle-like excitations in the quantized electromagnetic field with photon energy, $\hbar \omega$, and momentum, $\hbar \mathbold{k}$ \cite{Townsend00,Shankar04}. 
Equivalently, electromagnetic radiation can be understood from a classical perspective to be coupled oscillations in the \emph{electric field}, $\mathbold{E}(\mathbold{r},t)$, and \emph{the magnetic field}, $\mathbold{B}(\mathbold{r},t)$, that propagate together at the speed of light, $c_0$ \cite{Jackson75,Born80}. 
Across the electromagnetic spectrum, radiation tends to interact with matter on size scales comparable to its wavelength [\emph{cf.\@} \cref{tab:EM_spectrum}]. 
\begin{table}[]
 \centering
 \caption{Approximate wavelength ranges for all types of electromagnetic radiation}\label{tab:EM_spectrum}
 \begin{tabular}{@{}lll@{}} 
 \toprule
 spectral band & wavelength range & comparison \\ \midrule 
 radio waves & $\gtrapprox$ \SI{1}{\metre} & large objects \\ 
 microwaves & \SI{1}{\metre} to \SI{1}{\milli\metre} & meter stick \\ 
 terahertz radiation & \SI{1}{\milli\metre} to \SI{100}{\micro\metre} & grain of salt \\ 
 infrared radiation & \SI{100}{\micro\metre} to \SI{700}{\nano\metre} & human hair thickness, biological cells \\ 
 visible light & \SI{700}{\nano\metre} to \SI{400}{\nano\metre} & bacteria \\ 
 ultraviolet radiation & \SI{400}{\nano\metre} to \SI{5}{\nano\metre} & viruses, large molecules \\ 
 x-rays & \SI{5}{\nano\metre} to \SI{1}{\pico\metre} & atoms \\
 gamma rays &  $\lessapprox$ \SI{1}{\pico\metre} & atomic nuclei \\ 
 \bottomrule
 \end{tabular}
 \end{table}
Falling at the red end of the x-ray spectrum, soft x-rays have $\lambda$ approaching the atomic scale with high frequency $\omega = 2 \pi c_0 / \lambda$ comparable to atomic resonances in relatively light elements \cite{Attwood17}. 
To provide physics background for \cref{ch:introduction,ch:diff_eff,ap:quantum_spectral,app:x-ray_materials,ap:grating_basics}, this appendix outlines basics of classical-wave and photon descriptions for soft x-rays and their interaction with atomic electrons under the framework of non-relativistic quantum mechanics. 

\section{X-ray Nomenclature}\label{sec:histroical}
%%%%%%%%%%%%%%%%%%%%%%%%%%%%%%%%%%%%%%%%%-------------------------------------------------
The defining feature of x-rays that led to their discovery is their ability to penetrate through certain materials.
In R\"ontgen's experiments of the 1890s \cite{Rontgen1898a,Rontgen1898b,Rontgen1896}, x-rays were mainly produced by the deceleration of electrons as \emph{bremmstrahlung}, or ``braking radiation'', in an early electric discharge tube \cite{Compton35,Rybicki86,Attwood17,Als-Nielsen11}.\footnote{X-ray spectral lines characteristic of fluorescence occurring in the anode are also produced.} 
Such a device consists of a glass bulb that houses two electrodes held under partial vacuum while an applied high voltage (the \emph{tube voltage}, $V_T$) produces electric discharge from the negatively-biased cathode \cite{Coolidge1913,Compton35,Als-Nielsen11}. 
When tube voltages on the order of tens of \si{\kilo\volt} up to \SI{100}{\kilo\volt} were applied to the apparatus, it was found that a \emph{new kind of ray}, or \emph{x-ray}, with a unique penetrating ability was produced \cite{Rontgen1898a,Rontgen1896}. 
This then-unknown form of radiation was detected through observing fluorescence in \emph{barium platinocyanide} (Ba[Pt(CN)$_4$]), which was coated on a screen nearby the electric discharge tube, where it was discovered that even with the apparatus covered in light-blocking cardboard and the room darkened, Ba[Pt(CN)$_4$] produced a fluorescent glow when tube voltages up to \SI{100}{\kilo\volt} were applied. 

R\"ontgen discovered that radiation produced by $V_T \lessapprox \SI{100}{\kilo\volt}$ was able to penetrate though relatively low-$\mathcal{Z}$ materials such as cardboard whereas it was easily absorbed by heavier materials, such as pieces of metal lab equipment. %\footnote{On this note, R\"ontgen discovered that human bone, being rich in calcium ($\mathcal{Z}=20$), tends to absorb hard x-rays significantly more than the surrounding soft tissue. This enabled the first x-ray photograph of a human body part, which effectively gave birth to the radiographic techniques used in fields such as medicine and dentistry today \cite{Mould1995}.}   
With the maximum photon energy generated by bremmstrahlung being $q_e V_T$, x-rays with $\hbar \omega$ ranging from tens of \si{\kilo\electronvolt} to \SI{100}{\kilo\electronvolt} are referred to as \emph{hard x-rays} for this historical reason \cite{Als-Nielsen11}. 
On the other hand, radiation produced by $V_T \sim \SI{1}{\kilo\volt}$ was observed to be easily absorbed by virtually any material in early experiments, and as a result, x-rays with $ \SI{200}{\electronvolt}\lessapprox \hbar \omega \lessapprox \SI{2}{\kilo\electronvolt}$ are referred to as \emph{soft x-rays} \cite{Attwood17}. %[\emph{cf.\@} \cref{sec:SXR_med}]
However, these definitions vary in the literature and additionally, intermediate x-rays with $\SI{1}{\kilo\electronvolt} \lessapprox \hbar \omega \lessapprox \SI{5}{\kilo\electronvolt}$ are sometimes called \emph{tender x-rays} \cite{Northrup16,Senf16}.

\section{Photons and Classical Electromagnetic Waves}\label{sec:EM_waves_vac}
%%%%%%%%%%%%%%%%%%%%%%%%%%%%%%%%%%%%%%%%%--------------------------------------------------
The behavior of electromagnetic waves is governed by \emph{Maxwell's equations}, which are the foundation for classical electrodynamics \cite{Landau60,Jackson75,Born80,Rybicki86,Kahn02,Griffiths17,Attwood17}. 
In their microscopic form, these four equations are:\footnote{Throughout this thesis, $\cdot$ indicates a scalar (\emph{dot}) product while $\times$ indicates a vector (\emph{cross}) product such that $\divergence$ and $\curl$ represent \emph{divergence} and \emph{curl} vector operators, respectively.} 
\begin{subequations}
\begin{align}
 \div \mathbold{E}(\mathbold{r},t) &= \frac{\rho(\mathbold{r},t)}{\epsilon_0} \label{eq:Maxwell_vac_start_1} \\
 \curl \mathbold{E}(\mathbold{r},t) &= - \pdv{\mathbold{B}(\mathbold{r},t)}{t} \label{eq:Maxwell_vac_start_2}\\
 \div \mathbold{B}(\mathbold{r},t) &= 0 \label{eq:Maxwell_vac_start_3} \\
 \curl \mathbold{B}(\mathbold{r},t) &= \mu_0 \mathbfcal{J}(\mathbold{r},t) + \frac{1}{c_0^2} \pdv{\mathbold{E}(\mathbold{r},t)}{t}, \label{eq:Maxwell_vac_start_4}
 \end{align}
\end{subequations}
where $\epsilon_0$ is the vacuum permittivity and $\mu_0$ is the vacuum permeability with the speed of light given by $c_0 \equiv \left( \epsilon_0 \mu_0 \right)^{-1/2}$ [\emph{cf.\@} \cref{tab:fundamental_constants}]. 
Additionally, $\rho(\mathbold{r},t)$ is the volume density of electric charge and $\mathbfcal{J}(\mathbold{r},t)$ is a directional electric current density; these quantities are related to each other through the following \emph{continuity equation}:
\begin{equation} \label{eq:continuity}
 \divergence \mathbfcal{J}(\mathbold{r},t) + \pdv{\rho(\mathbold{r},t)}{t} = 0 , 
 \end{equation}
which ensures that electric charge is conserved \cite{Landau60,Jackson75,Born80,Rybicki86,Griffiths17,Attwood17}. 
The basic laws of electromagnetism are described by \cref{eq:Maxwell_vac_start_1,eq:Maxwell_vac_start_2,eq:Maxwell_vac_start_3,eq:Maxwell_vac_start_4}: 
\begin{enumerate}[noitemsep]
	\item \cref{eq:Maxwell_vac_start_1} is \emph{Gauss's law}, which describes how electric field lines diverge from a source of electric charge $\rho(\mathbold{r},t)$ (\emph{i.e.}, $\div \mathbold{E}(\mathbold{r},t) \neq 0$) %\cite{Lagrange1869,Gauss1877}
	\item \cref{eq:Maxwell_vac_start_2} is \emph{Faraday's law of induction}, which states that a dynamic magnetic field (\emph{i.e.}, $\pdv{\mathbold{B}(\mathbold{r},t)}{t} \neq 0$) generates a curled electric field (\emph{i.e.}, $\curl \mathbold{E}(\mathbold{r},t) \neq 0$)
	\item \cref{eq:Maxwell_vac_start_3} is a statement that magnetic charge does not exist and hence magnetic field lines are never divergent (\emph{i.e.}, $\div \mathbold{B}(\mathbold{r},t) = 0$ always)
	\item \cref{eq:Maxwell_vac_start_4} is the \emph{Amp\`ere-Maxwell equation}, an extension of \emph{Amp\`ere's circuital law}, which describes how an electric current with a density $\mathbfcal{J}(\mathbold{r},t)$ generates a curled magnetic field (\emph{i.e.}, $\curl \mathbold{B}(\mathbold{r},t) \neq 0$); Maxwell's correctional term, $\left( 1 / c_0^2 \right) \pdv*{\mathbold{E}(\mathbold{r},t)}{t}$, states that a dynamic electric field also generates a curled magnetic field
\end{enumerate}

In vacuum, electric charge and electric current are null but it can be shown that wave behavior arises from Maxwell's equations provided that there is some far-away source that generates a dynamic electric or magnetic field. 
Maxwell's equations given by \cref{eq:Maxwell_vac_start_1,eq:Maxwell_vac_start_2,eq:Maxwell_vac_start_3,eq:Maxwell_vac_start_4} for this situation read as
\begin{subequations}
\begin{align}
 \div \mathbold{E}(\mathbold{r},t) &= 0 \label{eq:Maxwell_vac_1}  \\
 \curl \mathbold{E}(\mathbold{r},t) &= - \mu_0 \pdv{\mathbold{H}(\mathbold{r},t)}{t}  \label{eq:Maxwell_vac_2} \\
 \div \mathbold{H}(\mathbold{r},t) &= 0 \label{eq:Maxwell_vac_3}  \\
 \curl \mathbold{H}(\mathbold{r},t) &= \epsilon_0 \pdv{\mathbold{E}(\mathbold{r},t)}{t} . \label{eq:Maxwell_vac_4}
 \end{align}
\end{subequations}
For convenience of notation, the \emph{auxiliary magnetic field}, $\mathbold{H}(\mathbold{r},t)$, has been substituted for the \emph{fundamental magnetic field}, $\mathbold{B}(\mathbold{r},t)$, through the constitutive relation $\mathbold{B}(\mathbold{r},t) = \mu_0 \mathbold{H}(\mathbold{r},t)$ \cite{Landau60,Jackson75,Griffiths17}.\footnote{This constitutive relation holds in vacuum and also in non-magnetized media, which need not be discussed in this context. Because of this, $\mathbold{B}(\mathbold{r},t)$ and $\mathbold{H}(\mathbold{r},t)$ are both referred to as \emph{the magnetic field} in this thesis.} 
Qualitatively, \cref{eq:Maxwell_vac_1,eq:Maxwell_vac_3} state that field lines are non-divergent while \cref{eq:Maxwell_vac_2,eq:Maxwell_vac_4} describe how a time-varying $\mathbold{H}(\mathbold{r},t)$ leads to a curled $\mathbold{E}(\mathbold{r},t)$ and vice-versa. 
These conditions can be seen to lead to wave behavior by combining \cref{eq:Maxwell_vac_1,eq:Maxwell_vac_2,eq:Maxwell_vac_3,eq:Maxwell_vac_4}\footnote{In particular, this can be done by first taking curls of \cref{eq:Maxwell_vac_2,eq:Maxwell_vac_4}: 
\begin{align*}
 \begin{split}
 \curl \curl \mathbold{E}(\mathbold{r},t) &= \grad \left[\div \mathbold{E}(\mathbold{r},t)\right] - \laplacian \mathbold{E}(\mathbold{r},t) = - \mu_0 \pdv{\left[ \curl \mathbold{H} (\mathbold{r},t) \right]}{t} = - \epsilon_0 \mu_0 \pdv[2]{\mathbold{E}(\mathbold{r},t)}{t}\\
 \curl \curl \mathbold{H}(\mathbold{r},t) &= \grad \left[\divergence \mathbold{H}(\mathbold{r},t)\right] - \laplacian \mathbold{H}(\mathbold{r},t) = \epsilon_0 \pdv{\left[ \curl \mathbold{E} (\mathbold{r},t) \right]}{t} = - \epsilon_0 \mu_0 \pdv[2]{\mathbold{H}(\mathbold{r},t)}{t}
 \end{split}
 \end{align*}
and then inserting \cref{eq:Maxwell_vac_1,eq:Maxwell_vac_3}. 
 \label{footnote:combine}} to arrive at the following identical expressions for $\mathbold{E}(\mathbold{r},t)$ and $\mathbold{H}(\mathbold{r},t)$:
\begin{equation} \label{eq:wave_vector_vac}
 \left( \laplacian - \epsilon_0 \mu_0 \pdv[2]{t} \right) \left\{
  \begin{array}{lr}
  \mathbold{E}(\mathbold{r},t)\\
  \mathbold{H}(\mathbold{r},t)
  \end{array}
 \right\} = \mathbf{0} ,
 \end{equation}
where $\mathbf{0}$ is the null vector. 
Considering each component of $\mathbold{E}(\mathbold{r},t)$ and $\mathbold{H}(\mathbold{r},t)$ separately, this is equivalent to six scalar differential equations that each take the form of the scalar \emph{wave equation} \cite{Born80,Griffiths17}: 
\begin{equation} \label{eq:scalar_wave_general}
 \left( \laplacian + \frac{1}{v_p^2} \pdv[2]{t} \right) u (\mathbold{r},t) = 0 ,
 \end{equation}
where the scalar field $u (\mathbold{r},t)$ represents the wave medium and $v_p = c_0 \equiv \left( \epsilon_0 \mu_0 \right)^{-1/2}$ is the wave propagation speed. 
While other types of waves are described by this scalar wave equation (\emph{e.g.}, water waves, sound waves, etc.), classical electromagnetic waves are oscillations in the vector fields $\mathbold{E}(\mathbold{r},t)$ and $\mathbold{H}(\mathbold{r},t)$ and thus a full vector treatment is generally required to describe their behavior.\footnote{However, in situations where it can be assumed that $\mathbold{E}(\mathbold{r},t)$ continuously oscillates along the same axis (\emph{i.e.}, if the light is \emph{linearly polarized} to a high degree), only one scalar component need be considered. 
Then, \cref{eq:scalar_wave_general} with $v_p = c_0$ can be used to evaluate wave behavior while the other components of $\mathbold{E}(\mathbold{r},t)$ and $\mathbold{H}(\mathbold{r},t)$ follow from \cref{eq:Maxwell_vac_1,eq:Maxwell_vac_2,eq:Maxwell_vac_3,eq:Maxwell_vac_4}. See also \cref{ap:grating_basics}.}

\subsection{Time-Harmonic Classical Wave Modes}\label{sec:time_harmonic}
%%%%%%%%%%%%%%%%%%%%%%%%%%%%%%%%%%%%%%%%%--------------------------------------------------
Oscillatory solutions to \cref{eq:scalar_wave_general} are commonly discussed as normal modes with some singular value for $\omega$ such that if the scalar field $u (\mathbold{r},t)$ were continuously measured at a fixed position $\mathbold{r}_0$, a sinusoidal function would be generated: 
\begin{subequations}
\begin{equation}\label{eq:cosine_mode}
 u (\mathbold{r}_0,t) \equiv u (t) = u_0 \cos \left( \Phi - \omega t \right) ,
 \end{equation}
where $u_0$ is the amplitude of the wave and $\Phi$ is its phase. 
Alternatively, this can be expressed as a complex phasor: 
\begin{equation}\label{eq:phasor_mode}
 u (t) = u_0 \mathrm{e}^{i \left( \Phi - \omega t \right)} ,
 \end{equation} 
\end{subequations}
with $i \equiv \sqrt{-1}$ as the imaginary unit. 
However, realistic electromagnetic waves tend to exist in superpositions of these normal modes such that there is always some spread in $\omega$ \cite{Born80}. 
This can be gleaned from the temporal Fourier transforms of the electromagnetic fields:
\begin{equation}\label{eq:temporal_fourier}
 \left\{
 \begin{array}{lr}
 \mathbold{E} \left( \mathbold{r} , \omega \right) \\
 \mathbold{H} \left( \mathbold{r} , \omega \right)
 \end{array}
 \right\} = \int_{-\infty}^{\infty} \left\{
 \begin{array}{lr}
 \mathbold{E} \left( \mathbold{r} , t ' \right) \\
 \mathbold{H} \left( \mathbold{r} , t ' \right)
 \end{array}
 \right\} \mathrm{e}^{i \omega  t'} \dd{t'} ,
 \end{equation}  
where $\mathbold{E} \left( \mathbold{r} , \omega \right)$ and $\mathbold{H} \left( \mathbold{r} , \omega \right)$ can have singular values only if $\mathbold{E} \left( \mathbold{r} , t \right)$ and $\mathbold{H} \left( \mathbold{r} , t \right)$ are pure sinusoids lasting for a formally infinite amount of time. 
Conversely, the most general solution to \cref{eq:wave_vector_vac} is a superposition of normal wave modes spanning all values of $\omega$, which can be represented using the following inverse Fourier transforms of the electromagnetic fields:
\begin{equation}
 \left\{
 \begin{array}{lr}
 \mathbold{E} \left( \mathbold{r} , t \right) \\
 \mathbold{H} \left( \mathbold{r} , t \right)
 \end{array}
 \right\} = \frac{1}{2 \pi} \int_{-\infty}^{\infty} \left\{
 \begin{array}{lr}
 \mathbold{E} \left( \mathbold{r} , \omega ' \right) \\
 \mathbold{H} \left( \mathbold{r} , \omega ' \right)
 \end{array}
 \right\} \mathrm{e}^{-i \omega ' t} \dd{\omega} ' . 
 \end{equation} 
For simplicity, however, it is useful to imagine a wave that extends infinitely in time and space so that just one frequency mode is present.\footnote{This can be written by making the following replacement:
\begin{equation*}
 \left\{
 \begin{array}{lr}
 \mathbold{E} \left( \mathbold{r} , \omega ' \right) \\
 \mathbold{H} \left( \mathbold{r} , \omega ' \right)
 \end{array} 
 \right\} \to \left\{
 \begin{array}{lr}
 \mathbold{E} \left( \mathbold{r} \right) \\
 \mathbold{H} \left( \mathbold{r} \right)
 \end{array} 
 \right\} \delta_D \left( \omega ' - \omega \right) 
 \end{equation*} 
with $\delta_D (\omega ' - \omega)$ being a \emph{Dirac delta function}, which is defined as \begin{equation*}
\delta_D \left( x \right) \equiv
   \begin{cases}
     \infty, & \text{if}\ x = 0 \\
     0, & \text{if}\ x \neq 0 .
   \end{cases} 
\end{equation*}
\label{footnote:dirac_delta}}  %with $\ell_{\text{coh}} \to \infty$
In this case, the electromagnetic fields can be written as 
\begin{equation}\label{eq:single-mode}
 \left\{
 \begin{array}{lr}
 \mathbold{E} \left( \mathbold{r} , t \right) \\
 \mathbold{H} \left( \mathbold{r} , t \right)
 \end{array} 
 \right\} =  \left\{
 \begin{array}{lr}
 \mathbold{E} \left( \mathbold{r} \right) \\
 \mathbold{H} \left( \mathbold{r} \right)
 \end{array} 
 \right\} \mathrm{e}^{-i \omega t} , 
 \end{equation} 
where $\mathbold{E}(\mathbold{r})$ and $\mathbold{H}(\mathbold{r})$ are the \emph{time-harmonic} fields, which implicitly assume a $\mathrm{e}^{-i \omega t}$ time dependence. 

Electromagnetic fields oscillating in time necessarily have a corresponding spatial frequency. 
In other words, a normal mode of frequency $\omega$ at a fixed time $t_0$ exhibits periodicity as a function of $\mathbold{r}$; essentially, this is the wavelength of the radiation.  
This can be deduced by inserting \cref{eq:single-mode} into \cref{eq:Maxwell_vac_1,eq:Maxwell_vac_2,eq:Maxwell_vac_3,eq:Maxwell_vac_4} so that the $\mathrm{e}^{-i \omega t}$ terms drop out, resulting in Maxwell's equations in time-harmonic form: 
\begin{subequations}
\begin{align} %eq:Maxwell_vac_TH 1-4
 \div \mathbold{E} \left( \mathbold{r} \right) &= 0 \label{eq:Maxwell_vac_TH_1} \\
 \curl \mathbold{E} \left( \mathbold{r} \right) &= i \omega \mu_0 \mathbold{H} \left( \mathbold{r} \right) \label{eq:Maxwell_vac_TH_3} \\
 \div \mathbold{H} \left( \mathbold{r} \right) &= 0 \label{eq:Maxwell_vac_TH_2} \\
 \curl \mathbold{H} \left( \mathbold{r} \right) &= -i \omega \epsilon_0 \mathbold{E} \left( \mathbold{r} \right) \label{eq:Maxwell_vac_TH_4} .
 \end{align}
\end{subequations}
These equations can be combined in a manner similar to footnote~\ref{footnote:combine} to arrive at a wave equation analogous to \cref{eq:wave_vector_vac}. 
Alternatively, this can be done by inserting \cref{eq:single-mode} into \cref{eq:wave_vector_vac} to yield the \emph{Helmholtz equation} for $\mathbold{E}(\mathbold{r})$ and $\mathbold{H}(\mathbold{r})$: % in vacuum: 
\begin{equation} \label{eq:Helmholtz}
 \left( \laplacian + \frac{\omega^2}{c_0^2} \right) \left\{
 \begin{array}{lr}
 \mathbold{E} \left( \mathbold{r} \right) \\
 \mathbold{H} \left( \mathbold{r} \right)
 \end{array}
 \right\} = \mathbf{0} .
 \end{equation} 

A general solution to \cref{eq:Helmholtz}\footnote{Once the Helmholtz equation has been solved for the fields, time dependences for single modes can be recovered by multiplying $\mathbold{E} \left( \mathbold{r} \right)$ and $\mathbold{H} \left( \mathbold{r} \right)$ by $\mathrm{e}^{-i \omega t}$; the real part of the phasors correspond to the physical fields.} can be expressed as a superposition of spatial wave modes using inverse Fourier transforms: %
\begin{equation} \label{eq:fourier_fields}
 \left\{
  \begin{array}{lr}
  \mathbold{E} \left( \mathbold{r} \right) \\
  \mathbold{H} \left( \mathbold{r} \right)
  \end{array}
 \right\} = \frac{1}{(2 \pi)^3} \int_{-\mathbold{\infty}}^{\mathbold{\infty}} \left\{
  \begin{array}{lr}
  \mathbold{E} \left( \mathbold{k} \right) \\
  \mathbold{H} \left( \mathbold{k} \right)
  \end{array}
 \right\} \mathrm{e}^{i \mathbold{k} \cdot \mathbold{r}} \dd[3]{\mathbold{k}},
 \end{equation}
where $\mathbold{k}$ is a 3D spatial frequency vector (\emph{i.e.}, the frequency analog of $\mathbold{r}$). 
This vector, which points in the direction of wave propagation, can be assumed to be purely real because there is no mechanism for wave attenuation in vacuum \cite{Landau60,Jackson75,Griffiths17,Born80}. 
Inserting this into the \cref{eq:Helmholtz} shows that $k_0^2 \equiv | \mathbold{k} |^2 = \omega^2 / c_0^2$: 
\begin{equation}
 \laplacian \left\{
  \begin{array}{lr}
  \mathbold{E} \left( \mathbold{r} \right) \\
  \mathbold{H} \left( \mathbold{r} \right)
  \end{array}
 \right\} = \frac{1}{(2 \pi)^3} \laplacian \int_{-\mathbold{\infty}}^{\mathbold{\infty}} \left\{
  \begin{array}{lr}
  \mathbold{E} \left( \mathbold{k} \right) \\
  \mathbold{H} \left( \mathbold{k} \right)
  \end{array}
 \right\} \mathrm{e}^{i \mathbold{k} \cdot \mathbold{r}} \dd[3]{\mathbold{k}}
  = - k_0^2 \left\{
  \begin{array}{lr}
  \mathbold{E} \left( \mathbold{r} \right) \\
  \mathbold{H} \left( \mathbold{r} \right)
  \end{array}
 \right\} ,
 \end{equation}
where $k_0 = 2 \pi / \lambda$ is the \emph{wave number in vacuum}. 
The corresponding \emph{dispersion relation in vacuum} is  
\begin{equation} \label{eq:disp_rel_vac}
 \omega = c_0 k_0 , 
 \end{equation}
which states that a wave mode of frequency $\omega$ and wavelength $\lambda$ propagates at a speed $c_0$ in vacuum, without attenuation. 
Further, inserting \cref{eq:fourier_fields} into \cref{eq:Maxwell_vac_TH_1,eq:Maxwell_vac_TH_2,eq:Maxwell_vac_TH_3,eq:Maxwell_vac_TH_4} shows that electromagnetic waves in vacuum are transverse. 
\begin{figure}
 \centering
 \includegraphics[scale=1.25]{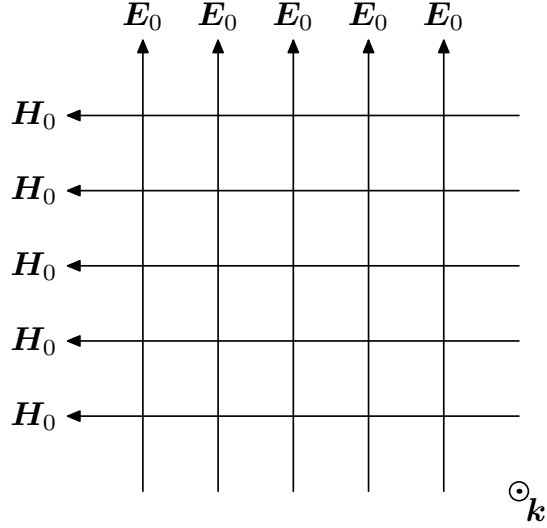}
 \caption{Orientation of vectors in a transverse electromagnetic wavefront}\label{fig:wavefront}
 \end{figure} 
That is, the electromagnetic fields are perpendicular to both each other and the wave vector $\mathbold{k}$;\footnote{In other words, surfaces defined by fields of constant phase (called \emph{wavefronts}) are everywhere perpendicular to lines that trace the direction of propagation (known as \emph{rays}).} this is shown in \cref{fig:wavefront}, where $\mathbold{E}_0$ and $\mathbold{H}_0 = (Z_0 k_0)^{-1} \mathbold{k} \times \mathbold{E}_0$ are their vector amplitudes:
\begin{subequations}
\begin{align} 
 \mathbold{k} \cdot \mathbold{E} \left( \mathbold{r} \right) &= 0 \label{eq:Maxwell_vac_FT_1} \\
 \mathbold{k} \times \mathbold{E} \left( \mathbold{r} \right) &= Z_0 k_0 \mathbold{H} \left( \mathbold{r} \right) \label{eq:Maxwell_vac_FT_3} \\
 \mathbold{k} \cdot \mathbold{H} \left( \mathbold{r} \right) &= 0 \label{eq:Maxwell_vac_FT_2} \\
 \mathbold{k} \times \mathbold{H} \left( \mathbold{r} \right) &= - \frac{k_0}{Z_0} \mathbold{E} \left( \mathbold{r} \right) \label{eq:Maxwell_vac_FT_4} ,
 \end{align}
\end{subequations}
with $Z_0 \equiv \mu_0 c_0$ as the impedance of free space. 

\subsection{Treatment of Photons}\label{sec:photon_intro}
%%%%%%%%%%%%%%%%%%%%%%%%%%%%%%%%%%%%%%%%%--------------------------------------------------
Quantum-mechanically, electromagnetic radiation can be represented using a set of abstract vectors residing in \emph{Fock space} \cite{Fock1932} that indicate the number of photons present in a given normal wave mode defined by a wave vector $\mathbold{k}$ and a linear polarization unit vector $\hat{\mathbold{e}}_{\mathbold{k},\upsilon}$ indexed by $\upsilon = 1,2$, which parameterizes the polarization state \cite{Shankar04,Townsend00,Fontana_82,kapitell4,wysin2011quantization}.  
These polarization unit vectors\footnote{Here, these basis vectors are chosen to indicate directions associated with linear polarization (as opposed to circular polarization) so that the unit vectors are real ($\hat{\mathbold{e}}_{\mathbold{k},\upsilon} = \hat{\mathbold{e}}^*_{\mathbold{k},\upsilon}$).} represent the mutually orthogonal directions of the electromagnetic fields [\emph{cf.\@} \cref{fig:wavefront}]:\footnote{Here, $\delta$ indicates a \emph{Kronecker delta function} defined by: 
\begin{equation}\label{eq:Kronecker_delta}
  \delta_{\upsilon,\upsilon'} =
  \begin{cases}
    1, & \text{if } \upsilon = \upsilon' \\
    0, & \text{if } \upsilon \neq \upsilon' . 
  \end{cases}
 \end{equation}} 
\begin{subequations}
\begin{equation}
 \hat{\mathbold{e}}_{\mathbold{k},\upsilon} \cdot \hat{\mathbold{e}}_{\mathbold{k}',\upsilon'} = \delta_{\mathbold{k},\mathbold{k}'} \delta_{\upsilon,\upsilon'}
 \end{equation}
and are both orthogonal to the direction of wave propagation defined by $\mathbold{k}$: 
\begin{equation}
 \mathbold{k} \cdot \hat{\mathbold{e}}_{\mathbold{k},\upsilon} = 0 \quad \text{for } \upsilon=1,2 .
 \end{equation}
\end{subequations}

\subsubsection{Fock States}\label{sec:fock_states}
%%%%%%%%%%%%%%%%%%%%%%%%%%%%%%%%%%%%%%%%%--------------------------------------------------
A single photon in a mode described by $\mathbold{k}$ and $\upsilon$ is represented by the Fock state $\ket{1_{\mathbold{k},\upsilon}}$. 
Similar to the treatment of quantum harmonic oscillators \cite{Shankar04,Griffiths05,Townsend00}, these Fock states change excitation levels through the application of an \emph{annihilation operator}, $\underline{a}_{\mathbold{k},\upsilon}$, or its adjoint, the \emph{creation operator}, $\underline{a}^{\dagger}_{\mathbold{k},\upsilon}$, (here ${}^{\dagger}$ indicates a conjugate transpose). 
These operators decrease the number of photons in a given mode by one:
\begin{subequations}
\begin{equation}\label{eq:annihilation_one}
 \underline{a}_{\mathbold{k},\upsilon} \ket{1_{\mathbold{k},\upsilon}} = \ket{0_{\mathbold{k},\upsilon}} \equiv \ket{0}, 
 \end{equation}
where $\ket{0}$ is the \emph{vacuum state} with $\underline{a}_{\mathbold{k},\upsilon} \ket{0} \equiv 0$, or increase the number of photons by one:
\begin{equation}\label{eq:creation_one}
 \underline{a}^{\dagger}_{\mathbold{k},\upsilon} \ket{1_{\mathbold{k},\upsilon}} = \sqrt{2} \ket{2_{\mathbold{k},\upsilon}} .
 \end{equation}
\end{subequations} 
Such a state with two photons, $\ket{2_{\mathbold{k},\upsilon}}$, is possible because photons are \emph{bosons}; there is no limit to the number of photons that can exist in a single state. 
More generally, a state of $N_{\mathbold{k},\upsilon}$ photons in a single mode is defined as 
\begin{equation}\label{eq:N_photons}
 \ket{N_{\mathbold{k},\upsilon}} = \frac{\left( \underline{a}^{\dagger}_{\mathbold{k},\upsilon} \right)^{N_{\mathbold{k},\upsilon}}}{\sqrt{N_{\mathbold{k},\upsilon} ! }} \ket{0}
 \end{equation}
with 
\begin{subequations}
\begin{equation}\label{eq:N_photon_destruction}
 \underline{a}_{\mathbold{k},\upsilon} \ket{N_{\mathbold{k},\upsilon}} = \sqrt{N_{\mathbold{k},\upsilon}} \ket{N_{\mathbold{k},\upsilon} - 1} 
 \end{equation}
and 
\begin{equation}\label{eq:N_photon_creation}
 \underline{a}^{\dagger}_{\mathbold{k},\upsilon} \ket{N_{\mathbold{k},\upsilon}} = \sqrt{N_{\mathbold{k},\upsilon} + 1} \ket{N_{\mathbold{k},\upsilon} + 1} .
 \end{equation}
\end{subequations} 
These photon states are taken to be orthonormal to each other such that \cite{Shankar04,Townsend00,Fontana_82,kapitell4,wysin2011quantization} 
\begin{equation}\label{eq:photon_orthonormal}
  \braket{N_{\mathbold{k}',\upsilon'}'}{N_{\mathbold{k},\upsilon}} =
  \begin{cases}
    1, & \text{for states with identical } N, \mathbold{k}, \text{and } \upsilon \\
    0, & \text{otherwise}
  \end{cases}
 \end{equation}
and moreover, the \emph{photon-number operator} $\underline{N}_{\mathbold{k},\upsilon} \equiv \underline{a}^{\dagger}_{\mathbold{k},\upsilon} \underline{a}_{\mathbold{k},\upsilon}$ returns the number of photons in a pure Fock state \cite{Shankar04}:
\begin{equation}\label{eq:number_operator}
 \underline{N}_{\mathbold{k},\upsilon} \ket{N_{\mathbold{k},\upsilon}} = \underline{a}^{\dagger}_{\mathbold{k},\upsilon} \underline{a}_{\mathbold{k},\upsilon} \ket{N_{\mathbold{k},\upsilon}} = N_{\mathbold{k},\upsilon} \ket{N_{\mathbold{k},\upsilon}} .
 \end{equation}

\subsubsection{Quantum Operators for the Electromagnetic Field}\label{sec:QEM_operators}
%%%%%%%%%%%%%%%%%%%%%%%%%%%%%%%%%%%%%%%%%--------------------------------------------------
In the framework of non-relativistic quantum mechanics, an operator $\underline{O}$ that represents an observable quantity such as position, momentum or energy must be Hermitian such that $\underline{O}^{\dagger} = \underline{O}$ \cite{Shankar04}. 
Operators that represent electromagnetic quantities therefore should also be Hermitian as they act on Fock states with the non-Hermitian operators $\underline{a}_{\mathbold{k},\upsilon}$ and $\underline{a}^{\dagger}_{\mathbold{k},\upsilon}$. 
To formulate this, it is useful to start by defining an electromagnetic vector potential, $\mathbold{A}  ( \mathbold{r} , t)$, in the \emph{Coulomb gauge}, where $\div \mathbold{A}  ( \mathbold{r} , t) = 0$, while the scalar potential that normally describes static electric fields is null; the electric and magnetic fields are then determined from \cite{Jackson75,Rybicki86,Griffiths17} 
\begin{subequations}
\begin{align}
 \mathbold{E} ( \mathbold{r} , t) &= - \pdv{\mathbold{A} ( \mathbold{r} , t)}{t} \label{eq:E-field_from_A} \\ 
 \text{and} \quad \mathbold{B} ( \mathbold{r} , t) &= \curl \mathbold{A} ( \mathbold{r} , t) \label{eq:B-field_from_A} . 
 \end{align}
 \end{subequations}
Using this notation, a classical electromagnetic wave is described by solutions to the following differential equation similar to \cref{eq:wave_vector_vac}:
\begin{subequations}
\begin{equation}
 \laplacian \mathbold{A} ( \mathbold{r} , t) - \frac{1}{c_0^2} \pdv[2]{\mathbold{A} ( \mathbold{r} , t)}{t} = 0 \label{eq:potential_wave}
 \end{equation}
with constraints 
\begin{align}
 \divergence \pdv{\mathbold{A} ( \mathbold{r} , t)}{t}&= 0 \label{eq:potential_E-field} \\
 \text{and} \quad \div \mathbold{A} ( \mathbold{r} , t) &= 0 \label{eq:potential_B-field} .
 \end{align}
\end{subequations}
A general, real solution to \cref{eq:potential_wave} can be written as a summation of discrete wave modes indexed by $\mathbold{k}$ and $\upsilon$ with a temporal frequency $\omega_{\mathbold{k}} \equiv \abs{\mathbold{k}} c_0$:
\begin{equation}\label{eq:potential_wave_solution}
 \mathbold{A} ( \mathbold{r} , t) = \sum_{\mathbold{k}} \sum_{\upsilon = 1}^2 \frac{1}{\sqrt{\mathcal{V}}} \hat{\mathbold{e}}_{\mathbold{k},\upsilon} \left[ C_{\mathbold{k},\upsilon} \mathrm{e}^{i (\mathbold{k} \cdot \mathbold{r} - \omega_{\mathbold{k}} t)} + C^*_{\mathbold{k},\upsilon} \mathrm{e}^{-i (\mathbold{k} \cdot \mathbold{r} - \omega_{\mathbold{k}} t)} \right] , 
 \end{equation}
where it is assumed, as a mathematical trick, that there are boundaries imposed by a large box of volume $\mathcal{V}$ explained in the following paragraph. 

A Hermitian operator for the quantized electromagnetic field can be written in an analogous fashion to \cref{eq:potential_wave_solution}, where the coefficients are replaced by terms involving $\underline{a}_{\mathbold{k},\upsilon}$ and $\underline{a}^{\dagger}_{\mathbold{k},\upsilon}$ \cite{Townsend00}. %[chapter 14]
In the \emph{Schr\"odinger picture} of quantum mechanics, where operators carry no time dependence, this is given as \cite{Shankar04,Townsend00} 
\begin{equation}
 \underline{\mathbold{A}} ( \mathbold{r} ) = \sum_{\mathbold{k}} \sum_{\upsilon = 1}^2  \sqrt{\frac{\hbar}{2 \mathcal{V} \epsilon_0 \omega_{\mathbold{k}}}} \hat{\mathbold{e}}_{\mathbold{k},\upsilon} \left[ \underline{a}_{\mathbold{k},\upsilon} \mathrm{e}^{i \mathbold{k} \cdot \mathbold{r} } + \underline{a}^{\dagger}_{\mathbold{k},\upsilon} \mathrm{e}^{-i \mathbold{k} \cdot \mathbold{r} } \right] , \label{eq:vector_potential_operator_fixedtime} % \tag{\ref{eq:vector_potential_operator_fixed}} ,
 \end{equation}
where $\mathcal{V}$ is the \emph{normalization volume} for field quantization \cite[\emph{cf.\@} \cref{eq:potential_wave_solution}]{Townsend00,kapitell4,wysin2011quantization}.\footnote{Following Townsend~\cite{Townsend00}, the use of this normalization volume is motivated by the \emph{particle in a box} solutions of a square potential well that are discussed in introductory quantum mechanics \cite{Griffiths05}. Like the eigenfunctions for a particle in a box, the quantized field modes are discrete, as indicated by the summation over $\mathbold{k}$ in \cref{eq:vector_potential_operator_fixedtime}. It should be emphasized that this is just a mathematical trick and $\mathcal{V}$ should be taken to approach infinity to describe real systems.\label{footnote:normalization_volume}}
Further, in direct analogy with \cref{eq:E-field_from_A,eq:B-field_from_A}, the \emph{electric-field operator} is taken as
\begin{equation}\label{eq:E-field_operator_fixed_linear}
 \underline{\mathbold{E}} ( \mathbold{r} ) = \sum_{\mathbold{k}} \sum_{\upsilon = 1}^2  i \sqrt{\frac{\hbar \omega_{\mathbold{k}}}{2 \mathcal{V} \epsilon_0}} \hat{\mathbold{e}}_{\mathbold{k},\upsilon} \left[ \underline{a}_{\mathbold{k},\upsilon} \mathrm{e}^{i \mathbold{k} \cdot \mathbold{r} } - \underline{a}^{\dagger}_{\mathbold{k},\upsilon} \mathrm{e}^{-i \mathbold{k} \cdot \mathbold{r}} \right] 
 \end{equation}
and the \emph{magnetic-field operator} as
\begin{equation}\label{eq:B-field_operator_fixed_linear}
 \underline{\mathbold{B}} ( \mathbold{r} ) = \sum_{\mathbold{k}} \sum_{\upsilon = 1}^2 i \sqrt{\frac{\hbar}{2 \mathcal{V} \epsilon_0 \omega_{\mathbold{k}}}} \mathbold{k} \times \hat{\mathbold{e}}_{\mathbold{k},\upsilon} \left[ \underline{a}_{\mathbold{k},\upsilon} \mathrm{e}^{i \mathbold{k} \cdot \mathbold{r} } + \underline{a}^{\dagger}_{\mathbold{k},\upsilon} \mathrm{e}^{-i \mathbold{k} \cdot \mathbold{r} } \right] .
 \end{equation} 
The Hamiltonian operator for the quantized electromagnetic field can then be formulated as the quantum analog of the energy contained within the classical field, given by 
\begin{equation}\label{eq:EM_energy_density}
 U = \frac{1}{2} \int_{\mathcal{V}} \left( \epsilon_0 \norm{\mathbold{E} \left( \mathbold{r} , t \right)}^2 + \mu_0 \norm{\mathbold{H} \left( \mathbold{r} , t \right)}^2  \right) \dd[3]{\mathbold{r}} 
 \end{equation}
over a volume $\mathcal{V}$. 

Using the operators defined in \cref{eq:E-field_operator_fixed_linear,eq:B-field_operator_fixed_linear} in place of $\mathbold{E} \left( \mathbold{r} , t \right)$ and $\mathbold{B} \left( \mathbold{r} , t \right) = \mu_0 \mathbold{H} \left( \mathbold{r} , t \right)$, this Hamiltonian operator comes out to \cite{Townsend00,Fontana_82,kapitell4,wysin2011quantization}:
\begin{subequations}
\begin{equation}\label{eq:photon_Hamiltonian_operator}
 \underline{\mathcal{H}}_{EM} = \sum_{\mathbold{k}} \sum_{\upsilon = 1}^2 \hbar \omega_{\mathbold{k}} \left( \underline{a}^{\dagger}_{\mathbold{k},\upsilon} \underline{a}_{\mathbold{k},\upsilon} + \frac{1}{2} \right) .
 \end{equation}
Applying this operator to a single-photon state $\ket{1_{\mathbold{k},\upsilon}}$ shows that the energy added to the system is equal to the energy of a photon, $\mathcal{E}_{\gamma} \equiv \hbar \omega_{\mathbold{k}}$:
\begin{equation}\label{eq:photon_energy}
 \underline{\mathcal{H}}_{EM} \ket{1_{\mathbold{k},\upsilon}} = \sum_{\mathbold{k}} \sum_{\upsilon = 1}^2 \hbar \omega_{\mathbold{k}} \left( \underline{a}^{\dagger}_{\mathbold{k}',\upsilon'} \underline{a}_{\mathbold{k}',\upsilon'} + \frac{1}{2} \right) \ket{1_{\mathbold{k},\upsilon}} = \left( \mathcal{E}_0 + \hbar \omega_{\mathbold{k}} \right) \ket{1_{\mathbold{k},\upsilon}} ,
 \end{equation}
where only terms with $\mathbold{k}' = \mathbold{k}$ and $\upsilon' = \upsilon$ survive the summation. 
Here, the energy of a photon, $\hbar \omega_{\mathbold{k}}$, is measured relative to the \emph{vacuum energy}, $\mathcal{E}_0$:
\begin{equation}\label{eq:vacuum_energy}
 \bra{0} \underline{\mathcal{H}}_{EM} \ket{0} = \bra{0} \sum_{\mathbold{k}} \sum_{\upsilon = 1}^2 \hbar \omega_{\mathbold{k}} \left( \underline{a}^{\dagger}_{\mathbold{k},\upsilon} \underline{a}_{\mathbold{k},\upsilon} + \frac{1}{2} \right) \ket{0} = \bra{0} \underbrace{\frac{1}{2} \sum_{\mathbold{k}} \sum_{\upsilon = 1}^2 \hbar \omega_{\mathbold{k}}}_{\mathcal{E}_0 \to \infty} \ket{0} ,
 \end{equation}
\end{subequations}
which is formally infinite \cite{Shankar04}. 
Similarly, it can be shown that the operator for the momentum of the electromagnetic field\footnote{Classically, this is given by
\begin{equation*}\label{eq:wave_momentum}
 \mathbold{p}_{EM} \equiv \epsilon_0 \mu_0 \int_{\mathcal{V}} \mathbold{S}(\mathbold{r},t) \dd[3]{\mathbold{r}} , 
 \end{equation*}
where integration is performed over some volume $\mathcal{V}$ and $\mathbold{S}(\mathbold{r},t) \equiv \mathbold{E}(\mathbold{r},t) \times \mathbold{H}(\mathbold{r},t)$ is \emph{Poynting's vector}, which describes the directional energy flux carried by an electromagnetic wave\cite{Landau60,Jackson75}.} applied to $\ket{1_{\mathbold{k},\upsilon}}$ yields $\hbar \mathbold{k}$ for the momentum of a single photon: 
\begin{equation}\label{eq:photon_momentum}
 \underline{\mathbold{p}}_{EM} \ket{1_{\mathbold{k},\upsilon}} = \sum_{\mathbold{k}'} \sum_{\upsilon' = 1}^2 \hbar \mathbold{k}' \underline{a}^{\dagger}_{\mathbold{k}',\upsilon'} \underline{a}_{\mathbold{k}',\upsilon'} \ket{1_{\mathbold{k},\upsilon}} = \hbar \mathbold{k} \ket{1_{\mathbold{k},\upsilon}} .
 \end{equation}
Together, \cref{eq:photon_energy,eq:photon_momentum} indicate that photons can be described as having definite energy, $\hbar \omega_{\mathbold{k}}$, and definite momentum, $\hbar \mathbold{k}$, where using \cref{eq:disp_rel_vac}, it is seen that $\hbar \omega = c_0 \abs{\hbar \mathbold{k}}$.  
This is consistent with the relativistic energy-momentum relation:
\begin{equation}
 \mathcal{E}_{\gamma} = \sqrt{\left( m_0 c_0^2 \right)^2 + \left( p c_0 \right)^2}
 \end{equation}
with $\mathcal{E}_{\gamma} = \hbar \omega_{\mathbold{k}}$, rest mass $m_0=0$ and $p = \abs{\hbar \mathbold{k}}$, which describes photons as massless particles that carry momentum. 

\subsection{Coherent States and Classical Wave Modes}\label{sec:coherent_states}
%%%%%%%%%%%%%%%%%%%%%%%%%%%%%%%%%%%%%%%%%--------------------------------------------------
Quantum-mechanically, an electromagnetic wave of a single mode with a wave vector $\mathbold{k}$ and a polarization index $\upsilon$ can be understood as a \emph{coherent state} consisting of a large but uncertain number of photons. 
This is described by a superposition of Fock states with different values of photon number, $N_{\mathbold{k},\upsilon}$, such that the overall state is characterized by the mean number of photons. 
These coherent states, written as $\ket{\alpha_{\mathbold{k},\upsilon}}$, are defined by their property that the application of $\underline{a}^{\dagger}_{\mathbold{k},\upsilon}$ and $\underline{a}_{\mathbold{k},\upsilon}$ does not have a significant effect on the overall photon state. 
That is, they are eigenstates of the annihilation operator \cite{Fontana_82}:
\begin{subequations}
\begin{equation}
 \underline{a}_{\mathbold{k},\upsilon} \ket{\alpha_{\mathbold{k},\upsilon}} = \alpha_{\mathbold{k},\upsilon} \ket{\alpha_{\mathbold{k},\upsilon}} ,
 \end{equation}
where $\alpha_{\mathbold{k},\upsilon}$ is a complex eigenvalue, and the mean number of photons is determined from the following expectation value of the photon-number operator given by \cref{eq:number_operator}: 
\begin{equation}\label{eq:mean_photons}
 \expval{\underline{N_{\mathbold{k},\upsilon}}} \equiv \expval{\underline{N_{\mathbold{k},\upsilon}}}{\alpha_{\mathbold{k},\upsilon}} = \expval{\underline{a}^{\dagger}_{\mathbold{k},\upsilon} \underline{a}_{\mathbold{k},\upsilon} }{\alpha_{\mathbold{k},\upsilon}} =  \norm{\alpha_{\mathbold{k},\upsilon}}^2 
 \end{equation}
while the standard deviation can be shown to be \cite{Fontana_82}:
\begin{equation}\label{eq:uncertain_photons}
 \sigma_N \equiv \sqrt{\expval{\underline{N_{\mathbold{k},\upsilon}}^2} - \expval{\underline{N_{\mathbold{k},\upsilon}}}^2} = \norm{\alpha_{\mathbold{k},\upsilon}} .
 \end{equation}
\end{subequations}
With these properties, coherent states yield a result for $\expval{\underline{\mathbold{E}} ( \mathbold{r} )}$ that is consistent with a classical electromagnetic wave, where $\underline{\mathbold{E}} ( \mathbold{r} )$ is the electric-field operator defined by \cref{eq:E-field_operator_fixed_linear}. 
Dropping the subscripts in $\ket{\alpha_{\mathbold{k},\upsilon}}$ and $\ket{N_{\mathbold{k},\upsilon}}$, a coherent state can be written explicitly as \cite{Fontana_82,kapitell4} 
\begin{equation}\label{eq:coherent_quantum_state}
 \ket{\alpha} = \mathrm{e}^{- \frac{1}{2} \norm{\alpha}^2} \sum_{N=0}^{\infty} \frac{\alpha^{N}}{\sqrt{N!}} \ket{N} \quad \text{with } \alpha \equiv \alpha_{\mathbold{k},\upsilon} \text{ and } N \equiv N_{\mathbold{k},\upsilon}, 
 \end{equation}
where $\norm{\alpha}^2$ is the mean number of photons and $\norm{\alpha}$ is the uncertainty [\emph{cf.\@} \cref{eq:mean_photons,eq:uncertain_photons}]. 
As the number of photons becomes very large, the fractional uncertainty $\sigma_N / \expval{\underline{N_{\mathbold{k},\upsilon}}} = \norm{\alpha}^{-1}$ becomes very small and the coherent state may be treated as a classical wave of a single mode. 

The utility of treating soft x-rays like classical waves or photons depends on the context of the physical scenario. 
Due to their relatively high photon energy, $\mathcal{E}_{\gamma} \equiv \hbar \omega$, soft x-rays tend to be emitted from relatively dim sources on a photon-by-photon basis in a non-classical fashion. 
As alluded to in \cref{sec:astro_plasmas}, this is especially relevant in x-ray astronomy, where the number of photons collected during a typical observation tends to be low. 
The interaction between soft x-rays and atomic electrons is also best described in terms of photons, which is introduced in \cref{sec:atomic_interaction} for the purpose of treating spectral line phenomena in \cref{ap:quantum_spectral}. 
On the other hand, beamline 6.3.2 of the Advanced Light Source \cite[\emph{cf.\@} \cref{sec:als_beam}]{ALS_632} provides a bright, nearly-monochromatic beam of soft x-rays for the diffraction-efficiency experiments described in \cref{ch:master_fab,ch:grating_replication}. 
Such a beam can be regarded as a highly-coherent state with a large number of photons that can be treated as a classical wave for most intents and purposes. 
\begin{figure}
 \centering
 \includegraphics[scale=1.25]{Appendix-A/Figures/diffraction_arc12.mps}
 \caption[{Illustration of coherence length}]{Illustration of coherence length, $\ell_{\text{coh}}$, given by \cref{eq:coh_len}.}\label{fig:coherence_length}
 \end{figure}
However, as alluded to at the start of \cref{sec:EM_waves_vac}, real sources are always composed of some small spread of wave modes so that there is some finite \emph{coherence length} [\emph{cf.\@} \cref{fig:coherence_length}] that can be approximated the distance it takes for two superimposed modes of wavelength $\lambda$ and $\lambda + \Delta \lambda$ to become $\pi$ radians out of phase \cite{Attwood17}:
\begin{equation}\label{eq:coh_len}
 \ell_{\text{coh}} \approx \frac{\lambda^2}{2 \Delta \lambda} .
 \end{equation}

\section{Quantum Interaction with Atomic Electrons}\label{sec:atomic_interaction}
%%%%%%%%%%%%%%%%%%%%%%%%%%%%%%%%%%%%%%%%%--------------------------------------------------
The way that soft x-ray photons interact with atomic electrons is of interest in this thesis for describing the following phenomena: 
\begin{enumerate}[noitemsep]
  \item Spectral lines produced from bound-bound transitions in highly-charged ions present in cosmic plasmas (discussed  in \cref{ch:introduction,ap:quantum_spectral})
  \item Scattering from electrons bound in neutral atoms as soft x-rays interact with optical materials (discussed in \cref{app:x-ray_materials} and referenced in \cref{ch:diff_eff})
 \end{enumerate}
Here, non-relativistic quantum mechanics is used to treat electron-photon interactions so that, strictly speaking, results are only valid in atoms of relatively low $\mathcal{Z}$, where electrons have classical velocities much smaller than the speed of light, $c_0$.\footnote{Stated differently, it is considered that electronic binding energies are negligible compared to the electron rest energy, $m_e c_0^2 \approx \SI{511}{\kilo\electronvolt}$.} 
\begin{table}[]
 \centering
 \caption[Summary of notation used for atomic electron states]{Summary of notation used for atomic electron states. Shells K, L, M, etc.\ are tied to the \emph{principal quantum number}, $n$. The inner-most shell of any atom features only the s orbital, which can house just two electrons. As $n$ increases, the p, d, f and g orbitals are introduced, which correspond to quantized angular momentum states with a \emph{azimuthal quantum number}, $\ell = 0, 1 , 2 \dotsc n-1$, and a \emph{magnetic quantum number}, $m_{\ell} = -\ell, -\ell + 1, \dotsc 0, \dotsc \ell -1, \ell$. Owing to their intrinsic spin states given by the \emph{spin quantum number} taking on $m_s = \pm 1/2$, two electrons can exist in an orbital with specified $n$, $\ell$ and $m_{\ell}$.}\label{tab:shell_orbitals} 
 \begin{tabular}{@{}llllll@{}} 
 \toprule
 atomic shell & K & L & M & N & O \\ \midrule
 %orbital designation & $s$ & $p$ & $d$ & $f$ \\
 principal quantum number ($n$) & 1 & 2 & 3 & 4 & 5 \\ 
 maximum orbital angular momentum ($\ell$) & 0 & 1 & 2 & 3 & 4  \\
 possible number of electrons in shell & 2 & 8 & 18 & 32 & 50 \\ \midrule
 introduced orbital (subshell) & s & p & d & f & g \\
 possible number of electrons in subshell & 2 & 6 & 10 & 14 & 18 \\
 \bottomrule
 \end{tabular}
 \end{table}
While this framework suffices for discussion of basic physics, relativistic corrections are needed for treating:
\begin{enumerate}[noitemsep]
  \item Fine-structure and electron spin effects in spectral lines 
  \item Scattering from inner-shell electrons in high-$\mathcal{Z}$ atoms 
 \end{enumerate}
In the present discussion, a non-relativistic electron is represented by an abstract \emph{wave vector}, $\ket{\Psi_e}$, in \emph{Hilbert space} \cite{Hilbert1928}, where vectors describing different bound or free electron states are all taken to be normalized and orthogonal to one another:
\begin{equation}\label{eq:electron_orthonormal}
  \braket{\Psi_e'}{\Psi_e} =
  \begin{cases}
    1, & \text{for states with identical quantum numbers} \\ 
    0, & \text{otherwise.}
  \end{cases}
 \end{equation}
For bound electrons, $\ket{\Psi_e}$ depends only on the principal, azimuthal and magnetic quantum numbers $n$, $\ell$ and $m_{\ell}$ [\emph{cf.\@} \cref{tab:shell_orbitals}] while free electrons are described as plane waves that depend on the particle's kinetic energy \cite{Shankar04,Griffiths05,Townsend00}. 

Bound-bound transitions, and often times scattering events, involve a change in electronic state where the wave vector, $\ket{\Psi_e (t)}$, evolves in time according to the \emph{Schr\"odinger equation} \cite{Schrodinger1926}:
\begin{equation}\label{eq:Schrodinger_TD}
 \underline{\mathcal{H}}_{\text{atom}} \ket{\Psi_e (t)} = i \hbar \dv{t} \ket{\Psi_e (t)} ,
\end{equation}
where $\underline{\mathcal{H}}_{\text{atom}}$ is an arbitrary Hamiltonian operator that describes the total non-relativistic energy of an electron bound to a highly-charged ion or a neutral atom \cite{Griffiths05,Shankar04,Townsend00,Bethe57}. 
To start, a generic case is considered where the Hamiltonian is assumed to depend only on the phase-space coordinates of a single transitioning electron. 
While this condition is already fulfilled in hydrogen-like ions where there is only one bound electron, the \emph{self-consistent field approximation} \cite{Rybicki86} is invoked to describe also the approximate behavior of bound-bound transitions in a helium-like ion as well as scattering in neutral atoms. 
For instance, the transitioning electron in the case of a helium-like ion is, in principle, subject to the attractive force from the nucleus in addition to an averaged repulsive force generated by the second electron, which stays stationary in a bound state [\emph{cf.\@} \cref{sec:selection_rules_helium}]; in the case of an electron bound in a general neutral atom, the self-consistent field approximation takes into account the averaged repulsive force from the other $\mathcal{Z} - 1$ bound electrons. 
With $\underline{\mathbold{p}}$ and $\underline{\mathbold{r}}$ being the momentum and position operators for a single electron wave vector, $\ket{\Psi_e}$, and $\underline{V}(\underline{\mathbold{r}})$, an arbitrary potential-energy operator that describes a hydrogen-like ion or approximations of other highly-charged  ions and neutral atoms, the Hamiltonian operator is written generally as
\begin{equation}\label{eq:generic_ion_Hamiltonian}
 \underline{\mathcal{H}}_{\text{atom}} = \frac{\underline{\mathbold{p}}^2}{2 m_e} + \underline{V}(\underline{\mathbold{r}}) .
 \end{equation}

\subsection{Bound Electrons Coupled to the Photon Field}\label{sec:ions_EM}
%%%%%%%%%%%%%%%%%%%%%%%%%%%%%%%%%%%%%%%%%--------------------------------------------------
Describing bound-bound transitions or scattering in terms of photons requires coupling the electronic Hamiltonian operator, $\underline{\mathcal{H}}_{\text{atom}}$ [\emph{cf.\@} \cref{eq:generic_ion_Hamiltonian}], to the quantized electromagnetic field, which is associated with the Hamiltonian operator $\underline{\mathcal{H}}_{EM}$ defined by \cref{eq:photon_Hamiltonian_operator} \cite{Shankar04,Townsend00,Fontana_82,kapitell4,wysin2011quantization}. 
As a first step, it is useful to consider a single bound electron as a classical particle of charge $-q_e$ subject a force characterized by $V(\mathbold{r})$, the classical potential energy analog to $\underline{V}(\underline{\mathbold{r}})$ introduced above. 
In a purely classical picture of a one-electron atom, the electron orbits the nucleus a distance $r$ away with a velocity $v$ but because the charge experiences a centripetal acceleration, $a = v^2 / r$, its orbit should decay as it emanates electromagnetic radiation.\footnote{For an electron with charge $-q_e$ and acceleration $a$, the total power radiated in all directions is given by $q_e^2 a^2 /6 \pi \epsilon_0 c_0^3$ from the \emph{Larmor radiation formula} \cite{Jackson75,Rybicki86}.} 
Starting from this framework however, a quantum picture can be developed to explain how electrons interact with photons as they transition between bound states. 
To start, the \emph{Lagrangian} for such a particle of mass $m_e$ coupled to the classical field represented by $\mathbold{A} ( \mathbold{r} , t )$ can be taken as \cite{Landau75,Landau76,Kahn02,Goldstein02}\footnote{For simplicity, the mass of the nucleus here is treated as being infinite so that center-of-mass corrections are ignored.}
\begin{equation}\label{eq:charged_particle_lagrangian}
 \mathcal{L} \left( \dot{\mathbold{r}}, \mathbold{r} , t \right) = \frac{1}{2} m_e \dot{\mathbold{r}}^2 - V (\mathbold{r}) - q_e \mathbold{A} ( \mathbold{r} , t ) \cdot \dot{\mathbold{r}} ,
 \end{equation}
where $\dot{\mathbold{r}} \equiv \dv*{\mathbold{r}}{t}$ and $\mathbold{A} ( \mathbold{r} , t )$ is defined by \cref{eq:potential_wave_solution}. 
Classically, a Hamiltonian for a charged particle subject to the electromagnetic field is generated by carrying out a \emph{Legendre transformation} of \cref{eq:charged_particle_lagrangian} \cite{Landau75,Landau76,Goldstein02}: 
\begin{subequations}
\begin{equation}
 \mathcal{H} \left( \mathbold{p} , \mathbold{r} , t \right) = \mathbold{p} \cdot \dot{\mathbold{r}} - \mathcal{L} \left( \dot{\mathbold{r}}, \mathbold{r} , t \right) ,
 \end{equation}
where $\mathbold{p}$ is the \emph{canonical momentum}:
\begin{equation}
 \mathbold{p} \equiv \pdv{\mathcal{L} \left( \dot{\mathbold{r}}, \mathbold{r} , t \right)}{\dot{\mathbold{r}}} = m_e \dot{\mathbold{r}} - q_e \mathbold{A} ( \mathbold{r} , t ) .
 \end{equation}
This comes out to 
\begin{equation}\label{eq:coupled_Hamiltonian_classical}
 \mathcal{H} \left( \mathbold{p} , \mathbold{r} , t \right) = \frac{\left[ \mathbold{p} + q_e \mathbold{A} ( \mathbold{r} , t ) \right]^2}{2 m_e} + V (\mathbold{r}) , 
 \end{equation}
\end{subequations}
which suggests that $\underline{\mathcal{H}}_{\text{atom}}$ can be coupled to the electromagnetic field by replacing the usual momentum with the canonical momentum: $\mathbold{p} \to \mathbold{p} + q_e \mathbold{A} ( \mathbold{r} , t )$. 

Now, $\underline{\mathcal{H}}_{\text{atom}}$ is coupled to the quantized electromagnetic field by replacing quantities $\mathbold{p}$, $\mathbold{r}$, $\mathbold{A} ( \mathbold{r} , t )$ and $\mathcal{H}$ in \cref{eq:coupled_Hamiltonian_classical} by their quantum operator counterparts, $\underline{\mathbold{r}}$, $\underline{\mathbold{p}}$ and $\underline{\mathbold{A}} ( \mathbold{r} )$ (with the latter given by \cref{eq:vector_potential_operator_fixedtime}), while also adding in $\underline{\mathcal{H}}_{EM}$ \cite{Shankar04,Townsend00,Fontana_82,kapitell4,wysin2011quantization}.  
Noting that the operators $\underline{\mathbold{p}}$ and $\underline{\mathbold{A}} ( \mathbold{r} )$ do not in general commute, 
\begin{subequations} 
\begin{align}\label{eq:coupled_Hamiltonian1}
 \begin{split}
 \underline{\mathcal{H}} &= \frac{ \left[ \underline{\mathbold{p}} + q_e \underline{\mathbold{A}} ( \mathbold{r} ) \right]^2}{2 m_e} + \underline{V}(\underline{\mathbold{r}}) + \underline{\mathcal{H}}_{EM} \\
 &= \underbrace{\frac{\underline{\mathbold{p}}^2}{2 m_e} + \underline{V}(\underline{\mathbold{r}})}_{\underline{\mathcal{H}}_{\text{atom}}} + \underline{\mathcal{H}}_{EM} + \frac{q_e \underline{\mathbold{p}} \cdot \underline{\mathbold{A}} ( \mathbold{r} )}{2 m_e} + \frac{q_e \underline{\mathbold{A}} ( \mathbold{r} ) \cdot \underline{\mathbold{p}}}{2 m_e} + \frac{q_e^2 \underline{\mathbold{A}}^2 ( \mathbold{r} )}{2 m_e} . 
 \end{split}
 \end{align}
However, this can be simplified by examining how the $\underline{\mathbold{p}} \cdot \underline{\mathbold{A}} ( \mathbold{r} )$ term operates in the position basis and then exploiting the condition on $\underline{\mathbold{A}} ( \mathbold{r} )$ put in place by the Coulomb gauge, where $\div \underline{\mathbold{A}} ( \mathbold{r} ) = 0$ \cite{Shankar04,Townsend00,Fontana_82,Rybicki86}. 
With the momentum operator taking on the form $\underline{\mathbold{p}} \to - i \hbar \grad$ in the position basis, the $\div \underline{\mathbold{A}}(\mathbold{r})$ piece of $\underline{\mathbold{p}} \cdot \underline{\mathbold{A}}(\mathbold{r})$ acts on a wave function $\Psi(\mathbold{r})$ in the following way:  
\begin{align}
 \begin{split}
 &\div \underline{\mathbold{A}}(\mathbold{r}) \, \Psi(\mathbold{r}) = \underbrace{\div \underline{\mathbold{A}}(\mathbold{r} ) }_{0 \text{ in Coulomb gauge}} \Psi(\mathbold{r}) + \underline{\mathbold{A}}(\mathbold{r}) \cdot \grad \Psi(\mathbold{r}) \\
 &\text{therefore, } \underline{\mathbold{p}} \cdot \underline{\mathbold{A}}(\mathbold{r}) = \underline{\mathbold{A}}(\mathbold{r}) \cdot \underline{\mathbold{p}} .
 \end{split} 
 \end{align}
After making this simplification using \cref{eq:potential_B-field}, \cref{eq:coupled_Hamiltonian1} becomes 
\begin{equation}\label{eq:coupled_Hamiltonian3}
 \underline{\mathcal{H}} = \underbrace{\underline{\mathcal{H}}_{\text{atom}} + \underline{\mathcal{H}}_{EM}}_{\underline{\mathcal{H}}^{(0)}} + \underbrace{\frac{q_e \underline{\mathbold{A}} ( \mathbold{r} ) \cdot \underline{\mathbold{p}}}{m_e}}_{\underline{\mathcal{H}}^{(1)}} + \underbrace{\frac{q_e^2 \underline{\mathbold{A}}^2 ( \mathbold{r} )}{2 m_e}}_{\underline{\mathcal{H}}^{(2)}} .
 \end{equation}
\end{subequations}
This shows that the total Hamiltonian, $\underline{\mathcal{H}}$, is equal to the sum of the Hamiltonians for an atom and the quantized electromagnetic field separately (with the latter denoted by $\underline{\mathcal{H}}^{(0)}$), plus two additional terms involving the operator for the quantized electromagnetic field $\underline{\mathbold{A}} ( \mathbold{r} )$, denoted by $\underline{\mathcal{H}}^{(1)}$ and $\underline{\mathcal{H}}^{(2)}$ \cite{Shankar04,Townsend00,Fontana_82}. 
Now that the momentum and position operators for the electron, $\underline{\mathbold{p}}$ and $\underline{\mathbold{r}}$, and the annihilation and creation operators for photons, $\underline{a}_{\mathbold{k},\upsilon}$ and $\underline{a}^{\dagger}_{\mathbold{k},\upsilon}$, are included in the same Hamiltonian, the operator $\underline{\mathcal{H}}$ acts on a state defined by the direct product of a particle wave function and a photon state. 
For example, an atomic electron in a state defined by a wave vector $\ket{\Psi_e}$ combined with $N$ photons existing in a mode described by $\mathbold{k}$ and $\upsilon$ is written as $\ket{\Psi_e} \otimes \ket{N_{\mathbold{k},\upsilon}}$ \cite{Townsend00}. 

\subsection{Time-Dependent Perturbation Theory}\label{sec:quantum_perturbations} %of Radiative Transitions
%%%%%%%%%%%%%%%%%%%%%%%%%%%%%%%%%%%%%%--------------------------------------------------
A bound electron that stays stationary in a particular state for a sufficient amount of time can be treated using the time-independent form of the Schr\"odinger equation:
\begin{equation}\label{eq:TI_Schrodinger}
 \underline{\mathcal{H}}_{\text{atom}} \ket{\Psi_e} = \mathcal{E}_e \ket{\Psi_e} ,
 \end{equation}
where $\ket{\Psi_e}$ is the wave vector of the stationary bound state and $\mathcal{E}_e$ is the electron binding energy.\footnote{Essentially, this is an \emph{eigenvalue problem}, which is often formulated in terms of traditional vectors as $\underline{\mathbold{M}} \mathbold{v} = \lambda \mathbold{v}$, where $\underline{\mathbold{M}}$ is a square matrix (in place of an abstract operator), $\mathbold{v}$ is a column vector and $\lambda$ is the scalar eigenvalue \cite{Shankar04}.}  
Such an electron is associated with a time-dependent wave vector given by $\ket{\Psi_e (t)} = \ket{\Psi_e} \mathrm{e}^{-i \mathcal{E} t / \hbar}$, where its probability density is independent of time:
\begin{equation}
 \norm{\braket{\mathbold{r}}{\Psi_e (t)}}^2 = \norm{\braket{\mathbold{r}}{\Psi_e}}^2 \equiv \norm{\Psi (\mathbold{r})}^2 .
 \end{equation}
As an electron makes a bound-bound transition, however, there is an associated probability that evolves with time and this behavior directly determines how the electron interacts with the quantized electromagnetic field to absorb, emit or scatter a photon of a particular energy \cite{Townsend00,Fontana_82}. 
In principle, this can be handled using a \emph{time-evolution operator}, $\underline{U} (t)$, that acts on an initial wave function condition, $\ket{\Psi (0)}$, to describe how the state evolves with time, $t$ \cite{Townsend00,Shankar04}:
\begin{subequations}
\begin{equation}
 \underline{U} (t) \ket{\Psi (0)} = \ket{\Psi (t)} . 
 \end{equation}
Inserting this into the time-dependent Schr\"odinger equation [\emph{cf.\@} \cref{eq:Schrodinger_TD}] yields the following relation that defines the functional form of $\underline{U} (t)$:
\begin{equation}\label{eq:evolution_operator_Schro}
 i \hbar \dv{t} \underline{U} (t) = \underline{\mathcal{H}} \, \underline{U} (t) ,
 \end{equation}
where $\underline{\mathcal{H}}_{\text{atom}}$ has been replaced by the full Hamiltonian, $\underline{\mathcal{H}}$ [\emph{cf.\@} \cref{eq:coupled_Hamiltonian3}]. % defined in \cref{eq:coupled_Hamiltonian3}. %the full Hamiltonian,
\end{subequations}
However, \cref{eq:evolution_operator_Schro} cannot be solved exactly for this Hamiltonian and hence \emph{time-dependent perturbation theory} must be used to examine how $\underline{\mathcal{H}}^{(1)}$ and $\underline{\mathcal{H}}^{(2)}$ contribute to electron-photon interactions \cite{Townsend00,Shankar04,Bethe57}. % and hence spectral lines in the measured spectrum \cite{Townsend00,Shankar04,Bethe57}. 

So far in this appendix, quantum mechanics has been formulated in the Schr\"odinger picture, where operators do not depend on time but the wave vectors in general do [\emph{cf.\@} appendix~\ref{sec:QEM_operators}]. %as mentioned in \cref{sec:QEM_operators}, 
To carry out time-dependent perturbation theory for radiative transitions and scattering, however, it is useful to switch from the Schr\"odinger picture to the \emph{interaction picture}, where both operators and wave functions carry time dependence with the former given by \cite{Townsend00,Shankar04} 
\begin{equation}\label{eq:interaction_operator}
 \underline{O}_I (t) = \mathrm{e}^{\frac{i}{\hbar} t \underline{\mathcal{H}}^{(0)}} \underline{O} \, \mathrm{e}^{-\frac{i}{\hbar} t \underline{\mathcal{H}}^{(0)}} .
 \end{equation}
In a similar manner to \cref{eq:evolution_operator_Schro}, the time-evolution operator in the interaction picture, $\underline{U}_I (t)$, is defined by the following relation \cite{Townsend00,Shankar04}: 
\begin{subequations}
\begin{equation}\label{eq:evolution_operator}
 i \hbar \dv{t} \underline{U}_I (t) = \underline{\mathcal{H}}^{(\text{pert})}_I (t) \, \underline{U}_I (t) ,
 \end{equation} 
where $\underline{\mathcal{H}}^{(\text{pert})}_I (t)$ can be considered to be either $\underline{\mathcal{H}}^{(1)}$ or $\underline{\mathcal{H}}^{(2)}$ inserted into \cref{eq:interaction_operator}.
For an initial state $\ket{A}$ and a final state $\ket{B}$, which designate both the electron and photon states (\emph{e.g.}, $\ket{\Psi_e} \otimes \ket{N_{\mathbold{k},\upsilon}}$), this time-evolution operator is used to calculate the probability of the transition occurring as function of time:
%this probability is the norm squared of the following matrix element:
\begin{equation}\label{eq:prob_matrix_element}
 \mathscr{P}_{A \to B} (t) = \norm{\matrixel{B}{\underline{U}_I (t)}{A}}^2 . 
 \end{equation}
To first order, a solution to \cref{eq:evolution_operator} for $\underline{U}_I (t)$ can be calculated from 
\begin{equation}\label{eq:first_order_evolution}
 \underline{U}_I (t) \approx 1 - \frac{i}{\hbar} \int_0^t \underline{\mathcal{H}}^{(\text{pert})}_I (t') \dd{t'}
 \end{equation}
with $\underline{U}_I (0) = 1$ \cite{Townsend00,Shankar04}. 
\end{subequations}
This operator, which contains $\underline{\mathcal{H}}^{(1)}$ and $\underline{\mathcal{H}}^{(2)}$ terms in first order, now can be used to calculate the probability of various electron-photon processes to occur as a function of time. 

\subsection{Framework for a First-Order Transition}\label{sec:gen_first_order}
%%%%%%%%%%%%%%%%%%%%%%%%%%%%%%%%%%%%%%%%%--------------------------------------------------
A general first-order transition between two electron-photon states can be described as a single-state transition between one arbitrary state, $\ket{A} = \ket{\Psi_A} \otimes \ket{N_{\mathbold{k},\upsilon}}$, to another, $\ket{B} = \ket{\Psi_B} \otimes \ket{N_{\mathbold{k}',\upsilon'}'}$, which are associated with initial and final electronic binding energies, $\mathcal{E}_A$ and $\mathcal{E}_B$, as well as initial and final wave vectors, $\mathbold{k}$ and $\mathbold{k}'$, respectively \cite{Shankar04,kapitell4,wysin2011quantization}. 
As indicated by \cref{eq:prob_matrix_element}, the probability as a function of time for this transition to occur is determined by evaluating the norm squared of the following matrix element: 
\begin{align}\label{eq:general_evolution1}
 \begin{split}
 \bra{\Psi_B} &\otimes \bra{N_{\mathbold{k}',\upsilon'}'} \underline{U}_I (t) \ket{\Psi_A} \otimes \ket{N_{\mathbold{k},\upsilon}} \\
 &= - \frac{i}{\hbar} \int_0^t \bra{\Psi_B} \otimes \bra{N_{\mathbold{k}',\upsilon'}'} \underline{\mathcal{H}}^{(\text{pert})}_{I} (t') \ket{\Psi_A} \otimes \ket{N_{\mathbold{k},\upsilon}} \dd{t'} , 
 \end{split}
 \end{align}
where from \cref{eq:interaction_operator}, 
\begin{equation}\label{eq:H_pert_general}
 \underline{\mathcal{H}}^{(\text{pert})}_I (t) = \mathrm{e}^{\frac{i}{\hbar} t \underline{\mathcal{H}}^{(0)}} \underline{\mathcal{H}}^{(\text{pert})} \mathrm{e}^{-\frac{i}{\hbar} t \underline{\mathcal{H}}^{(0)}} 
 \end{equation}
with $\underline{\mathcal{H}}^{(0)} = \underline{\mathcal{H}}_{\text{atom}} + \underline{\mathcal{H}}_{EM}$ [\emph{cf.\@} \cref{eq:coupled_Hamiltonian3,eq:generic_ion_Hamiltonian,eq:photon_Hamiltonian_operator}]. 
As a consequence of the assumed eigenvalue relation for the unperturbed electron state [\emph{cf.\@} \cref{eq:TI_Schrodinger}], it holds true that 
\begin{subequations}
\begin{equation}\label{eq:exponential_electron_eigenvalue}
 \bra{\Psi_B} \mathrm{e}^{ \frac{i}{\hbar} t \underline{\mathcal{H}}_{\text{atom}}} = \bra{\Psi_B} \mathrm{e}^{ \frac{i}{\hbar} \mathcal{E}_B t} \quad \text{and} \quad \mathrm{e}^{- \frac{i}{\hbar} t \underline{\mathcal{H}}_{\text{atom}}} \ket{\Psi_A} = \mathrm{e}^{- \frac{i}{\hbar} \mathcal{E}_A t} \ket{\Psi_A} ,
 \end{equation}
while from the properties of Fock states defined in \cref{eq:N_photons,eq:N_photon_destruction}, 
\begin{align}\label{eq:exponential_photon_eigenvalue}
 \begin{split}
 \bra{N_{\mathbold{k}',\upsilon'}'} \mathrm{e}^{\frac{i}{\hbar} t \underline{\mathcal{H}}_{EM}} &= \bra{N_{\mathbold{k}',\upsilon'}'} \mathrm{e}^{\frac{i \mathcal{E}_0}{\hbar} t} \mathrm{e}^{i N_{\mathbold{k}',\upsilon'}' \omega_{\mathbold{k}'} t} \\
 &\text{and} \quad \mathrm{e}^{-\frac{i}{\hbar} t \underline{\mathcal{H}}_{EM}} \ket{N_{\mathbold{k},\upsilon}} = \mathrm{e}^{-\frac{i \mathcal{E}_0}{\hbar} t} \mathrm{e}^{-i N_{\mathbold{k},\upsilon} \omega_{\mathbold{k}} t} \ket{N_{\mathbold{k},\upsilon}} ,
 \end{split}
 \end{align}
\end{subequations}
where $\mathcal{E}_0$ is the vacuum energy [\emph{cf.\@} \cref{eq:vacuum_energy}]. 
The matrix element from \cref{eq:general_evolution1} then becomes 
\begin{subequations}
\begin{align}\label{eq:general_evolution2a}
 \begin{split}
 &\bra{\Psi_B} \otimes \bra{N_{\mathbold{k}',\upsilon'}'} \underline{U}_I (t) \ket{\Psi_A} \otimes \ket{N_{\mathbold{k},\upsilon}} \\
 &\quad = - \frac{i}{\hbar} \int_0^t \mathrm{e}^{\frac{i}{\hbar} \left( \Delta \mathcal{E}_e - \Delta \gamma \right) t'} \bra{\Psi_B} \otimes \bra{N_{\mathbold{k}',\upsilon'}'} \underline{\mathcal{H}}^{(\text{pert})} \ket{\Psi_A} \otimes \ket{N_{\mathbold{k},\upsilon}} \dd{t'}
 \end{split}
 \end{align}
with $\Delta \mathcal{E}_e \equiv \mathcal{E}_B - \mathcal{E}_A$ and $\Delta \gamma \equiv N_{\mathbold{k}',\upsilon'}' \, \hbar \omega_{\mathbold{k}'} - N_{\mathbold{k},\upsilon} \, \hbar \omega_{\mathbold{k}}$. 

Evaluating the integral in \cref{eq:general_evolution2a} gives 
\begin{equation}\label{eq:general_evolution2b}
 \bra{\Psi_B} \otimes \bra{N_{\mathbold{k}',\upsilon'}'} \underline{U}_I (t) \ket{\Psi_A} \otimes \ket{N_{\mathbold{k},\upsilon}} = \mathcal{M} \, \left( \frac{\mathrm{e}^{\frac{i}{\hbar} \left( \Delta \mathcal{E}_e - \Delta \gamma \right) t} - 1}{\Delta \mathcal{E}_e - \Delta \gamma} \right) ,
 \end{equation}
where $\mathcal{M} \equiv \bra{\Psi_B} \otimes \bra{N_{\mathbold{k},\upsilon}'} \underline{\mathcal{H}}^{(\text{pert})} \ket{\Psi_A} \otimes \ket{N_{\mathbold{k},\upsilon}}$ is the matrix element for a time-independent perturbation arising from $\underline{\mathcal{H}}^{(\text{pert})}$ \cite{Shankar04,Townsend00}. 
\end{subequations}
Using $\varpi \equiv \left( \Delta \mathcal{E}_e - \Delta \gamma \right) / \hbar$, the time-dependent piece of \cref{eq:general_evolution2b} 
can be rewritten as 
\begin{equation}\label{eq:complex_exponential_sinc}
 \frac{\mathrm{e}^{i \varpi t} - 1}{i \varpi} = \frac{\mathrm{e}^{i \varpi t / 2}}{i \varpi} \underbrace{\left( \mathrm{e}^{i \varpi t / 2} - \mathrm{e}^{-i \varpi t / 2}  \right)}_{2 i \sin \left( \frac{\varpi t}{2} \right)} = t \, \mathrm{e}^{i \varpi t / 2} \sinc \left( \frac{\varpi t}{2} \right) ,
 \end{equation}
where $\sinc (x) \equiv \sin (x) / x$ is the \emph{sinc function}, and the probability of absorption occurring as a function of time according to \cref{eq:prob_matrix_element} is \cite{Shankar04,Townsend00} 
\begin{equation}\label{eq:general_prob_trans}
 %\begin{split}
 \mathscr{P}_{A \to B} (t) \equiv \norm{\bra{\Psi_B} \otimes \bra{N_{\mathbold{k}',\upsilon'}' } \underline{U}_I (t) \ket{\Psi_A} \otimes \ket{N_{\mathbold{k},\upsilon}}}^2 = \frac{t^2}{\hbar^2} \norm{\mathcal{M}}^2 \sinc^2 \left( \frac{\varpi t}{2} \right) .
 \end{equation}
From \cref{eq:general_prob_trans}, it can be stated that the probability of a general first-order transition between electron-photon states $\ket{A}$ and $\ket{B}$ depends on $\norm{\mathcal{M}}^2$ while there is spread of possible photon energies that becomes narrower as $t$ associated with the transition progresses. 
The long-time limit of this distribution can be represented as a Dirac delta function \cite{Townsend00}: % \cite[\emph{cf.\@} footnote~\ref{footnote:dirac_delta}]{Townsend00}: 
\begin{equation}\label{eq:Dirac_limit}
 \lim_{t \to \infty} \frac{t}{\pi} \sinc^2 \left( \frac{\varpi t}{2} \right) = \delta_D \left( \frac{\varpi}{2} \right) \equiv
   \begin{cases}
     \infty, & \text{if}\ \frac{\varpi}{2} = 0 \\
     0, & \text{if}\ \frac{\varpi}{2} \neq 0 .
   \end{cases} 
 \end{equation}
Inserting \cref{eq:Dirac_limit} into \cref{eq:absorption_evolution4} and recovering the definition of $\varpi$ indicates that the relation $\Delta \gamma = \Delta \mathcal{E}_e$ is only necessarily satisfied in the limit that $t \to \infty$:%\footnote{Here, the following property of Dirac delta functions has been used:
%\begin{equation*}
% \delta_D \left( \frac{\Delta \mathcal{E}_e - \Delta \gamma }{2 \hbar} \right) = 2 \hbar \delta_D \left( \Delta \mathcal{E}_e - \Delta \gamma \right) .
% \end{equation*}}
\begin{align}\label{eq:general_evolution5}
 \begin{split}
 \mathscr{P}_{A \to B} (t) &= \frac{t^2}{\hbar^2} \norm{\mathcal{M}}^2 \sinc^2 \left( \frac{\left( \Delta \mathcal{E}_e - \Delta \gamma \right) t}{2 \hbar} \right) \\
 \mathscr{P}_{A \to B}^{\infty} (t) \equiv \lim_{t \to \infty} \mathscr{P}_{A \to B} (t) &= \frac{2 \pi t}{\hbar} \norm{\mathcal{M}}^2 \delta_D \left( \Delta \mathcal{E}_e - \Delta \gamma \right) \propto t ,
 \end{split}
 \end{align}
which states that under first-order perturbation theory, the probability for the transition to occur increases in proportion to $t$ in the long-time limit while the spread of possible photon energies that can be absorbed becomes infinitesimally narrow with $t \to \infty$.  
\begin{figure}
 \centering
 \includegraphics[scale=1]{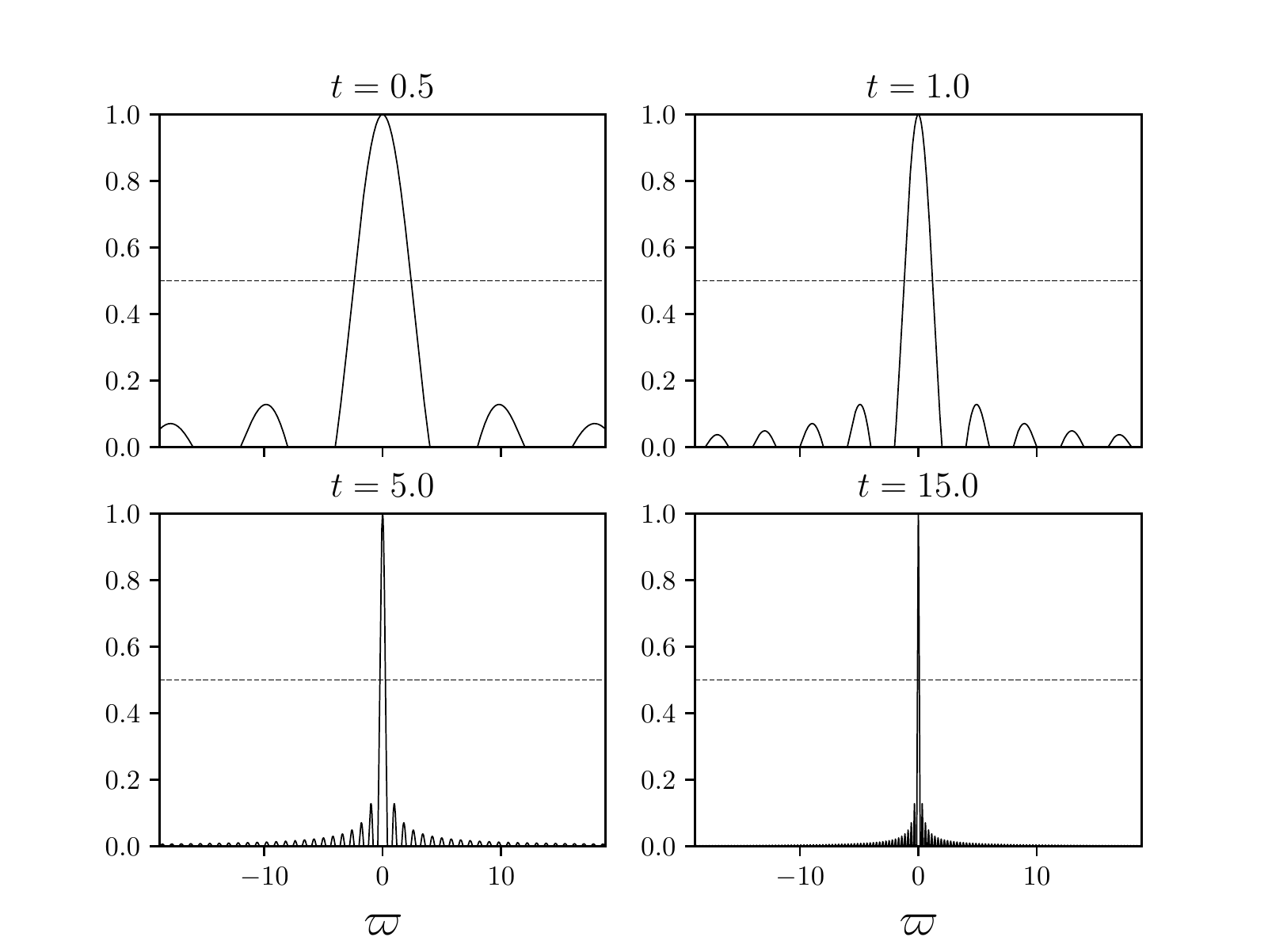}
 \caption[Plot of sinc-squared functions from quantum perturbation theory]{Plot of $\sinc^2 \left( \varpi t /2 \right)$ with $\varpi \equiv \left( \Delta \mathcal{E}_e - \Delta \gamma \right) / \hbar$ for four values of time $t$ using arbitrary units. Dashed lines mark half maximum of the central peak.}\label{fig:sinc_squared} %show that the central-peak width decreases with increasing $t$.
 \end{figure}
This is shown graphically in \cref{fig:sinc_squared}, where it is seen that the spectrum of possible photon energies for a first-order transition becomes narrower as time increases.\footnote{Note, however, that the factor of $t^2$ in \cref{eq:general_evolution5} is left out for the sake of comparison.} 

Depending on $\underline{\mathcal{H}}^{(\text{pert})}$, the quantity $\varpi = \left( \Delta \mathcal{E}_e - \Delta \gamma \right) / \hbar$ reduces to different forms. 
Processes that involve an electronic transition with a net change in binding energies, and either the absorption or emission of a single photon, can be described using first-order perturbations of the operator $\underline{\mathcal{H}}^{(1)} = (q_e / m_e) \, \underline{\mathbold{A}} ( \mathbold{r} ) \cdot \underline{\mathbold{p}}$ \cite[\emph{cf.\@} \cref{sec:ions_EM}]{Townsend00,Shankar04,Fontana_82,kapitell4,wysin2011quantization}. 
This can be seen by noting that $\underline{\mathcal{H}}^{(1)}$ depends linearly on both the Hilbert-space momentum operator $\underline{\mathbold{p}}$ and the Fock-space operators $\underline{a}_{\mathbold{k},\upsilon}$ and $\underline{a}^{\dagger}_{\mathbold{k},\upsilon}$ that are contained within the operator $\underline{\mathbold{A}} ( \mathbold{r} )$: 
\begin{align}\label{eq:single_photon_Hamiltonian}
 \begin{split}
 \underline{\mathcal{H}}^{(1)} = \frac{q_e}{m_e} \underline{\mathbold{A}} ( \mathbold{r} ) \cdot \underline{\mathbold{p}} &= \underbrace{\frac{q_e}{m_e} \sum_{\mathbold{k}} \sum_{\upsilon = 1}^2  \sqrt{\frac{\hbar}{2 \mathcal{V} \epsilon_0 \omega_{\mathbold{k}}}} \left( \hat{\mathbold{e}}_{\mathbold{k},\upsilon} \cdot \underline{\mathbold{p}} \right) \, \underline{a}_{\mathbold{k},\upsilon} \mathrm{e}^{i \mathbold{k} \cdot \mathbold{r} }}_{\text{absorption}} \\
 &+  \underbrace{\frac{q_e}{m_e} \sum_{\mathbold{k}} \sum_{\upsilon = 1}^2  \sqrt{\frac{\hbar}{2 \mathcal{V} \epsilon_0 \omega_{\mathbold{k}}}} \left( \hat{\mathbold{e}}_{\mathbold{k},\upsilon} \cdot \underline{\mathbold{p}} \right) \, \underline{a}^{\dagger}_{\mathbold{k},\upsilon} \mathrm{e}^{-i \mathbold{k} \cdot \mathbold{r}}}_{\text{emission}} . 
 \end{split}
 \end{align} 
In this case, $\varpi \to \left(\Delta \mathcal{E}_e - \hbar \omega_{\mathbold{k}} \right) / \hbar$ with $\mathcal{E}_B \neq \mathcal{E}_A$ and a photon number change of $\abs{N_{\mathbold{k}',\upsilon'}' - N_{\mathbold{k},\upsilon}} = 1$ [\emph{cf.\@} \cref{sec:single_absorption_emission}]. 
Therefore, the distribution in \cref{fig:sinc_squared} is a function of the absorbed or emitted photon energy with a centroid at $\Delta \mathcal{E}_e \equiv \hbar \omega_{\text{trans}}$, where $\omega_{\text{trans}}$ is the \emph{transition frequency}. 
On the other hand, $\underline{\mathcal{H}}^{(2)} = \left( q_e^2 / 2 m_e \right) \underline{\mathbold{A}}^2 \left( \mathbold{r} \right)$ contains cross-terms of the annihilation and creation operators: 
\begin{align}\label{eq:two_photon_Hamiltonian}
 \begin{split}
 &\underline{\mathcal{H}}^{(2)} = \frac{q_e^2}{2 m_e} \underline{\mathbold{A}}^2 \left( \mathbold{r} \right) \\
 &= \sum_{\mathbold{k},\mathbold{k}'} \sum_{\upsilon,\upsilon' = 1}^2 \left( \frac{\hbar q_e^2 \hat{\mathbold{e}}_{\mathbold{k},\upsilon} \cdot \hat{\mathbold{e}}_{\mathbold{k}',\upsilon'} }{4 \mathcal{V} \epsilon_0 m_e \omega_{\mathbold{k}} \, \omega_{\mathbold{k}'}} \right) \left[ \underline{a}_{\mathbold{k},\upsilon} \mathrm{e}^{i \mathbold{k} \cdot \mathbold{r} } + \underline{a}^{\dagger}_{\mathbold{k},\upsilon} \mathrm{e}^{-i \mathbold{k} \cdot \mathbold{r} } \right] \left[ \underline{a}_{\mathbold{k}',\upsilon'} \mathrm{e}^{i \mathbold{k}' \cdot \mathbold{r} } + \underline{a}^{\dagger}_{\mathbold{k}',\upsilon'} \mathrm{e}^{-i \mathbold{k}' \cdot \mathbold{r} } \right] 
 \end{split}
 \end{align} 
and so describes, among other two-photon processes, scattering phenomena, where $\underline{a}_{\mathbold{k},\upsilon}$ annihilates the initial photon and $\underline{a}^{\dagger}_{\mathbold{k}',\upsilon'}$ creates the scattered photon. 
In such a \emph{coherent scattering} process [\emph{cf.\@} \cref{sec:SXR_med}], the initial and final electron states are the same ($\mathcal{E}_B = \mathcal{E}_A$) while there is no change in the number of photons $\left( N_{\mathbold{k}',\upsilon'}' = N_{\mathbold{k},\upsilon} \right)$. 
As a result, the distribution in \cref{fig:sinc_squared} is in this case a function of scattered photon energy with a peak corresponding to the energy of the incident photon and $\varpi \to \omega_{\mathbold{k}'} - \omega_{\mathbold{k}}$.  

The first-order behavior just described is somewhat analogous to the spectrum of a finite, undamped sinusoidal wave in classical electrodynamics. 
To demonstrate this, let the following scalar function represent the strength of the electric field in such a wave as it passes by some fixed position over some duration, $\mathcal{T}$ \cite{Fontana82}:
\begin{subequations}
\begin{equation}\label{eq:wave}
    u(t) =
    \begin{cases}
      u_0 \cos \left( \Phi - \omega_0 t \right) , & 0 \leq t \leq \mathcal{T} \\
      0, & \text{otherwise.} 
    \end{cases}
  \end{equation}
The temporal Fourier transform of this function is 
\begin{align}
 \begin{split}
 u (\omega) &= \int_{-\infty}^{\infty} u(t) \, \mathrm{e}^{i \omega t} \dd{t} = u_0 \int_0^{\mathcal{T}} \cos \left( \Phi - \omega_0 t \right) \mathrm{e}^{i \omega t} \dd{t} \\
 &= \frac{i u_0}{2} \left( \mathrm{e}^{-i \Phi} \frac{1 - \mathrm{e}^{i \left( \omega + \omega_0 \right) \mathcal{T}}}{\left( \omega + \omega_0 \right)} + \mathrm{e}^{i \Phi} \frac{1 - \mathrm{e}^{i \left( \omega - \omega_0 \right) \mathcal{T}}}{\left( \omega - \omega_0 \right)} \right) ,
 \end{split}
 \end{align}
which describes two distributions in the frequency domain centered around $\omega = \pm \omega_0$ with full width at half maximum roughly equal to $2 \pi / \mathcal{T}$. 
Assuming that $\mathcal{T} \gg \omega_0^{-1}$, the contribution from the $\omega + \omega_0$ term can be neglected and then $\norm{u (\omega)}^2$, which is proportional to the energy density of the field, is \cite{Fontana82} 
\begin{equation}
 \norm{u (\omega)}^2 = \frac{u_0^2 \mathcal{T}^2}{4} \sinc^2 \left( \frac{\Delta \omega \mathcal{T}}{2} \right) ,
 \end{equation}
\end{subequations}
where $\Delta \omega \equiv \omega - \omega_0$.
Here, the parameter $\varpi \equiv \left( \Delta \mathcal{E}_e - \Delta \gamma \right) / \hbar$ used in this subsection is analogous to $\Delta \omega$ for a classical wave, where in \cref{fig:sinc_squared}, the spread of $\omega$ would become increasingly narrower as $\mathcal{T} \to \infty$. 

\subsection{Single-Photon Absorption and Emission}\label{sec:single_absorption_emission} 
%%%%%%%%%%%%%%%%%%%%%%%%%%%%%%%%%%%%%%--------------------------------------------------
Depending on the initial and final electron states, first-order perturbations of $\underline{\mathcal{H}}^{(1)}$ can describe bound-bound as well as bound-free absorption and emission processes. 
First, \emph{photo-absorption} can be described as a bound-bound state transition between an electron-photon states $\ket{A} = \ket{\Psi_G} \otimes \ket{N_{\mathbold{k},\upsilon}}$ and $\ket{B} = \ket{\Psi_E} \otimes \ket{\left( N_{\mathbold{k},\upsilon} - 1 \right)}$, where in the latter there is one photon missing from the original $N_{\mathbold{k},\upsilon}$ photons that are assumed to exist in a mode described by $\mathbold{k}$ and $\upsilon$ \cite{Townsend00}. 
Here, the transition is considered to take place in a hypothetical two-level atom with a ground state $\ket{\Psi_G}$ and an excited state $\ket{\Psi_E}$, with binding energies $\mathcal{E}_G$ and $\mathcal{E}_E$ (assuming $\mathcal{E}_E > \mathcal{E}_G$).
In the context of astrophysical soft x-ray spectroscopy, this can be taken to represent one mode of the x-ray continuum source generated by an active galactic nucleus that is absorbed by a particular transition in a highly-charged ion [\emph{cf.\@} \cref{sec:astro_plasmas}]. 
The probability for absorption to occur as a function of time according to \cref{eq:prob_matrix_element} is \cite{Shankar04,Townsend00,kapitell4,wysin2011quantization} 
\begin{align}\label{eq:absorption_evolution4}
 \begin{split}
 &\mathscr{P}_{G \to E} (t) \equiv \norm{\bra{\Psi_E} \otimes \bra{\left( N_{\mathbold{k},\upsilon} -1 \right)} \underline{U}_I (t) \ket{\Psi_G} \otimes \ket{N_{\mathbold{k},\upsilon}}}^2 \\
 &= \frac{t^2}{\hbar^2} \norm{\mathcal{M}_{\text{abso}}}^2 \sinc^2 \left( \frac{\varpi t}{2} \right) \quad \text{where } \varpi \equiv \omega_{\text{trans}} - \omega_{\mathbold{k}} .
 \end{split}
 \end{align}
With $\underline{\mathcal{H}}^{(1)} = \left( q_e / m_e \right) \underline{\mathbold{A}} ( \mathbold{r} ) \cdot \underline{\mathbold{p}}$ given by \cref{eq:single_photon_Hamiltonian} as the perturbing operator, $\norm{\mathcal{M}_{\text{abso}}}^2$ can be evaluated by carrying out the following Fock-space operations: 
\begin{align} 
\begin{split}\label{eq:Fock_operations}
 &\bra{\Psi_E} \otimes \bra{\left( N_{\mathbold{k},\upsilon} -1 \right)} \underline{\mathbold{A}} ( \mathbold{r} ) \cdot \underline{\mathbold{p}} \ket{\Psi_G} \otimes \ket{N_{\mathbold{k},\upsilon}} \\
 &\: = \bra{\Psi_E} \otimes \bra{\left( N_{\mathbold{k},\upsilon} -1 \right)} \sum_{\mathbold{k}'} \sum_{\upsilon' = 1}^2 \sqrt{\frac{\hbar}{2 \mathcal{V} \epsilon_0 \omega_{\mathbold{k}'}}} \left( \hat{\mathbold{e}}_{\mathbold{k}',\upsilon'} \cdot \underline{\mathbold{p}} \right) \underline{a}_{\mathbold{k}',\upsilon'} \mathrm{e}^{i \mathbold{k}' \cdot \mathbold{r} } \ket{\Psi_G} \otimes \ket{N_{\mathbold{k},\upsilon}} \\
 &\; + \bra{\Psi_E} \otimes \bra{\left( N_{\mathbold{k},\upsilon} -1 \right)} \sum_{\mathbold{k}'} \sum_{\upsilon' = 1}^2 \sqrt{\frac{\hbar}{2 \mathcal{V} \epsilon_0 \omega_{\mathbold{k}'}}} \left( \hat{\mathbold{e}}_{\mathbold{k}',\upsilon'} \cdot \underline{\mathbold{p}} \right) \underline{a}^{\dagger}_{\mathbold{k}',\upsilon'} \mathrm{e}^{-i \mathbold{k}' \cdot \mathbold{r} } \ket{\Psi_G} \otimes \ket{N_{\mathbold{k},\upsilon}} \\
 &\: = \sqrt{\frac{\hbar N_{\mathbold{k},\upsilon}}{2 \mathcal{V} \epsilon_0 \omega_{\mathbold{k}}}} \hat{\mathbold{e}}_{\mathbold{k},\upsilon} \cdot \bra{\Psi_B} \underline{\mathbold{p}} \mathrm{e}^{i \mathbold{k} \cdot \mathbold{r} } \ket{\Psi_G} \underbrace{\braket{\left( N_{\mathbold{k},\upsilon} - 1 \right)}}_{1} \\
 &\; + \sqrt{\frac{\hbar \left( N_{\mathbold{k},\upsilon} + 1 \right)}{2 \mathcal{V} \epsilon_0 \omega_{\mathbold{k}}}} \hat{\mathbold{e}}_{\mathbold{k},\upsilon} \cdot \bra{\Psi_E} \underline{\mathbold{p}} \mathrm{e}^{-i \mathbold{k} \cdot \mathbold{r} } \ket{\Psi_G} \underbrace{\braket{\left( N_{\mathbold{k},\upsilon} - 1 \right)}{\left( N_{\mathbold{k},\upsilon} + 1 \right)}}_{0}
 \end{split}
 \end{align}
so that it becomes\footnote{The evaluation of $\hat{\mathbold{e}}_{\mathbold{k},\upsilon} \cdot \bra{\Psi_E} \underline{\mathbold{p}} \mathrm{e}^{ i \mathbold{k} \cdot \mathbold{r}} \ket{\Psi_G}$ is handled in \cref{ap:quantum_spectral}. }
\begin{align}
 \begin{split}
 \norm{\mathcal{M}_{\text{abso}}}^2 &= \norm{\bra{\Psi_E} \otimes \bra{\left( N_{\mathbold{k},\upsilon} -1 \right)} \underline{\mathcal{H}}^{(1)} \ket{\Psi_G} \otimes \ket{N_{\mathbold{k},\upsilon}}}^2 \\ 
 &= \frac{q_e^2}{m_e^2} \norm{\bra{\Psi_E} \otimes \bra{\left( N_{\mathbold{k},\upsilon} -1 \right)} \underline{\mathbold{A}} ( \mathbold{r} ) \cdot \underline{\mathbold{p}} \ket{\Psi_G} \otimes \ket{N_{\mathbold{k},\upsilon}}}^2 \\
 &= \frac{q_e^2 \hbar N_{\mathbold{k},\upsilon}}{2 m_e^2 \mathcal{V} \epsilon_0 \omega_{\mathbold{k}}} \norm{\hat{\mathbold{e}}_{\mathbold{k},\upsilon} \cdot \bra{\Psi_E} \underline{\mathbold{p}} \mathrm{e}^{ i \mathbold{k} \cdot \mathbold{r}} \ket{\Psi_G}}^2 .
 \end{split}
 \end{align}
Single-photon absorption can also occur as a bound-free process known as \emph{photo-ionization}, or the \emph{photo-electric effect}, where a photon with $\mathcal{E}_{\gamma}$ just exceeding $\mathcal{E}_e$ is absorbed as the electron is ejected with kinetic energy equal to $\mathcal{E}_{\gamma} - \mathcal{E}_e$ \cite{Kahn02}.\footnote{\emph{e.g.}, from a highly-charged ion in a cosmic plasma or a neutral atom in a material} 
In this case, the initial state is taken to be the ground state with $\ket{A} = \ket{\Psi_G} \otimes \ket{N_{\mathbold{k},\upsilon}}$ but the final state is $\ket{B} = \ket{\Psi_C} \otimes \ket{\left( N_{\mathbold{k},\upsilon} - 1 \right)}$, where $\ket{\Psi_C}$ is a continuum state for a free electron. 
This phenomenon can occur in cosmic plasmas with a strong soft x-ray field as well as in materials where \emph{absorption edges} are formed [\emph{cf.\@} \cref{sec:SXR_med}]. 

Conversely, bound-bound \emph{stimulated emission} can be described as being precisely the opposite of the photo-absorption process where there is an electron-photon state change from $\ket{A} = \ket{\Psi_E} \otimes \ket{\left( N_{\mathbold{k},\upsilon} -1 \right)}$ to $\ket{B} = \ket{\Psi_G} \otimes \ket{N_{\mathbold{k},\upsilon}}$. 
Their matrix elements are therefore complex conjugates of one another:
\begin{equation}
\bra{\Psi_G} \otimes \bra{N_{\mathbold{k},\upsilon}} \underline{U}_I (t) \ket{\Psi_E} \otimes \ket{\left( N_{\mathbold{k},\upsilon} -1 \right)} = \bra{\Psi_E} \otimes \bra{\left( N_{\mathbold{k},\upsilon} -1 \right)} \underline{U}_I (t) \ket{\Psi_G} \otimes \ket{N_{\mathbold{k},\upsilon}}^* 
 \end{equation}
so that from the Fock-state operations of \cref{eq:Fock_operations}, it follows that
\begin{align}\label{eq:abso_stim_balance}
 \begin{split}
 \mathcal{M}_{\text{stim}} &\equiv \bra{\Psi_G} \otimes \bra{N_{\mathbold{k},\upsilon}} \underline{\mathcal{H}}^{(1)} \ket{\Psi_E} \otimes \ket{\left( N_{\mathbold{k},\upsilon} - 1 \right)} = \mathcal{M}_{\text{abso}}^* \\
 &= \hat{\mathbold{e}}_{\mathbold{k},\upsilon} \cdot \bra{\Psi_G} \underline{\mathbold{p}} \mathrm{e}^{ -i \mathbold{k} \cdot \mathbold{r}} \ket{\Psi_E}
 \end{split}
 \end{align}
and as a result, $\norm{\mathcal{M}_{\text{abso}}}^2 = \norm{\mathcal{M}_{\text{stim}}}^2$. 
This implies that for either photo-absorption or stimulated emission, the probability as a function of time for any transition between two electron-photon states $\ket{A}$ and $\ket{B}$ is of the form given by \cref{eq:general_evolution5}, where assuming that the photon changes from $\ket{N_{\mathbold{k},\upsilon}}$ to $\ket{\left( N_{\mathbold{k},\upsilon} - 1 \right)}$ or vice-versa, $\mathcal{M}$ is either $\mathcal{M}_{\text{abso}}$ or $\mathcal{M}_{\text{stim}}$. 
In either case, after all Fock-state operations have been carried out, the norm squared of this matrix element is 
\begin{equation}\label{eq:abso_stim_matrix_element}
 \norm{\mathcal{M}}^2 = \frac{q_e^2 \hbar N_{\mathbold{k},\upsilon}}{2 m_e^2 \mathcal{V} \epsilon_0 \omega_{\mathbold{k}}} \norm{\hat{\mathbold{e}}_{\mathbold{k},\upsilon} \cdot \bra{\Psi_B} \underline{\mathbold{p}} \mathrm{e}^{\pm i \mathbold{k} \cdot \mathbold{r}} \ket{\Psi_A}}^2 ,
 \end{equation}
where in the complex exponential, the $+$ and $-$ signs correspond to emission and absorption, respectively. 
This demonstrates that, in addition to the effect that $\underline{\mathbold{p}} \mathrm{e}^{\pm i \mathbold{k} \cdot \mathbold{r}}$ has as a perturbing operator, the probability of a photon with energy $\mathcal{E}_{\gamma} = \hbar \omega_{\mathbold{k}}$ being absorbed depends on $N_{\mathbold{k},\upsilon}$, the number of photons present in a particular mode defined by $\mathbold{k}$ and $\upsilon$. 
Moreover, \emph{radiative recombination} can occur as a free-bound process, where a free electron (\emph{e.g.}, one in a cosmic plasma) with speed $v_e$ and kinetic energy $\frac{1}{2} m_e v_e^2$ fills a vacancy in an ion with a binding energy $\mathcal{E}_e$ and a photon of energy $\frac{1}{2} m_e v_e^2 - \mathcal{E}_e$ is emitted in the process. 
This process can be thought of as the opposite of photo-ionization, where an electron is absorbed by an ion as a photon is emitted. 
Taking the vacancy to be filled as the ground state, this can be described as a change in state from $\ket{A} = \ket{\Psi_C} \otimes \ket{N_{\mathbold{k},\upsilon}}$ (where $\ket{\Psi_C}$ is a continuum state) to $\ket{B} = \ket{\Psi_G} \otimes \ket{\left( N_{\mathbold{k},\upsilon} + 1 \right)}$. 

Both photo-absorption and stimulated emission can also be understood as arising from the presence of a classical electromagnetic wave, where the classical vector potential $\mathbold{A} ( \mathbold{r} , t)$ [\emph{cf.\@} \cref{eq:potential_wave_solution}] is used in place of the vector potential operator $\underline{\mathbold{A}} ( \mathbold{r} )$ \cite[\emph{cf.\@} \cref{eq:vector_potential_operator_fixedtime}]{Rybicki86,Shankar04}. 
Although this is justified for large photon numbers, where the radiation can be treated as a coherent state [\emph{cf.\@} \cref{sec:coherent_states}], the classical-wave approximation does not necessarily hold in x-ray astrophysics, where photon counts typically are low. 
%As discussed in \cref{sec:coherent_states}, this is justified for large photon numbers where the radiation can be treated as a coherent state. 
%However, in x-ray astrophysics where photon counts are low, the classical-wave approximation does not necessarily hold 
Moreover, a photon description of radiation is needed to explain the phenomenon of \emph{spontaneous emission}, where there is no stimulating radiation field. 
This process can be described as a transition from $\ket{A} = \ket{\Psi_E} \otimes \ket{0}$ to $\ket{B} = \ket{\Psi_G} \otimes \ket{1_{\mathbold{k},\upsilon}}$, where a photon is emitted from the vacuum state \cite{Fontana_82,Townsend00,Shankar04}. 
In this case, \cref{eq:abso_stim_matrix_element} holds with $N_{\mathbold{k},\upsilon} = 1$ so that the matrix element is
\begin{align}\label{eq:emission_matrix_element}
 \begin{split}
 \mathcal{M}_{\text{spon}} &\equiv \bra{\Psi_G} \otimes \bra{1_{\mathbold{k},\upsilon}} \underline{\mathcal{H}}^{(1)} \ket{\Psi_E} \otimes \ket{0} \\
 &= \frac{q_e}{m_e} \sqrt{\frac{\hbar}{2 \mathcal{V} \epsilon_0 \omega_{\mathbold{k}}}} \hat{\mathbold{e}}_{\mathbold{k},\upsilon} \cdot \bra{\Psi_G} \underline{\mathbold{p}} \mathrm{e}^{-i \mathbold{k} \cdot \mathbold{r}} \ket{\Psi_E}  .
 \end{split}
 \end{align}
This is often the prominent mechanism for discrete photon emission in diffuse plasmas where the soft x-ray field is too weak to induce stimulated emission appreciably. 
Two-photon processes, including second-order perturbations of $\underline{\mathcal{H}}^{(1)}$ as well as first-order perturbations of $\underline{\mathcal{H}}^{(2)}$, can also occur in bound-bound transitions.\footnote{\emph{e.g.}, in helium-like ions as described in \cref{sec:selection_rules_helium}} 
However, since the probability for a transition to occur depends on the norm squared of a matrix element involving $\underline{\mathcal{H}}^{(\text{pert})}_I (t)$, single-photon processes are proportional to the fine-structure constant, $\alpha_f \equiv q^2_e / 4 \pi \epsilon_0 \hbar c_0$, while two-photon process are proportional to $\alpha_f^2$ [\emph{cf.\@} \cref{eq:coupled_Hamiltonian3}]. 
Because of this, transitions associated with first-order $\underline{\mathcal{H}}^{(1)}$ are significantly more likely to occur than those associated with first-order $\underline{\mathcal{H}}^{(2)}$ or second-order $\underline{\mathcal{H}}^{(1)}$ \cite{Shankar04,Townsend00}. 
For this reason, single-photon transitions are what contribute the most to soft x-ray spectra from the hot, diffuse plasmas described in \cref{sec:astro_plasmas}, where scattering can also typically be neglected \cite{Paerels03}. 

\subsection{Transition Rates}\label{sec:transition_rate_sub}
%%%%%%%%%%%%%%%%%%%%%%%%%%%%%%%%%%%%%%%%%--------------------------------------------------
The $t \to \infty$ first-order behavior described by \cref{eq:general_evolution5} can be used to derive a typical transition rate for a given process characterized by a time-independent matrix element, $\mathcal{M}$. 
Using $\Delta \mathcal{E}_e$ and $\Delta \gamma \equiv N_{\mathbold{k}',\upsilon'}' \, \hbar \omega_{\mathbold{k}'} - N_{\mathbold{k},\upsilon} \, \hbar \omega_{\mathbold{k}}$ introduced in \cref{sec:gen_first_order}, this transition rate can be determined from the following expression:
\begin{subequations}
\begin{equation}\label{eq:trans_rate_sigle_mode_general}
 \frac{\mathscr{P}_{A \to B}^{\infty} (t)}{t} = \frac{2 \pi}{\hbar} \norm{\mathcal{M}}^2 \delta_D \left( \Delta \mathcal{E}_e - \Delta \gamma \right) , 
 \end{equation}
which is independent of time. 
While this approach can be used to calculate transition rates for any first-order process, the main interest here is to determine the relative transition rates of various bound-bound transitions in highly-charged ions as addressed in \cref{ap:quantum_spectral}. 
In this case, \cref{eq:trans_rate_sigle_mode_general} with $\Delta \gamma = \hbar \omega_{\mathbold{k}}$ becomes 
\begin{equation}\label{eq:trans_rate_sigle_mode}
 \Gamma_{\upsilon} (\hbar \omega_{\mathbold{k}}) \equiv  \frac{\mathscr{P}_{A \to B}^{\infty} (t)}{t} = \frac{2 \pi}{\hbar} \norm{\mathcal{M}}^2 \delta_D \left( \Delta \mathcal{E}_e - \hbar \omega_{\mathbold{k}} \right) , 
 \end{equation}
\end{subequations}
a time-independent function of $\mathcal{E}_{\gamma} = \hbar \omega_{\mathbold{k}}$. 
However, a realistic spectral line is formed from a large number of ions undergoing the same process with probabilistic values for $\mathcal{E}_{\gamma}$ and photon-propagation direction, $\mathbold{k} / k_0$. 

To take into account a range of $\mathcal{E}_{\gamma}$ and $\mathbold{k} / k_0$, recall that the electromagnetic field has been quantized using boundary conditions imposed by a large box of volume $\mathcal{V}$ [\emph{cf.\@} \cref{sec:photon_intro}]. 
This means that, mathematically, the possible photon modes are restricted to a lattice where there are a finite number of these states with wave vector between $\mathbold{k}$ and $\mathbold{k}+ \mathbold{\Delta k}$ to consider. 
Following Townsend~\cite{Townsend00}, the number of modes with wave number in between $k_0$ and $k_0 + \Delta k_0$, and propagation-direction solid angle between $\Omega$ and $\Omega + \Delta \Omega$, can be written as  
\begin{subequations}
\begin{equation}\label{eq:photon_states1}
 \frac{\mathcal{V}}{\left( 2 \pi \right)^3} k_0^2 \Delta k_0 \Delta \Omega \to \frac{\mathcal{V}}{\left( 2 \pi \right)^3} k_0^2 \dd{k_0} \dd{\Omega} , 
 \end{equation}
where infinitesimal intervals are used in the limit that $\mathcal{V} \to \infty$ to describe a physical system. 
In terms of photon energy, this is written as
\begin{equation}\label{eq:photon_states2}
 \frac{\mathcal{V}}{\left( 2 \pi \right)^3} k_0^2 \dd{k_0} \dd{\Omega} = \frac{\mathcal{V}}{\left( h c_0 \right)^3} \mathcal{E}^2_{\gamma} \dd{\mathcal{E}_{\gamma}} \dd{\Omega} \equiv \rho_{\text{state}} \left( \mathcal{E}_{\gamma} \right) \dd{\mathcal{E}_{\gamma}} \dd{\Omega} , 
 \end{equation}
where
\begin{equation}\label{eq:photon_state_density}
 \rho_{\text{state}} \left( \mathcal{E}_{\gamma} \right) \equiv  \frac{\mathcal{V}}{\left( h c_0 \right)^3} \mathcal{E}^2_{\gamma} \quad \text{or} \quad \rho_{\text{state}} \left( \omega_{\mathbold{k}} \right) \equiv \frac{\mathcal{V}}{\left( 2 \pi \right)^3} \frac{\omega_{\mathbold{k}}^2}{\hbar c_0^3}
 \end{equation}
is the \emph{density of photon states} per photon energy, per solid angle \cite{Townsend00}.  
\end{subequations}

With the transition probability for a single photon mode being $\norm{\bra{B} \underline{U}_I (t) \ket{A}}^2$, the overall transition probability taking into account the range of photon modes is calculated from summing $\norm{\bra{B} \underline{U}_I (t) \ket{A}}^2$ over all possible states. 
In the limit that $\mathcal{V} \to \infty$, the summation of these probabilities for a single polarization mode becomes an integral using \cref{eq:photon_states1,eq:photon_states2,eq:photon_state_density}:
\begin{align}\label{eq:absorption_evolution6}
 \begin{split}
 \mathscr{P}_{\upsilon} (t) \equiv \sum_{\mathbold{k}}^{\mathbold{k} + \mathbold{\Delta k}} \norm{\bra{B} \underline{U}_I (t) \ket{A}}^2 &\to \frac{\mathcal{V}}{\left( 2 \pi \right)^3} \int k_0^2 \dd{k_0} \int \dd{\Omega} \norm{\bra{B} \underline{U}_I (t) \ket{B}}^2 \\
 &= \int \rho_{\text{state}} \left( \mathcal{E}_{\gamma} \right) \dd{\mathcal{E}_{\gamma}} \int \dd{\Omega} \norm{\bra{B} \underline{U}_I (t) \ket{A}}^2 
 \end{split}
 \end{align}
and from \cref{eq:trans_rate_sigle_mode}, it is known that 
\begin{equation*}
 \mathscr{P}_{A \to B}^{\infty} (t) \equiv \lim_{t \to \infty} \norm{\bra{B} \underline{U}_I (t) \ket{A}}^2 = \frac{2 \pi t}{\hbar} \norm{\mathcal{M}}^2 \delta_D \left( \Delta \mathcal{E}_e - \mathcal{E}_{\gamma} \right) = t \, \Gamma_{\upsilon} (\mathcal{E}_{\gamma}) .
 \end{equation*}
Therefore, using \cref{eq:absorption_evolution6}, the probability of the transition occurring in the long-time limit (for a single polarization mode) is 
\begin{align}
 \begin{split}\label{eq:long_time_trans_prob}
 \mathscr{P}_{\upsilon}^{\infty} &\equiv \lim_{t \to \infty} \mathscr{P}_{\upsilon} (t) = \frac{2 \pi t}{\hbar} \int \rho_{\text{state}} \left( \mathcal{E}_{\gamma} \right) \dd{\mathcal{E}_{\gamma}} \int \norm{\mathcal{M}}^2 \delta_D \left( \Delta \mathcal{E}_e - \mathcal{E}_{\gamma} \right) \dd{\Omega} \\
 &= \frac{2 \pi t}{\hbar} \int \rho_{\text{state}} \left( \Delta \mathcal{E}_e \right) \norm{\mathcal{M}}^2 \Big|_{\mathcal{E}_{\gamma} = \Delta \mathcal{E}_e} \dd{\Omega} = t \int \rho_{\text{state}} \left( \Delta \mathcal{E}_e \right) \Gamma_{\upsilon} (\Delta \mathcal{E}_e) \dd{\Omega} ,
 \end{split}
 \end{align}
where, along with $\rho_{\text{state}} \left( \mathcal{E}_{\gamma} \right)$, $\mathcal{M}$, and $\Gamma_{\upsilon} (\mathcal{E}_{\gamma})$ are evaluated at $\mathcal{E}_{\gamma} = \Delta \mathcal{E}_e$ \cite{Shankar04,Townsend00,Griffiths05}. 
While this is an approximation, $\mathscr{P}_{\upsilon}^{\infty}$ can be used to obtain a probabilistic rate for a certain transition. 
Per solid angle, for a single polarization state, this can be defined as 
\begin{subequations}
\begin{align}
 \begin{split}\label{eq:diff_transition_rate}
 \Gamma_{\upsilon \Omega} \equiv \dv{\Omega}(\frac{\mathscr{P}_{\upsilon}^{\infty}}{t}) &= \frac{2 \pi}{\hbar} \rho_{\text{state}} \left( \Delta \mathcal{E}_e \right) \norm{\mathcal{M}}^2 \Big|_{\mathcal{E}_{\gamma} = \Delta \mathcal{E}_e} \\
 &= \frac{\mathcal{V}}{4 \pi^2 \hbar^2 c_0^3} \left( \frac{\Delta \mathcal{E}_e}{\hbar} \right)^2  \norm{\mathcal{M}}^2 \Big|_{\mathcal{E}_{\gamma} = \Delta \mathcal{E}_e} \propto \omega_{\text{trans}}^2 ,
 \end{split}
 \end{align}
while the total transition rate is obtained from summing over both polarization states and integrating over the solid angle of interest:
\begin{equation}\label{eq:transition_rate}
 \Gamma \equiv \sum_{\upsilon = 1}^2 \int \Gamma_{\upsilon \Omega} \dd{\Omega} .
 \end{equation}
\end{subequations}
The idea that the probability rate for a given transition is approximately independent of time according to \cref{eq:diff_transition_rate,eq:transition_rate} is known as \emph{Fermi's golden rule} \cite{Townsend00,Kahn02,Shankar04,kapitell4,wysin2011quantization}. 
The fact that $\Gamma_{\upsilon \Omega} \propto \omega_{\text{trans}}^2$ implies that transition rates for processes involving soft x-rays are expected to be large compared to those associated with lower-energy radiation due to their relatively high photon energy. % $\mathcal{E}_{\gamma} = \Delta \mathcal{E}_e$. 
An important consequence of this is that highly-charged ions undergoing soft x-ray transitions require high rates of excitation for local thermodynamic equilibrium to be established in a cosmic plasma \cite{Paerels03}. 

\section{Summary}\label{sec:apAsummary}
%%%%%%%%%%%%%%%%%%%%%%%%%%%%%%%%%%%%%%%%%--------------------------------------------------
This appendix outlines mathematical framework for discussions relating to soft x-ray interaction with matter throughout this dissertation. 
In particular, the classical description of electromagnetic waves laid out in \cref{sec:EM_waves_vac} is used as a starting point for x-ray optics discussion in \cref{ch:diff_eff,app:x-ray_materials}. 
Photons are described quantum-mechanically as vectors in Fock space that are acted upon by annihilation and creation operators contained within Hermitian operators for the electromagnetic field. 
Their interaction with atomic electrons, which are described quantum-mechanically as vectors in Hilbert space, is treated in \cref{sec:atomic_interaction} as a segue into \cref{ap:quantum_spectral}, which addresses spectral lines produced from highly-charged ions found in cosmic plasmas. 
% !TEX root = ../McCoy-Dissertation.tex
\Appendix{On X-ray Spectral Lines}\label{ap:quantum_spectral}
%%%%%%%%%%%%%%%%%%%%%%%%%%%%%%%%%%%%%%%%%--------------------------------------------------
Chemical abundances in the solar photosphere provide an estimate for which atomic nuclei are expected to be the most abundant in the Universe overall as a result of supernovae and other baryon-dispersing processes that have taken place in the Milky Way Galaxy \cite{Rolfs88,Carroll07}. 
The top ten most abundant isotopes inferred from multi-wavelength spectral observations \cite{Grevesse98,Asplund09} are those listed in \cref{tab:astro_element_nuclei_mass}: after hydrogen and helium, the most abundant is element is oxygen followed by carbon, neon, nitrogen, magnesium, silicon, iron and sulfur in decreasing proportions. 
These are the primary atomic nuclei generated by stars with mass $M_{*} \gtrapprox 8 M_{\odot}$. 
\begin{table}[]
 \centering
 \caption[Nuclear masses for stable isotopes of astrophysically abundant elements]{Nuclear masses for stable isotopes of astrophysically abundant elements compared to the mass of the electron, $m_e$ [\emph{cf.\@} \cref{tab:fundamental_constants}]. Data are listed in order of standard solar abundance \cite{Grevesse98,Asplund09}.}\label{tab:astro_element_nuclei_mass}
 \begin{tabular}{@{}lllll@{}} 
 \toprule
 isotope & nucleus & mass \\ \midrule
 hydrogen-1 & 1 proton & $\SI{1.673e-27}{\kilogram} \approx \num{1800} m_e$ \\
 helium-4 & 2 protons + 2 neutrons & $\SI{6.646e-27}{\kilogram} \approx \num{7300} \, m_e$ \\
 \midrule
 oxygen-16 & 8 protons + 8 neutrons & $\SI{2.656e-26}{\kilogram} \approx \num{29200} \, m_e$ \\
 carbon-12 & 6 protons + 6 neutrons &  $\SI{1.993e-26}{\kilogram} \approx \num{21900} \, m_e$ \\
 neon-20 & 10 protons + 10 neutrons & $\SI{3.3209e-26}{\kilogram} \approx \num{36400} \, m_e$ \\
 nitrogen-14 & 7 protons + 7 neutrons & $\SI{2.325e-26}{\kilogram} \approx \num{25500} \, m_e$ \\
 magnesium-24 & 12 protons + 12 neutrons & $\SI{3.983e-26}{\kilogram} \approx \num{43700} \, m_e$ \\
 silicon-28 & 14 protons + 14 neutrons & $\SI{4.646e-26}{\kilogram} \approx \num{51000} \, m_e$ \\
 iron-56 & 26 protons + 30 neutrons & $\SI{9.288e-26}{\kilogram} \approx \num{102000} \, m_e$ & \\
 sulfur-32 & 16 protons + 16 neutrons & $\SI{5.309e-26}{\kilogram} \approx \num{58200} \, m_e$ & \\
 \bottomrule
 \end{tabular}
 \end{table}  
To summarize briefly, following the \emph{main sequence} lifespan of such a massive star, where hydrogen ($\mathcal{Z}=1$) is converted into helium ($\mathcal{Z}=2$) in its core for tens to hundreds of millions of years, the newly-generated helium nuclei (also known as \emph{alpha particles}) through the \emph{triple-alpha} process\footnote{Note that an alpha particle (\emph{i.e.} a helium-4 nucleus) is composed of two protons and two neutrons. In the triple-alpha process, two alpha particles fuse to form the unstable isotope beryllium-8 ($\mathcal{Z}=4$) that then fuses with a third alpha particle to form a stable carbon-12 nucleus.} begin to fuse into nuclei of carbon ($\mathcal{Z}=6$), which in turn fuse further with alpha particles to synthesize nuclei of oxygen ($\mathcal{Z}=8$) and neon ($\mathcal{Z}=10$) \cite{Rolfs88,Carroll07}. 
Due to the sufficiently high core temperature achieved in high-mass stars, carbon nuclei ultimately are able to fuse together to generate more oxygen and neon in addition to sodium and magnesium ($\mathcal{Z}=11,12$). 
As density and temperature increase further, oxygen nuclei also fuse to produce a core consisting mostly of silicon ($\mathcal{Z}=14$) in addition to other byproducts such as magnesium, phosphorous and sulfur ($\mathcal{Z}=12,15,16$). 
Lastly, a series of reactions starting with silicon fusing with helium commences to produce a core dominated by iron ($\mathcal{Z}=26$) prior to the occurrence of a \emph{core-collapse supernova} that ejects these materials into the interstellar medium \cite{Rolfs88,Carroll07,Ryden10}.
During the explosion, other elements with $\mathcal{Z} \lessapprox 26$ are synthesized in relative proportions that depend on the mass of the star, $M_{*}$, as well as other heavier elements in much lower abundances \cite{Woosley86,Rolfs88}. 
Left behind is a neutron star for $M_{*} \lessapprox 25 M_{\odot}$ or a black hole for still more massive stars. 
Meanwhile, the majority of elements heavier than iron are thought to be generated in rare quantities only during extremely energetic events such as the collision of orbiting neutron stars \cite{Kasen17}. 

With the exception of iron due to its relatively high $\mathcal{Z}$, the inner-most electrons (\emph{i.e., K-shell} electrons; see \cref{tab:shell_orbitals}) in the abundant elements listed in \cref{tab:astro_element_nuclei_mass}, save hydrogen and helium, have binding energies characteristic of the soft x-ray spectrum, where photon energy ranges from $\SI{250}{\electronvolt} \lessapprox \mathcal{E}_{\gamma} \lessapprox \SI{2}{\kilo\electronvolt}$. 
Additionally, $\mathcal{E}_e$ for electrons in the next higher-laying shell of iron (\emph{i.e.,} the \emph{L-shell}) fall within this range [\emph{cf.\@} \cref{tab:astro_element_binding}]. 
\begin{table}[]
 \centering
 \caption[Electronic binding energies for the inner-most electrons in neutral atoms of astrophysically abundant elements]{Electronic binding energies, $\mathcal{E}_e$, and associated electromagnetic wavelengths for the inner-most electrons in neutral atoms of the astrophysically abundant elements listed in \cref{tab:astro_element_nuclei_mass}. While the lightest elements are characteristic of vacuum and extreme ultraviolet (UV) radiation, those with $\mathcal{Z} \geq 6$ have K-shell electronic binding energies falling in the soft x-ray and beyond [\emph{cf.\@} \cref{tab:first_element_binding}]. Additionally, L-shell binding energies approach soft x-ray energies for elements heavier than sulfur (not listed are L-shell binding energies for $\mathcal{Z} < 14$) \cite{Attwood17}.}\label{tab:astro_element_binding}
 \begin{tabular}{@{}lllll@{}} 
 \toprule
 chemical element & shell & binding energy & wavelength & spectral band \\ \midrule
 hydrogen ($\mathcal{Z}=1$) & K & \SI{13.6}{\electronvolt} & \SI{91.16}{\nano\metre} & vacuum UV \\
 helium ($\mathcal{Z}=2$) & K & \SI{24.6}{\electronvolt} & \SI{50.4}{\nano\metre} & vacuum UV \\
 \midrule
 oxygen ($\mathcal{Z}=8$) & K & \SI{543.1}{\electronvolt} & \SI{2.28}{\nano\metre} & soft x-ray \\
 carbon ($\mathcal{Z}=6$) & K & \SI{284.2}{\electronvolt} & \SI{4.36}{\nano\metre} & soft x-ray \\
 neon ($\mathcal{Z}=10$) & K & \SI{870.2}{\electronvolt} & \SI{1.42}{\nano\metre} & soft x-ray \\
 nitrogen ($\mathcal{Z}=7$) & K & \SI{409.9}{\electronvolt} & \SI{3.03}{\nano\metre} & soft x-ray \\
 magnesium ($\mathcal{Z}=12$) & K & \SI{1303.0}{\electronvolt} & \SI{0.952}{\nano\metre} & soft x-ray \\ \midrule
 silicon ($\mathcal{Z}=14$) & K & \SI{1838.9}{\electronvolt} & \SI{0.674}{\nano\metre} & soft x-ray \\
 \hspace{5mm}'' \hspace{12mm}'' & L & \SI{149.7}{\electronvolt} & \SI{8.28}{\nano\metre} & extreme UV \\ 
 \hspace{5mm}'' \hspace{12mm}'' & L & \SI{99.8}{\electronvolt} & \SI{12.4}{\nano\metre} & extreme UV \\
 \hspace{5mm}'' \hspace{12mm}'' & L & \SI{99.2}{\electronvolt} & \SI{12.5}{\nano\metre} & extreme UV \\ \midrule
 iron ($\mathcal{Z}=26$) & K & \SI{7112.0}{\electronvolt} & \SI{0.174}{\nano\metre} & x-ray \\ 
 \hspace{3mm}'' \hspace{10mm}'' & L & \SI{844.6}{\electronvolt} & \SI{1.47}{\nano\metre} & soft x-ray \\
 \hspace{3mm}'' \hspace{10mm}'' & L & \SI{719.9}{\electronvolt} & \SI{1.72}{\nano\metre} & soft x-ray \\
 \hspace{3mm}'' \hspace{10mm}'' & L & \SI{706.8}{\electronvolt} & \SI{1.75}{\nano\metre} & soft x-ray \\ \midrule
 sulfur ($\mathcal{Z}=16$) & K & \SI{2472}{\electronvolt} & \SI{0.502}{\nano\metre} & soft x-ray \\
 \hspace{5mm}'' \hspace{11mm}'' & L & \SI{230.9}{\electronvolt} & \SI{5.37}{\nano\metre} & extreme UV \\
 \hspace{5mm}'' \hspace{11mm}'' & L & \SI{163.6}{\electronvolt} & \SI{7.58}{\nano\metre} & extreme UV \\
 \hspace{5mm}'' \hspace{11mm}'' & L & \SI{162.5}{\electronvolt} & \SI{7.63}{\nano\metre} & extreme UV \\
 \bottomrule
 \end{tabular}
 \end{table}
Electrons in these inner-most shells listed in \cref{tab:astro_element_binding} are also the outer-most electrons in highly-charged ions by definition. 
However, the listed data pertain to neutral atoms \cite{Attwood17} whereas K-shell and L-shell electrons in highly-charged ions have binding energies that are somewhat larger due to a lack of outer-shell electrons; this has the consequence that for elements such as silicon and sulfur, L-shell binding energies move into the soft x-ray spectrum as the degree of ionization increases \cite{Kahn02,Paerels03}. 
Therefore, overall, atoms with $6 \leq \mathcal{Z} \lessapprox 14$ ionized down to the K-shell and atoms with $16 \lessapprox \mathcal{Z} \lessapprox 26$ ionized down to the L-shell have their outer-most electrons bound with soft x-ray energies. 
With cosmic abundance weighted toward lower $\mathcal{Z}$, \emph{hydrogen-like} and \emph{helium-like} ions of $6 \leq \mathcal{Z} \lessapprox 14$, where just one or two bound K-shell electrons remain, respectively, are of particular interest for soft x-ray spectroscopy. 
Before prominent spectral lines from these ions are discussed in \cref{sec:selection_rules_hydrogen,sec:selection_rules_helium}, basics of spectral line formation and \emph{electromagnetic-multipole transitions} are outlined in \cref{sec:form_spectral,sec:multipole}, respectively. 
A summary is then given in \cref{sec:apBsummary}. 

\section{Formation of Spectral Lines}\label{sec:form_spectral}
%%%%%%%%%%%%%%%%%%%%%%%%%%%%%%%%%%%%%%--------------------------------------------------
Until \cref{sec:selection_rules_hydrogen,sec:selection_rules_helium}, where details of specific ions are considered, the quantum-mechanical Hamiltonian $\underline{\mathcal{H}}_{\text{atom}}$ defined by \cref{eq:generic_ion_Hamiltonian} is taken to describe a hypothetical two-level atom, where a single electron is subject to an arbitrary force as it makes a jump between two quantum energy levels, and in doing so, interacts with the quantized electromagnetic field to absorb or emit radiation [\emph{cf.\@} \cref{sec:single_absorption_emission}]. 
According to Fermi's golden rule [\emph{cf.\@} \cref{sec:transition_rate_sub}], the probability for such a bound-bound transition to occur increases indefinitely as $\mathscr{P} (t) = \Gamma t$, where $\Gamma$ is given by \cref{eq:transition_rate}. 
While $\Gamma^{-1}$ provides a typical timescale for a certain transition to occur, this does not hold for long timescales because the fact that the probability for an electron to be in its initial state decreases as $t$ progresses must be taken into account \cite{Fontana_82,Griffiths05,kapitell4}. 
If it is assumed that the electron is undergoing a bound-bound transition from some electron-photon state, $\ket{A}$ to another state, $\ket{B}$, the probability of the electron being in the initial state $\ket{A}$ is $1- \mathscr{P} (t)$, where $\mathscr{P} (t) \neq \Gamma t$ is the probability for the transition to occur. 
Then, the rate of change for $\mathscr{P} (t)$ is proportional to the probability of the electron being in state $\ket{A}$:
\begin{subequations}
\begin{equation}
 \dv{\mathscr{P} (t)}{t} = \Gamma \left[ 1- \mathscr{P} (t) \right] .
 \end{equation}
This differential equation can be solved to give 
\begin{equation}\label{eq:probability_decay}
 \mathscr{P} (t) = 1 - \mathrm{e}^{- \Gamma t} ,
 \end{equation}
\end{subequations}
which states that the probability for the transition to have occurred by a time $t$ exhibits asymptotic behavior with $\Gamma$ as a damping parameter. 
Alternatively, this can be argued by considering a distribution of $\mathcal{N}_0$ ions that are all supposed to be in state $\ket{A}$ at some time $t=0$ \cite{Townsend00,kapitell4}. 
The number of ions that remain in state $\ket{A}$ is a function of time, $\mathcal{N}_A (t)$, defined by the following differential equation:
\begin{subequations}
\begin{equation}
 \dv{\mathcal{N}_A (t)}{t} = - \Gamma \mathcal{N}_A (t) ,
 \end{equation}
which can be solved to give $\mathcal{N}_A (t) = \mathcal{N}_0 \mathrm{e}^{- \Gamma t}$. 
Therefore, the probability that $\mathcal{N}_B (t)$ ions have transitioned to state $\ket{B}$ after a time $t$ can be written as  
\begin{equation}\label{eq:number_decay}
 \mathcal{N}_B (t) = \mathcal{N}_0 - \mathcal{N}_A (t) =  \mathcal{N}_0 \left( 1 - \mathrm{e}^{- \Gamma t} \right) ,
 \end{equation}
\end{subequations}
which is equivalent to the probability given by \cref{eq:probability_decay} \cite{Townsend00}. 

In contrast to the \emph{sinc-squared} behavior of first-order transitions [\emph{cf.\@} \cref{fig:sinc_squared}], the decaying transition probability considered here is characteristic of a damped oscillator. 
That is, it is analogous to the spectrum of a classical electromagnetic wave that starts at time $t=0$ and decays over a time interval $\mathcal{T}$, which can be taken to approach infinity \cite{Fontana82}.  
In direct analogy to \cref{eq:wave}, let the magnitude of the electric field for this wave be represented using
\begin{equation}\label{eq:wave_pulse_decay}
    u(t) =
    \begin{cases}
      u_0 \cos \left( \Phi - \omega_0 t \right) \mathrm{e}^{- \frac{1}{2} \Gamma t} , & 0 \leq t \leq \mathcal{T} \\
      0, & \text{otherwise,} 
    \end{cases}
  \end{equation}
where $\omega_0$ effectively plays the role of the transition frequency of a quantum system, $\omega_{\text{trans}} \equiv \Delta \mathcal{E}_e / \hbar$ [\emph{cf.\@} \cref{sec:gen_first_order}]. 
Here, a factor of $1/2$ appears in the exponential so that the intensity of the wave (which is proportional to $\norm{u (t)}^2$) depends on $\mathrm{e}^{- \Gamma t}$; in quantum mechanics, it is the wave function that oscillates at $\omega_{\text{trans}}$ and decays according to $\mathrm{e}^{- \frac{1}{2} \Gamma t}$ so that the transition probability has a $\mathrm{e}^{- \Gamma t}$ dependence \cite{Rybicki86}. 

The temporal Fourier transform of \cref{eq:wave_pulse_decay} is \cite{Fontana82}
\begin{align}
 \begin{split}
 u (\omega) &= \int_{-\infty}^{\infty} u(t) \, \mathrm{e}^{i \omega t} \dd{t} = u_0 \int_0^{\mathcal{T}} \cos \left( \Phi - \omega_0 t \right) \mathrm{e}^{- \frac{1}{2} \Gamma t} \mathrm{e}^{i \omega t} \dd{t} \\
 &= \frac{u_0}{2} \left( \mathrm{e}^{-i \Phi} \frac{1 - \mathrm{e}^{i \left( \omega + \omega_0 \right) \mathcal{T}} \mathrm{e}^{- \frac{1}{2} \Gamma \mathcal{T}}}{\frac{1}{2} \Gamma - i \left( \omega + \omega_0 \right)} + \mathrm{e}^{i \Phi} \frac{1 - \mathrm{e}^{i \left( \omega - \omega_0 \right) \mathcal{T}} \mathrm{e}^{- \frac{1}{2} \Gamma \mathcal{T}}}{\frac{1}{2} \Gamma - i \left( \omega - \omega_0 \right)} \right) 
 \end{split}
 \end{align}
while $\norm{u (t)}^2$ is proportional to the wave intensity. 
Assuming $\mathcal{T} \gg \omega_0^{-1}$ so that the $\omega + \omega_0$ term can be neglected, $\norm{u (t)}^2$ can be written as 
\begin{subequations}
\begin{equation}\label{eq:pulse_intensity}
 \norm{u (\omega)}^2 = \frac{u_0^2 \mathcal{T}^2}{4} \left( \frac{1 + \mathrm{e}^{- \Gamma \mathcal{T}} - 2 \cos \left( \Delta \omega \mathcal{T} \right) \mathrm{e}^{- \frac{1}{2} \Gamma \mathcal{T}}}{\left( \Delta \omega \mathcal{T} \right)^2 + \frac{1}{4} \left( \Gamma \mathcal{T} \right)^2 } \right) 
 \end{equation}
and in the limit that $\mathcal{T} \to \infty$, this becomes
\begin{equation}
 \norm{u (\omega)}^2 = \frac{u_0^2}{4} \left( \frac{1}{\Delta \omega^2 + \frac{1}{4} \Gamma^2} \right) ,
 \end{equation}
which can be normalized\footnote{A normalization coefficient, $\mathcal{A}$, can be determined from
\begin{equation*}
 \mathcal{A} \int_{- \infty}^{\infty} \frac{\dd{\omega}}{\left( \omega - \omega_0 \right)^2 + \frac{1}{4} \Gamma^2} = \mathcal{A} \frac{2 \pi}{\Gamma} = 1, 
 \end{equation*}
where, with $a \equiv \Gamma / 2$:
\begin{equation*}
 \int \frac{\dd{\omega}}{\left( \omega - \omega_0 \right)^2 + a^2} = \frac{\arctan \left( \frac{\omega - \omega_0}{a} \right)}{a} + \text{ a constant} .
 \end{equation*}} 
to yield a \emph{Lorentzian} distribution \cite{Fontana82}:
\begin{equation}\label{eq:normalized_lorentzian}
 \norm{u (\omega)}^2 \to \frac{1}{2 \pi} \left( \frac{\Gamma}{\Delta \omega^2 + \frac{1}{4} \Gamma^2} \right) .
 \end{equation}
\end{subequations}
As mentioned in \cref{sec:gen_first_order}, the quantum analog of $\Delta \omega$ for a bound-bound transition is the difference between the nominal energy associated with the transition and the absorbed or emitted photon energy, $\mathcal{E}_{\gamma} \equiv \hbar \omega$:
\begin{equation*}
 \varpi \equiv \frac{\left( \Delta \mathcal{E}_e - \hbar \omega \right)}{\hbar} = \omega_{\text{trans}} - \omega .
 \end{equation*}
Therefore, the radiation absorbed or emitted by a collection of highly-charged ions undergoing a certain transition with $\omega_{\text{trans}}$ and $\Gamma$ is expected, under ideal conditions, to appear as a Lorentzian distribution with a line profile of the form %proportional to:
\begin{equation}\label{eq:normalized_lorentzian_QM}
 \phi_{\text{nat}} \equiv \frac{1}{2 \pi} \left( \frac{\Gamma}{\left( \omega_{\text{trans}} - \omega \right)^2 + \frac{1}{4} \Gamma^2} \right) .
 \end{equation}
In other words, a \emph{naturally-broadened} spectral line is expected to appear as a Lorentzian distribution if it can be assumed that the ions are virtually stationary and non-interacting; other line profiles are discussed in \cref{sec:astro_plasmas}.

The \emph{full width at half maximum (FWHM)} of such a spectral line is equal to $\Gamma$, which is shown graphically in \cref{fig:lorentzian}. 
\begin{figure}
 \centering
 \includegraphics[scale=1]{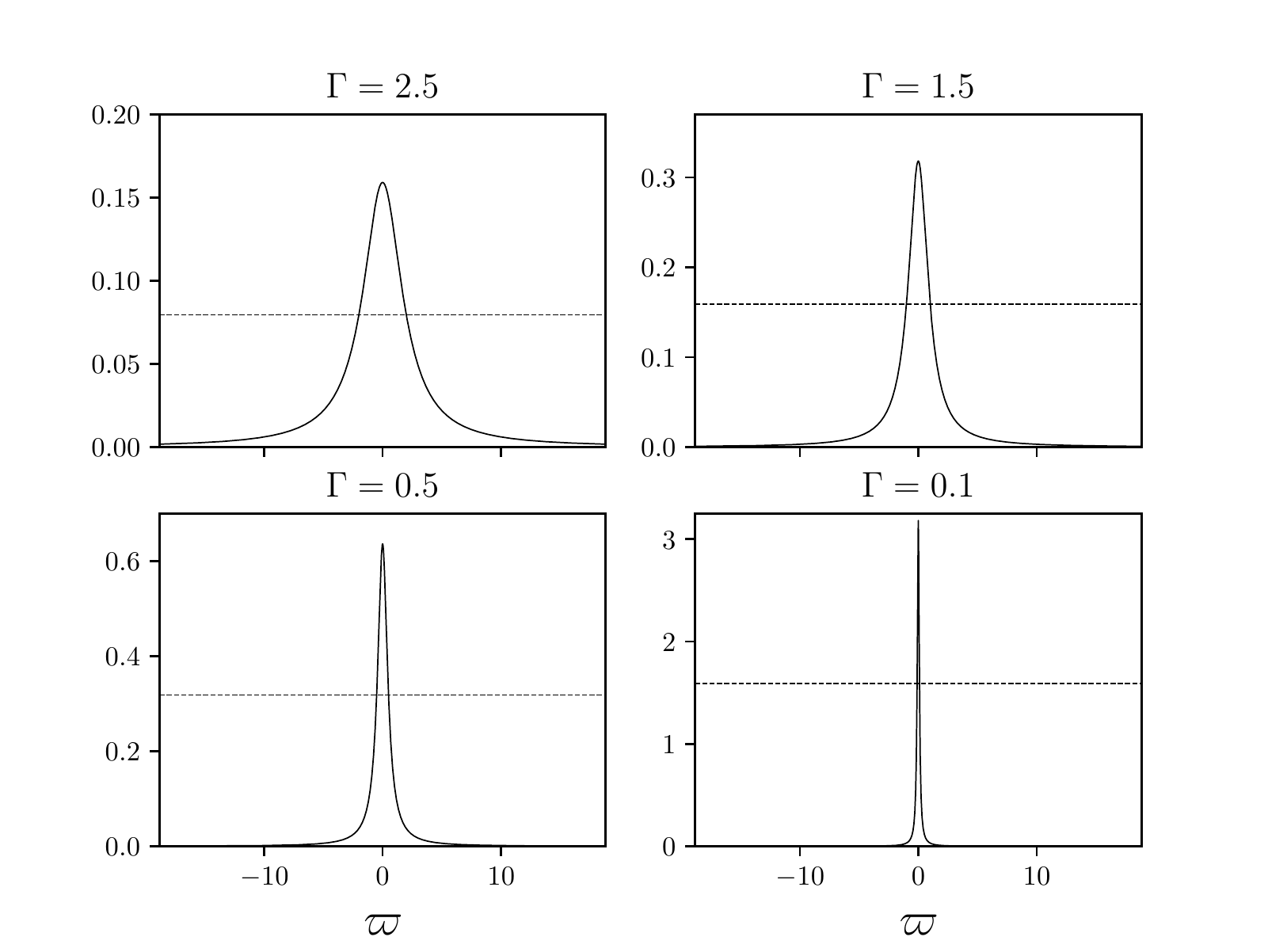}
 \caption[Plots of Lorentzian functions for naturally-broadened spectral lines]{Plots of Lorentzian functions defined by \cref{eq:normalized_lorentzian_QM} with $\varpi = \omega_{\text{trans}} - \omega$ for four values of the damping term $\Gamma$ using arbitrary units. Dashed lines indicate half maximum with $\text{FWHM}=\Gamma$.}\label{fig:lorentzian}
 \end{figure}
However, this assumes that only the initial state, $\ket{A}$, is broadened, or equivalently, that the electron spends an infinite amount of time in the final state, $\ket{B}$.  
In such a case where this assumption is not justified, $1 / \Gamma_A$ and $1 / \Gamma_B$, the timescales for both states $\ket{A}$ and $\ket{B}$, must be taken into account by making the replacement $\Gamma \to \Gamma_A + \Gamma_B$ \cite{Carroll07,Rybicki86}. 

\section{Electromagnetic-Multipole Transitions}\label{sec:multipole}
%%%%%%%%%%%%%%%%%%%%%%%%%%%%%%%%%%%%%%%%%--------------------------------------------------
In a given species of highly-charged ion, there exist many possible bound states for an inner-shell electron to transition to or from in the process of forming a spectral line [\emph{cf.\@} \cref{sec:form_spectral}]. 
While this depends on how electronic states are populated in a plasma, each of these potential transitions has an associated probabilistic rate, $\Gamma$, that depends on $\norm{\mathcal{M}}^2$ as described by Fermi's golden rule [\emph{cf.\@} \cref{eq:diff_transition_rate,eq:transition_rate}]. 
This transition rate has been formulated in \cref{sec:transition_rate_sub} by first considering the probability for a single transition from one electron-photon state, $\ket{A}$, to another, $\ket{B}$, as a function of time (given by $\norm{\bra{B} \underline{U}_I (t) \ket{A}}^2$) and then integrating this quantity over a range of wave modes. 
To recap, $\underline{U}_I (t)$ is the time-evolution operator from \cref{sec:quantum_perturbations}, given by \cref{eq:first_order_evolution}:
\begin{equation}\label{eq:first_order_evolution_recap}
 \underline{U}_I (t) \approx 1 - \frac{i}{\hbar} \int_0^t \mathrm{e}^{\frac{i}{\hbar} t' \underline{\mathcal{H}}^{(0)}} \underline{\mathcal{H}}^{(1)} \mathrm{e}^{-\frac{i}{\hbar} t' \underline{\mathcal{H}}^{(0)}} \dd{t'} ,
 \end{equation}
where $\underline{\mathcal{H}}^{(0)} = \underline{\mathcal{H}}_{\text{ion}} + \underline{\mathcal{H}}_{\text{EM}}$ is the known Hamiltonian and $\underline{\mathcal{H}}^{(1)}$ is the perturbing Hamiltonian that describes single-photon, bound-bound transitions [\emph{cf.\@} \cref{eq:single_photon_Hamiltonian}]. 
Additionally, \cref{sec:gen_first_order} shows that $\norm{\bra{B} \underline{U}_I (t) \ket{A}}^2$ has a sinc-squared time dependence under first-order perturbation theory, and for a photon state change from $\ket{N_{\mathbold{k},\upsilon}}$ to $\ket{N_{\mathbold{k},\upsilon} - 1}$ or vice-versa, it ends up being proportional to the following, equal quantities [\emph{cf.\@} \cref{eq:abso_stim_matrix_element}] at any given time, $t$:
\begin{subequations}
\begin{align}\label{eq:general_norm_squared_transition}
 \begin{split}
 \text{absorption:} \quad \norm{\mathcal{M}}^2 & = \norm{\bra{\Psi_B} \otimes \bra{\left( N_{\mathbold{k},\upsilon} -1 \right)} \underline{\mathcal{H}}^{(1)} \ket{\Psi_A} \otimes \ket{N_{\mathbold{k},\upsilon}}}^2 \\ 
 \text{emission:} \quad \norm{\mathcal{M}}^2 & = \norm{\bra{\Psi_B} \otimes \bra{N_{\mathbold{k},\upsilon} } \underline{\mathcal{H}}^{(1)} \ket{\Psi_A} \otimes \ket{\left( N_{\mathbold{k},\upsilon} -1 \right) }}^2 .
 \end{split}
 \end{align}
In either case, if $\underline{\mathcal{H}}^{(1)} = (q_e / m_e) \underline{\mathbold{A}} ( \mathbold{r} ) \cdot \underline{\mathbold{p}}$ is inserted and all Fock-space operations are carried out, this quantity reduces to
\begin{equation}\label{eq:general_norm_squared_transition_reduced}
  \norm{\mathcal{M}}^2 = \frac{q_e^2 \hbar N_{\mathbold{k},\upsilon}}{2 m_e^2 \mathcal{V} \epsilon_0 \omega_{\mathbold{k}}} \norm{\hat{\mathbold{e}}_{\mathbold{k},\upsilon} \cdot \bra{\Psi_B} \underline{\mathbold{p}} \mathrm{e}^{\pm i \mathbold{k} \cdot \mathbold{r}} \ket{\Psi_A}}^2 ,
 \end{equation}
\end{subequations}
where $\ket{\Psi_A}$ and $\ket{\Psi_B}$ are the initial and final electron states, respectively, that depend only on the quantum numbers $n$, $\ell$ and $m_{\ell}$ [\emph{cf.\@} \cref{tab:shell_orbitals}] while $\hat{\mathbold{e}}_{\mathbold{k},\upsilon}$ indicates the direction of the photon electric field [\emph{cf.\@} \cref{sec:photon_intro}]. 
Moreover, $-$ in the complex exponential corresponds to emission while $+$ corresponds to absorption. 
Evaluating $\norm{\mathcal{M}}^2$ then gives an indication of the most probable transitions in a particular ion and hence which spectral lines are expected to appear the strongest in a measured spectrum \cite{Kahn02}. 

To motivate the idea that each possible transition in an ion contributes to the spectrum with a certain probability weight, the following classical scenario is considered as a rough analogy. 
The electron states $\ket{\Psi_A}$ and $\ket{\Psi_B}$ describe the spatial probability distributions for the initial and final states of the transitioning electron; these are the so-called \emph{electron clouds} defined by functions of position $\norm{\Psi_A (\mathbold{r})}^2 \equiv \norm{\braket{\mathbold{r}}{\Psi_A}}^2$ and $\norm{\Psi_B (\mathbold{r})}^2 \equiv \norm{\braket{\mathbold{r}}{\Psi_B}}^2$. 
In this analogy, each of these electron clouds described by $\norm{\Psi_{n \ell m} (\mathbold{r})}^2$, which depend on quantum numbers $n$, $\ell$ and $m_{\ell}$, can be thought to represent a continuous distribution of electric charge with a volume density, $\rho(\mathbold{r})$. 
Additionally, the orbital motion of the electron can be thought of as a loop of atomic current so that there is an associated electric current density, $\mathbfcal{J} (\mathbold{r})$, whose vector direction is directly related to $m_{\ell}$. 
From a quantum-mechanical viewpoint, this corresponds to the \emph{probability current} \cite{Griffiths05}:
\begin{equation}
 \mathbfcal{J} (\mathbold{r}) \to \frac{i \hbar}{2 m_e} \left[ \Psi_{n \ell m} (\mathbold{r}) \grad \Psi_{n \ell m}^* (\mathbold{r}) - \Psi_{n \ell m}^* (\mathbold{r}) \grad \Psi_{n \ell m} (\mathbold{r}) \right] \quad \text{with } m \equiv m_{\ell} .
 \end{equation}

As an electron makes a bound-bound-transition from an excited state to the ground state and emits radiation in this classical analogy, these charge and current distributions, $\rho(\mathbold{r},t)$ and $\mathbfcal{J} (\mathbold{r},t)$, oscillate in time between the initial configuration with $\rho_A (\mathbold{r})$ and $\mathbfcal{J}_A (\mathbold{r})$, and the final configuration with $\rho_B (\mathbold{r})$ and $\mathbfcal{J}_B (\mathbold{r})$.\footnote{Note that these two quantities are related to each other through the continuity equation defined by \cref{eq:continuity} \cite{Landau60,Jackson75,Griffiths17,Griffiths05}:
\begin{equation*}
 \div \mathbfcal{J} (\mathbold{r},t) + \pdv{\rho(\mathbold{r},t)}{t} = 0 .
 \end{equation*}}
These oscillating functions, $\rho(\mathbold{r},t)$ and $\mathbfcal{J} (\mathbold{r},t)$, then lead to emission and absorption of electromagnetic waves according to Maxwell's equations [\emph{cf.\@} \cref{sec:EM_waves_vac}]. 
Depending on $\rho (\mathbold{r})$ and $\mathbfcal{J} (\mathbold{r})$ of the initial and final states as well as the frequency of oscillation $(\omega_{\text{trans}} = \Delta \mathcal{E}_e / \hbar$), the radiation pattern associated with such an oscillation may be described as being characteristic of an \emph{electric dipole}, a \emph{magnetic dipole} or more generally, any type of \emph{electromagnetic multipole}, each of which is expected to radiate with a different intensity pattern \cite{Kahn02}. 
The analogy to quantum mechanics is that the probability for a certain transition to occur is related to which type of multipole radiation it is characteristic of. 
There are, however, crucial quantum effects that come into play that weaken this analogy. 
Besides the fact that radiation is discretized into photons, particularly important examples of this include relativistic and spin-related aspects of the electron that are not explained directly by the non-relativistic theory of quantum mechanics.  
While these effects must be taken into account to explain some spectral lines, much of this classical analogy holds for bound-bound radiative processes in highly-charged ions and moreover, different types of transitions are often classified in terms of electromagnetic-multipole radiation [\emph{cf.\@} \cref{sec:allowed_transitions,sec:forbidden_transitions}]. 
Through analyzing these processes, so-called \emph{selection rules} arise that dictate which of these transitions are mostly likely to occur. 

\subsection{Electric-Dipole Transitions}\label{sec:allowed_transitions}
%%%%%%%%%%%%%%%%%%%%%%%%%%%%%%%%%%%%%%%%%--------------------------------------------------
From a classical perspective, the force that an electromagnetic wave exerts on a freely-moving particle at a position $\mathbold{r}$ with charge $-q_e$, mass $m_e$ and velocity $\mathbold{v} \equiv \dv*{\mathbold{r}}{t}$ is the \emph{Lorentz force} \cite{Griffiths17,Jackson75,Rybicki86,Kahn02}:\footnote{This can be shown by carrying out the \emph{Euler-Lagrange equation}:
\begin{equation*} 
\dv{t} \left( \pdv{\mathcal{L} \left( \dot{\mathbold{r}}, \mathbold{r} , t \right)}{\dot{\mathbold{r}}} \right) = \pdv{\mathcal{L} \left( \dot{\mathbold{r}}, \mathbold{r} , t \right)}{\mathbold{r}} ,
 \end{equation*}
where $\mathcal{L} \left( \dot{\mathbold{r}}, \mathbold{r} , t \right)$ is the Lagrangian for a charged particle given by \cref{eq:charged_particle_lagrangian} with $V (r) \to 0$.} 
\begin{align}
\begin{split}\label{eq:Lorentz_force}
 \mathbold{F}(\mathbold{r},\mathbold{v},t) &= - q_e \left( \mathbold{E}(\mathbold{r},t) + \mathbold{v} \times \mathbold{B}(\mathbold{r},t) \right) \\
 &= - q_e \left( \mathbold{E}(\mathbold{r},t) + \frac{\mathbold{v}}{c_0} \times \frac{\mathbold{k}}{k_0} \times \mathbold{E}(\mathbold{r},t) \right) .
 \end{split}
 \end{align}
With the condition $c_0 k_0 \mathbold{B}(\mathbold{r},t) = \mathbold{k} \times \mathbold{E}(\mathbold{r},t)$ for a transverse wave [\emph{cf.\@} \cref{sec:time_harmonic}], the magnetic-field component of the force depends on $\mathbold{v} /c_0$ and hence its magnitude is expected to be small for non-relativistic motion. 
In spirit of the the classical-quantum analogy outlined above, it might then be expected that radiative bound-bound transitions in ions tend to be dominated by the electric-field component, at least in situations where relativistic effects can be neglected. 
Moreover, for bound electrons with small velocities and localized motion, the electric field appears spatially uniform; radiative transitions meeting these criteria should then be characteristic of electric-dipole radiation \cite{deBergevin09}. 
To investigate this, an electric-dipole Hamiltonian term is introduced as the analog to the energy of a classical, point electric dipole: 
\begin{subequations}
\begin{equation}\label{eq:E1_classical}
 U_{\text{E}1} = - \mathbold{\mu}_{\text{E}1} \cdot \mathbold{E}_0 = q_e \mathbold{r} \cdot \mathbold{E}_0 ,
 \end{equation}
where $\mathbold{E}_0$ is a constant electric field. 
Similar to $\underline{\mathcal{H}}^{(1)} = (q_e / m_e) \underline{\mathbold{A}} ( \mathbold{r} ) \cdot \underline{\mathbold{p}}$, this operator is written as 
\begin{equation}\label{eq:E1_hamiltonian}
 \underline{\mathcal{H}}^{(\text{E}1)} \equiv q_e \underline{\mathbold{E}}_0 \cdot \underline{\mathbold{r}} ,
 \end{equation}
where $\underline{\mathbold{E}}_0$ is the electric-field operator defined by \cref{eq:E-field_operator_fixed_linear} with $\mathrm{e}^{\pm i \mathbold{k} \cdot \mathbold{r}} \to 1$ to invoke a constant electric field:
\begin{equation}\label{eq:E-field_operator_fixed_linear_constant}
 \underline{\mathbold{E}}_0 \equiv \sum_{\mathbold{k}} \sum_{\upsilon = 1}^2  i \sqrt{\frac{\hbar \omega_{\mathbold{k}}}{2 \mathcal{V} \epsilon_0}} \hat{\mathbold{e}}_{\mathbold{k},\upsilon} \left[ \underline{a}_{\mathbold{k},\upsilon} - \underline{a}^{\dagger}_{\mathbold{k},\upsilon} \right] ,
 \end{equation}
\end{subequations}
where $\underline{a}_{\mathbold{k},\upsilon}$ and $\underline{a}^{\dagger}_{\mathbold{k},\upsilon}$ are annihilation and creation operators for a Fock state defined by a wave vector $\mathbold{k}$ and a polarization index $\upsilon$ [\emph{cf.\@} \cref{sec:photon_intro}]. 

Noting that bound-bound absorption and emission are opposite processes [\emph{cf.\@} \cref{sec:single_absorption_emission}], the effect that $\underline{\mathcal{H}}^{(\text{E}1)}$ has as a perturbing operator can be examined by considering just one of these cases under the \emph{self-consistent field approximation} [\emph{cf.\@} \cref{sec:atomic_interaction}]. 
To demonstrate this for spontaneous emission, \cref{eq:general_norm_squared_transition} is used to write $\norm{\mathcal{M}}^2$ with $N_{\mathbold{k},\upsilon} = 1$: 
\begin{equation}
 \norm{\mathcal{M}}^2 = \norm{\bra{\Psi_B} \otimes \bra{1_{\mathbold{k},\upsilon} } \underline{\mathcal{H}}^{(1)} \ket{\Psi_A} \otimes \ket{0}}^2 .
 \end{equation}
Inserting $\underline{\mathcal{H}}^{(\text{E}1)}$ [\emph{cf.\@} \cref{eq:E1_hamiltonian}] in place of $\underline{\mathcal{H}}^{(1)}$ gives 
\begin{subequations}
\begin{align}
 \begin{split}
 \norm{\mathcal{M}_{\text{E}1}}^2 &\equiv \norm{\bra{\Psi_B} \otimes \bra{1_{\mathbold{k},\upsilon} } \underline{\mathcal{H}}^{(\text{E}1)} \ket{\Psi_A} \otimes \ket{0}}^2 \\
 &= q_e^2 \norm{\bra{\Psi_B} \otimes \bra{1_{\mathbold{k},\upsilon} } \left( \underline{\mathbold{E}}_0 \cdot \underline{\mathbold{r}} \right) \ket{\Psi_A} \otimes \ket{0}}^2 \\
 &= q_e^2 \frac{\hbar \omega_{\mathbold{k}}}{2 \mathcal{V} \epsilon_0} \norm{\bra{\Psi_B} \left( \hat{\mathbold{e}}_{\mathbold{k},\upsilon} \cdot \underline{\mathbold{r}} \right) \ket{\Psi_A}}^2 , 
 \end{split} 
 \end{align}
while the transition rate, $\Gamma$, is calculated using $\hbar \omega_{\mathbold{k}} = \Delta \mathcal{E}_e$ [\emph{cf.\@} \cref{sec:transition_rate_sub}]: 
\begin{equation}\label{eq:E1_spon_term}
 \norm{\mathcal{M}_{\text{E}1}}^2 \to \frac{q_e^2 \Delta \mathcal{E}_e }{2 \epsilon_0 \mathcal{V}} \norm{\hat{\mathbold{e}}_{\mathbold{k},\upsilon} \cdot \bra{\Psi_B} \underline{\mathbold{r}} \ket{\Psi_A} }^2 .
 \end{equation} 
\end{subequations}
From \cref{eq:emission_matrix_element}, it is known that $\norm{\mathcal{M}}^2$ for a general case of spontaneous emission, using $\underline{\mathcal{H}}^{(1)} = (q_e / m_e) \underline{\mathbold{A}} ( \mathbold{r} ) \cdot \underline{\mathbold{p}}$, is 
\begin{subequations}
\begin{align}
 \begin{split}
 \norm{\mathcal{M}_{\text{spon}} }^2 &\equiv \norm{\bra{\Psi_G} \otimes \bra{1_{\mathbold{k},\upsilon}} \underline{\mathcal{H}}^{(1)} \ket{\Psi_E} \otimes \ket{0}}^2 \\
 &= \frac{q_e^2}{m_e^2} \left( \frac{\hbar}{2 \mathcal{V} \epsilon_0 \omega_{\mathbold{k}}} \right) \norm{\hat{\mathbold{e}}_{\mathbold{k},\upsilon} \cdot \bra{\Psi_G} \underline{\mathbold{p}} \mathrm{e}^{-i \mathbold{k} \cdot \mathbold{r}} \ket{\Psi_E}}^2 ,
 \end{split}
 \end{align}
which, when evaluated at $\hbar \omega_{\mathbold{k}} = \Delta \mathcal{E}_e$ under the approximation $\mathrm{e}^{- i \mathbold{k} \cdot \mathbold{r}} \approx 1$, becomes 
\begin{align}
 \begin{split}\label{eq:normal_spon_term}
 \norm{\mathcal{M}_{\text{spon}} }^2 &\to \norm{\bra{\Psi_G} \otimes \bra{1_{\mathbold{k},\upsilon}} \underline{\mathcal{H}}^{(1)} \ket{\Psi_E} \otimes \ket{0}}^2 \\
 &= \frac{q_e^2}{m_e^2} \left( \frac{\hbar^2}{2 \mathcal{V} \epsilon_0 \Delta \mathcal{E}_e } \right) \norm{\hat{\mathbold{e}}_{\mathbold{k},\upsilon} \cdot \bra{\Psi_G} \underline{\mathbold{p}} \ket{\Psi_E}}^2 .
 \end{split}
 \end{align}
\end{subequations}
Then, using the following quantum operator commutation relation:\footnote{This makes use of the following fundamental commutation relation between the position and momentum operators $\underline{\mathbold{r}}$ and $\underline{\mathbold{p}}$ \cite{Griffiths05,Shankar04}:
\begin{equation}\label{eq:pos_mom_commuation}
 \left[ \underline{r}_{\alpha}, \underline{p}_{\beta} \right] \equiv \underline{r}_{\alpha} \underline{p}_{\beta} - \underline{p}_{\beta} \underline{r}_{\alpha} = i \hbar \delta_{\alpha,\beta} ,
 \end{equation}
where $\alpha, \beta = 1,2,3$ index single components of $\underline{\mathbold{r}}$ and $\underline{\mathbold{p}}$ and $\delta_{\alpha,\beta}$ is a Kronecker delta [\emph{cf.\@} \cref{eq:Kronecker_delta}].}
\begin{subequations}
\begin{equation}
 \left[ \underline{\mathcal{H}}^{(0)} , \underline{\mathbold{r}} \right] \equiv \underline{\mathcal{H}}^{(0)} \underline{\mathbold{r}} - \underline{\mathbold{r}} \underline{\mathcal{H}}^{(0)} = - \frac{i \hbar}{m_e} \underline{\mathbold{p}} , \label{eq:commutation1}
 \end{equation}
where $\underline{\mathcal{H}}^{(0)} = \underline{\mathcal{H}}_{\text{ion}}$ is the unperturbed Hamiltonian, $\norm{\mathcal{M}_{\text{E}1}}^2$ and $\norm{\mathcal{M}_{\text{spon}} }^2$ in \cref{eq:E1_spon_term,eq:normal_spon_term} can be shown to be equivalent by solving \cref{eq:commutation1} for $\underline{\mathbold{p}}$ to write the matrix element in \cref{eq:normal_spon_term} as \cite{Townsend00,Rybicki86} 
\begin{align}\label{eq:dipole_inner_product2}
 \begin{split}
 \hat{\mathbold{e}}_{\mathbold{k},\upsilon} \cdot \matrixel{\Psi_B}{\underline{\mathbold{p}}}{\Psi_A} &= \frac{i m_e}{\hbar} \hat{\mathbold{e}}_{\mathbold{k},\upsilon} \cdot  \left( \matrixel{\Psi_B}{\underline{\mathcal{H}}^{(0)} \underline{\mathbold{r}}}{\Psi_A} - \matrixel{\Psi_B}{\underline{\mathbold{r}} \underline{\mathcal{H}}^{(0)}}{\Psi_A} \right) \\
 &= \frac{i m_e}{\hbar} \underbrace{\left( \mathcal{E}_{B} - \mathcal{E}_{A} \right)}_{\Delta \mathcal{E}_e} \hat{\mathbold{e}}_{\mathbold{k},\upsilon} \cdot \matrixel{\Psi_B}{ \underline{\mathbold{r}} }{\Psi_A} .
 \end{split}
 \end{align}
Carrying out a squared norm on both sides shows that $\norm{\mathcal{M}_{\text{E}1}}^2 = \norm{\mathcal{M}_{\text{spon}} }^2$ under the approximation $\mathrm{e}^{- i \mathbold{k} \cdot \mathbold{r}} \approx 1$ with
\begin{equation}
 \norm{\hat{\mathbold{e}}_{\mathbold{k},\upsilon} \cdot \matrixel{\Psi_B}{\underline{\mathbold{p}}}{\Psi_A}}^2 = \left( \frac{m_e \Delta \mathcal{E}_e}{\hbar} \right)^2 \norm{\hat{\mathbold{e}}_{\mathbold{k},\upsilon} \cdot \matrixel{\Psi_B}{ \underline{\mathbold{r}} }{\Psi_A}}^2 .
 \end{equation}
\end{subequations}

The above analysis indicates that electric-dipole transitions are associated with the $0^{\text{th}}$-order approximation of the complex exponential in \cref{eq:general_norm_squared_transition}, which is valid for either absorption or emission: 
\begin{equation}\label{eq:Taylor_complex_exponential}
 \mathrm{e}^{\pm i \mathbold{k} \cdot \mathbold{r}} = 1 \pm i \mathbold{k} \cdot \mathbold{r} + \frac{\left( \pm i \mathbold{k} \cdot \mathbold{r} \right)^2}{2} \dotsc 
 \end{equation} 
Because a factor of unity enters into $\norm{\mathcal{M}_{\text{E}1}}^2$ in this approximation, these transitions are expected to have relatively large values for the transition rate, $\Gamma$.   
Physically, the condition $\mathbold{k} \cdot \mathbold{r} \ll 1$ implies that the electron undergoing the transition stays localized to within regions much smaller than the wavelength of the absorbed or emitted photon, $\lambda = 2 \pi / k_0 = h c_0 / \mathcal{E}_{\gamma}$  \cite{Shankar04,Townsend00}. 
In other words, the electric field associated with the photon appears spatially uniform to the transitioning electron and this is the case for many transitions in highly-charged ions, or more generally, for transitions involving K-shell electrons in a low-to-mid $\mathcal{Z}$ atom, which have typical radii $\sim a_0 / \mathcal{Z}$ (with $a_0$ as the \emph{Bohr radius}\footnote{This is the distance between the nucleus and the most probable location of a ground-state electron in a hydrogen atom, which can be obtained using the semi-classical \emph{Bohr model of the atom}, where the centripetal force felt by the electron orbiting the proton with a velocity $v$ is balanced by electrostatic attraction between the two charges \cite{Shankar04,Griffiths05}: %\cite{Bohr13a,Bohr13b} ,Tipler07
\begin{equation*}\label{eq:Bohr_model}
 \frac{m_e v^2}{r} = \frac{q_e^2}{4 \pi \epsilon_0 r^2} .
 \end{equation*}
Then, assuming the orbital velocity $v$ takes on only discrete values $v = n \hbar / m_e r$ with $n=1,2,3\dotsc$, the above expression and solving for $r$ with $n=1$ then gives the Bohr radius defined in \cref{tab:fundamental_constants}.}) \cite{Attwood17}. 
For these reasons, $\mathrm{e}^{\pm i \mathbold{k} \cdot \mathbold{r}} \approx 1$ is referred to as the \emph{electric-dipole approximation} and transitions that give $\Gamma \neq 0$ under these conditions tend to be most dominant in a given ion. 

For a transitioning electron, a set of selection rules for the electric-dipole approximation can be determined by first by projecting the electric-dipole matrix element into the position basis:
\begin{equation}
 \hat{\mathbold{e}}_{\mathbold{k},\upsilon} \cdot \bra{\Psi_B} \underline{\mathbold{r}} \ket{\Psi_A} = \hat{\mathbold{e}}_{\mathbold{k},\upsilon} \cdot \int_{-\infty}^{\infty} \Psi_B^* (\mathbold{r}) \, \mathbold{r} \, \Psi_A (\mathbold{r}) \dd[3]{\mathbold{r}} ,
 \end{equation}
and noting that the integral comes out to zero unless the integrand is an even function. 
Because the operator $\underline{\mathbold{r}}$ has \emph{odd parity},\footnote{Broadly, this is defined as the sign change that an operator or wave function incurs from switching the sign on all phase space coordinates: $\underline{\mathbold{p}} \to -\underline{\mathbold{p}}$ and $\underline{\mathbold{r}} \to -\underline{\mathbold{r}}$ \cite{Rybicki86}.} one of the wave functions must be even while the other must be odd to ensure that the integrand is an even function. 
With parity in a one-electron system defined as $(-1)^{\ell}$ (\emph{i.e.}, $+1$ for even and $-1$ for odd), this is verified by \emph{Laporte's rule}, which states that \emph{parity must change during electric-dipole transitions} \cite{Rybicki86,Bethe57}. 
Meanwhile, the spin quantum number, $m_s$, is not expected to change under a pure electric-dipole transition because the \emph{spin-angular-momentum operator}, $\underline{\mathbold{S}}$, does not appear in the perturbing Hamiltonian.
From these considerations, selection rules for an electric-dipole transition between a state with $\ell = \ell_A, m_{\ell} = m_A$ to one with $\ell = \ell_B, m_{\ell} = m_B$ can be stated as 
\begin{align}
 \begin{split}\label{eq:hydrogen_E1_selection_rules}
 \Delta \ell = \ell_B - \ell_A &= \pm 1 \\
 \Delta m_{\ell} = m_B - m_A &= 0, \pm 1 \quad (\text{with } \Delta m_s =0),
  \end{split}
 \end{align}
where the latter follows from the fact that for any given $\ell$, $m_{\ell}$ can take on values $-\ell, -\ell + 1, \dotsc 0, \dotsc \ell -1, \ell$ [\emph{cf.\@} \cref{tab:shell_orbitals}]. 

In a system with two electrons, the parity is given by $(-1)^{\ell_1 + \ell_2}$, where $\ell_{1,2}$ are the azimuthal quantum numbers for each electron. 
Assuming that one electron is in the ground state with $\ell_1 = 0$ while the other is left arbitrary as $\ell_2 = \ell$, the quantum number for the total orbital angular momentum has possible values $L = 0,1$ in the excited state and therefore the selection rules are given by \cite{Rybicki86,Bethe57} 
\begin{align}\label{eq:helium_E1_selection_rules}
 \begin{split}
 \Delta S &= 0 \quad \text{and} \quad \Delta L = 0, \, \pm 1 , \\
 &\text{except for } L = 0 \to L = 0 ,
  \end{split}
 \end{align}
where $\Delta S = 0$ indicates that the total spin of the system does not change in an electric-dipole transition while the $L = 0 \to L = 0$ rules follows from the fact that there must a transfer of angular momentum between the transitioning electron and the interacting photon. 
In any case, the appropriate selection rules dictate which transitions are \emph{allowed} under the electric-dipole approximation while all others are referred to being \emph{forbidden}. 
However, as demonstrated in \cref{sec:forbidden_transitions}, these processes are not truly forbidden but rather occur with much lower probability. 
Electric-dipole transitions can also occur along with a spin magnetic dipole interaction; such a transition is referred to as being \emph{semi-forbidden} and occurs with probability intermediate between allowed and forbidden transitions. 

\subsection{Magnetic-Dipole and Forbidden Transitions}\label{sec:forbidden_transitions}
%%%%%%%%%%%%%%%%%%%%%%%%%%%%%%%%%%%%%%%%%--------------------------------------------------
Bound-bound transitions that yield $\Gamma \neq 0$ under the electric-dipole approximation [\emph{cf.\@} \cref{sec:allowed_transitions}] are expected to occur most frequently and hence contribute to the strongest spectral lines in a given ion, which are also known as \emph{resonance lines}. 
In contrast, transitions that yield $\norm{\mathcal{M}}^2 \neq 0$ only with more terms beyond $\mathrm{e}^{\pm i \mathbold{k} \cdot \mathbold{r}} \approx 1$ are said to be electric-dipole forbidden to indicate their relatively low transition rate, $\Gamma$. 
Based on the arguments given at the start of \cref{sec:allowed_transitions}, magnetic-dipole interactions are expected to be a contributor to these forbidden transitions. 
In analogy to \cref{eq:E1_classical,eq:E1_hamiltonian}, the energy associated with a classical, point magnetic dipole is given by 
\begin{subequations}
\begin{equation}\label{eq:M1_classical}
 U_{\text{M}1} = - \mathbold{\mu}_{\text{M}1} \cdot \mathbold{B}_0 ,
 \end{equation}
where $\mathbold{B}_0$ is a constant magnetic field. 
Then, the corresponding Hamiltonian operator is taken as 
\begin{equation}\label{eq:M1_hamiltonian}
 \underline{\mathcal{H}}^{(\text{M}1)} \equiv - \underline{\mathbold{\mu}}_{\text{M}1} \cdot \underline{\mathbold{B}}_0 ,
 \end{equation}
where $\underline{\mathbold{B}}_0$ is the magnetic-field operator defined in \cref{eq:B-field_operator_fixed_linear} with $\mathrm{e}^{\pm i \mathbold{k} \cdot \mathbold{r}} \to 1$:
\begin{equation}\label{eq:B-field_operator_fixed_linear_constant}
 \underline{\mathbold{B}}_0 = \sum_{\mathbold{k}} \sum_{\upsilon = 1}^2 i \sqrt{\frac{\hbar}{2 \mathcal{V} \epsilon_0 \omega_{\mathbold{k}}}} \mathbold{k} \times \hat{\mathbold{e}}_{\mathbold{k},\upsilon} \left[ \underline{a}_{\mathbold{k},\upsilon} + \underline{a}^{\dagger}_{\mathbold{k},\upsilon} \right] 
 \end{equation}
\end{subequations}
and $\underline{\mathbold{\mu}}_{\text{M}1}$ is the operator for the magnetic-dipole moment. 
This operator includes contributions from both the orbital angular momentum and the intrinsic angular momentum associated with electron spin; these components are considered separately in the following two subsections. 

\subsubsection{Orbital Magnetic-Dipole and Electric-Quadrupole}
%%%%%%%%%%%%%%%%%%%%%%%%%%%%%%%%%%%%%%%%%--------------------------------------------------
In analogy to a classical magnetic dipole, the component of $\underline{\mathbold{\mu}}_{\text{M}1}$ that arises from orbital angular momentum is expressed as \cite{Townsend00,Shankar04,Griffiths05} 
\begin{equation}\label{eq:orbital_moment_operator}
 \underline{\mathbold{\mu}}_{\ell} \equiv - \frac{q_e}{2 m_e} \underline{\mathbold{L}} ,
 \end{equation}
where $\underline{\mathbold{L}}$ is the \emph{orbital-angular-momentum operator}, which is defined as
\begin{subequations} 
\begin{equation}\label{eq:eq:orbit_ang_oper}
 \underline{\mathbold{L}} \equiv \underline{\mathbold{r}} \times \underline{\mathbold{p}} = \hat{\mathbold{x}} \underline{L}_x + \hat{\mathbold{y}} \underline{L}_y + \hat{\mathbold{z}} \underline{L}_z .
 \end{equation}
 Neglecting spin, the quantum numbers $\ell$ and $m_{\ell}$ define eigenvalue relations for $\underline{\mathbold{L}}^2$, as well as for an arbitrary vector component of $\underline{\mathbold{L}}$ to indicate the direction of the magnetic moment. 
Taking this direction to be oriented along the $z$-axis and using the notation $\ket{\Psi_e} \equiv \ket{n, \ell, \, m_{\ell} }$, these relations are \cite{Townsend00,Shankar04,Griffiths05} 
\begin{align}
 \begin{split}\label{eq:orbital_eigenvalues}
 \underline{L}^2 \, \ket{n, \, \ell, \, m_{\ell}} &= \ell (\ell + 1) \hbar^2 \, \ket{n, \, \ell, \, m_{\ell}} \\
 \underline{L}_z \, \ket{n, \, \ell, \, m_{\ell}} &=  m_{\ell} \hbar \, \ket{n, \, \ell, \, m_{\ell}} .
 \end{split}
 \end{align}
Meanwhile, $\underline{L}_x$ and $\underline{L}_y$ can be expressed in terms of $\underline{L}_+$ and $\underline{L}_-$, the raising and lowering operators that serve to change the value of $m_{\ell}$ by $\pm 1$:
\begin{equation}
 \underline{L}_x = \frac{1}{2} \left( \underline{L}_+ + \underline{L}_- \right) \quad \text{and} \quad \underline{L}_y = \frac{i}{2} \left( \underline{L}_- - \underline{L}_+ \right) 
 \end{equation}
with 
\begin{equation}\label{eq:orbital_raise_lower}
 \underline{L}_{\pm} \ket{n, \, \ell, \, m_{\ell}} = \hbar \sqrt{\ell \left( \ell + 1 \right) - m_{\ell} \left( m_{\ell} \pm \right)} \ket{n, \ell, \, \left( m_{\ell} \pm 1 \right)}.
 \end{equation}
\end{subequations}

Defining the orbital-angular-momentum piece of the magnetic-dipole Hamiltonian operator as 
\begin{equation}
 \underline{\mathcal{H}}^{(\text{M}1)}_{\ell} \equiv - \underline{\mathbold{\mu}}_{\ell} \cdot \underline{\mathbold{B}}_0 ,
 \end{equation}
the example of spontaneous emission discussed in \cref{sec:allowed_transitions} is now returned to in order to investigate properties of orbital magnetic-dipole transitions. 
In this situation, the relevant form of $\norm{\mathcal{M}}^2$ with $\underline{\mathcal{H}}^{(\text{M}1)}_{\ell}$ inserted in place of the general expression, $\underline{\mathcal{H}}^{(1)} = (q_e / m_e) \underline{\mathbold{A}} ( \mathbold{r} ) \cdot \underline{\mathbold{p}}$, is 
\begin{align}
 \begin{split}
 \norm{\mathcal{M}_{\text{M}1 \ell}}^2 &\equiv \norm{\bra{\Psi_B} \otimes \bra{1_{\mathbold{k},\upsilon} } \underline{\mathcal{H}}^{(\text{M}1)}_{\ell} \ket{\Psi_A} \otimes \ket{0}}^2 \\
 &= \left( \frac{q_e}{2 m_e} \right)^2 \norm{\bra{\Psi_B} \otimes \bra{1_{\mathbold{k},\upsilon} } \left( \underline{\mathbold{B}}_0 \cdot \underline{\mathbold{L}} \right) \ket{\Psi_A} \otimes \ket{0}}^2 \\
 &= \left( \frac{q_e}{2 m_e} \right)^2 \frac{\hbar}{2 \mathcal{V} \omega_{\mathbold{k}} \epsilon_0} \norm{\bra{\Psi_B} \left[ \left( \mathbold{k} \cross \hat{\mathbold{e}}_{\mathbold{k},\upsilon} \right) \cdot \underline{\mathbold{L}} \right] \ket{\Psi_A}}^2 , 
 \end{split} 
 \end{align}
where it is seen that the perturbing operator takes the form $\left( \mathbold{k} \cross \hat{\mathbold{e}}_{\mathbold{k},\upsilon} \right) \cdot \underline{\mathbold{L}}$. 
As always, this term is evaluated at $\hbar \omega_{\mathbold{k}} = \Delta \mathcal{E}_e$ to arrive at a transition rate given by Fermi's golden rule [\emph{cf.\@} \cref{sec:transition_rate_sub}], where $\Delta \mathcal{E}_e$ is the electronic binding energy difference associated with the transition. 
Then, using the following vector relation \cite{Townsend00}:
\begin{equation}\label{eq:mag_dipole_vector_trick}
 \left( \mathbold{k} \times \hat{\mathbold{e}}_{\mathbold{k},\upsilon} \right) \cdot \underbrace{\left( \underline{\mathbold{r}} \times \underline{\mathbold{p}} \right)}_{\underline{\mathbold{L}}} =  \left( \hat{\mathbold{e}}_{\mathbold{k},\upsilon} \cdot \underline{\mathbold{p}} \right) \left( \mathbold{k} \cdot \underline{\mathbold{r}} \right) - \left( \hat{\mathbold{e}}_{\mathbold{k},\upsilon} \cdot \underline{\mathbold{r}} \right) \left( \mathbold{k} \cdot \underline{\mathbold{p}} \right) ,
 \end{equation}
it can be shown that this magnetic-dipole term appears in $\underline{\mathcal{H}}^{(1)} = (q_e / m_e) \underline{\mathbold{A}} ( \mathbold{r} ) \cdot \underline{\mathbold{p}}$ when the approximation $\mathrm{e}^{\pm i \mathbold{k} \cdot \mathbold{r}} \approx 1 \pm i \mathbold{k} \cdot \mathbold{r}$ is invoked. 
While the first term results in $\bra{\Psi_B} \left( \hat{\mathbold{e}}_{\mathbold{k},\upsilon} \cdot \underline{\mathbold{p}} \right) \ket{\Psi_A}$ for an electric dipole, the next term in the series yields $\bra{\Psi_B} \left( \hat{\mathbold{e}}_{\mathbold{k},\upsilon} \cdot \underline{\mathbold{p}} \right) \left( \mathbold{k} \cdot \underline{\mathbold{r}} \right) \ket{\Psi_A}$, where $\mathbold{r}$ is now treated as the position operator for the electron, $\underline{\mathbold{r}}$. 

By expressing the perturbing operator as 
\begin{align}\label{eq:term_split_trick}
 \begin{split}
 \left( \hat{\mathbold{e}}_{\mathbold{k},\upsilon} \cdot \underline{\mathbold{p}} \right) \left( \mathbold{k} \cdot \underline{\mathbold{r}} \right) &= \frac{1}{2} \underbrace{\left[ \left( \hat{\mathbold{e}}_{\mathbold{k},\upsilon} \cdot \underline{\mathbold{p}} \right) \left( \mathbold{k} \cdot \underline{\mathbold{r}} \right) - \left( \hat{\mathbold{e}}_{\mathbold{k},\upsilon} \cdot \underline{\mathbold{r}} \right) \left( \mathbold{k} \cdot \underline{\mathbold{p}} \right) \right]}_{\text{orbital magnetic dipole}}\\
 &+ \frac{1}{2} \underbrace{ \left[ \left( \hat{\mathbold{e}}_{\mathbold{k},\upsilon} \cdot \underline{\mathbold{p}} \right) \left( \mathbold{k} \cdot \underline{\mathbold{r}} \right) + \left( \hat{\mathbold{e}}_{\mathbold{k},\upsilon} \cdot \underline{\mathbold{r}} \right) \left( \mathbold{k} \cdot \underline{\mathbold{p}} \right) \right] }_{\text{electric quadrupole}},
 \end{split}
 \end{align}
it is seen that terms for both orbital-magnetic-dipole and \emph{electric-quadrupole} transitions are present. 
While the former is evident from \cref{eq:mag_dipole_vector_trick}, the latter term in \cref{eq:term_split_trick} can be recognized as an electric-quadrupole moment by using the following commutation relation:  
\begin{subequations}
\begin{align}
 \begin{split}
 \left[ \underline{\mathcal{H}}^{(0)} , \left( \hat{\mathbold{e}}_{\mathbold{k},\upsilon} \cdot \underline{\mathbold{r}} \right) \left( \mathbold{k} \cdot \underline{\mathbold{r}} \right) \right] &\equiv \underline{\mathcal{H}}^{(0)} \left( \hat{\mathbold{e}}_{\mathbold{k},\upsilon} \cdot \underline{\mathbold{r}} \right) \left( \mathbold{k} \cdot \underline{\mathbold{r}} \right) - \left( \hat{\mathbold{e}}_{\mathbold{k},\upsilon} \cdot \underline{\mathbold{r}} \right) \left( \mathbold{k} \cdot \underline{\mathbold{r}} \right) \underline{\mathcal{H}}^{(0)} \\ 
 &= - \frac{i \hbar }{m_e} \left[ \left( \hat{\mathbold{e}}_{\mathbold{k},\upsilon} \cdot \underline{\mathbold{r}} \right) \left( \mathbold{k} \cdot \underline{\mathbold{p}} \right) + \left( \mathbold{k} \cdot \underline{\mathbold{r}} \right) \left( \hat{\mathbold{e}}_{\mathbold{k},\upsilon} \cdot \underline{\mathbold{p}} \right) \right] ,
 \end{split}
 \end{align}
where $\underline{\mathcal{H}}^{(0)} = \underline{\mathcal{H}}_{\text{ion}}$.\footnote{Motivated by Fitzpatrick \cite{Fitzpatrick_QM}, this result is obtained from carrying out the following operations, where $\hat{\mathbold{e}}$ is shorthand for $\hat{\mathbold{e}}_{\mathbold{k},\upsilon}$:
\begin{align*}
 &\left[ \underline{\mathcal{H}}^{(0)} , \left( \hat{\mathbold{e}} \cdot \underline{\mathbold{r}} \right) \left( \mathbold{k} \cdot \underline{\mathbold{r}} \right) \right] = \sum_{\alpha,\beta} \hat{e}_{\alpha} k_{\beta} \left[ \underline{\mathcal{H}}^{(0)}, \underline{r}_{\alpha} \underline{r}_{\beta} \right] = \sum_{\alpha,\beta} \hat{e}_{\alpha} k_{\beta} \underline{r}_{\alpha} \underbrace{\left[ \underline{\mathcal{H}}^{(0)}, \underline{r}_{\beta} \right]}_{\text{\cref{eq:commutation1}}} + \sum_{\alpha,\beta} \hat{e}_{\alpha} k_{\beta} \underbrace{\left[ \underline{\mathcal{H}}^{(0)}, \underline{r}_{\alpha} \right]}_{\text{\cref{eq:commutation1}}} \underline{r}_{\beta} \\ 
 &\quad = - \frac{i \hbar }{m_e} \sum_{\alpha,\beta} \hat{e}_{\alpha} k_{\beta} \left( r_{\alpha} p_{\beta} + \underbrace{p_{\alpha} r_{\beta}}_{\text{use \cref{eq:pos_mom_commuation}}} \right) = - \frac{i \hbar }{m_e} \sum_{\alpha,\beta} \hat{e}_{\alpha} k_{\beta} \left( r_{\alpha} p_{\beta} + r_{\beta} p_{\alpha} - i \hbar \delta_{\alpha,\beta} \right) \\
 &\quad = - \frac{i \hbar }{m_e} \left[ \left( \hat{\mathbold{e}} \cdot \mathbold{r} \right) \left( \mathbold{k} \cdot \mathbold{p} \right) + \left( \mathbold{k} \cdot \mathbold{r} \right) \left( \hat{\mathbold{e}} \cdot \mathbold{p} \right) - i \hbar \underbrace{\left( \hat{\mathbold{e}} \cdot \mathbold{k} \right)}_0 \right] ,
 \end{align*}
where $\alpha$ and $\beta$ index single components of $\underline{\mathbold{r}}$, $\underline{\mathbold{p}}$, $\mathbold{k}$ and $\hat{\mathbold{e}}$ while $\delta_{\alpha,\beta}$ is a Kronecker delta.} 
Using this relation, the effect that this latter term has as a perturbing operator can be determined through evaluating the following matrix element:
\begin{align}
 \begin{split}
 &\matrixel{\Psi_B}{\left[ \left( \hat{\mathbold{e}}_{\mathbold{k},\upsilon} \cdot \underline{\mathbold{r}} \right) \left( \mathbold{k} \cdot \underline{\mathbold{p}} \right) + \left( \mathbold{k} \cdot \underline{\mathbold{r}} \right) \left( \hat{\mathbold{e}}_{\mathbold{k},\upsilon} \cdot \underline{\mathbold{p}} \right) \right]}{\Psi_A}  \\
 &= \frac{i m_e}{\hbar} \left( \matrixel{\Psi_B}{\underline{\mathcal{H}}^{(0)} \left( \hat{\mathbold{e}}_{\mathbold{k},\upsilon} \cdot \underline{\mathbold{r}} \right) \left( \mathbold{k} \cdot \underline{\mathbold{r}} \right)}{\Psi_A} - \matrixel{\Psi_B}{\left( \hat{\mathbold{e}}_{\mathbold{k},\upsilon} \cdot \underline{\mathbold{r}} \right) \left( \mathbold{k} \cdot \underline{\mathbold{r}} \right) \underline{\mathcal{H}}^{(0)}}{\Psi_A} \right) \\
 &= \frac{i m_e}{\hbar} \underbrace{\left( \mathcal{E}_{B} - \mathcal{E}_{A} \right)}_{\Delta \mathcal{E}_e} \matrixel{\Psi_B}{\underbrace{\left( \hat{\mathbold{e}}_{\mathbold{k},\upsilon} \cdot \underline{\mathbold{r}} \right) \left( \mathbold{k} \cdot \underline{\mathbold{r}} \right)}_{\text{electric quadrupole}}}{\Psi_A} ,
 \end{split}
 \end{align}
\end{subequations}
where as in \cref{eq:dipole_inner_product2}, $\Delta \mathcal{E}_e$ is the difference in electronic binding energies for the initial and final states. 
With $\alpha,\beta =1,2,3$ indexing tensor components, the energy associated with a classical electric quadrupole is 
\begin{subequations}
\begin{equation}\label{eq:electric_quad_energy}
 \frac{1}{6} \sum_{\alpha,\beta} \mathcal{Q}_{\alpha,\beta} \pdv{E_{\beta}}{r_{\alpha}} ,
 \end{equation}
with being $\mathcal{Q}_{\alpha,\beta}$ the \emph{quadrupole-moment tensor} \cite{Jackson75}.
For a set of point charges with $-q_e$, this can be written as 
\begin{equation}\label{eq:electric_quad_moment}
 \mathcal{Q}_{\alpha,\beta} = -q_e \left( 3 r_{\alpha} r_{\beta} - r^2 \delta_{\alpha,\beta} \right) .
 \end{equation}
\end{subequations}

With the approximation $\mathrm{e}^{\pm i \mathbold{k} \cdot \mathbold{r}} \approx 1 \pm i \mathbold{k} \cdot \mathbold{r}$, the spatial derivative of the electric field with a vector direction $\hat{e}_{\beta}$ gives $k_{\alpha} \hat{e}_{\beta}$ so that \cref{eq:electric_quad_energy} becomes 
\begin{subequations}
\begin{equation}\label{eq:electric_quad_energy_approx}
 \frac{1}{6} \sum_{\alpha,\beta} \mathcal{Q}_{\alpha,\beta} k_{\alpha} \hat{e}_{\beta} . 
 \end{equation}
Then, it can be shown using \cref{eq:electric_quad_energy,eq:electric_quad_moment} that the term $\left(\hat{\mathbold{e}} \cdot \mathbold{r} \right) \left( \mathbold{k} \cdot \mathbold{r} \right)$ is proportional to the energy of an electric quadrupole:
\begin{equation}
 \sum_{\alpha,\beta} \mathcal{Q}_{\alpha,\beta} k_{\alpha} \hat{e}_{\beta} \propto \sum_{\alpha,\beta} \left( 3 r_{\alpha} r_{\beta} \hat{e}_{\beta} - r^2 \delta_{\alpha,\beta} k_{\alpha} \hat{e}_{\beta} \right) = \left( 3 \hat{\mathbold{e}} \cdot \mathbold{r} \right) \left( \mathbold{k} \cdot \mathbold{r} \right) - r^2 \underbrace{\mathbold{k} \cdot \hat{\mathbold{e}}}_0 . 
 \end{equation}
\end{subequations}
This indicates that, unlike an electric-dipole transition, an electric-quadrupole transition is subject to a photon electric field that varies spatially to a small degree. 
Like an electric-dipole transition, however, such a transition does not depend on angular momentum and hence there should be no effect on electron spin. 
Ultimately, the relevant matrix element depends on the square of the position operator, $\underline{\mathbold{r}}^2$, instead of $\underline{\mathbold{r}}$ as it does in \cref{eq:dipole_inner_product2} for the case of electric-dipole transitions. 

Because orbital magnetic-dipole transitions depends on the operator $\underline{\mathbold{L}}$ and electric-quadrupole transitions depends on $\underline{\mathbold{r}}^2$, both are associated with even parity. 
This implies that the initial and final wave function in a one-electron system must have the same parity, $(-1)^{\ell}$. 
In the magnetic-dipole case, $\underline{\mathbold{L}}$ can at most change $m_{\ell}$ by one due to the presence of $\mathbold{L}_{\pm}$ [\emph{cf.\@} \cref{eq:orbital_raise_lower}] to first order, while in any case, $\ell$ remains unchanged. 
Therefore, selection rules for such a transition (neglecting spin) can be taken as
\begin{align}
 \begin{split}\label{eq:hydrogen_M1_selection_rules}
 \Delta \ell = \ell_B - \ell_A &= 0 \\
 \Delta m_{\ell} = m_B - m_A &= 0, \pm 1 .
  \end{split}
 \end{align}
For electric-quadrupole transitions, $\ell$ can either stay the same or change by $\pm 2$ to provide even parity overall and thus these selection rules are 
\begin{align}
 \begin{split}\label{eq:hydrogen_E2_selection_rules}
 \Delta \ell = \ell_B - \ell_A &= 0, \pm 2 \\
 \Delta m_{\ell} = m_B - m_A &= 0, \pm 1, \pm 2 ,
  \end{split}
 \end{align}
where it is assumed that there is no change in the spin quantum number, $m_s$. 
Either of these situations can be generalized to two-electron systems, where in particular, it holds true that \emph{the configuration does not change} for magnetic-dipole transitions \cite{Rybicki86}. 
Explained next, this condition can also be fulfilled in the event that the total spin of the system changes while the orbital angular momentum is unaffected. 

\subsubsection{Spin Magnetic-Dipole and Semi-Forbidden}
%%%%%%%%%%%%%%%%%%%%%%%%%%%%%%%%%%%%%%%%%--------------------------------------------------
So far in this appendix, non-relativistic quantum mechanics has been employed and therefore electron spin has not been properly taken into account. 
Because of this, a second component to $\underline{\mathbold{\mu}}_{\text{M}1}$ that depends on electron spin must be added to the perturbing Hamiltonian manually to describe certain phenomena. 
With the total operator written as $\underline{\mathbold{\mu}}_{\text{M}1} \equiv \underline{\mathbold{\mu}}_{\ell} + \underline{\mathbold{\mu}}_{s}$, the spin term is defined as
\begin{subequations}
\begin{equation}\label{eq:spin_moment_operator}
 \underline{\mathbold{\mu}}_{s} \equiv - \frac{q_e g_e}{2 m_e} \underline{\mathbold{S}} ,
 \end{equation} 
where $g_e \approx \num{2.002319}$ is the \emph{g-factor} of electron that arises in relativistic quantum mechanics and 
\begin{equation}\label{eq:spin_ang_oper}
 \underline{\mathbold{S}} = \hat{\mathbold{x}} \underline{S}_x + \hat{\mathbold{y}} \underline{S}_y + \hat{\mathbold{z}} \underline{S}_z
 \end{equation}
is the spin-angular-momentum operator that acts on electron spin states represented by \emph{spinors} of the form $\ket{\chi_e} \equiv \ket{s, \, m_s}$ \cite{Shankar04,Townsend00}. 
In the case of a single electron, $s=1/2$ while $m_s = \pm 1/2$ but more generally, $s$ is the total spin of the system and $m_s$ takes on values $-s, -s + 1, \dotsc 0, \dotsc s -1, s$ that indicate the net spin direction of the system. 
Just as in \cref{eq:orbital_eigenvalues} for orbital angular momentum, the following eigenvalue relations hold: 
\begin{align}
 \begin{split}\label{eq:spin_eigenvalues}
 \underline{S}^2 \, \ket{s, \, m_{s}} &= s (s + 1) \hbar^2 \, \ket{s, \, m_{s}} \\
 \underline{S}_z \, \ket{s, \, m_{s}} &=  m_{s} \hbar \, \ket{s, \, m_{s}} ,
 \end{split}
 \end{align}
while $\underline{S}_x$ and $\underline{S}_y$ can be expressed in terms of $\underline{S}_+$ and $\underline{S}_-$, the raising and lowering operators for spin states that change the value of $m_s$ by $\pm 1$:
\begin{equation}
 %\begin{split}
 \underline{S}_x = \frac{1}{2} \left( \underline{S}_+ + \underline{S}_- \right) \quad \text{and} \quad \underline{S}_y = \frac{i}{2} \left( \underline{S}_- - \underline{S}_+ \right)
 %\end{split}
 \end{equation}
with 
\begin{equation}
 \underline{S}_{\pm} \ket{s, \, m_s} = \hbar \sqrt{\ell \left( \ell + 1 \right) - m_{\ell} \left( m_{\ell} \pm \right)} \ket{s, \, \left( m_s \pm 1 \right)}.
 \end{equation}
\end{subequations}
Considering this piece of the Hamiltonian alone for a one-electron system:
\begin{equation}
 \underline{\mathcal{H}}^{(\text{M}1)}_{s} \equiv - \underline{\mathbold{\mu}}_{s} \cdot \underline{\mathbold{B}}_0 ,
 \end{equation}
it is expected that like in the case of an orbital magnetic-dipole transition, only the direction of the angular momentum, $m_s$, is affected while $s$ cannot change due to parity considerations. 
On the other hand, parity can change if the electric-dipole term $\underline{\mathcal{H}}^{(\text{E}1)} = q_e \underline{\mathbold{B}}_0 \cdot \underline{\mathbold{r}}$ is included with the perturbing Hamiltonian. %from \cref{sec:allowed_transitions}
In this case, the selection rules defined in \cref{eq:hydrogen_E1_selection_rules,eq:helium_E1_selection_rules} are still satisfied but with the added flexibility that the total spin of the system can change by one quantum number. 
Such a \emph{semi-forbidden} transition that requires both electric-dipole and spin-magnetic-dipole interaction is said to produce \emph{intercombination lines}. 

\section{Hydrogen-like Ions}\label{sec:selection_rules_hydrogen}
%%%%%%%%%%%%%%%%%%%%%%%%%%%%%%%%%%%%%%%%%--------------------------------------------------
The single electron bound in a hydrogen-like ion is subject to a central potential established by the nuclear charge, $\mathcal{Z} q_e$. 
In non-relativistic quantum mechanics, where spin is neglected, this is described with the following Hamiltonian operator \cite{Bethe57,Townsend00,Kahn02,Shankar04}: 
\begin{equation}\label{eq:hydrogen_ion_Hamiltonian}
 \underline{\mathcal{H}}_{\text{H}} = \frac{\underline{\mathbold{p}}^2}{2 m_R} - \hbar c_0 \alpha_f \frac{\mathcal{Z}}{\abs{\underline{\mathbold{r}}}} ,
 \end{equation}
where $\alpha_f$ is the fine-structure constant, $\underline{\mathbold{p}}$ and $\underline{\mathbold{r}}$ are the momentum and position operators for the electron and $m_R$ is the two-body reduced mass:
\begin{equation}\label{eq:reduced_mass}
 m_R \equiv \frac{m_e \, m_{\text{nucleus}}}{m_e + m_{\text{nucleus}}} \approx m_{\text{nucleus}}
 \end{equation}
with $m_{\text{nucleus}}$ provided in \cref{tab:astro_element_nuclei_mass} for various astrophysically abundant elements. 
\begin{subequations}
Without spin, wave functions $\braket{\mathbold{r}}{\Psi_e} \equiv \Psi_{n \ell m} (\mathbold{r})$ only depends on $n$, $\ell$ and $m_{\ell}$; they are solutions to the time-independent Schr\"ondiger equation, which can written in spherical coordinates as \cite{Townsend00,Shankar04,Griffiths05} 
\begin{equation}\label{eq:TI_Schrodinger_hydrogen}
 - \left( \frac{\hbar^2}{2 m_R} \laplacian + \hbar c_0 \alpha_f \frac{\mathcal{Z}}{r} \right) \Psi_{n \ell m} (r, \theta, \phi) = \mathcal{E}_{n \ell m} \Psi_{n \ell m} (r, \theta, \phi) \quad \text{with } m \equiv m_{\ell},
 \end{equation}
where the momentum operator takes on the form $\underline{\mathbold{p}} = - i \hbar \grad$ in the position basis and the eigenvalue $\mathcal{E}_{n \ell m}$ is the binding energy of the electron. 
Solutions to \cref{eq:TI_Schrodinger_hydrogen} exist in closed form, where the eigenvalues depend only on the principal quantum number, $n$ \cite{Townsend00,Shankar04,Griffiths05}:
\begin{align}\label{eq:simple_hydrogen_energy_Z}
 \begin{split}
 \mathcal{E}_{n \ell m} \to \mathcal{E}_n &= - \frac{1}{2} m_R c_0^2 \alpha^2_f \frac{\mathcal{Z}^2}{n^2} \\
 \text{with ground state} \quad \mathcal{E}_1 &= - \frac{1}{2} m_R c_0^2 \alpha^2_f \mathcal{Z}^2 \approx \left( \SI{-13.61}{\electronvolt} \right) \mathcal{Z}^2.
 \end{split}
 \end{align}
\end{subequations}

Due to the central-potential symmetry of $\underline{\mathcal{H}}_{\text{H}}$ [\emph{cf.\@} \cref{eq:hydrogen_ion_Hamiltonian}], the eigenfunctions can be split into radial and angular parts \cite{Townsend00,Bethe57,Kahn02,Shankar04}: 
%\begin{subequations} 
\begin{equation}\label{eq:separable_wave_function} 
 \Psi_{n \ell m} (r, \theta, \phi) \equiv \mathcal{R}_{n \ell}(r) \, Y_{\ell m} (\theta,\phi) \quad \text{with } m \equiv m_{\ell}. 
 \end{equation}
Defining $a_R \equiv m_e a_0 / m_R$, the radial component can be written explicitly as
\begin{subequations} 
\begin{equation}\label{eq:radial_solution_Schrodinger}
 \mathcal{R}_{n \ell}(r) = \sqrt{\left( \frac{2 \mathcal{Z}}{n a_R} \right)^3 \frac{(n - \ell -1)!}{2 n \left[ (n + \ell)! \right]^3}} \, \mathrm{e}^{- \mathcal{Z} r / n a_R} \left( \frac{2 \mathcal{Z} r}{n a_R} \right)^{\ell + 1} \mathcal{L}_{n-\ell-1}^{2 \ell + 1} \left( \frac{2 \mathcal{Z} r}{n a_R} \right) 
 \end{equation}  
with 
\begin{equation}\label{eq:radial_solution_Schrodinger_polynomial1}
 \mathcal{L}_{n-\ell-1}^{2 \ell + 1} ( X ) \equiv (-1)^{2 \ell + 1} \left( \dv{X} \right)^{2 \ell + 1} \mathcal{L}_{n-\ell-1} ( X ) ,
 \end{equation}
as an \emph{associated Laguerre polynomial} and 
\begin{equation}\label{eq:radial_solution_Schrodinger_polynomial2}
 \mathcal{L}_{n-\ell-1} ( X ) \equiv \mathrm{e}^{X} \left( \dv{X} \right)^{n-\ell-1} \left( \mathrm{e}^{-X} X^{n-\ell-1} \right) . 
 \end{equation}
\end{subequations}
as a \emph{Laguerre polynomial} of order $n-\ell-1$ \cite{Griffiths05,Shankar04}. 
%\footnote{Defining $X \equiv 2 \mathcal{Z} r / n a_R$, $\mathcal{L}_{n-\ell-1}^{2 \ell + 1} ( X )$ is an \emph{associated Laguerre polynomial}: 
%\begin{equation*}\label{eq:radial_solution_Schrodinger_polynomial1}
% \mathcal{L}_{n-\ell-1}^{2 \ell + 1} ( X ) \equiv (-1)^{2 \ell + 1} \left( \dv{X} \right)^{2 \ell + 1} \mathcal{L}_{n-\ell-1} ( X ) ,
% \end{equation*}
%where $\mathcal{L}_{n-\ell-1} ( X )$ is a \emph{Laguerre polynomial} of order $n-\ell-1$: 
%\begin{equation*}\label{eq:radial_solution_Schrodinger_polynomial2}
% \mathcal{L}_{n-\ell-1} ( X ) \equiv \mathrm{e}^{X} \left( \dv{X} \right)^{n-\ell-1} \left( \mathrm{e}^{-X} X^{n-\ell-1} \right) . 
% \end{equation*} } 
The angular component can be defined in terms of \emph{spherical harmonics}, which form a complete set of orthogonal functions on a unit sphere: 
\begin{subequations}
\begin{equation}\label{eq:angular_solution_Schrodinger}
 Y_{\ell m} (\theta,\phi) = \sqrt{\frac{2 \ell + 1}{4 \pi} \frac{(\ell - m)!}{(\ell + m)!}} \, P_{\ell}^{m} \left[ \cos (\theta) \right] \mathrm{e}^{i m \phi} \quad \text{with } m \equiv m_{\ell} ,
 \end{equation}
where 
\begin{equation}\label{eq:angular_solution_Schrodinger_polynomial1}
 P_{\ell}^{m} \left( X \right) \equiv \left( 1 - X^2 \right)^{|m|/2}  \left( \dv{X} \right)^{|m|} P_{\ell} \left( X \right) 
 \end{equation}
is an \emph{associated Legendre function} with $P_{\ell} \left( X \right)$ being a \emph{Legendre polynomial} given by the \emph{Rodrigues formula} \cite{Griffiths05,Shankar04}: 
\begin{equation}\label{eq:angular_solution_Schrodinger_polynomial2}
 P_{\ell} \left( X \right) \equiv \frac{1}{2^{\ell} \ell !} \left( \dv{X} \right)^{\ell} \left( X^2 - 1 \right)^{\ell} . 
 \end{equation}
\end{subequations}
The eigenfunctions of a hydrogen-like atom constructed under this framework thus are known exactly and can be expressed analytically. 

%\footnote{Defining $X \equiv \cos (\theta) $, $P_{\ell}^{m} \left( X \right)$ is an \emph{associated Legendre function} defined by: 
%\begin{equation*}\label{eq:angular_solution_Schrodinger_polynomial1}
% P_{\ell}^{m} \left( X \right) \equiv \left( 1 - X^2 \right)^{|m|/2}  \left( \dv{X} \right)^{|m|} P_{\ell} \left( X \right) ,
% \end{equation*}
%with $P_{\ell} \left( X \right)$ being a \emph{Legendre polynomial} given by the \emph{Rodrigues formula}: 
%\begin{equation*}\label{eq:angular_solution_Schrodinger_polynomial2}
% P_{\ell} \left( X \right) \equiv \frac{1}{2^{\ell} \ell !} \left( \dv{X} \right)^{\ell} \left( X^2 - 1 \right)^{\ell} . 
% \end{equation*} }

%which form a complete set of orthogonal functions on a unit sphere. 
%\end{subequations}

\subsection{Fine-Structure Corrections}
%%%%%%%%%%%%%%%%%%%%%%%%%%%%%%%%%%%%%%%%%--------------------------------------------------
Together, the functions $\mathcal{R}_{n \ell}(r)$ and $Y_{\ell m} (\theta,\phi)$ [\emph{cf.\@} \cref{eq:radial_solution_Schrodinger,eq:angular_solution_Schrodinger}] effectively describe the shape of an electron cloud given by the probability distribution, $\norm{\Psi_{n \ell m} (r, \theta, \phi)}^2$. 
Additionally, $\mathcal{E}_n$ given by \cref{eq:simple_hydrogen_energy_Z} can be used to arrive at approximate values for centroids of spectral lines that arise from transitioning atomic electrons. 
For example, using $\mathcal{E}_n$ to evaluate the difference in electronic binding energies:
\begin{equation}\label{eq:approx_lyman}
 \Delta \mathcal{E}_e = \mathcal{E}_2 - \mathcal{E}_1 = \frac{3}{8} m_R c_0^2 \alpha^2_f \mathcal{Z}^2 ,
 \end{equation}
shows that transitions between the K-shell and the L-shell fall comfortably within the soft x-ray range for abundant hydrogen-like ions with $6 \lessapprox \mathcal{Z} \lessapprox 14$ [\emph{cf.\@} \cref{tab:astro_element_lyman_approx}].  
\begin{table}[]
 \centering
 \caption[Nominal photon energies for Lyman-alpha transitions in abundant hydrogen-like ions]{Nominal photon energies and associated electromagnetic wavelengths for $n=1$ to $n=2$ transitions in abundant hydrogen-like ions according to non-relativistic quantum mechanics using \cref{eq:approx_lyman}.}\label{tab:astro_element_lyman_approx}
 \begin{tabular}{@{}lllll@{}} 
 \toprule
 chemical element & ion & transition energy & wavelength & spectral band \\ \midrule
 hydrogen ($\mathcal{Z}=1$) & \ion{H}{i} & \SI{10.2}{\electronvolt} & \SI{121.4}{\nano\metre} & far/vacuum UV \\
 helium ($\mathcal{Z}=2$) & \ion{He}{ii} & \SI{40.8}{\electronvolt} & \SI{30.4}{\nano\metre} & vacuum UV \\
 oxygen ($\mathcal{Z}=8$) & \ion{O}{viii} & \SI{653.4}{\electronvolt} & \SI{1.90}{\nano\metre} & soft x-ray \\
 carbon ($\mathcal{Z}=6$) & \ion{C}{vi} & \SI{367.5}{\electronvolt} & \SI{3.37}{\nano\metre} & soft x-ray \\
 neon ($\mathcal{Z}=10$) & \ion{Ne}{x} & \SI{1021.0}{\electronvolt} & \SI{1.21}{\nano\metre} & soft x-ray \\
 nitrogen ($\mathcal{Z}=7$) & \ion{N}{vii} & \SI{500.3}{\electronvolt} & \SI{2.48}{\nano\metre} & soft x-ray \\
 magnesium ($\mathcal{Z}=12$) & \ion{Mg}{xii} & \SI{1470.2}{\electronvolt} & \SI{0.843}{\nano\metre} & soft x-ray \\
 silicon ($\mathcal{Z}=14$) & \ion{Si}{xiv} & \SI{2001.1}{\electronvolt} & \SI{0.620}{\nano\metre} & soft x-ray \\
 iron ($\mathcal{Z}=26$) & \ion{Fe}{xxvi} & \SI{6901.7}{\electronvolt} & \SI{0.180}{\nano\metre} & x-ray \\
 \bottomrule
 \end{tabular}
 \end{table}
However, there are relativistic and fermionic effects of the electron that have not been accounted for with non-relativistic treatment of 
quantum mechanics but nonetheless have an effect on electronic binding energies and, therefore, spectral line centroids \cite{Bethe57,Kahn02}. 
While this is more accurately described by relativistic quantum mechanics using the \emph{Dirac equation} or techniques in \emph{quantum field theory}, perturbative corrections to $\underline{\mathcal{H}}_{\text{H}}$ can be introduced to arrive at a corrected expression for electronic binding energy \cite{Griffiths05,Townsend00}: 
\begin{subequations}
\begin{equation}\label{eq:rel_corrections}
 \left( \underline{\mathcal{H}}_{\text{H}} + \text{corrections} \right) \ket{\Psi_e} = \mathcal{E}_{\text{corrected}} \ket{\Psi_e} .
 \end{equation}
In particular, these corrections include higher-order terms to relativistic kinetic energy:
\begin{equation}\label{eq:rel_kinetic_expansion}
 \sqrt{ \underline{\mathbold{p}}^2 c_0^2 + m_e^2 c_0^4} - m_e c_0^2 = \frac{\underline{\mathbold{p}}^2}{2 m_e} - \frac{\underline{\mathbold{p}}^4}{8 m_e^3 c_0^2} + \cdots
 \end{equation}
as well as the \emph{spin-orbit interaction} \cite{Griffiths05,Shankar04}. 
This latter correction arises from the fact that in the reference frame of a bound electron with orbital angular momentum, there is a magnetic field generated by the apparent motion of the nucleus that exerts a force on the spin-magnetic moment; the perturbing Hamiltonian for this interaction is given by 
\begin{equation}\label{eq:spin-orbit}
 \underline{\mathcal{H}}_{\text{S-O}} = \frac{\mathcal{Z} q_e^2}{2 m_e^2 c_0^2 \underline{\mathbold{r}}^3} \, \underline{\mathbold{L}} \cdot \underline{\mathbold{S}} ,
 \end{equation} 
\end{subequations}
which involves the orbital and spin-angular-momentum operators, $\underline{\mathbold{L}}$ and $\underline{\mathbold{S}}$ given by \cref{eq:eq:orbit_ang_oper,eq:spin_ang_oper} \cite{Townsend00}. 

\subsection{Lyman Doublet Splitting}
%%%%%%%%%%%%%%%%%%%%%%%%%%%%%%%%%%%%%%%%%--------------------------------------------------
Taking into account the fine-structure effects described by \cref{eq:rel_corrections,eq:rel_kinetic_expansion,eq:spin-orbit}, the result is that the full expression for electronic binding energy depends on $n$ but also the quantum number for total angular momentum, $j$ \cite{Griffiths05,Townsend00}:
\begin{align}\label{eq:fine_hydrogen_energy_Z}
 \begin{split}
 \mathcal{E}_{\text{corrected}} \to \mathcal{E}_{n,j} &= - \frac{1}{2} m_R c_0^2 \alpha^2_f \frac{\mathcal{Z}^2}{n^2} \left[ 1 + \frac{\alpha_f^2}{n^2} \left( \frac{n}{j + 1/2} - \frac{3}{4} \right) \right] \\
 \text{with ground state} \quad \mathcal{E}_{1,1/2} &= - \frac{1}{2} m_R c_0^2 \alpha^2_f \left( 1 + \frac{1}{4} \alpha_f^2 \right) \mathcal{Z}^2 . 
 \end{split}
 \end{align}
With the electron-spin number being $s = 1/2$, the number $j$ can take on values $\left( \ell + 1/2 \right), \, \left( \ell - 1/2 \right), \, \left( \ell -3/2 \right), \, \ldots, \, \abs{\ell - 1/2}$: in a state with $n=2$ and $\ell = 1$, for example, $j$ can either be $3/2$ or $1/2$ and hence there are two possible transitions to the ground state with $n=1$, $\ell=0$ and $j = 1/2$. 
These can be written in \emph{spectroscopic notation} as \cite{Rybicki86}
\begin{align}\label{eq:hydrogen_doublet}
 \begin{split}
 2 p \; {}^2 \! P_{3/2} &\leftrightarrow 1 s \; {}^2 \! S_{1/2} \quad \text{(resonance)} \\ 
 2 p \; {}^2 \! P_{1/2} &\leftrightarrow 1 s \; {}^2 \! S_{1/2} \quad \text{(intercombination)}.
 \end{split}
 \end{align}
The former transition satisfies the selection rules for a pure electric-dipole transition [\emph{cf.\@} \cref{eq:hydrogen_E1_selection_rules}], where $\Delta \ell =1$ so that $j$ transitions in between $3/2$ and $1/2$. 
In contrast, the latter transition also involves $\Delta \ell =1$ but additionally, the fact that $j$ is unchanged implies that a spin flip must occur. 
With a spin violation to the usual selection rules, this means that this latter transition is semi-forbidden under the electric-dipole approximation [\emph{cf.\@} \cref{sec:allowed_transitions}]. 

These transitions described by \cref{eq:hydrogen_doublet} give rise to resonance and intercombination lines that are revealed to be closely spaced when \cref{eq:fine_hydrogen_energy_Z} is used to calculate $\Delta \mathcal{E}_e$:
\begin{subequations}
\begin{align}
 \text{resonance: } \Delta \mathcal{E}_R &\equiv \mathcal{E}_{2,3/2} - \mathcal{E}_{1,1/2} = \frac{3}{8} m_R c_0^2 \alpha^2_f \left( 1 + \frac{5}{16} \alpha_f^2 \right) \mathcal{Z}^2  \\ 
 \text{intercombination: } \Delta \mathcal{E}_I &\equiv \mathcal{E}_{2,1/2} - \mathcal{E}_{1,1/2} = \frac{3}{8} m_R c_0^2 \alpha^2_f \left( 1 + \frac{11}{48} \alpha_f^2 \right) \mathcal{Z}^2 , 
 \end{align}
where the difference between the two is
\begin{equation}
 \Delta \mathcal{E}_R - \Delta \mathcal{E}_I = \frac{1}{32} m_R c_0^2 \alpha^2_f \mathcal{Z}^2 \approx \left( \SI{0.85}{\electronvolt} \right) \mathcal{Z}^2 ,
 \end{equation}
\end{subequations}
which shows that the resonance transition has a slightly higher energy than the intercombination line. 
Whether or not this doublet of spectral lines are resolved, these \emph{Lyman-alpha} transitions between the K-shell and the L-shell are among the most prominent spectral lines for hydrogen-like ions \cite{Kahn02}. 
Similar Lyman-series transitions exist for other shells with $n > 2$ and $\ell > 1$ but only some values of $j$ obey the electric-dipole selection rules.  
In pure magnetic-dipole transitions, only $m_{\ell}$ or $m_s$ are affected while there cannot be a change in $\ell$. 
Therefore, an electron involved in such a forbidden transition is confined to a single shell so that by \cref{eq:fine_hydrogen_energy_Z}, $\Delta \mathcal{E}_e$ is much smaller than $\mathcal{E}_{\gamma}$ for soft x-rays. % and hence not directly relevant to x-ray spectroscopy.   
An example of an electric-quadrupole transition is a jump between ($n=3$, $\ell=2$) to ($n=1$, $\ell=0$) but such processes have very low transition rates and hence contribute little to measured spectra \cite{Townsend00}. 

\section{Helium-like Ions}\label{sec:selection_rules_helium}
%%%%%%%%%%%%%%%%%%%%%%%%%%%%%%%%%%%%%%%%%--------------------------------------------------
In a helium-like ion, a transitioning electron feels attraction from the central nuclear charge but additionally, it is subject to repulsive force from the second bound electron. 
While important spin-related effects come into play, a non-relativistic Hamiltonian that neglects spin can be written as \cite{Griffiths05,Townsend00,Bethe57,Kahn02} 
\begin{subequations}
\begin{equation}\label{eq:helium_ion_Hamiltonian}
 \underline{\mathcal{H}}_{\text{He}} = \frac{1}{2 m_e} \left( \underline{\mathbold{p}}_1^2 + \underline{\mathbold{p}}_2^2 \right) - \hbar c_0 \alpha_f \left( \frac{\mathcal{Z}}{ \abs{ \underline{ \mathbold{r} }_1 } } + \frac{\mathcal{Z}}{  \abs{ \underline{ \mathbold{r} }_2 } }  - \frac{1}{\abs{ \underline{\mathbold{r}}_2 -  \underline{ \mathbold{r} }_1  }} \right) ,
 \end{equation}
where subscripts$_{1,2}$ indicate the phase-space operators, $\underline{\mathbold{p}}$ and $\underline{\mathbold{r}}$, that act on the wave vectors $\ket{\Psi_{e1}}_1$ and $\ket{\Psi_{e2}}_2$ for each electron.\footnote{These wave vectors can be projected into their respective position bases:
\begin{equation*}
 \braket{\mathbold{r}_1}{\Psi_{e1}}_1 = \Psi_{e1} (\mathbold{r}_1) \quad \text{and} \quad \braket{\mathbold{r}_2}{\Psi_{e2}}_2 = \Psi_{e2} (\mathbold{r}_2) .
 \end{equation*}
With the coordinates interchanged, this is written as 
\begin{equation*}
 \braket{\mathbold{r}_1}{\Psi_{e2}}_1 = \Psi_{e2} (\mathbold{r}_1) \quad \text{and} \quad \braket{\mathbold{r}_2}{\Psi_{e1}}_2 = \Psi_{e1} (\mathbold{r}_2) .
 \end{equation*}
} 
Here, the nucleus is treated as being stationary for simplicity so that the reduced-mass term with $m_R$ [\emph{cf.\@} \cref{eq:reduced_mass}] is omitted. 
Although the Schr\"ondiger equation cannot be solved exactly for $\underline{\mathcal{H}}_{\text{He}}$, approximate solutions can be found by treating the electron-electron repulsion term as a perturbation to two hydrogen-like Hamiltonians for each electron given by \cref{eq:hydrogen_ion_Hamiltonian} with $m_R = m_e$:
\begin{equation}\label{eq:helium_ion_Hamiltonian2}
 \underline{\mathcal{H}}_{\text{He}} = \underline{\mathcal{H}}_{\text{H}1} + \underline{\mathcal{H}}_{\text{H}2} + \frac{\hbar c_0 \alpha_f}{\abs{ \underline{\mathbold{r}}_2 -  \underline{ \mathbold{r} }_1 }} .
 \end{equation}
\end{subequations}

Separately, the two electrons in a helium-like ion have quantum numbers $n_{1,2}$, $\ell_{1,2}$ and $m_{\ell \, 1,2}$ but as discussed in \cref{sec:allowed_transitions}, it is assumed that at least one of the electrons is in the ground state so that $n_1 = 1, \, \ell_1 = 1 \text{ and } m_{\ell \, 1} = 0$ while the other is left arbitrary with $n_2 \to n, \, \ell_2 \to \ell \text{ and } m_{\ell \, 2} \to m_{\ell}$. 
However, with two overlapping electron wave vectors, \emph{identical-particle} and spin effects must also be taken into account to determine how the electrons exist in superposition with a combined wave function $\Psi_e \left( \mathbold{r}_1, \mathbold{r}_2 \right)$. 
These phenomena hinge on the rules of \emph{Fermi-Dirac statistics}, which require that the overall state of the system be antisymmetric under exchange of particle coordinates \cite{Shankar04,Griffiths05}. 
This overall state includes the combined wave vector $\ket{\Psi_e}$ as well as a spinor $\ket{\chi_e} \equiv \ket{S, \, m_S}$ that describes the total spin of the system. 

\subsection{The Ground State}\label{sec:helium_ground}
%%%%%%%%%%%%%%%%%%%%%%%%%%%%%%%%%%%%%%%%%--------------------------------------------------
In the ground state of a helium-like ion with $n = 1, \, \ell = 0 \text{ and } m_{\ell} = 0$ for both $\ket{\Psi_{e1}}_1$ and $\ket{\Psi_{e2}}_2$, the \emph{Pauli exclusion principle} \cite{Shankar04,Griffiths05} dictates that the two electrons must have opposite spin such that the total spin quantum number is $S=0$ and the configuration is written in spectroscopic notation as
\begin{equation}
 1 s^2 \; {}^1 \! S_{0} .
 \end{equation}
Using the following shorthand for spin-up and spin-down states $\ket{s, \, m_s}$ of the individual electrons: 
\begin{align}
\begin{split}
  \ket{1/2, \, +1/2}_1 &\equiv \ket{\uparrow}_1 \quad \ket{1/2, \, +1/2}_2 \equiv \ket{\uparrow}_2 \\
 \ket{1/2, \, -1/2}_1 &\equiv \ket{\downarrow}_1 \quad \ket{1/2, \, -1/2}_2 \equiv \ket{\downarrow}_2 , 
 \end{split}
 \end{align}
the total spinor $\ket{\chi_e}$ is an antisymmetric \emph{singlet state} \cite{Griffiths05,Townsend00}:
\begin{subequations}
\begin{equation}
 \ket{0, \, 0} = \frac{1}{\sqrt{2}} \left[ \ket{\uparrow}_1 \otimes \ket{\downarrow}_2 - \ket{\downarrow}_1 \otimes \ket{\uparrow}_2 \right] .
 \end{equation}
This allows the wave vector to be a symmetric superposition of $\ket{\Psi_{e1}}$ and $\ket{\Psi_{e2}}$ known as \emph{parahelium}:
\begin{align}
 \begin{split}
 \ket{\Psi_e} &= \frac{1}{\sqrt{2}} \left[ \ket{\Psi_{e1}}_1 \otimes \ket{\Psi_{e2}}_2 + \ket{\Psi_{e2}}_1 \otimes \ket{\Psi_{e1}}_2 \right] \\
 \text{or} \quad \Psi_e ( \mathbold{r}_1, \mathbold{r}_2 ) &= \frac{1}{\sqrt{2}} \left[ \Psi_{e1} (\mathbold{r}_1) \, \Psi_{e2} (\mathbold{r}_2) + \Psi_{e2} (\mathbold{r}_1) \, \Psi_{e1} (\mathbold{r}_2) \right] .
 \end{split}
 \end{align} 
In the special case of the ground-state configuration, both electrons have identical quantum numbers so this is written as $\ket{1, \, 0, \, 0}_1 \otimes \ket{1, \, 0, \, 0}_2$ using $\ket{n, \, \ell, \, m_{\ell}}$ for a single-electron state. 
The overall state that combines the symmetric wave vector and the antisymmetric spinor then is written as 
\begin{equation}
 \ket{1 s^2 \; {}^1 \! S_{0}} \equiv \ket{1, \, 0, \, 0}_1 \otimes \ket{1, \, 0, \, 0}_2 \otimes \ket{0, \, 0} .
 \end{equation}
\end{subequations}

In principle, the total energy of the ground state is determined from the following expectation value:
\begin{subequations}
\begin{align}
 \begin{split}
 \mathcal{E}_{(1 s^2 \; {}^1 \! S_{0})} &\equiv \expval{ \underbrace{ \underline{\mathcal{H}}_{\text{He}} }_{ \text{ \cref{eq:helium_ion_Hamiltonian2} } } }{1 s^2 \; {}^1 \! S_{0}} \\
 &= \underbrace{ \expval{\underline{\mathcal{H}}_{\text{H}1}}{1 s^2 \; {}^1 \! S_{0}} }_{\mathcal{E}_1} + \underbrace{\expval{\underline{\mathcal{H}}_{\text{H}2}}{1 s^2 \; {}^1 \! S_{0}}}_{\mathcal{E}_1} \\ 
 &\quad \quad+ \underbrace{\expval{ \left( \frac{\hbar c_0 \alpha_f}{\abs{ \mathbold{r}_2 - \mathbold{r}_1 }} \right) }{1 s^2 \; {}^1 \! S_{0}} }_{\text{electron-electron}} ,
 \end{split}
 \end{align}
where $\mathcal{E}_1 = - \frac{1}{2} m_e c_0^2 \alpha^2_f \mathcal{Z}^2$ is the ground-state energy of a hydrogen-like ion\footnote{Neglecting spin and relativistic effects, $\mathcal{E}_1$ is given by \cref{eq:simple_hydrogen_energy_Z} with $n=1$ and $m_R = m_e$.} while the last term is the energy contributed by the electron-electron repulsion. %[chapter 12]
Following Townsend~\cite{Townsend00}, this piece of the expectation value can be evaluated in the position basis using the following matrix element:\footnote{Note that because $\underline{\mathcal{H}}_{\text{He}}$ does not depend on the spin operator, $\underline{\mathbold{S}}$, 
\begin{equation*}
 \expval{ \underline{\mathcal{H}}_{\text{He}} }{1 s^2 \; {}^1 \! S_{0}} = \bra{1, \, 0, \, 0}_1 \otimes \bra{1, \, 0, \, 0}_2 \underline{\mathcal{H}}_{\text{He}} \ket{1, \, 0, \, 0}_1 \otimes \ket{1, \, 0, \, 0}_2 \underbrace{ \braket{0, \, 0} }_1 .
 \end{equation*}} 
\begin{align}\label{eq:perturbing_helium_term}
 \begin{split}
 &\expval{ \left( \frac{\hbar c_0 \alpha_f}{\abs{ \underline{\mathbold{r}}_2 -  \underline{ \mathbold{r} }_1 }} \right) }{1 s^2 \; {}^1 \! S_{0}} \\
 &\quad \quad \quad \quad = \iint \underbrace{\norm{\braket{\mathbold{r}_1}{1, \, 0, \, 0}_1}^2}_{\norm{\Psi_e (\mathbold{r}_1)}^2} \underbrace{\norm{\braket{\mathbold{r}_2}{1, \, 0, \, 0}_2}^2}_{\norm{\Psi_e (\mathbold{r}_2)}^2}  \left( \frac{\hbar c_0 \alpha_f}{\abs{ \underline{\mathbold{r}}_2 -  \underline{ \mathbold{r} }_1 }} \right) \dd[3]{\mathbold{r}_1} \dd[3]{\mathbold{r}_2}   .
 \end{split}
 \end{align}
\end{subequations} 
For both sets of coordinates, $\mathbold{r}_1$ and $\mathbold{r}_2$, the wave function takes the form of a hydrogen-like ion as in \cref{eq:separable_wave_function,eq:radial_solution_Schrodinger,eq:angular_solution_Schrodinger}:
\begin{equation}
 \braket{\mathbold{r}}{n, \, \ell, \, m } = \Psi_e (\mathbold{r}) = \mathcal{R}_{n \ell}(r) \, Y_{\ell m} (\theta,\phi) \quad \text{with } m \equiv m_{\ell} ,
 \end{equation}
where in the case of the ground state, 
\begin{equation}
 \braket{\mathbold{r}}{1, \, 0, \, 0 } = \mathcal{R}_{1 0}(r) \, Y_{0 0} (\theta,\phi) = \frac{1}{\sqrt{\pi}} \left( \frac{\mathcal{Z}}{a_0} \right)^{3/2} \mathrm{e}^{-\frac{\mathcal{Z}}{a_0} r} .
 \end{equation}
\begin{subequations}
While the mathematics are not carried here, the integral in \cref{eq:perturbing_helium_term} comes out to \cite{Townsend00} %by Townsend~\cite{Townsend00}
\begin{equation} %\footnote{See \cite[chapter 7]{Griffiths05} and \cite[chapter 12]{Townsend00}.}
 \expval{ \left( \frac{\hbar c_0 \alpha_f}{\abs{ \underline{\mathbold{r}}_2 -  \underline{ \mathbold{r} }_1 }} \right) }{1 s^2 \; {}^1 \! S_{0}} = \frac{5}{8} m_e c_0^2 \alpha_f^2 \mathcal{Z} \approx \left( \SI{34}{\electronvolt} \right) \mathcal{Z} 
 \end{equation}
and the total energy of the ground state then is %according to this then is 
\begin{equation}
 \mathcal{E}_{(1 s^2 \; {}^1 \! S_{0})} = \underbrace{- m_e c_0^2 \alpha^2_f \mathcal{Z}^2}_{2 \mathcal{E}_1} + \frac{5}{8} m_e c_0^2 \alpha_f^2 \mathcal{Z} = m_e c_0^2 \alpha_f^2 \mathcal{Z} \left( \frac{5}{8} - \mathcal{Z} \right) .
 \end{equation}
\end{subequations} 
However, this approximation is several percentage points off from the experimentally-determined value. 
%cannot be solved exactly. 
A more accurate result for the ground state energy comes from the \emph{variational method}, where the wave function for each electron has an effective value for $\mathcal{Z}$ that is allowed to vary; this is written as $\tilde{\mathcal{Z}}$ so that the modified wave function for both sets of coordinates is \cite{Griffiths05,Townsend00}
\begin{subequations}
\begin{equation}
 \tilde{\Psi}_e (\mathbold{r}) = \frac{1}{\sqrt{\pi}} \left( \frac{\tilde{\mathcal{Z}}}{a_0} \right)^{3/2} \mathrm{e}^{-\frac{ \tilde{\mathcal{Z}}}{a_0} r} .
 \end{equation}
By expressing the Hamiltonian in \cref{eq:helium_ion_Hamiltonian} as
\begin{align}\label{eq:helium_ion_Hamiltonian_variational}
 \begin{split}
 \underline{\mathcal{H}}_{\text{He}} &= \frac{1}{2 m_e} \left( \underline{\mathbold{p}}_1^2 + \underline{\mathbold{p}}_2^2 \right) - \hbar c_0 \alpha_f \left( \frac{ \tilde{\mathcal{Z}}}{ \abs{ \underline{ \mathbold{r} }_1 } } + \frac{ \tilde{\mathcal{Z}}}{  \abs{ \underline{ \mathbold{r} }_2 } } \right)  \\
 &\quad + \hbar c_0 \alpha_f \left( \frac{ \left( \tilde{\mathcal{Z}} - \mathcal{Z} \right)}{ \abs{ \underline{ \mathbold{r} }_1 } } + \frac{ \left( \tilde{\mathcal{Z}} - \mathcal{Z} \right) }{ \abs{ \underline{ \mathbold{r} }_2 } } + \frac{1}{\abs{ \underline{\mathbold{r}}_2 -  \underline{ \mathbold{r} }_1 }} \right) 
 \end{split}
 \end{align}
and then carrying out the expectation value $\expval{\underline{\mathcal{H}}_{\text{He}}}{1 s^2 \; {}^1 \! S_{0}}$ gives a function of $\tilde{\mathcal{Z}}$ that minimizes at $\tilde{\mathcal{Z}} = \mathcal{Z} - \frac{5}{16}$ so that 
\begin{equation}
 \mathcal{E}_{(1 s^2 \; {}^1 \! S_{0})} \approx \underbrace{- m_e c_0^2 \alpha^2_f}_{\SI{-27.2}{\electronvolt}} \left( \mathcal{Z} - \frac{5}{16}  \right)^2 .
 \end{equation}
\end{subequations}
This result yields $\mathcal{E}_{(1 s^2 \; {}^1 \! S_{0})} = \SI{-77.5}{\electronvolt}$ for $\mathcal{Z} = 2$ whereas the most accepted experimentally-determined value is $\mathcal{E}_{(1 s^2 \; {}^1 \! S_{0})} = \SI{-79.0}{\electronvolt}$ \cite{Griffiths05,Townsend00}.\footnote{Note that this binding energy lays in the extreme UV spectrum while analogous binding energies for helium-like ions of low-to-mid $\mathcal{Z}$ exist in the soft x-ray spectrum.} 

\subsection{L-shell Excited States and Prominent Transitions}\label{eq:helium_Lshell}
%%%%%%%%%%%%%%%%%%%%%%%%%%%%%%%%%%%%%%%%%--------------------------------------------------
In contrast to the ground state of helium [\emph{cf.\@} \cref{sec:helium_ground}], the total electronic spin can be either $S=0$ or $S=1$ if one of the two bound electrons is in an excited state, in which case there are three possible \emph{triplet states}, $\ket{1, \, m_S}$, that give the same total spin \cite{Griffiths05,Townsend00}: 
\begin{subequations}
\begin{align}
 \begin{split}
 \ket{1, \, +1} = \ket{\uparrow}_1 \otimes \ket{\uparrow}_2, \quad \ket{1, \, -1} = \ket{\downarrow}_1 \otimes \ket{\downarrow}_2 \\
 \text{and} \quad \ket{1, \, 0} = \frac{1}{\sqrt{2}} \left( \ket{\uparrow}_1 \otimes \ket{\downarrow}_2 + \ket{\downarrow}_1 \otimes \ket{\uparrow}_2 \right) .
 \end{split}
 \end{align}
Because each of these spinors are symmetric with respect to particle exchange, the spatial part of the overall wave vector then must be antisymmetric to satisfy Fermi-Dirac statistics: 
\begin{align}
 \begin{split}
 \ket{\Psi_e} &= \frac{1}{\sqrt{2}} \left[ \ket{\Psi_{e1}}_1 \otimes \ket{\Psi_{e2}}_2 - \ket{\Psi_{e2}}_1 \otimes \ket{\Psi_{e1}}_2 \right] \\
 \text{or} \quad \Psi_e ( \mathbold{r}_1, \mathbold{r}_2 ) &= \frac{1}{\sqrt{2}} \left[ \Psi_{e1} (\mathbold{r}_1) \, \Psi_{e2} (\mathbold{r}_2) - \Psi_{e2} (\mathbold{r}_1) \, \Psi_{e1} (\mathbold{r}_2) \right] ;
 \end{split}
 \end{align}
this is referred to as \emph{orthohelium} \cite{Griffiths05,Townsend00}. 
\end{subequations}
While the ground state of a helium-like ion is necessarily parahelium, configurations where one electron is an excited state can either be parahelium or orthohelium. 

Despite the fact that spin does not appear in $\underline{\mathcal{H}}_{\text{He}}$ [\emph{cf.\@} \cref{eq:helium_ion_Hamiltonian}], parahelium and orthohelium tend to have different electronic binding energies that arise from identical-particle and spin effects that have no clear classical analog. 
In particular, with the wave vector being symmetric in parahelium, the spatial part of the system exhibits boson-like properties such that electrons tend to be closer together than they do in orthohelium, where the antisymmetric nature of the wave vector causes the opposite phenomenon \cite{Griffiths05,Townsend00}. 
Because of this, electrons in parahelium experience slightly more repulsion from each other and hence are expected to have higher energy (\emph{i.e.}, binding energy that is less negative) than electrons in orthohelium.  
This breaks degeneracy in the spin configuration and therefore it is useful to consider each possible state with $S=0$ and $S=1$ as being a contributor to measured spectral lines. 
For example, if $n=2$ and $S=0$ in an excited parahelium state, the total angular momentum is either $J=1$ or $J=0$ depending on if the excited electron is associated with $\ell=1$ or $\ell=0$. 
These two configurations can be written in spectroscopic notation as
\begin{subequations}
\begin{equation}\label{eq:he_singlets}
 1s \, 2 p \; {}^1 \! P_{1} \quad \text{and} \quad 1s \, 2 s \; {}^1 \! S_{0} ,
 \end{equation}
with the following wave vectors:
\begin{align}
 \ket{1s \, 2 p \; {}^1 \! P_{1} } &= \frac{1}{\sqrt{2}} \left[ \ket{1, \, 0, \, 0}_1 \otimes \ket{2, \, 1, \, m_{\ell}}_2 + \ket{2, \, 1, \, m_{\ell}}_1 \otimes \ket{1, \, 0, \, 0}_2 \right] \otimes \ket{0, \, 0} \\
 \ket{1s \, 2 s \; {}^1 \! S_{0} } &= \frac{1}{\sqrt{2}} \left[ \ket{1, \, 0, \, 0}_1 \otimes \ket{2, \, 0, \, 0 }_2 + \ket{2, \, 0, \, 0 }_1 \otimes \ket{1, \, 0, \, 0}_2 \right] \otimes \ket{0, \, 0} .
 \end{align}
\end{subequations}
%\begin{subequations}
For $n=2$ and $S=1$, as another example, the total angular momentum takes on values $J=2,1,0$ from the addition of $L=\ell$ and $S$. 
If $\ell = 0$, it must be that $J=1$ and the configuration is
\begin{subequations}
\begin{equation}\label{eq:he_triplet1}
 1s \, 2 s \; {}^3 \! S_{1} 
 \end{equation}
with 
\begin{equation}
 \ket{1s \, 2 s \; {}^3 \! S_{1}} = \frac{1}{\sqrt{2}} \left[ \ket{1, \, 0, \, 0}_1 \otimes \ket{2, \, 0, \, 0 }_2 - \ket{2, \, 0, \, 0 }_1 \otimes \ket{1, \, 0, \, 0}_2 \right] \otimes \ket{1, \, m_S} .
 \end{equation}
\end{subequations}
In the event that $\ell = 1$, all three possible values of $J$ are possible: 
\begin{subequations}
\begin{equation}\label{eq:he_triplet2}
 1s \, 2 s \; {}^3 \! P_{2} , \quad 1s \, 2 s \; {}^3 \! P_{1} \quad \text{and} \quad 1s \, 2 s \; {}^3 \! P_{0} ,
 \end{equation}
%\end{subequations}
with
\begin{align}
\begin{split}
 \ket{1s \, 2 s \; {}^3 \! P_{2}} &= \frac{1}{\sqrt{2}} \left[ \ket{1, \, 0, \, 0}_1 \otimes \ket{2, \, 1, \, \pm 2}_2 - \ket{2, \, 1, \, \pm 2}_1 \otimes \ket{1, \, 0, \, 0}_2 \right] \otimes \ket{1, \, \pm 2} \\
 \ket{1s \, 2 s \; {}^3 \! P_{1}} &= \frac{1}{\sqrt{2}} \left[ \ket{1, \, 0, \, 0}_1 \otimes \ket{2, \, 1, \, 0}_2 - \ket{2, \, 1, \, 0}_1 \otimes \ket{1, \, 0, \, 0}_2 \right] \otimes \ket{1, \, \pm 1} \\
 \ket{1s \, 2 s \; {}^3 \! P_{0}} &= \frac{1}{\sqrt{2}} \left[ \ket{1, \, 0, \, 0}_1 \otimes \ket{2, \, 1, \, \mp 1}_2 - \ket{2, \, 1, \, \mp 1}_1 \otimes \ket{1, \, 0, \, 0}_2 \right] \otimes \ket{1, \, \pm 1} .
 \end{split}
 \end{align}
\end{subequations}
%and $\mathcal{E}_{(1s \, 2 s \; {}^3 \! P_{0})} > \mathcal{E}_{(1s \, 2 s \; {}^3 \! P_{1})} > \mathcal{E}_{(1s \, 2 s \; {}^3 \! P_{2})}$ due to hyperfine splitting \cite{Townsend00}.
%\subsection{Prominent Transitions}\label{sec:prom_heli}
%%%%%%%%%%%%%%%%%%%%%%%%%%%%%%%%%%%%%%%%%--------------------------------------------------
%The two electrons bound in a helium-like ion can exist either in antisymmetric singlet states or symmetric triplet states, also referred to as parahelium and orthohelium, respectively [\emph{cf.\@} \cref{eq:helium_Lshell}]. 
Excited states thus can be either be singlet or triplet states, with the former giving rise to slightly higher $\Delta \mathcal{E}$ to the parahelium ground state, $1 s^2 \; {}^1 \! S_{0}$ \cite[\emph{cf.\@} \cref{sec:helium_ground}]{Griffiths05,Townsend00}. 

Assuming that one electron stays in its ground state, there are six possible L-shell excited states for the other electron. 
Two of these are singlet states, where the total spin of the system is null ($S=0$), and therefore the total angular momentum quantum number, $J$, can either be \num{0} or \num{1}. 
These two configurations [\emph{cf.\@} \cref{eq:he_singlets}] give rise to the following transitions to the ground state:
\begin{align}\label{eq:he_trans1}
 \begin{split}
 1s \, 2 p \; {}^1 \! P_{1} &\rightarrow 1 s^2 \; {}^1 \! S_{0} \quad \text{(electric dipole)} \\ 
 1s \, 2 s \; {}^1 \! S_{0} &\rightarrow 1 s^2 \; {}^1 \! S_{0} \quad \text{(two-photon)} .
 \end{split}
 \end{align}
The first transition listed in \cref{eq:he_trans1} satisfies the electric-dipole selection rules outlined in \cref{sec:allowed_transitions} and hence is a resonance line that occurs most frequently for a given ion [\emph{cf.\@} \cref{tab:astro_element_he_res}]. 
\begin{table}[]
 \centering
 \caption[Nominal photon energies for L-shell resonance transitions in abundant helium-like ions]{Nominal photon energies and associated electromagnetic wavelengths for L-shell resonance transitions in abundant helium-like ions \cite{Porquet2010}.}\label{tab:astro_element_he_res}
 \begin{tabular}{@{}lllll@{}} 
 \toprule
 chemical element & ion & transition energy & wavelength & spectral band \\ \midrule
 oxygen ($\mathcal{Z}=8$) & \ion{O}{vii} & \SI{574.0}{\electronvolt} & \SI{2.1602}{\nano\metre} & soft x-ray \\
 carbon ($\mathcal{Z}=6$) & \ion{C}{v} & \SI{307.9}{\electronvolt} & \SI{4.0267}{\nano\metre} & soft x-ray \\
 neon ($\mathcal{Z}=10$) & \ion{Ne}{ix} & \SI{922.0}{\electronvolt} & \SI{1.3447}{\nano\metre} & soft x-ray \\
 nitrogen ($\mathcal{Z}=7$) & \ion{N}{vi} & \SI{430.7}{\electronvolt} & \SI{2.8787}{\nano\metre} & soft x-ray \\
 magnesium ($\mathcal{Z}=12$) & \ion{Mg}{xi} & \SI{1352.2}{\electronvolt} & \SI{0.91688}{\nano\metre} & soft x-ray \\
 silicon ($\mathcal{Z}=14$) & \ion{Si}{xiii} & \SI{1865.0}{\electronvolt} & \SI{0.66479}{\nano\metre} & soft x-ray \\
 iron ($\mathcal{Z}=26$) & \ion{Fe}{xxv} & \SI{6700.4}{\electronvolt} & \SI{0.18504}{\nano\metre} & x-ray \\
 \bottomrule
 \end{tabular}
 \end{table}
The second, on the other hand, is a two-photon transition that necessarily occurs with a much lower probability [\emph{cf.\@} \cref{sec:quantum_perturbations}]. 
Meanwhile, the other four possible excited states are triplet states with $S=1$, including one state with null orbital angular momentum, $1s \, 2 s \; {}^3 \! S_{1}$ [\emph{cf.\@} \cref{eq:he_triplet1}], which gives rise to a magnetic-dipole transition:
\begin{equation}
 1s \, 2 s \; {}^3 \! S_{1} \rightarrow 1 s^2 \; {}^1 \! S_{0} \quad \text{(forbidden/magnetic dipole).} 
 \end{equation}
Such a transition is said to be relativistic in the sense that it is the electron spin that changes, not the orbital angular momentum. 
\begin{table}[]
 \centering
 \caption[Nominal photon energies for L-shell forbidden transitions in abundant helium-like ions]{Nominal photon energies and associated electromagnetic wavelengths for L-shell forbidden transitions in abundant helium-like ions \cite{Porquet2010}.}\label{tab:astro_element_he_forb}
 \begin{tabular}{@{}lllll@{}} 
 \toprule
 chemical element & ion & transition energy & wavelength & spectral band \\ \midrule
 %hydrogen ($\mathcal{Z}=1$) & \ion{H}{i} & \SI{10.2}{\electronvolt} & \SI{121.4}{\nano\metre} & far/vacuum UV \\
 %helium ($\mathcal{Z}=2$) & \ion{He}{ii} & \SI{40.8}{\electronvolt} & \SI{30.4}{\nano\metre} & vacuum UV \\
 oxygen ($\mathcal{Z}=8$) & \ion{O}{vii} & \SI{561.1}{\electronvolt} & \SI{2.2098}{\nano\metre} & soft x-ray \\
 carbon ($\mathcal{Z}=6$) & \ion{C}{v} & \SI{299.0}{\electronvolt} & \SI{4.1472}{\nano\metre} & soft x-ray \\
 neon ($\mathcal{Z}=10$) & \ion{Ne}{ix} & \SI{905.1}{\electronvolt} & \SI{1.3699}{\nano\metre} & soft x-ray \\
 nitrogen ($\mathcal{Z}=7$) & \ion{N}{vi} & \SI{419.8}{\electronvolt} & \SI{2.9535}{\nano\metre} & soft x-ray \\
 magnesium ($\mathcal{Z}=12$) & \ion{Mg}{xi} & \SI{1331.1}{\electronvolt} & \SI{0.93143}{\nano\metre} & soft x-ray \\
 silicon ($\mathcal{Z}=14$) & \ion{Si}{xiii} & \SI{1839.4}{\electronvolt} & \SI{0.67403}{\nano\metre} & soft x-ray \\
 iron ($\mathcal{Z}=26$) & \ion{Fe}{xxv} & \SI{6636.6}{\electronvolt} & \SI{0.18682}{\nano\metre} & x-ray \\
 \bottomrule
 \end{tabular}
 \end{table}
Because it is forbidden according the electric-dipole selection rules given by \cref{eq:helium_E1_selection_rules}, this transition occurs much less frequently than the resonance transition and hence such a spectral line appears relatively dim in the spectrum [\emph{cf.\@} \cref{tab:astro_element_he_forb}]. 

The other three triplet states [\emph{cf.\@} \cref{eq:he_triplet2}] have the same orbital-angular-momentum state but different values of $J$ that yield $\mathcal{E}_{(1s \, 2 s \; {}^3 \! P_{0})} > \mathcal{E}_{(1s \, 2 s \; {}^3 \! P_{1})} > \mathcal{E}_{(1s \, 2 s \; {}^3 \! P_{2})}$ due to fine-structure splitting [\emph{cf.\@} \cref{sec:selection_rules_hydrogen}]. 
These are associated with the following transitions to the ground state:
\begin{align}
 \begin{split}\label{eq:he_trans2}
 %\mathscr{W}: \, 1s \, 2 p \; {}^1 \! P_{1} &\rightarrow 1 s^2 \; {}^1 \! S_{0} \quad \text{(resonance/electric dipole)} \\ 
 %1s \, 2 s \; {}^1 \! S_{0} &\leftrightarrow 1 s^2 \; {}^1 \! S_{0} \quad \text{(two-photon)} \\  
 1s \, 2 s \; {}^3 \! P_{2} &\rightarrow 1 s^2 \; {}^1 \! S_{0} \quad \text{(magnetic quadrupole)} \\ 
 1s \, 2 s \; {}^3 \! P_{1} &\rightarrow 1 s^2 \; {}^1 \! S_{0} \quad \text{(semi-forbidden electric dipole)} \\ 
 %\mathscr{Z}: \, 1s \, 2 s \; {}^3 \! S_{1} &\rightarrow 1 s^2 \; {}^1 \! S_{0} \quad \text{(forbidden/magnetic dipole)} 
 1s \, 2 s \; {}^3 \! P_{0} &\rightarrow 1 s^2 \; {}^1 \! S_{0} \quad \text{(strictly forbidden)} .
 \end{split}
 \end{align}
With $\Delta J = 2$ and a change in parity, the first of the ground-state transitions listed in \cref{eq:he_trans2} is an electric-dipole forbidden, magnetic-quadrupole transition that occurs with low probability and becomes intense only for ions with $\mathcal{Z} \gtrapprox 16$ \cite{Porquet2010}. 
The transition from the $J=1$ state is an electric-dipole transition that also occurs with a spin flip and hence is semi-forbidden [\emph{cf.\@} \cref{tab:astro_element_he_interx,tab:astro_element_he_intery}]. %with a probability intermediate between that of resonance and electric dipole forbidden transitions
\begin{table}[]
 \centering
 \caption[Nominal photon energies for L-shell intercombination transitions in abundant helium-like ions]{Nominal photon energies and associated electromagnetic wavelengths for L-shell intercombination transitions in abundant helium-like ions \cite{Porquet2010}.}\label{tab:astro_element_he_interx}
 \begin{tabular}{@{}lllll@{}} 
 \toprule
 chemical element & ion & transition energy & wavelength & spectral band \\ \midrule
 oxygen ($\mathcal{Z}=8$) & \ion{O}{vii} & \SI{568.7}{\electronvolt} & \SI{2.1801}{\nano\metre} & soft x-ray \\
 carbon ($\mathcal{Z}=6$) & \ion{C}{v} & \SI{304.4}{\electronvolt} & \SI{4.0728}{\nano\metre} & soft x-ray \\
 neon ($\mathcal{Z}=10$) & \ion{Ne}{ix} & \SI{915.0}{\electronvolt} & \SI{1.3550}{\nano\metre} & soft x-ray \\
 nitrogen ($\mathcal{Z}=7$) & \ion{N}{vi} & \SI{426.3}{\electronvolt} & \SI{2.9082}{\nano\metre} & soft x-ray \\
 magnesium ($\mathcal{Z}=12$) & \ion{Mg}{xi} & \SI{1343.1}{\electronvolt} & \SI{0.92282}{\nano\metre} & soft x-ray \\
 silicon ($\mathcal{Z}=14$) & \ion{Si}{xiii} & \SI{1854.7}{\electronvolt} & \SI{0.66850}{\nano\metre} & soft x-ray \\
 iron ($\mathcal{Z}=26$) & \ion{Fe}{xxv} & \SI{6682.3}{\electronvolt} & \SI{0.18554}{\nano\metre} & x-ray \\
 \bottomrule
 \end{tabular}
 \end{table}
\begin{table}[]
 \centering
 \caption[Nominal photon energies for L-shell intercombination transitions in abundant helium-like ions]{Nominal photon energies and associated electromagnetic wavelengths for L-shell intercombination transitions in abundant helium-like ions \cite{Porquet2010}.}\label{tab:astro_element_he_intery}
 \begin{tabular}{@{}lllll@{}} 
 \toprule
 chemical element & ion & transition energy & wavelength & spectral band \\ \midrule
 oxygen ($\mathcal{Z}=8$) & \ion{O}{vii} & \SI{568.6}{\electronvolt} & \SI{2.1804}{\nano\metre} & soft x-ray \\
 carbon ($\mathcal{Z}=6$) & \ion{C}{v} & \SI{304.4}{\electronvolt} & \SI{4.0730}{\nano\metre} & soft x-ray \\
 neon ($\mathcal{Z}=10$) & \ion{Ne}{ix} & \SI{914.8}{\electronvolt} & \SI{1.3553}{\nano\metre} & soft x-ray \\
 nitrogen ($\mathcal{Z}=7$) & \ion{N}{vi} & \SI{426.3}{\electronvolt} & \SI{2.9084}{\nano\metre} & soft x-ray \\
 magnesium ($\mathcal{Z}=12$) & \ion{Mg}{xi} & \SI{1343.1}{\electronvolt} & \SI{0.92312}{\nano\metre} & soft x-ray \\
 silicon ($\mathcal{Z}=14$) & \ion{Si}{xiii} & \SI{1853.8}{\electronvolt} & \SI{0.66882}{\nano\metre} & soft x-ray \\
 iron ($\mathcal{Z}=26$) & \ion{Fe}{xxv} & \SI{6667.6}{\electronvolt} & \SI{0.18595}{\nano\metre} & x-ray \\
 \bottomrule
 \end{tabular}
 \end{table}
Finally, the transition from the $J=0$ state to the ground state (also $J=0$) is \emph{strictly forbidden} because angular momentum must be carried away by the emitted photon. 
While resonance lines are of most interest for absorption spectroscopy of extended galactic halos and the intergalactic medium [\emph{cf.\@} \cref{sec:astro_plasmas}], helium-like ions play a special role in collisional plasma diagnostics, where ratios of their prominent emission lines [\emph{cf.\@} \cref{tab:astro_element_he_res,tab:astro_element_he_forb,tab:astro_element_he_interx,tab:astro_element_he_intery}] provide a way for physical parameters such as density and temperature to be measured \cite{Porquet2010,Porquet01,Porquet01b,Pradhan85,Gabriel69,Kahn02,Paerels03,Vink2011}. 

\section{Summary}\label{sec:apBsummary}
%%%%%%%%%%%%%%%%%%%%%%%%%%%%%%%%%%%%%%%%%--------------------------------------------------
A spectral line is formed from a large number of identical ions undergoing a bound-bound transition that occurs with a decaying probability dependent on the quantum-mechanical transition rate, $\Gamma$ [\emph{cf.\@} \cref{sec:atomic_interaction,sec:form_spectral}]. 
Approximating the perturbing Hamiltonian, $\underline{\mathcal{H}}^{(1)}$, using $\mathrm{e}^{\pm i \mathbold{k} \cdot \mathbold{r}} \approx 1$ yields electric-dipole transitions with large $\Gamma$, which give rise to resonance lines [\emph{cf.\@} \cref{sec:multipole}]. 
Orbital magnetic-dipole and electric-quadrupole transitions result from the approximation $\mathrm{e}^{\pm i \mathbold{k} \cdot \mathbold{r}} \approx 1 \pm i \mathbold{k} \cdot \mathbold{r}$; in either case, $\Gamma$ is proportional to the small quantity $\left( \mathbold{k} \cdot \mathbold{r} \right)^2$ and as a result, forbidden transitions\footnote{This includes higher-order electromagnetic-multipole transitions such as \emph{magnetic-quadrupole} and \emph{electric-octupole} transitions.} occur much less frequently than electric-dipole transitions. 
Additionally, a spin-magnetic-dipole piece of $\underline{\mathcal{H}}^{(1)}$ must be included manually to take into account transitions involving spin flips. 
When such a transition occurs in the absence of an electric-dipole interaction, it is equivalent to an orbital-magnetic-dipole transition in the sense that it is forbidden under the electric-dipole approximation and hence should occur with a comparatively low $\Gamma$. 
Alternatively, an electric-dipole transition can occur with a spin flip, in which case the interaction has both electric-dipole and spin magnetic-dipole attributes; the result is an intercombination line that is said to be semi-forbidden with typical values for $\Gamma$ in between those of resonance lines and forbidden lines. 
Overall, the spectral lines most relevant to sensitive absorption spectroscopy in the soft x-ray are resonance lines associated hydrogen-like and helium-like ions of astrophysically-abundant, low-to-mid $\mathcal{Z}$ elements [\emph{cf.\@} \cref{tab:astro_element_nuclei_mass}].

% !TEX root = ../McCoy-Dissertation.tex
\Appendix{On X-ray Reflection}\label{app:x-ray_materials} 
%%%%%%%%%%%%%%%%%%%%%%%%%%%%%%%%%%%%%%%%%--------------------------------------------------
Motivated in \cref{ch:introduction}, x-ray reflection gratings are a technology suitable for sensitive absorption spectroscopy provided that they perform with high diffraction efficiency and high spectral resolving power. 
Although \cref{sec:grating_fab_intro} outlines methods for the manufacture of surface reliefs for x-ray reflection gratings, what is not addressed therein is that achieving an efficient, blazed response from a grating requires a reflective overcoat that takes into account the physics of how soft x-rays interact with matter. 
The principles behind these physical phenomena give insight into which materials and film thicknesses are appropriate for these overcoats under \emph{total external reflection} and additionally, considerations for surface roughness on the grating groove facets. 
Ultimately, reviewing this physics is prudent to understanding how fabricated x-ray reflection gratings can be tested empirically for diffraction efficiency [\emph{cf.\@} \cref{sec:als_testing}] and compared to theoretical models that are based on vector theories of diffraction [\emph{cf.\@} \cref{sec:integral_method}]. 
The goal of this appendix is to provide an overview for the physics of x-ray reflection starting with how soft x-rays propagate through a medium as they interact with atomic electrons in \cref{sec:SXR_med}. 
With an \emph{index of refraction} for soft x-rays formulated from these phenomena, \cref{sec:planar_interface} treats reflection from a mirror flat and discusses how the choice of material for a reflective overcoat and the level of surface roughness produced impacts specular reflectivity; a summary is provided in \cref{sec:appC_summary}. 

\section{Soft X-rays in Materials}\label{sec:SXR_med}
%%%%%%%%%%%%%%%%%%%%%%%%%%%%%%%%%%%%%%%%%--------------------------------------------------
Electromagnetic radiation in vacuum is described classically as coupled oscillations in the electric field $\mathbold{E}(\mathbold{r},t)$ and the magnetic field $\mathbold{B}(\mathbold{r},t)$ that propagate as a wave at the speed of light, $c_0$ [\emph{cf.\@} \cref{sec:EM_waves_vac}]. 
As such an electromagnetic wave propagates inside a material on the other hand, $\mathbold{E}(\mathbold{r},t)$ and $\mathbold{B}(\mathbold{r},t)$ interact with the bound charge density, $\rho_b(\mathbold{r},t)$, and the bound current density, $\mathbfcal{J}_b(\mathbold{r},t)$, associated with the atomic electrons in a manner that varies substantially across the electromagnetic spectrum \cite{Jackson75,Landau60,Griffiths17}. 
Semi-classically, the dominant interactions can be summarized as: 
\begin{enumerate}[noitemsep]
  \item Atomic nuclei and their electrons are forced slightly in opposite directions by $\mathbold{E}(\mathbold{r},t)$, generating an \emph{electric-dipole moment} that is said to be polarized toward the direction of the positive charge. 
  \item The total orbital angular momentum and quantum-mechanical spin of atomic electrons have a small tendency to align with $\mathbold{B}(\mathbold{r},t)$, which induces a net \emph{magnetic-dipole moment} that would otherwise not be there in a non-magnetic material.
\end{enumerate} 
These induced dipole moments are described macroscopically, per unit volume by the \emph{electric polarization field}, $\mathbold{P}(\mathbold{r},t)$, and the \emph{magnetization field}, $\mathbold{M}(\mathbold{r},t)$, which are directly related to $\rho_b(\mathbold{r},t)$ and $\mathbfcal{J}_b(\mathbold{r},t)$ \cite{Jackson75,Landau60,Griffiths17}: 
\begin{subequations}
\begin{equation}\label{eq:electric_polarization_field}
 \rho_b (\mathbold{r},t) \equiv - \div \mathbold{P}(\mathbold{r},t) 
 \end{equation}
\begin{equation}\label{eq:magnetization_field}
 \mathbfcal{J}_b (\mathbold{r},t) \equiv \curl \mathbold{M}(\mathbold{r},t) . 
 \end{equation} 
\end{subequations}
Thus, as $\mathbold{P}(\mathbold{r},t)$ and $\mathbold{M}(\mathbold{r},t)$ are driven by $\mathbold{E}(\mathbold{r},t)$ and $\mathbold{B}(\mathbold{r},t)$ of an electromagnetic wave, $\rho_b(\mathbold{r},t)$ and $\mathbfcal{J}_b(\mathbold{r},t)$ act as source terms in Maxwell's equations [\emph{cf.\@} \cref{eq:Maxwell_vac_start_1,eq:Maxwell_vac_start_2,eq:Maxwell_vac_start_3,eq:Maxwell_vac_start_4}], which describe how more radiation is produced from these interactions. %in a manner similar in concept to a basic radio antenna. 
Then, the overall electromagnetic wave that is \emph{refracted} and \emph{attenuated} as it propagates in the medium with a group velocity smaller than $c_0$ is the superposition of the original wave with these radiated waves, which generally have a phase shift associated with them \cite{Jackson75,Rybicki86,Griffiths17}. 

The above scenario describes the classical, macroscopic effect that scattering from atomic electrons has on wave propagation inside a  material. 
However, it was implicitly assumed that the material could be treated as a continuous medium, or equivalently, that the wavelength of the radiation, $\lambda$, is much larger than the atomic scale, which can be taken as the Bohr radius $a_0 \approx \SI{0.05}{\nm}$ [\emph{cf.\@} \cref{sec:allowed_transitions}]. 
Although this approximation is often valid, for example, in the visible light spectrum, where $\SI{700}{\nm}\gtrapprox \lambda \gtrapprox \SI{400}{\nm}$, it breaks down for soft x-rays and other radiation with $\lambda \lessapprox a_0$  \cite{Landau60,Attwood17,Als-Nielsen11}. 
While this condition comes with its own set of consequences, which are described throughout this appendix, another aspect to consider is the highly frequency-dependent nature of scattering in materials. 
For reference, the frequency, $\omega = 2 \pi c_0 / \lambda$, of visible light is large enough to neglect the magnetic-dipole processes that may come into play for radio waves and other relatively low-frequency radiation but still much smaller than virtually all atomic resonances, which are given by $\omega_e \equiv \mathcal{E}_e / \hbar$, where $\mathcal{E}_e$ is an electronic binding energy \cite{Landau60,Jackson75}. 
That is, with timescales for magnetic-dipole processes being much longer than the timescale $\omega^{-1}$, electric-dipole interactions are the dominant contribution to the scattered radiation from bound electrons with $\omega \ll \omega_e$. 
As frequency increases into the hard x-ray spectrum however, $\omega^{-1}$ is so short that even electric-dipole processes do not have time to respond to the oscillation of the wave and as a result, scattering is characteristic of a free electron with $\omega \gg \omega_e$ \cite{Attwood17,Als-Nielsen11}. 
Importantly, soft x-rays exist in between these two limiting cases, where $\omega$ is in the neighborhood of $\omega_e$ for atomic electrons and hence a more detailed description of scattering should be considered. 

For scattered radiation to interfere constructively and contribute to the overall electromagnetic wave that propagates in a material, the scattering processes must be \emph{coherent} in the sense that $\omega$ for the driving and radiating waves are nominally identical. 
This therefore excludes inelastic processes such as \emph{Compton scattering}, where the incoming photon\footnote{Typically, this is a hard x-ray photon with substantial quantized momentum [\emph{cf.\@} \cref{ap:x-ray_intro}].} transfers momentum to an atomic electron as it ejects it and scatters as a new photon with reduced $\omega$ and increased $\lambda$ \cite{Als-Nielsen11}. 
There then are two main types of coherent scattering alluded to above that are relevant in different scenarios depending on how photon energy, $\mathcal{E}_{\gamma} \equiv \hbar \omega$, compares to $\mathcal{E}_e$ for a given atomic electron: 
\begin{enumerate}[noitemsep]
 \item \emph{Rayleigh scattering} for $\mathcal{E}_{\gamma} \ll \mathcal{E}_e$, where the electron binding force has a strong effect on the interaction (\emph{e.g.}, visible light scattering from virtually any atom). 
 \item \emph{Thomson scattering} for $\mathcal{E}_{\gamma} \gg \mathcal{E}_e$, where the electron binding force is unimportant in the interaction (\emph{e.g.}, hard x-rays scattering from electrons in low-to-mid $\mathcal{Z}$ atoms). 
 \end{enumerate} 
These coherent-scattering processes can be treated quantum-mechanically as the absorption of a photon with $\abs{\mathbold{k}} = k_0 \equiv 2 \pi / \lambda$ coupled with the emission of a new photon with $\abs{\mathbold{k}'} = k_0$ that propagates in a different, probabilistic direction. 

Rayleigh scattering can be described as a second-order perturbation of $\underline{\mathcal{H}}^{(1)} \equiv  \left( q_e / m_e \right) \underline{\mathbold{A}} ( \mathbold{r} ) \cdot \underline{\mathbold{p}}$ [\emph{cf.\@} \cref{sec:ions_EM}], where the relevant time evolution operator for an atomic electron under the \emph{self-consistent field approximation} can be written as\footnote{Here, operators are expressed in the interaction picture using \cref{eq:interaction_operator}.}
\begin{subequations}
\begin{equation}
 \underline{U}_I (t) \approx 1 - \frac{i}{\hbar} \int_0^t \underline{\mathcal{H}}^{(1)}_I (t') \dd{t'} + \left(- \frac{i}{\hbar} \right)^2 \int_0^t \underline{\mathcal{H}}^{(1)}_I (t') \dd{t'} \int_0^{t'} \underline{\mathcal{H}}^{(1)}_I (t'') \dd{t''} ,
 \end{equation}
where the first term describes first-order processes such as bound-bound transitions and \emph{photo-ionization} [\emph{cf.\@} \cref{sec:single_absorption_emission}] while the second-order term can be expanded in terms of a complete set of intermediate electron-photon states ,$\sum_{I} \ket{I} \bra{I}$, where $\ket{I}$ is any possible intermediate electron-photon state. 
Quantum-mechanically, this describes all the possible transitions that the electron can make as it interacts with photons on its way from the initial state $\ket{A}$ to the final state $\ket{B}$ \cite{Townsend00,deBergevin09}. 
The corresponding matrix element (neglecting the first term) then is 
\begin{equation}%\label{eq:second_order_evolution_term}
 \matrixel{B}{\underline{U}_I (t)}{A} = \left(- \frac{i}{\hbar} \right)^2 \int_0^t \dd{t'} \int_0^{t'} \dd{t''} \sum_{I} \matrixel{B}{\underline{\mathcal{H}}^{(1)}_I (t')}{I} \matrixel{I}{\underline{\mathcal{H}}^{(1)}_I (t'')}{A} ,
 \end{equation}
\end{subequations}
where electron changes state from $\ket{A}$ to $\ket{I}$ but reverts to its original state after the transition from $\ket{I}$ to $\ket{B}$. 
On the other hand, Thomson scattering can be described as a first-order perturbation of $\underline{\mathcal{H}}^{(2)} \equiv \left( q_e^2 / 2 m_e \right) \underline{\mathbold{A}}^2 \left( \mathbold{r} \right)$ [\emph{cf.\@} \cref{sec:ions_EM}], where the relevant time-evolution operator is
\begin{subequations}
\begin{equation}%\label{eq:first_order_evolution}
 \underline{U}_I (t) \approx 1 - \frac{i}{\hbar} \int_0^t \underline{\mathcal{H}}^{(2)}_I (t') \dd{t'}
 \end{equation}
so that the matrix element is determined from 
\begin{equation}%\label{eq:first_order_evolution}
 \matrixel{B}{\underline{U}_I (t)}{A} = - \frac{i}{\hbar} \int_0^t \bra{A} \underline{\mathcal{H}}^{(2)}_I (t') \ket{B} \dd{t'} .
 \end{equation}
\end{subequations}
However, with $\mathcal{E}_{\gamma} \sim \mathcal{E}_e$ for the inner-most (\emph{i.e.}, K-shell; see \cref{tab:shell_orbitals}) electrons in low-to-mid $\mathcal{Z}$ atoms, dispersion corrections to scattering are required for accurately describing how soft x-rays interact with materials and additionally, there is a high chance for radiation to be absorbed via photo-ionization for $\mathcal{E}_{\gamma} \gtrapprox \mathcal{E}_e$ [\emph{cf.\@} \cref{sec:single_absorption_emission}]. 
This can be gleaned from \cref{tab:first_element_binding}, where the binding energies of K-shell electrons for the first \num{15} atoms are listed in order of $\mathcal{Z}$. 
\begin{table}[]
 \centering
 \caption[Binding energies for the inner-most electrons of the first fifteen elements in order of atomic number]{Binding energies, $\mathcal{E}_e$, and associated electromagnetic wavelengths, $\lambda$, for the inner-most (\emph{i.e.}, K-shell) electrons of the first fifteen elements in order of atomic number $\mathcal{Z}$ \cite{Attwood17}. The lightest elements have K-shell electrons characteristic of high-energy ultraviolet (UV) radiation while these binding energies move into the soft x-ray starting with $\mathcal{Z}=6$.}\label{tab:first_element_binding}
 \begin{tabular}{@{}lllll@{}} 
 \toprule
 chemical element & shell & binding energy & wavelength & spectrum \\ \midrule
 hydrogen ($\mathcal{Z}=1$) & K & \SI{13.6}{\electronvolt} & \SI{91.16}{\nano\metre} & vacuum UV \\
 helium ($\mathcal{Z}=2$) & K & \SI{24.6}{\electronvolt} & \SI{50.4}{\nano\metre} & vacuum UV \\
 lithium ($\mathcal{Z}=3$) & K & \SI{54.7}{\electronvolt} & \SI{22.67}{\nano\metre} & extreme UV \\
 beryllium ($\mathcal{Z}=4$) & K & \SI{111.5}{\electronvolt} & \SI{11.12}{\nano\metre} & extreme UV \\
 boron ($\mathcal{Z}=5$) & K & \SI{188}{\electronvolt} & \SI{6.6}{\nano\metre} & extreme UV \\
 carbon ($\mathcal{Z}=6$) & K & \SI{284.2}{\electronvolt} & \SI{4.36}{\nano\metre} & soft x-ray \\
 nitrogen ($\mathcal{Z}=7$) & K & \SI{409.9}{\electronvolt} & \SI{3.03}{\nano\metre} & soft x-ray \\
 oxygen ($\mathcal{Z}=8$) & K & \SI{543.1}{\electronvolt} & \SI{2.28}{\nano\metre} & soft x-ray \\
 fluorine ($\mathcal{Z}=9$) & K & \SI{696.7}{\electronvolt} & \SI{1.78}{\nano\metre} & soft x-ray \\
 neon ($\mathcal{Z}=10$) & K & \SI{870.2}{\electronvolt} & \SI{1.42}{\nano\metre} & soft x-ray \\
 sodium ($\mathcal{Z}=11$) & K & \SI{1070.8}{\electronvolt} & \SI{1.16}{\nano\metre} & soft x-ray \\
 magnesium ($\mathcal{Z}=12$) & K & \SI{1303.0}{\electronvolt} & \SI{0.952}{\nano\metre} & soft x-ray \\
 aluminum ($\mathcal{Z}=13$) & K & \SI{1559.6}{\electronvolt} & \SI{0.795}{\nano\metre} & soft x-ray \\
 silicon ($\mathcal{Z}=14$) & K & \SI{1838.9}{\electronvolt} & \SI{0.674}{\nano\metre} & soft x-ray \\
 phosphorous ($\mathcal{Z}=15$) & K & \SI{2145.5}{\electronvolt} & \SI{0.578}{\nano\metre} & soft x-ray \\
 \bottomrule
 \end{tabular}
 \end{table}
As seen in the table, atoms with $\mathcal{Z} \geq 6$ starting with carbon have K-shell binding energies that fall in the soft x-ray range while in contrast, the outer-most electrons in atoms that partake in chemical bonds have relatively weak binding energies characteristic of UV radiation. 
Moreover, heavy atoms have K-shell electrons characteristic of hard x-rays with $\mathcal{E}_{\gamma}$ on the order of tens of \si{\kilo\electronvolt} and intermediate-shell binding energies characteristic of soft x-rays and other smaller-$\mathcal{E}_{\gamma}$ radiation. 
It is for these reasons that while hard x-rays are known for their ability to penetrate through materials of relatively low $\mathcal{Z}$ [\emph{cf.\@} \cref{sec:histroical}], soft x-rays are easily absorbed by relatively light materials such as carbon, nitrogen and oxygen that are ubiquitous in nature as well as all heavier atoms. 

Overall, the following two properties of soft x-rays in particular have a drastic effect on their scattering behavior: 
\begin{enumerate}[noitemsep]
 \item The photon energy range $\SI{250}{\electronvolt} \lessapprox \mathcal{E}_{\gamma} \lessapprox \SI{2}{\kilo\electronvolt}$ coincides with the spectrum of K-shell binding energies in low-to-mid $\mathcal{Z}$ atoms.
 \item The wavelength range $\SI{5}{\nm} \gtrapprox \lambda \gtrapprox \SI{0.5}{\nm}$ approaches the atomic scale. 
\end{enumerate}
As explained in the following subsections, this scattering behavior leads to the need for grazing-incidence angles on materials with relatively high $\mathcal{Z}$ and \si{\nm}-scale surface roughness to achieve substantial soft x-ray reflection from an optical component, such as a mirror or a reflection grating, which is represented as a boundary between vacuum and the material considered. 

\subsection{Coherent Scattering in the Born Approximation}\label{sec:scattering_atoms}
%%%%%%%%%%%%%%%%%%%%%%%%%%%%%%%%%%%%%%%%%-------------------------------------------------- 
Motivated by textbooks on x-ray physics \cite{Attwood17,Als-Nielsen11,deBergevin09}, coherent scattering is here treated semi-classically by first considering an isolated, neutral atom with $\mathcal{Z}$ electrons to behave as a collection of $\mathcal{Z}$ harmonic oscillators\footnote{It is assumed that the atomic nucleus of charge $+ \mathcal{Z} q_e$ remains stationary due to its relatively large mass; only the $\mathcal{Z}$ electrons are imagined to behave as harmonic oscillators.} that are driven by the fields of an incident electromagnetic wave with frequency $\omega$ and wave vector $\mathbold{k}$ with $\abs{\mathbold{k}} = k_0 \equiv 2 \pi / \lambda$. 
Additionally, the \emph{Born approximation} is invoked to make the assumption that radiation is sufficiently weak such that the primary electromagnetic wave alone is responsible for driving atomic electrons while the contribution from scattered radiation is ignored. %extinction length....
This driving force comes from the \emph{Lorentz force}, $\mathbold{F}(\mathbold{r},\mathbold{v},t)$, for an electron at a position $\mathbold{r}$ with charge $-q_e$, mass $m_e$ and velocity $\mathbold{v} \equiv \dv*{\mathbold{r}}{t}$ [\emph{cf.\@} \cref{eq:Lorentz_force}]. 
For atoms of relatively low $\mathcal{Z}$, the effect of the magnetic field can be neglected with $|\mathbold{v}| \ll c_0$ \cite{Attwood17,Als-Nielsen11} and the driving force at a given position $\mathbold{r}$ reduces to 
\begin{equation}
 \mathbold{F}(\mathbold{r},t) \approx -q_e \mathbold{E}_0 \mathrm{e}^{i \left( \mathbold{k} \cdot \mathbold{r} - \omega t \right)} ,
 \end{equation}
where $\mathbold{E}_0$ describes the amplitude and polarization of its electric field \cite{Attwood17,deBergevin09}. 
Stated differently, the magnitude of electron binding energies in these atoms (\emph{e.g.}, those listed in \cref{tab:first_element_binding}) are non-relativistic in the sense that they are negligible compared to the electron rest energy, $m_e c_0^2 \approx \SI{511}{\kilo\electronvolt}$. 

Assuming very small oscillation amplitudes,\footnote{Since the approximation $v=|\mathbold{v}| \ll c_0$ has been made for low-to-mid $\mathcal{Z}$ atoms, the distance that oscillating electrons transverse is $\sim v /\omega$, which is much smaller than the soft x-ray wavelength $\lambda \sim c_0 /\omega$ \cite{Landau60}. Therefore, this assumption is justified for non-relativistic electron motion.\label{footnote:non-rel_assumption}} the position of an electron can be considered to oscillate about its average position, $\Delta \mathbold{r}$, with the same time dependence as the driving electric field, along the direction defined by $\mathbold{E}_0$. 
Ignoring restoring and damping forces for the moment, this can be expressed using \emph{Newton's second law}, which states that the acceleration $\mathbold{a} \equiv \dv*{\mathbold{v}}{t}$ induced in one of these electrons is proportional to this force \cite{Landau76,Goldstein02}: 
\begin{equation}\label{eq:acceleration_electron}
 \mathbold{a} \equiv \dv[2]{\mathbold{r}}{t} = \frac{\mathbold{F}(\mathbold{r},\mathbold{v},t)}{m_e} \approx \frac{-q_e}{m_e} \mathbold{E}_0 \cos \left( \mathbold{k} \cdot \Delta \mathbold{r} - \omega t \right) . 
 \end{equation}
Due to this acceleration, such an electric charge radiates its own electromagnetic waves in a toroidal pattern characteristic of \emph{Larmor radiation} [\emph{cf.\@} \cref{sec:ions_EM}], where the magnitude of the radiated electric field as observed at a far-field distance $R \gg c_0 / \omega$, at the \emph{retarded time},\footnote{Evaluating \cref{eq:single_electron_Larmor} at the retarded time takes into account the fact that changes to the electromagnetic fields as observed in the far-field are delayed by a time $R/c_0$ due to the finite speed of light \cite{Landau60,Jackson75,Born80,Rybicki86,Griffiths17}.} $t - R/c_0$, is given by \cite{Jackson75,Rybicki86}:
\begin{equation}\label{eq:single_electron_Larmor}
 E_{\text{rad}} \left( R, t \right) = \frac{q_e \sin \left( \vartheta \right)}{4 \pi \epsilon_0 c_0^2 R} \abs{\mathbold{a} (t - R/c_0)} 
 \end{equation}
with $\epsilon_0$ as the vacuum permittivity. 
Here, $\vartheta$ is the polar angle between $\mathbold{a}$ and a direction representing an arbitrary outgoing ray indicated by a wave vector $\mathbold{k}'$. 
\begin{figure}
 \centering
 \includegraphics[scale=0.5]{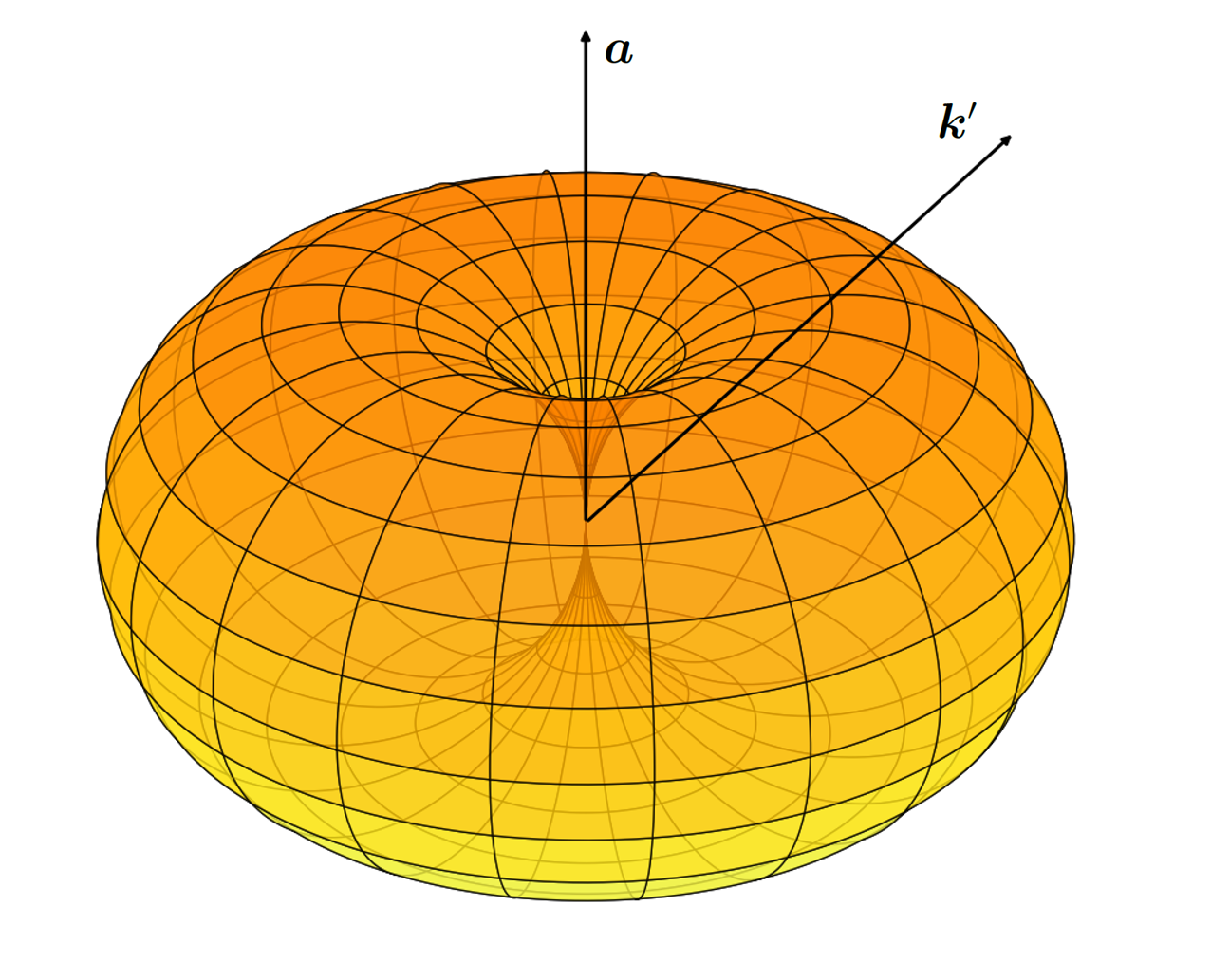}
 \caption[Characteristic torus shape of electric dipole radiation in coherent scattering]{Characteristic $\sin^2 \left( \vartheta \right)$ torus shape of Larmor radiation, where $\vartheta$ is the angle between the charge acceleration $\mathbold{a}$ and the propagation direction of scattered ray, which is represented by $\mathbold{k}'$.}\label{fig:torus_radition} 
 \end{figure} 
The toroidal shape seen in \cref{fig:torus_radition} represents the magnitude of the radiated intensity, which depends on $E^2_{\text{rad}} \left( R, t \right)$ and hence is proportional to $\sin^2 \left( \vartheta \right)$. 

The above scenario outlines the process of Thomson scattering for a single electron from a classical viewpoint: an incident electromagnetic wave of frequency $\omega$ induces the charge to oscillate, which in turn produces radiation of the same frequency in a characteristic toroidal pattern. 
From a quantum-mechanical perspective, the distribution seen in \cref{fig:torus_radition} represents the probability for a single photon to be scattered in a particular direction defined by $\mathbold{k}'$ \cite{Shankar04,deBergevin09,Attwood17,Als-Nielsen11}. 
However, because the approximation $\mathcal{E}_{\gamma} \gg \mathcal{E}_e$ is generally not valid for soft x-rays, restoring and damping forces for atomic electrons must be taken into account. 
That is, in analogy to the concept of vibrational resonance in a macroscopic object, the $j^{\text{th}}$ electron in an atom can be thought of semi-classically as having a resonance frequency, $\omega_j$, that it is equivalent to $\omega_e \equiv \mathcal{E}_{e} / \hbar$ introduced at the start of \cref{sec:SXR_med}. 
Simultaneously, the oscillation is damped as the electron radiates away its energy from a classical perspective; this is accounted for with a damping term, $\Gamma_j$.\footnote{Quantum-mechanically, $\Gamma_j$ is associated with the transition rate [\emph{cf.\@} \cref{sec:transition_rate_sub}] and hence damping of the probability for a scattering event to occur, which ultimately results in a small spread of possible energies for the scattered photon. 
That is, in a similar manner to the damping behavior exhibited in bound-bound transitions that result in naturally-broadened spectral lines [\emph{cf.\@} \cref{sec:form_spectral}], $\Gamma_j$ characterizes the width of this distribution, which can be assumed to be very small such that $\Gamma_j \ll \omega$.} 
While determining values for $\omega_j$ and $\Gamma_j$ from theory requires the use of rigorous quantum mechanics, arguments familiar in classical physics in conjunction with tabulated data are sufficient to study how coherent scattering depends on $\omega$ and the particular type of atom under question \cite{Shankar04,deBergevin09,Attwood17,Als-Nielsen11}. 

With knowledge of the range of binding frequencies, $\omega_j$, for the elements and the assumption that $\Gamma_j / \omega \ll 1$, coherent scattering that takes into account atomic resonances can be treated under the Born approximation framework discussed above \cite{Attwood17,deBergevin09}.  
However, unlike relatively long-$\lambda$ radiation such as visible light, $\lambda$ for soft x-rays is comparable to the spatial distribution of electrons in an atom as alluded to previously. 
Because of this, each of the $\mathcal{Z}$ electrons in an atom generally experiences a different phase of the incident wave, which affects how each electron oscillates about its average position, $\Delta \mathbold{r}_j$, and in turn, the overall scattering event. 
Illustrated for a single electron in \cref{fig:atomic_scattering}, this vector $\Delta \mathbold{r}_j$ originates at the location of the atomic nucleus while the oscillation is described by some function $\mathbold{X}_j (t)$, which is to be determined. 
\begin{figure}
 \centering
 \includegraphics[scale=1.25]{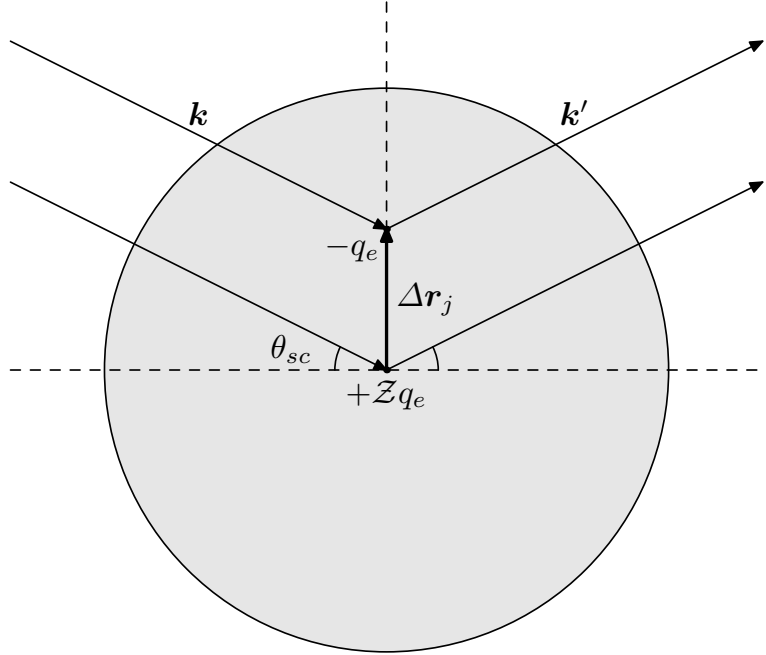}
 \caption{Coherent scattering from the $j^{\text{th}}$ electron in a neutral atom with $\mathcal{Z}$ electrons}\label{fig:atomic_scattering}
 \end{figure}
The electric field felt by the $j^{\text{th}}$ electron then is $\mathbold{E}_j(\Delta \mathbold{r}_j,t) = \mathbold{E}_{0} \mathrm{e}^{i \left( \mathbold{k} \cdot \Delta \mathbold{r}_j - \omega t \right)}$ and therefore each electron in general radiates with a different phase as observed in the far-field. 
As a generalization of \cref{eq:acceleration_electron}, equating forces gives the following differential equation that describes the position $\mathbold{X}_j (t)$ of the $j^{\text{th}}$ electron \cite{Attwood17,deBergevin09}: 
\begin{subequations}
\begin{equation}\label{eq:osc_diff_eq}
 \dv[2]{\mathbold{X}_j (t)}{t} + \Gamma_j \dv{\mathbold{X}_j (t)}{t} + \omega_j^2 \mathbold{X}_j (t) = - \frac{q_e}{m_e} \mathbold{E}_0 \mathrm{e}^{i \left( \mathbold{k} \cdot \Delta \mathbold{r}_j - \omega t \right)} .
 \end{equation}
With the assumption that $v=|\mathbold{v}| \ll c_0$ for non-relativistic atomic electrons stated at the start of \cref{sec:SXR_med}, the position of each electron is considered to oscillate about $\Delta \mathbold{r}_j$ with the same time dependence of the driving field: $\mathbold{X}_j (t) = \Delta \mathbold{X}_j \mathrm{e}^{-i \omega t}$. 
Factors of $\mathrm{e}^{-i \omega t}$ drop out when this is inserted into \cref{eq:osc_diff_eq} and solving for the oscillation amplitude $\Delta \mathbold{X}_j$ gives \cite{Attwood17,deBergevin09}:
\begin{equation}\label{eq:electron_osc_amp}
 \Delta \mathbold{X}_j \equiv \frac{\mathbold{X}_j (t)}{\mathrm{e}^{-i \omega t}} = \frac{q_e}{m_e} \frac{\mathbold{E}_0 \mathrm{e}^{i \mathbold{k} \cdot \Delta \mathbold{r}_j}}{\left( \omega^2 - \omega_j^2 + i \omega \Gamma_j \right)} . 
 \end{equation}
\end{subequations}

For a point of observation $\mathbold{r}$,\footnote{To clarify, this is an arbitrary position in the line of sight for a scattered wave (given by $\mathbold{k}'$), where the electric field radiated by the atomic electrons is supposed to be measured.} the distance from the $j^{\text{th}}$ electron is $R_j \equiv | \mathbold{r} -\Delta \mathbold{r}_j |$ and the magnitude of the radiated electric field in the far-field (\emph{i.e.}, $R_j \gg c_0 / \omega$) is
%\begin{subequations}
\begin{equation}\label{eq:single_electron_radiate}
 E_{\text{rad},j} \left( R_j, t \right) = \frac{q_e \sin \left( \vartheta \right)}{4 \pi \epsilon_0 c_0^2 R_j} \dv[2]{X_j (t - R_j/c_0)}{t} = - \frac{q_e \omega^2 \sin \left( \vartheta \right)}{4 \pi \epsilon_0 c_0^2} \frac{\abs{\Delta \mathbold{X}_j} \mathrm{e}^{i \omega \left( \frac{R_j}{c_0} - t \right)}}{R_j} ,
 \end{equation}
where $X_j (t) \equiv \abs{\mathbold{X}_j (t)}$ and $\vartheta$ is the angle between the direction of charge oscillation along $\mathbold{E}_0$ and the direction of the outgoing scattered wave denoted by $\mathbold{k}'$ [\emph{cf.\@} \cref{fig:atomic_radiating}]. 
\begin{figure}
 \centering
 \includegraphics[scale=1.25]{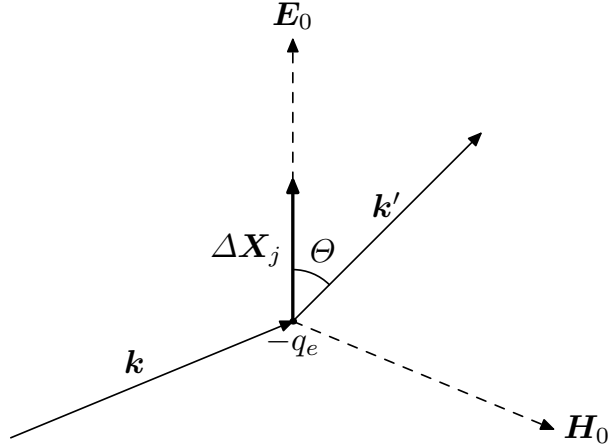}
 \caption[Scattering geometry for the $j^{\text{th}}$ electron in an atom]{Scattering geometry for the $j^{\text{th}}$ electron in an atom that oscillates along a direction defined by $\Delta \mathbold{X}_j$, which is parallel to the electric field of the incident electromagnetic wave with vector amplitude $\mathbold{E}_0$ (and perpendicular to the magnetic field with vector amplitude $\mathbold{H}_0$, as in \cref{fig:wavefront}). By \cref{eq:single_electron_radiate}, the magnitude of the radiated electric field (at a far-field distance from the electron $R_j$) depends on the angle $\vartheta$.}\label{fig:atomic_radiating}
 \end{figure} 
The magnitude of the radiated electric field at a far-field position $\mathbold{r}$ for an atom with $\mathcal{Z}$ electrons then is
\begin{align}
 \begin{split} \label{eq:charge_radiation1}
 E_{\text{rad}} \left( \mathbold{r} , t \right) &= \sum_{j=1}^{\mathcal{Z}} E_{\text{rad},j} \left( R_j, t \right) = - \frac{q_e \omega^2 \sin \left( \vartheta \right)}{4 \pi \epsilon_0 c_0^2} \mathrm{e}^{-i \omega t} \sum_{j=1}^{\mathcal{Z}} \frac{|\Delta \mathbold{X}_j|}{R_j} \mathrm{e}^{i k_0 R_j} \\
 &= - \abs{\mathbold{E}_0}  \mathrm{e}^{-i \omega t} r_e \sin \left( \vartheta \right) \sum_{j=1}^{\mathcal{Z}} \frac{\omega^2 \mathrm{e}^{i \left( k_0 R_j + \mathbold{k} \cdot \Delta \mathbold{r}_j \right)}}{R_j \left( \omega^2 - \omega_j^2 + i \omega \Gamma_j \right)} ,
 \end{split}
 \end{align} 
where \cref{eq:electron_osc_amp} was inserted for $\abs{\Delta \mathbold{X}_j}$ and $r_e \equiv q_e^2 / 4 \pi \epsilon_0 m_e c_0^2$ is the classical electron radius. 
For $\mathbold{r}$ with $\abs{\mathbold{r}} \equiv r \gg \abs{\Delta \mathbold{r}_j}$, the following approximations for $R_j \equiv | \mathbold{r} -\Delta \mathbold{r}_j |$ can be made: 
\begin{subequations}
\begin{equation}
 R_j^2 \approx r^2 - 2 \mathbold{r} \cdot \Delta \mathbold{r}_j \quad \text{and} \quad R_j \approx r \sqrt{\left( 1 - \frac{2 \mathbold{r} \cdot \Delta \mathbold{r}_j}{r^2} \right)} \approx r - \frac{\mathbold{r}}{r} \cdot \Delta \mathbold{r}_j 
 \end{equation}
and moreover, $\mathbold{r}/r$ is very close to the direction of the scattered wave given by $\mathbold{k}'/k_0$ such that the $R_j^{-1}$ term in \cref{eq:charge_radiation1} is roughly equal to $r^{-1}$ \cite{Attwood17}. 
Then, 
\begin{equation}
 k_0 R_j \approx k_0 \left( r - \frac{\mathbold{r}}{r} \cdot \Delta \mathbold{r}_j \right) \approx k_0 r - \mathbold{k}' \cdot \Delta \mathbold{r}_j
 \end{equation}
\end{subequations}
and the term in the exponential of \cref{eq:charge_radiation1}, $k_0 R_j + \mathbold{k} \cdot \Delta \mathbold{r}_j$, can be approximated as $k_0 r + \mathbold{Q} \cdot \Delta \mathbold{r}_j$, with $\mathbold{Q} \equiv \mathbold{k} - \mathbold{k}'$ as the \emph{scattering vector}.\footnote{It may help to point out that this vector, also called the \emph{wave-vector transfer} \cite{deBergevin09,Als-Nielsen11} is sometimes defined with the opposite sign (\emph{i.e.}, $\mathbold{k}' - \mathbold{k}$) \cite{Landau60,Sinha88,deBergevin09,Attwood17} and written using other symbols such as $\mathbold{q}$ \cite{Landau60,Sinha88,deBergevin09} or $\Delta \mathbold{k}$ \cite{Attwood17}.} 
Finally, \cref{eq:charge_radiation1} becomes 
\begin{equation}\label{eq:charge_radiation2}
 E_{\text{rad}} \left( \mathbold{r} , t \right) = - \abs{\mathbold{E}_0}  \mathrm{e}^{i \left(k_0 r -\omega t \right)} \frac{r_e}{r} \sin \left( \vartheta \right) \sum_{j=1}^{\mathcal{Z}} \frac{\omega^2 \mathrm{e}^{i \mathbold{Q} \cdot \Delta \mathbold{r}_j}}{\left( \omega^2 - \omega_j^2 + i \omega \Gamma_j \right)} 
 \end{equation}
for a point of observation in the far-field, where $r \gg \abs{\Delta \mathbold{r}_j}$. 
%\end{subequations}

At this point, \cref{eq:charge_radiation2} describes the radiation produced from scattering in a single atom, where the spacing in between electrons cannot be neglected due to the condition $\lambda \gg a_0$ not being fulfilled. 
Dividing this expression by the magnitude of the incident electric field highlights the fact that the scattered field emerges as a spherical wavefront that is modulated by $\sin \left( \vartheta \right)$ and a complex function $f^a(\mathbold{Q},\omega)$ known as the \emph{atomic scattering factor} \cite{deBergevin09,Als-Nielsen11,Attwood17}: 
\begin{subequations}
\begin{equation}\label{eq:spherical_scatter}
 \frac{E_{\text{rad}} \left( \mathbold{r} , t \right)}{\abs{\mathbold{E}_0} \mathrm{e}^{-i \omega t}} = - r_e \sin \left( \vartheta \right) \left( \frac{\mathrm{e}^{i k_0 r}}{r} \right) f^a \left( \mathbold{Q}, \omega \right)  
 \end{equation}
with\footnote{This quantity can be formulated in a few different ways. In some texts \cite{deBergevin09,Als-Nielsen11}, a function of $\mathbold{Q}$ alone, referred to as the \emph{atomic structure factor}, is defined as the Fourier transform (with respect to $\mathbold{Q}$) of the electron distribution in an atom. Then, $\omega$-dependent (complex) dispersion corrections are introduced to account for the resonant behavior of bound electrons. These terms depending on $\mathbold{Q}$ and $\omega$ separately are added together to arrive at an expression for $f^a(\mathbold{Q},\omega)$, also called the \emph{atomic form factor}.} 
\begin{equation}\label{eq:atomic_scattering_factor}
 f^a \left( \mathbold{Q}, \omega \right) \equiv \sum_{j=1}^{\mathcal{Z}} \frac{\omega^2 \mathrm{e}^{i \mathbold{Q} \cdot \Delta \mathbold{r}_j}}{\left( \omega^2 - \omega_j^2 + i \omega \Gamma_j \right)} ,
 \end{equation} 
\end{subequations}
which describes how radiation of frequency $\omega$ is scattered by a particular type of atom in a direction given by the scattering vector, $\mathbold{Q}$. 
Using the scattering angle defined in \cref{fig:atomic_scattering}, $\theta_{sc}$, the magnitude of this vector is
\begin{subequations}
\begin{equation}\label{eq:scattering_angle}
 \abs{\mathbold{Q}} \equiv \abs{\mathbold{k} - \mathbold{k}'} = 2 k_0 \sin \left( \theta_{sc} \right) = \frac{4 \pi}{\lambda} \sin \left( \theta_{sc} \right) 
 \end{equation}
and assuming $\abs{\Delta \mathbold{r}_j} \sim a_0$, the phase term $\mathbold{Q} \cdot \Delta \mathbold{r}_j$ in \cref{eq:charge_radiation2,eq:atomic_scattering_factor} has a magnitude bounded by 
\begin{equation}\label{eq:atomic_phase_inequality}
 \abs{\mathbold{Q} \cdot \Delta \mathbold{r}_j} \leq \frac{4 \pi a_0}{\lambda} \sin \left( \theta_{sc} \right) . 
 \end{equation}
Therefore, the phase difference felt by each of the $j$ electrons can be ignored when the following condition is met \cite{Attwood17}: 
\begin{equation}\label{eq:scatter_condition}
 \sin \left( \theta_{sc} \right)  \ll \frac{\lambda}{a_0}, 
 \end{equation}
\end{subequations}
which occurs for near-forward scattering with $\mathbold{Q} \approx \mathbf{0}$ such that $f^a(\mathbold{Q},\omega)$ is at a maximum for a given $\omega$, where $\mathbf{0}$ is the null vector. 
Physically, this suggests that all electrons oscillate virtually in phase due to $\mathbold{k}$ and $\mathbold{k}'$ being close to parallel. 
In contrast, $f^a(\mathbold{Q},\omega)$ decreases with increasing $\mathbold{Q}$ and these scattered waves of different phases tend to cancel each other out \cite{Als-Nielsen11}. 

For radiation with $\lambda \lessapprox a_0$, \cref{eq:scatter_condition} is only satisfied for a particular range of $\theta_{sc}$. 
Hard x-rays with large $a_0 / \lambda$, for example, only meet this condition for very small $\theta_{sc}$. 
However, as $\lambda$ becomes large compared to $a_0$ (\emph{e.g.}, the red end of the soft x-ray spectrum), this condition for $\theta_{sc}$ is loosened and as a result, scattering with all electrons in phase can occur over a slightly wider range of angles \cite{Attwood17}. %double-check this...
As $\lambda$ increases further to the point of becoming characteristic of visible light, $a_0 / \lambda$ is very small and thus coherent scattering occurs over all angles. 
For any of these cases, \cref{eq:atomic_scattering_factor} can be simplified to
\begin{equation}\label{eq:atomic_scattering_factor_reduced1}
 f^a \left( \mathbf{0}, \omega \right) \equiv f^0 \left( \omega \right) = \sum_{j=1}^{\mathcal{Z}} \frac{\omega^2}{\left( \omega^2 - \omega_j^2 + i \omega \Gamma_j \right)} , 
 \end{equation}
where the superscript$^0$ indicates that the \emph{forward-scattering approximation} with $\mathbold{Q} \to \mathbf{0}$ has been made \cite{Attwood17}. 

Equivalently, $f^0 \left( \omega \right)$ can be defined using scattering \emph{oscillator strengths} denoted by $g_s$, which can loosely be understood as the number electrons with a binding frequency $\omega_s$ with $\sum_s g_s = \mathcal{Z}$:\footnote{Quantum-mechanically, $g_s$ for a given atomic scattering event is related to the transition rate, $\Gamma$, while $\sum_s g_s = \mathcal{Z}$ follows from the \emph{Thomas-Reiche-Kuhn sum rule} \cite{Attwood17,Rybicki86}.}
\begin{subequations}
\begin{equation}\label{eq:atomic_scattering_factor_reduced2}
 f^0 \left( \omega \right) \equiv \sum_s \frac{g_s \, \omega^2}{\left( \omega^2 - \omega_s^2 + i \omega \Gamma_s \right)} .
 \end{equation} 
This quantity is directly related to the effective scattering cross-section for radiation of frequency $\omega$ \cite{Als-Nielsen11,Attwood17}:
\begin{equation}\label{eq:scattering_cross-section}
 \sigma_a \left( \omega \right) = \frac{8 \pi}{3} r_e^2 \, \norm{f^0 \left( \omega \right)}^2 = \sigma_e \, \norm{\sum_s \frac{g_s \omega^2}{\left( \omega^2 - \omega_s^2 + i \omega \Gamma_s \right)}}^2 ,
 \end{equation}
where $\sigma_e \equiv 8 \pi r_e^2 / 3$ is the \emph{Thomson cross-section} for a free electron. 
Note that for a scenario where $\omega$ is much higher than all $\omega_s$ in a particular atom, $f^0(\omega) \to \mathcal{Z}$ and $\sigma_a \left( \omega \right) \to \mathcal{Z}^2 \sigma_e$, which is characteristic of Thomson scattering for $\mathcal{Z}$ electrons. 
Mentioned previously, hard x-rays scattering from electrons in relatively light atoms can be described in this way while on the other hand, the opposite case with $\omega \ll \omega_s$ has \cref{eq:scattering_cross-section} reducing to the well-known expression for the Rayleigh scattering cross-section:
\begin{equation}
 \sigma_a \left( \omega \right) \to \sigma_e \sum_s g_s^2 \left( \frac{\omega}{\omega_s} \right)^{4} ,
 \end{equation}
\end{subequations}
which indicates that the amplitude of scattering by a single bound electron has a $\omega^4$ dependence \cite{Jackson75,deBergevin09,Attwood17}. 

Complex atomic scattering factors in the forward-scattering approximation, $f^0(\omega)$, are plotted as a function of photon energy, $\mathcal{E}_{\gamma} = \hbar \omega$, across the soft x-ray spectrum in \cref{fig:atomic_scattering_data,fig:atomic_scattering_data2,fig:atomic_scattering_data3} for a few light elements using data tabulated by the Center for X-ray Optics (CXRO) at Lawrence Berkeley National Laboratory \cite{CXRO_database,LBNL_web}. 
\begin{figure}
 \centering
 \includegraphics[scale=0.95]{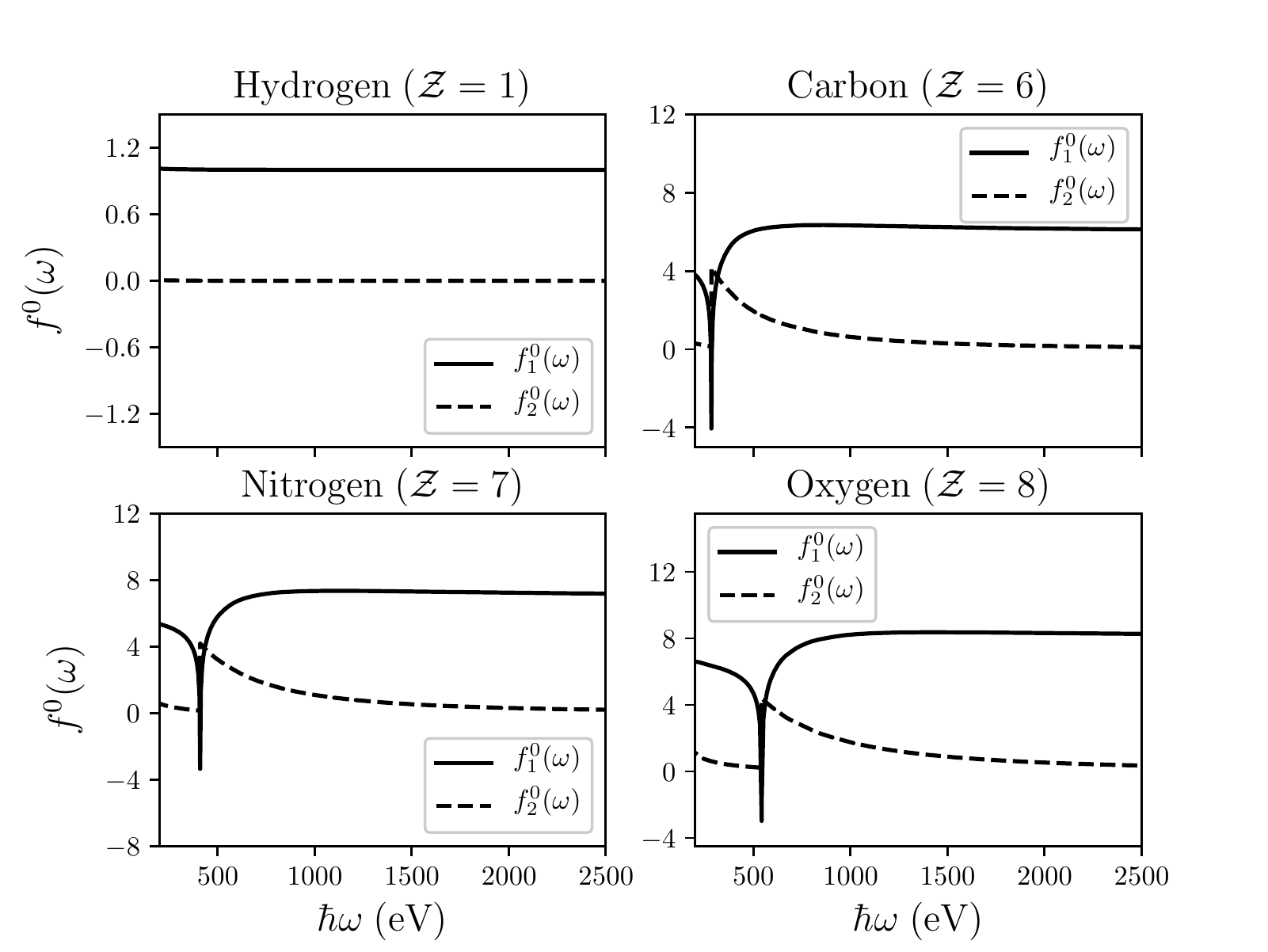}
 \caption[Atomic scattering factors for hydrogen, carbon, nitrogen and oxygen across the soft x-ray spectrum]{Atomic scattering factors for hydrogen, carbon, nitrogen and oxygen in the forward-scattering approximation across the soft x-ray spectrum \cite{CXRO_database}}\label{fig:atomic_scattering_data}
 \end{figure}
\begin{figure}
 \centering
 \includegraphics[scale=0.95]{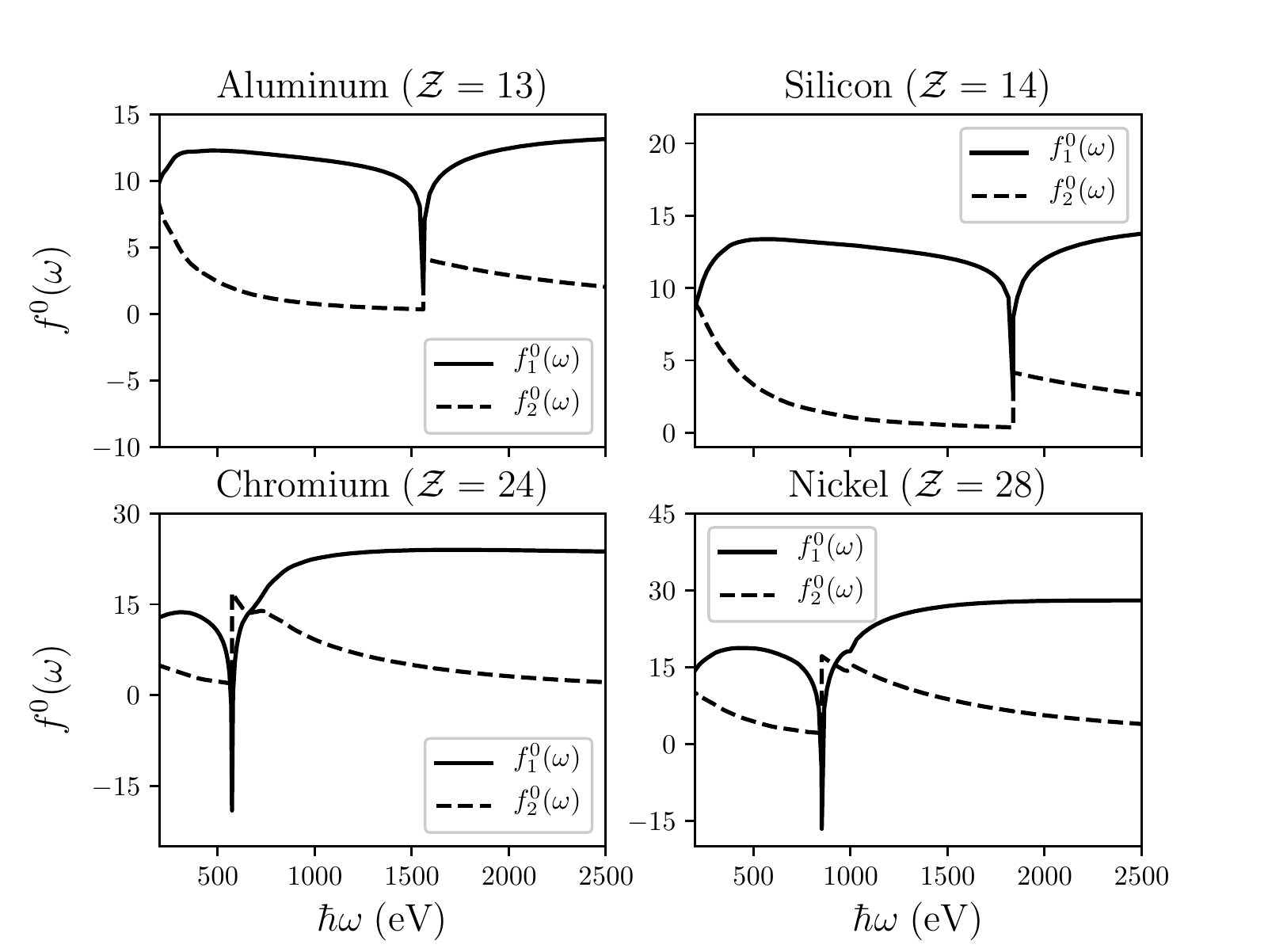}
 \caption[Atomic scattering factors for aluminum, silicon, chromium and nickel across the soft x-ray spectrum]{Atomic scattering factors for aluminum, silicon, chromium and nickel in the forward-scattering approximation across the soft x-ray spectrum \cite{CXRO_database}}\label{fig:atomic_scattering_data2}
 \end{figure}
\begin{figure}
 \centering
 \includegraphics[scale=0.95]{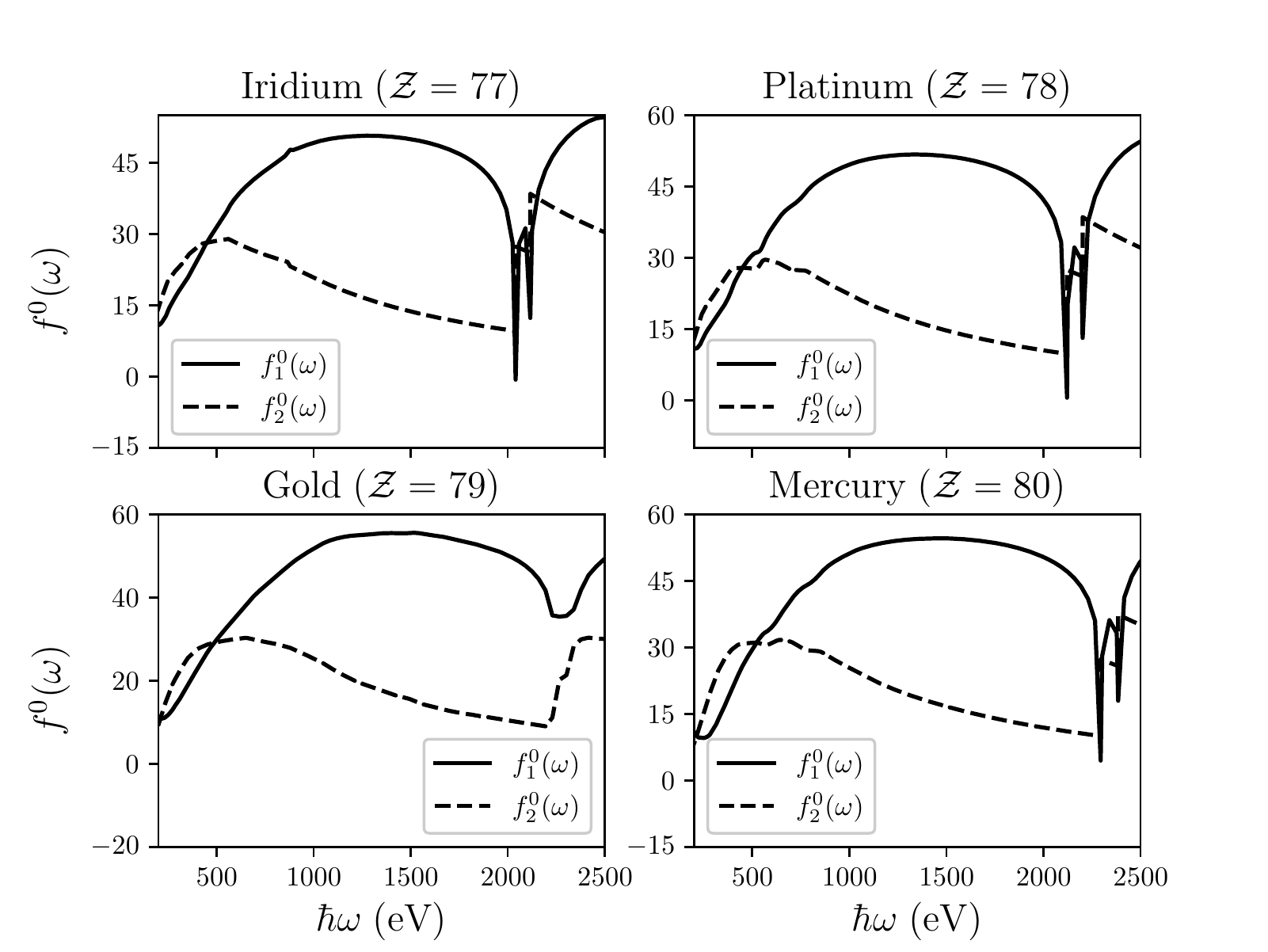}
 \caption[Atomic scattering factors for iridium, platinum, gold and mercury across the soft x-ray spectrum]{Atomic scattering factors for iridium, platinum, gold and mercury in the forward-scattering approximation across the soft x-ray spectrum \cite{CXRO_database}}\label{fig:atomic_scattering_data3}
 \end{figure}
In the plots, $f^0(\omega)$ is decomposed into real and imaginary parts: 
\begin{equation}\label{eq:atomic_scattering_factor_reduced3}
 f^0 \left( \omega \right) = \Re \left[ f^0 \left( \omega \right) \right] + i \Im \left[ f^0 \left( \omega \right) \right] \equiv f_1^0 \left( \omega \right) - i f_2^0 \left( \omega \right) ,
 \end{equation} 
where $f_1^0(\omega)$ is associated with the amplitude of the scattered wave while $f_2^0(\omega)$ is related to attenuation and absorption. 
As can be gleaned from these figures, $f_1^0(\omega)$ tends to increase with $\mathcal{Z}$, which represents the number of electrons that are radiating together in phase under the forward-scattering approximation. 
Meanwhile, $f_2^0(\omega)$ also increases with $\mathcal{Z}$, where the effect is most pronounced near prominent \emph{absorption edges} that are caused by photo-ionization. 

K-shell absorption edges are seen clearly in \cref{fig:atomic_scattering_data} for carbon, nitrogen and oxygen ($\mathcal{Z}=$ \numrange{6}{8}) whereas there is a flat response from the lone electron in hydrogen ($\mathcal{Z}=1$) that has $\mathcal{E}_e \approx \SI{13.6}{\electronvolt}$, which is much smaller than $\mathcal{E}_{\gamma} \equiv \hbar \omega$ for soft x-rays.
As a result, soft x-rays experience a high degree of absorption and attenuation in the elements with $\mathcal{Z} \geq 6$ while on the other hand, hydrogen and other very light elements are virtually transparent to soft x-rays with $f_1^0(\omega) \approx 1$ and $f_2^0(\omega)$ not departing significantly from zero, which is analogous to the behavior exhibited by hard x-rays scattering from low-to-mid $\mathcal{Z}$ atoms. 
Further, \cref{fig:atomic_scattering_data2} shows higher-energy K-shell absorption edges in aluminum and silicon ($\mathcal{Z}=$ \numlist{13;14}) while L-shell absorption lines at lower $\mathcal{E}_{\gamma}$ are seen in chromium and nickel ($\mathcal{Z}=$ \numlist{24;28}).\footnote{Note, however, that each of these four atoms tend to bond with oxygen so that in practice, these materials oxidize and hence some degree of oxygen absorption should also be expected.}
Although relativistic corrections are needed to treat scattering from high-$\mathcal{Z}$ atoms such as iridium, platinum, gold and mercury ($\mathcal{Z}=$ \numrange{77}{80}), experimentally-determined values\footnote{CXRO tabulates data for chemical elements up to uranium ($\mathcal{Z} = 92$) \cite{Henke93,Gullikson98,CXRO_database,Attwood17}. In practice, $f_2^0(\omega)$ is experimentally determined by measuring absorption while $f_1^0(\omega)$ follows from the \emph{Kramers-Kronig relations} for causal consistency \cite{Kronig26,Als-Nielsen11,Attwood17,Jackson75,Landau60}.} of $f_1^0(\omega)$ and $f_2^0(\omega)$ for these atoms are plotted in \cref{fig:atomic_scattering_data3}. 
Signatures of M-shell absorption in these heavy atoms are seen at $\mathcal{E}_{\gamma} > \SI{2}{\kilo\electronvolt}$ while N-shell absorption contributes to attenuation for $\mathcal{E}_{\gamma} \lessapprox \SI{800}{\electronvolt}$. 
Due to this broadband response free of sharp absorption edges below \SI{2}{\kilo\electronvolt}, these materials are desirable for soft x-ray reflectivity; this is discussed further in the next section. %\cref{sec:x-ray_index}. 

\subsection{Soft X-ray Index of Refraction}\label{sec:x-ray_index}
%%%%%%%%%%%%%%%%%%%%%%%%%%%%%%%%%%%%%%%%%--------------------------------------------------
Typically, the \emph{index of refraction} of a material is defined macroscopically for radiation of a given frequency $\omega$ assuming $\lambda \gg a_0$. 
For visible light, where $\omega$ is large enough to neglect contributions from magnetic-dipole interactions as discussed at the start of \cref{sec:SXR_med}, the index of refraction depends only on the electric polarization field, $\mathbold{P}(\mathbold{r},t)$ [\emph{cf.\@} \cref{eq:electric_polarization_field}], as it enters through the \emph{electric displacement field}, $\mathbold{D}(\mathbold{r},t)$, which is defined as \cite{Landau60,Jackson75,Griffiths17,Born80}: 
\begin{subequations}
\begin{equation}\label{eq:displacement_field_def}
 \mathbold{D}(\mathbold{r},t) \equiv \epsilon_0 \mathbold{E}(\mathbold{r},t) + \mathbold{P}(\mathbold{r},t) .
 \end{equation}
Because it takes a finite amount of time for atomic electric-dipole moments to respond to $\mathbold{E}(\mathbold{r},t)$, $\mathbold{D}(\mathbold{r},t)$ depends on $\mathbold{E}(\mathbold{r},t)$ not only at a time $t$ like it does in quasi-static situations,\footnote{For example, in electrostatics and in situations where $\mathbold{E}(\mathbold{r},t)$ varies slowly in time, the relation $\mathbold{D}(\mathbold{r},t) = \epsilon \mathbold{E}(\mathbold{r},t)$ (with $\epsilon$ as a constant) can often be used.} but also generally on all previous times leading up to $t$. 
This can be written as:%define linear material
\begin{equation}
 \mathbold{D}(\mathbold{r},t) = \epsilon_0 \mathbold{E}(\mathbold{r},t) + \int_0^{\infty} f_{\epsilon} (t') \, \mathbold{E}(\mathbold{r},t-t') \dd{t'} ,
 \end{equation}
where $f_{\epsilon} (t')$ is, for now, an unknown function of time that depends on the properties of a given material \cite{Landau60,Jackson75}. 
Expressing the fields as inverse Fourier transforms shows that $\mathbold{D}(\mathbold{r},t)$ and $\mathbold{E}(\mathbold{r},t)$ are related to each other through a complex, material-dependent function of frequency:
\begin{align}
 \begin{split}\label{eq:epsilon_omega_def}
 \mathbold{D}(\mathbold{r},t) &= \frac{1}{2 \pi} \int_{-\infty}^{\infty} \mathbold{D}(\mathbold{r},\omega) \, \mathrm{e}^{-i \omega t} \dd{\omega} \\
 &= \frac{\epsilon_0}{2 \pi} \int_{-\infty}^{\infty} \mathbold{E}(\mathbold{r},\omega) \, \mathrm{e}^{-i \omega t} \dd{\omega} + \frac{1}{2 \pi} \int_{-\infty}^{\infty} \int_{0}^{\infty} f_{\epsilon} (t') \, \mathrm{e}^{-i \omega (t - t')} \mathbold{E}(\mathbold{r},\omega) \dd{t'} \dd{\omega} \\
 &= \left[ \epsilon_0 + \int_{0}^{\infty} f_{\epsilon} (t') \, \mathrm{e}^{i \omega t'} \dd{t'} \right] \mathbold{E}(\mathbold{r},t) \equiv \epsilon (\omega) \, \mathbold{E}(\mathbold{r},t) . %\\ %&\quad \quad \quad
 \end{split}
 \end{align}
\end{subequations}
Using this function $\epsilon (\omega)$, which describes permittivity of the medium for a given frequency, the complex index of refraction is defined as 
\begin{equation}\label{eq:refractive_index}
 \tilde{\nu} (\omega) \equiv \sqrt{\frac{\epsilon (\omega)}{\epsilon_0}} .
 \end{equation}
However, as emphasized in the preceding discussion, the condition $\lambda \gg a_0$ is not fulfilled for soft x-rays. 
Notwithstanding this, an effective index of refraction can be formulated by using an approximate expression for $\mathbold{P}(\mathbold{r},t)$ that takes into account the density of atoms constituting a material. 

In principle, describing the behavior of soft x-rays in a medium requires solving the microscopic form of Maxwell's equations [\emph{cf.\@} \cref{eq:Maxwell_vac_start_1,eq:Maxwell_vac_start_2,eq:Maxwell_vac_start_3,eq:Maxwell_vac_start_4}] using the electric charge and current present in individual atoms as source terms \cite{Attwood17,deBergevin09}. 
On the other hand, the macroscopic version of Maxwell's equations in a dispersive medium can be expressed in a time-harmonic form similar to \cref{eq:Maxwell_vac_TH_1,eq:Maxwell_vac_TH_2,eq:Maxwell_vac_TH_3,eq:Maxwell_vac_TH_4} for electromagnetic waves in vacuum. 
In the absence of free charge and current, these can be written as:
\begin{subequations}
\begin{align} 
 \mathbold{\nabla} \cdot \mathbold{E} \left( \mathbold{r} \right) &= 0 \label{eq:Maxwell_med_TH_1} \\
 \mathbold{\nabla} \cdot \mathbold{H} \left( \mathbold{r} \right) &= 0 \label{eq:Maxwell_med_TH_2} \\
 \mathbold{\nabla} \times \mathbold{E} \left( \mathbold{r} \right) &= i \omega \mu_0 \mathbold{H} \left( \mathbold{r} \right) \label{eq:Maxwell_med_TH_3} \\
 \mathbold{\nabla} \times \mathbold{H} \left( \mathbold{r} \right) &= -i \omega \epsilon (\omega) \mathbold{E} \left( \mathbold{r} \right) \label{eq:Maxwell_med_TH_4} ,
 \end{align}
\end{subequations}
with $\mu_0$ as the vacuum permeability. 
These equations can be rearranged\footnote{That is, by taking curls of \cref{eq:Maxwell_med_TH_3,eq:Maxwell_med_TH_4}: 
\begin{align*}
 \begin{split}
 \mathbold{\nabla} \times \mathbold{\nabla} \times \mathbold{E}(\mathbold{r}) &= \mathbold{\nabla}\left[\mathbold{\nabla} \cdot \mathbold{E}(\mathbold{r})\right] - \laplacian \mathbold{E}(\mathbold{r}) = i \omega \mu_0 \mathbold{\nabla} \times \mathbold{H} (\mathbold{r}) = \mu_0 \epsilon (\omega) \omega^2 \mathbold{E} (\mathbold{r}) \\
 \mathbold{\nabla} \times \mathbold{\nabla} \times \mathbold{H}(\mathbold{r}) &= \mathbold{\nabla}\left[\mathbold{\nabla} \cdot \mathbold{H}(\mathbold{r})\right] - \laplacian \mathbold{H}(\mathbold{r}) = - i \epsilon (\omega) \mathbold{\nabla} \times \mathbold{E} (\mathbold{r}) = \mu_0 \epsilon (\omega) \omega^2 \mathbold{H} (\mathbold{r})
 \end{split}
 \end{align*}
and inserting \cref{eq:Maxwell_med_TH_1,eq:Maxwell_med_TH_2}.} to give the following expression: 
\begin{equation} \label{eq:Helmholtz_med}
 \left( \laplacian + \mu_0 \epsilon (\omega) \omega^2 \right) \left\{
  \begin{array}{lr}
  \mathbold{E}(\mathbold{r})\\
  \mathbold{H}(\mathbold{r})
  \end{array}
 \right\} = \mathbf{0} ,
 \end{equation}
which is recognized as the \emph{Helmholtz equation} for the medium \cite{Landau60,Jackson75}. 
Similar to what is done in \cref{sec:EM_waves_vac} for electromagnetic waves in vacuum, solutions to \cref{eq:Helmholtz_med} can be studied by expressing the time-harmonic fields as a superposition of spatial wave modes using inverse Fourier transforms:
\begin{equation} \label{eq:fourier_fields_complex}
 \left\{
  \begin{array}{lr}
  \mathbold{E} \left( \mathbold{r} \right) \\
  \mathbold{H} \left( \mathbold{r} \right)
  \end{array}
 \right\} = \frac{1}{(2 \pi)^3} \int_{-\mathbold{\infty}}^{\mathbold{\infty}} \left\{
  \begin{array}{lr}
  \mathbold{E} \left( \mathbold{\tilde{k}} \right) \\
  \mathbold{H} \left( \mathbold{\tilde{k}} \right)
  \end{array}
 \right\} \mathrm{e}^{i \mathbold{\tilde{k}} \cdot \mathbold{r}} \dd[3]{\mathbold{\tilde{k}}},
 \end{equation}
where $\mathbold{\tilde{k}}$ is a wave vector that characterizes the direction of wave propagation and the spatial frequency of the electromagnetic fields.

Because $\epsilon(\omega)$ is generally complex, $\mathbold{\tilde{k}}$ is considered to be a complex wave vector:
%\begin{subequations}
\begin{equation}
 \mathbold{\tilde{k}} = \mathbold{k} + i \mathbold{\kappa} , \label{eq:k_tilde_def}
 \end{equation}
where $\mathbold{k}$ and $\mathbold{\kappa}$ are real vectors denoting the direction and magnitude of wave propagation and wave attenuation, respectively. 
Inserting \cref{eq:fourier_fields_complex} into \cref{eq:Helmholtz_med} gives 
\begin{equation}
 \laplacian \left\{
  \begin{array}{lr}
  \mathbold{E} \left( \mathbold{r} \right) \\
  \mathbold{H} \left( \mathbold{r} \right)
  \end{array}
 \right\} = \frac{1}{(2 \pi)^3} \laplacian \int_{-\mathbold{\infty}}^{\mathbold{\infty}} \left\{
  \begin{array}{lr}
  \mathbold{E} \left( \mathbold{\tilde{k}} \right) \\
  \mathbold{H} \left( \mathbold{\tilde{k}} \right)
  \end{array}
 \right\} \mathrm{e}^{i \mathbold{\tilde{k}} \cdot \mathbold{r}} \dd[3]{\mathbold{\tilde{k}}}
  = - \tilde{k}^2 \left\{
  \begin{array}{lr}
  \mathbold{E} \left( \mathbold{r} \right) \\
  \mathbold{H} \left( \mathbold{r} \right)
  \end{array}
 \right\} 
 \end{equation}
with
\begin{equation}\label{eq:disp_rel_med}
 \tilde{k}^2 \equiv \mathbold{\tilde{k}} \cdot \mathbold{\tilde{k}} = \mathbold{k} \cdot \mathbold{k} + 2 i \mathbold{k} \cdot \mathbold{\kappa} - \mathbold{\kappa} \cdot \mathbold{\kappa} = \mu_0 \, \epsilon (\omega) \, \omega^2 
 \end{equation}
%\end{subequations}
as the \emph{dispersion relation in the material}. 
Dividing this expression by $k_0^2 = \omega^2 \epsilon_0 \mu_0$ [\emph{cf.\@} \cref{eq:disp_rel_vac}] shows that $\tilde{k} = k_0 \tilde{\nu} (\omega)$  while inserting \cref{eq:fourier_fields_complex} into \cref{eq:Maxwell_med_TH_1,eq:Maxwell_med_TH_2,eq:Maxwell_med_TH_3,eq:Maxwell_med_TH_4} yields the following expressions:
\begin{subequations}
\begin{align} %eq:Maxwell_med_TH 1-4
 \left( \mathbold{k} + i \mathbold{\kappa} \right) \cdot \mathbold{E} \left( \mathbold{r} \right) &= 0 \label{eq:Maxwell_med_FT_1} \\
 \left( \mathbold{k} + i \mathbold{\kappa} \right) \cdot \mathbold{H} \left( \mathbold{r} \right) &= 0 \label{eq:Maxwell_med_FT_2} \\
 \left( \mathbold{k} + i \mathbold{\kappa} \right) \times \mathbold{E} \left( \mathbold{r} \right) &= Z_0 k_0 \mathbold{H} \left( \mathbold{r} \right) \label{eq:Maxwell_med_FT_3} \\
 \left( \mathbold{k} + i \mathbold{\kappa} \right) \times \mathbold{H} \left( \mathbold{r} \right) &= - \tilde{\nu}^2 (\omega) \frac{k_0}{Z_0} \mathbold{E} \left( \mathbold{r} \right) \label{eq:Maxwell_med_FT_4} ,
 \end{align}
\end{subequations}
with $Z_0$ as the impedance of free space. 
These equations indicate that, in contrast to the behavior of electromagnetic radiation in vacuum, the intensity of a wave drops off as $\mathrm{e}^{-2 \mathbold{\kappa} \cdot \mathbold{r}}$ as it propagates through a material and moreover, waves are only transverse in situations where $\mathbold{k}$ and $\mathbold{\kappa}$ are aligned; otherwise, wavefronts are distorted as they propagate and decay in different directions. 

Now, with $\lambda \gg a_0$ not fulfilled for soft x-rays, formulating $\tilde{\nu} (\omega)$ starts with considering the classical treatment of coherent scattering outlined in \cref{sec:scattering_atoms}, where the $j^{\text{th}}$ electron in an atom responds to an incident electromagnetic wave by oscillating about its average position, $\Delta \mathbold{r}_j$ [\emph{cf.\@} \cref{eq:osc_diff_eq,eq:electron_osc_amp}]. 
As this electrons oscillates along the direction of $\mathbold{E}_0$ with the function $\mathbold{X}_j (t) = \Delta \mathbold{X}_j \mathrm{e}^{-i \omega t}$ given by 
\begin{subequations}
\begin{equation}\label{eq:electron_osc}
 \mathbold{X}_j (t) = \frac{q_e}{m_e} \frac{\mathbold{E}_0 \mathrm{e}^{i \left( \mathbold{k} \cdot \Delta \mathbold{r}_j - \omega t \right)}}{\left( \omega^2 - \omega_j^2 + i \omega \Gamma_j \right)} ,
 \end{equation}
it radiates a wave of frequency $\omega$ according to \cref{eq:single_electron_radiate}. 
The most significant contributors to the modification of the incident wave, however, are forward-scattering events, where the phase term can be neglected [\emph{cf.\@} \cref{sec:scattering_atoms}]. 
In this case, each of the $\mathcal{Z}$ electrons in an atom oscillates in phase: 
\begin{equation}\label{eq:atomic_osc}
 \mathbold{X}_a (t) \equiv \sum_{j=1}^{\mathcal{Z}} \mathbold{X}_j (t) \approx \frac{q_e}{m_e} \mathbold{E}_0 \mathrm{e}^{-i \omega t} \underbrace{\sum_{s} \frac{g_s}{\left( \omega^2 - \omega_j^2 + i \omega \Gamma_j \right)}}_{f^0 (\omega) / \omega^2} ,
 \end{equation}
where $f^0 (\omega)$ is the atomic scattering factor in the forward-scattering approximation [\emph{cf.\@} \cref{eq:atomic_scattering_factor_reduced2}]. 
Therefore, the scattered intensity is at its maximum in this direction for a given $\omega$ as the atom behaves as a single, oscillating electric dipole:
\begin{equation}\label{eq:atomic_dipole_moment}
 \mathbold{p}_a (t) \equiv - q_e \mathbold{X}_a (t) .
 \end{equation}
\end{subequations}

For a large collection of atoms constituting a solid material with an average density of $\mathcal{N}_a$ atoms per unit volume, each atomic dipole is radiating its own wavefront but due to the relative positions of these atoms, partial destructive interference occurs unless the scattering angle illustrated in \cref{fig:atomic_scattering}, $\theta_{sc}$, is close to zero \cite{Attwood17}. 
In analogy with \cref{eq:atomic_phase_inequality}, this condition for $\theta_{sc}$ is approximately 
\begin{equation}
 \sin \left( \theta \right) \ll \frac{\lambda}{a} ,
 \end{equation}
where $a \equiv \mathcal{N}_a^{-1/3}$ is a typical spacing between atoms, which is usually several times larger than the Bohr radius, $a_0 \approx \SI{0.05}{\nm}$. 
Considering only scattered waves with $\theta_{sc} \to 0$ and a quasi-uniform medium with a near constant $\mathcal{N}_a$, \cref{eq:atomic_dipole_moment} can be used to express the electric polarization field as: 
\begin{equation}\label{eq:electric_polarization_model}
 \mathbold{P}(\mathbold{r},t) = - \mathcal{N}_a \frac{q_e^2}{m_e} \frac{f^0 (\omega)}{\omega^2} \mathbold{E} (\mathbold{r}, t) ,
 \end{equation}
where values for $\mathcal{N}_a$ for PMMA [\emph{cf.\@} \cref{sec:binary_ebeam}], silica (\ce{SiO2}), nickel and gold are listed in \cref{tab:element_densities}. 
\begin{table}[]
 \centering
 \caption[Mass density, atomic number density and typical atomic spacing for PMMA, silica, nickel and gold.]{Mass density, atomic number density and typical atomic spacing for PMMA (\ce{C5H8O2})$_n$, silica (\ce{SiO2}), nickel and gold \cite{CXRO_database}.}\label{tab:element_densities} 
 \begin{tabular}{@{}lllll@{}}
 \toprule  %calculated using CXRO value for g/cm^3 density, divided by the appropriate atomic mass (averages used for PMMA and silica)
 material & mass density & number density ($\mathcal{N}_a$) & typical atomic spacing \\ \midrule
 PMMA & \SI{1.19}{\gram\per\centi\metre\tothe{3}} & \num{1.08e29} atoms \si{\per\metre\tothe{3}} & $\mathcal{N}_a^{-1/3} \approx \SI{0.21}{\nm}$ \\ 
 silica & \SI{2.20}{\gram\per\centi\metre\tothe{3}} & \num{6.62e28} atoms \si{\per\metre\tothe{3}} & $\mathcal{N}_a^{-1/3} \approx \SI{0.25}{\nm}$ \\ 
 nickel & \SI{8.90}{\gram\per\centi\metre\tothe{3}} & \num{9.14e28} atoms \si{\per\metre\tothe{3}} & $\mathcal{N}_a^{-1/3} \approx \SI{0.22}{\nm} $\\ 
 gold & \SI{19.32}{\gram\per\centi\metre\tothe{3}} & \num{5.90e28} atoms \si{\per\metre\tothe{3}} & $\mathcal{N}_a^{-1/3} \approx \SI{0.27}{\nm}$ \\  \bottomrule
 \end{tabular}
 \end{table}
Then, using $\mathbold{P}(\mathbold{r},t) = \left[ \epsilon (\omega) - \epsilon_0  \right] \mathbold{E}(\mathbold{r},t)$ [\emph{cf.\@} \cref{eq:displacement_field_def,eq:epsilon_omega_def}] and $k_0^2 = \omega^2 / c_0^2 = (2 \pi / \lambda)^2$ [\emph{cf.\@} \cref{eq:disp_rel_vac}] yields the following expression for $\epsilon (\omega)$: % from \cref{eq:electric_polarization_model}:
\begin{equation}\label{eq:epsilon_omega_model}
 \epsilon (\omega)  = \epsilon_0 - \frac{\mathcal{N}_a q_e^2}{m_e} \frac{f^0 (\omega)}{\omega^2} = \epsilon_0 \left( 1 - \frac{4 \pi}{k_0^2} \mathcal{N}_a b^0 (\omega) \right) ,
 \end{equation}
where $b^0 (\omega) \equiv r_e f^0 (\omega)$ is the \emph{scattering length} in the forward-scattering approximation with $r_e \approx \SI{2.8e-15}{\metre}$ as the classical electron radius \cite{Als-Nielsen11,deBergevin09}.\footnote{In the limit of very high frequency, $\left( \omega^2 - \omega_s^2 + i \omega \Gamma_s \right) \approx \omega^2$ and therefore the electrons behave as if they were unbound and \cref{eq:epsilon_omega_model} becomes 
\begin{equation*}\label{eq:epsilon_omega_approx}
 \tilde{\nu}^2 (\omega) = \frac{\epsilon (\omega)}{\epsilon_0} \approx 1 - \frac{\omega_p^2}{\omega^2} \quad \text{with } \omega_p^2 \equiv \frac{\mathcal{N}_a \mathcal{Z} q_e^2}{\epsilon_0 m_e} ,
 \end{equation*}
where $\omega_p$ is the \emph{plasma frequency} of the medium. 
Physically, this reflects the fact that electron polarization processes cannot keep up when $\omega$ is sufficiently high, as alluded to previously \cite{Landau60,Jackson75}.} 

The complex index of refraction, which is valid for soft x-rays and other radiation satisfying the forward-scattering approximation can be expressed as [\emph{cf.\@} \cref{eq:refractive_index,eq:epsilon_omega_model}]
\begin{subequations}
\begin{equation}\label{eq:index_model}
 \tilde{\nu} (\omega) \equiv \left( \frac{\epsilon (\omega)}{\epsilon_0} \right)^{1/2} = \left( 1 - \frac{\lambda^2}{\pi} \mathcal{N}_a b^0 (\omega) \right)^{1/2} . 
 \end{equation}
From \cref{fig:atomic_scattering_data,fig:atomic_scattering_data2,fig:atomic_scattering_data3}, however, $\Re [b^0 (\omega)] \lessapprox 50 r_e \approx \SI{1.4e-13}{\metre}$ while, for soft x-rays, $\lambda^2 \sim \left( \SI{1}{\nm} \right)^2 = \SI{1e-18}{\metre\squared}$; with a typical value of $\mathcal{N}_a \sim \SI{e29}{\per\metre\tothe{3}}$ [\emph{cf.\@} \cref{tab:element_densities}], the quantity $\lambda^2 \mathcal{N}_a \Re \left[ b^0 (\omega) \right] / \pi$ is $\sim \num{e-3}$. 
Thus, it is justified to approximate \cref{eq:index_model} using the Taylor-series expansion of a square root, to first order:\footnote{\emph{i.e.}, $\sqrt{1 + x} \approx 1 + x/2$ for small $x$} 
\begin{equation}\label{eq:index_approx_alt}
 \tilde{\nu} (\omega) \approx \frac{\lambda^2}{2 \pi} \mathcal{N}_a b^0 (\omega) = 1 - 2 \pi \frac{c_0^2}{\omega^2} \mathcal{N}_a r_e f^0 (\omega) .
 \end{equation}
\end{subequations}
Similar to the treatment of $f^0 (\omega)$ in \cref{eq:atomic_scattering_factor_reduced3}, it is useful to split up $\tilde{\nu} (\omega)$ into real and imaginary components, $\nu(\omega)$ and $\xi(\omega)$, also known as the \emph{optical constants} [\emph{cf.\@} \cref{tab:optical_constants}]. 
\begin{table}[]
 \centering
 \caption[Notation for complex index of refraction]{Notation for complex index of refraction, $\tilde{\nu} (\omega)$}\label{tab:optical_constants}
 \begin{tabular}{@{}lll@{}} 
 \toprule
  & real part & imaginary part \\ \midrule
 this dissertation & $\nu (\omega) \equiv 1 - \delta_{\nu} (\omega)$  & $\xi (\omega)$ \\
 common optical notation  & $n$ & $k$ or $\kappa$ \\
 common x-ray notation & $1-\delta$ & $\beta$ \\ \bottomrule
 \end{tabular}
 \end{table}
This can be written as 
\begin{subequations}
\begin{equation}
 \tilde{\nu} (\omega) = \Re \left[ \tilde{\nu} (\omega) \right] + i \Im \left[ \tilde{\nu} (\omega) \right] \equiv \nu(\omega) + i \xi(\omega)
 \end{equation}
with 
\begin{align}
 \nu(\omega) &= 1 - 2 \pi \frac{c_0^2}{\omega^2} \mathcal{N}_a r_e f^0_1 \left( \omega \right) \equiv 1 - \delta_{\nu} (\omega) \quad \text{and} \\
 \xi(\omega) &= 2 \pi \frac{c_0^2}{\omega^2} \mathcal{N}_a r_e f^0_2 \left( \omega \right) ,
 \end{align}
where $f^0 \left( \omega \right) \equiv f^0_1 \left( \omega \right) - i f^0_2 \left( \omega \right)$ [\emph{cf.\@} \cref{eq:atomic_scattering_factor_reduced3}]. 
\end{subequations}
These expressions indicate that $\nu(\omega) < 1$ over the soft x-ray spectrum, with possible exceptions near resonances, where $\omega \approx \omega_s$. 
This implies that in an opposite fashion to visible light and other relatively long-$\lambda$ radiation, the phase velocity, $c_p \equiv c_0 / \nu (\omega)$, can be greater than $c_0$ for soft x-rays with the wavelength in the medium, $\lambda' \equiv \lambda / \nu(\omega)$, increasing relative to $\lambda$ rather than decreasing. 
Moreover, with $\nu(\omega) \lessapprox 1$, the effect of refraction in the soft x-ray bandpass is expected to be small [\emph{cf.\@} \cref{sec:planar_interface}]. 

The optical constants defined above are plotted in \cref{fig:X-ray_index} for the materials listed in \cref{tab:element_densities} using data obtained from CXRO \cite{CXRO_database}. 
\begin{figure}
 \centering
 \includegraphics[scale=0.95]{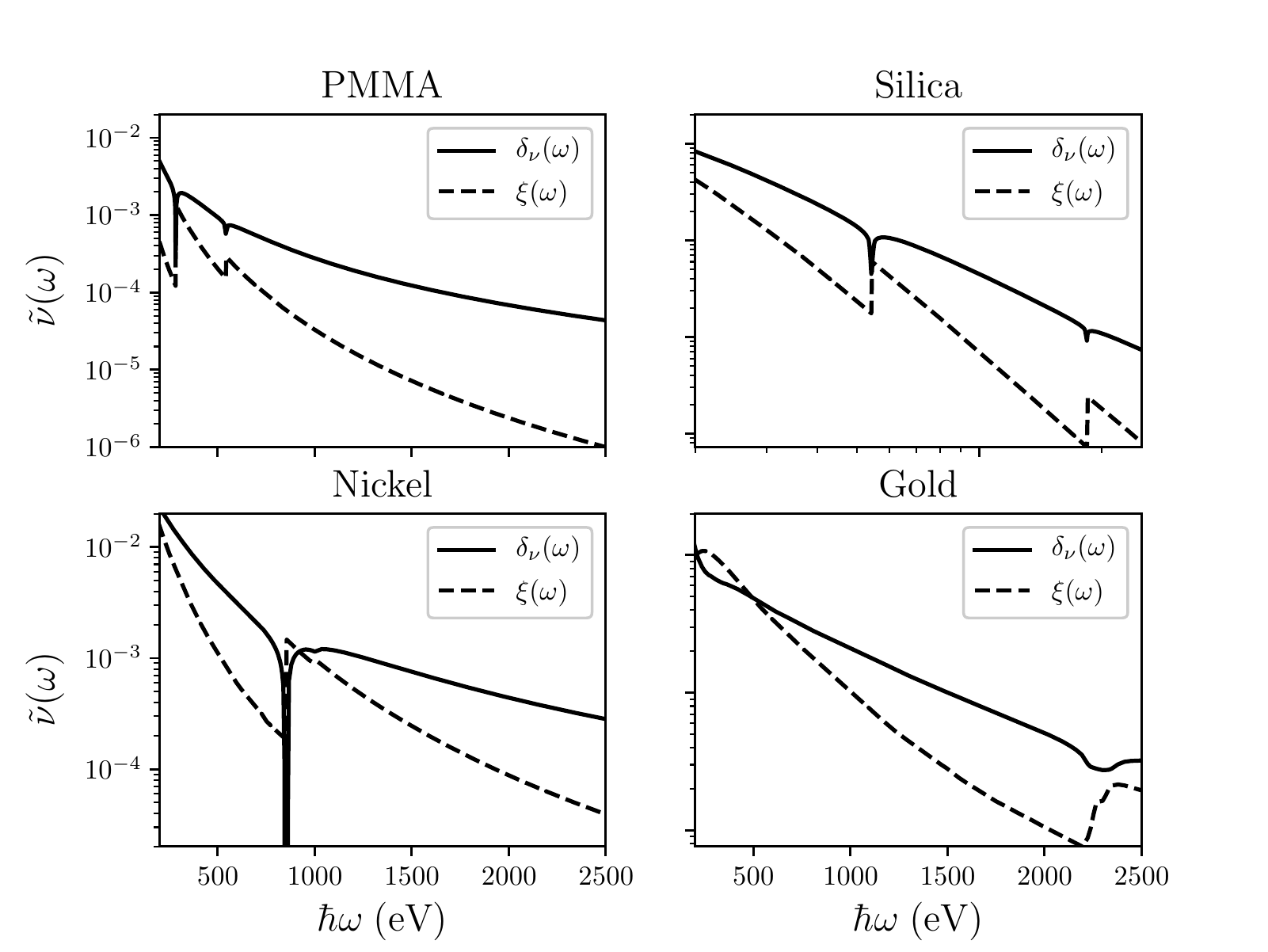}
 \caption[Soft x-ray index of refraction for PMMA, silica, nickel and gold]{Soft x-ray index of refraction for PMMA (\ce{C5H8O2})$_n$, silica (\ce{SiO2}), nickel (\ce{Ni}) and gold (\ce{Au}) on a logarithmic scale \cite{CXRO_database}}\label{fig:X-ray_index} 
 \end{figure} 
PMMA and silica, as materials with molecular composition, have optical constants that depend on an average of the atomic scattering factors for the constituent atoms and as a result, multiple absorption edges are evident in these plots. 
Using these optical constants, the wave number in the medium now can be written as 
\begin{subequations}
\begin{equation}
 \tilde{k} \equiv \tilde{\nu} (\omega) k_0 = \frac{\omega}{c_0} \left[ 1 - \delta_{\nu} (\omega) + i \xi(\omega) \right] . 
 \end{equation}
For the special case of a wave decaying in the direction of propagation with $\mathbold{\tilde{k}} \cdot \mathbold{r} = \tilde{k} r$, the electric field is of the form
\begin{equation}
 \mathbold{E} \left( \mathbold{r} , t \right) = \mathbold{E}_0 \mathrm{e}^{i (\tilde{k} r - \omega t)} = \mathbold{E}_0 \mathrm{e}^{i k_0 \left( r - c_0 t \right)} \mathrm{e}^{i k_0 r \delta_{\nu} (\omega)} \mathrm{e}^{-k_0 r \xi(\omega)} ,
 \end{equation}
which indicates the medium induces a phase shift $k_0 r \delta_{\nu} (\omega)$ and causes the wave to decay exponentially with wave intensity dropping off as $\mathrm{e}^{- 2 k_0 r \xi(\omega)}$. 
The scattered wave therefore becomes \SI{1}{\radian} out of phase with the original wave over a distance known as the \emph{extinction length} \cite{deBergevin09}:
\begin{equation}\label{eq:extinction_length}
 k_0 r \delta_{\nu} (\omega) = 1 \implies \ell_{\text{ext}} \equiv \frac{\lambda}{2 \pi \delta_{\nu} (\omega)} ,
 \end{equation}
while on the other hand, the $1/\mathrm{e}$ length scale for absorption is \cite{Attwood17}
\begin{equation}\label{eq:eabsorption_length}
 2 k_0 r \xi(\omega) = 1 \implies \ell_{\text{abs}} \equiv \frac{\lambda}{4 \pi \xi(\omega)} .
 \end{equation}
\end{subequations}
These two characteristic length scales are plotted in \cref{fig:X-ray_lengths} using the optical constants from \cref{fig:X-ray_index}. 
\begin{figure}
 \centering
 \includegraphics[scale=0.95]{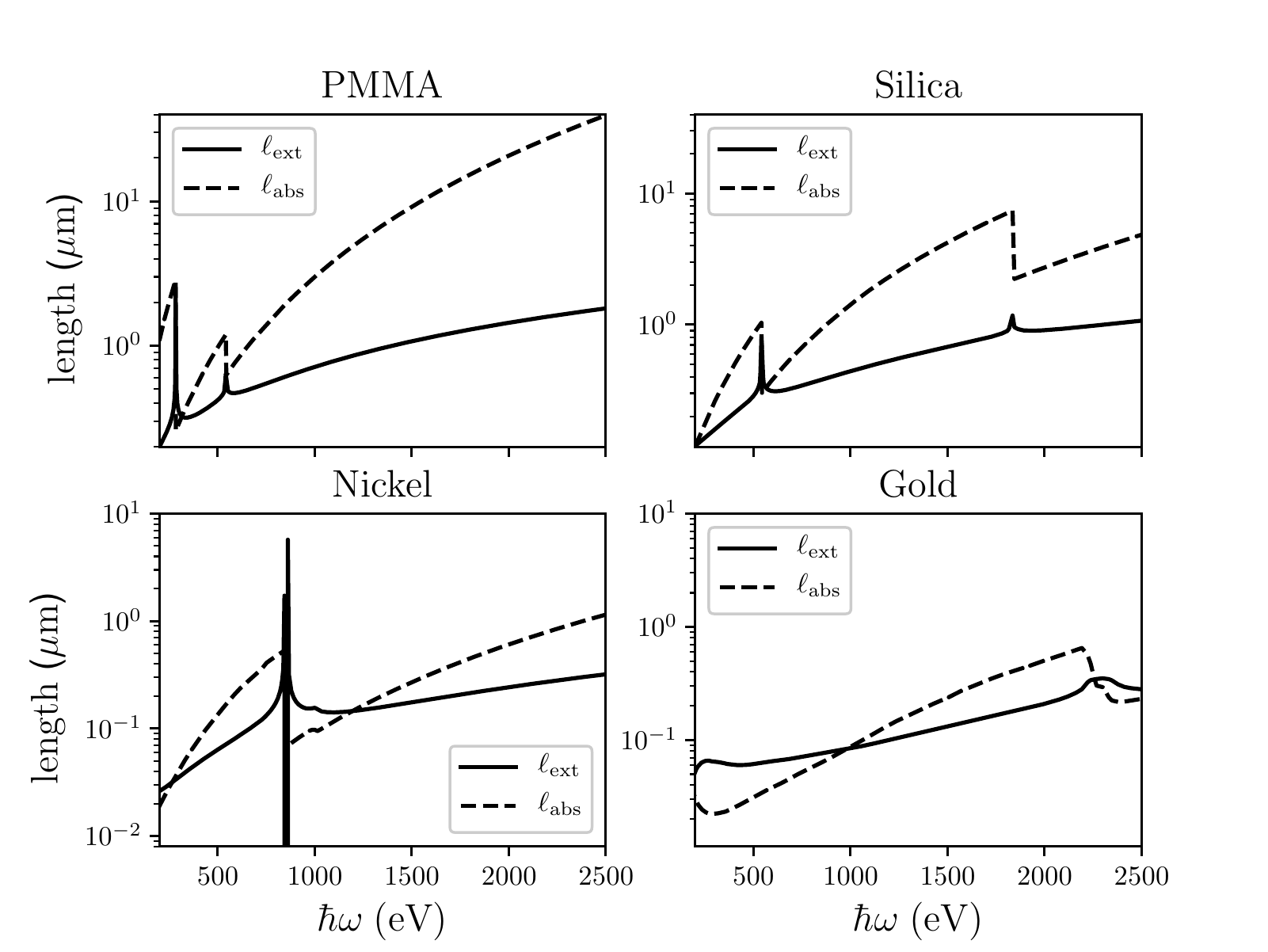}
 \caption[Characteristic length scales for soft x-rays in PMMA, silica, nickel and gold]{Characteristic length scales for soft x-rays in PMMA (\ce{C5H8O2})$_n$, silica (\ce{SiO2}), nickel (\ce{Ni}) and gold (\ce{Au}). The extinction length, $\ell_{\text{ext}}$, arises from a phase shift due to the real index while $\ell_{\text{abs}}$ describes the length scale of absorption due to the imaginary index \cite{CXRO_database}}\label{fig:X-ray_lengths} 
 \end{figure} 
This demonstrates that soft x-rays are attenuated over short, sub-\si{\um} distances that decrease as $\mathcal{Z}$ increases, especially near prominent absorption edges. 

\section{Reflection from a Mirror Flat}\label{sec:planar_interface}
%%%%%%%%%%%%%%%%%%%%%%%%%%%%%%%%%%%%%%%%%--------------------------------------------------
Several important properties of reflective overcoats for x-ray reflection gratings can be gleaned from considering how a single mode of soft x-rays with frequency $\omega$ reflects and refracts at a planar interface between vacuum, with index of refraction equal to unity (\emph{i.e.}, $\tilde{\nu} (\omega) = \nu (\omega) = 1$) and a material that represents a thick mirror with a complex index of refraction, $\tilde{\nu} (\omega) \equiv \nu (\omega) + i \xi(\omega)$, where $\nu (\omega) \equiv 1 - \delta_{\nu} (\omega) \lessapprox 1$ and $\xi(\omega) \ll 1$ [\emph{cf.\@} \cref{sec:x-ray_index}]. 
To draw an analogy with radiation illuminating an idealized sawtooth facet of an x-ray reflection grating [\emph{cf.\@} \cref{fig:conical_reflection_edit}], the angle of incidence measured relative to the surface of the mirror is taken to be $\zeta$. 
Using the coordinate system defined in \cref{fig:2Dmirror_angles}, the incident wave vector given by 
\begin{subequations}
\begin{equation}\label{eq:2D_incident_wave}
 \mathbold{k} = k_x \mathbold{\hat{x}} + k_y \mathbold{\hat{y}} = k_0 \left[ \cos \left( \zeta \right) \mathbold{\hat{x}} - \sin \left( \zeta \right) \mathbold{\hat{y}} \right] 
 \end{equation}
with $\abs{\mathbold{k}} = k_0 \equiv 2 \pi / \lambda = \sqrt{k_x^2 + k_y^2}$ and $k_z = 0$. 
\begin{figure}
 \centering
 \includegraphics[scale=1.25]{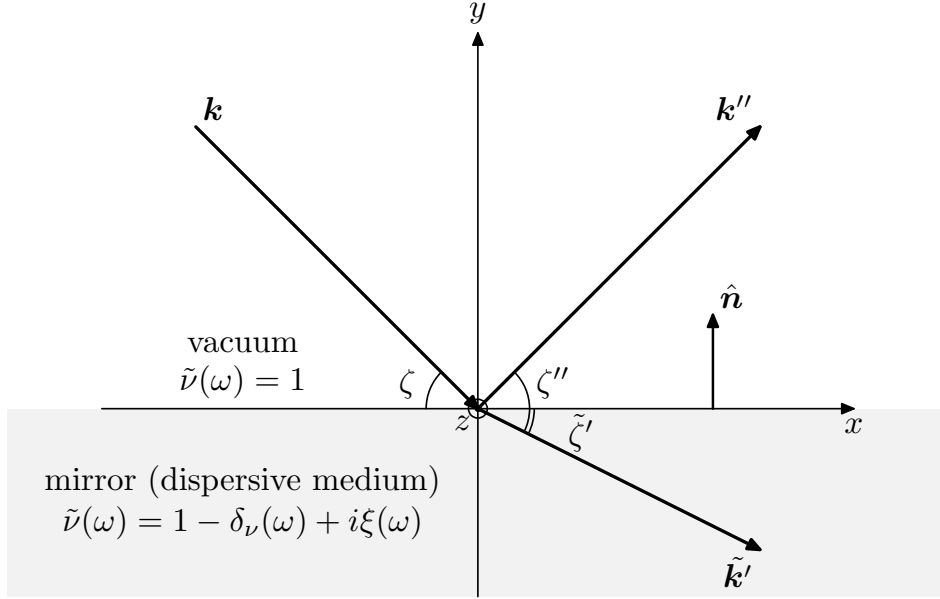}
 \caption[Geometry for in-plane reflection and refraction]{Geometry for in-plane reflection and refraction, where $\mathbold{k}$ is the incident wave vector, $\mathbold{k}''$ is the reflected wave vector and $\mathbold{\tilde{k}}'$ is the refracted wave vector.}\label{fig:2Dmirror_angles}
 \end{figure}
The corresponding electromagnetic wavefront is represented by 
\begin{equation} \label{eq:incident_fields}
 \mathbold{E} \left( x, y, z \right) = \mathbold{E}_0 \mathrm{e}^{i \left( k_x x + k_y y \right)} \text{ and } \mathbold{H} \left( x, y, z \right) = \frac{1}{Z_0 k_0} \mathbold{k} \times \mathbold{E} \left( x, y, z \right) ,
 \end{equation}
where $\mathbold{E}_0$ is a vector describing the polarization and amplitude of the incident electric field [\emph{cf.\@} \cref{sec:EM_waves_vac}]. 
\end{subequations}
Kept in arbitrary form, the reflected ray is also associated with a real wave vector: 
\begin{subequations}
\begin{equation}\label{eq:3D_reflected_wave_vector}
 \mathbold{k''} = k_x'' \mathbold{\hat{x}} + k_y'' \mathbold{\hat{y}} + k_z'' \mathbold{\hat{z}}
 \end{equation}
with $\abs{\mathbold{k''}} = k_0 = \sqrt{k_x^2 + k_y^2 + k_z^2}$ while the reflected wavefront can be written using 
\begin{equation}\label{eq:reflected_fields}
\mathbold{E''} \left( x, y, z \right) = \mathbold{E}_0'' \mathrm{e}^{i \left( k_x'' x + k_y'' y + k_z'' z \right)} \text{ and } \mathbold{H''} \left( x, y, z \right) = \frac{1}{Z_0 k_0} \mathbold{k''} \times \mathbold{E''} \left( x, y, z \right) ,
 \end{equation} 
where $\mathbold{E}_0''$ describes the polarization and amplitude of the reflected electric field.  
\end{subequations}
On the other hand, the refracted ray is described with a complex wave vector [\emph{cf.\@} \cref{eq:k_tilde_def}]:
\begin{subequations}
\begin{equation}\label{eq:3D_refracted_wave_vector}
 \mathbold{\tilde{k}'} = \mathbold{k'} + i \mathbold{\kappa'} = \tilde{k}_x' \mathbold{\hat{x}} + \tilde{k}_y' \mathbold{\hat{y}} + \tilde{k}_z' \mathbold{\hat{z}}
 \end{equation}
with $\abs{\mathbold{\tilde{k}'}} = \tilde{k}' \equiv \tilde{\nu} (\omega) \, k_0 = \sqrt{\tilde{k}_x'^2 + \tilde{k}_y'^2 + \tilde{k}_z'^2}$ and the refracted wavefront is associated with complex electromagnetic fields, where $\mathbold{E}_0'$ describes the polarization and initial amplitude of the refracted electric field: 
\begin{equation}\label{eq:refracted_fields}
 \mathbold{E'} \left( x, y, z \right) = \mathbold{E}_0' \mathrm{e}^{i \left( \tilde{k_x}' x + \tilde{k_y}' y + \tilde{k_z}' z \right)} \text{ and } \mathbold{H'} \left( x, y, z \right) = \frac{1}{Z_0 k_0} \mathbold{\tilde{k}'} \times \mathbold{E'} \left( x, y, z \right) .
 \end{equation} 
\end{subequations}

To describe rigorously how the reflected and refracted fields are related to the incident wave, the Helmholtz equation [\emph{cf.\@} \cref{eq:Helmholtz}] is solved above the mirror surface, in vacuum:
\begin{subequations}
\begin{equation} 
 \left( \laplacian + k_0^2 \right) \left\{
 \begin{array}{lr}
 \mathbold{E} \left( x, y, z \right) + \mathbold{E''} \left( x, y, z \right) \\
 \mathbold{H} \left( x, y, z \right) + \mathbold{H''} \left( x, y, z \right) \label{eq:Helmholtz_above_mirror}
 \end{array}
 \right\} = \mathbf{0} ,
 \end{equation}
and below the surface, in a dispersive medium representing the mirror material [\emph{cf.\@} \cref{eq:Helmholtz_med}]:
\begin{equation} 
 \left( \laplacian + \tilde{k}'^2 \right) \left\{
 \begin{array}{lr}
 \mathbold{E'} \left( x, y, z \right) \\
 \mathbold{H'} \left( x, y, z \right) \label{eq:Helmholtz_below_mirror}
 \end{array}
 \right\} = \mathbf{0} . 
 \end{equation}
\end{subequations}
Additionally, the incident, reflected and refracted fields must be phase-matched across the boundary, which is initially taken to be a perfectly smooth plane representing the surface of the mirror at $y=0$. 
This can be handled using the general boundary conditions for electrodynamics in the absence of surface current and surface charge \cite{Landau60,Jackson75}: 
\begin{subequations}
\begin{align}
 \mathbold{\hat{n}} \times \left[ \mathbold{E} \left( x, 0, z \right) + \mathbold{E''} \left( x, 0, z \right) \right] &= \mathbold{\hat{n}} \times \mathbold{E'} \left( x, 0, z \right) \label{eq:general_boundary_1} \\
 \mathbold{\hat{n}} \times \left[\mathbold{H} \left( x, 0, z \right)  + \mathbold{H''} \left( x, 0, z \right) \right] &= \mathbold{\hat{n}} \times \mathbold{H'} \left( x, 0, z \right) \label{eq:general_boundary_2} \\
 \mathbold{\hat{n}} \cdot \left[ \mathbold{E} \left( x, 0, z \right) + \mathbold{E''} \left( x, 0, z \right) \right] &= \tilde{\nu}^2 (\omega) \, \mathbold{\hat{n}} \cdot \mathbold{E'} \left( x, 0, z \right) \label{eq:general_boundary_3} \\
 \mathbold{\hat{n}} \cdot \left[ \mathbold{H} \left( x, 0, z \right) + \mathbold{H''} \left( x, 0, z \right) \right] &= \mathbold{\hat{n}} \cdot \mathbold{H'} \left( x, 0, z \right) \label{eq:general_boundary_4} , 
 \end{align}
\end{subequations}
where $\mathbold{\hat{n}} = \mathbold{\hat{y}}$ is a unit vector pointing from the dispersive medium to vacuum [\emph{cf.\@} \cref{fig:2Dmirror_angles}]. 
Inserting \cref{eq:incident_fields,eq:reflected_fields,eq:refracted_fields} into these relations verifies that reflection and refraction occur in-plane with the requirement that $k_z \equiv 0 = k_z'' = \tilde{k_z}'$. 
These boundary conditions also demand that $k_x = k_x'' = \tilde{k_x}'$, which leads to the \emph{law of reflection} and \emph{Snell's law of refraction}. 
Explicitly, with $k_x'' \equiv k_0 \cos \left( \zeta '' \right)$ the former reads as 
\begin{equation}\label{eq:reflection}
 k_x'' = k_x \implies \zeta'' = \zeta ,
 \end{equation}
where $k_y'' \equiv k_0 \sin \left( \zeta '' \right)$. 
Meanwhile, $\tilde{k_x}' \equiv \tilde{k}' \cos \left( \tilde{\zeta}' \right)$ is associated with a complex angle, $\tilde{\zeta}'$, so that the latter reads as
%\begin{subequations}
\begin{subequations}
\begin{equation}\label{eq:refraction}
  \tilde{k_x}' = k_x \implies \tilde{\nu}(\omega) \cos \left( \tilde{\zeta}' \right) = \cos \left( \zeta \right) 
 \end{equation}
with $\tilde{k_y}' \equiv - \tilde{k}' \sin \left( \tilde{\zeta}' \right)$. 
Defining $\tilde{\zeta}' \equiv \zeta'_{\Re} + i \zeta'_{\Im}$ and using a trigonometric identity involving complex angles,\footnote{$\cos \left( \zeta'_{\Re} + i \zeta'_{\Im}\right) = \cos \left( \zeta'_{\Re} \right) \cosh \left( \zeta'_{\Im} \right) - i \sin \left( \zeta'_{\Re} \right) \sinh \left( \zeta'_{\Im} \right)$} \cref{eq:refraction} yields the following relations:
\begin{align}
 \nu (\omega) \cos \left( \zeta'_{\Re} \right) \cosh \left( \zeta'_{\Im} \right) + \xi (\omega) \sin \left( \zeta'_{\Re} \right) \sinh \left( \zeta'_{\Im} \right) &= \cos \left( \zeta \right) \label{eq:refractionA} \\
 \xi (\omega) \cos \left( \zeta'_{\Re} \right) \cosh \left( \zeta'_{\Im} \right) - \nu (\omega) \sin \left( \zeta'_{\Re} \right) \sinh \left( \zeta'_{\Im} \right) &=0 \label{eq:refractionB} ,
 \end{align}
\end{subequations}
which imply that $\zeta'_{\Im} = 0$ if $\xi (\omega) = 0$. 

Crucially, appreciable reflectivity from an interface at soft x-ray frequencies can only be achieved in the regime of \emph{total external reflection (TER)}, where $\zeta$ is small enough such that the refracted ray travels virtually parallel to the surface of the mirror with $\zeta'_{\Re} \approx 0$ instead of propagating down into the dispersive medium substantially. 
\begin{figure}
 \centering
 \includegraphics[scale=0.95]{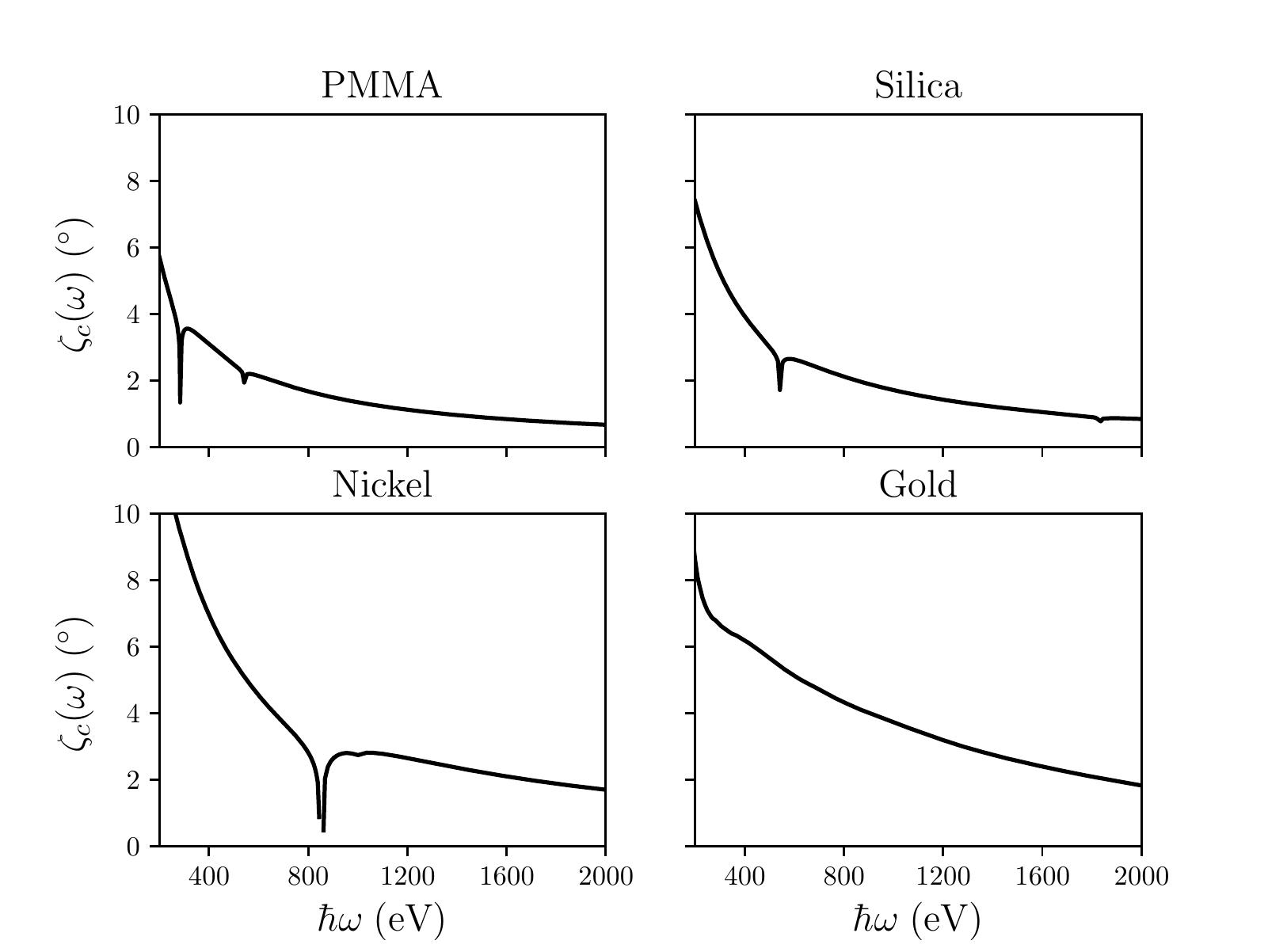}
 \caption[Critical angle for total external reflection (TER) of PMMA, silica, nickel and gold across the soft x-ray spectrum]{Critical angle for total external reflection (TER) of PMMA (\ce{C5H8O2})$_n$, silica (\ce{SiO2}), nickel (\ce{Ni}) and gold (\ce{Au}) across the soft x-ray spectrum \cite{CXRO_database}}\label{fig:critical_angle}
 \end{figure} 
While this is shown explicitly in \cref{sec:reflectivity_polarization,sec:total_refl}, it can be gleaned by neglecting $\xi (\omega)$ for the moment so that $\tilde{\nu} (\omega) = \nu (\omega) \lessapprox 1$ and $\zeta'_{\Im} = 0$ with \cref{eq:refractionA} reducing to 
\begin{equation}\label{eq:Snell_real}
 \cos \left( \zeta \right) = \nu (\omega) \cos \left( \zeta'_{\Re} \right) . 
 \end{equation}
From this expression, the condition for TER is met when $\zeta$ is smaller than a certain \emph{critical angle}, $\zeta_c (\omega)$, that corresponds to $\zeta'_{\Re} = 0$ with $\cos \left( \zeta \right) = \nu (\omega) \equiv 1 - \delta_{\nu} (\omega)$. 
This can be written as
\begin{equation}\label{eq:critical_angle}
 \zeta_c (\omega) \equiv \arccos \left[ 1 - \delta_{\nu} (\omega) \right] \approx \sqrt{2 \delta_{\nu} (\omega)} ,
 \end{equation}
where the approximation is valid for $\delta_{\nu} (\omega) \ll 1$ \cite{Attwood17}. 
Using data for $\delta_{\nu} (\omega)$ obtained from CXRO \cite{CXRO_database}, $\zeta_c (\omega)$ is plotted as a function of photon energy in \cref{fig:critical_angle} for the materials from \cref{fig:X-ray_index,fig:X-ray_lengths}, where it is seen that $\zeta \lessapprox 2^{\circ}$ is typically required to achieve TER across the soft x-ray bandpass. 

The term \emph{total external reflection} by name implies that all incident radiation is reflected from a mirror with $\nu (\omega) < 1$. 
Although this is approximately true, a real material with a complex index of refraction, $\tilde{\nu} (\omega)$, has an imaginary part, $\xi(\omega) \neq 0$, that takes into account losses in reflectivity and additionally, causes the refracted ray to penetrate slightly into the mirror rather than travel parallel to the surface as deduced above \cite{Attwood17}. 
The former is quantified by analyzing the \emph{Fresnel equations} for specular reflectivity in \cref{sec:reflectivity_polarization} while the latter [\emph{cf.\@} \cref{sec:total_refl}] manifests as a \emph{penetration depth} that informs thickness requirements for reflective overcoats. 
Moreover, the presence of surface roughness leads to losses in specular reflectivity in a manner that depends on the size scale of surface features and the wave vector of incident radiation [\emph{cf.\@} \cref{sec:rough_surface}]. % this is the subject of \cref{sec:rough_surface}. 

\subsection{Fresnel Reflectivity in Orthogonal Polarizations}\label{sec:reflectivity_polarization}
%%%%%%%%%%%%%%%%%%%%%%%%%%%%%%%%%%%%%%%%%--------------------------------------------------
The reflectivity of a mirror flat that can be regarded as being infinitely thick\footnote{In practice, this means that a slab of material is assumed to be thick enough for the refracted ray to be completely attenuated or absorbed before reaching another interface.} is defined as the ratio of the specularly-reflected and incident waves in terms of flux through a unit area of a surface parallel to the boundary \cite{Born80}. 
Using $\mathbold{S}(\mathbold{r}) = \mathbold{E}(\mathbold{r}) \times \mathbold{H}(\mathbold{r})$ as the time-harmonic form of \emph{Poynting's vector} \cite{Griffiths17,Jackson75}, this can be written as 
\begin{equation}\label{eq:fresnel_refl_def}
 \mathcal{R}_F \equiv \frac{\mathbold{S}''(\mathbold{r}) \cdot \mathbold{\hat{y}}}{\mathbold{S}(\mathbold{r}) \cdot \mathbold{\hat{y}}} = \frac{k_y''}{k_y} \frac{\abs{\mathbold{E}_0''}^2}{\abs{\mathbold{E}_0}^2} = \frac{\abs{\mathbold{E}_0''}^2}{\abs{\mathbold{E}_0}^2} ,
 \end{equation} 
which reduces to the ratio of field strengths for the incident and reflected waves. 
By symmetry, the boundary conditions [\emph{cf.\@} \cref{eq:general_boundary_1,eq:general_boundary_2,eq:general_boundary_3,eq:general_boundary_4}] demand that the time-harmonic electromagnetic fields are constant across the surface of the mirror, except for a phase shift that depends on the parallel component of the wave number, $k_{\parallel}$. 
Using the coordinate system defined in \cref{fig:2Dmirror_angles}, where $k_z = 0$ by construction, $k_{\parallel} \equiv k_x = k_0 \cos \left( \zeta \right)$ so that translations in $x$ on the boundary come with a $\mathrm{e}^{i k_x x}$ dependence while translations in $z$ introduce no phase shift. 
This allows the electromagnetic fields at the surface of the mirror to be described as being \emph{space-harmonic} such that the time-harmonic incident fields can be written as
\begin{equation}
 \left\{
 \begin{array}{lr}
 \mathbold{E} \left( \mathbold{r} \right) \\
 \mathbold{H} \left( \mathbold{r} \right)
 \end{array} 
 \right\} =  \left\{
 \begin{array}{lr}
 \mathbold{E} \left( y \right) \\
 \mathbold{H} \left( y \right)
 \end{array} 
 \right\} \mathrm{e}^{i k_x x} \underbrace{\mathrm{e}^{i k_z z}}_1 
 \end{equation} 
with similar expressions for the reflected and refracted fields. 
The Helmholtz equations given by \cref{eq:Helmholtz_above_mirror,eq:Helmholtz_below_mirror} then reduce to 
\begin{subequations}
\begin{equation} 
 \left( \frac{d^2}{d y^2} + k_y^2 \right) \left\{
 \begin{array}{lr}
 \mathbold{E} \left( y \right) + \mathbold{E''} \left( y \right) \\
 \mathbold{H} \left( y \right) + \mathbold{H''} \left( y \right) \label{eq:Helmholtz_above_mirror_simplified}
 \end{array}
 \right\} = \mathbf{0} 
 \end{equation}
in vacuum with $k_y^2 = k_0^2 - k_x^2 = k_0^2 \sin^2 \left( \zeta \right)$ and 
\begin{equation} 
 \left( \frac{d^2}{d y^2} + \tilde{k_y}'^2 \right) \left\{
 \begin{array}{lr}
 \mathbold{E'} \left( y \right) \\
 \mathbold{H'} \left( y\right) \label{eq:Helmholtz_below_mirror_simplified}
 \end{array}
 \right\} = \mathbf{0} 
 \end{equation}
inside the material of the mirror with $\tilde{k_y}'^2 = \tilde{k}'^2 - k_x^2 = k_0^2 \left[ \tilde{\nu}^2(\omega) - \cos^2 \left( \zeta \right) \right]$. 
\end{subequations}

The incident wave is assumed to be traveling toward the mirror surface with a $\mathrm{e}^{i k_y y}$ dependence, where $k_y \equiv -k_0 \sin \left( \zeta \right)$ [\emph{cf.\@} \cref{eq:2D_incident_wave,fig:2Dmirror_angles}]. 
Meanwhile, the reflected wave travels away with a $\mathrm{e}^{-i k_y y}$ dependence so that the total electric field in vacuum can be written as 
\begin{subequations}
\begin{equation}
 \mathbold{E} \left( y \right) + \mathbold{E''} \left( y \right) = \mathbold{E}_0 \mathrm{e}^{i k_y y} + \mathbold{E}_0'' \mathrm{e}^{-i k_y y} ,
 \end{equation}
which is a solution to \cref{eq:Helmholtz_above_mirror}. 
On the other hand, a physically-realistic solution to \cref{eq:Helmholtz_below_mirror_simplified} is one which propagates away from the surface, into the mirror with
\begin{equation}
 \mathbold{E'} \left( y \right) = \mathbold{E}_0' \mathrm{e}^{i \tilde{k_y}' y} 
 \end{equation}
\end{subequations}
so that using \cref{eq:incident_fields,eq:reflected_fields,eq:refracted_fields} for the corresponding magnetic fields, the boundary conditions given by \cref{eq:general_boundary_1,eq:general_boundary_2,eq:general_boundary_3,eq:general_boundary_4} reduce to
\begin{subequations}
\begin{align} 
 \mathbold{\hat{y}} \times \left[ \mathbold{E}_0 + \mathbold{E}_0'' \right] &= \mathbold{\hat{y}} \times \mathbold{E}_0' \label{eq:mirror_boundary_1} \\
 \mathbold{\hat{y}} \times \left[ \mathbold{k} \times \mathbold{E}_0 + \mathbold{k''} \times \mathbold{E}_0'' \right] &= \mathbold{\hat{y}} \times \left[ \mathbold{\tilde{k}'} \times \mathbold{E}_0' \right] \label{eq:mirror_boundary_2} \\
 \mathbold{\hat{y}} \cdot \left[ \mathbold{E}_0 + \mathbold{E}_0'' \right] &= \tilde{\nu}^2 (\omega) \, \mathbold{\hat{y}} \cdot \mathbold{E}_0' \label{eq:mirror_boundary_3} \\
 \mathbold{\hat{y}} \cdot \left[ \mathbold{k} \times \mathbold{E}_0 + \mathbold{k''} \times \mathbold{E}_0'' \right] &= \mathbold{\hat{y}} \cdot \left[ \mathbold{\tilde{k}'} \times \mathbold{E}_0' \right] \label{eq:mirror_boundary_4} .
 \end{align} %This shows that polarization is the same for incident, refracted and reflected rays?
\end{subequations}
To evaluate how polarization, given by the orientation of the electric field vector, affects the reflectivity of a mirror, the Helmholtz equations given by \cref{eq:Helmholtz_above_mirror_simplified,eq:Helmholtz_below_mirror_simplified} are solved with the boundary conditions defined in \cref{eq:mirror_boundary_1,eq:mirror_boundary_2,eq:mirror_boundary_3,eq:mirror_boundary_4}, in two orthogonal, linear polarization states:
\begin{itemize}[noitemsep]
  \item \emph{s-polarization}: the electric field is perpendicular to the plane of incidence and parallel to the surface, and 
  \item \emph{p-polarization}: the electric field is parallel to the plane of incidence and perpendicular to the the surface, 
\end{itemize}
with $u(y)$ as a scalar function representing the magnitude of the field in each case. 
Similarly, from \cref{eq:mirror_boundary_1,eq:mirror_boundary_2,eq:mirror_boundary_3,eq:mirror_boundary_4}, the reflected and refracted fields can also be represented with scalar functions $u''(y)$ and $u'(y)$, respectively, and \cref{eq:Helmholtz_above_mirror_simplified,eq:Helmholtz_below_mirror_simplified} are reduced to their scalar form:
\begin{subequations}
\begin{align}
 \left( \dv[2]{y} + k_y^2 \right) \left[ u(y) + u''(y)  \right] &= 0 \quad \text{for $y>0$} \\
 \left( \dv[2]{y} + \tilde{k_y}'^2 \right) u'(y) &= 0 \quad \text{for $y<0$} .
  \end{align}
 \end{subequations}
Solving these equations for each orthogonal polarization allows the reflectivity in any arbitrary polarization to determined through superposition. 

The incident, reflected and refracted electric fields in s-polarization, which are parallel to the surface of the mirror, can be written as
\begin{align}
\begin{split}
 \mathbold{E}(y) &= \mathcal{A}_s \mathbold{\hat{z}} \mathrm{e}^{i k_y y} \quad \text{(incident)} \quad \quad \mathbold{E''}(y) = \mathcal{A}_s'' \mathbold{\hat{z}} \mathrm{e}^{-i k_y y} \quad \text{(reflected)} \\
 \mathbold{E'}(y) &= \mathcal{A}_s' \mathbold{\hat{z}} \mathrm{e}^{i \tilde{k_y}' y} \quad \text{(refracted/attenuated)} ,
 \end{split}
 \end{align}
where $\mathcal{A}_s$ is the known amplitude of the incident electric field while $\mathcal{A}_s''$ and $\mathcal{A}_s'$ are the unknown amplitudes of the reflected and refracted electric fields. 
From \cref{eq:mirror_boundary_1,eq:mirror_boundary_2,eq:mirror_boundary_3,eq:mirror_boundary_4}, the boundary conditions for this polarization give\footnote{Here, the following vector triple-product rules have been used: $\mathbold{A} \times \left( \mathbold{B} \times \mathbold{C} \right) = \left( \mathbold{A} \cdot \mathbold{C} \right) \mathbold{B} - \left( \mathbold{A} \cdot \mathbold{B} \right) \mathbold{C}$ and $\mathbold{A} \cdot \left( \mathbold{B} \times \mathbold{C} \right) = \mathbold{B} \cdot \left( \mathbold{C} \times \mathbold{A} \right) = \mathbold{C} \cdot \left( \mathbold{A} \times \mathbold{B} \right)$.}  
\begin{subequations}
\begin{align}
%\begin{split}
 \mathbold{\hat{y}} \times \left[ \mathcal{A}_s \mathbold{\hat{z}} + \mathcal{A}_s'' \mathbold{\hat{z}} \right] = \mathbold{\hat{y}} \times \mathcal{A}_s' \mathbold{\hat{z}} &\implies \mathcal{A}_s + \mathcal{A}_s'' = \mathcal{A}_s' \\
 \mathbold{\hat{y}} \times \left[ \mathbold{k} \times \mathcal{A}_s \mathbold{\hat{z}}  + \mathbold{k''} \times \mathcal{A}_s'' \mathbold{\hat{z}} \right] = \mathbold{\hat{y}} \times \left[ \mathbold{\tilde{k}'} \times \mathcal{A}_s' \mathbold{\hat{z}} \right] &\implies \left( \mathcal{A}_s - \mathcal{A}_s'' \right) k_y = \mathcal{A}_s' \tilde{k_y}' \\ 
 \mathbold{\hat{y}} \cdot \left[ \mathcal{A}_s \mathbold{\hat{z}'} + \mathcal{A}_s'' \mathbold{\hat{z}'} \right] = \tilde{\nu}^2 (\omega) \, \mathbold{\hat{y}} \cdot \mathcal{A}_s' \mathbold{\hat{z}'} &\implies 0 = 0 \\
 \mathbold{\hat{y}} \cdot \left[ \mathbold{k} \times \mathcal{A}_s \mathbold{\hat{z}} + \mathbold{k''} \times \mathcal{A}_s'' \mathbold{\hat{z}} \right] = \mathbold{\hat{y}} \cdot \left[ \mathbold{\tilde{k}'} \times \mathcal{A}_s' \mathbold{\hat{z}} \right] &\implies \mathcal{A}_s + \mathcal{A}_s'' = \mathcal{A}_s' ,
 %\end{split}
 \end{align}
\end{subequations}
which can be rearranged to yield complex coefficients for reflection and transmission, respectively:
\begin{equation}\label{eq:s_coeff}
 \tilde{r}_s \equiv \frac{\mathcal{A}_s''}{\mathcal{A}_s} = \frac{k_y - \tilde{k_y}'}{k_y + \tilde{k_y}'} \quad \text{and} \quad \tilde{t}_s \equiv \frac{\mathcal{A}_s'}{\mathcal{A}_s} = \frac{2 k_y}{k_y + \tilde{k_y}'} 
 \end{equation}
with $k_y = - k_0 \sin (\zeta)$ and $\tilde{k_y}' = - \sqrt{\tilde{\nu}^2 (\omega) - \cos^2 \left( \zeta \right)}$. 
While the \emph{transmissivity} of a mirror in s-polarization depends on $\tilde{t}_s$, the reflectivity is determined from
\begin{equation}\label{eq:s_reflectivity}
 \mathcal{R}_s (\zeta) = \norm{\tilde{r}_s}^2 = \norm{\left( \frac{\sin \left( \zeta \right) - \sqrt{\tilde{\nu}^2 (\omega) - \cos^2 \left( \zeta \right)}}{\sin \left( \zeta \right) + \sqrt{\tilde{\nu}^2 (\omega) - \cos^2 \left( \zeta \right)}} \right)}^2 .
 \end{equation}
In p-polarization on the other hand, the incident, reflected and refracted electric fields are perpendicular to the surface of the mirror. 
The corresponding magnetic fields are then parallel to the surface, which can be written as 
\begin{subequations}
\begin{align}
\begin{split}
 \mathbold{H} (y) &= \frac{\mathcal{A}_p}{Z_0} \mathbold{\hat{z}} \mathrm{e}^{i k_y y} \quad \text{(incident)} \quad \quad \mathbold{H''} (y) = \frac{\mathcal{A''}_p}{Z_0} \mathbold{\hat{z}} \mathrm{e}^{-i k_y y} \quad \text{(reflected)} \\
 \mathbold{H'} (y) &= \frac{\mathcal{A'}_p}{Z_0} \mathbold{\hat{z}} \mathrm{e}^{i \tilde{k_y}' y} \quad \text{(refracted/attenuated)} ,
 \end{split}
 \end{align} 
\end{subequations}
where $\mathcal{A}_p$ is the known amplitude of the incident electric field, while $\mathcal{A}_p''$ and $\mathcal{A}_p'$ are the unknown amplitudes of the reflected and refracted electric fields. 
Using \cref{eq:Maxwell_vac_FT_4,eq:Maxwell_med_FT_4} for the electric field, the boundary conditions for this polarization yield 
\begin{subequations}
\begin{align}
%\begin{split}
 \mathbold{\hat{y}} \times \left[\mathbold{k} \times \mathcal{A}_p \mathbold{\hat{z}} + \mathbold{k''} \times \mathcal{A}_p'' \mathbold{\hat{z}} \right] = \mathbold{\hat{y}} \times \left[ \frac{\mathbold{\tilde{k}'}  \times \mathcal{A}_p' \mathbold{\hat{z}}}{\tilde{\nu}^2(\omega)} \right] &\implies \left( \mathcal{A}_p - \mathcal{A}_p'' \right) k_y = \frac{\mathcal{A}_p' \tilde{k_y}'}{\tilde{\nu}^2(\omega)} \\ 
 \mathbold{\hat{y}} \times \left[ \mathcal{A}_p \mathbold{\hat{z}} + \mathcal{A}_p'' \mathbold{\hat{z}} \right] = \mathbold{\hat{y}} \times \mathcal{A}_p' \mathbold{\hat{z}} &\implies \mathcal{A}_p + \mathcal{A}_p'' = \mathcal{A}_p' \\
 \mathbold{\hat{y}} \cdot \left[ \mathbold{k} \times \mathcal{A}_p \mathbold{\hat{z}} + \mathbold{k''} \times \mathcal{A}_p'' \mathbold{\hat{z}} \right] = \mathbold{\hat{y}} \cdot \left[\mathbold{\tilde{k}'}  \times \mathcal{A}_p' \mathbold{\hat{z}} \right] &\implies \mathcal{A}_p + \mathcal{A}_p'' = \mathcal{A}_p' \\
 \mathbold{\hat{y}} \cdot \left[ \mathcal{A}_p \mathbold{\hat{z}} + \mathcal{A}_p'' \mathbold{\hat{z}} \right] = \mathbold{\hat{y}} \cdot \mathcal{A}_p' \mathbold{\hat{z}} &\implies 0 = 0 .
 %\end{split}
 \end{align}
\end{subequations}
These expressions can be rearranged as
\begin{equation}\label{eq:p_coeff}
 \tilde{r}_p \equiv \frac{\mathcal{A}_p''}{\mathcal{A}_p} = \frac{\tilde{\nu}^2(\omega) k_y  - \tilde{k_y}'}{\tilde{k_y}' + \tilde{\nu}^2(\omega) k_y} \quad \text{and} \quad \tilde{t}_p \equiv \frac{\mathcal{A}_p'}{\mathcal{A}_p} = \frac{2 \tilde{\nu}^2(\omega) k_y}{\tilde{k_y}' + \tilde{\nu}^2(\omega) k_y} , 
 \end{equation}
with the former giving the following expression for reflectivity in p-polarization:
\begin{equation}\label{eq:p_reflectivity}
 \mathcal{R}_p (\zeta) = \norm{\tilde{r}_p}^2 = \norm{\left( \frac{\tilde{\nu}^2 (\omega) \sin \left( \zeta \right) - \sqrt{\tilde{\nu}^2 (\omega) - \cos^2 \left( \zeta \right)}}{\tilde{\nu}^2 (\omega) \sin \left( \zeta \right) + \sqrt{\tilde{\nu}^2 (\omega) - \cos^2 \left( \zeta \right)}} \right)}^2 .
 \end{equation}

The expressions for $\mathcal{R}_s$ and $\mathcal{R}_p$ in \cref{eq:s_reflectivity,eq:p_reflectivity} are the reflection coefficients found in the \emph{Fresnel equations} \cite{Landau60,Jackson75,Born80,Griffiths17,Attwood17,Fresnel1823}, which assume a perfectly smooth surface. 
However, these expressions can be simplified in the soft x-ray regime with $\nu (\omega) \lessapprox 1$ and $\xi(\omega) \ll 1$ using 
\begin{subequations}
\begin{equation}\label{eq:index_approx}
 \tilde{\nu}^2 (\omega) = \nu^2 (\omega) + 2 i \nu(\omega) \xi(\omega) - \xi^2(\omega) \approx \nu^2 (\omega) + 2 i \xi(\omega) ,
 \end{equation}
where $\nu^2 (\omega) = \left[ 1 - \zeta^2_c (\omega) / 2 \right]^2 \approx 1 - \zeta^2_c (\omega)$ from the approximate definition of $\zeta_c (\omega)$ given by \cref{eq:critical_angle} \cite{Attwood17}. 
Then, \cref{eq:index_approx} becomes
\begin{equation}
 \tilde{\nu}^2 (\omega) \approx 1 - \zeta^2_c (\omega) + 2 i \xi(\omega) ,
 \end{equation}
\end{subequations}
and additionally, at grazing-incidence angles with $\zeta \ll 1$ it is justified to use the approximations $\sin \left( \zeta \right) \approx \zeta$ and $\cos^2 \left( \zeta \right) \approx 1 - \zeta^2$ so that the reflection coefficients can be written as  
\begin{subequations}
\begin{equation}\label{eq:s_refl_approx}
 \tilde{r}_s \approx \left( \frac{\zeta - \sqrt{\zeta^2 - \zeta^2_c (\omega) + 2 i \xi(\omega)}}{\zeta + \sqrt{\zeta^2 - \zeta^2_c (\omega) + 2 i \xi(\omega)}} \right)\quad \text{and} 
 \end{equation}
\begin{equation}\label{eq:p_refl_approx}
 \tilde{r}_p \approx \left( \frac{\zeta \left[ 1 - \zeta^2_c (\omega) + 2 i \xi(\omega) \right] - \sqrt{\zeta^2 - \zeta^2_c (\omega) + 2 i \xi(\omega)}}{\zeta \left[ 1 - \zeta^2_c (\omega) + 2 i \xi(\omega) \right] + \sqrt{\zeta^2 - \zeta^2_c (\omega) + 2 i \xi(\omega)}} \right) . 
 \end{equation}
\end{subequations}
Knowing that $\xi(\omega) \, \zeta  \ll  1$ and $\zeta^2_c (\omega) \zeta  \ll  1$ with $\zeta  \ll  1$, these relations indicate that reflectivity in s- and p-polarizations are virtually equal to each other at grazing-incidence angles (\emph{i.e.}, $\mathcal{R}_F \approx \mathcal{R}_s \approx \mathcal{R}_p$). 
\begin{figure}
 \centering
 \includegraphics[scale=0.95]{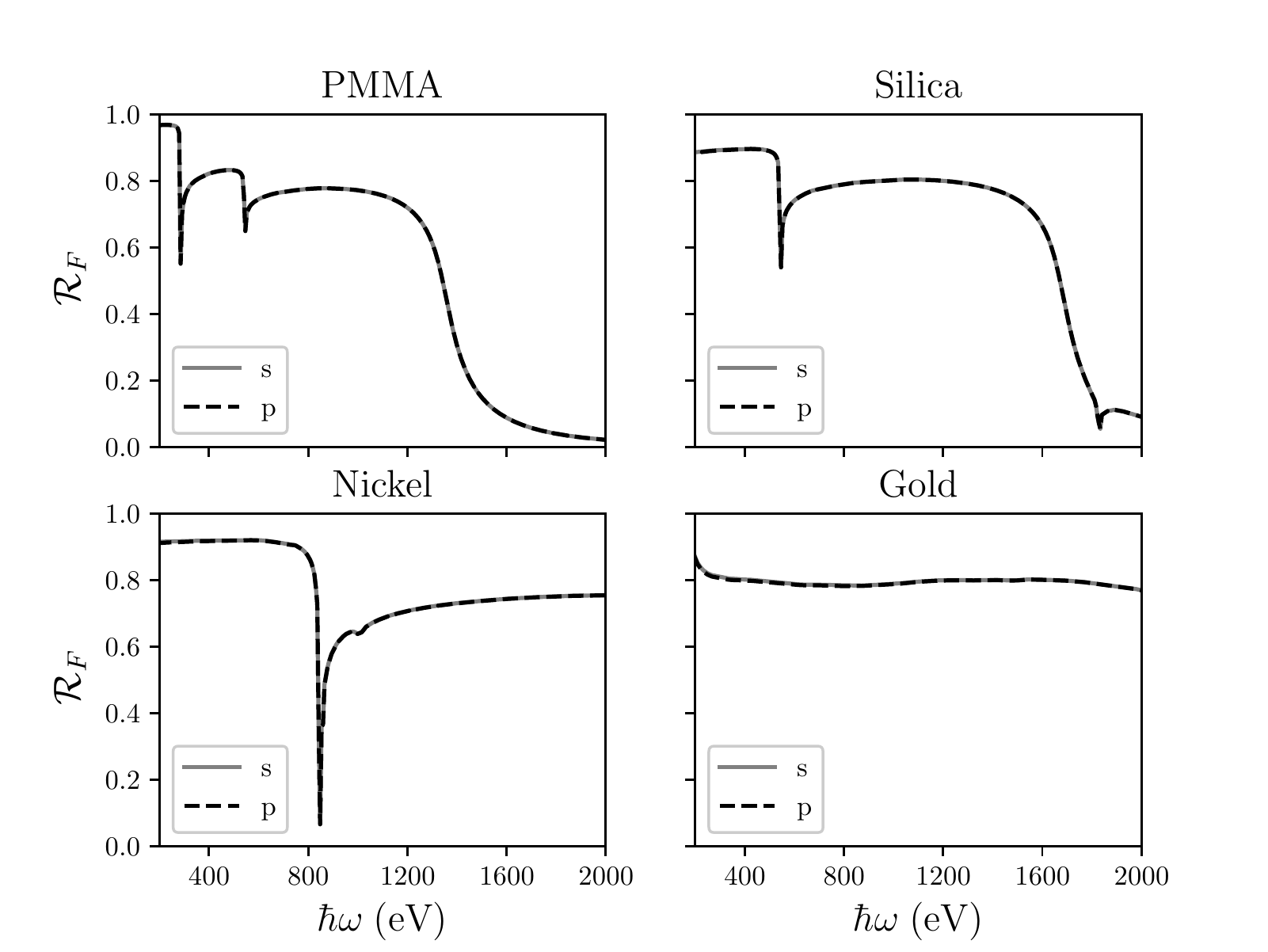}
 \caption[Fresnel soft x-ray reflectivity in orthogonal polarizations at a $1^{\circ}$ graze angle for PMMA, silica, nickel and gold]{Fresnel soft x-ray reflectivity in orthogonal polarizations at $\zeta = 1^{\circ}$ for PMMA (\ce{C5H8O2})$_n$, silica (\ce{SiO2}), nickel (\ce{Ni}) and gold (\ce{Au}), demonstrating lack of polarization sensitivity \cite{CXRO_database}}\label{fig:X-ray_refl}
 \end{figure} 
This is demonstrated graphically in \cref{fig:X-ray_refl}, where reflectivity data gathered from CXRO \cite{CXRO_database} are plotted for the materials from \cref{fig:X-ray_index,fig:X-ray_lengths,fig:critical_angle}.\footnote{Note, however, that the condition for TER, $\zeta = 1^{\circ} < \zeta_c (\omega)$, is not satisfied toward the blue end of the soft x-ray spectrum for PMMA and silica, which causes reflectivity to approach very small values.} 

With an insensitivity to polarization at grazing incidence, the expressions for $\mathcal{R}_s$ formulated above can be used to describe Fresnel reflectivity for a substrate or a thick slab of material (\emph{i.e.}, $\mathcal{R}_{\text{slab}} \equiv \mathcal{R}_F \approx \mathcal{R}_s$). 
For oxidizing materials, however, a thin film of oxide can form on a surface exposed to air. 
As alluded to in \cref{sec:crystal_etching}, this is the case for a silicon substrate with a layer of native oxide (\ce{SiO2}) that is a few \si{\nm} thick; another example is nickel oxide (NiO) forming on a layer of nickel. 
In any such case, taking $\tilde{\nu}_1 (\omega)$ as the complex index of refraction for the oxide layer, the complex reflection coefficient for the first interface in s-polarization is given by 
\begin{subequations}
\begin{equation}\label{eq:refl_coeff01}
 \tilde{r}_{0,1} = \frac{k_{y,0} - \tilde{k}_{y,1}}{k_{y,0} + \tilde{k}_{y,1}} = \frac{\sin \left( \zeta \right) - \sqrt{\tilde{\nu}_1^2 (\omega) - \cos^2 \left( \zeta \right)}}{\sin \left( \zeta \right) + \sqrt{\tilde{\nu}_1^2 (\omega) - \cos^2 \left( \zeta \right)}} ,
 \end{equation}
where $k_{y,0} = k_0 \sin \left( \zeta \right)$ and $\tilde{k}_{y,1} = k_0 \sqrt{\tilde{\nu}_1^2 (\omega) - \cos^2 \left( \zeta \right)}$ are the components of the wave vector normal to the boundary, in vacuum and in the oxide layer, respectively. 
Similarly, with $\tilde{\nu}_2 (\omega)$ as the complex index of refraction for the bulk material,\footnote{\emph{e.g.}, a silicon substrate or a relatively thick nickel layer} the complex reflection coefficient for the second interface, in the same polarization is 
\begin{equation}\label{eq:refl_coeff12}
 \tilde{r}_{1,2} = \frac{\tilde{k}_{y,1} - \tilde{k}_{y,2}}{\tilde{k}_{y,1} + \tilde{k}_{y,2}} = \frac{\sqrt{\tilde{\nu}_1^2 (\omega) - \cos^2 \left( \zeta \right)} - \sqrt{\tilde{\nu}_2^2 (\omega) - \cos^2 \left( \zeta \right)}}{\sqrt{\tilde{\nu}_1^2 (\omega) - \cos^2 \left( \zeta \right)} + \sqrt{\tilde{\nu}_2^2 (\omega) - \cos^2 \left( \zeta \right)}} , 
 \end{equation}
\end{subequations}
where $\tilde{k}_{y,2} = k_0 \sqrt{\tilde{\nu}_2^2 (\omega) - \cos^2 \left( \zeta \right)}$ is the component of the wave vector normal to the surface in the bulk material. 

\subsection{Penetration Depth for Total External Reflection}\label{sec:total_refl}
%%%%%%%%%%%%%%%%%%%%%%%%%%%%%%%%%%%%%%%%%--------------------------------------------------
As described at the start of \cref{sec:planar_interface}, the refracted ray in the regime of TER only travels exactly parallel to the surface of the mirror if the imaginary part of the refractive index, $\xi (\omega)$, is zero.  
In such a hypothetical scenario where $\zeta < \zeta_c (\omega)$ and $\xi (\omega) = 0$, the approximate reflection coefficients given by \cref{eq:s_refl_approx,eq:p_refl_approx} become
\begin{subequations}
\begin{equation}\label{eq:s_refl_approx_TER}
 \tilde{r}_s \approx \left( \frac{\zeta - i \sqrt{\zeta^2_c (\omega) - \zeta^2}}{\zeta + i \sqrt{\zeta^2_c (\omega) - \zeta^2}} \right)\quad \text{and} 
 \end{equation}
\begin{equation}\label{eq:p_refl_approx_TER}
 %\begin{split}
 \tilde{r}_p \approx \left( \frac{\zeta \left[ 1 - \zeta^2_c (\omega) \right] - i \sqrt{\zeta^2_c (\omega) - \zeta^2}}{\zeta \left[ 1 - \zeta^2_c (\omega) \right] + i \sqrt{\zeta^2_c (\omega) - \zeta^2}} \right) , 
 \end{equation}
\end{subequations}
which yield $\mathcal{R} = \norm{\tilde{r}}^2 =1$ for both polarizations. 
While this is assumed for the purposes of defining $\zeta_c (\omega)$ in \cref{eq:critical_angle}, it is necessary to consider the effect of $\xi (\omega) \neq 0$ to describe how the radiation penetrates slightly into the bulk of a mirror or reflection grating with a complex index of refraction $\tilde{\nu} (\omega)$ \cite{Attwood17,Vinogradov85}. 
To do this, the refracted ray is treated as having a complex wave vector $\mathbold{\tilde{k}}' \equiv \mathbold{k}' + i \mathbold{\kappa}'$ with $\tilde{k_z}'=0$ [\emph{cf.\@} \cref{sec:x-ray_index,fig:2Dmirror_angles}]:
\begin{equation}\label{eq:refacted_ray_TER}
 \mathbold{\tilde{k}}' = \tilde{k_x}' \mathbold{\hat{x}} + \tilde{k_y}' \mathbold{\hat{y}} = \tilde{k}' \left[ \cos \left( \tilde{\zeta}' \right) \mathbold{\hat{x}} - \sin \left( \tilde{\zeta}' \right) \mathbold{\hat{y}} \right] , 
 \end{equation}
where $\tilde{\zeta}' \equiv \zeta'_{\Re} + i \zeta'_{\Im}$ is a complex angle, $\tilde{k}' \equiv \tilde{\nu} (\omega) k_0$ is the complex wave number in the material and $k_0 \equiv 2 \pi / \lambda$ is the wave number in vacuum. 

According to Snell's law [\emph{cf.\@} \cref{eq:refraction}], $\tilde{k_x}' = k_x$, with $k_x \equiv k_0 \cos \left( \zeta \right)$ as the $x$-component of the incident wave vector from \cref{eq:2D_incident_wave} and hence
\begin{equation}\label{refract_ky_gen}
 \tilde{k_y}' \equiv - \sqrt{\tilde{k}'^2 - k_x^2} = - k_0 \sqrt{\tilde{\nu}^2 (\omega) - \cos^2 \left( \zeta \right)} .
 \end{equation}
Because $\Im \left[ \tilde{k_x}' \right] \equiv \kappa_x' = 0$ and $\Im \left[ \tilde{k_y}' \right] \equiv \kappa_y' \neq 0$ by Snell's law, $\mathbold{k}' = k_x \mathbold{\hat{x}} + k_y' \mathbold{\hat{y}}$ and $\mathbold{\kappa}' = \kappa_y' \mathbold{\hat{y}}$ point in different directions with the refracted wave being distorted and non-planar rather than transverse \cite{Attwood17}. 
The electromagnetic fields of such a wave are then attenuated by a factor of $\mathrm{e}^{- \mathbold{\kappa}' \cdot \mathbold{r}} = \mathrm{e}^{- \kappa_y' y}$ as they penetrate into the material, where $\kappa_y' < 0$ and $y < 0$.  
The corresponding intensity then drops to $1/\mathrm{e} \approx \SI{37}{\percent}$ of its initial value over a $y$-distance $y_{1/\mathrm{e}} < 0$: 
\begin{subequations}
\begin{equation}
 \mathrm{e}^{-1} \equiv \mathrm{e}^{- 2 \kappa_y' y_{1/\mathrm{e}}} \implies y_{1/\mathrm{e}} \equiv \frac{1}{2} \kappa_y'^{-1} .
 \end{equation}
This \emph{penetration depth} can be written generally as \cite{Attwood17,Vineyard82,Als-Nielsen11}
\begin{equation}\label{eq:penetration_depth}
 \mathcal{D}_{\perp} \equiv \abs{y_{1/\mathrm{e}}} = \frac{1}{2} \abs{\Im \left[ \tilde{k_{\perp}}' \right]}^{-1} = \frac{\lambda}{4 \pi \sqrt{\tilde{\nu}^2 (\omega) - \cos^2 \left( \zeta \right)}} ,
 \end{equation}
\end{subequations}
where $\tilde{k_{\perp}}' \equiv \tilde{k_y}'$ is component of the complex wave vector perpendicular to the surface of the mirror with $\Im \left[ \tilde{k_{\perp}}' \right] \equiv \kappa_y'$ in this coordinate system. 

A simpler, approximate expression for $\mathcal{D}_{\perp}$ can be derived\footnote{The following analysis follows closely from Attwood and Sakdinawat~\cite{Attwood17}, chapter 3.} by first using trigonometric identities similar to \cref{eq:refractionA,eq:refractionB} to rewrite $\tilde{k_x}'$ and $\tilde{k_y}'$ as 
\begin{subequations}
\begin{align}
 \tilde{k_x}' \equiv k_x' + i \kappa_x &= \tilde{k}' \left[ \cos \left( \zeta'_{\Re} \right) \cosh \left( \zeta'_{\Im} \right) - i \sin \left( \zeta'_{\Re} \right) \sinh \left( \zeta'_{\Im} \right) \right] \label{eq:refract_x_1} \\ 
 \tilde{k_y}' \equiv k_y' + i \kappa_y &= - \tilde{k}' \left[ \sin \left( \zeta'_{\Re} \right) \cosh \left( \zeta'_{\Im} \right) + i \cos \left( \zeta'_{\Re} \right) \sinh \left( \zeta'_{\Im} \right) \right] \label{eq:refract_x_2} 
 \end{align}
\end{subequations}
but because $\zeta'_{\Re}$ deviates only slightly from zero for $\zeta < \zeta_c (\omega)$, it is justified to make the small-angle approximations $\cos \left( \zeta'_{\Re} \right) \approx 1 - \zeta'^2_{\Re} / 2$ and $\sin \left( \zeta'_{\Re} \right) \approx \zeta'_{\Re}$. 
Further, with $\xi (\omega) \ll 1$ [\emph{cf.\@} \cref{sec:x-ray_index}], $\zeta'_{\Im}$ is expected to be a small quantity and hence the hyperbolic trigonometric functions in \cref{eq:refract_x_1,eq:refract_x_2} can be approximated as $\sinh \left( \zeta'_{\Im} \right) \approx \zeta'_{\Im}$ and $\cosh \left( \zeta'_{\Im} \right) \approx 1 + \zeta'^2_{\Im}/2$ so that $\tilde{k_x}'$ and $\tilde{k_y}'$ reduce to \cite{Attwood17}
\begin{subequations}
\begin{align}
 \tilde{k_x}' &\approx \tilde{k}' \left[\left( 1 - \frac{\zeta'^2_{\Re}}{2} \right) \left( 1 + \frac{\zeta'^2_{\Im}}{2} \right) - i \zeta'_{\Re} \zeta'_{\Im} \right] \label{eq:refract_x_1_approx} \\
 \tilde{k_y}' &\approx - \tilde{k}' \left[ \zeta'_{\Re} \left( 1 + \frac{\zeta'^2_{\Im}}{2} \right) + i \left( 1 - \frac{\zeta'^2_{\Re}}{2} \right) \zeta'_{\Im} \right]\label{eq:refract_x_2_approx} .
 \end{align} 
\end{subequations}
It is then useful to split \cref{eq:refract_x_1_approx,eq:refract_x_2_approx} into real and imaginary components with the complex wave number expanded as $\tilde{k}' = k_0 \left[ \nu(\omega) + i \xi (\omega) \right]$: 
\begin{subequations}
\begin{align}
 k_x' \equiv \Re \left[ \tilde{k_x}' \right] &= k_0 \left[ \nu(\omega) \left( 1 - \frac{\zeta'^2_{\Re}}{2} \right) \left( 1 + \frac{\zeta'^2_{\Im}}{2} \right) + \xi (\omega) \zeta'_{\Re} \zeta'_{\Im} \right] \label{eq:real_refract_x} \\
 \kappa_x' \equiv \Im \left[ \tilde{k_x}' \right] &= k_0 \left[ \xi(\omega) \left( 1 - \frac{\zeta'^2_{\Re}}{2} \right) \left( 1 + \frac{\zeta'^2_{\Im}}{2} \right) - \nu (\omega) \zeta'_{\Re} \zeta'_{\Im} \right] \label{eq:imag_refract_x} \\ 
 k_y' \equiv \Re \left[ \tilde{k_y}' \right] &= - k_0 \left[ \nu(\omega) \zeta'_{\Re} \left( 1 + \frac{\zeta'^2_{\Im}}{2} \right) - \xi(\omega) \left( 1 - \frac{\zeta'^2_{\Re}}{2} \right) \zeta'_{\Im} \right] \label{eq:real_refract_y} \\
 \kappa_y' \equiv \Im \left[ \tilde{k_y}' \right] &= - k_0 \left[ \xi(\omega) \zeta'_{\Re} \left( 1 + \frac{\zeta'^2_{\Im}}{2} \right) + \nu(\omega) \left( 1 - \frac{\zeta'^2_{\Re}}{2} \right) \zeta'_{\Im} \right] \label{eq:imag_refract_y} .
 \end{align}
\end{subequations}
Applying the small-angle approximation $k_x \approx k_0 ( 1 - \zeta^2 / 2 )$ and $k_x' = k_x$ from Snell's law to \cref{eq:real_refract_x} yields 
\begin{subequations}
\begin{equation}\label{eq:real_x_condition}
 \nu(\omega) \left( 1 - \frac{\zeta'^2_{\Re}}{2} \right) \left( 1 + \frac{\zeta'^2_{\Im}}{2} \right) + \xi (\omega) \zeta'_{\Re} \zeta'_{\Im} \approx 1 - \zeta^2 / 2 
 \end{equation}
while \cref{eq:imag_refract_x} with $\kappa_x' = 0$ gives
\begin{equation}\label{eq:imag_x_condition}
 \xi(\omega) \left( 1 - \frac{\zeta'^2_{\Re}}{2} \right) \left( 1 + \frac{\zeta'^2_{\Im}}{2} \right) - \nu (\omega) \zeta'_{\Re} \zeta'_{\Im} = 0 . 
 \end{equation} 
\end{subequations}
Keeping only first-order $\zeta'_{\Re}$ and $\zeta'_{\Im}$ terms, \cref{eq:imag_x_condition,eq:imag_refract_x} can be used to arrive at an approximate expression for $\zeta'_{\Im}$: 
\begin{equation}\label{eq:approx_relation_real_imag_angles}
 \zeta'_{\Im} \approx \frac{\xi(\omega)}{\zeta'_{\Re} \nu (\omega)} 
 \end{equation}
\begin{subequations}
and after dividing by $\nu(\omega)$ and inserting \cref{eq:approx_relation_real_imag_angles} for $\zeta'_{\Im}$, \cref{eq:real_x_condition} becomes 
\begin{equation}
 1 + \frac{\xi^2 (\omega)}{2 \zeta'^2_{\Re}} - \frac{\zeta'^2_{\Re}}{2} - \underbrace{\frac{\xi^2 (\omega)}{3 \nu^2 (\omega)}}_{\approx 0} \approx \nu^{-1} (\omega) \left( 1 - \frac{\zeta^2}{2} \right) , \label{eq:real_x_condition_approx} %\left[ 1 + \delta_{\nu} (\omega) \right] 
 \end{equation}
where the term with an under-brace can be neglected since $\xi (\omega) \ll 1$ while $\nu (\omega) \approx 1$. 
Using $\nu^{-1} (\omega) \equiv \left[ 1 - \delta_{\nu} (\omega) \right]^{-1} \approx 1 + \delta_{\nu} (\omega)$ for $\delta_{\nu} (\omega) \ll 1$, \cref{eq:real_x_condition_approx} can be further approximated as 
\begin{equation}
 \frac{\xi^2 (\omega)}{2 \zeta'^2_{\Re}} - \frac{\zeta'^2_{\Re}}{2} \approx \delta_{\nu} (\omega) - \frac{\zeta^2}{2} - \underbrace{\frac{\zeta^2 \delta_{\nu} (\omega)}{2}}_{\approx 0} , \label{eq:real_x_condition_approx2} 
 \end{equation}
where, because $\zeta \ll 1$ and $\delta_{\nu} (\omega) \ll 1$, the term with an underbrace is comparatively small and hence can be ignored. 
Using $\delta_{\nu} (\omega) \approx \zeta_c^2 (\omega) / 2$ from the approximate definition of $\zeta_c (\omega)$ [\emph{cf.\@} \cref{eq:critical_angle}], \cref{eq:real_x_condition_approx2} yields a quadratic formula for $\zeta'^2_{\Re}$:
\begin{equation}\label{eq:real_x_condition_mod_quad}
 \zeta'^4_{\Re} - \left[ \zeta^2 - \zeta^2_c (\omega)  \right] \zeta'^2_{\Re} - \xi^2 (\omega) = 0 .
 \end{equation}
\end{subequations}

With solutions to \cref{eq:real_x_condition_mod_quad} of the form
\begin{equation}\label{eq:real_x_condition_mod_quad_sol}
 \zeta'^2_{\Re} = \frac{1}{2} \left( \zeta^2 - \zeta^2_c (\omega) \pm \sqrt{\left[ \zeta^2_c (\omega) - \zeta^2 \right]^2 + 4 \xi^2 (\omega)} \right) , 
 \end{equation}
an approximate, real expression for $\zeta'_{\Re}$ can be written using the $+$ sign solution: 
\begin{equation}
 \zeta'_{\Re} = \frac{1}{\sqrt{2}} \sqrt{\zeta^2 - \zeta^2_c (\omega) + \left[ \zeta^2_c (\omega) - \zeta^2  \right] \sqrt{1 + 4 \frac{\xi^2 (\omega)}{\left[ \zeta^2_c (\omega) - \zeta^2  \right]^2}}} \label{eq:real_refr_angle_sol} ,
 \end{equation}
where in the limit of small $\zeta$, it is justified to use the following Taylor-series expansion:
\begin{equation}
 \sqrt{1 + 4 X^2} \approx 1 + 2 X^2 \quad \text{with} \quad X \equiv \frac{\xi (\omega)}{\zeta^2_c (\omega) - \zeta^2} .
 \end{equation}
Then, $\zeta'_{\Re}$ in \cref{eq:real_refr_angle_sol} can be further approximated as 
\begin{subequations}
\begin{equation}\label{eq:zetaprime_real}
 \zeta'_{\Re} \approx \frac{\xi(\omega)}{\sqrt{\zeta^2_c (\omega) - \zeta^2}} 
 \end{equation}
while from \cref{eq:approx_relation_real_imag_angles} using $\nu (\omega) \approx 1$, 
\begin{equation}\label{eq:zetaprime_imag}
 \zeta'_{\Im} \approx \frac{\xi(\omega)}{\zeta'_{\Re} \nu (\omega)} \approx \sqrt{\zeta^2_c (\omega) - \zeta^2} .
 \end{equation} 
\end{subequations}
Together, \cref{eq:zetaprime_real,eq:zetaprime_imag} give an approximate expression for the complex angle associated with the refracted ray in \cref{eq:refacted_ray_TER}, $\tilde{\zeta}'$, assuming $\zeta \ll \zeta_c (\omega)$:
\begin{equation}
 \tilde{\zeta}' \equiv \zeta'_{\Re} + i \zeta'_{\Im} \approx \frac{1}{\sqrt{\zeta^2_c (\omega) - \zeta^2}} \left( \xi(\omega) + i \left[ \zeta^2_c (\omega) - \zeta^2 \right] \right) ,
 \end{equation}
which can be used to arrive at an approximate expression for $\kappa_y'$. 
From \cref{eq:imag_refract_y}, 
\begin{subequations}
\begin{equation}
 \kappa_y' \approx - k_0  \nu(\omega) \left( 1 - \frac{\zeta'^2_{\Re}}{2} \right) \zeta'_{\Im} = - k_0 \nu(\omega) \left(1 - \frac{\xi^2(\omega)}{\zeta^2_c (\omega) - \zeta^2} \right) \sqrt{\zeta^2_c (\omega) - \zeta^2} ,
 \end{equation} 
where the term multiplied by $\xi(\omega)$ has been neglected since $\xi(\omega) \ll \nu(\omega)$ while moreover, taking $\nu (\omega) \approx 1$ and neglecting the small $\xi^2(\omega)$ term gives 
\begin{equation}
 \kappa_y' \approx - k_0 \sqrt{\zeta^2_c (\omega) - \zeta^2} . 
 \end{equation}
\end{subequations}
The penetration depth from \cref{eq:penetration_depth} then becomes 
\begin{equation}\label{eq:penetration_depth_approx}
 \mathcal{D}_{\perp} = \frac{1}{2} \abs{\kappa_y'}^{-1} \approx \frac{\lambda}{4 \pi \sqrt{\zeta^2_c (\omega) - \zeta^2}} ,
 \end{equation}
which suggests that materials with larger critical angles are associated with smaller values of $\mathcal{D}_{\perp}$ for fixed $\zeta$. 
However, as $\zeta$ approaches $\zeta_c (\omega)$, \cref{eq:penetration_depth_approx} no longer holds and the fields penetrate further into the material \cite{Attwood17,Gibaud2009}. 
\begin{figure}
 \centering
 \includegraphics[scale=0.95]{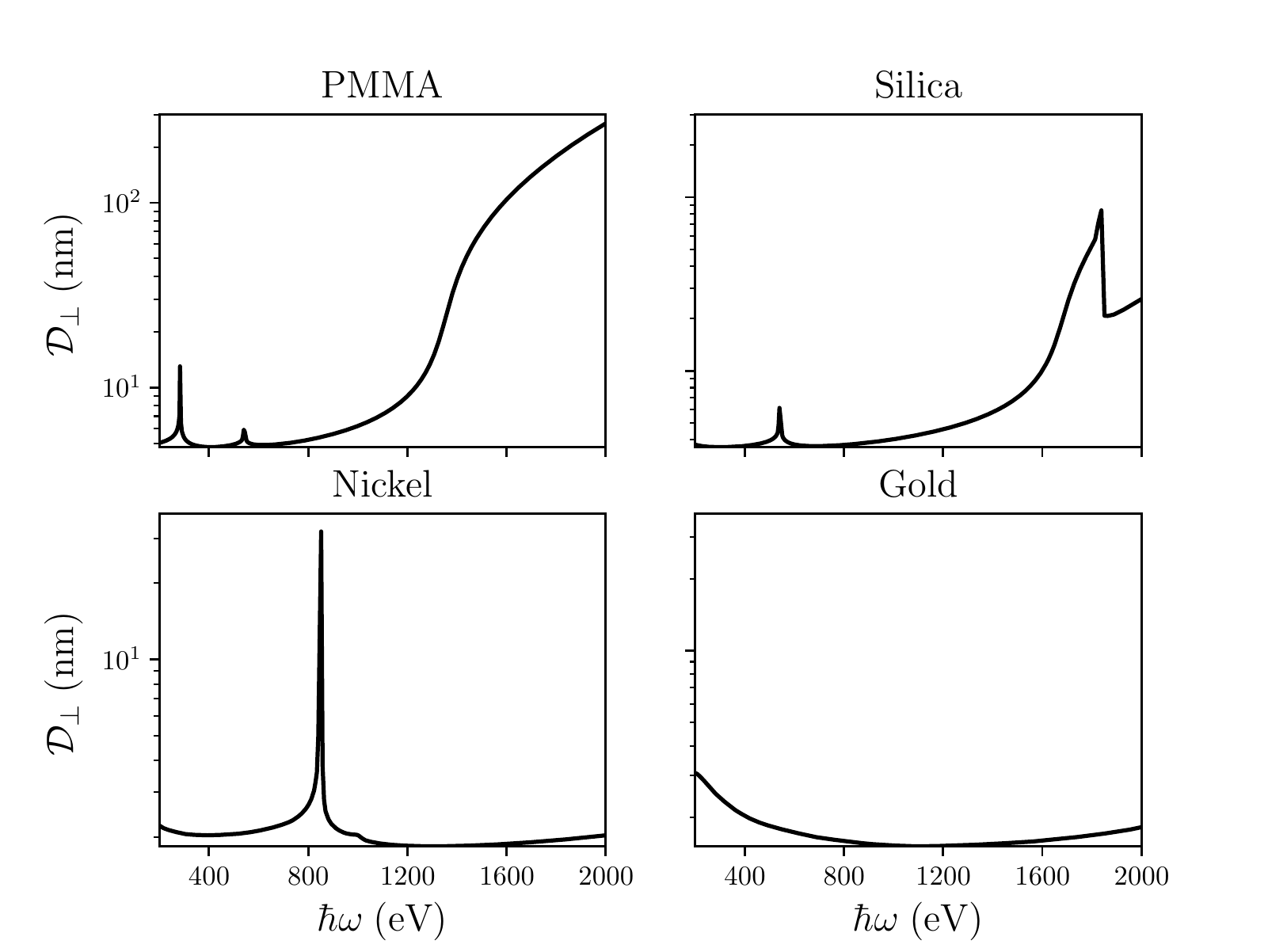}
 \caption[TER penetration depth for PMMA, silica, nickel and gold]{TER penetration depth, $\mathcal{D}_{\perp}$, at $\zeta = 1^{\circ}$ for PMMA (\ce{C5H8O2})$_n$, silica (\ce{SiO2}), nickel (\ce{Ni}) and gold (\ce{Au}) \cite{CXRO_database}}\label{fig:X-ray_atten}
 \end{figure} 
This is demonstrated in \cref{fig:X-ray_atten}, where the exact expression for $\mathcal{D}_{\perp}$ using \cref{eq:penetration_depth} with $\zeta = 1^{\circ}$ is plotted across the soft x-ray spectrum for the materials from \cref{fig:X-ray_index,fig:X-ray_lengths,fig:critical_angle,fig:X-ray_refl}. 
It is seen in these plots that $\mathcal{D}_{\perp}$ sharply increases at absorption edges and over other spectral regions where $\zeta \gtrapprox \zeta_c (\omega)$, which is the case for PMMA and silica toward the blue end of the soft x-ray spectrum. 

A mirror flat designed for soft x-rays under TER in practice can be a layer of an appropriate material \numrange{4}{5} $\mathcal{D}_{\perp}$ thick coated on a substrate such as silicon or fused silica. 
Over such a depth, \SIrange{98}{99}{\percent}  of the incident wave intensity is attenuated and as a result, reflections that occur at the interface of the coating and the substrate can be neglected. 
This principle also holds for x-ray reflection gratings, where, as alluded to in \cref{sec:grating_tech_dev}, only a relatively thin layer of a reflective material need be coated on a grating surface relief. 
The penetration depth data from \cref{fig:X-ray_atten} combined with the reflectivity data from \cref{fig:X-ray_refl} indicate that a gold coating provides an efficient, broadband response for grazing-incidence soft x-rays with a relatively thin deposited layer (typically \SIrange{10}{15}{\nm}). 
This is beneficial for x-ray reflection gratings with groove spacings of a few hundred \si{\nm} in terms of diffraction efficiency, both for the overall reflectivity and for maintaining the sawtooth groove shape that affects how efficiency is distributed among propagating orders [\emph{cf.\@} \cref{ch:diff_eff,ap:grating_basics}]. 
Additionally, nickel has similar properties for $\hbar \omega \lessapprox \SI{800}{\electronvolt}$ and may be appropriate in situations where the prominent L-shell absorption edge can be avoided. 

In principle, a thin, reflective overcoat for an x-ray reflection grating can be produced through a variety of \emph{chemical vapor} or \emph{physical vapor} deposition techniques \cite{Franssila10}. 
While chemical vapor deposition (especially \emph{atomic layer deposition} \cite{George10}) is advantageous for producing conformal coatings on corrugated surfaces, these processes are very time consuming and therefore become impractical in terms of throughput for spectrometers that require many grating replicas. 
Physical-vapor-deposition techniques such as \emph{plasma sputtering} or \emph{electron-beam vapor deposition} \cite{Singh05} instead are commonly used for this application with the level of coating conformity depending on the geometry inside the processing chamber. 
The nature of the surface to be coated, however, must be taken into account to ensure that a quality film is produced. 
That is, for a grating mold etched in silicon with native oxide or patterned in resist that contains oxygen, the overcoat material must also be oxidizing to promote wetting and adhesion \cite{Trolier-McKinstry17}. 
While nickel satisfies this condition, gold does not oxidize and therefore a thin film of an oxidizing metal such as chromium or titanium must be first deposited to provide a wetted, metallic surface for a gold layer to adhere to. 
Moreover, the wettability of a reflective overcoat material also contributes to the level of surface roughness on a mirror or similarly, the blazed groove facets of a reflection grating. 

\subsection{Surface Roughness}\label{sec:rough_surface}
%%%%%%%%%%%%%%%%%%%%%%%%%%%%%%%%%%%%%%%%%--------------------------------------------------
Any realistic surface features some level of roughness that, depending on the incident wave vector and the surface profile, serves to reduce specular reflectivity relative to what is predicted by the Fresnel equations [\emph{cf.\@} \cref{sec:reflectivity_polarization}]. 
In principle, a rough surface can be described with a pseudo-random surface profile, $y = Y(x,z)$, where $x$, $y$ and $z$ are the coordinates defined in \cref{fig:2Dmirror_angles} with the average value $\langle Y(x,z) \rangle = 0$ corresponding to a perfectly smooth surface in the plane defined by $y = 0$. 
While the exact analytical form of the function $Y(x,z)$ is unknown in practice, its statistical properties can be used to describe approximately the effect that a rough surface has on incident radiation. 
First, this function fluctuates in value to a level characterized by the variance of $Y(x,z)$:
\begin{equation}\label{eq:surface_variance}
 \sigma^2 \equiv \langle Y^2(x,z) \rangle - \underbrace{\langle Y(x,z) \rangle^2}_0 ,
 \end{equation} 
where $\sigma$ is the \emph{root mean square (RMS) roughness} that describes the typical spread in surface feature heights \cite{Beckmann63,Ogilvy87,Als-Nielsen11,Sentenac2009}.   
Additionally, information on the spread of lateral sizes on the surface is contained within the \emph{autocorrelation function} associated with $Y(x,z)$ \cite{Beckmann63,Hogrefe87,Ogilvy87,Sentenac2009}: 
\begin{equation}\label{eq:auto_correlation}
 C (\mathbold{\tau}) = \frac{\langle Y(\mathbold{\tau}) Y(\mathbf{0}) \rangle}{\sigma^2} ,
 \end{equation}
where $\mathbold{\tau} \equiv \tau_x  \mathbold{\hat{x}} + \tau_z  \mathbold{\hat{z}}$ is a position vector confined to the $y = 0$ plane that represents the directional distance between an arbitrary $(x,z)$ point and an origin $(0,0) \equiv \mathbf{0}$. 

Taking the Fourier transform of \cref{eq:auto_correlation} gives a \emph{power spectral density (PSD) function}, which describes the distribution of spatial frequency components making up the rough surface \cite{Beckmann63,Wen15,Ogilvy87}:
\begin{subequations}
\begin{equation}\label{eq:PSD_general}
 PSD \left( \mathbold{K} \right) = \int_{-\infty}^{\infty} C (\mathbold{\tau}) \, \mathrm{e}^{i \mathbold{K} \cdot \mathbold{\tau}} \dd[2]{\mathbold{\tau}} ,
 \end{equation}
where $\mathbold{K} \equiv K_x  \mathbold{\hat{x}} + K_z  \mathbold{\hat{z}}$ is a spatial analog to $\mathbold{\tau}$. 
The exponential term in the case of an isotropic surface becomes $\mathbold{K} \cdot \mathbold{\tau} = K \tau \cos (\phi)$ with $\tau \equiv |\mathbold{\tau}|$, $K \equiv |\mathbold{K}|$ and $\tan (\phi) \equiv \tau_z / \tau_x$ so that \cref{eq:PSD_general} becomes 
\begin{equation}\label{eq:PSD_isotropic}
 PSD \left( K \right) = \int_{0}^{2 \pi} \int_{0}^{\infty} C (\tau) \, \mathrm{e}^{i K \tau \cos (\phi)} \tau \dd{\tau} \dd{\phi} = 2 \pi \int_{0}^{\infty} C (\tau) \, J_0 \left( K \tau \right) \tau \dd{\tau} ,
 \end{equation}
\end{subequations}
where $J_0 \left(K \tau \right)$ is the $0^{\text{th}}$-order \emph{Bessel function of the first kind}\footnote{Here, \emph{Bessel's first integral} is used with $n=0$: 
\begin{equation*}
 J_n (w) = \frac{1}{2 \pi i^n} \int_0^{2 \pi} \mathrm{e}^{i w \cos (\phi)} \mathrm{e}^{i n \phi} \dd{\phi} .
 \end{equation*}} 
and under this assumption of isotropy, $C (\mathbold{\tau}) = C (\tau)$. 
Moreover, this autocorrelation function is often taken to be a normal distribution: 
\begin{equation}\label{eq:auto_correlation_normal}
 C (\tau) = \mathrm{e}^{- \tau^2 / \ell_{\text{corr}}^2} ,
 \end{equation}
where $\ell_{\text{corr}}$ is the \emph{correlation length}, which is defined by the value of $\tau$ that corresponds to $C (\tau)$ reaching $1/\mathrm{e} \approx \SI{37}{\percent}$ of its maximum value \cite{Beckmann63,Sentenac2009,Ogilvy87,Hogrefe87}. 
Physically, $\ell_{\text{corr}}$ represents a typical size scale of a rough surface feature: $\ell_{\text{corr}} \to 0$ represents very small features that approach white noise while very large $\ell_{\text{corr}}$ indicates the presence of low-frequency surface distortion \cite{Beckmann63,Wen15,Sentenac2009}.

The effect that surface roughness has on incident radiation depends strongly on the typical lateral size scale of surface features, which is characterized by $\ell_{\text{corr}}$. 
If $\sigma$ is small enough however, a surface can be treated as being smooth such that the Fresnel equations for specular reflection given [\emph{cf.\@} \cref{eq:s_reflectivity,eq:p_reflectivity}] are valid. 
While pure specular reflection occurs when the phase of an incident plane wave varies linearly and continuously across a perfectly smooth surface \cite{Jackson75,Born80,Attwood17}, surface roughness with relatively large $\ell_{\text{corr}}$ introduces local phase variations from the differing path lengths that rays take to reflect from the random surface features associated with the roughness \cite{Beckmann63,Ogilvy87}. 
For rays striking two points on a surface that differ in depth by $y = \sigma$ [\emph{cf.\@} \cref{fig:smooth_criterion}], the path-length difference for a specularly reflected wavefront is $\Delta s = 2 \sigma \sin \left( \zeta \right)$, where $\zeta$ is the incidence angle measured relative to the tangent plane of the surface. 
\begin{figure}
 \centering
 \includegraphics{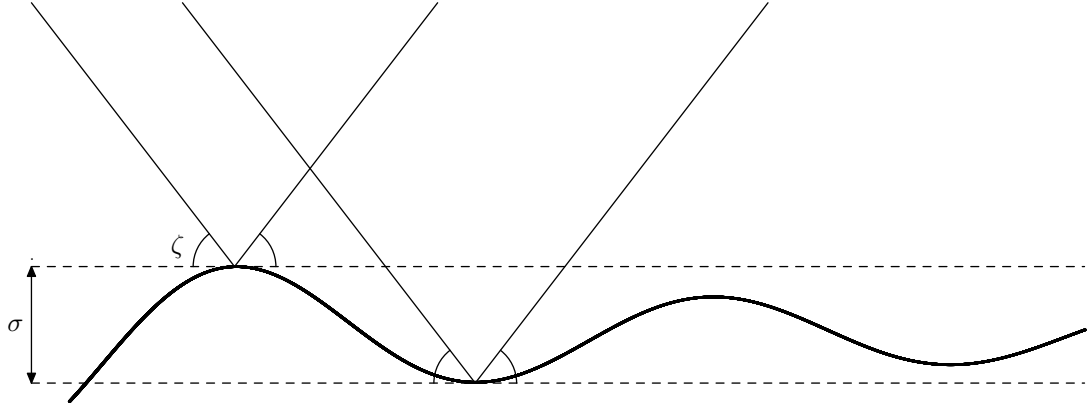}
 \caption{2D representation of rays incident on a rough surface}\label{fig:smooth_criterion}
 \end{figure} 
The phase shift associated with $\Delta s$ is \cite{Beckmann63,Hogrefe87} 
\begin{subequations}
\begin{equation}\label{eq:roughness_phase}
 \Delta \Phi = k_0 \Delta s = \frac{4 \pi \sigma}{\lambda} \sin \left( \zeta \right) ,
 \end{equation}
which provides a measure of how in-phase specularly reflected light is. 

The level of smoothness for a surface can be quantified by comparing $\Delta \Phi$ to some fixed value, $\Delta \Phi_0$, so that \cref{eq:roughness_phase} can be rearranged to give a requirement on $\sigma$: 
\begin{equation}\label{eq:roughness_inequality_general}
 \sigma < \frac{\Delta \Phi_0 \lambda}{4 \pi \sin \left( \zeta \right)} .
 \end{equation}
\end{subequations}
While the choice of $\Delta \Phi_0$ is somewhat arbitrary, commonly quoted values are $\Delta \Phi_0 = \pi /2$ (the \emph{Rayleigh criterion}) and $\Delta \Phi_0 = \pi /8$ (the \emph{Fraunhofer criterion}) \cite{Beckmann63}. 
Regardless of the exact definition for $\Delta \Phi_0$, the basic principle behind \cref{eq:roughness_inequality_general} can be understood qualitatively by considering the everyday example of the glare that the Sun produces as it reflects from pavement. 
That is, when the Sun is overhead, $\sin \left( \zeta \right) \approx 1$ so that $\sigma < \Delta \Phi_0 \lambda / 4 \pi$ may not be fulfilled for $\SI{700}{\nm} \gtrapprox \lambda \gtrapprox \SI{400}{\nm}$ with typical pavement RMS roughness values. 
On the other hand, pavement can appear smooth to sunlight with sufficiently small $\zeta$ when the Sun is low in the sky, which results in a glare that comes from specular reflection. 
For soft x-rays with $\SI{5}{\nm} \gtrapprox \lambda \gtrapprox \SI{0.5}{\nm}$, these conditions have much tighter constraints but at grazing-incidence angles in the regime of TER, high-quality surfaces with sufficiently small $\sigma$ can be regarded as being smooth to incident radiation. 
Using the Fraunhofer criterion and taking $\zeta = 1^{\circ}$ as a typical graze angle, \cref{eq:roughness_inequality_general} becomes $\sigma \lessapprox 1.8 \lambda$, which suggests that nanoscale values of $\sigma$ are needed to realize a smooth surface for grazing-incidence soft x-rays. 

\subsubsection{Debye-Waller Regime}\label{sec:DW_factor}
%%%%%%%%%%%%%%%%%%%%%%%%%%%%%%%%%%%%%%%%%--------------------------------------------------
The scenario described in the preceding paragraph relies on the assumption of $\ell_{\text{corr}}$ being large enough such that the electromagnetic fields at the boundary of the surface can be defined locally in a precise way. 
In the context of soft x-rays and other radiation with $\nu (\omega) = 1 - \delta_{\nu} (\omega) \lessapprox 1$, this condition is considered fulfilled when $\ell_{\text{corr}} \gg \ell_{\text{ext}}$, with $\ell_{\text{ext}} \equiv \lambda / 2 \pi \delta_{\nu} (\omega)$ being the extinction length \cref{eq:extinction_length} \cite[\emph{cf.\@} \cref{eq:extinction_length}]{deBergevin09,Gibaud2009}. 
Determining how the incident electromagnetic fields respond to a rough surface in principle requires solving the Helmholtz equations given by \cref{eq:Helmholtz_above_mirror,eq:Helmholtz_below_mirror} using boundary conditions that are a generalization of \cref{eq:general_boundary_1,eq:general_boundary_2,eq:general_boundary_3,eq:general_boundary_4} for a perfectly smooth surface: 
\begin{subequations}
\begin{align}
 \mathbold{\hat{n}} \times \left[ \mathbold{E} \left( x, Y(x,z), z \right) + \mathbold{E''} \left( x, Y(x,z), z \right) \right] &= \mathbold{\hat{n}} \times \mathbold{E'} \left( x, Y(x,z), z \right) \label{eq:rough_boundary_1} \\
 \mathbold{\hat{n}} \times \left[\mathbold{H} \left( x, Y(x,z), z \right)  + \mathbold{H''} \left( x, Y(x,z), z \right) \right] &= \mathbold{\hat{n}} \times \mathbold{H'} \left( x, Y(x,z), z \right)  \label{eq:rough_boundary_2} \\
 \mathbold{\hat{n}} \cdot \left[ \mathbold{E} \left( x, Y(x,z), z \right) + \mathbold{E''} \left( x, Y(x,z), z \right) \right] &= \tilde{\nu}^2 (\omega) \, \mathbold{\hat{n}} \cdot \mathbold{E'} \left( x, Y(x,z), z \right) \label{eq:rough_boundary_3} \\
 \mathbold{\hat{n}} \cdot \left[ \mathbold{H} \left( x, Y(x,z), z \right) + \mathbold{H''} \left( x, Y(x,z), z \right) \right] &= \mathbold{\hat{n}} \cdot \mathbold{H'} \left( x, Y(x,z), z \right) \label{eq:rough_boundary_4} , 
 \end{align}
\end{subequations}
where $Y(x,z)$ describes the rough surface profile and $\mathbold{\hat{n}} \equiv \mathbold{\hat{n}} (x,z)$ is a unit vector normal to the tangent plane of a roughness feature located at a point $(x,z)$. 
While this is in general a very complicated problem, the \emph{tangent-plane approximation} can be invoked for $\ell_{\text{corr}} \gg \ell_{\text{ext}}$ such that $Y(x,z)$ varies slow enough to make the assumption $\mathbold{\hat{n}} (x,z) \approx \mathbold{\hat{y}}$. 

In the tangent-plane approximation, \cref{eq:rough_boundary_1,eq:rough_boundary_2,eq:rough_boundary_3,eq:rough_boundary_4} for s-polarization\footnote{Because the expressions for Fresnel reflectivity in orthogonal polarizations, $\mathcal{R}_s$ and $\mathcal{R}_p$, are virtually equal at grazing incidence [\emph{cf.\@} \cref{fig:X-ray_refl}], only s-polarization need be considered here.} yield the following two relations after taking into account Snell's law that mandates $k_x = k_x'' = \tilde{k_x}' = k_0 \cos (\zeta)$ and $k_z = k_z'' = \tilde{k_z}' \equiv 0$: 
\begin{subequations}
\begin{align}
 \mathcal{A}_s \mathrm{e}^{i k_y Y(x,z)} + \mathcal{A}_s'' \mathrm{e}^{-i k_y Y(x,z)} &= \mathcal{A}_s' \mathrm{e}^{i \tilde{k_y}' Y(x,z)} \label{eq:tangent_plane1} \\
 \mathcal{A}_s k_y \mathrm{e}^{i k_y Y(x,z)} - \mathcal{A}_s'' k_y \mathrm{e}^{-i k_y Y(x,z)} &= \mathcal{A}_s' \tilde{k_y}' \mathrm{e}^{i \tilde{k_y}' Y(x,z)} \label{eq:tangent_plane2}
 \end{align}
\end{subequations}
with $\mathcal{A}_s$, $\mathcal{A}_s''$ and $\mathcal{A}_s'$ being the amplitudes of the incident, reflected and refracted wavefronts, respectively, while $k_y = -k_0 \sin (\zeta)$ and $\tilde{k_y}' = -k_0 \sqrt{\tilde{\nu}^2 (\omega) - \cos^2 (\zeta)}$. 
These two equations can be solved for the reflected amplitude: 
\begin{equation}
 \mathcal{A}_s'' = \mathcal{A}_s \mathrm{e}^{2 i k_y Y(x,z)} \left( \frac{k_y - \tilde{k_y}'}{\tilde{k_y}' + k_y} \right)
 \end{equation}
and using this relation, a reduced reflectivity can be defined as 
\begin{subequations}
\begin{equation}\label{eq:DW_refl_s}
 \mathcal{R}_{DW} \equiv \norm{\frac{\langle \mathcal{A}_s'' \rangle}{\mathcal{A}_s}}^2 = \norm{\langle \mathrm{e}^{2 i k_y Y(x,z)} \rangle}^2 \underbrace{\norm{\left( \frac{k_y - \tilde{k_y}'}{\tilde{k_y}' + k_y} \right) }^2 }_{\mathcal{R}_s} , %= \mathrm{e}^{-2 k_y^2 \sigma^2} ,
 \end{equation}
where, assuming that the randomness of the surface profile, $Y(x,z)$, is described by a normal distribution\footnote{For a random variable $X$ that is described by a Gaussian distribution, the \emph{Baker-Hausdorff theorem} states that $\langle \mathrm{e}^{i X} \rangle = \mathrm{e}^{-\frac{1}{2} \langle X^2 \rangle}$ \cite{Als-Nielsen11}. Moreover, $\sigma^2 \equiv \langle Y^2 (x,z) \rangle$ from \cref{eq:surface_variance}.} and using the definition of $\sigma^2$ in \cref{eq:surface_variance},
\begin{equation}\label{eq:DW_factor}
 \langle \mathrm{e}^{2 i k_y Y(x,z)} \rangle = \mathrm{e}^{-2 k_y^2 \langle Y^2(x,z) \rangle} = \mathrm{e}^{-2 k_y^2 \sigma^2}
 \end{equation}
\end{subequations}
is known as the \emph{Debye-Waller factor} \cite{Gibaud2009}. 
With the under-braced term in \cref{eq:DW_refl_s} recognized as $\mathcal{R}_F$ for s-polarization, \cref{eq:DW_factor} is equivalent to the fraction of radiation lost to surface roughness for large $\ell_{\text{corr}}$, $\mathcal{R}_{DW} / \mathcal{R}_F = \mathrm{e}^{-4 k_y^2 \sigma^2}$. 

\subsubsection{Intermediate Regime}\label{sec:intermediate_rough}
%%%%%%%%%%%%%%%%%%%%%%%%%%%%%%%%%%%%%%%%%--------------------------------------------------
For a surface assumed to be dominated by features with a large $\ell_{\text{corr}}$ such that the Debye-Waller factor is valid, the reduction in specular reflectivity is attributed to diffuse scatter in vacuum with the total integrated scatter being equal to $1 - \mathrm{e}^{-4 k_y^2 \sigma^2}$ \cite{Wen15}. 
While this expression breaks down as $\ell_{\text{corr}}$ decreases\footnote{This occurs when the tangent-plane approximation is no longer valid such that the boundary conditions given by \cref{eq:rough_boundary_1,eq:rough_boundary_2,eq:rough_boundary_3,eq:rough_boundary_4} must be applied for an arbitrary normal vector $\mathbold{\hat{n}} (x,z)$.} and does not describe how the intensity of scattered light is distributed, this behavior can be understood in part by considering a rough surface to be a superposition of sinusoidal components with various periods and orientations described by the vector $\mathbold{K}$ introduced for the the PSD function [\emph{cf.\@} \cref{eq:PSD_general}]. 
Defining $\abs{\mathbold{K}} \equiv K \equiv 2 \pi / \Lambda$, a single mode of a rough surface in principle acts as a sinusoidal reflection grating with a groove spacing equal to $\Lambda$, which is an effective \emph{surface wavelength} \cite{Cash87,Sentenac2009,Aschenbach85}. 
Approximate scattering behavior then can be gleaned by considering how an electromagnetic wave interacts with such a structure for 
\begin{enumerate}[noitemsep]
  \item the exact \emph{in-plane} case, where the plane defined by the incident wave vector, $\mathbold{k}$, and $\mathbold{K}$ is perpendicular to the surface defined by $y=0$, and 
  \item the totally \emph{off-plane} case, where $\mathbold{k}$ is perpendicular to $\mathbold{K}$.
 \end{enumerate}

The locations of diffracted orders for a grating for an oblique incidence angle are described by the \emph{generalized grating equation} [\emph{cf.\@} \cref{eq:off-plane_incidence_orders,eq:off-plane_orders}]: 
\begin{equation} \label{eq:grating_rough}
 \sin \left( \alpha \right) + \sin \left( \beta \right) = \frac{n \lambda}{\Lambda \sin \left( \gamma \right)} ,
 \end{equation}
where $\alpha$ is the polar incidence angle, $\beta$ is the polar diffracted angle of the $n^{\text{th}}$ order and $\gamma$ is the cone-opening angle of the diffraction pattern [\emph{cf.\@} \cref{fig:conical_reflection_edit,fig:conical_reflection}]. 
For the in-plane scenario [\emph{cf.\@} \cref{fig:in_plane_scatter}], all diffracted orders are confined to this same plane with $\sin (\gamma) = 1$ so that the incidence angle measured relative to the surface is $\zeta = \pi / 2 - \alpha$. 
\begin{figure}
 \centering
 \includegraphics[scale=1.0]{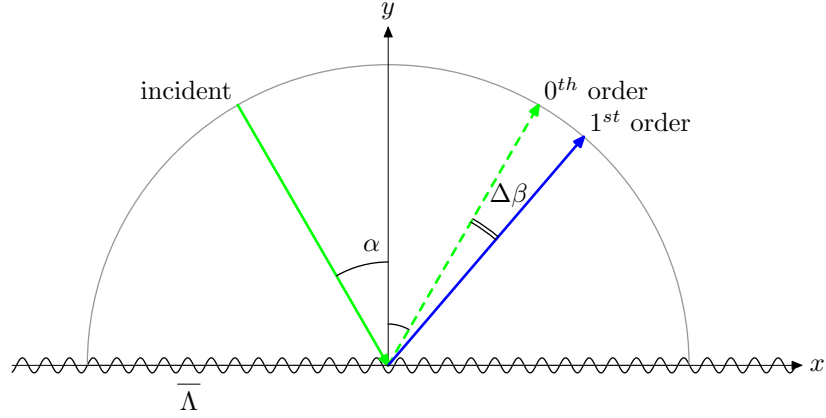}
 \caption{In-plane scatter from a single mode of surface roughness}\label{fig:in_plane_scatter}
 \end{figure}
The diffracted angle moreover is taken to be $\beta = \Delta \beta - \alpha$, where $\Delta \beta$ is the angular deviation from specular for the scattered ray, so that 
\cref{eq:grating_rough} becomes 
\begin{subequations}
\begin{equation}
 \cos \left( \zeta \right) - \cos \left( \Delta \beta + \zeta \right) = \frac{n \lambda}{\Lambda} .
 \end{equation}
Using a trigonometric identity\footnote{$\cos \left( \Delta \beta + \zeta \right) = \cos \left( \Delta \beta \right) \cos \left( \zeta \right) - \sin \left( \Delta \beta \right) \sin \left( \zeta \right)$} and taking $n=1$ for first-order diffraction\footnote{According to the scalar treatment of diffraction described in \cref{sec:sinusoid_principle}, diffracted orders with $n = \pm 1$ are expected to dominate over those with larger $\abs{n}$.} gives 
\begin{equation}
 \frac{\lambda}{\Lambda} = \underbrace{\cos \left( \zeta \right)}_{\approx 1} \left[ 1 - \underbrace{\cos \left( \Delta \beta \right)}_{\approx 1} \right] + \underbrace{\sin \left( \zeta \right)}_{\approx \zeta} \underbrace{\sin \left( \Delta \beta \right)}_{\approx \Delta \beta} , %\approx \zeta \Delta \beta ,  
 \end{equation}
where, using the under-braced small-angle approximations for a grazing-incidence angle $\zeta \ll 1$, the scattered angle is 
\begin{equation}\label{eq:in-plane_scatter_angle}
 \Delta \beta \approx \frac{\lambda}{\zeta \Lambda} .
 \end{equation}
\end{subequations}

The scenario for the off-plane case [\emph{cf.\@} \cref{fig:off_plane_scatter}] corresponds to $\alpha = 0$, $\beta =  \Delta \beta $ and $\gamma = \zeta$ with all diffracted orders confined to the surface of a cone so that \cref{eq:grating_rough} becomes 
\begin{subequations}
\begin{equation} 
 \underbrace{\sin \left( \zeta \right)}_{\approx \zeta} \underbrace{\sin \left( \Delta \beta \right)}_{\approx \Delta \beta} = \frac{n \lambda}{\Lambda} , 
 \end{equation}
where the under-braced approximations hold for small angles. 
\begin{figure}
 \centering
 \includegraphics[scale=1.0]{Appendix-C/Figures/diffraction_arc11.mps}
 \caption{Off-plane scatter from a single mode of surface roughness}\label{fig:off_plane_scatter}
 \end{figure}
In this case, however, the off-plane scattered angle is $\varphi''$ that is labeled in \cref{fig:off_plane_scatter} and defined by the following relation:
\begin{equation}
 \sin \left( \varphi'' \right) = \tan \left( \zeta \right) \tan \left( \Delta \beta \right) ,
 \end{equation}
which can be approximated as $\varphi'' \approx \zeta \Delta \beta$. 
Taking $n=1$ for first-order diffraction, this diffracted angle can be expressed as 
\begin{equation}\label{eq:off-plane_scatter_angle}
 \varphi'' \approx \frac{\lambda}{\Lambda} .
 \end{equation}
\end{subequations}
Comparing \cref{eq:in-plane_scatter_angle,eq:off-plane_scatter_angle} shows that the angular spread of in-plane scattering is $\zeta^{-1}$ times larger than that of off-plane scattering assuming $\zeta \ll 1$ for grazing-incidence soft x-rays \cite{Aschenbach85}.
This suggests that diffuse scattering produced by a rough surface is most detrimental for an x-ray reflection grating in the plane defined by the incident wave vector and the direction normal to the active blazed groove facets \cite{Cash87}. 
However, determining the intensity of scattered radiation requires the use of more rigorous techniques such as the \emph{distorted-wave Born approximation} \cite{Vineyard82,Sinha88,Vinogradov85,Daillant2009}. 

\subsubsection{Nevot-Croce Regime}\label{sec:NC_factor}
%%%%%%%%%%%%%%%%%%%%%%%%%%%%%%%%%%%%%%%%%--------------------------------------------------
Beyond diffuse scattering vacuum, specular reflectivity can also be reduced due to absorption, as radiation scatters into the depth of the reflective material that makes up a rough surface. 
This predominately occurs when $\ell_{\text{corr}}$ is so small that typical values for $\Lambda$ do not produce diffraction by \cref{eq:grating_rough} for the generalized grating equation \cite{deBoer95}. 
In the case of a surface with $\ell_{\text{corr}} \ll \ell_{\text{ext}}$, there is no precise phase relationship at the boundary $Y (x,z)$ and instead, the boundary conditions defined by \cref{eq:rough_boundary_1,eq:rough_boundary_2,eq:rough_boundary_3,eq:rough_boundary_4} are met only on average so that \cref{eq:tangent_plane1,eq:tangent_plane2} for the tangent-plane approximation are again valid, but only for \emph{effective amplitudes} \cite{Gibaud2009}. 
To show this, multiplying \cref{eq:tangent_plane1} by $k_y$ gives 
\begin{subequations}
\begin{align}
 \mathcal{A}_s k_y \mathrm{e}^{i k_y Y(x,z)} + \mathcal{A}_s'' k_y \mathrm{e}^{-i k_y Y(x,z)} &= \mathcal{A}_s' k_y \mathrm{e}^{i \tilde{k_y}' Y(x,z)} \label{eq:tangent_plane1_eff} \\
 \mathcal{A}_s k_y \mathrm{e}^{i k_y Y(x,z)} - \mathcal{A}_s'' k_y \mathrm{e}^{-i k_y Y(x,z)} &= \mathcal{A}_s' \tilde{k_y}' \mathrm{e}^{i \tilde{k_y}' Y(x,z)} \label{eq:tangent_plane2_eff} 
 \end{align}
\end{subequations}
while adding the two relations and subtracting \cref{eq:tangent_plane2_eff} from \cref{eq:tangent_plane2_eff} yields 
\begin{subequations}
\begin{align}
 2 \mathcal{A}_s k_y \mathrm{e}^{i k_y Y(x,z)} &= \mathcal{A}_s' \left( k_y + \tilde{k_y}' \right) \mathrm{e}^{i \tilde{k_y}' Y(x,z)}  \label{eq:tangent_plane1_eff_comb} \\
 2 \mathcal{A}_s'' k_y \mathrm{e}^{-i k_y Y(x,z)} &= \mathcal{A}_s' \left( k_y - \tilde{k_y}' \right) \mathrm{e}^{i \tilde{k_y}' Y(x,z)}  \label{eq:tangent_plane2_eff_comb} .
 \end{align}
\end{subequations}

The effective amplitudes on the boundary correspond to average values with \cref{eq:tangent_plane1_eff_comb,eq:tangent_plane2_eff_comb} rewritten as
\begin{subequations}
\begin{align}
 \langle \mathcal{A}_s \rangle &= \frac{1}{2 k_y} \langle \mathcal{A}_s' \rangle \left( k_y + \tilde{k_y}' \right) \langle \mathrm{e}^{i \left( \tilde{k_y}' - k_y \right) Y(x,z)} \rangle \label{eq:tangent_plane1_eff_comb_alt} \\
 \langle \mathcal{A}_s'' \rangle  &= \frac{1}{2 k_y} \langle \mathcal{A}_s' \rangle \left( k_y - \tilde{k_y}' \right) \langle \mathrm{e}^{i \left( \tilde{k_y}' + k_y \right) Y(x,z)} \rangle \label{eq:tangent_plane2_eff_comb_alt} 
 \end{align}
\end{subequations}
with the reduced reflectivity given by\footnote{Similar to the derivation of the Debye-Waller factor, it is again assumed that $Y(x,z)$ is a random variable described by Gaussian distribution so that the following relation holds: $\langle \mathrm{e}^{i X} \rangle = \mathrm{e}^{-\frac{1}{2} \langle X^2 \rangle}$ \cite{Als-Nielsen11}.} 
\begin{subequations}
\begin{equation}
 \mathcal{R}_{NC} \equiv \norm{\frac{\langle \mathcal{A}_s'' \rangle}{\langle \mathcal{A}_s \rangle}}^2 = \norm{\frac{\langle \mathrm{e}^{i \left( \tilde{k_y}' + k_y \right) Y(x,z)} \rangle}{\langle \mathrm{e}^{i \left( \tilde{k_y}' - k_y \right) Y(x,z)} \rangle}}^2 \underbrace{\norm{\left( \frac{k_y - \tilde{k_y}'}{\tilde{k_y}' + k_y} \right)}^2}_{\mathcal{R}_s} , \label{eq:NC_refl_s}
 \end{equation}
where
\begin{equation}
 \frac{\langle \mathrm{e}^{i \left( \tilde{k_y}' + k_y \right) Y(x,z)} \rangle}{\langle \mathrm{e}^{i \left( \tilde{k_y}' - k_y \right) Y(x,z)} \rangle} = \frac{\mathrm{e}^{- \frac{1}{2} \left( \tilde{k_y}' + k_y \right)^2 \langle Y^2 (x,z) \rangle}}{\mathrm{e}^{- \frac{1}{2} \left( \tilde{k_y}' - k_y \right)^2 \langle Y^2 (x,z) \rangle}} = \mathrm{e}^{- 2 k_y \tilde{k_y}' \sigma^2}
 \end{equation}
is a complex quantity known as the \emph{Nevot-Croce factor} \cite{Gibaud2009,deBoer95,Wen15,Nevot80}. 
With $\mathcal{R}_s$ for Fresnel reflectivity recognized in \cref{eq:NC_refl_s}, the fraction of specularly reflected radiation lost due to absorption can be written as 
\begin{equation}\label{eq:NC_refl_ratio}
 \frac{\mathcal{R}_{NC}}{\mathcal{R}_F} = \norm{\mathrm{e}^{- 2 k_y \tilde{k_y}' \sigma^2}}^2 = \mathrm{e}^{-2 k_y \left( \tilde{k_y}' + \tilde{k_y}'^* \right) \sigma^2} = \mathrm{e}^{-4 k_0^2 \sin \left( \zeta \right) \Re \left[ \sqrt{\tilde{\nu}^2 (\omega) - \cos^2 \left( \zeta \right) } \right] \sigma^2}
 \end{equation}
for a thick slab, where again, $\mathcal{R}_F \approx \mathcal{R}_s \approx \mathcal{R}_p$ for grazing-incidence soft x-rays of any polarization [\emph{cf.\@} \cref{sec:reflectivity_polarization}]. 
\end{subequations}
Treating reduced specular reflectivity using Nevot-Croce factors is considered a decent approximation if surface features with relatively large lateral sizes can be ignored while atomic-scale features with small $\sigma$ dominate surface roughness. 
In any case, the effect that a rough surface has on the specular reflectivity of a mirror flat is analogous to the losses in diffracted orders that result from surface roughness on the blazed groove facets of an x-ray reflection grating. % and therefore, this phenomenon must be kept under control to achieve high diffraction efficiency. 

\section{Summary}\label{sec:appC_summary}
%%%%%%%%%%%%%%%%%%%%%%%%%%%%%%%%%%%%%%%%%--------------------------------------------------
Due to the fact that their photon energy range, \SI{250}{\electronvolt}~$\lessapprox \mathcal{E}_{\gamma} \lessapprox$~\SI{2}{\kilo\electronvolt}, coincides with the spectrum of K-shell binding energies in low-to-mid $\mathcal{Z}$ atoms while their wavelength range, \SI{5}{\nm}~$\gtrapprox \lambda \gtrapprox$~\SI{0.5}{\nm}, approaches the atomic scale, soft x-rays are easily absorbed by $\mathcal{Z} \geq 6$ materials and additionally, appreciable reflectivity from a surface can only be achieved at grazing-incidence angles, $\zeta$, that are smaller than the critical angle for total external reflection (TER), $\zeta_c (\omega)$. 
Although materials with moderate $\mathcal{Z}$ (\emph{e.g.}, nickel) may provide high reflectivity over a limited bandpass, those with high $\mathcal{Z}$ (\emph{e.g.}, gold) tend to offer broadband reflectivity at soft x-ray wavelengths. 
These phenomena can be gleaned from the complex index of refraction for a given material, $\tilde{\nu} (\omega) = \nu (\omega) + i \xi (\omega)$, 
where $\nu (\omega) \equiv 1 - \delta_{\nu} (\omega) \lessapprox 1$ leads to $\zeta_c (\omega) \approx \sqrt{2 \delta_{\nu} (\omega)}$ with $\delta_{\nu} (\omega) \ll 1$ while $\xi (\omega) \neq 0$ gives rise to reflectivity losses and in turn, a nonzero penetration depth, $\mathcal{D}_{\perp}$, that informs thickness requirements for deposited, reflective films used in the soft x-ray. 
A film \numrange{4}{5} $\mathcal{D}_{\perp}$ thick (typically $\sim \SI{15}{\nm}$) effectively functions as a thick slab, and with Fresnel reflectivity being virtually polarization-insensitive in the regime of TER, the quantity need only be calculated for a single polarization. 
Finally, effects of nanoscale surface roughness in the limit of small and large correlation length, $\ell_{\text{corr}}$, can be modeled using Nevot-Croce and Debye-Waller factors, respectively, while the intermediate region requires more advanced treatment. 
% !TEX root = ../McCoy-Dissertation.tex
\Appendix{Scalar Treatment of Gratings}\label{ap:grating_basics}
%%%%%%%%%%%%%%%%%%%%%e%%%%%%%%%%%%%%%%%%%%--------------------------------------------------
As motivated in \cref{sec:grating_tech_dev}, diffraction gratings are the instrument of choice for sensitive, high-resolution spectroscopy in the soft x-ray spectrum, particularly for the photon-energy range $\SI{250}{\electronvolt} \lessapprox \mathcal{E}_{\gamma} \lessapprox \SI{1}{\kilo\electronvolt}$, where energy-dispersive detectors intended for higher-energy x-rays face limitations in terms of spectral resolving power, $\mathscr{R}$. 
While x-rays are often associated with their particle-like behavior [\emph{cf.\@} \cref{ap:x-ray_intro}], like all types of waves they experience the phenomenon of diffraction as they bend around obstacles, spread outward from narrow openings or otherwise superpose to produce interference patterns.\footnote{An everyday example of this phenomenon is waves on the surface of a shallow pond, which undergo diffraction as they pass through a narrow opening created by protruding rocks, or when multiple waves from splashes overlap to produce interference fringes.} 
Fundamentally, this behavior arises because according to the \emph{Huygens-Fresnel principle}, any given wavefront can be understood as being composed of a set of point sources that emanate secondary spherical waves \cite{Born80}. 
This implies that when a wave encounters an aperture, the scenario is equivalent to there being a new distribution of point sources that interfere with one another as they propagate, producing diffraction patterns. 
As these secondary waves move forward, they interfere with one another to form the diffracted wavefront. 
In the case of multiple obstacles or openings, an additional diffraction pattern is produced as the wavefronts from each aperture overlap and produce interference fringes. 
For a large number of evenly-spaced apertures, the system behaves like a diffraction grating, where an ordered interference pattern is produced in the far field that depends on the spatial periodicity, the wavelength and the angle of incidence \cite{Loewen97}. 

Considering wave mechanics alone, a diffraction grating for soft x-rays can be realized as a periodic structure with a very fine pitch to accommodate the wavelength of radiation $\SI{5}{\nm} \gtrapprox \lambda = h c_0 / \mathcal{E}_{\gamma} \gtrapprox \SI{1.25}{\nm}$. 
However, when the interaction of radiation with matter is taken into account [\emph{cf.\@} \cref{app:x-ray_materials}], soft x-rays and other short-$\lambda$ radiation are found to reflect appreciably only at grazing incidence angles in the regime of \emph{total external reflection}. 
This restriction introduces practical challenges and drives the consideration of alternative grating geometries that produce conical diffraction patterns from oblique angles of incidence. 
To introduce these concepts and ultimately motivate the specifics of what goes into the design of efficient reflection gratings for soft x-ray spectroscopy, this appendix outlines basic physics of diffraction gratings according to scalar theories of diffraction, which, by construction, ignore effects related to the polarization of electromagnetic radiation and the atomic structure of materials. 
In this case, scalar waves are imagined to diffract from idealized structures that scale equally with any wavelength so that according to the Huygens-Fresnel principle, a grating can be described as a continuous surface of point sources with periodic variations in amplitude, phase or both. 
While the electromagnetic vector treatment of gratings outlined in \cref{sec:integral_method} is needed to describe these phenomena most accurately, scalar theories of diffraction can be used to describe roughly the behavior of diffraction gratings, including the effect that groove shape has on the dispersed radiation. 

\section{Wave Interference From an Array of Slits}\label{sec:slit_array}
%%%%%%%%%%%%%%%%%%%%%%%%%%%%%%%%%%%%%%%%%--------------------------------------------------
As a first step toward describing diffraction gratings, a generic, scalar plane wave of wavelength $\lambda$ is imagined to be incident on a barrier with two narrow slits\footnote{The lengths of these slits are taken to be formally infinite so that this scenario can be treated two-dimensionally. In other words, each aperture is considered to be sufficiently long such that diffraction from edge effects along this direction can be neglected.} separated by a distance $d$, each with a width $W \ll \lambda$ so that both can be thought of as the source of a true circular wave according to the Huygens-Fresnel principle.
This is illustrated in \cref{fig:double_slit_experiment}, where the barrier lays in the plane defined by the $x$-axis and the direction pointing out of the page.
\begin{figure}
 \centering
 \includegraphics[scale=1.07]{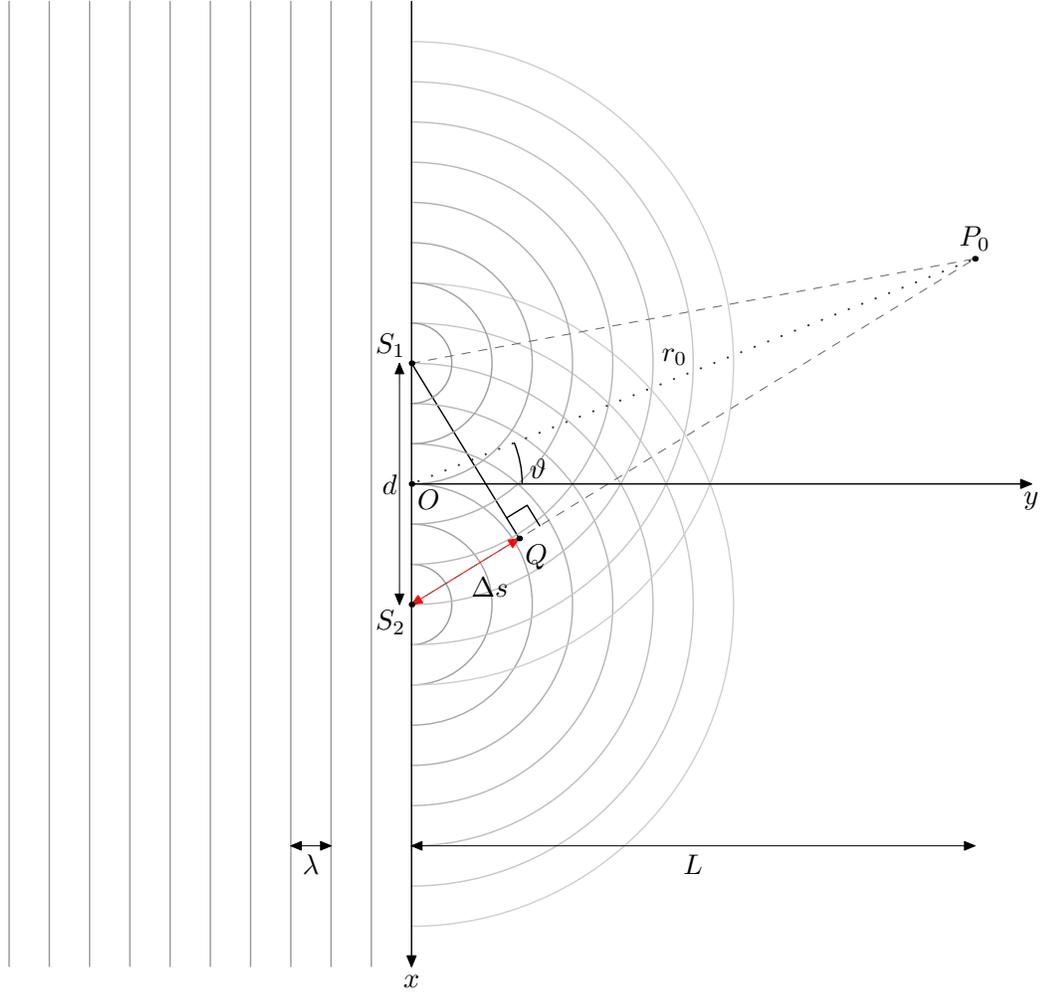}
 \caption{Near-field interference from two very small slits}\label{fig:double_slit_experiment}
 \end{figure}
Initially, it is assumed that the incident wave vector, $\mathbold{k}$ (with $\abs{\mathbold{k}} \equiv k = 2 \pi / \lambda$) is oriented perpendicular to this barrier so that the incident wave propagates in the $y$-direction, $\mathbold{\hat{y}}$, with a scalar field satisfying \cref{eq:scalar_wave_general} of the form 
\begin{equation}
 u(y,t) \propto \mathrm{e}^{i \left( k y - \omega t \right)} ,
 \end{equation}
where $\omega = 2 \pi c / \lambda$ is the frequency of the wave and $c$ is the speed of the wave. 
Meanwhile, each slit emanates circular waves that can be expressed in polar coordinates, $\mathbold{s} \equiv (s, \vartheta)$, as 
\begin{equation}\label{eq:circular_wave}
 u(s,t) \propto \frac{1}{\sqrt{s}} \, \mathrm{e}^{i \left( k s - \omega t \right)} ,
 \end{equation}
where $s \equiv \sqrt{x^2 + y^2}$ is a 2D radial distance and $\vartheta \equiv - \arccos \left( x /s \right)$ for $y \geq 0$. 
As indicated in \cref{fig:double_slit_experiment}, the incident planar wave gives rise to two point sources, $S_1$ and $S_2$, at the location of each slit: $(x,y) = (\pm d/2,0)$ relative to the $x-y$ origin, $O$. 
These two sources separately define scalar waves $u_1 (s_1,t)$ and $u_2 (s_2,t)$ of the form given in \cref{eq:circular_wave}, where $s_1$ and $s_2$ are the radial distances from each slit. 

The interference pattern caused by the placement of these sources is determined by probing the intensity of the combined scalar wave $u ( \mathbold{s} , t ) = u_1 ( \mathbold{s} , t ) + u_2 ( \mathbold{s} , t )$ at some fixed point, $P_0$, away from the barrier. 
With the magnitude of \emph{Poynting's vector}, $\mathbold{S}$, being proportional to $\norm{u}^2$, the double-slit \emph{interference function} is taken to be $I_2 \propto \norm{u}^2$, where the maximum of the function is normalized to unity. 
Shown in \cref{fig:double_slit_experiment}, this arbitrary point where wave intensity\footnote{As described in \cref{sec:amplitude_phase}, the wave intensity, $\mathcal{I}$, that passes through a plane parallel to the barrier depends on $\vartheta$ such that $\mathcal{I} \equiv \mathbold{S} \cdot \mathbold{\hat{y}} \propto \norm{u}^2 \cos \left( \vartheta \right)$.} is imagined to measured is located a distance $L$ from the barrier so that radial distance from $O$ to $P_0$ is $r_0 = L \sec (\vartheta)$. 
Because the incident wavefronts are assumed to be oriented parallel to the barrier, sources $S_1$ and $S_2$ generate waves with identical amplitude and phase. 
However, by the time their wavefronts reach $P_0$, there will generally be differences in amplitude and phase due to the fact that there is a \emph{path-length difference}, $\Delta s \equiv s_2 - s_1 \equiv \overline{S_2 Q}$, where $s_1 \equiv \overline{S_1 P_0}$ and $s_2 \equiv \overline{S_2 P_0}$ in \cref{fig:double_slit_experiment}. 
These path lengths $s_1$ and $s_2$ can be determined using the \emph{law of cosines}: 
\begin{align}
 \begin{split}
 s_1^2 &= \left( \frac{d}{2} \right)^2 + r_0^2 - r_0 \, d \cos \left(\frac{\pi}{2} - \vartheta \right) = \left( \frac{d}{2} \right)^2 + L^2 \sec^2 \left( \vartheta \right) - L \, d \tan \left( \vartheta \right) \\
 s_2^2 &= \left( \frac{d}{2} \right)^2 + r_0^2 - r_0 \, d \cos \left(\frac{\pi}{2} + \vartheta \right) = \left( \frac{d}{2} \right)^2 + L^2 \sec^2 \left( \vartheta \right) + L \, d \tan \left( \vartheta \right) . \label{eq:path_length_exact}
 \end{split}
 \end{align} 
In terms of \cref{fig:double_slit_experiment}, sources $S_1$ and $S_2$ have radial distances $s_1$ and $s_2$ to $P_0$ while the amplitudes of the $S_1$ and $S_2$ wavefronts are defined to be $\mathcal{A}_1$ and $\mathcal{A}_2$, respectively. 
Especially for small $L$, the difference between these amplitudes depends on $s_1$ and $s_2$ separately, instead of $\Delta s$. 
On the other hand, the $S_2$ wavefront is shifted by a phase $\Phi = - k \Delta s$ relative to the $S_1$ wavefront. 
Using this notation, the $S_1$ scalar field at $P_0$ is $u_1 (t) \equiv \mathcal{A}_1 \mathrm{e}^{- i \omega t}$ and the $S_2$ scalar field at $P_0$ is $u_2 (t) = \mathcal{A}_2 \mathrm{e}^{i (\Phi -  \omega t)}$ with the overall wave at this point being described by 
\begin{equation}\label{eq:scalar_field_double_slit}
 u (t) = u_1 (t) + u_2 (t) = \mathrm{e}^{- i \omega t} \left( \mathcal{A}_1 + \mathcal{A}_2 \mathrm{e}^{i \Phi} \right) \quad \text{for 2 slits} .
 \end{equation}

As a next step toward describing a diffraction grating, the \emph{far-field approximation} is invoked such that the distance to $P_0$ is taken to be very large compared to the spacing between the slits so that $L \gg d$ and rays $\overrightarrow{S_1 P_0}$ and $\overrightarrow{S_2 P_0}$ are approximately parallel to each other. 
\begin{figure}
\centering
\includegraphics[scale=1.07]{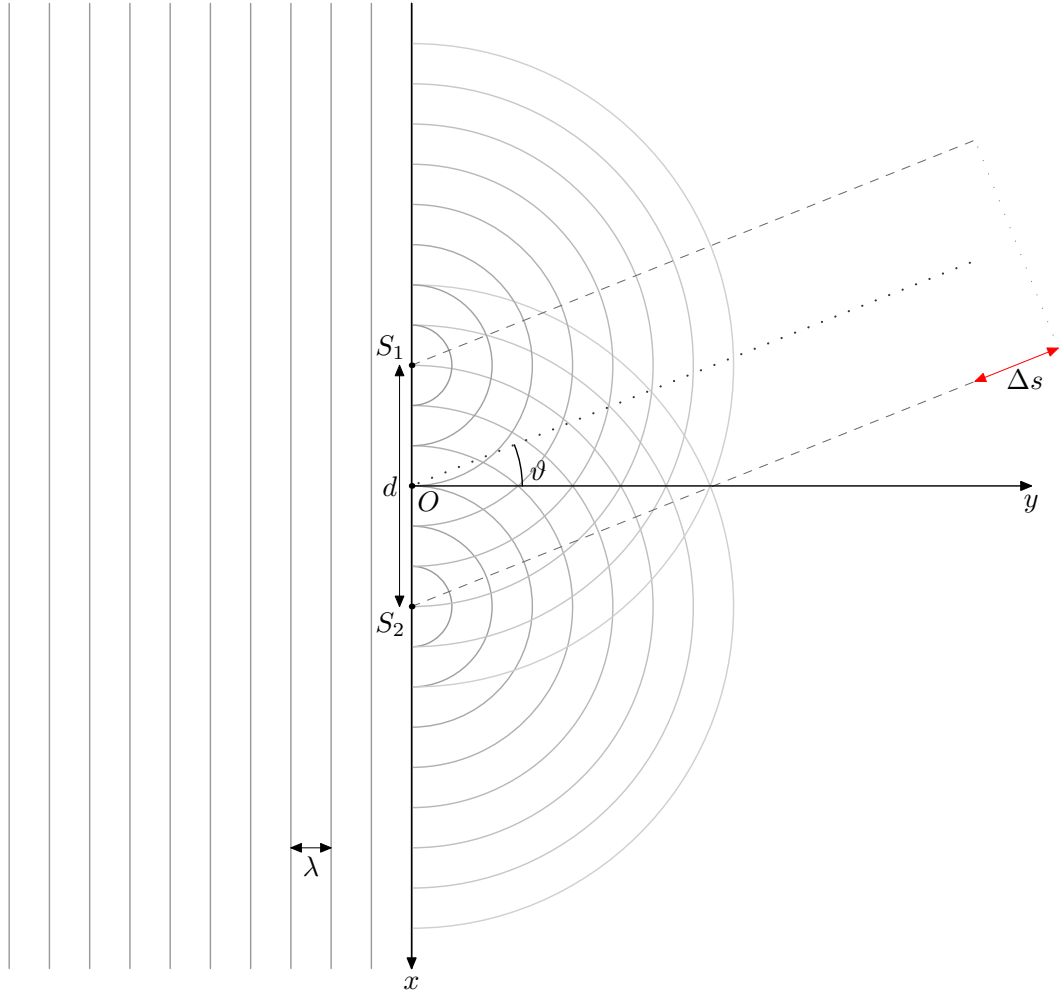}
\caption{Far-field interference from two very small slits}\label{fig:double_slit_experiment_approx}
\end{figure}
This scenario is illustrated in \cref{fig:double_slit_experiment_approx}, where $P_0$ is effectively at an infinite distance from the barrier.  
In this case, $r_0 \gg \Delta s$ and the diffracted wavefronts are roughly planar with $\mathcal{A}_1 \approx \mathcal{A}_2 \equiv \mathcal{A}_0$, where $\mathcal{A}_0$ is the amplitude measured at $P_0$. 
Using the approximation $s_2 + s_1 \approx 2 r_0$, the difference of the two terms in \cref{eq:path_length_exact} becomes 
\begin{equation}
 s_2^2 - s_1^2 = 2 L \, d \tan \left( \vartheta \right) = \left( s_2 + s_1 \right) \left( s_2 - s_1 \right) \approx 2 L \sec \left( \vartheta \right) \Delta s 
 \end{equation}
and, therefore, the path-length difference in the far field is $\Delta s = d \sin (\vartheta)$ assuming that the incident wave is normally incident on the barrier. 
The phase shift is $\Phi = - k \Delta s$ with the double-slit interference function given by
\begin{equation}\label{eq:double-slit_intensity}
 I_2 \left( \Phi \right) \propto \norm{u ( t )}^2 = 2 \mathcal{A}_0^2 \left[ 1 + \cos \left( \Phi \right) \right] \quad \text{for 2 slits,}
 \end{equation}
where $u (t)$ follows from \cref{eq:scalar_field_double_slit} using $\mathcal{A}_1 = \mathcal{A}_2 = \mathcal{A}_0$. 
Maxima of this function correspond to $\Phi = 2 n \pi$ or, equivalently, $\Delta s = - n \lambda$ with $n = 0, \pm 1, \pm 2, \pm 3 \dotsc$  
Physically, these maxima represent constructive interference fringes with angular locations given by 
\begin{equation}\label{eq:fringes}
 \sin \left( \vartheta_{\text{max}} \right) = - \frac{n \lambda}{d} \quad \text{for } n = 0, \pm 1, \pm 2, \pm 3 \dotsc 
 \end{equation}
provided that $-1 < n \lambda / d < 1$. 
\begin{figure}
\centering
\includegraphics[scale=0.99]{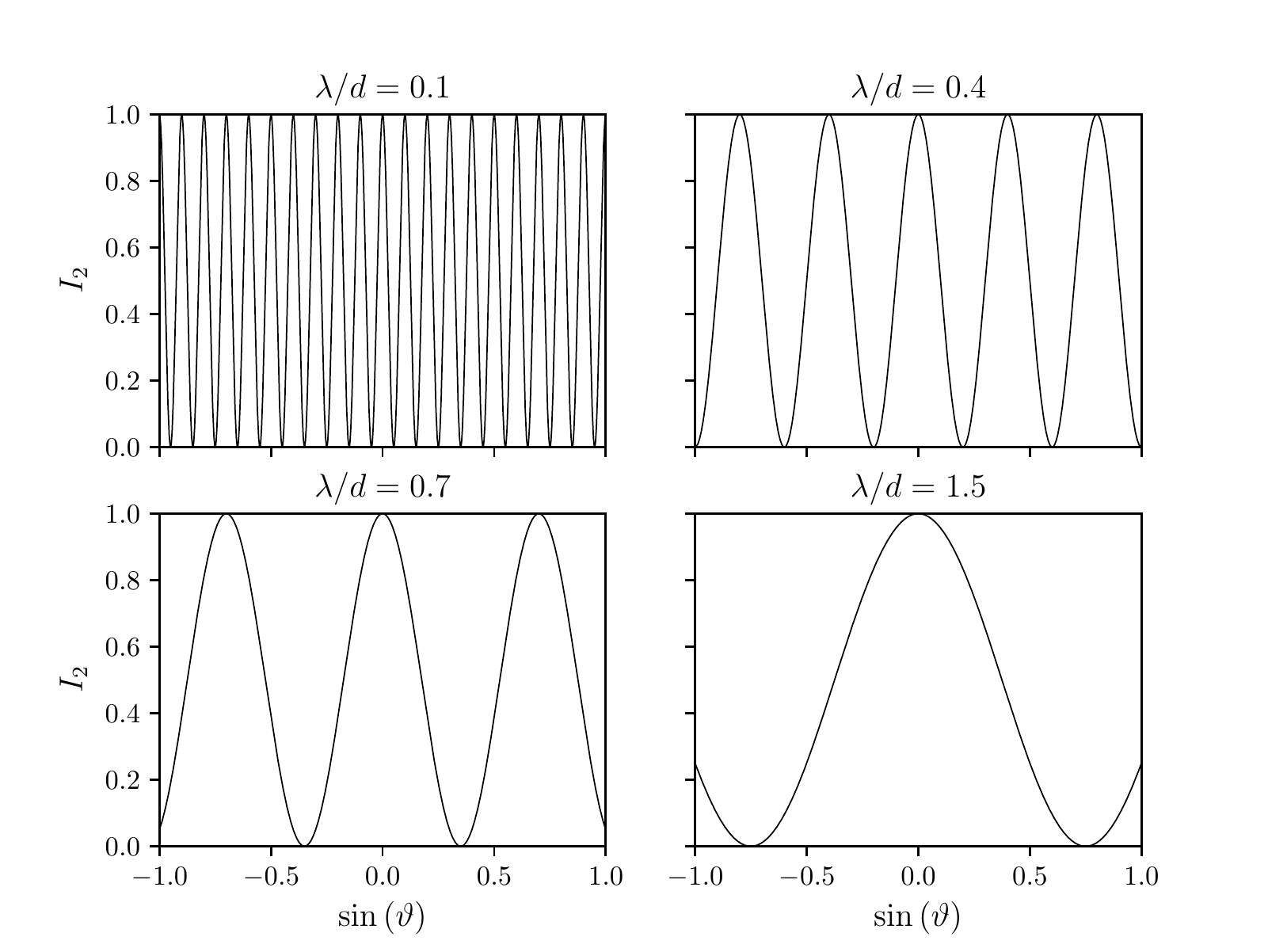}
\caption[Normalized interference function for two point sources for various wavelengths]{Normalized interference function for two point sources as a function of $\sin (\vartheta)$ for various $\lambda / d$ ratios.}\label{fig:double_slit_plot}
\end{figure} 
The sinusoidal dependence of the far-field double-slit interference pattern on $k d$, and hence the ratio $\lambda / d$, is demonstrated in \cref{fig:double_slit_plot}. 

The double-slit scenario outlined above can be generalized to $N$ equally-spaced slits, each an identical source $S_m$ (for $m = 1, 2, \dotsc , N$). 
In this case, $P_0$ is considered to be a distance $L \gg (N - 1) d$ away from the barrier so that far-field approximations are valid. 
At $P_0$, the amplitude associated with each slit is virtually the same in this approximation so that the scalar field at this point due to the $m^{\text{th}}$ source $S_m$ is 
\begin{subequations} 
\begin{equation}
 u_m (t) = u_1 (t) \mathrm{e}^{i \left( m - 1 \right) \Phi} = \mathcal{A}_0 \, \mathrm{e}^{i \left[ \left( m -1 \right) \Phi - \omega t \right]} \quad \text{for }  m = 1 , 2, 3 \dotsc
 \end{equation}
The total scalar field at $P_0$ is given by 
\begin{equation}
 u (t) = \sum_{m=1}^{N} u_m (t) = u_1 (t) \mathrm{e}^{-i \Phi} \sum_{m=1}^{N} \mathrm{e}^{i m \Phi} \quad \text{for $N$ slits} ,
 \end{equation}
\end{subequations}
where the following exponential sum-formula is used:
\begin{equation}\label{eq:exp_sum}
 \mathrm{e}^{-i \Phi} \sum_{m=1}^{N} \mathrm{e}^{i m \Phi} = \sum_{m=0}^{N-1} \mathrm{e}^{i m \Phi} = \frac{1 - \mathrm{e}^{i N \Phi}}{1 - \mathrm{e}^{i \Phi}} 
 \end{equation}
and the resulting interference function is %intensity at $P_0$ is: 
\begin{subequations}
\begin{equation}\label{eq:N-slit_intensity}
 I_N \left( \Phi \right) \propto \norm{u (t)}^2 = \mathcal{A}_0^2 \frac{1 - \cos \left( N \Phi \right)}{1 - \cos \left( \Phi \right)}  \quad \text{for $N$ slits} .
 \end{equation} 
Using various trigonometric identities,\footnote{In particular, $ 2 \sin^2 \left( \Phi / 2 \right) = 1 - \cos \left( \Phi \right)$ and $\sin^2 \left( \Phi \right) = \left[ 1 - \cos \left( \Phi \right) \right] \left[ 1 + \cos \left( \Phi \right) \right]$.} \cref{eq:double-slit_intensity} is recovered in the case of $N = 2$ and \cref{eq:N-slit_intensity} can be normalized\footnote{First, defining the constant of proportionality in \cref{eq:N-slit_intensity} to be $I_0$, the following trigonometric relation holds: 
\begin{equation*}
 I_N \left( \Phi \right) = I_0 \frac{1 - \cos \left( N \Phi \right)}{1 - \cos \left( \Phi \right)} = I_0 \frac{\sin^2 \left( \frac{N}{2} \Phi \right)}{\sin^2 \left( \frac{1}{2} \Phi \right)}. 
 \end{equation*}
Then, using \emph{L'H\^opital's rule} twice shows that the maxima of \cref{eq:N-slit_intensity} are given by $$\lim_{\Phi \to 2 n \pi} I_N \left( \Phi \right) = \frac{1}{2} I_0 N^2 .$$} by setting $I_0 N^2 \to 2$ to give the unitless \emph{multi-slit interference function} \cite{Loewen97}: 
\begin{equation}\label{eq:multi-slit_interference}
 I_{N} \left( \Phi \right) = \left( \frac{\sin \left( \frac{N}{2} \Phi \right)}{N \sin \left( \frac{1}{2} \Phi \right)} \right)^2 \quad \text{for $N$ slits, normalized} .
 \end{equation} 

Since it has been assumed that the plane wave is normally incident on the barrier, the phase shift is $\Phi = \Phi^{\text{normal}} \equiv - k d \sin \left( \vartheta \right)$ and \cref{eq:multi-slit_interference} becomes 
\begin{equation}\label{eq:multi-slit_interference_normal_inc}
 I_{N} \left( k d,  \sin \left( \vartheta \right) \right) = \left( \frac{\sin \left( \frac{N}{2} k d \sin \left( \vartheta \right) \right)}{N \sin \left( \frac{1}{2} k d \sin \left( \vartheta \right) \right)} \right)^2 \quad \text{for $N$ slits, normalized}
 \end{equation} 
\end{subequations}
with $k d \equiv 2 \pi \left( \lambda / d \right)^{-1}$. 
This interference pattern is plotted in \cref{fig:multi_slit_plot} for various values of $N$ with $\lambda / d = 0.4$. 
\begin{figure}
\centering
\includegraphics[scale=0.99]{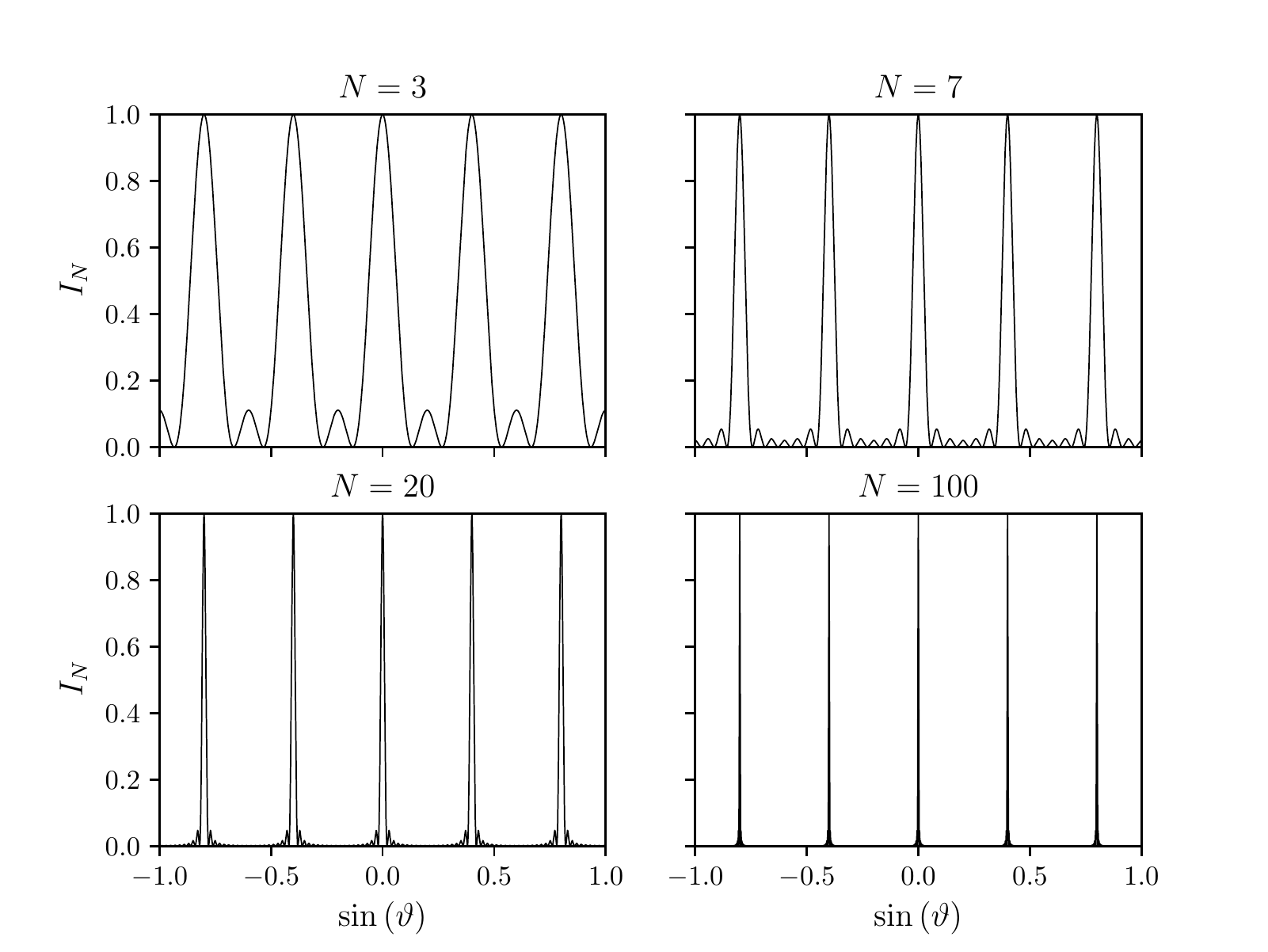}%see intensity_plots.py
\caption[Normalized, far-field interference functions for $N$ sources]{Normalized, far-field interference functions for $N$ sources as a function of $\sin (\vartheta)$ with $\lambda / d = 0.4$.}\label{fig:multi_slit_plot}
\end{figure}
In these plots, it is seen that the number of primary fringes is the same as the double-slit experiment in \cref{fig:double_slit_plot}. 
Given by \cref{eq:fringes}, there are five of these fringes corresponding to a value of $\vartheta_{\text{max}}$ with $n = 0, \pm 1, \pm 2$. 
For $N > 2$, there also exist secondary fringes with maxima separated by points of zero intensity. 
From \cref{eq:N-slit_intensity}, this condition is for normal incidence is 
\begin{equation}\label{eq:fringe_min}
 \begin{split}
 \cos \left( N \Phi \right) = 1 &\implies N \Phi = 2 n \pi \\
 \Phi = \Phi^{\text{normal}} \equiv - k d \sin (\vartheta) &\implies \sin \left( \vartheta_{\text{min}} \right) = - \frac{n}{N} \frac{\lambda}{d} ,
 \end{split}
 \end{equation}
where $\Phi^{\text{normal}}$ is the multi-slit phase shift for this scenario of normal incidence. 
Seen in \cref{fig:multi_slit_plot}, the number of these minima increases with $N$, which causes the number of secondary maxima to increase while their intensities become suppressed. 
Simultaneously, the angular width of the main fringes decreases due to these closely spaced minima, especially as $N$ becomes very large. 

In the limit that $N \to \infty$, the structure can be treated as periodic with $K$ being the \emph{grating wave number} used in \cref{sec:grating_boundary}: 
\begin{equation}\label{eq:grating_number}
 K \equiv \frac{2 \pi}{d} .
 \end{equation}
This causes each fringe to be concentrated into an angular direction $\vartheta_{\text{max}}$ given by \cref{eq:fringes} for $n = 0, \pm 1, \pm 2$ and the secondary fringes vanish. 
The $N$ sources in this scenario start to behave like a diffraction grating, where far-field fringes (now referred to as \emph{diffracted orders}) propagate as planar waves. 
For an arbitrary $\lambda / d$ ratio, the \emph{diffracted angle}, $\vartheta_{\text{max}} \to - \beta_n$ from \cref{eq:fringes} is given by 
\begin{subequations}
\begin{equation}\label{eq:normal_incidence_orders}
 \sin \left( \beta_n \right) = \frac{n \lambda}{d} \quad \text{for } n = 0, \pm 1, \pm 2, \pm 3 \dotsc ,
 \end{equation}
which is a special case of the grating equation for normal incidence. 
For each order number, $n$, such that $- \pi / 2 < \Re \left[ \beta_n \right] < \pi /2$, \emph{propagating orders} exist in the far field. 
Otherwise, when $\abs{n \lambda / d} > 1$ the orders are \emph{evanescent} with $\Re \left[ \beta_n \right] \pm \pi / 2$ and $\Im \left[ \beta_n \right] \neq 0$. 

The expression in \cref{eq:normal_incidence_orders} can also be viewed in terms of wave numbers involving the grating periodicity along the $x$-direction and the $x$-component of the diffracted wave. 
That is, the diffracted wave vector has an $x$-component $k'_{x,n} \equiv k \sin \left( \beta_n \right)$ equal to an integer number of grating wave vectors:
\begin{equation}\label{eq:normal_incidence_orders_wave_number}
 k'_{x,n} = n K \quad \text{for } n = 0, \pm 1, \pm 2, \pm 3 \dotsc 
 \end{equation}
In other words, dispersion takes place in the direction orthogonal to the groove direction and to the grating normal (\emph{i.e.}, the $x$-axis in \cref{fig:double_slit_plot}). 
\end{subequations}
It can be gleaned from \cref{eq:normal_incidence_orders,eq:normal_incidence_orders_wave_number}  that for $\lambda$ characteristic of hard x-rays, diffraction is produced from $d$ on the order of atomic scales \cite{Bragg1913a,Compton35,Als-Nielsen11}. 
In this way, crystalline materials featuring regularly-spaced atoms (\emph{e.g.}, silicon shown in \cref{fig:crystal_structure}) produce x-ray diffraction patterns that can be used for a variety of applications including x-ray spectroscopy \cite{Giacconi79}. 
However, there is little versatility for crystal spectrometers in the soft x-ray band, where $\lambda$ is comparatively long and instead, custom diffraction gratings are better suited for the task.

\section{Off-Plane Geometry}\label{sec:off-plane_geo}
%%%%%%%%%%%%%%%%%%%%%%%%%%%%%%%%%%%%%%%%%--------------------------------------------------
The grazing-incidence requirement of soft x-rays [\emph{cf.\@} \cref{app:x-ray_materials}] requires generalizing the multi-slit interference scenario outlined in \cref{sec:slit_array} to include extra components to the phase shift $\Phi = - k \Delta s$ that arise from using oblique incidence angles in 3D geometries. 
To start, the incident plane wave can be imagined to have a wave vector $\mathbold{k}$ that is confined to the $x-y$ plane, but at an angle $\alpha$ as illustrated in the top panel of \cref{fig:oblique_off-plane}. 
Instead of $\mathbold{k} = k \mathbold{\hat{y}}$ as in \cref{fig:double_slit_experiment,fig:double_slit_experiment_approx}, the wave vector for this scenario is written as 
\begin{subequations}
\begin{equation}\label{eq:in-plane_wave_vector}
 \mathbold{k} = - k \left[ \sin (\alpha) \mathbold{\hat{x}} + \cos(\alpha) \mathbold{\hat{y}} \right] ,
 \end{equation}
so that the total path-length difference becomes $\Delta s = - d \left[ \sin (\beta_n) + \sin(\alpha) \right]$ in the far field, for $N \to \infty$. 
In this case, the diffraction pattern can be described two-dimensionally such that the incident ray is \emph{in-plane} with the orientations of all diffracted fringes. 
The multi-slit phase shift for this scenario is given by 
\begin{equation}\label{eq:in-plane_phase_shift}
 \Phi^{\text{in-plane}} = - k \Delta s = k d \left[ \sin (\beta_n) + \sin(\alpha) \right] = \Phi^{\text{normal}} + k d \sin(\alpha), 
 \end{equation}
where $\Phi^{\text{normal}} \equiv - k d \sin (\vartheta) \to k d \sin (\beta_n)$ is the phase shift for normal incidence [\emph{cf.\@} \cref{sec:slit_array,fig:double_slit_plot}]. 
\begin{figure}
 \hspace{-2cm}  
 \includegraphics[scale=0.85]{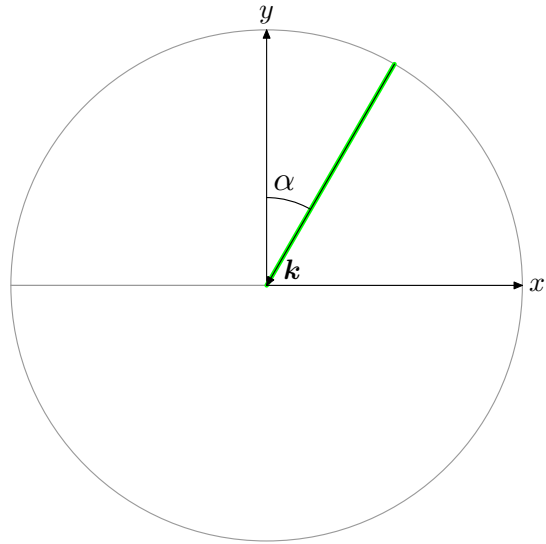}
 \centering 
 \rule{\textwidth}{.4pt}
 \includegraphics[scale=0.85]{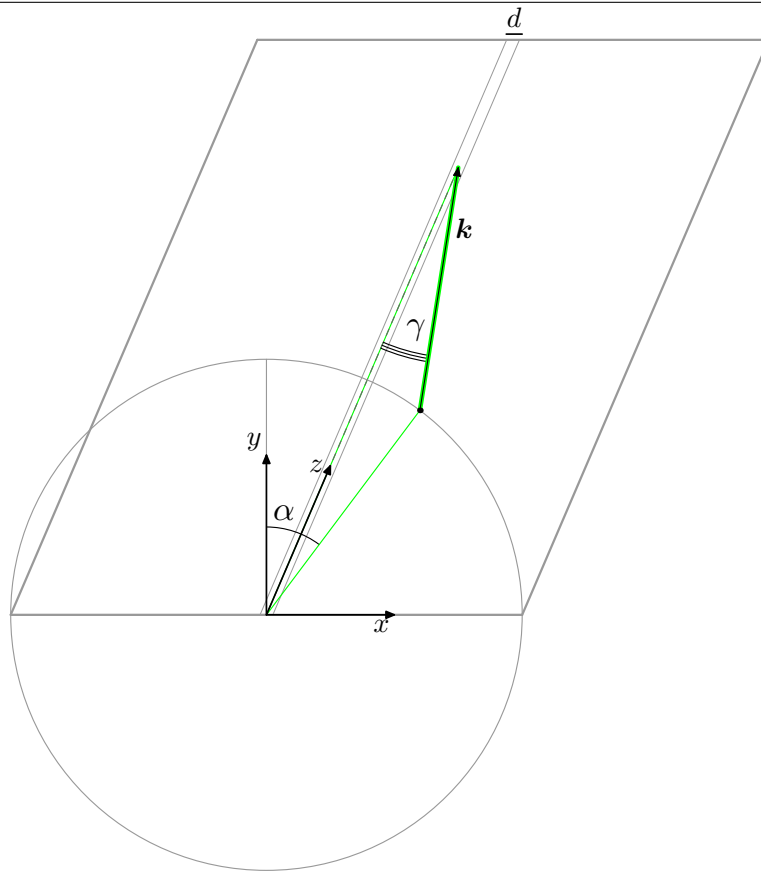}
 \caption[Orientation of a representative wave vector for a generic oblique angle of incidence]{Orientation of a representative wave vector for a generic oblique angle of incidence. The top panel shows the vector confined to the $x-y$ plane, where in-plane diffraction occurs. The bottom panel shows the vector with a $z$-component, in which case off-plane diffraction occurs. The bottom panel appears the same as the top panel in projection when viewed directly down the $z$-axis.}\label{fig:oblique_off-plane}
 \end{figure}

This relation given by \cref{eq:in-plane_phase_shift} implies that with choice of large $\sin(\alpha)$, the phase shift is increased substantially. 
Noting that \cref{eq:normal_incidence_orders} in this case becomes 
\begin{equation}\label{eq:in-plane_orders}
 \sin \left( \alpha \right) + \sin \left( \beta_n \right) = \frac{n \lambda}{d} \quad \text{for } n = 0, \pm 1, \pm 2, \pm 3 \dotsc ,
 \end{equation}
the in-plane geometry under grazing incidence with $\sin(\alpha)$ approaching unity allows for large diffracted angles as compared to normal incidence for the same value of $d$. 
\end{subequations} 
This allows $d \gg \lambda$ provided that there is a very shallow angle of incidence confined to the $x-y$ plane as in \cref{fig:double_slit_plot} and the top panel of \cref{fig:oblique_off-plane}. 
For example, the soft x-ray reflection gratings on \emph{XMM-Newton}, which are designed for an in-plane geometry with $\alpha \approx 88.5^{\circ}$, have a groove spacing of $d \sim \SI{1.5}{\um}$ \cite{denHerder01}. 
On the other hand, \emph{critical angle transmission gratings} use $\alpha$ on the order of just a few degrees and in this case the aperture spacing is commonly $\SI{200}{\nm}$ \cite{Heilmann11}. 

In a more general scenario to that just described, the wave vector $\mathbold{k}$ has a $z$ component characterized by the angle $\gamma$ illustrated in the bottom panel of \cref{fig:oblique_off-plane} such that \cref{eq:in-plane_wave_vector} for the wave vector becomes [\emph{cf.\@} \cref{fig:grating_angles}] 
\begin{subequations}
\begin{equation}
 \mathbold{k} =  - k \left[ \sin (\alpha) \sin (\gamma) \mathbold{\hat{x}} + \cos(\alpha) \sin (\gamma) \mathbold{\hat{y}} - \cos (\gamma) \mathbold{\hat{z}} \right] ,
 \end{equation}
which will be shown to cause \emph{off-plane} diffraction [\emph{cf.\@} \cref{sec:grating_boundary}]. 
The multi-slit phase shift between the sources as observed in the far field is 
\begin{equation}\label{eq:off-plane_phase}
 \Phi^{\text{off-plane}} = - k \Delta s = k d \sin (\gamma) \left[ \sin (\beta_n) + \sin(\alpha) \right] = \sin (\gamma) \, \Phi^{\text{in-plane}}  
 \end{equation}
as a generalization of \cref{eq:in-plane_phase_shift}, which indicates a reduction relative to the in-plane case by a factor of $\sin (\gamma)$. 
This has the consequence that for extreme off-plane geometries with $\gamma$ of a few degrees, $d$ must be smaller than what it would be for a typical in-plane grazing incidence geometry [\emph{cf.\@} \cref{eq:inplane_limit,eq:offplane_limit}]. 
For example, off-plane gratings for soft x-rays currently under study commonly range from $\SI{400}{\nm}$ down to $\SI{160}{\nm}$ depending on the design of the spectrometer \cite{McEntaffer13,Miles18,McCurdy20,McCoy20}. 
This scenario of off-plane diffraction can also be described using framework for in-plane diffraction, where $\mathbold{k}$ is projected onto the the $x-y$ plane so that its magnitude, $\bar{k} \equiv k \sin (\gamma)$, can be thought of as an effective wave number with an increased wavelength, $\bar{\lambda} = \lambda \csc (\gamma)$. 
\end{subequations}

As the number of slits become very large and the fringes behave as collimated diffracted orders, equating $\Phi^{\text{off-plane}} = 2 n \pi$ with $n = 0, \pm 1, \pm 2, \pm 3 \dotsc$ gives the \emph{generalized grating equation}:
\begin{subequations}
\begin{equation}\label{eq:off-plane_incidence_orders}
 \sin \left( \alpha \right) + \sin \left( \beta_n \right) = \frac{n \lambda}{d \sin(\gamma)} \quad \text{for } n = 0, \pm 1, \pm 2, \pm 3 \dotsc 
 \end{equation}
Similar to \cref{eq:normal_incidence_orders_wave_number}, this can be expressed in terms of wave numbers [\emph{cf.\@} \cref{sec:reflected_diffracted}]: 
\begin{equation}\label{eq:off-plane_orders_wave_number}
  k'_{x,n} - k_x = n K \quad \text{for } n = 0, \pm 1, \pm 2, \pm 3 \dotsc ,
 \end{equation}
where $k'_{x,n} \equiv k \sin (\gamma) \sin(\beta_n)$ and $k_x \equiv - k \sin (\gamma) \sin(\alpha)$ so that it can be seen explicitly that grating dispersion takes places entirely in the $x$-direction (the \emph{grating-dispersion direction}).  
In this geometry, diffracted orders by definition have the following $y$-component to the wave vector: 
\begin{equation}
 k'_{y,n} = \pm \sqrt{k^2 - k'^2_{x,n} - k_z^2} \quad \text{with } k_z \equiv k \cos (\gamma) .
 \end{equation}
\end{subequations}
Propagating orders that correspond to those values of $n$ for which $k'_{y,n}$ is real lay on the surface of a cone with an opening angle of $2 \gamma$ while still being equally spaced along the $x$-axis \cite{Neviere78}. 
\begin{figure}
 \centering
 \includegraphics[scale=0.85]{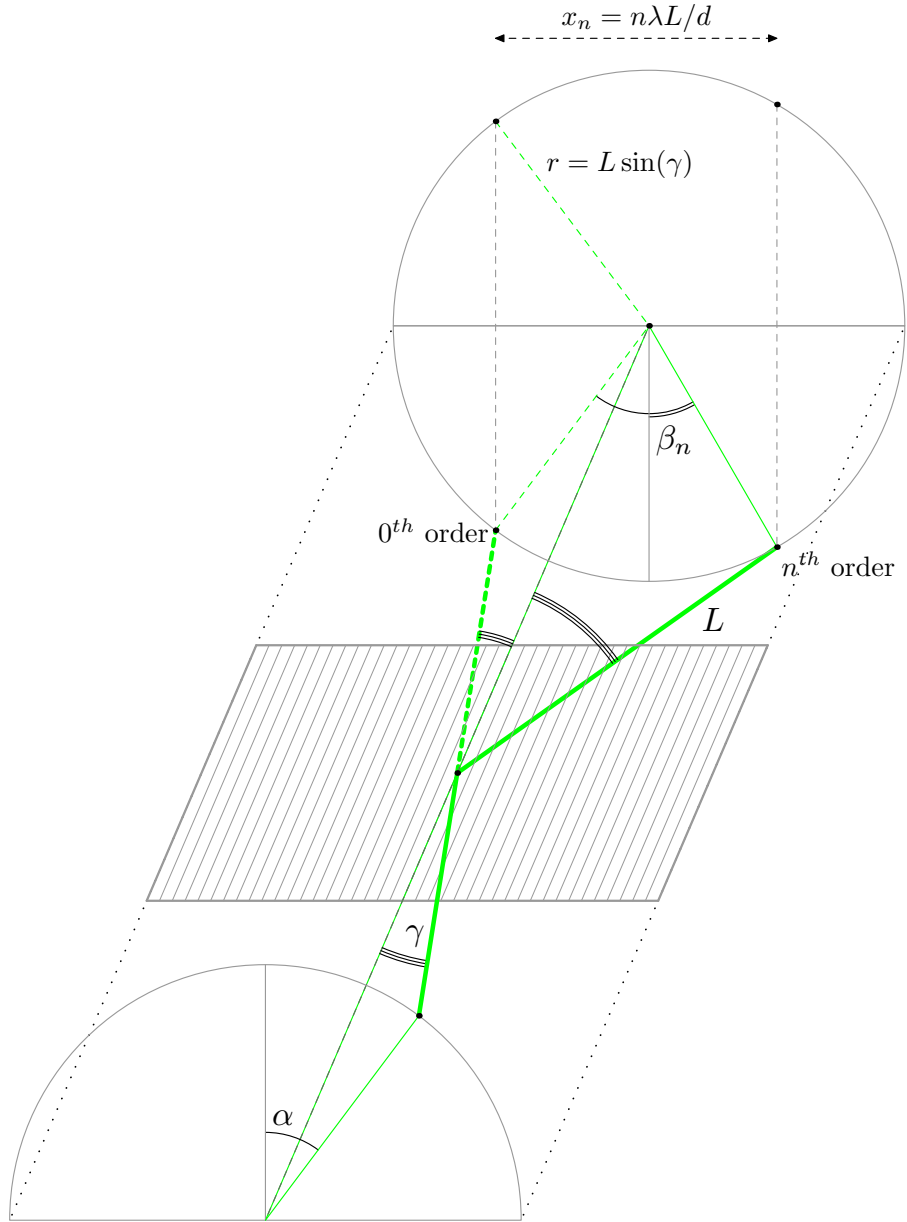}
 \caption{Geometry for a transmission grating producing conical diffraction}\label{fig:conical_transmission}
 \end{figure}
This is visualized for the case of a transmission grating in \cref{fig:conical_transmission}, where the $0^{\text{th}}$ order passes straight through the structure while any given propagating order intersects with the circle drawn in the figure. 

A reflection grating in this conical geometry is shown in \cref{fig:conical_reflection}, where the $0^{\text{th}}$ order is equivalent to the reflected beam from a mirror and all propagating orders lay on the upper-half of the cone. 
\begin{figure}
 \centering
 \includegraphics[scale=0.85]{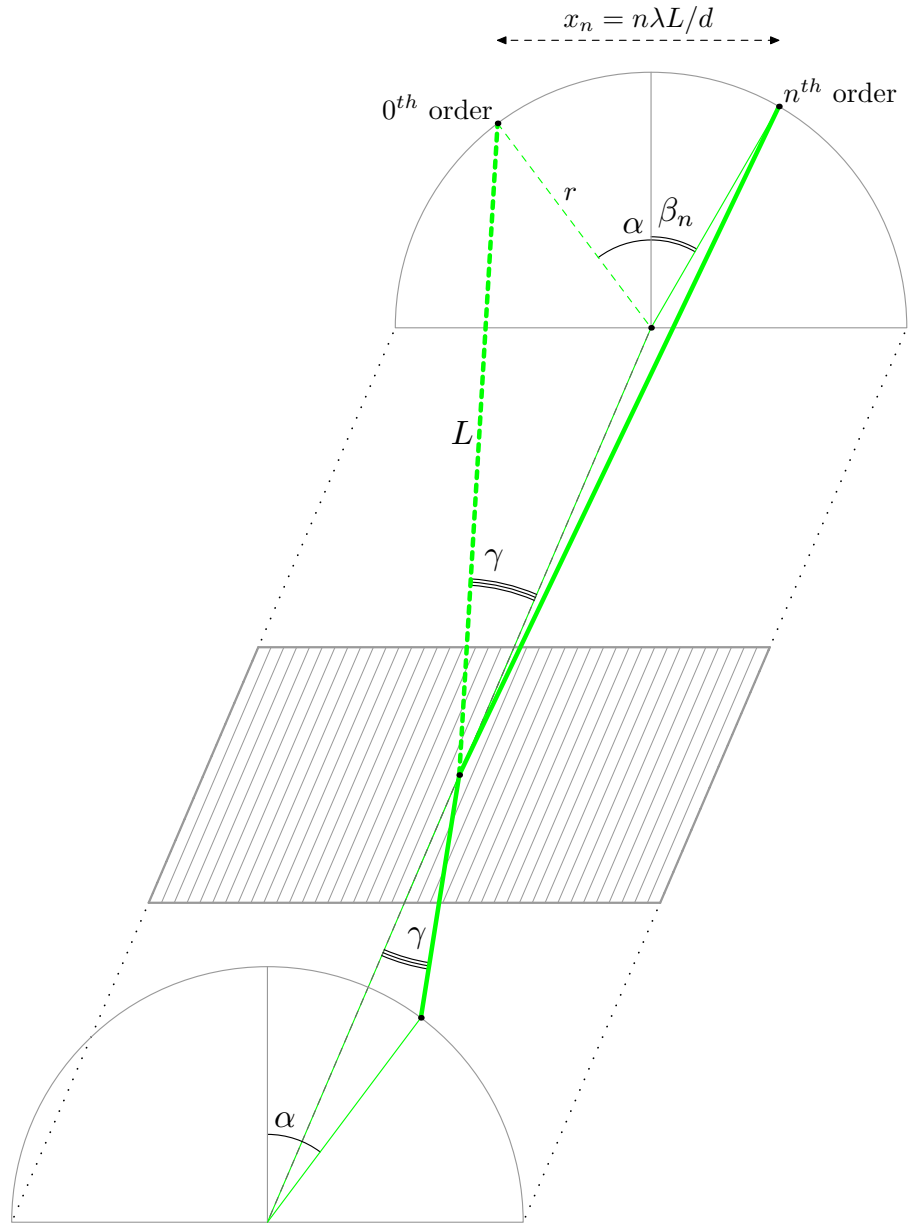}
 \caption{Geometry for a reflection grating producing conical diffraction}\label{fig:conical_reflection}
 \end{figure}
At a distance $L \gg (N-1) d$ away from the grating for either case, the radius of this circle is $r = L \sin (\gamma)$ [\emph{cf.\@} \cref{eq:arc_radius}] and the $x$-distance between the $n^{\text{th}}$ and $0^{\text{th}}$ orders is given by 
\begin{subequations}
\begin{equation}
 x_n = r \left[ \sin \left( \alpha \right) + \sin \left( \beta_n \right) \right] .
 \end{equation}
Using \cref{eq:off-plane_incidence_orders} this becomes [\emph{cf.\@} \cref{eq:linear_dispersion}] 
\begin{equation}\label{eq:dispersion}
 x_n = \frac{n \lambda L}{d} ,
 \end{equation}
\end{subequations}
as labeled in \cref{fig:conical_transmission,fig:conical_reflection}. 
This shows that no matter the incidence angle, the linear dispersion between propagating orders on an imaging detector a distance $L$ away from the grating is proportional to $\lambda /d$ as well as $\abs{n}$. 

\section{On Spectral Resolving Power}\label{sec:resolving_power}
%%%%%%%%%%%%%%%%%%%%%%%%%%%%%%%%%%%%%%%%%--------------------------------------------------
In spectroscopy, the purpose of a diffraction grating is to disperse white light according to color so that the intensity of its constituent wavelength components can be measured and plotted as a spectrum. 
From \cref{eq:off-plane_incidence_orders}, this is based on the principle that a monochromatic plane wave incident on a barrier with $N$ infinitesimal slits produces a number of primary fringe maxima depending on the ratio $\lambda /d$; in the limit that $N$ is very large, these maxima become diffracted orders that are observed as plane waves in the far field [\emph{cf.\@} \cref{sec:slit_array}]. 
On a screen a distance $L \gg (N - 1) d$ away from a grating with $N$ slits or grooves, the $n^{\text{th}}$ diffracted order exists at a position relative to $0^{\text{th}}$ order given by \cref{eq:dispersion}. 
The secondary fringes are separated by a distance $\Delta x = L \lambda / N d$ [\emph{cf.\@} \cref{eq:fringe_min}], which implies that each primary fringe has a width $\Delta x$ on the screen. 
Based on these considerations alone, the spectral resolving power, $\mathscr{R}$, of such a grating is \cite{Loewen97} 
\begin{equation}\label{eq:diffraction_limit_res}
 \frac{\lambda}{\Delta \lambda} = \abs{\frac{x_n}{\Delta x}} = \abs{n} N ,
 \end{equation}
which applies only to a \emph{diffraction-limited} scenario, where all other aberrations arising from optical imperfections can be neglected. 
Essentially, this states that the maximum $\mathscr{R}$ that can possibly be attained is dependent on the number of slit sources and that spectra become increasingly easier to resolve as the magnitude of the order number increases. 
However, \cref{eq:diffraction_limit_res} is also dependent on the angles of incidence, which can be seen using \cref{eq:off-plane_incidence_orders} to write 
\begin{equation}
 \abs{n} = \sin \left( \gamma \right) \abs{\sin \left( \alpha \right) + \sin \left( \beta_n \right)} \frac{d}{\lambda}
 \end{equation} 
so that \cref{eq:diffraction_limit_res} becomes 
\begin{subequations}
\begin{equation}\label{eq:diffraction_limit_res_angle}
 \frac{\lambda}{\Delta \lambda} = \sin \left( \gamma \right) \abs{\sin \left( \alpha \right) + \sin \left( \beta_n \right)} \frac{N d}{\lambda} ,
 \end{equation}
which states that higher $\mathscr{R}$ is achieved with large angles of incidence and diffraction \cite{Loewen97}. 
The maximum $\mathscr{R}$ attainable for a grating with $N$ perfectly spaced apertures is then 
\begin{equation}\label{eq:diffraction_limit_res_angle_max}
 \frac{\lambda}{\Delta \lambda} = \frac{2 N d}{\lambda} ,
 \end{equation}
\end{subequations}
where $N d$ is the ruled width of the grating. 

As a simple example that demonstrates the concept of diffraction-limited $\mathscr{R}$, it is supposed that collimated white light composed of representative red, green and blue wavelengths in equal parts is normally incident on the barrier drawn in \cref{fig:double_slit_plot}, but with $N$ regularly-spaced, infinitesimal slits. 
Shown in \cref{fig:multi_slit_plot_color}, each color is dispersed by a different amount proportional to the ratio $\lambda / d$ so that effectively, a spectrum is produced around each primary fringe maximum for $n = \pm 1, \pm 2$.  
\begin{figure}
\centering
\includegraphics[scale=0.99]{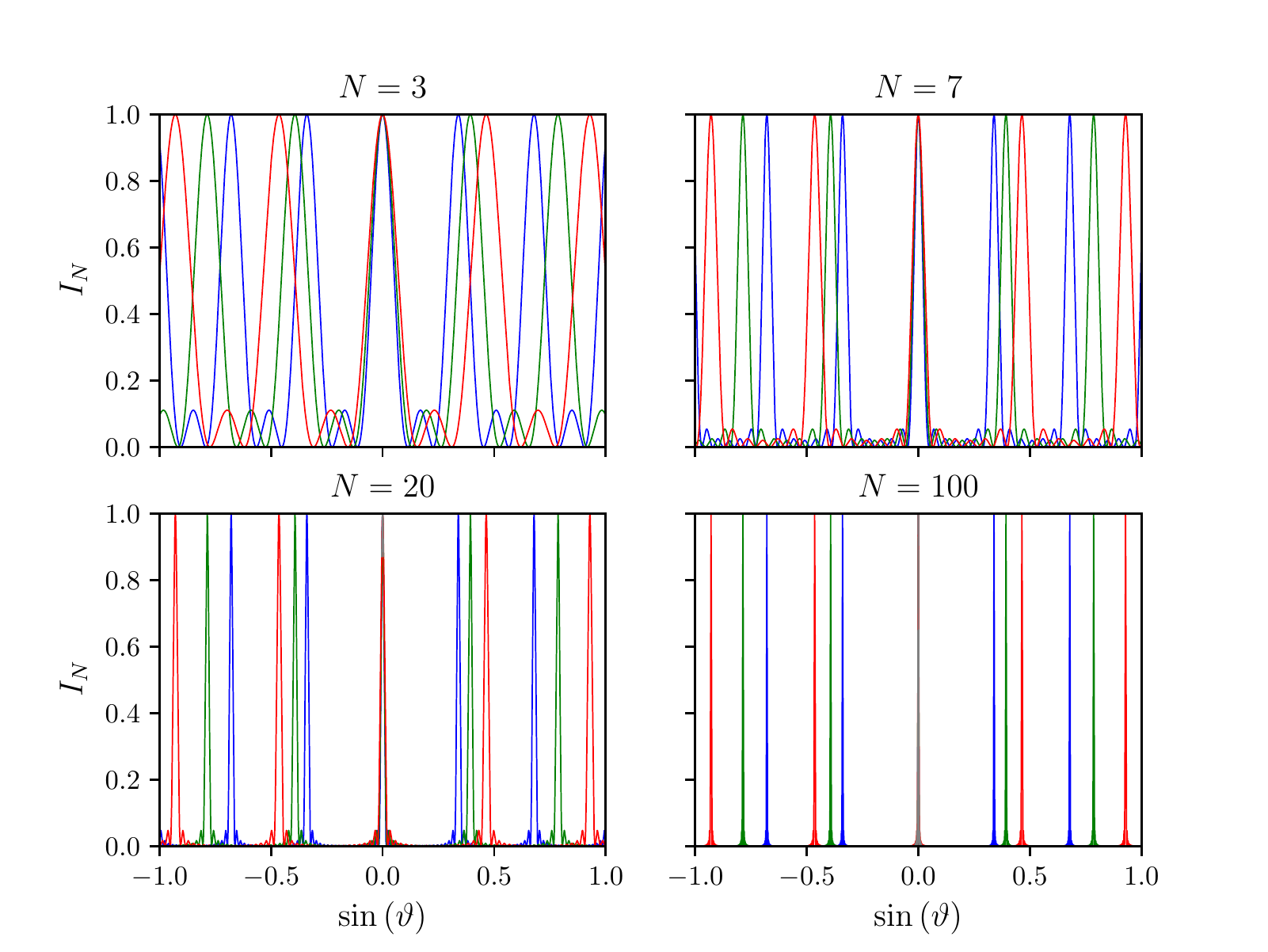}
\caption[Far-field interference of polychromatic light for $N$ sources]{Far-field interference for $N$ sources as a function of $\sin (\vartheta)$ for wavelengths corresponding to red ($\lambda = \SI{650}{\nm}$), green ($\lambda = \SI{550}{\nm}$) and blue ($\lambda = \SI{475}{\nm}$) with $d = \SI{1.4}{\micro\metre}$.}\label{fig:multi_slit_plot_color}
\end{figure}  
It is clear from the figure that unless $N$ is large enough, these fringes overlap with each other and hence the colors cannot be distinguished through grating dispersion. 
Although only a modest value of $N$ is needed to distinguish between the red, green and blue wavelengths in the case of \cref{fig:multi_slit_plot_color}, generally a very large number of slits is needed to separate colors closer together in wavelength. 
However, as the number of these fringes increases with decreasing $\lambda / d$, inevitably there will be some spectral overlap. 
For this reason, diffracted orders generally need to be separated for useful spectra to be extracted. 
\begin{figure}
\centering
\includegraphics[scale=0.99]{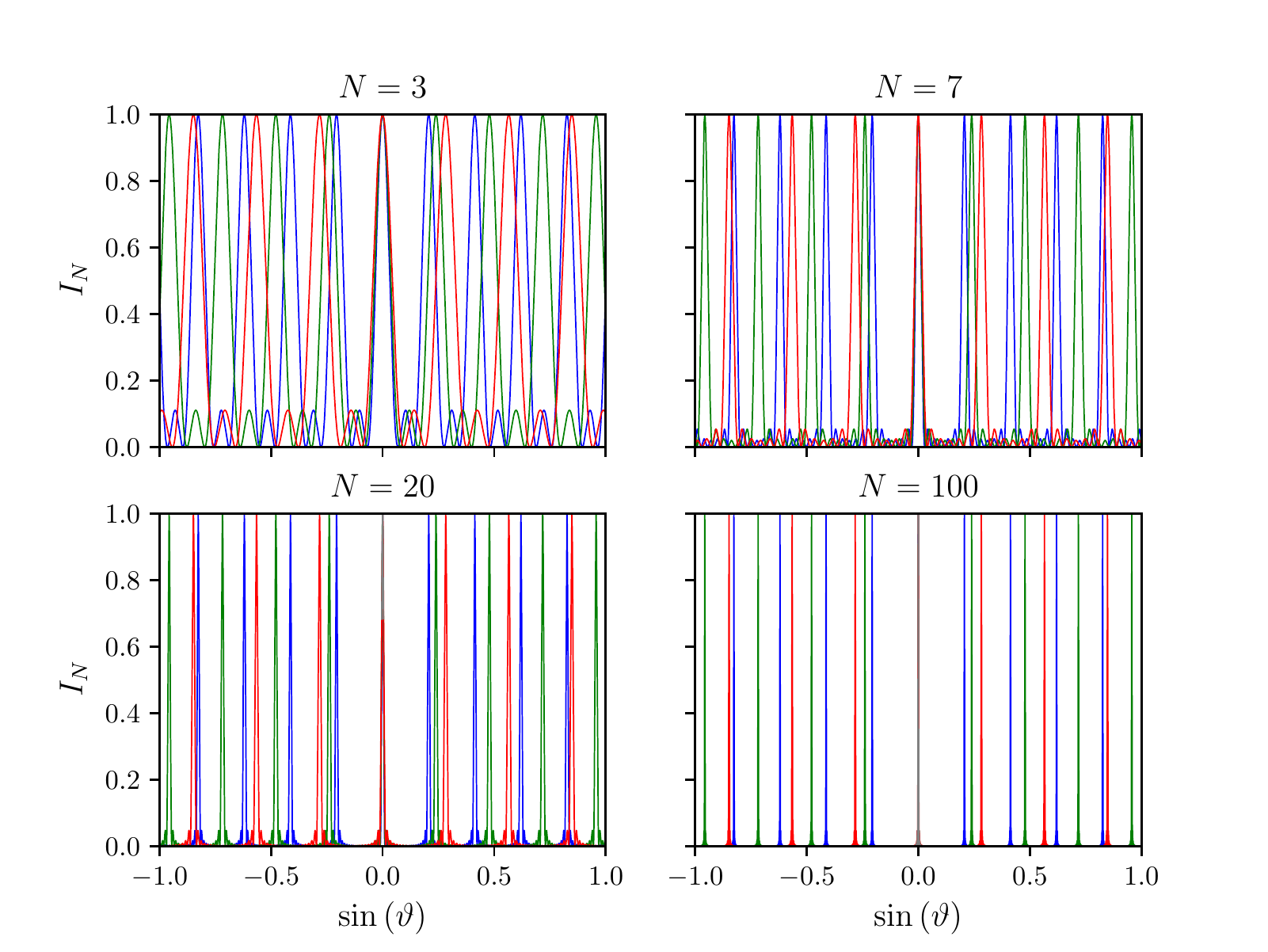}
\caption[Far-field interference of polychromatic light for $N$ sources with order confusion]{Far-field interference for $N$ sources as a function of $\sin (\vartheta)$ for wavelengths corresponding to red ($\lambda = \SI{650}{\nm}$), green ($\lambda = \SI{550}{\nm}$) and blue ($\lambda = \SI{475}{\nm}$) with $d = \SI{2.3}{\micro\metre}$, demonstrating the concept of order confusion.}\label{fig:multi_slit_plot_color_confusion}
\end{figure}
This is demonstrated in \cref{fig:multi_slit_plot_color_confusion}, where the red, green and blue wavelengths from \cref{fig:multi_slit_plot_color} are dispersed by $N$ slits with a different value for $d$.

Spectral resolving power can also be treated using the framework of \emph{Fraunhofer diffraction} such that an array of $N$ in-phase, infinitesimal slits, each spaced by the same distance $d$, can be described as a sum of \emph{Dirac delta functions}:\footnote{A Dirac delta function is defined as 
\begin{equation*}
\delta_D \left( x \right) \equiv
   \begin{cases}
     \infty, & \text{if}\ x = 0 \\
     0, & \text{if}\ x \neq 0 .
   \end{cases} 
\end{equation*}}
\begin{subequations}
\begin{equation}\label{eq:amplitude_dirac_finite}
 \mathcal{A} (x) = \mathcal{A}_0 \sum_{m=1}^{N} \delta_D \left( \frac{x}{d} - m \right) .
 \end{equation}
Taking the Fourier transform and using the exponential-sum formula given by \cref{eq:exp_sum} with $\Phi \to - k'_x d$ yields 
\begin{align}
 \begin{split}\label{eq:DC_FT}
 \mathcal{A} \left( k'_x \right) &= \mathcal{A}_0 \, d \int_{-\infty}^{\infty} \sum_{m=1}^{N} \delta_D \left( x - m d \right) \mathrm{e}^{-i k'_x x} \dd{x} \\
 &= \mathcal{A}_0 \, d \sum_{m=1}^{N} \mathrm{e}^{-i k'_x m d} = \mathrm{e}^{-i k'_x d} \frac{1 - \mathrm{e}^{- i N k'_x d}}{1 - \mathrm{e}^{- i k'_x d}} .
 \end{split}
 \end{align}
The corresponding far-field interference pattern is the squared norm of this function: 
\begin{equation}
 I_N = \norm{\mathcal{A} \left( k'_x \right)}^2 = \mathcal{A}_0^2 \frac{1 - \cos \left( N k'_x d \right)}{1 - \cos \left( k'_x d \right)} ,
 \end{equation}
\end{subequations}
which is equivalent to the multi-slit intensity function given by \cref{eq:N-slit_intensity}. 
This indicates that a grating with a perfect but finite slit periodicity is associated with diffraction-limited $\mathscr{R}$ [\emph{cf.\@} \cref{eq:diffraction_limit_res,eq:diffraction_limit_res_angle,eq:diffraction_limit_res_angle_max}]. 

In the limit that $N \to \infty$, \cref{eq:amplitude_dirac_finite} becomes 
\begin{equation}\label{eq:amplitude_dirac}
 \mathcal{A} (x) = \mathcal{A}_0 \sum_{n=-\infty}^{\infty} \delta_D \left( \frac{x}{d} - n \right) \equiv \mathcal{A}_0 \, \Sha \left( \frac{x}{d} \right) = \mathcal{A}_0 \, d \, \Sha \left( x \right) ,
 \end{equation}
where $\Sha \left( x \right)$ is known as the \emph{Dirac comb}. 
The Fourier transform of \cref{eq:amplitude_dirac} is
\begin{subequations}
\begin{align}\label{eq:eq:amplitude_dirac_FT}
 \begin{split}
 \mathcal{A} \left( k'_x \right) &= \mathcal{A}_0 \int_{-\infty}^{\infty} \Sha \! \left( \frac{x}{d} \right) \mathrm{e}^{-i k'_x x} \dd{x} \\
 &= \mathcal{A}_0 \int_{-\infty}^{\infty} \sum_{n=-\infty}^{\infty} \delta_D \left( \frac{x}{d} - n \right) \mathrm{e}^{-i k'_x x} \dd{x} = \mathcal{A}_0 \sum_{n=-\infty}^{\infty} \mathrm{e}^{-i k'_x n d} , 
 \end{split}
 \end{align}
which is recognized up to a factor of $2 \pi / d \equiv K$ as the Fourier-series expansion of a Dirac comb:
\begin{equation}\label{eq:sha_FS}
 \Sha \! \left( - \frac{k'_x}{2 \pi} \right) \equiv d \sum_{n=-\infty}^{\infty} \mathrm{e}^{-i k'_x n d} = 2 \pi \, \Sha \! \left( k'_x \right)  
 \end{equation}
so that 
\begin{equation}\label{eq:sha_FT}
 \int_{-\infty}^{\infty} \Sha \! \left( \frac{x}{d} \right) \mathrm{e}^{-i k'_x x} \dd{x} = \Sha \! \left( \frac{k'_x}{K} \right) .
 \end{equation}
\end{subequations}
Here, $\mathcal{A} \left( k'_x \right)$ is the far-field amplitude of a grating with $\mathscr{R} \to \infty$, where each diffracted order is associated with a singular value of $k'_x$ as described by the sum of Dirac delta functions. 

In practice, $N$ can be considered to approach infinity for an x-ray reflection of substantial size.  
The silicon grating with $d \lessapprox \SI{160}{\nm}$ described in \cref{sec:crystal_etching}, for example, has $N \gtrapprox \num{468750}$ so that diffraction-limited $\mathscr{R}$ is on the order of \num{100000} or more by \cref{eq:diffraction_limit_res}. 
On the other hand, the spectral resolving power of an x-ray reflection grating, is limited (typically to a few thousand) by how well the radially-ruled profile of the grating grooves matches the focal length of a Wolter-I telescope [\emph{cf.\@} \cref{sec:grating_tech_intro}]. 
While beamline testing for $\mathscr{R}$ is beyond the scope of this dissertation, it is noted that an imperfect $\mathscr{R}$ effectively serves to broaden the angular spread of each propagating order so that, in principle, the total flux associated with each diffracted beam is unaffected. 
Because of this, $\mathscr{R} \to \infty$ with $N \to \infty$ is assumed for the examination of scalar diffraction efficiency in \cref{sec:amplitude_phase}. 

\section{Groove Shape Impact on Diffraction Efficiency}\label{sec:amplitude_phase}
%%%%%%%%%%%%%%%%%%%%%%%%%%%%%%%%%%%%%%%%%--------------------------------------------------
A diffraction grating broadly can be considered to be any sort of structure that causes an incident wave to undergo periodic shifts in amplitude, phase, or possibly both, across its surface. 
The example of water waves passing through a set of evenly-spaced obstacles is analogous to a transmission grating for electromagnetic radiation that features a finely-pitched periodic array of absorbing slabs. %mentioned previously
In the simplest of scenarios, such an optical device functions as an \emph{amplitude grating}, where, by the Huygens-Fresnel principle, the point sources of wavelength $\lambda$ that constitute its surface can be regarded to vary in amplitude alone. 
According to the framework of Fraunhofer diffraction [\emph{cf.\@} \cref{sec:resolving_power}], the far-field interference pattern then is the Fourier transform of this periodic amplitude function \cite{Born80}. 
That is, if an array of slits can be described with an amplitude function $\mathcal{A} (x)$, then the Fourier transform is a function of the $x$-component of the diffracted wave vector in the far field with $k'_x \equiv k \sin \left( \gamma \right) \sin \left( \beta_n \right)$: 
\begin{subequations}
\begin{equation}
 \mathcal{A} \left( k'_x \right) = \int_{-\infty}^{\infty} \mathcal{A} (x) \, \mathrm{e}^{-i k'_x x} \dd{x} ,
 \end{equation}
so that taking the inverse-Fourier transform recovers $\mathcal{A} (x)$: 
\begin{equation}\label{eq:amplitude_Fourier}
 \mathcal{A} (x) = \frac{1}{2 \pi} \int_{-\infty}^{\infty} \mathcal{A} \left( k'_x \right) \mathrm{e}^{i k'_x x} \dd{k'_x} .
 \end{equation}
\end{subequations}

On the other hand, a reflection grating realized as a mirror with regularly-spaced grooves etched into its surface (also described as a \emph{surface-relief grating}) may not provide any variation in amplitude at all. 
Rather, as light traverses the depth of the grooves and reflects back under ideal conditions, the net result is a periodic phase shift and the surface can be thought of as a \emph{phase grating} that determines the far-field interference pattern.  
To study this, a more general version of \cref{eq:amplitude_Fourier} is defined as the grating \emph{transmittance function} \cite{Sanchez-Lopez09,Harvey19,Casini14}:
\begin{subequations}
\begin{equation} \label{eq:grating_transmittance}
 G (x) \equiv \mathcal{A} (x) \mathrm{e}^{i \Phi (x)} = \frac{1}{2 \pi} \int_{-\infty}^{\infty} G \left( k'_x \right) \mathrm{e}^{i k'_x x} \dd{k'_x} 
 \end{equation}
with a Fourier transform describing the far-field interference pattern given by 
\begin{equation}
 G \left( k'_x \right) = \int_{-\infty}^{\infty} G (x) \, \mathrm{e}^{-i k'_x x} \dd{x} .
 \end{equation}
If $G(x)$ is periodic with $G (x+d) = G (x)$, it can be expressed as a Fourier series: 
\begin{equation} \label{eq:grating_transmittance_series}
 G (x) = \sum_{n=-\infty}^{\infty} G_n \mathrm{e}^{i n K x} ,
 \end{equation}
where $K \equiv 2 \pi /d$ [\emph{cf.\@} \cref{eq:grating_number}] and the Fourier coefficients are
\begin{equation} \label{eq:transmit_fourier_coeff}
 G_n = \frac{1}{d} \int_{0}^{d} G \left( x \right) \mathrm{e}^{- i n K x} \dd{x} \quad \text{for } n = 0, \pm 1, \pm 2, \pm 3 \dotsc
 \end{equation} 
\end{subequations}
Far-field intensity in the $n^{\text{th}}$ diffracted order, $\mathcal{I}_n$, as is it passes through a plane parallel to the grating surface depends on the following projection of Poynting's vector:
\begin{equation}\label{eq:order_intensity_scalar}
 \mathcal{I}_n = \mathbold{S}_n \cdot \mathbold{\hat{y}} \propto \sin \left( \gamma \right) \cos \left( \beta_n \right) \norm{G_n}^2 ,
 \end{equation}
where $\mathbold{S}_n$ describes the directional energy flux of the $n^{\text{th}}$ propagating order. 

The assumption of $G (x+d) = G (x)$ for all $x$ implies a grating with $\mathscr{R} \to \infty$ so that in this case, the form of $G (x)$ only impacts diffraction efficiency [\emph{cf.\@} \cref{sec:resolving_power}]. 
Although the interaction of the incident wave with the grating structure is not treated under this framework, its intensity can be taken as $\mathcal{I}_{\text{inc}} \propto \sin \left( \gamma \right) \cos \left( \alpha \right)$ in analogy to \cref{eq:transmit_fourier_coeff} so that scalar diffraction efficiency is defined as 
\begin{equation}\label{eq:scalar_efficiency}
 \mathscr{E}_n \equiv \frac{\mathcal{I}_n}{\mathcal{I}_{\text{inc}}} \propto \frac{\cos \left( \beta_n \right)}{\cos \left( \alpha \right)} \norm{G_n}^2 .
 \end{equation}
Approximate behavior for \emph{relative diffraction efficiency} can be gleaned using \cref{eq:scalar_efficiency}, where the constant of proportionality is determined from requiring that $\sum_n \mathscr{E}_n = 1$ for all $n$ corresponding to propagating orders [\emph{cf.\@} \cref{sec:rel_eff_perf}]. 
In the discussion that follows, the efficiency behavior of common groove shapes are considered below with emphasis on phase gratings, where a function $\Phi (x+d) = \Phi (x)$ is used to describe surface-relief reflection gratings with various groove profiles. 

To formulate descriptions for amplitude and phase gratings, an array of $N \to \infty$ infinitesimal slits given by \cref{eq:amplitude_dirac} with $\mathcal{A} (x+d) = \mathcal{A} (x)$ is again considered. 
Inserting \cref{eq:amplitude_dirac} as $G(x)$ with $\Phi (x) = 0$ into \cref{eq:transmit_fourier_coeff} shows that the amplitude of each diffracted order in the far field is identical:
\begin{align} 
 \begin{split}\label{eq:amplitude_fourier_coeff}
 G_n &= \frac{\mathcal{A}_0}{d} \int_{0}^{d} \Sha \left( \frac{x}{d} \right) \mathrm{e}^{- i n K x} \dd{x} = \mathcal{A}_0 \int_{0}^{d} \sum_{m=-\infty}^{\infty} \delta_D \left( x - m d \right) \mathrm{e}^{- i n K x} \dd{x} \\
 &= \mathcal{A}_0 \int_{0}^{d} \delta_D \left( x - d \right) \mathrm{e}^{- i n K x} \dd{x} = \mathcal{A}_0 \mathrm{e}^{- i n \left( 2 \pi \right)} = \mathcal{A}_0 ,
 \end{split}
 \end{align} 
which is the same result obtained from \cref{eq:DC_FT} and the multi-slit interference function given by \cref{eq:N-slit_intensity} in the limit that $N \to \infty$. 
\begin{figure}
 \centering
 \includegraphics[scale=0.62]{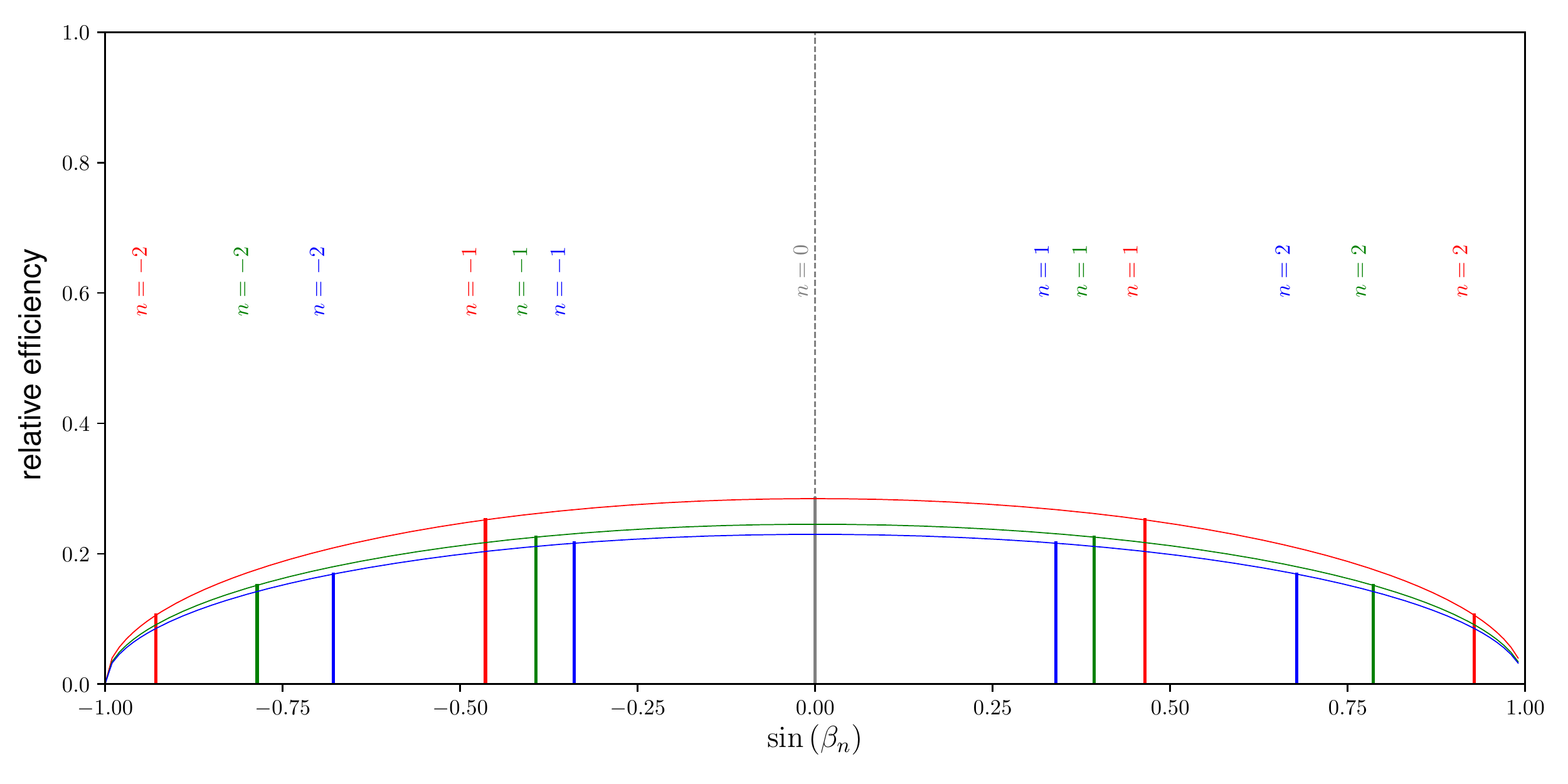}
 \caption[Relative efficiency of a Dirac-comb amplitude grating.]{Relative efficiency of a Dirac-comb amplitude grating at normal incidence with $\lambda / d$ for red, green and blue.}\label{fig:dirac_amplitude}
 \end{figure} 
Relative diffraction efficiency, $\mathscr{E}_n$, in this case is modulated only by the trigonometric factors involving $\beta_n$ and $\alpha$ that appear in \cref{eq:scalar_efficiency} such that for $\alpha = 0$,
\begin{equation}
 \mathscr{E}_n \propto \cos \left( \beta_n \right) = \sqrt{1 - \left( \frac{n \lambda}{d} \right)^2} ,
 \end{equation}
which is plotted as relative efficiency in \cref{fig:dirac_amplitude} for three values of $\lambda / d$ that are representative of red, green and blue wavelengths [\emph{cf.\@} \cref{fig:multi_slit_plot_color}]. 
Unlike the Dirac comb [\emph{cf.\@} \cref{eq:amplitude_dirac,eq:amplitude_fourier_coeff}], real gratings have apertures with finite size or more generally, apertures where amplitude $\mathcal{A} (x)$ and phase $\Phi (x)$ vary across the periodic distance $d$, which results in a set of $G_n$ that vary between diffracted orders.  
The Dirac comb, however, can be viewed as a descriptor for the locations of diffracted order and it can be convolved with the diffraction pattern from a single slit to describe a grating with a finite aperture \cite{Loewen97,DeRoo16thesis}. 
Using a similar approach, the Fourier components for square wave, sinusoidal and sawtooth transmittance functions with $G (x+d) = G (x)$ are described in \cref{eq:square_principle,sec:sinusoid_principle,sec:sawtooth_principle}. 
Each example, as in \cref{fig:dirac_amplitude}, uses $\lambda =$~\SIlist{650;550;475}{\nm} for red, green and blue, respectively, with a groove spacing of $d = \SI{1.4}{\micro\metre}$. 
While these examples explicitly assume in-plane diffraction with $\sin \left( \gamma \right) = 1$, the more general off-plane case can recovered by substituting $\lambda \to \bar{\lambda} \equiv \lambda \csc \left( \gamma \right)$ [\emph{cf.\@} \cref{sec:off-plane_geo}]. 

\subsection{The Square Wave}\label{eq:square_principle}
%%%%%%%%%%%%%%%%%%%%%%%%%%%%%%%%%%%%%%%%%--------------------------------------------------
The first example to consider is a binary amplitude grating with a slit width $\mathscr{W}$, which is not assumed to be small compared to $\lambda$ as in the case of the Dirac comb. 
This parameter $\mathscr{W}$ sets the duty cycle, $\mathscr{W}/d$, of the piece-wise transmittance function for a square wave: 
\begin{subequations}
\begin{equation}
    G (x) = 
                \begin{cases}
                  \mathcal{A}_0 , &\quad \text{for } 0 < x \leq \mathscr{W} \\
                  0 , &\quad \text{for } \mathscr{W} < x \leq d ,
                \end{cases}
 \end{equation}
where \cref{eq:transmit_fourier_coeff} gives the following Fourier coefficients: 
\begin{equation}\label{eq:G_terms_square}
 G_n = \frac{\mathcal{A}_0}{d} \int_0^{\mathscr{W}} \mathrm{e}^{- i n K x} \dd{x} = \frac{i \mathcal{A}_0}{2 n \pi} \left( \mathrm{e}^{- i n K \mathscr{W}} - 1 \right) \quad \text{for } n = 0, \pm 1, \pm 2, \pm 3 \dotsc
 \end{equation}
Physically, this describes the behavior of a transmission grating featuring an array of absorbing slabs with scalar diffraction efficiency [\emph{cf.\@} \cref{eq:scalar_efficiency}] given by 
\begin{align}\label{eq:binary_amplitude_intensity}
 \begin{split}
 \mathscr{E}_n &\propto \frac{\cos \left( \beta_n \right)}{\cos \left( \alpha \right)} \norm{G_n}^2 \\
 \text{with} \quad \norm{G_n}^2 &\propto \frac{1}{2} \left( \frac{1}{n \pi} \right)^2 \left[ 1 - \cos \left( n K \mathscr{W} \right) \right] = \left( \frac{\mathscr{W}}{d} \right)^2 \sinc^2 \left( \frac{n \pi \mathscr{W}}{d} \right) ,
 \end{split}
\end{align}
\end{subequations}
where $\sinc (x) \equiv \sin(x) / x$ is the \emph{sinc function} [\emph{cf.\@} \cref{fig:sinc_squared}]. 
This shows that the intensity of propagating orders is dependent on the the ratio $n \mathscr{W} / d$ such that when it is an integer, intensity reaches zero. 

For example, a grating with a \SI{50}{\percent} duty cycle with $\mathscr{W} = d/2$ has all even orders other than $n=0$ vanish since $\norm{G_n}^2 \propto \sinc^2 \left( n \pi / 2 \right)$ for a fixed geometry; a \SI{25}{\percent} duty cycle with $\mathscr{W} = d/4$ yields $\norm{G_n}^2 \propto \sinc^2 \left( n \pi / 4 \right)$ and therefore every $4^{\text{th}}$ order is suppressed. 
\begin{figure}
 \centering
 \includegraphics[scale=0.62]{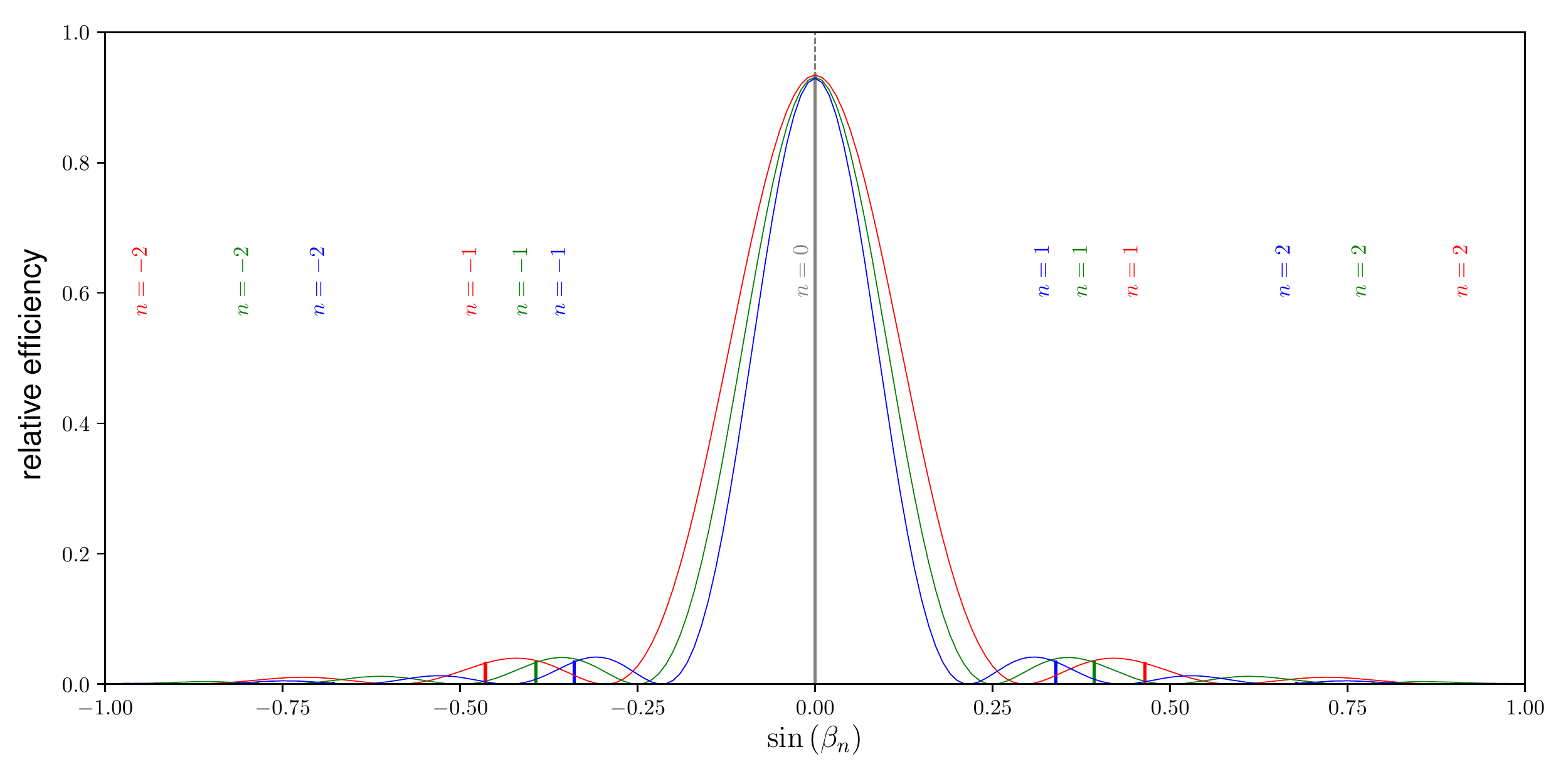}
 \caption[Relative efficiency of a square-wave amplitude grating.]{Relative efficiency of a $\mathscr{W}/d = \SI{50}{\percent}$ square-wave amplitude grating at normal incidence with $\lambda / d$ for red, green and blue.}\label{fig:square_amplitude}
 \end{figure} 
Relative efficiency for a grating at normal incidence with $\mathscr{W} = d/2$ is plotted in \cref{fig:square_amplitude}, where it is seen that propagating orders with $n= \pm 2$ have $\mathscr{E}_n \approx 0$ as a result of $\mathscr{W}/d = \SI{50}{\percent}$. % a \SI{50}{\percent} duty cycle. 
In the figure, it is also seen that $\mathscr{E}_n$ for $n= \pm 1$ comprise a small percentage of the total efficiency. 
While this can be improved with smaller $\mathscr{W}/d$, efficiency is ultimately spread relatively evenly amongst unsuppressed propagating orders. 

A binary phase grating can be taken to represent a reflection grating used near normal incidence, where radiation picks up a phase shift $\Phi_g$ as radiation reflects from the the top and bottom of rectangular grooves\footnote{$\Phi_g$ is not to be confused with the multi-slit phase shift, $\Phi^{\text{off-plane}}$, given by \cref{eq:off-plane_phase}.} of width $\mathcal{W}$ and depth $h_0$. 
For an in-plane geometry, this phase shift from the groove facet can be written explicitly as\footnote{However, it should be noted that if radiation illuminates significant portions of the groove sidewalls, then this analysis does not hold.}
\begin{equation}\label{eq:laminar_phase}
 \Phi_g = \frac{2 \pi}{\lambda} h_0 \left[ \cos \left( \alpha \right) + \cos \left( \beta_n \right) \right] . 
 \end{equation}
With amplitude constant everywhere, a binary phase grating can be described using
\begin{subequations}
\begin{equation}
    G (x) = 
                \begin{cases}
                  \mathcal{A}_0 \mathrm{e}^{i \Phi_g} , &\quad \text{for } 0 < x \leq \mathscr{W} \\
                  \mathcal{A}_0 , &\quad \text{for } \mathscr{W} < x \leq d 
                \end{cases}
 \end{equation}
with Fourier coefficients coming out to
\begin{align}\label{eq:G_terms_square_phase}
 \begin{split}
 G_n &= \frac{\mathcal{A}_0}{d} \left( \mathrm{e}^{i \Phi_g} \int_0^{\mathscr{W}} \mathrm{e}^{- i n K x} \dd{x} + \int_{\mathscr{W}}^d \mathrm{e}^{- i n K x} \dd{x} \right) \\
 &= \frac{i \mathcal{A}_0}{2 n \pi} \left( \mathrm{e}^{i \Phi_g} - 1 \right) \left( \mathrm{e}^{- i n K \mathscr{W}} - 1 \right) \quad \text{for } n = 0, \pm 1, \pm 2, \pm 3 \dotsc  
 \end{split}
 \end{align}
so that the scalar diffraction efficiency is  
\begin{align}\label{eq:binary_phase_intensity}
 \begin{split}
 \mathscr{E}_n &\propto \frac{\cos \left( \beta_n \right)}{\cos \left( \alpha \right)} \norm{G_n}^2 \\
 \text{with} \quad \norm{G_n}^2 &\propto \left( \frac{1}{n \pi} \right)^2 \left[ 1 - \cos \left( n K \mathscr{W} \right) \right] \left[ 1 - \cos \left( \Phi_g \right) \right] \\
 &= \left( \frac{\mathscr{W} \Phi_g}{d} \right)^2 \sinc^2 \left( \frac{n \pi \mathscr{W}}{d} \right) \sinc^2 \left( \frac{\Phi_g}{2} \right) .
 \end{split}
 \end{align}
\end{subequations}
Therefore, similar to the results obtained from the amplitude amplitude grating considered above given by \cref{eq:binary_amplitude_intensity}, it is evident from \cref{eq:binary_phase_intensity} that the intensity of diffracted orders in this case is dependent on $n \mathscr{W} / d$ as well as $\Phi_g / 2 \pi$. 
\begin{figure}
 \centering
 \includegraphics[scale=0.62]{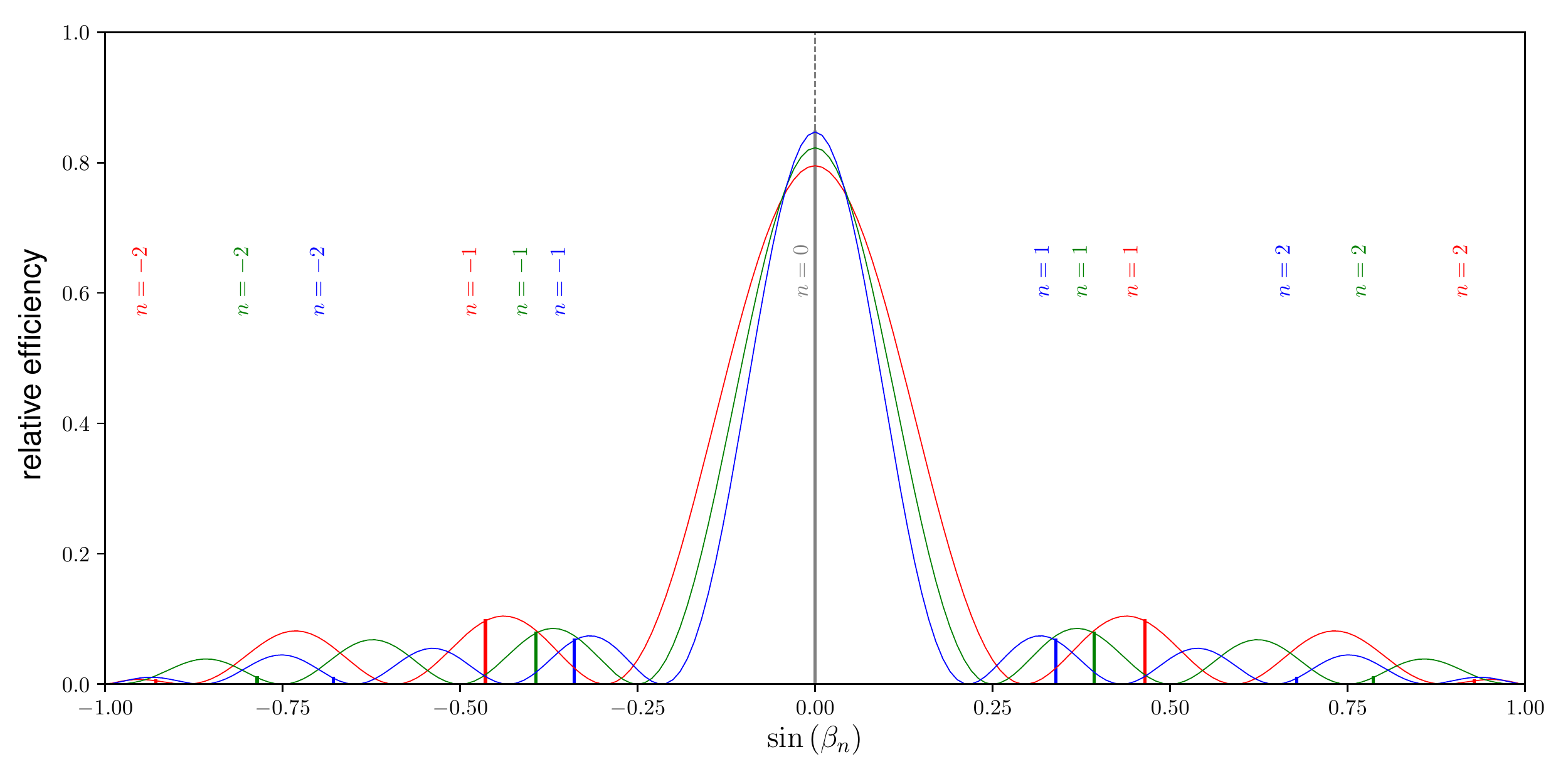}
 \caption[Relative efficiency of a square-wave phase grating.]{Relative efficiency of a $\mathscr{W}/d = \SI{50}{\percent}$ square-wave phase grating at normal incidence with $h_0 / \lambda = 0.3$.}\label{fig:square_phase}
 \end{figure} 
This is illustrated in \cref{fig:square_phase} for a grating with $\mathscr{W} = d/2$, $\alpha = 0$ and $h_0 / \lambda = 0.3$, showing that the depth of the grating grooves, $h_0$, has an effect on the diffracted intensity in an analogous fashion to the aperture duty cycle, $\mathscr{W} / d$. 

\subsection{The Sinusoid}\label{sec:sinusoid_principle}
%%%%%%%%%%%%%%%%%%%%%%%%%%%%%%%%%%%%%%%%%--------------------------------------------------
As a next example, a sinusoidal amplitude grating is considered to have the following transmittance function: 
\begin{subequations}
\begin{equation}\label{eq:sin_amplitude_grating}
 G (x) = \mathcal{A} (x) = \frac{\mathcal{A}_0}{2} \left[ 1 + \sin (K x) \right] .
 \end{equation}
Because $\mathcal{A} (x)$ is constructed purely from $0^{\text{th}}$ and $\pm 1^{\text{st}}$ order spatial harmonics, only diffracted orders corresponding to Fourier coefficients with $n = 0, \pm 1$ should have a non-zero amplitude. 
This can be verified by inserting \cref{eq:sin_amplitude_grating} into \cref{eq:transmit_fourier_coeff} shows that $G_n = 0$ unless $n = 0, \pm 1$:
\begin{align}
\begin{split}
 G_n &= \frac{\mathcal{A}_0}{2 d} \int_{0}^{d} \left[ 1 + \sin (K x) \right] \mathrm{e}^{- i n K x} \dd{x} \quad \text{for } n = 0, \pm 1, \pm 2, \pm 3 \dotsc \\
 &= \frac{\mathcal{A}_0}{2 d} \int_{0}^{d} \left[ \mathrm{e}^{- i n K x} + \sin \left( K x \right) \cos \left( n K x \right) - i\sin \left( K x \right) \sin \left( n K x \right) \right] \dd{x} \\
 &= \frac{\mathcal{A}_0}{2 d} \left[ \underbrace{ \int_{0}^{d} \mathrm{e}^{- i n K x} \dd{x}}_\text{$d$ for $n=0$} + \overbrace{ \int_{0}^{d} \sin \left( K x \right) \cos \left( n K x \right) \dd{x} }^\text{always 0} - i \underbrace{ \int_{0}^{d} \sin \left( K x \right) \sin \left( n K x \right) \dd{x} }_\text{$\pm \frac{d}{2}$ for $n = \pm 1$} \right]
 \end{split}
 \end{align}
and therefore, the Fourier coefficients can be written as 
\begin{equation}\label{eq:sin_amplitude_grating_coef}
    G_n =
                \begin{cases}
                  -\frac{i \mathcal{A}_0}{4} , &\quad \text{for } n = 1 \\
                  \frac{\mathcal{A}_0}{2} , &\quad \text{for } n = 0  \\
                  \frac{i \mathcal{A}_0}{4} , &\quad \text{for } n = -1 \\
                  0 , &\quad \text{otherwise.}
                \end{cases}
 \end{equation}
This implies that the intensity of $0^{\text{th}}$ order should be four times the intensity of the $\pm 1^{\text{st}}$ orders while all other orders vanish: 
\begin{equation}\label{eq:sin_amplitude_grating_inten}
 \mathscr{E}_n \propto \frac{\cos \left( \beta_n \right)}{\cos \left( \alpha \right)} \norm{G_n}^2 \propto \frac{\cos \left( \beta_n \right)}{\cos \left( \alpha \right)}
                \begin{cases}
                  \frac{1}{16} , &\quad \text{for } n = \pm 1 \\
                  \frac{1}{4} , &\quad \text{for } n = 0  \\
                  0 , &\quad \text{otherwise}.
                \end{cases}
 \end{equation}
\end{subequations}
Although \cref{eq:sin_amplitude_grating} does not describe a physical grating in a general sense, this framework is useful to describe gratings with only one propagating order.  

A sinusoidal surface-relief grating with a groove-depth profile described by 
\begin{equation}
 h (x) = \frac{h_0}{2} \left[ 1 + \sin (K x) \right] ,
 \end{equation}
with $h_0$ as the maximum depth of the grating grooves, can be considered as a phase grating for a geometry where all portions of the groove are illuminated.  
In this case, a result different from \cref{eq:sin_amplitude_grating,eq:sin_amplitude_grating_coef,eq:sin_amplitude_grating_inten} is obtained with a phase function given by 
\begin{equation}
 \Phi (x) = \frac{\Phi_g}{2} \left[ 1 + \sin \left( K x \right) \right] = \Phi (x + d) ,
 \end{equation}
where $\Phi_g$ is the maximum phase shift that radiation picks up as it reflected from the groove structure [\emph{cf.\@} \cref{eq:laminar_phase}]. 
With amplitude assumed to be constant everywhere, the transmittance function is 
\begin{subequations}
\begin{equation}
 G (x) = \mathcal{A}_0 \mathrm{e}^{i \frac{\Phi_g}{2} \left[ 1 + \sin \left( K x \right) \right]}
 \end{equation} 
and then the following integral must be evaluated to determine the Fourier coefficients:
\begin{equation}\label{eq:sine_coef_integral}
 G_n = \frac{\mathcal{A}_0}{d} \mathrm{e}^{i \frac{\Phi_g}{2}} \int_{0}^{d} \mathrm{e}^{i \frac{\Phi_g}{2} \sin \left( K x \right)} \mathrm{e}^{- i n K x} \dd{x} \quad \text{for } n = 0, \pm 1, \pm 2, \pm 3 \dotsc .
 \end{equation}
\end{subequations}
The exponential sine term can be handled using a property of \emph{Bessel functions of the first kind}:
\begin{equation}\label{eq:Bessel_property}
 \mathrm{e}^{i \frac{\Phi_g}{2} \sin \left( K x \right)} = \sum_{m=-\infty}^{\infty} J_m \left( \frac{\Phi_g}{2} \right) \mathrm{e}^{i m K x} ,
 \end{equation}
where $J_m (\Phi_g / 2)$ is the $m^{\text{th}}$-order Bessel function. 
\begin{figure}
 \centering
 \includegraphics[scale=0.62]{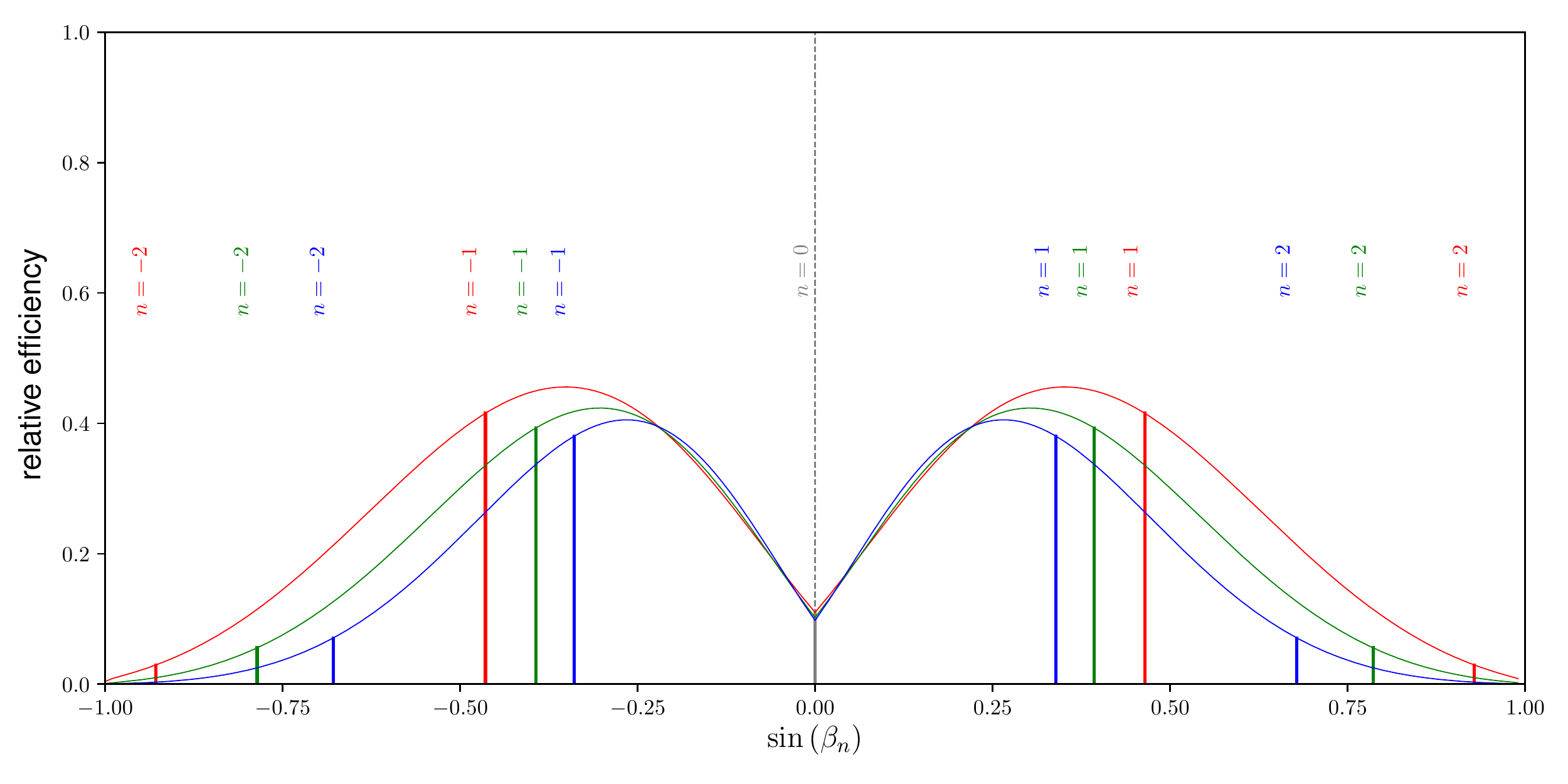}
 \caption[Relative efficiency of a sinusoidal phase grating.]{Relative efficiency of a sinusoidal phase grating at normal incidence with $h_0 / \lambda = 0.3$.}\label{fig:sine_phase} 
 \end{figure} 

Combining \cref{eq:sine_coef_integral,eq:Bessel_property} gives
\begin{align}\label{eq:G_terms_sine_phase}
 \begin{split}
 G_n &= \frac{\mathcal{A}_0}{d} \mathrm{e}^{i \frac{\Phi_g}{2}} \int_{0}^{d} \sum_{m=-\infty}^{\infty} J_m \left( \frac{\Phi_g}{2} \right) \mathrm{e}^{i \left( m - n \right) K x} \dd{x} \\
 &= \mathcal{A}_0 \mathrm{e}^{i \frac{\Phi_g}{2} } J_n \left( \frac{\Phi_g}{2} \right) \quad \text{for } n = 0, \pm 1, \pm 2, \pm 3 \dotsc ,
 %= \frac{\mathcal{A}_0}{d} \mathrm{e}^{i \chi} \int_{0}^{d} J_n \left( \chi \right) \dd{x} \\
 %= \mathcal{A}_0 \mathrm{e}^{i \chi} J_n \left( \chi \right)
 \end{split}
 \end{align}
which shows that orders of $\left| n \right| > 1$ can have a non-zero intensity depending on the value of $\Phi_g$, with the relative intensity of the $n^{\text{th}}$ order being 
\begin{equation}
 \mathscr{E}_n \propto \frac{\cos \left( \beta_n \right)}{\cos \left( \alpha \right)} \norm{G_n}^2 \quad \text{with} \quad \norm{G_n}^2 \propto \norm{J_n \left( \frac{\Phi_g}{2} \right)}^2. 
 \end{equation}
For a grating used at normal incidence, the diffraction efficiency becomes, using \cref{eq:laminar_phase} for $\Phi_g$, 
\begin{equation}
 \mathscr{E}_n \propto \frac{\cos \left( \beta_n \right)}{\cos \left( \alpha \right)} \norm{J_n \left( \pi \frac{h_0}{\lambda} \sqrt{1 - \left( \frac{n \lambda}{d} \right)^2 } \right)}^2  .
 \end{equation}
This is illustrated in \cref{fig:sine_phase} for $h_0 / \lambda = 0.3$, where it is seen that most diffracted efficiency is contained within orders $n = \pm 1$. 

\subsection{The Sawtooth}\label{sec:sawtooth_principle}
%%%%%%%%%%%%%%%%%%%%%%%%%%%%%%%%%%%%%%%%%--------------------------------------------------
The final grating groove shape to consider is the sawtooth, which is defined as a periodic function that increases with some slope over the duration of the period and then falls with infinite slope to start the cycle over again. 
If a hypothetical sawtooth amplitude grating is considered, the transmittance function is 
\begin{subequations}
\begin{equation}
 G (x) = \mathcal{A} (x) = \frac{\mathcal{A}_0}{d} x \quad \text{for } 0 < x \leq d ,
 \end{equation}
where $\mathcal{A}_0$ is the amplitude and $\mathcal{A}_0 / d$ is the slope of the sawtooth. 
Inserting this expression into \cref{eq:transmit_fourier_coeff} gives: 
\begin{equation}\label{eq:G_terms_square}
 G_n = \frac{\mathcal{A}_0}{d^2} \int_0^d x \mathrm{e}^{- i n K x} \dd{x} = \frac{\mathcal{A}_0 i}{2 \pi n} \quad \text{for } n = 0, \pm 1, \pm 2, \pm 3 \dotsc
 \end{equation}
so that $\mathscr{E}_n$ drops off as $n^{-2}$. 
\end{subequations}
Although this does not describe a physical diffraction grating, an ideal, blazed reflection grating can be treated as a surface where phase varies as a sawtooth: 
\begin{equation}
 \Phi (x) = \Phi_g \frac{x}{d} \quad \text{for } 0 < x \leq d , %\quad \text{and } \\
 %G \left( x \right) &= \mathcal{A}_0 \mathrm{e}^{i \Phi_g \frac{x}{d}} \quad \text{for } 0 < x \leq d ,
 \end{equation} 
where $\Phi_g$ is the phase shift difference between the top and bottom of a groove with depth $h_0$ [\emph{cf.\@} \cref{eq:laminar_phase}]. 
With Fourier coefficients for $G(x) = \mathcal{A}_0 \mathrm{e}^{i \Phi (x)}$ in this case given by 
\begin{align}
 \begin{split}
 G_n &= \frac{\mathcal{A}_0}{d} \int_0^d \mathrm{e}^{i \left( \frac{\Phi_g}{2 \pi} - n \right) K x} \dd{x} \\
 &= \frac{\mathcal{A}_0 i}{\left( \Phi_g - 2 \pi n \right)} \left( 1 - \mathrm{e}^{i \left( \Phi_g - 2 \pi n \right)} \right) \quad \text{for } n = 0, \pm 1, \pm 2, \pm 3 \dotsc ,
 \end{split}
 \end{align}
the scalar diffraction efficiency of the $n^{\text{th}}$ order is 
\begin{align}\label{eq:sawtooth_phase}
\begin{split}
 \mathscr{E}_n &\propto \frac{\cos \left( \beta_n \right)}{\cos \left( \alpha \right)} \norm{G_n}^2 \\
 \text{with} \quad \norm{G_n}^2  &\propto \frac{2 \left[ 1 - \cos \left( \Phi_g - 2 \pi n  \right) \right]}{\left( \Phi_g - 2 \pi n \right)^2} \propto \sinc^2 \left( \frac{\Phi_g}{2} - n \pi \right) . 
 \end{split}
 \end{align}

The sinc-squared function in \cref{eq:sawtooth_phase} indicates that the intensity of orders varies in a similar way to single-slit interference, where the maximum intensity corresponds to $\Phi_g = 2 \pi n$. 
However, in this case there exists a maximum for each diffracted order and $\Phi_g$ depends on the slope of the sawtooth with the groove depth given by $h_0 = d \tan \left( \delta \right)$:
%\begin{subequations}
\begin{equation}\label{eq:sawtooth_path}
 \Phi_g = \frac{2 \pi}{\lambda} d \tan \left( \delta \right) \left[ \cos (\alpha) + \cos (\beta_n) \right] .
 \end{equation}
Interference maxima then correspond to set of $n$ and $\lambda$ that satisfy the following relation:
\begin{equation}\label{eq:blaze_condition}
 n \pi = \frac{\Phi_g}{2} \implies \frac{n \lambda}{d} = \tan \left( \delta \right) \left[ \cos (\alpha) + \cos (\beta_n) \right] .
\end{equation}
Equating this with the generalized grating equation given by \cref{eq:off-plane_incidence_orders}, solving for $\tan \left( \delta \right)$ and recognizing the \emph{tangent of an average} trigonometric identity yields 
\begin{equation}
 \tan \left( \delta \right) = \frac{\sin (\alpha) + \sin (\beta_n)}{\cos (\alpha) + \cos (\beta_n)} \equiv \tan \left( \frac{\alpha + \beta_n}{2} \right),
 \end{equation}
%\end{subequations}
which shows that diffracted angles corresponding to maximum intensity are all given by $\beta_n = 2 \delta - \alpha$ and suggests that the slope of the sawtooth can be thought of as being similar to a mirror flat, where the reflected angle corresponds to the grating's blaze response. 
Inserting $\beta_n = 2 \delta - \alpha$ into \cref{eq:off-plane_incidence_orders} and then recovering the off-plane case using $\lambda \to \bar{\lambda} = \lambda \csc (\gamma)$ gives the equation for the \emph{blaze wavelength}: 
\begin{equation}\label{eq:blaze_wavelength_general}
 %\begin{split}
 \lambda_b = \frac{d \sin \left( \gamma \right)}{n} \left[ \sin \left( \alpha \right) + \sin \left( 2 \delta - \alpha \right)  \right] = \frac{2 d \sin \left( \gamma \right) \sin \left( \delta \right)}{n} \cos \left( \delta - \alpha \right)  ,
 %\end{split}
 \end{equation}
which describes a set of wavelengths and order numbers where diffraction efficiency is maximized. 
Using $\beta_n = 2 \delta - \alpha$ inserted into \cref{eq:sawtooth_path}, the phase shift corresponding to the blazed diffracted angle is
\begin{equation}\label{eq:sawtooth_path_blaze}
 \Phi_b  \equiv \frac{2 \pi}{\lambda} d \sin (\gamma) \left[ \sin (\alpha) + \sin (2 \delta - \alpha) \right] 
 \end{equation}
so that the diffraction efficiency is a special case of \cref{eq:sawtooth_phase} with $\Phi_g = \Phi_b$:
\begin{subequations}
\begin{align}\label{eq:sawtooth_phase_blaze1}
 \begin{split}
 \mathscr{E}_n &\propto \frac{\cos \left( 2 \delta - \alpha \right)}{\cos \left( \alpha \right)} \norm{G_n}^2 \\
 \text{with} \quad \norm{G_n}^2 &\propto \sinc^2 \left( \frac{d}{\lambda} \pi \tan \left( \delta \right) \sin (\gamma) \left[ \cos (\alpha) + \cos (2 \delta - \alpha) \right]  - n \pi \right) . 
 \end{split}
 \end{align}

For a scenario parameterized by fixed values of $\lambda$, $d$, $\delta$ and $\gamma$, the term $\norm{G_n}^2$ in \cref{eq:sawtooth_phase_blaze1} can be considered to behave as a function of $\alpha$, 
\begin{figure}
 \centering
 \includegraphics[scale=0.62]{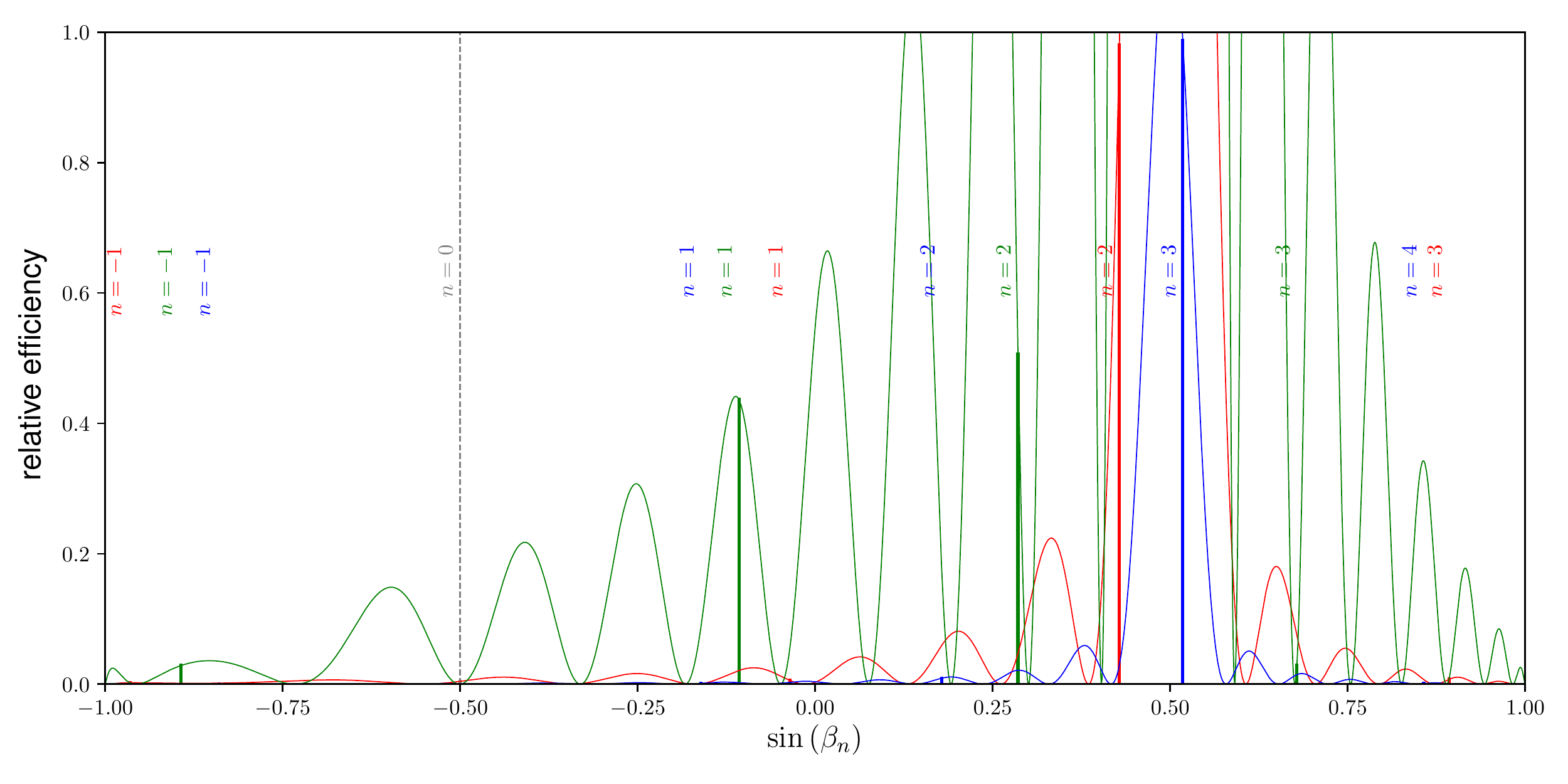}
 \caption[Relative efficiency of a sawtooth phase grating.]{Relative efficiency of a sawtooth phase grating in a Littrow configuration with $\alpha = \delta = 30^{\circ}$.}\label{fig:sawtooth_phase} 
 \end{figure} 
which has local extrema characterized by values for $\alpha$ that satisfy the following partial derivative condition:
\begin{equation}\label{eq:blaze_partial1}
 \pdv{\alpha}(\norm{G_n}^2) \propto \pdv{f_n}{\Phi_b} \pdv{\Phi_b}{\alpha} = 0 \quad \text{with} \quad f_n \equiv \sinc^2 \left( \frac{\Phi_b}{2} - n \pi \right) .
 \end{equation}
Noting that 
\begin{equation}\label{eq:blaze_partial2}
 \pdv{f_n}{\Phi_b} = \pdv{f_n}{\Phi'} \underbrace{ \pdv{\Phi'}{\Phi_b}}_{1/2} = \frac{\sinc \left( \Phi' \right) \left[ \Phi' \cos \left( \Phi' \right) - \sin \left( \Phi' \right)  \right]}{\left( \Phi' \right)^2} \quad \text{with} \quad \Phi' \equiv \frac{\Phi_b}{2} - n \pi
 \end{equation}
and 
\begin{equation}\label{eq:blaze_partial3}
 \pdv{\Phi_b}{\alpha} = \frac{2 \pi}{\lambda} \sin (\gamma) \left[ \cos \left( \alpha \right) - \cos \left( 2 \delta - \alpha \right) \right] ,
 \end{equation}
the condition defined in \cref{eq:sawtooth_phase_blaze1} is guaranteed to be satisfied if $\pdv*{\Phi_b}{\alpha} = 0$ with $\alpha = \delta$ 
\end{subequations}
so that \cref{eq:blaze_wavelength_general} for the blaze wavelength reduces to 
\begin{equation}\label{eq:blaze_wavelength_Litt}
 \lambda_b = \frac{2 d \sin \left( \delta \right) \sin \left( \gamma \right)}{n} . 
 \end{equation}

An example of an efficiency response in a Littrow configuration with $\alpha = \delta = 30^{\circ}$ and $\sin \left( \gamma \right) = 1$ is plotted in \cref{fig:sawtooth_phase}. 
\begin{figure}
 \centering
 \includegraphics[scale=0.62]{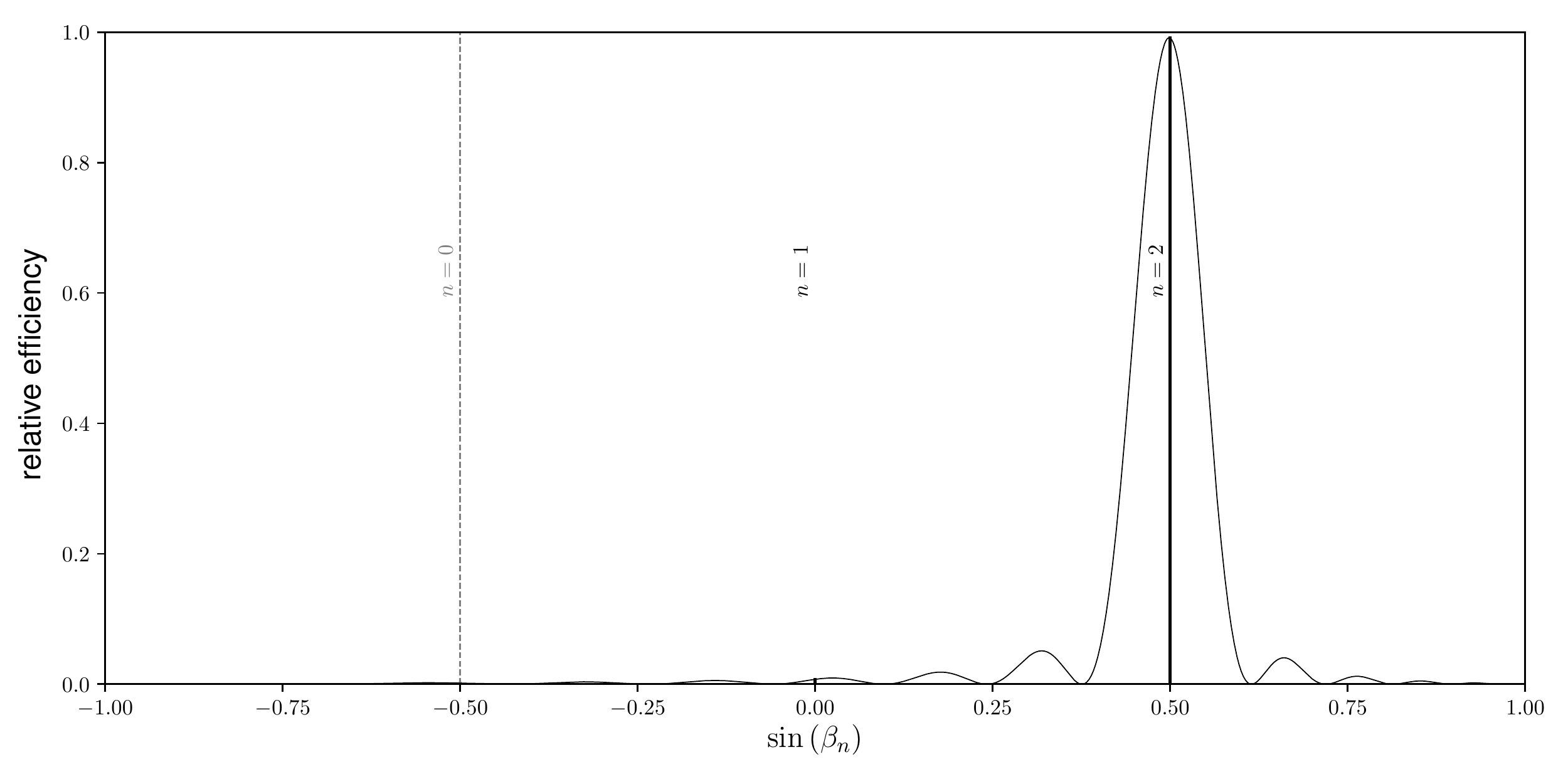}
 \caption[Relative efficiency of a sawtooth phase grating at the blaze wavelength in second order.]{Relative efficiency of a sawtooth phase grating in a Littrow configuration with $\alpha = \delta = 30^{\circ}$, at the blaze wavelength for $n=2$.}\label{fig:sawtooth_phase_blaze} 
 \end{figure} 
Because $\lambda / d$ for red, green and blue do not satisfy exactly \cref{eq:blaze_wavelength_Litt} with $2 \sin \left( \delta \right) = 1$, diffraction efficiency is spread between orders rather than being concentrated at $\sin \left( \beta_n \right) = 0.5$. 
Red and blue values of $\lambda / d$, however, come close to satisfying the blaze condition for $n = 2$ and $n = 3$ and as a result, their diffraction efficiency is more concentrated near the blaze angle, as opposed to the case for green.  
A value of $\lambda / d = 0.5$, on the other hand, satisfies \cref{eq:blaze_wavelength_Litt} for $n=2$ and the blaze response of the grating is maximized, as illustrated in \cref{fig:sawtooth_phase_blaze}. 
Due to these properties, blazed gratings used in a Littrow configuration are optimal for maximizing $\mathscr{E}_n$ for a given $\beta_n$ and hence a given bandpass of interest [\emph{cf.\@} \cref{sec:grating_tech_dev}].

\section{Summary}\label{sec:appD_summary}
%%%%%%%%%%%%%%%%%%%%%%%%%%%%%%%%%%%%%%%%%-------------------------------------------------- 
Basic physics of diffraction gratings can be gleaned by considering Huygens-Fresnel principle applied to an array of periodically spaced slit sources, where grating behavior arises as the number of slits, $N$, becomes very large. 
According to the framework of Fraunhofer diffraction, the far-field diffraction pattern produced by a grating can be determined from the Fourier transform of its transmittance function, $G(x)$, which describes periodic variations in amplitude and phase caused by the grating. 
Reflection gratings can often be modeled as phase gratings determined by the path-length differences picked up as light reflects from a groove facet. 
While scalar this approach to modeling diffraction efficiency is limited due to its neglect of electromagnetic polarization and radiation interaction with materials, it is found that gratings with sawtooth-shaped groove facets have the ability to concentrate diffraction efficiency in the $n^{\text{th}}$ order over a particular spectral range. 

%\include{Appendix-E/Appendix-E}
%%%%%%%%%%%%%%%%%%%%%%%%%%%%%%%%%%%%%%%%%%%%%%%%%%%%%%%%%%%%%%%
% ESM students need to include a Nontechnical Abstract as the %
% last appendix.                                              %
%%%%%%%%%%%%%%%%%%%%%%%%%%%%%%%%%%%%%%%%%%%%%%%%%%%%%%%%%%%%%%%
% This \include command should point to the file containing
% that abstract.
%\include{nontechnical-abstract}
%%%%%%%%%%%%%%%%%%%%%%%%%%%%%%%%%%%%%%%%%%%
} % End of the \allowdisplaybreak command %
%%%%%%%%%%%%%%%%%%%%%%%%%%%%%%%%%%%%%%%%%%%

%%%%%%%%%%%%%%%%
% BIBLIOGRAPHY %
%%%%%%%%%%%%%%%%
% You can use BibTeX or other bibliography facility for your
% bibliography. LaTeX's standard stuff is shown below. If you
% bibtex, then this section should look something like:
	\begin{singlespace}
	\bibliographystyle{GLG-bibstyle}
	\addcontentsline{toc}{chapter}{Bibliography}
	\bibliography{report}
	\end{singlespace}

%\begin{singlespace}
%\begin{thebibliography}{99}
%\addcontentsline{toc}{chapter}{Bibliography}
%\frenchspacing

%\bibitem{Wisdom87} J. Wisdom, ``Rotational Dynamics of Irregularly Shaped Natural Satellites,'' \emph{The Astronomical Journal}, Vol.~94, No.~5, 1987  pp. 1350--1360.

%\bibitem{G&H83} J. Guckenheimer and P. Holmes, \emph{Nonlinear Oscillations, Dynamical Systems, and Bifurcations of Vector Fields}, Springer-Verlag, New York, 1983.

%\end{thebibliography}
%\end{singlespace}

\backmatter

% Vita
% Place your vita below.

{\noindent \large \textbf{Education}}

\begin{itemize}[noitemsep]
 \item \emph{Doctor of Philosophy} in Astronomy \& Astrophysics (Anticipated May 2021) at The Pennsylvania State University, University Park, PA
 \item Graduate Student in Physics (2012-2016) at The University of Iowa, Iowa City, IA 
 \item \emph{Bachelor of Science} in Astronomy, Physics (May 2012) at The University of Massachusetts, Amherst, MA 
 \end{itemize}

{\noindent \large \textbf{Academic Employment}}

\begin{itemize}[noitemsep]
 \item \emph{Postdoctoral Scholar} at the Department of Astronomy \& Astrophysics, The Pennsylvania State University (2021-present)
 \item \emph{Graduate Research Assistant} at the Department of Astronomy \& Astrophysics, The Pennsylvania State University (2019-2020)
 \item \emph{NASA Space Technology Research Fellow} at the Department of Astronomy \& Astrophysics, The Pennsylvania State University (2016-2019)
 \item \emph{NASA Space Technology Research Fellow} at the Department of Physics \& Astronomy, The University of Iowa (2015-2016)
 \item \emph{Graduate Research Assistant} at the Department of Physics \& Astronomy, The University of Iowa (2013-2015)
 \item \emph{Graduate Teaching Assistant} at the Department of Physics \& Astronomy, The University of Iowa (2012-2013)
 \end{itemize}

{\noindent \large \textbf{Peer-Reviewed Journal Articles}}

\begin{itemize}[noitemsep]
 \item McCoy, Vershuuren, Miles, and McEntaffer (2020) ``X-ray verification of sol-gel resist shrinkage in substrate-conformal imprint lithography for a replicated blazed reflection grating" in \emph{OSA Continuum} \cite{McCoy20b}
 %\item ``X-ray verification of sol-gel resist shrinkage in substrate-conformal imprint lithography for a replicated blazed reflection grating" in \emph{OSA Continuum}, 3, 11 (2020) \cite{McCoy20b}
 \item McCoy, McEntaffer, and Miles (2020) ``Extreme Ultraviolet and Soft X-Ray Diffraction Efficiency of a Blazed Reflection Grating Fabricated by Thermally Activated Selective Topography Equilibration" in \emph{The Astrophysical Journal (ApJ)} \cite{McCoy20}
 %\item ``Extreme Ultraviolet and Soft X-Ray Diffraction Efficiency of a Blazed Reflection Grating Fabricated by Thermally Activated Selective Topography Equilibration" in \emph{The Astrophysical Journal (ApJ)}, 891, 2 (2020) \cite{McCoy20}
 \item McCoy, McEntaffer, and Eichfeld (2018) ``Fabrication of astronomical x-ray reflection gratings using thermally activated selective topography equilibration" in \emph{The Journal of Vacuum Science \& Technology B} \cite{McCoy18}
 %\item ``Fabrication of astronomical x-ray reflection gratings using thermally activated selective topography equilibration" in \emph{The Journal of Vacuum Science \& Technology B}, 36,6 (2018) \cite{McCoy18}
 \item McCoy, Schultz, Tutt, Rogers, Miles, and McEntaffer (2016) ``A Primer for Telemetry Interfacing in Accordance with NASA Standards Using Low Cost FPGAs" in \emph{The Journal of Astronomical Instrumentation} \cite{McCoy_telemetry}
 %\item ``A Primer for Telemetry Interfacing in Accordance with NASA Standards Using Low Cost FPGAs" in \emph{The Journal of Astronomical Instrumentation}, 05, 01, 1640002 (2016) \cite{McCoy_telemetry}
 \item five co-authored papers in \emph{Experimental Astronomy}, \emph{ApJ} and \emph{The Journal of Astronomical Telescopes, Instruments and Systems} (2017-2020) \cite{Rogers17,Miles18,Miles19,Rogers20,McCurdy20}
 \end{itemize}

\end{document}